\documentclass[reprint,aps,amsmath,amssymb,rmp,floatfix,longbibliography]{revtex4-1}

\usepackage[utf8]{inputenc}
\usepackage[english]{babel}

\usepackage{import}
\usepackage{graphicx} 

\usepackage{bm} 
\usepackage{siunitx} 
\usepackage{textcomp} 
\newcommand{\del}{\partial}

\usepackage{xcolor}
\definecolor{light-gray}{gray}{0.8}
\definecolor{apsblue}{rgb}{0.176, 0.152, 0.57}
\usepackage[colorlinks=true, linkcolor=black, citecolor=apsblue, filecolor=black, urlcolor=apsblue]{hyperref}

\usepackage[mathlines]{lineno}

\newcommand*\dif{\mathop{}\!\mathrm{d}}

\begin{document}

\title{Beam-driven plasma-wakefield acceleration}

\author{C. A. Lindstr{\o}m}
\thanks{C.~A.~Lindstr{\o}m and S.~Corde contributed equally; \href{mailto:c.a.lindstrom@fys.uio.no}{c.a.lindstrom@fys.uio.no}, \href{mailto:sebastien.corde@polytechnique.edu}{sebastien.corde@polytechnique.edu}.}
\affiliation{\mbox{Department of Physics, University of Oslo, Oslo, Norway}}
\affiliation{\mbox{Deutsches Elektronen-Synchrotron DESY, Hamburg, Germany}}

\author{S. Corde}
\thanks{C.~A.~Lindstr{\o}m and S.~Corde contributed equally; \href{mailto:c.a.lindstrom@fys.uio.no}{c.a.lindstrom@fys.uio.no}, \href{mailto:sebastien.corde@polytechnique.edu}{sebastien.corde@polytechnique.edu}.}
\affiliation{\mbox{Laboratoire d'Optique Appliquée (LOA), CNRS, \'Ecole Polytechnique, ENSTA}, \mbox{Institut Polytechnique de Paris, Palaiseau, France}}
\affiliation{\mbox{SLAC National Accelerator Laboratory, Menlo Park, CA, USA}}

\author{R. D'Arcy}
\affiliation{\mbox{John Adams Institute for Accelerator Science, Department of Physics}, \mbox{University of Oxford, Oxford, UK}}
\affiliation{\mbox{Deutsches Elektronen-Synchrotron DESY, Hamburg, Germany}}

\author{S. Gessner}
\affiliation{\mbox{SLAC National Accelerator Laboratory, Menlo Park, CA, USA}}

\author{M. Gilljohann}
\affiliation{\mbox{Laboratoire d'Optique Appliquée (LOA), CNRS, \'Ecole Polytechnique, ENSTA}, \mbox{Institut Polytechnique de Paris, Palaiseau, France}}

\author{M. J. Hogan}
\affiliation{\mbox{SLAC National Accelerator Laboratory, Menlo Park, CA, USA}}

\author{J. Osterhoff}
\email{josterhoff@lbl.gov}
\affiliation{\mbox{Lawrence Berkeley National Laboratory, Berkeley, California, USA}}
\affiliation{\mbox{Deutsches Elektronen-Synchrotron DESY, Hamburg, Germany}}


\begin{abstract}
    Beam-driven plasma-wakefield acceleration (PWFA) has emerged as a transformative technology with the potential to revolutionize the field of particle acceleration, especially toward compact accelerators for high-energy and high-power applications. Charged particle beams are used to excite density waves in plasma with accelerating fields reaching up to \SI{100}{GV/m}, thousands of times stronger than the fields provided by radio-frequency cavities. Plasma-wakefield-accelerator research has matured over the span of four decades from basic concepts and proof-of-principle experiments to a rich and rapidly progressing sub-field with dedicated experimental facilities and state-of-the-art simulation codes. We review the physics, including theory of linear and nonlinear plasma wakefields as well as beam dynamics of both the wakefield driver and trailing bunches accelerating in the plasma wake, and address challenges associated with energy efficiency and preservation of beam quality. Advanced topics such as positron acceleration, self-modulation, internal injection, long-term plasma evolution and multistage acceleration are discussed. Simulation codes and major experiments are surveyed, spanning the use of electron, positron and proton bunches as wakefield drivers. Finally, we look ahead to future particle colliders and light sources based on plasma technology.
\end{abstract}

\maketitle

\tableofcontents



\section{Introduction}
\label{sec:introduction}

\begin{figure}[t]
    \centering\includegraphics[width=\linewidth]{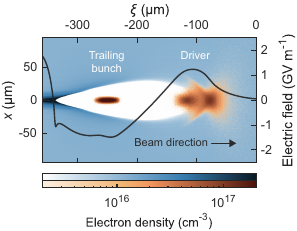}
    \caption{An electron bunch (right; orange color map) in a plasma (blue color map) drives a nonlinear plasma wake, here simulated using the particle-in-cell method. Strong longitudinal electric fields (black line) cause the driver to rapidly decelerate and lose energy to the wake. A second bunch (left) trailing behind the driver can acquire a significant fraction of this energy, leading to rapid acceleration, while being focused by the exposed plasma ions in the wake (white region). Since both bunches are ultrarelativistic, hence moving at close to light speed, their longitudinal separation remains approximately constant until the driver electrons are depleted of their energy. Here, $\xi$ and $x$ are the longitudinal and transverse directions. From \textcite{Lindstrom2024} (CC-BY 4.0).}
    \label{fig:plasma-wake-schematic}
\end{figure}

The particle accelerator has evolved from its inception in the early 1900's to become a major work horse of scientific progress, ranging from subatomic physics and material science to cancer therapy and semi-conductor ion implantation. Paralleled only by the success of the laser and the computer, particle accelerators have experienced an exponential increase in capabilities over time \cite{Shiltsev2011,Shiltsev2020}, but at the expense of increasing size and cost. At the energy frontier, particle accelerators based on radio-frequency (rf) structures are fast approaching their fundamental limits \cite{Shiltsev2021}: magnetic-field strength and synchrotron radiation limit circular accelerators \cite{FCC2019a,FCC2019b}, while material breakdown in metallic accelerating cavities limits linear accelerators \cite{Grudiev2009,Aicheler2012}. Circumventing these problems for future generations of high-energy-physics experiments will require a technological leap. One of the most promising ideas to this end is plasma-wakefield acceleration (PWFA), whereby an ionized gas is used as an accelerating medium. Immune to further breakdowns, the accelerating fields in a plasma can reach several orders of magnitude higher than in rf cavities, promising compact and potentially cheaper linear particle accelerators \cite{Joshi2003}. The field of PWFA has previously been reviewed by \textcite{Joshi2007}, \textcite{Muggli2016} and \textcite{Hogan2016}, and more recently by \textcite{Chen2026} and \textcite{Saberi2026}.

The basic principle behind plasma-wakefield acceleration is the following (see Fig.~\ref{fig:plasma-wake-schematic}): an intense bunch of relativistic charged particles traveling through a plasma will repel electrons (if negatively charged) or attract electrons (if positively charged), which creates a region of charge separation behind the bunch. This region, called a \textit{plasma wake}, has strong electric and magnetic fields that tend to rapidly restore quasi-neutrality. Its typical scale is the plasma wavelength, $\lambda_p$, and the bunch needs to be shorter than this to effectively excite the wake. If another particle bunch is traveling at an appropriate distance behind the wake-driving bunch---typically within one plasma wavelength---it can be accelerated by the electric field in the wake. This second bunch is sometimes referred to as the ``trailing" (used in this Review), ``accelerating", ``main" or ``witness" bunch; the latter referring to its observation of the field inside the plasma wake, also known as the \textit{plasma wakefield}. If the particle density of the driver is similar to that of the plasma, $n_0$, the characteristic scale of the wakefield is $E_0 = m_e c \omega_p/e$, called the \textit{wavebreaking field} \cite{Akhiezer1956,Dawson1959}, where $\omega_p  = 2\pi c/\lambda_p = \sqrt{n_0 e^2/m_e\epsilon_0}$ is the plasma-electron frequency, $m_e$ and $e$ are the electron mass and charge, respectively, and $c$ is the speed of light in vacuum. Its numerical value is given by 
\begin{equation}
    \label{eq:wavebreaking-field-simple}
    E_0\hspace{1pt}[\mathrm{V/m}] \approx 96  \sqrt{n_0[\mathrm{cm}^{-3}]}.
\end{equation}
A plasma density of order $10^{14}$--$10^{18}$~cm$^{-3}$, the range typically considered in experiments, corresponds to an accelerating field of 1--\SI{100}{GV/m}; one to three orders of magnitude higher than in an rf accelerator (10--\SI{100}{MV/m}). The corresponding plasma wavelength is in the range \SI{3}{mm}--\SI{30}{\micro\m}, one or more orders of magnitude smaller than the wavelength of rf accelerators (i.e., 30--\SI{300}{mm}).

Plasma wakefields can be driven in many different ways. While this Review discusses plasma wakefields driven by beams of charged particles, the wakefield can also be driven by lasers---often referred to as \textit{laser-wakefield acceleration} (LWFA) or \textit{laser--plasma acceleration}. This closely related field of research was reviewed by \textcite{Esarey2009}. Within beam-driven plasma-wakefield acceleration, several different particles can be used: electron bunches are the most common \cite{Chen1985,Ruth1985}, and can drive either weak perturbations (\textit{linear regime}) or strong density perturbations (\textit{nonlinear regime}), each with their own advantages and drawbacks, as discussed in Sec.~\ref{sec:plasma-wakefields}. The nonlinear \textit{blowout} regime \cite{Rosenzweig1991} is seen as particularly attractive for acceleration of electrons due to its large accelerating gradients and beam-quality-preserving focusing fields. Positively charged bunches---positrons or protons---can also be used to drive plasma wakes; for positrons this is similar to that of electrons in the linear regime \cite{Schroeder2010,Doche2017}, but different and less ideal in the nonlinear or \textit{suck-in} regime \cite{Lee2001, Corde2015}, as discussed in Sec.~\ref{sec:positron-acceleration}. Next, high-energy, high-charge proton bunches accelerated by synchrotrons store kilojoules of energy which can be used to drive high-amplitude wakes over long distances \cite{Caldwell2009}, but the proton bunch lengths are typically orders of magnitude longer than $\lambda_p$ for interesting plasma densities. For this reason, proton-driven plasma accelerators are operated in the \textit{self-modulated} regime \cite{Lotov1998,Kumar2010}, where a long bunch interacts with the plasma to form a train of short bunches---see Sec.~\ref{sec:self-modulated-wakefields} for more details. More exotic regimes, not covered in detail in this Review, include plasma-wakefield acceleration of muons \cite{Ahmed2015,Shiltsev2022} and channeling acceleration in solid-density plasmas within crystals and nanostructures \cite{Chattopadhyay2020,Gilljohann2023}.

\begin{figure}[t]
    \centering\includegraphics[width=\linewidth]{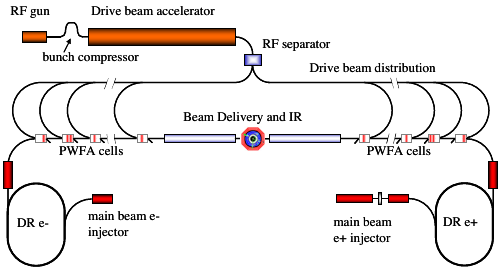}
    \caption{PWFA collider concept, where an rf-based linear accelerator (top left) produces electron drivers which are distributed to multiple PWFA stages. High-quality electron and positron bunches are produced using damping rings and subsequently accelerated in the PWFAs, and finally collided in the interaction region. From \textcite{Seryi2009} (CC-BY 3.0).}
    \label{fig:PWFA-collider-concept}
\end{figure}

The key motivation for using a particle beam to drive the plasma wakefield is the potential for large energy gain and high energy efficiency---both demonstrated in landmark experiments performed at SLAC \cite{Blumenfeld2007,Litos2014}. These two aspects make PWFA particularly well suited for compact, high-average-power accelerators for photon science, such as an x-ray free-electron laser (FEL) \cite{Assmann2020,Habib2023,Schroeder2025}, and for high-energy physics, such as a linear collider \cite{Rosenzweig1998,Seryi2009,Foster2023}, where the need for both high energy and high particle flux demands megawatts of beam power (see Fig.~\ref{fig:PWFA-collider-concept} and Sec.~\ref{sec:applications:colliders}). The wall-plug-to-beam energy efficiency of PWFA can be high because beam drivers can be produced with relatively high efficiency using klystrons \cite{Varian1939,Lien1970}, as utilized in rf-based linear-collider concepts like CLIC \cite{Aicheler2012}. Moreover, the energy contained in these drivers can be efficiently extracted over long propagation distances (until \textit{depletion}). This is because the beams are self-guided in the plasma wake (no \textit{diffraction}) and because the trailing bunch remains at a fixed distance behind the driver, since both bunches effectively travel at the speed of light (no \textit{dephasing}). If using drivers with large total energy content (i.e., joule-level electron/positron drivers or kilojoule-level proton drivers), beam-driven plasma accelerators can reach high particle energies in the GeV--TeV range.

While the benefits of PWFA are plentiful, any radical change in technology brings with it new challenges. Understanding and overcoming these issues has been the focal point of research---past and present. This includes questions concerning stability and long-distance propagation of beam drivers, which can be complicated by effects such as head erosion, mismatching, and the hosing instability---topics covered in Sec.~\ref{sec:driver-propagation}. Next, high beam quality, including low energy spread and small transverse emittance (i.e., area in phase space), is key to applications in high-energy physics and photon science. Accelerating a trailing bunch while simultaneously ensuring high energy efficiency and high beam quality can be challenging, for instance due to tight tolerances on synchronization, misalignment, and bunch structure, as well as effects such as Coulomb scattering \cite{Montague1984}, hosing and beam-breakup instabilities \cite{Whittum1991,Lebedev2017} and ion motion \cite{Rosenzweig2005}. These topics are covered extensively in Sec.~\ref{sec:evolution-trailing-bunch}. While solutions do exist for most issues regarding acceleration of electrons, the same is not true for positrons. This is because plasmas are inherently charge asymmetric; plasma ions are much heavier than plasma electrons and therefore the plasma responds differently to a positron beam compared to an electron beam. Positron acceleration is crucial for building an electron--positron collider, but is a challenging topic of research due to the complexity of its dynamics, as well as a lack of experimental facilities that provide suitable positron bunches. Nevertheless, recent progress in both experiments and theory shows promise---see Sec.~\ref{sec:positron-acceleration} or \textcite{Cao2024} for a review of positron acceleration in plasmas. Even if only electrons are needed, delivering the beam power required for PWFA-based FELs and linear colliders will be challenging. Accelerating particles at high repetition rate, typically kHz or higher, sets stringent demands on the plasma sources and the long-term plasma evolution that happen inside them---a topic covered in Sec.~\ref{sec:long-term-evolution}. Lastly, in order to reach very high energies, multiple acceleration stages are likely required. So-called \textit{staging} of plasma accelerators can be quite space consuming due to the complexity of injecting and extracting drivers while simultaneously preserving the transverse and longitudinal phase space of the accelerating bunch---this topic is covered in Sec.~\ref{sec:staging}.

\begin{figure}[t]
    \centering\includegraphics[width=\linewidth]{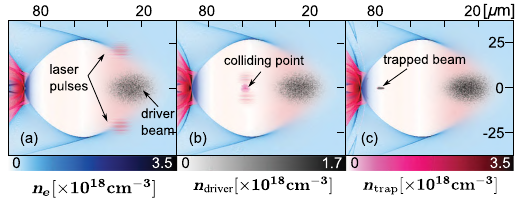}
    \caption{Generation of a new, high-quality electron bunch by injection of plasma electrons into the plasma wake (PIC simulation). In the plasma-photocathode scheme, additional electrons are released by an external laser entering transversely (a) from two sides and (b) constructively interfering on axis. The electrons, initially at rest, are (c) accelerated to near light speed by the wakefield before the wake has passed, locking them in phase with the beam driver. From \textcite{Li2013}.}
    \label{fig:plasma-photocathode-scheme}
\end{figure}

Beyond its use for rapidly accelerating particles, beam-driven plasma accelerators can also generate high-quality electron beams directly from the plasma, sometimes known as \textit{internal injection} as opposed to \textit{external injection} of bunches from external sources. This scheme promises so-called \textit{brightness transformation} (brightness being the density in phase space), since the energy stored in a lower-quality drive beam can be converted to accelerate a higher-quality injected beam \cite{Foerster2022,Zhang2025,Wood2026}. If plasma electrons can experience the strong accelerating gradient for long enough to become relativistic, they will be injected into the plasma wake and continue to be accelerated just like externally injected electrons. This does not usually happen for plasma electrons in a uniform plasma, but can be achieved using two main techniques: (1) new electrons can be released by ionization within the plasma wake, by the intense fields of a laser pulse or a high-charge-density particle beam; or (2) existing plasma electrons that are part of the expelled electron sheath can be captured by a sudden reduction in phase velocity of the wake, for instance by a sharp downramp in plasma density. Numerous schemes exist that perform this injection in various ways, as detailed in Sec.~\ref{sec:internal-injection}, some of which are common to laser-driven plasma accelerators and others that are only feasible in beam-driven plasma accelerators. One scheme in particular, the so-called \textit{plasma photocathode} injection \cite{Hidding2012}, colloquially known as ``Trojan Horse" injection, exploits the advantages of beam drivers to generate beams with extremely high brightness. Here, a laser pulse travels collinearly with the beam driver, releasing electrons from higher ionization levels than that used for the surrounding plasma---not normally possible with a laser driver, since that would itself ionize the higher levels. Importantly, these new electrons can be released directly on-axis, which means they can have very low transverse momentum and hence very low emittance. Such injected bunches can have normalized emittances at the nanometer scale (orders of magnitude lower than rf-photocathode sources), which combined with high current and small energy spread result in unprecedented beam brightness \cite{Habib2023}. Figure~\ref{fig:plasma-photocathode-scheme} illustrates an alternative scheme for plasma photocathodes where the laser pulse is coupled in transversely instead of collinearly \cite{Li2013}.

The earliest ideas of plasma-wakefield acceleration can be traced back to \textcite{Veksler1956} and \textcite{Fainberg1956}---referred to as ``coherent acceleration" or ``collective acceleration" \cite{Lawson1972}. A review by \textcite{Fainberg1968} details the foundational work performed in the USSR from 1950s onward, which included both theoretical \cite{Fainberg1960} and experimental advances \cite{Kharchenko1960,Kiselev1976}. While these developments were recognized in the US at the time \cite{AEC1972}, the research field in its modern form originated somewhat independently, starting with work led by John M.~Dawson and collaborators at UCLA \cite{Dawson2001}, who in the late 1970s discovered laser-driven plasma acceleration, as reported in a seminal paper by \textcite{Tajima1979}. Investigations into plasma acceleration driven instead by intense particle beams soon followed, both at UCLA, as reported by \textcite{Chen1985}, and at SLAC National Accelerator Laboratory, as reported by \textcite{Ruth1985}, driven in part by the possibility of performing such experiments in the SLAC linac \cite{Joshi2001}. At this time, the term ``plasma wakefield" was coined, inspired by contemporary work on wakefields in rf accelerators \cite{Voss1982,Bane1985b}; a change in nomenclature that may have contributed to the disconnect between modern and early literature.

From the start, two main methodologies have formed the pillars of the research field: theoretical studies augmented by numerical simulations, and experimental studies using beams from rf-based particle accelerators.

Particle-in-cell (PIC) simulations are the most commonly used numerical simulation for plasma accelerators. Such simulations place plasma and beam particles in a gridded simulation box, and move them in small time steps based on the electric and magnetic fields that evolve at each grid point (see Figs.~\ref{fig:plasma-wake-schematic} and \ref{fig:plasma-photocathode-scheme}). For high accuracy, billions of beam and plasma particles can be required, which means that simulations are often run on large supercomputers. While the most general PIC codes \cite{Fonseca2002} work equally well for beam-driven and laser-driven plasma accelerators, some codes are optimized specifically for beam drivers. The most common optimization is the \textit{quasistatic approximation} \cite{Sprangle1990,Mora1997}, which makes use of the fact that the relativistic beam particles evolve on a significantly longer timescale than the plasma electrons, which allows the simulation to be split into independent longitudinal slices \cite{Huang2006,Mehrling2014}. Further, cylindrical symmetry or near-symmetry allows the system to be \textit{Fourier decomposed} into its lowest azimuthal modes \cite{Lifschitz2009,Li2021}, which dramatically speeds up the simulation. These numerical simulations, discussed briefly in Sec.~\ref{sec:numerical-simulations}, also play a vital role in supporting the experimental research, both for planning of future experiments and as a way to interpret existing results.

\begin{figure}[t]
    \centering\includegraphics[width=0.90\linewidth]{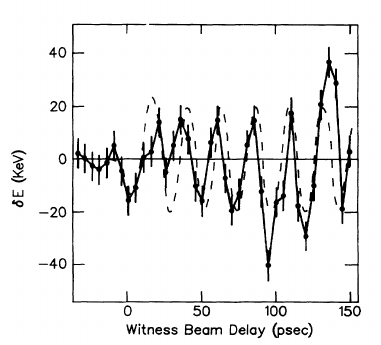}
    \caption{First observation of beam-driven plasma wakefields, achieved at Argonne National Laboratory. The energy change of a trailing bunch was measured at variable delays behind a leading drive bunch. From \textcite{Rosenzweig1988}.}
    \label{fig:first-observation}
\end{figure}

Experimental research was in many ways initiated by \textcite{Rosenzweig1988} with the first experimental observation of beam-driven plasma wakefields at Argonne National Laboratory. Two particle bunches were propagated through a \SI{30}{cm}-long plasma at a density of $n_0 \approx \SI{e13}{\per\cm\cubed}$. Varying the delay between the leading and trailing bunch, a small energy modulation was observed, as shown in Fig.~\ref{fig:first-observation}, proving the existence of an oscillating plasma wakefield in the linear regime. Later experiments at the same facility also demonstrated nonlinear plasma wakefields, first in the overdense ($n_b < n_0$) regime \cite{Rosenzweig1989} and then in the underdense ($n_b \gtrsim n_0$; i.e., blowout) regime  \cite{Barov1998,Barov2000}, where $n_b$ denotes the beam density.

\begin{figure}[t]
    \centering\includegraphics[width=0.95\linewidth]{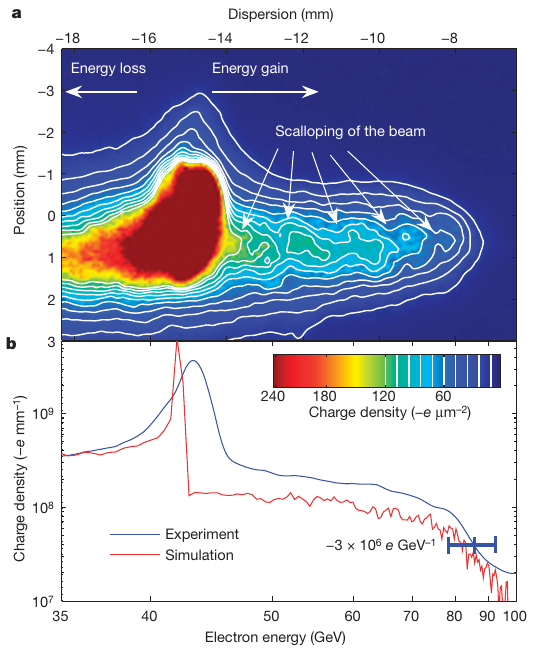}
    \caption{Energy doubling of 42-GeV electrons within an 85-cm-long plasma accelerator. In this landmark experiment, a 50-fs long bunch of $1.8\times10^{10}$ electrons focused to a beam size of approximately \SI{10}{\micro\m} propagated through lithium vapor of density \SI{2.7e17}{cm^{-3}}. The head of the bunch ionized the gas into a plasma and drove a plasma wakefield with fields up to \SI{52}{GV/m}, by which electrons in the tail of the bunch were accelerated. From \textcite{Blumenfeld2007}.}
    \label{fig:large-energy-gain}
\end{figure}

In the following decades, several important experimental milestones were reached. The first experiments to show the promise of beam-driven plasma acceleration were the demonstrations of large acceleration gradient \cite{Hogan2005} and energy gain \cite{Blumenfeld2007}, the latter by doubling the energy of \SI{42}{GeV} electrons to \SI{85}{GeV} in less than a meter (see Fig.~\ref{fig:large-energy-gain}). Later, high energy-transfer efficiency from the wake to the trailing bunch of up to 30\% was demonstrated in the first high-gradient plasma-wakefield accelerator using a distinct driver and trailing bunch \cite{Litos2014}. A similar, but more precise experiment demonstrated the flattening of the accelerating field via \textit{optimal beam loading} (see Sec.~\ref{sec:optimal-beam-loading-energy-spread}), which enabled the preservation of energy spread while simultaneously providing energy-transfer efficiencies beyond 40\% \cite{Lindstrom2021a}. Further control of the beam quality was then demonstrated with the preservation of emittance \cite{Lindstrom2024}. Large transformer ratio (peak accelerating versus decelerating field) as high as 4.6 and 7.8 \cite{Loisch2018,Roussel2020} have also been shown, by careful shaping of the driver's current profile. Studies of the evolution of the plasma after a beam--plasma interaction found an upper limit to the repetition rate of order MHz or higher \cite{DArcy2022}. Next, PWFAs have been used for photon-science applications (see Sec.~\ref{sec:applications:FEL}), specifically to boost the energy in an FEL \cite{Pompili2022}. Utilizing these electron-driven plasma wakes, internal injection and acceleration of bunches with low energy spread (of order 1\%) and low emittance (of order \SI{1}{\milli\m\milli\radian}) have been achieved using several different techniques, including ionization injection \cite{Oz2007,VafaeiNajafabadi2014,VafaeiNajafabadi2016}, as well as plasma photocathode and density-downramp injection \cite{Deng2019,Knetsch2021,CouperusCabada2021,Foerster2022,Wood2026,Zhang2025}. While all the above results were demonstrated with negatively charged electron beams, positively charged particle beams have also been used. Positrons have been both transported \cite{Hogan2003} and accelerated~\cite{Blue2003} in plasmas, first in the linear regime, and later also in the nonlinear regime~\cite{Corde2015} and in a hollow plasma channel~\cite{Gessner2016a}. Moreover, electrons were accelerated to \SI{2}{GeV} in less than \SI{10}{m} by a \SI{400}{GeV} proton driver \cite{Adli2018} that had been self-modulated into a train of shorter bunches~\cite{Gross2018,Adli2019,Turner2019}. Lastly, beam-driven plasma wakefields have been driven using electron bunches from a laser-wakefield accelerator, the so-called \textit{hybrid} scheme (see Sec.~\ref{sec:hybrid}), operating with a pair of drive--trailing bunches from the LWFA~\cite{Gotzfried2020, Kurz2021} or with internal injection in the PWFA~\cite{CouperusCabada2021} that enabled the first demonstration of brightness transformation~\cite{Foerster2022}. All these experiments, as well as the facilities in which they were performed, are discussed in Sec.~\ref{sec:experiments}.

Contemporary experimental research focuses on demonstrating the remaining milestones needed for large-scale applications. A particular focus is placed on preserving the beam quality---ultimately the 6D beam brightness---across long acceleration distances. Connected to this is also observation and suppression of the hosing or beam-breakup instability as well as ion motion. Demonstration of high overall energy-transfer efficiency is a stated goal at several facilities \cite{Joshi2018,DArcy2019a}, ideally in conjunction with beam-quality preservation. Work on demonstrating high-repetition-rate and high-average-power operation is ongoing, exploring both the physical limitations of the plasma-acceleration process as well as the technical limitations of the plasma-source design. Improvements in stability and reliability, of both plasmas and beams, is another aspect that is continuously maturing. Brightness transformation and ultrabright-beam generation based on internal injection is actively pursued, with a particular focus on the collinear plasma-photocathode technique. Such demonstrations are attempted both at facilities with rf-based electron sources as well as laser--plasma-based hybrid sources, where the latter may benefit from the intrinsic synchronization of its electron beam and injection laser. Lastly, positron experiments are planned to continue, but the lack of facilities providing high-energy positron beams means progress is uncertain.

Contemporary theoretical research is exploring the ultimate limitations of beam-driven plasma accelerators, focusing on at least six major topics: (1) identifying a viable positron-acceleration scheme, with numerous recent proposals~\cite{Cao2024}; (2) understanding how to maximize the energy efficiency while simultaneously suppressing transverse instabilities~\cite{Lebedev2017}; (3) staging of multiple plasma accelerators compactly and without significant emittance growth \cite{Lindstrom2021b}; (4) performing self-consistent simulations of very long plasma accelerators with many stages, both accurately and at reduced computational cost \cite{Vay2021,Chen2025}; (5) understanding the long-term evolution of plasmas after beam--plasma interaction, and the repetition-rate and heating limits of plasma sources \cite{Zgadzaj2020,Gilljohann2019,DArcy2022}; (6) lastly, a number of collider-specific requirements are being studied, such as spin polarization \cite{Reichwein2025} and asymmetric-emittance beams \cite{Diederichs2024}.

In short, beam-driven plasma-wakefield acceleration is progressing towards demonstration of its promise as a compact accelerator technology for large-scale and high-power accelerator facilities. This Review summarizes the history and state-of-the-art of this exciting field of research.
In the sections below, we first cover the basic physics of the beam--plasma interaction (Sec.~\ref{sec:physics}), as well as more advanced variations and aspects of PWFAs (Sec.~\ref{sec:advanced-topics}), followed by a review of the most important research methods and results, both theoretical and experimental (Sec.~\ref{sec:research-methods-and-results}), a detailed look at proposed applications, with a particular focus on high-energy physics and photon science (Sec.~\ref{sec:applications}), and finally some concluding remarks looking toward the future of the field (Sec.~\ref{sec:conclusion}).


\section{Physics}
\label{sec:physics}

This section describes the three basic components of a beam-driven plasma-wakefield accelerator: the plasma wakefield, the beam driver and the trailing bunch. In particular, Sec.~\ref{sec:plasma-wakefields} covers the short-timescale response of the plasma to a beam, while Secs.~\ref{sec:driver-propagation} and \ref{sec:evolution-trailing-bunch} describe the longer-timescale evolution of the driver and trailing bunch within the plasma wakefield, respectively.


\subsection{Plasma wakefields}
\label{sec:plasma-wakefields}

At the heart of PWFA is the excitation of a charge-density wave in the wake of a particle beam propagating in a plasma. This charge-density wave consists of a plasma-density perturbation accompanied by electromagnetic fields generated in the wake of the beam, referred to as plasma wakefields. In this section, we introduce the basic theory underlying plasma wakefields (Sec.~\ref{sec:physics:plasma-wakefields:beams-plasmas}) and discuss different regimes for their excitation: linear (Sec.~\ref{sec:lin_wake}), nonlinear (Sec.~\ref{sec:nonlin_wake}), high-transformer-ratio (Sec.~\ref{sec:high-transformer-ratio}) and hollow-channel wakefields (Sec.~\ref{sec:hc_linear}), as well as the effect of plasma temperature and the phenomenon of wavebreaking (Sec.~\ref{sec:plasma_temperature}). SI units are used throughout this Review.

\subsubsection{Beams and plasmas}
\label{sec:physics:plasma-wakefields:beams-plasmas}

Beams and plasmas are collections of charged particles which interact with one another via Maxwell's equations. Given a set of microscopic beam and plasma particle coordinates and momenta, we could in principle calculate the Lorentz force experienced by every particle and evolve the beam--plasma system forward in time. Yet this approach is not tractable in practice, and we are forced to use an approximate description of the beam--plasma system. The most generic approach to the problem, the kinetic description of the plasma and the beam, is to replace our set of particle coordinates and momenta with a distribution $f(\mathbf{x},\mathbf{p},t)$ which describes the probability of finding a particle at a given position and momentum $(\mathbf{x},\mathbf{p})$ and at a given time $t$. The six-dimensional space $(\mathbf{x},\mathbf{p})$ is referred to as the \textit{phase space}. In beam physics, we associate the volume of phase space occupied by beam particles with the beam emittance~\cite{Humphries1990, Reiser2008, Seryi2015} (defined in Sec.~\ref{sec:envelope-matching}).

In the absence of collisions between particles, the evolution of the distribution function for a given particle species (e.g., beam electrons or plasma electrons) is described by the Vlasov equation~\cite{Goldston1995, Kruer2003}
\begin{equation}
    \frac{\dif f}{\dif t} = \frac{\del f}{\del t} + \frac{\mathbf{p}}{\gamma m}\cdot \mathbf{\nabla} f + \mathbf{F} \cdot \mathbf{\nabla}_p f = 0, \label{eq:vlasov}
\end{equation}
where $m$ and $\gamma$ are the species particle mass and the relativistic Lorentz factor, respectively, $\mathbf{F}$ is the force acting on the particle species and $\mathbf{\nabla}_p=(\partial_{p_x},\partial_{p_y},\partial_{p_z})$ differentiates with respect to the momentum variable.
The assertion $\dif f/\dif t = 0$ is a statement of Liouville's theorem: under the action of conservative forces (i.e., no collisions or radiative damping) the phase-space volume occupied by the particles is conserved.

An even more simplified description can be obtained by adopting a fluid theory for the plasma, which can be derived by taking \textit{moments} of the Vlasov equation. The species density is given by
\begin{equation}
    n(\mathbf{x},t) = \int f(\mathbf{x},\mathbf{p},t) \dif \mathbf{p}.
\end{equation}
The first-order moment of the Vlasov equation is found by integrating Eq.~(\ref{eq:vlasov}) over momentum space, which yields the plasma-fluid continuity equation
\begin{equation}
\frac{\del n}{\del t} + \mathbf{\nabla} \cdot (n\mathbf{v}) = 0,\label{eq:contiuity}
\end{equation}
with $\mathbf{v}=\mathbf{p}/(\gamma m)$ the velocity. The second-order moment is found by multiplying Eq.~(\ref{eq:vlasov}) by $\mathbf{p}$ and again integrating over momentum space. This yields the plasma-fluid equation of motion
\begin{equation}
\left[\frac{\del }{\del t}+\mathbf{v}\cdot\nabla\right]\mathbf{p}=q(\mathbf{E}+\mathbf{v}\times\mathbf{B}), \label{eq:fluid_eom}
\end{equation}
where we have taken the cold fluid limit of zero temperature and identified $\mathbf{F}$ as the Lorentz force, with $\mathbf{E}$ and $\mathbf{B}$ the electric and magnetic fields, and $q$ the species particle charge. Equations~(\ref{eq:contiuity}) and~(\ref{eq:fluid_eom}), along with Maxwell's equations
\begin{align}
\nabla \cdot \mathbf{E} &= \frac{\rho}{\epsilon_0},\label{eq:gauss} \\
\nabla \cdot \mathbf{B} &= 0,\label{eq:gaussB} \\
\nabla \times \mathbf{E} & = -\frac{\del \mathbf{B}}{\del t},\label{eq:farad} \\
\nabla \times \mathbf{B} & = \mu_0\mathbf{j} + \frac{1}{c^2} \frac{\del \mathbf{E}}{\del t},\label{eq:ampere}
\end{align}
serve as the starting point for our discussion of linear plasma-wakefield theory. Here, $\rho = \sum_i n_i q_i$ and $\mathbf{j} = \sum_i n_i q_i \mathbf{v}_i$ are the charge and current densities (summed over all species; $i$), and $\epsilon_0$, $\mu_0$ and $c$ are the vacuum permittivity, permeability and speed of light, respectively.

\subsubsection{Linear wakefields} 
\label{sec:lin_wake}

In this section, we consider the effect of a low-density, ultrarelativistic particle beam propagating into a neutral, cold plasma with electron density $n_0$. Here low density means that the beam density $n_b$ is much less than the plasma density: $n_b \ll n_0$, or in other words, that the plasma is \textit{overdense}, denser than the beam. This implies, as we will see below, that the plasma response is electrostatic behind the beam and that the linear plasma wave does not have a magnetic field, while the beam fields are radially shielded by the plasma.

The plasma ion mass $m_i$ is much larger than the plasma electron mass $m_e$ ($m_i \geq 1836 m_e$) and plasma ions are thus assumed to be immobile on the short plasma timescale relevant for plasma-wakefield acceleration, which is the timescale of the plasma electron response $\omega_p^{-1}$, where
\begin{equation}
    \label{eq:plas_freq}
    \omega_p = \sqrt{\frac{n_0e^2}{m_e\epsilon_0}}
\end{equation}
is the plasma-electron frequency. With this assumption, the ions form a uniform stationary background and provide a restoring force for displaced plasma electrons (see Sec.~\ref{sec:ion-motion} for a review of the effect of ion motion for conditions where it cannot be neglected). Finally, we will also use the \textit{quasistatic approximation} \cite{Sprangle1990, Mora1997} where the beam is assumed not to evolve during the timescale of the plasma response, or in other words that the beam timescale is much larger than the plasma-electron timescale.

The beam creates a small perturbation in the plasma electron density as it propagates through it. To first order, the plasma density is described by $n(\mathbf{x},t) = n_0 + n_1(\mathbf{x},t)$, where the plasma perturbation $n_1 \ll n_0$. The plasma is assumed to be at rest and free from any static electric and magnetic fields prior to the arrival of the electron beam. Therefore, the plasma-fluid velocity $\mathbf{v} = \mathbf{v}_1(\mathbf{x},t)$ and electric field $\mathbf{E} = \mathbf{E}_1(\mathbf{x},t)$ are also first-order perturbative quantities. Following the derivation by \textcite{Ruth1985}, we start by linearizing Eqs.~(\ref{eq:contiuity}),~(\ref{eq:fluid_eom}), and~(\ref{eq:gauss}) to find
\begin{align}
    \frac{\del n_1}{\del t} &= -n_0 \nabla \cdot \mathbf{v}_1, \label{eq:pert_cont}\\
    \frac{\del\mathbf{v}_1}{\del t}&= -\frac{e}{m_e}\mathbf{E}_1, \label{eq:pert_eom}\\
    \nabla\cdot\mathbf{E}_1 &= - e\frac{n_1}{\epsilon_0}+q\frac{n_b}{\epsilon_0}, \label{eq:pert_gauss}
\end{align}
where the charge density $\rho = - en_1 + qn_b$ has been used, with $-en_1$ the charge density of the plasma and $qn_b$ the charge density of the beam, $e$ being the elementary charge and $q$ the charge of a beam particle. Taking the divergence of Eq.~(\ref{eq:pert_eom}) and combining the three equations yield the linear plasma wave equation
\begin{equation}
    \label{eq:plas_wave_t}
    \frac{\del^2 n_1}{\del t^2} + \omega_p^2 n_1 = \omega_p^2 \frac{q}{e} n_b.
\end{equation} 

We consider an ultrarelativistic drive bunch with velocity $v_b \simeq c$ and use the change of coordinates $\xi=z-ct,\tau=t$, where $z$ is the longitudinal coordinate along the beam propagation axis so that $\xi$ is the \textit{co-moving} coordinate. We also use the quasistatic approximation $\partial_\tau\ll c\partial_\xi$, these derivatives being applied to any physical quantity describing the plasma wakefields or the driver. Equation~(\ref{eq:plas_wave_t}) can then be re-expressed with the co-moving coordinate $\xi$ instead of $t$, using $\partial_t=-c\partial_\xi$,
\begin{equation}
\frac{\del^2 n_1}{\del \xi^2} + k_p^2 n_1 = k_p^2 \frac{q}{e} n_b,
\label{eq:plas_wave}
\end{equation}
where $k_p = \omega_p/c$ is the plasma wavenumber. Equation~(\ref{eq:plas_wave}) is a simple harmonic oscillator with a source term on the right-hand side given by the beam density. The solution for $n_1$ is a sinusoidal function of $k_p\xi$ behind the drive beam, thus taking the form of a wave in the wake of the driver with a phase velocity equal to the speed of the driver, here assumed to be approximately $c$. Using the Green's function of Eq.~(\ref{eq:plas_wave}), $-\frac{q}{e}\sin(k_p\xi)\Theta(-\xi)$ with $\Theta$ the Heaviside step function, the solution for $n_1$ is obtained by convolving it with the beam profile
\begin{equation}
    n_1(x,y,\xi)=-k_p\frac{q}{e}\int_\xi^{+\infty}n_b(x,y,\xi^\prime)\sin[k_p(\xi-\xi^{\prime})] \dif \xi^{\prime}.
    \label{eq:n1}
\end{equation}
Transversely, the plasma-density perturbation $n_1$ is local and follows directly the transverse profile of the beam. The transverse extent of the density perturbation thus matches that of the beam.

\begin{figure}[t]
    \centering
    \includegraphics[width=0.94\linewidth]{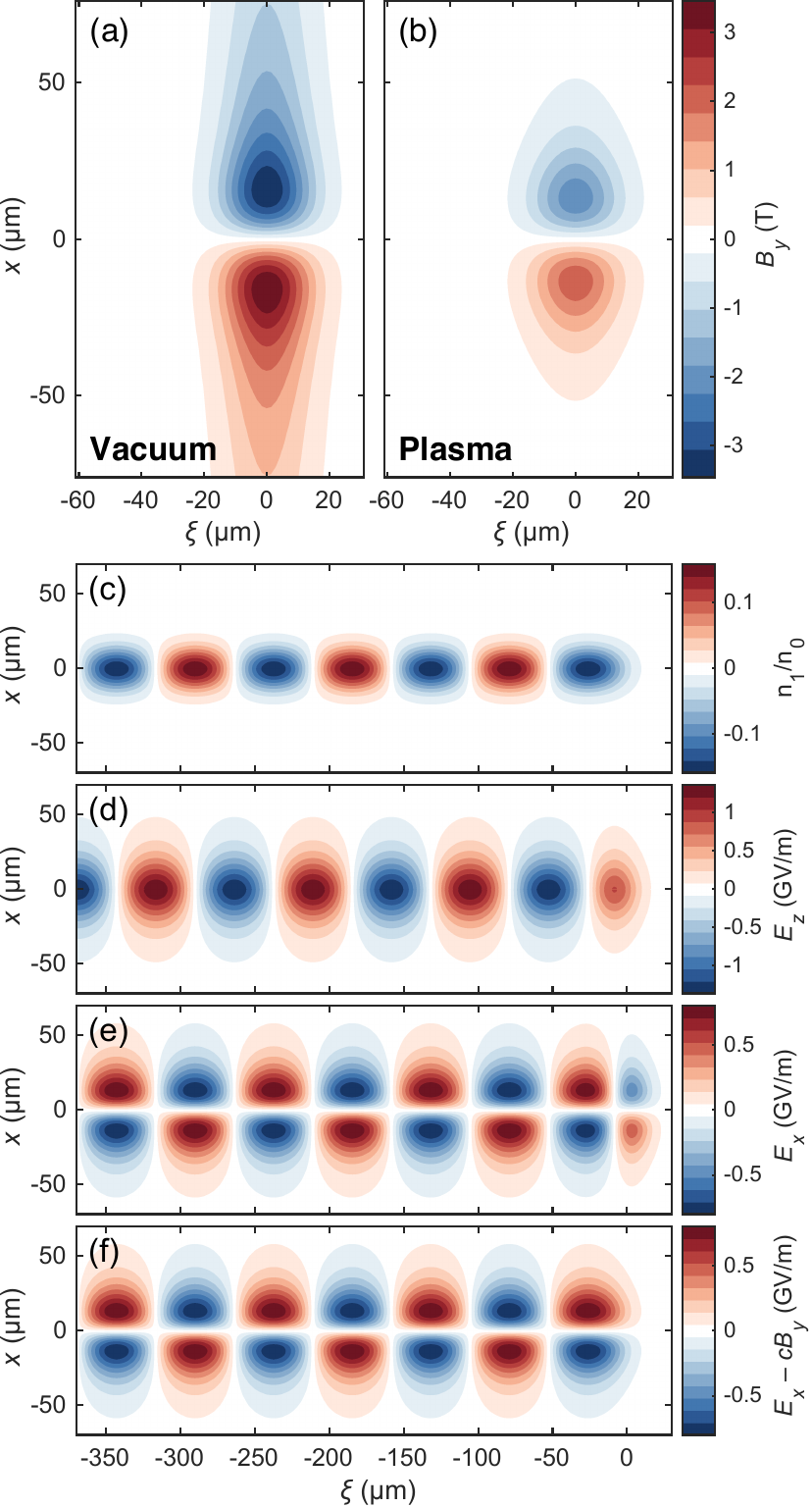}
    \caption{Linear plasma response to a relativistic, Gaussian beam with $N_b=2\times 10^8$ electrons and dimensions $\sigma_r=\sigma_z=\SI{10}{\micro\m}$ propagating to the right in a plasma with density $n_0=\SI{1e17}{cm^{-3}}$. The peak beam density is $n_b \simeq \SI{1.3e16}{cm^{-3}}$. Beam magnetic field in vacuum (a) and radially shielded by the plasma [(b), Eq.~(\ref{eq:B})], plasma density perturbation [(c), Eq.~(\ref{eq:n1})], longitudinal [(d), Eq.~(\ref{eq:Ez})] and transverse [(e), Eq.~(\ref{eq:Er})] electric field (including both terms from shielded beam electric field and from plasma wave), and transverse wakefield $W_x=E_x-cB_y$ (f) where the shielded self-field terms in $E_x$ and $B_y$ cancel out.}
    \label{fig:linear_response}
\end{figure}

The complete derivation for the fields $E_z$, $E_r$ and $B_\theta$ in the linear plasma wakefield can be found in \textcite{Keinigs1987} for the case with azimuthal symmetry where $n_b$ depends only on the radial cylindrical coordinate $r$ and on $\xi$. Using Maxwell's equations, going to the Fourier space of the $\xi$ coordinate and solving with respect to $r$ using the radial Green's function or with Hankel transforms, the solutions for the fields can be obtained and involve Bessel functions as follows
\begin{align}
\nonumber
&E_z(r,\xi) = -\frac{qk_p^2}{\epsilon_0}\int_0^{+\infty} r^\prime \dif r^\prime  K_0(k_pr_>)I_0(k_pr_<)\\
&\qquad\qquad\qquad\times\int_\xi^{+\infty} \dif \xi^\prime n_b(r^\prime,\xi^\prime) \cos{k_p(\xi-\xi^\prime)},\label{eq:Ez}\\
\nonumber
&E_r(r,\xi) = -\frac{q}{\epsilon_0}\int_0^{+\infty} r^\prime \dif r^\prime K_1(k_pr_>)I_1(k_pr_<)\\
&\times\left(\int_\xi^{+\infty} k_p \dif \xi^\prime \frac{\partial n_b(r^\prime,\xi^\prime)}{\partial r^\prime} \sin{k_p(\xi-\xi^\prime)}+\frac{\partial n_b(r^\prime,\xi)}{\partial r^\prime}\right),\label{eq:Er}\\
&B_\theta(r,\xi) = -\mu_0qv_b\int_0^{+\infty} r^\prime \dif r^\prime K_1(k_pr_>)I_1(k_pr_<) \frac{\partial n_b(r^\prime,\xi)}{\partial r^\prime}, 
\label{eq:B}
\end{align}
where $r_>$ (respectively $r_<$) is the larger (respectively smaller) of $r$ and $r^\prime$, and $I_n$ and $K_n$ are the \textit{n}th-order modified Bessel functions of the first and second kind, respectively. The terms involving $\del_{r^\prime} n_b(r^\prime,\xi)$ in $E_r$ and $B_\theta$ correspond to the self-generated fields of the beam that are radially shielded by the plasma over a plasma skin depth $k_p^{-1}$ [see Fig.~\ref{fig:linear_response}(b)], which reduce to the self fields of a bare beam in the vacuum limit $k_p\rightarrow0$ [see Fig.~\ref{fig:linear_response}(a)]. The sinusoidal terms in $E_z$ (cosine term) and $E_r$ (sine term) correspond to the plasma wave, with the characteristic property that $E_z$ and $E_r$ oscillations are 90{\textdegree} out of phase [see Figs.~\ref{fig:linear_response}(d) and (e)], and there is no $B$-field in the plasma wave behind the beam driver. The linear plasma wave is therefore of electrostatic nature. Similarly to the self fields, the plasma wakefields are also shielded radially over a plasma skin depth. In contrast to the density perturbation $n_1$, the fields are not necessarily local and can extend outside the beam. The transverse extent of the plasma wakefields matches the transverse extent of the shielded beam fields. For a small beam size, $k_p\sigma_r\ll1$, the plasma wakefields (and the energy in the plasma wave) thus extend radially over a plasma skin depth $k_p^{-1}$, while for a large beam size, $k_p\sigma_r\gg1$, they have the same transverse extent as that of the beam density $n_b$~\cite{Hue2021}. 

A short drive beam experiences a transverse force $F_r\simeq q(E_r-cB_\theta)$ that is focusing, and a longitudinal force $F_z=qE_z$ that is decelerating, consistent with a transfer of energy from the drive beam to the plasma wave. For the trailing bunch, due to the 90{\textdegree} phase shift between $E_z$ and $E_r$, exactly one quarter of the plasma wave period is simultaneously focusing and accelerating and thus suited for the acceleration and transport of the trailing bunch in the plasma. To maximize the energy-transfer efficiency from the plasma to the trailing bunch (see Sec.~\ref{sec:beam-loading}), there should be as little energy as possible left in the plasma behind the trailing bunch. Given that the energy is localized where the fields are, a mismatch in transverse size between the wakefields of the driver and trailing bunches is very detrimental to the energy efficiency. The solution is to use small beams, $k_p\sigma_r\ll1$, for both the driver and the trailing bunches, so that their wakefields have a similar transverse extent given by the plasma skin depth~\cite{Hue2021}.

\subsubsection{Nonlinear wakefields}
\label{sec:nonlin_wake}

When the ultrarelativistic particle beam has a bunch density that exceeds the plasma density, $n_b>n_0$, the plasma can no longer screen the particle beam and the linear perturbation theory used in Sec.~\ref{sec:lin_wake}, which assumes $n_1\ll n_0$, cannot be applied. Transversely, which is relevant for fully three-dimensional wakefields, plasma electrons experience large excursions from their initial positions, being either sucked in (for a positively charged beam) or blown out (for a negatively charged beam) by the particle bunch~\cite{Rosenzweig1991}. In the one-dimensional (1D) limit (i.e., an infinitely wide beam), the motion is purely longitudinal and in a nonlinear wakefield, plasma electrons oscillate along the propagation axis with a maximum speed approaching that of the particle bunch (i.e., near the speed of light) and with an excursion of the order of the plasma wavelength.

\begin{figure}[t]
    \centering
    \includegraphics[width=\linewidth]{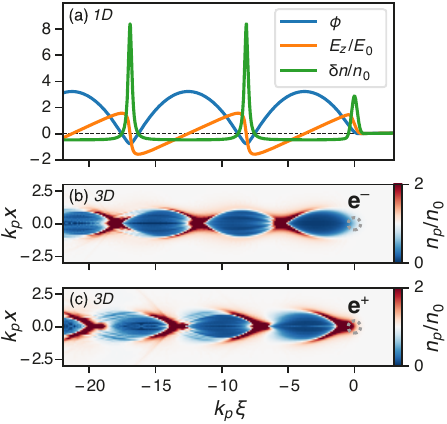}
    \caption{(a) 1D electron-driven nonlinear wakefield with normalized electric potential $\phi$, normalized electric field $E_z/E_0$ and plasma-density perturbation $\delta n/n_0$. (b) Snapshot of the plasma-electron density $n_p/n_0$ for a 3D nonlinear wakefield driven by an electron driver and (c) a positively charged (positron or proton) driver. In all cases $n_b/n_0=3$, while $k_p\sigma_z=0.21$ in (a) and $k_p\sigma_z=k_p\sigma_r=0.42$ in (b,c). Dashed lines in (b,c) represent the beam density contour at half maximum.}
  \label{fig:nonlinear_wakefields}
\end{figure}

In the quasistatic approximation where $n_b$ only depends on $\xi$, and with $v_b\simeq c$, the equation for nonlinear wakefields in 1D is given in terms of the normalized electric potential $\phi=e\Phi/m_ec^2$, where $\Phi$ is the electric potential, by~\cite{Rosenzweig1987,Krall1991}
\begin{align}
    \label{eq_1Dnonlinear_phi}
    &\frac{\dif ^2\phi}{\dif \xi^2}=\frac{k_p^2}{2}\left[
    \frac{2n_b}{n_0}+\frac{1}{(1+\phi)^2}-1 \right],\\
    \label{eq_1Dnonlinear_Ez}
    &\frac{E_z}{E_0}=-\frac{1}{k_p}\frac{\dif \phi}{\dif \xi},
\end{align}
with $E_0=m_ec\omega_p/e$ [Eq.~(\ref{eq:wavebreaking-field-simple})] the normalizing electric field, referred to as the nonrelativistic cold-plasma wavebreaking field (see Sec.~\ref{sec:plasma_temperature}). A 1D nonlinear wakefield obtained by solving Eqs.~(\ref{eq_1Dnonlinear_phi}--\ref{eq_1Dnonlinear_Ez}) is shown in Fig.~\ref{fig:nonlinear_wakefields}(a).

For fully three-dimensional (3D) nonlinear wakefields, there is no complete and self-consistent analytical solution. We must therefore rely on numerical modeling (see Sec.~\ref{sec:numerical-simulations}) and/or phenomenological models to provide a qualitative and quantitative description of the nonlinear wakefield. In addition to the dimensionless parameter $n_b/n_0$, in 3D the nonlinear wakefield also depends on the normalized current
\begin{equation}
    \label{eq:normalized-current}
    \Lambda \equiv 2I/I_A=k_p^2\sigma_r^2n_b/n_0
\end{equation}
or on the normalized bunch charge
\begin{equation}
    \label{eq:normalized-charge}
    \tilde{Q} \equiv k_p^3N_b/n_0=(2\pi)^{3/2}k_p\sigma_z\Lambda,
\end{equation}
where $I_A=4\pi\epsilon_0 m_ec^3/e\simeq\SI{17}{kA}$ is the Alfv{\'e}n current, $I$ is the peak current of the beam and $N_b$ the total number of particles in the beam. Beyond their definition in terms of $I$ and $N_b$, their expression in terms of the beam size $\sigma_r$ and the bunch length $\sigma_z$ are given for Gaussian beams. The relevant parameter for very short bunches $k_p\sigma_z\ll 1$ is $\tilde{Q}$~\cite{Barov2004,Rosenzweig2004}, while $\Lambda$ is more appropriate to describe wakefields driven by longer bunches [$k_p\sigma_z\gtrsim 0.2$~\cite{Lu2010}]. Based on these dimensionless parameters, the following regimes are defined:
\begin{itemize}
    \item $n_b/n_0\ll1$: the linear regime (Sec.~\ref{sec:lin_wake});
    \item $n_b/n_0\gtrsim1$ and ($\Lambda \ll 1$ or $\tilde{Q} \ll 1$): \\ the nonrelativistic and nonlinear regime;
    \item $n_b/n_0\gtrsim1$ and ($\Lambda\gtrsim 1$ or $\tilde{Q}\gtrsim 1$): \\ the relativistic and nonlinear regime.
\end{itemize}
For a negatively charged particle bunch (e.g., an electron beam) and $n_b/n_0\gtrsim1$, plasma electrons are expelled, or blown out, from the propagation axis by the particle bunch. Plasma electrons are then pulled back by the exposed plasma ions, overshooting the axis and setting up a nonlinear plasma oscillation. In this so-called blowout regime~\cite{Rosenzweig1991}, the plasma wave takes the form of ion cavities surrounded by thin electron sheaths, as shown in Fig.~\ref{fig:nonlinear_wakefields}(b)~\cite{Lee2000,Barov2004,Rosenzweig2004,Lotov2004,Lu2006a,Lu2006b}.

\begin{figure}[t]
    \centering
    \includegraphics[width=\linewidth]{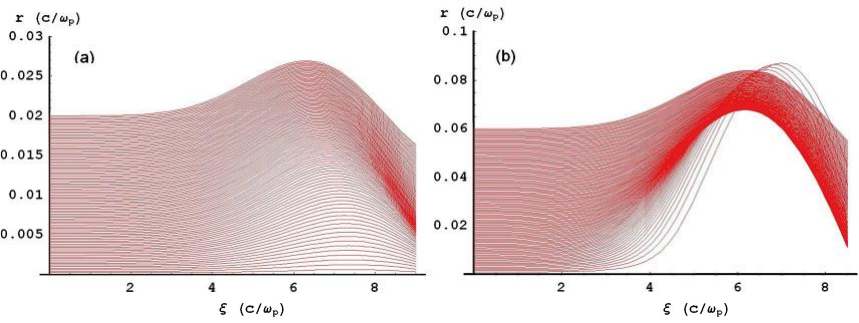}
    \caption{Plasma-electron trajectories for $n_b=n_0$ (a) and $n_b=10\:n_0$ (b). The electron bunch is located at $\xi=5\:k_p^{-1}$, with $k_p\sigma_z=\sqrt{2}$ and $k_p\sigma_r=0.01$. Note the crossing of trajectories in (b). Adapted from \textcite{Lu2006b}.}
  \label{fig:trajectory_crossing}
\end{figure}

While the linear regime (Sec.~\ref{sec:lin_wake}) and the 1D nonlinear regime [Eqs.~(\ref{eq_1Dnonlinear_phi}) and (\ref{eq_1Dnonlinear_Ez})] can be described by fluid theory, the blowout regime is characterized by transverse crossing of plasma-electron trajectories~\cite{Lu2006b}---a phenomenon sometimes referred to as \textit{transverse wavebreaking}. In this case, the fluid description must be abandoned in favor of a fully kinetic approach.
As shown in Fig.~\ref{fig:trajectory_crossing}, plasma electrons initially located closer to the axis can experience a stronger deflection than plasma electrons initially located further away from the axis, and their trajectories intersect each other. Trajectory crossing is in fact a necessary condition to reach the blowout regime~\cite{Lu2006b}, as otherwise a region void of plasma electrons cannot be established.

In addition to trajectory crossing, the blowout regime can also have another specific characteristic when the plasma response is relativistic ($\Lambda\gtrsim 1$ or $\tilde{Q}\gtrsim 1$): along the longitudinal axis, plasma electrons are initially pushed forward~\cite{Barov2004,Xu2020}, instead of backward as in the linear or 1D nonlinear regime. At the front of the particle bunch, the decelerating longitudinal electric field induces a backward motion for plasma electrons, but in a fully 3D configuration, the electric field of the bunch expels plasma electrons radially, while the bunch magnetic field pushes them longitudinally forward through $\bf{v}\times\bf{B}$; a forward push that overcomes the decelerating electric field.

Importantly, the blowout regime has different scalings depending on the relativistic or nonrelativistic nature of the plasma response. A phenomenological model can be used to describe quantitatively the fields in the ion cavity of the blowout regime, by considering that blown-out plasma electrons form a thin electron sheath circulating around the ion cavity~\cite{Lotov2004,Lu2006a,Lu2006b}. The dynamics can then be described in terms of two coupled systems: the electron sheath whose trajectory $r_b(\xi)$ depends on the fields through its equation of motion, and the fields that depend on the sheath trajectory (the latter characterizing the source terms in Maxwell's equations in this model). The plasma can then be considered as made of three parts: the ion cavity, the plasma-electron sheath, and the region outside with weak density perturbation, for which the response is linear~\cite{Lu2006b}. 

Transversely, the linear-response region outside the ion cavity typically extends over a plasma skin depth $k_p^{-1}$. For weak drivers ($\Lambda \ll 1$ or $\tilde{Q} \ll 1$), the blowout radius is smaller than the plasma skin depth and as a result, the linear-response region dominates over the contribution of the blowout cavity when calculating $E_z$. Thus, in this nonrelativistic blowout regime, despite the linear theory being invalid because $n_b/n_0\gtrsim1$, it still provides a useful estimate and scaling for $E_z$~\cite{Lu2010}. 

On the other hand, for the relativistic blowout regime ($\Lambda \gtrsim 1$ or $\tilde{Q} \gtrsim 1$), the blowout radius is larger than the plasma skin depth, and the linear theory predictions and scalings can fail. For ultrarelativistic blowout ($\Lambda \gg 1$ or $\tilde{Q} \gg 1$), the blowout cavity provides the dominant contribution and the coupled sheath--field system can be simplified and described by a single differential equation for the electron sheath trajectory $r_b(\xi)$~\cite{Lu2006a,Lu2006b}:
\begin{align}
    \label{eq:sheath-traj-lu}
    &\frac{k_p^2r_b^3}{4}\frac{\dif^2r_b}{\dif \xi^2}+\frac{k_p^2r_b^2}{2}\left[ \frac{\dif r_b}{\dif \xi}\right]^2+\frac{k_p^2r_b^2}{4} = \lambda(\xi,r_b),\\
    &\frac{E_z}{E_0}=-\frac{1}{2}k_pr_b\frac{\dif r_b}{\dif \xi},
\end{align}
where $\lambda(\xi,r_b)=k_p^2\int_0^{r_b}[n_b(\xi,r)/n_0]r\dif r$ is the normalized current profile of the electron beam when $r_b\gg\sigma_r$, and is related to the $\Lambda$ parameter by $\Lambda=\max\:\lambda(\xi)$. This model does not take into account ion motion, whose effect is discussed in Sec.~\ref{sec:ion-motion}. Near the maximum $R_b$ of $r_b(\xi)$ and for zero right-hand side in Eq.~(\ref{eq:sheath-traj-lu}), the trajectory resembles that of a circle~\cite{Lu2006a}; in this case $E_z(\xi)/E_0\simeq k_p\xi/2$ [$\xi=0$ being the location of the maximum of $r_b(\xi)$], consistent with expressions obtained for the so-called \textit{bubble regime} in laser-wakefield accelerators, where the bubble shape of the ion cavity is assumed~\cite{Kostyukov2004,Pukhov2004}. At the rear of the blowout cavity, the sheath-trajectory equation differs significantly from a circle and leads to the characteristic spike of $E_z$, a feature not reproduced in bubble models. 

A more accurate description than Eq.~(\ref{eq:sheath-traj-lu}) can be obtained by using a multi-sheath model~\cite{Dalichaouch2021}, an energy-conserving theory~\cite{Golovanov2023} or an adiabatic sheath model \cite{Liu2025}. The energy-conserving theory not only satisfies the energy-conservation law, but also provides much higher accuracy in a wide range of parameters, without the need to fit external parameters. It is not limited to the description of very large blowout cavities with $k_pR_b\gg1$, and can be used for $k_pR_b\sim1$. The energy-conserving sheath trajectory equation is given by~\cite{Golovanov2023}
\begin{align}
    \label{eq:sheath-traj}
    &\left(\frac{k_p^2r_b^3}{4}+r_b\right)\frac{\dif^2r_b}{\dif \xi^2}+\left(\frac{k_p^2r_b^2}{2}+1\right)\left[ \frac{\dif r_b}{\dif \xi}\right]^2+\frac{k_p^2r_b^2}{4} = \lambda(\xi,r_b),
\end{align}
which simplifies to Eq.~(\ref{eq:sheath-traj-lu}) for ultrarelativistic blowouts with $k_p r_b\gg1$. Importantly, while the seminal sheath model~\cite{Lu2006a, Lu2006b} with Eq.~(\ref{eq:sheath-traj-lu}) relies on the assumption that the plasma-electron sheath is described by a single plasma electron trajectory, which was shown to fail notably in the front half of the blowout cavity~\cite{Stupakov2016,2017Wang}, the energy-conserving theory with Eq.~(\ref{eq:sheath-traj}) does not rely on this single-electron-trajectory assumption and instead only requires that electrons in the sheath are moving
tangentially to the border of the blowout cavity.

For moderate bunch length ($k_p\sigma_z\gtrsim0.2$), the maximum blowout radius $R_b$ can be estimated by equating the repulsive space-charge force from the electron drive beam and the restoring force from the wakefield; the exact numerical coefficient in this estimate can be verified with simulations. It is given by~\cite{Lu2006b,Lu2010}
\begin{equation}
    \label{eq:blowout_radius}
    k_pR_b=2\sqrt{\Lambda}.
\end{equation}
This relation can also be obtained by considering the equilibrium blowout radius for a beam with a constant current $\lambda(\xi)=\Lambda$; the first two terms of the left-hand side of Eq.~(\ref{eq:sheath-traj}) go to zero. For short bunches with $k_p\sigma_z\ll1$, it is the initial kick given to plasma electrons by the drive bunch that determines the maximum blowout radius $R_b$. A similar relationship exists between $R_b$ and $\tilde{Q}$ of the form 
\begin{equation}
    \label{eq:blowout_radius_short}
    k_pR_b=C\sqrt{\tilde{Q}},
\end{equation}
with $C\simeq 2.8$ for short and narrow drivers~\cite{2017Wang}. This general scaling was shown to hold exactly in the nonrelativistic and nonlinear regime with $\tilde{Q}\ll1$, where the plasma response to a pointlike drive bunch can be exactly rescaled as $\tilde{Q}^{1/2}$~\cite{Stupakov2016}. Finally, for long bunches with $k_p\sigma_z\gg1$ (used to reach high transformer ratios; see Sec.~\ref{sec:high-transformer-ratio}), Eq.~(\ref{eq:blowout_radius}) will overestimate the blowout radius. In this case, an adiabatic sheath model is more accurate \cite{Liu2025}.

\begin{figure}[t]
    \centering
    \includegraphics[width=\linewidth]{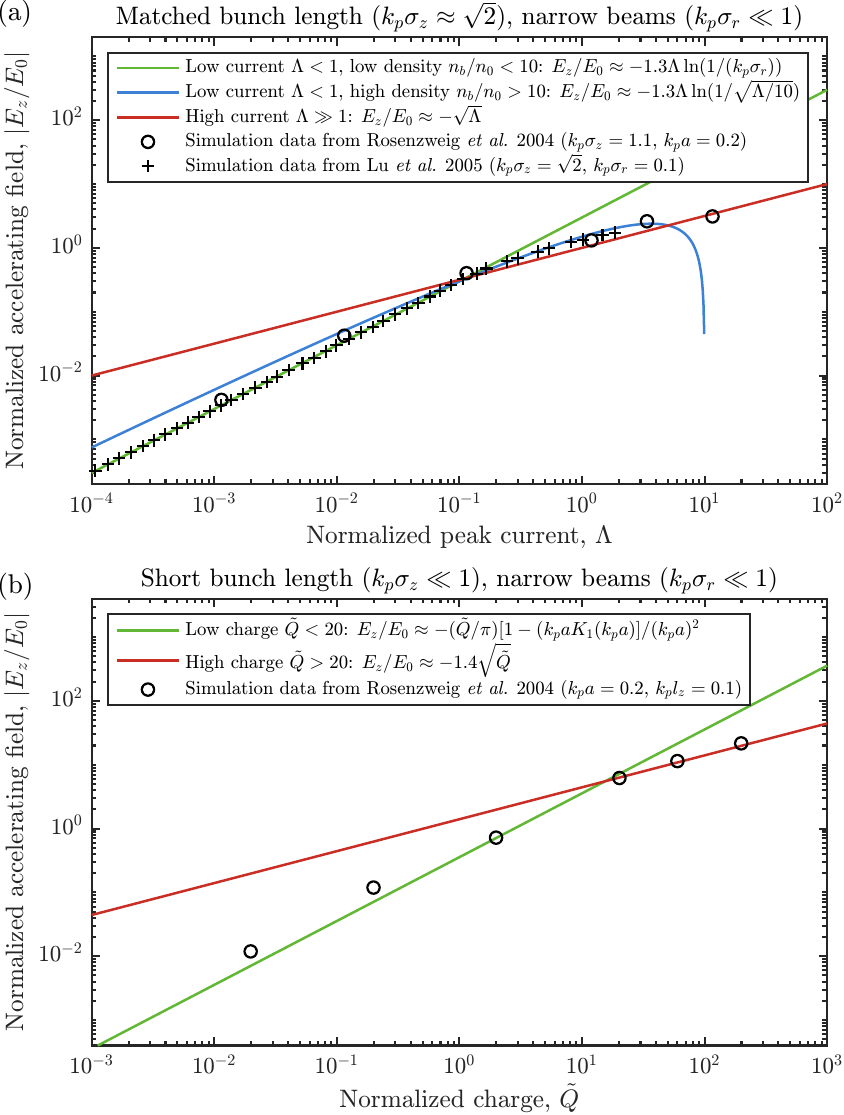}
    \caption{Accelerating field in linear and nonlinear plasma wakes driven by narrow electron bunches ($k_p\sigma_r \ll 1$, where $\sigma_r$ is the radial beam size). (a) For matched bunch lengths ($k_p\sigma_z\approx\sqrt{2}$), the field varies with the normalized peak current $\Lambda$ [Eq.~(\ref{eq:normalized-current})]. The relevant model in the area of validity (as indicated in the legend) agrees with simulation data from \textcite{Rosenzweig2004} and \textcite{Lu2005}, who both used Gaussian current profiles of lengths $k_p\sigma_z = 1.1$ and $\sqrt{2}$, respectively. The former simulated beams with a flat-top radial profile of radius $k_p a =0.2$, whereas the latter used a Gaussian radial profile of size $k_p\sigma_r=0.1$. (b) For short bunch lengths ($k_p\sigma_z\ll1$), the field instead varies with the normalized charge $\tilde{Q}$ [Eq.~(\ref{eq:normalized-charge})]. The low- and high-charge models agree with simulations from \textcite{Rosenzweig2004}, here using a flat-top current profile of length $k_pl_z=0.1$.}
  \label{fig:electric-field-prediction}
\end{figure}

The maximum accelerating field, near the rear of the blowout cavity, can be predicted analytically across a range of parameters, as shown in Fig.~\ref{fig:electric-field-prediction}. For narrow ($k_p\sigma_r < 0.3$) bunches with a length matched to the plasma skin depth ($k_p\sigma_z \approx \sqrt{2}$), operating in the nonrelativistic regime ($\Lambda \lesssim 1$), \textcite{Lu2005} found that for beam densities below $n_b/n_0 < 10$, the ``useful" field (i.e., avoiding the spike at the very end of the blowout cavity) is given by
\begin{equation}
    \label{eq:useful_field_matched}
    \frac{E_z}{E_0} \approx -1.3 \Lambda \ln\left(\frac{1}{k_p\sigma_r}\right).
\end{equation}
For higher densities ($n_b/n_0 > 10$ but still $\Lambda \lesssim1$), 
Eq.~(\ref{eq:useful_field_matched}) becomes 
\begin{equation}
    \label{eq:useful_field_matched_highdensity}
    \frac{E_z}{E_0} \approx -1.3 \Lambda \ln\left(\frac{1}{\sqrt{\Lambda/10}}\right).
\end{equation}
Similarly, for short and narrow bunches ($k_p\sigma_z \ll 1$, $k_p\sigma_r \ll 1$), \textcite{Barov2004} and \textcite{Rosenzweig2004} found that for normalized charges below $\tilde{Q} \lesssim 20$, the useful field is
\begin{equation}
    \label{eq:useful_field_short}
    \frac{E_z}{E_0} \approx -\frac{\tilde{Q}}{\pi}\frac{[1-(k_p a) K_1(k_p a)]}{(k_p a)^2},
\end{equation}
where $a \approx \sqrt{3}\sigma_r \ll 1$ is the radius of a uniform disk beam and $K_1$ is the first-order modified Bessel function of the second kind; the leading-order approximation to this is $E_z/E_0 \approx -(\tilde{Q}/2\pi)\ln(1.123/(k_p a))$. Equations~(\ref{eq:useful_field_matched}) and (\ref{eq:useful_field_short}) both scale linearly with charge, as expected in the linear regime, but surprisingly also partly hold in the nonlinear regime. In the strongly relativistic regime ($\Lambda \gg1$ or $\tilde{Q} \gg1$), the field grows linearly with $\xi$ as $E_z/E_0 \approx k_p\xi/2$ toward the rear of a spherical wake (for which $k_p\xi_\mathrm{rear} \approx -k_pR_b = -2\sqrt{\Lambda}$ or $-2.8\tilde{Q}^{1/2}$) \cite{Lu2006a}, such that the maximum useful field is $E_z/E_0 \approx -\sqrt{\Lambda}$ or $-1.4\tilde{Q}^{1/2}$, which scales sublinearly with charge. For moderate bunch lengths (i.e., for which $k_p R_b \approx 2 \sqrt{\Lambda}$), this approximation is consistent with \textcite{Golovanov2023} to within $\sim$10\% from $\Lambda\gtrsim3$.

The field properties in the blowout regime are ideal for preserving beam quality. Because the blowout cavity is devoid of plasma electrons and thus has a uniform charge density inside associated with the immobile ions (ignoring for now ion motion, see Sec.~\ref{sec:ion-motion}), the transverse focusing force, $F_r=q(E_r-cB_\theta$) for axisymmetric beams, is linear in $r$ and independent of $\xi$. This is a key property that allows for the preservation of transverse emittance~\cite{Clayton2016}. Furthermore, the Panofsky-Wenzel theorem~\cite{Panofsky1956} implies that 
\begin{equation}
    \label{eq:panofsky-wenzel-theorem}
    \partial_\xi F_r=\partial_r F_z,
\end{equation}
and thus the uniformity of $F_r$ along $\xi$ ($\partial_\xi F_r=0$) ensures that $F_z=qE_z$ is independent of $r$ and that the slice energy spread can be preserved: all particles from a given slice $\xi_0$ experience the same accelerating field $E_z$. These properties make the blowout regime a very promising candidate for highly efficient and high-quality plasma-based electron acceleration in PWFA (see Sec.~\ref{sec:evolution-trailing-bunch}).

In contrast to negatively charged drivers, the case of a bunch of positively charged particles (i.e., positrons or protons) exhibits some specific properties in the nonlinear regime $n_b/n_0\gtrsim1$. Plasma electrons are sucked in by a positively charged drive bunch instead of being expelled as in the blowout regime. As a result, plasma electrons flow through the drive particle bunch, and a sheath trajectory model as in Eq.~(\ref{eq:sheath-traj}) is not applicable to describe this phase of the wakefield formation. However, for bunches that are sufficiently short so that most plasma electrons cross the propagation axis behind the bunch, an ion cavity similar to the one of the blowout regime can be formed, as shown in Fig.~\ref{fig:nonlinear_wakefields}(c), because sucked-in plasma electrons are crossing and overshooting the axis in a region void of drive-bunch particles. Once plasma electrons have overshot the axis, an ion cavity is naturally formed with properties similar to that described above. This ion-cavity formation can occur in both nonlinear proton-driven PWFA~\cite{Caldwell2009} and positron-driven PWFA ~\cite{Corde2015}, the latter which was demonstrated experimentally. Here, it was shown that a longer positron drive bunch would lead to a more complicated wakefield where the ion cavity is not void of plasma electrons (see Sec.~\ref{sec:beam-loading} and Fig.~\ref{fig:beam_loading_posi}), opening the possibility for positron acceleration (see Sec.~\ref{sec:positron-acceleration}). In general, in the strongly nonlinear regime with $n_b/n_0\gg1$, a long positively charged drive bunch can lead to multiple oscillations of plasma electrons around the propagation axis if $\omega_{e,b}\sigma_z/c\gtrsim1$, where $\omega_{e,b}=(n_be^2/m_e\epsilon_0)^{1/2}$ is the plasma-electron frequency associated with the particle-bunch density. In this situation, the on-axis plasma electron density as well as the focusing force have strong oscillations along $\xi$, and the Panofsky-Wenzel theorem implies that $E_z$ is strongly non-uniform transversely. Such a non-ideal field structure can however be tolerable for a drive particle bunch, where preservation of beam quality is not important, as long as the drive bunch can efficiently transfer its energy to the plasma wakefields.

A critical quantity for the theoretical description of 3D nonlinear wakefields, underlying the previously discussed models, is the pseudo-potential $\psi=\phi- a_z$, with $a_z$ the longitudinal component of the normalized vector potential $\mathbf{a}=e\mathbf{A}/m_ec$ and $\mathbf{A}$ the vector potential. While the electric potential is sufficient to capture the physics of nonlinear plasma wakefields in 1D, as described by Eqs.~(\ref{eq_1Dnonlinear_phi}) and~(\ref{eq_1Dnonlinear_Ez}), $\mathbf{A}$ is required to account for the fully electromagnetic character of 3D nonlinear wakefields. In the quasistatic approximation, which represents a symmetry, the system is invariant under the transformation $z \rightarrow z+c\Delta T$, $t\rightarrow t+\Delta T$, and the gauge-dependent potential $\phi$ can be elevated to the gauge-independent pseudo-potential $\psi$ that fully determines the longitudinal and transverse wakefields~\cite{Blumenfeld2009}:
\begin{align}
    & E_z = -\frac{1}{k_p}\frac{\partial \psi}{\partial \xi}\ E_0,\\
    & E_r-cB_\theta = -\frac{1}{k_p}\frac{\partial \psi}{\partial r}\ E_0.
\end{align}
The variable $\psi$ is thus sufficient to predict the dynamics of the drive and trailing particle beams that propagate at nearly the speed of light $c$ along the $z$ axis. The quantity $\psi$ obeys a Poisson-like equation in the transverse plane, with $\rho-j_z/c$ as a source term~\cite{Lu2006b}. The mathematical identity $\partial_\xi\partial_r\psi=\partial_r\partial_\xi\psi$ provides an immediate proof of the Panofsky-Wenzel theorem [Eq.~(\ref{eq:panofsky-wenzel-theorem})] in the quasistatic case. In addition to being a powerful theoretical tool for modeling 3D nonlinear wakefields, $\psi$ also plays a fundamental role in the Hamiltonian dynamics of a test electron and thus on the physics of internal injection  (a test particle is assumed not to influence the rest of the system). Applying Noether's theorem, the quasistatic symmetry implies the existence of a constant of motion given by $H-cP_z$ for the test electron, where $H$ and $P_z=p_z-eA_z$ are the Hamiltonian and longitudinal component of the canonical momentum of the test electron, respectively. This conserved quantity provides a direct and fundamental relationship between the gauge-invariant $\psi$ and the kinetic energy and forward momentum of the test electron~\cite{Mora1997}:
\begin{equation}
\label{eq:qsa-constant-of-motion}
    \gamma-\frac{p_z}{m_ec}-\psi=\mathrm{constant}.
\end{equation}
Using initial conditions to determine the constant, this identity provides a condition for trapping a test electron into the wakefield, as discussed in Sec.~\ref{sec:internal-injection}. The wake phase velocity $v_\phi$, assumed to be $c$ until now, can also be accounted for in the quasistatic symmetry, in which case Eq.~(\ref{eq:qsa-constant-of-motion}) is replaced by Eq.~(\ref{Hamiltonian2}), $\xi=z-v_\phi t$ and $\psi=\phi-v_\phi a_z/c$.

\subsubsection{High-transformer-ratio wakefields}
\label{sec:high-transformer-ratio}

To maximize the energy that can be gained by the trailing bunch in a single plasma accelerator stage while keeping the driver energy at a reasonable level, the so-called transformer ratio $\mathcal{R}$, defined as
\begin{equation}
    \mathcal{R}=\frac{|E_\mathrm{acc}|}{|E_\mathrm{dec}|},
\end{equation}
has to be increased, where $E_\mathrm{acc}$ is the maximum accelerating field experienced by the trailing bunch and $E_\mathrm{dec}$ is the maximum decelerating field found in the driver. Indeed, for a drive-beam energy $E_d$, the maximum energy gain for the trailing bunch is limited to $\mathcal{R}E_d$. The transformer ratio is a property of the wakefield, and specific current profiles for the driver are required to generate high-transformer-ratio wakefields. In fact, the transformer ratio is a general characteristic for collinear wakefield acceleration~\cite{Bane1985a} and is relevant to wakefields driven also in metallic and dielectric structures, in addition to plasmas. Importantly, in the linear regime, it was found that $\mathcal{R}$ cannot be larger than 2 for a drive bunch whose current profile is symmetric about its midpoint~\cite{Bane1985b}. 

\begin{figure}[t]
    \centering
    \includegraphics[width=\linewidth]{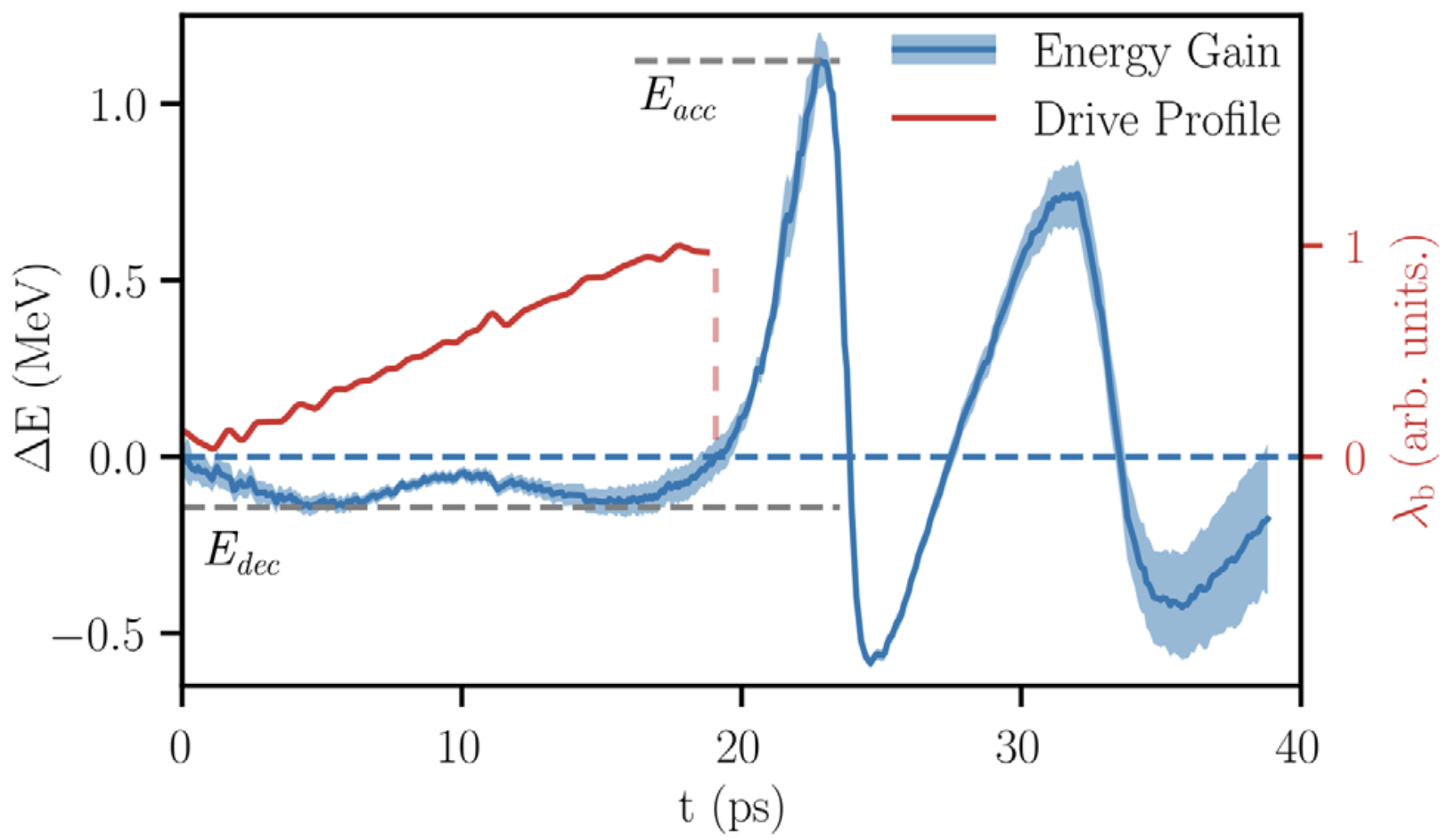}
    \caption{High-transformer-ratio plasma wakefields in a PWFA experiment. The energy change (blue), representing the longitudinal wakefield, and the drive-beam current profile (red) were measured as a function of time. A transformer ratio $\mathcal{R}$ of nearly 8 was achieved using a linearly-ramped drive-current profile. Adapted from \textcite{Roussel2020}.}
    \label{fig:high-transformer-ratio}
\end{figure}

Strategies to maximize $\mathcal{R}$ and overcome the $\mathcal{R}=2$ limit are therefore largely based on using longitudinally asymmetric drivers. In practice, the most effective way to increase $\mathcal{R}$ is to minimize the decelerating field $E_\mathrm{dec}$ within the drive beam for a given accelerating field $E_\mathrm{acc}$. This can be done by using a long, linearly-ramped current profile \cite{Bane1985a,Chen1986,Katsouleas1986,Rosenzweig1991,OShea2012,Massimo2014,Lemery2015,Su2023}, as shown in Fig.~\ref{fig:high-transformer-ratio}; a long adiabatic rise of the beam current to transfer energy slowly from the drive beam to the plasma wakefield, followed by a rapid fall to set the plasma oscillation with large amplitude. A precursor bunch (i.e., just ahead of the driver) or a modified bunch head \cite{Jiang2012} can also be used to improve the uniformity of $E_\mathrm{dec}$ along the drive bunch, as demonstrated experimentally by \textcite{Roussel2020}. Another scheme is using a train of multiple drive bunches with linearly increasing charge per bunch \cite{Kallos2007,Fang2011,Farmer2015}  and spaced by 1--1.5 plasma wavelengths. Wakefield excitation by a ramped train of electron bunches was demonstrated experimentally by \textcite{Verra2025}.

The first experimental demonstrations of high-transformer-ratio wakefields in advanced accelerators were obtained for dielectric wakefield accelerators~\cite{Jing2007, Jing2011} with $\mathcal{R}\approx5$~\cite{Gao2018}. In PWFA, two seminal results demonstrated experimentally transformer ratios of nearly 5~\cite{Loisch2018} and 8~\cite{Roussel2020} in the nonrelativistic and nonlinear regime. \textcite{Loisch2018} relied on the shaping of the photocathode laser temporal profile at the photoinjector to generate a ramped current profile for the drive electron beam, and used a short trailing electron bunch to probe $E_\mathrm{acc}$, with energy gain of \SI{0.43}{MeV} for the trailing bunch and maximum slice energy loss of \SI{0.09}{MeV} for the drive bunch. \textcite{Roussel2020} used an emittance exchange (EEX) process to map the transverse coordinate to the longitudinal coordinate, thereby shaping the longitudinal current profile when using a beam mask in the plane transverse to the beam axis just before the EEX beamline. With a long trailing electron bunch, the longitudinal wakefield was measured in a single shot (see Fig.~\ref{fig:high-transformer-ratio}), allowing the dependence of high-transformer-ratio wakefields with the shape of the drive beam to be uncovered. The observed transformer ratio of nearly 8 was well in excess of the expected value of $\mathcal{R}$ for such a drive-current profile and for single-mode linear wakefields~\cite{Bane1985a}, highlighting the potential of nonlinear (multimode) wakefields to enhance the transformer ratio further.

\subsubsection{Hollow-channel plasma wakefields}
\label{sec:hc_linear}

As discussed in Sec.~\ref{sec:nonlin_wake}, the nonlinear blowout regime is ideal for accelerating a trailing electron bunch in plasma. The transverse focusing forces are linear, the accelerating field is independent of transverse beam offset, and a properly loaded beam will be uniformly accelerated (see Sec.~\ref{sec:optimal-beam-loading-energy-spread}). By contrast, the region of the nonlinear blowout wakefield that is both focusing and accelerating for positrons is too small to load a trailing positron bunch unless significant modifications are made to the wake, as discussed in Secs.~\ref{sec:loaded} and \ref{sec:beam-plasma-shaping}.

\begin{figure}
    \centering
    \includegraphics[width=\linewidth]{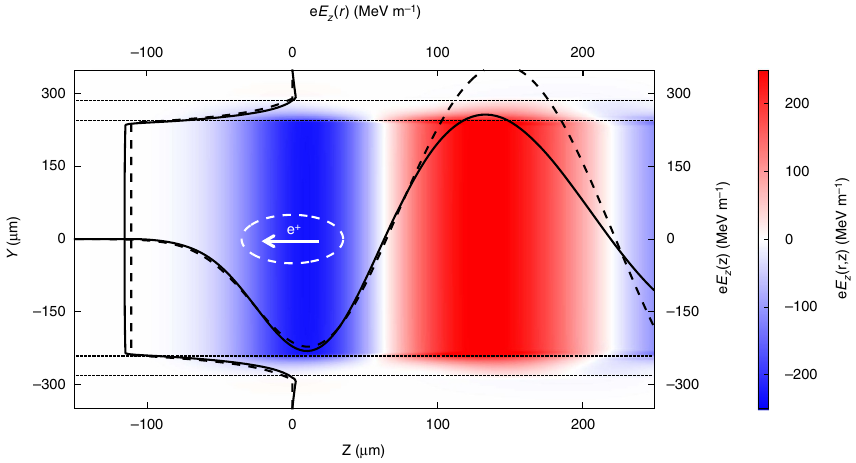}
    \caption{Longitudinal field in a hollow-channel plasma accelerator, driven by a positron beam moving toward the left. The plasma forms a cylindrical tube from radius 240 to \SI{290}{\micro\m} (dotted black line). The calculated/simulated field (dashed/solid lines) oscillates in the longitudinal direction ($Z$) but is uniform in the radial direction ($r$). There is no focusing force in the channel. From \textcite{Gessner2016a} (CC-BY 4.0).}
  \label{fig:hollow-channel-fields}
\end{figure}

The linear hollow-channel plasma wakefield accelerator is a concept combining the symmetrization of the plasma response to beams of opposite charge~\cite{Chiou1996,Chiou1998} and the absence of plasma constituents along the beam axis, resulting in no plasma focusing. In this scenario, an electron or positron beam propagates through a hollow tube of plasma. The beam drives a high-amplitude electromagnetic field inside the plasma channel. The electromagnetic fields are analogous to those excited by a beam transiting a structured waveguide~\cite{Chao1993} or a dielectric waveguide~\cite{OShea2016} and can be decomposed into components characterized by the azimuthal moment $m$.  For an on-axis beam driver, only the axisymmetric $m=0$ mode will be excited. In this case, the wakefield in the channel has desirable properties for accelerating a trailing bunch of electrons or positrons. Namely, the $m=0$ mode produces a radially uniform longitudinal field which extends to the plasma boundary (see Fig.~\ref{fig:hollow-channel-fields}), such that particles with different initial offsets with respect to the channel axis all gain energy at the same rate. The $E_r$ and $cB_\theta$ components are equal and opposite, and all other field components are zero, so there is no transverse force acting on the beam particles due to the electromagnetic wakefield. The absence of plasma in the channel implies that there are no transverse forces from on-axis plasma electrons or ions. If the trailing bunch also propagates on-axis, it will not experience a transverse force due to the wake. The longitudinal wakefield excited by a single particle of charge $q$ located at $\xi=0$ is given by
\begin{equation}
    \label{eq:hollow-channel-longitudinal-wakefield}
    W_z(\xi) = -q\frac{\mathcal{G}_0 k_p^2}{\pi \epsilon_0}\cos(\chi_0 k_p \xi)\Theta(-\xi),
\end{equation}
where $\mathcal{G}_0$ and $\chi_0$ are geometric factors from boundary conditions at the plasma inner radius and outer radius~\cite{Gessner2016b, Lindstrom2018}. If the inner radius is on the order of a few plasma skin depths, and the plasma wall is at least one skin depth thick, we find $\chi_0\approx0.5$ and $\mathcal{G}_0\approx0.1$.

If either the driver or trailing bunch is offset from the center of the channel, all higher azimuthal modes will be excited in addition to the $m=0$ mode. In particular, the beams will couple to the $m=1$ dipole mode which is always deflecting at short distances $|\xi| \ll 1/\chi k_p$. The transverse wakefield excited by a single particle of charge $q$ located at $\xi=0$ with a small transverse offset $\Delta x$ is 
\begin{equation}
    \label{eq:hollow-channel-transverse-wakefield}
    W_\perp(\xi) = -q\Delta x\frac{\mathcal{G}_1 k_p^3}{\pi \epsilon_0}\sin(\chi_1 k_p \xi)\Theta(-\xi),
\end{equation}
where $\chi_1$ is a wavelength modification factor for the $m=1$ mode. Wakefields from arbitrary beams can be obtained by convolving the single-particle wakefield with the current profile. Note that, at least for short-range wakes ($k_p\xi\ll1$), the transverse wakefield scales as $k_p^4$ whereas the longitudinal wakefield scales as $k_p^2$, implying that transverse wakefields are comparatively stronger at higher density. The main challenge of the hollow-channel plasma wakefield accelerator is therefore maintaining small transverse offsets; $\Delta x \ll k_p^{-1}$. In general, the transverse wakefield is deflecting away from the channel axis, and as the beam drifts toward the channel wall, it will excite a larger transverse wakefield, leading to stronger deflection. This is called the beam-breakup instability (BBU) and it has been studied extensively in the case of hollow-channel plasmas~\cite{Schroeder1999}.

Experimental studies of the linear hollow-channel plasma wakefield accelerator in the context of positron acceleration (see Sec.~\ref{sec:without-focusing}) have shown good agreement with these theoretical predictions and have highlighted the challenge associated with transverse wakefields~\cite{Lindstrom2018} and BBU instability. Strategies to mitigate BBU in the hollow-channel plasma wakefield often revert to the re-introduction of plasma focusing via on-axis plasma electrons from the channel walls [see e.g.~\textcite{Zhou2021} and Sec.~\ref{sec:beam-plasma-shaping}].

Going beyond the linear regime, nonlinear hollow-channel plasma wakefields in many ways resemble the blowout regime but without on-axis focusing, as described analytically by~\textcite{Golovanov2017}. These wakes can sustain strong, longitudinally uniform accelerating fields by means of beam loading, as well as by superimposing an accelerating positron bunch on a decelerating electron bunch~\cite{Zhou2022a}. Here, the linear response is of similar importance to the nonlinear contribution associated with the blowout effect in the resulting $E_z$ field. However, further work is needed to explore strategies ensuring stability and mitigating BBU.

In practice, the hollow-channel plasma accelerator is an idealized concept that assumes zero plasma density (i.e., a vacuum) on-axis with a sharp transition to full density at the inner radius $a$. Hollow channel plasmas that can be realized in the laboratory violate these assumptions (see Sec.~\ref{sec:positron-acceleration} and \ref{sec:experimental:facilities:slac}). For example, the channel may contain un-ionized, on-axis vapor~\cite{Kimura2011,Gessner2016a}, the channel may contain low-density plasma~\cite{Schroeder2013a}, or the plasma transition at the boundary is not sharp. In the latter case, there will be a region of the boundary where the wavenumber of the plasma matches the wavenumber of the hollow channel mode $k_{p(n<n_0)} = \chi_0 k_p.$~\cite{Shvets1996}. In this case, the electrostatic plasma mode is excited, which removes energy from the $m=0$ accelerating mode and further disrupts the plasma boundary. This issue is avoided if the transition region $\Delta a$ is narrow ($\Delta a < k_p^{-1}$).


\subsubsection{Wavebreaking and plasma temperature}
\label{sec:plasma_temperature}

Wavebreaking is sometimes used to refer to the breakdown of the fluid description for the plasma, which is associated with trajectory crossing, as discussed in Sec.~\ref{sec:nonlin_wake} and depicted in Fig.~\ref{fig:trajectory_crossing}. In this restrictive sense, the wavebreaking limit can be understood as the maximum amplitude of the plasma wave allowed in the fluid model. Wavebreaking is, however, more generally understood as the breaking of the regular structure of the wake, which can lead to trapping of some plasma electrons into the wake [see discussion in~\textcite{Lu2006c} on the definition of wavebreaking]. Longitudinal wavebreaking~\cite{Akhiezer1956} occurs when the longitudinal oscillation of plasma electrons becomes so large that these electrons can be injected into the accelerating portion of the wake---the wave breaks because a large population of plasma electrons no longer participates in maintaining the wave, and instead moves with it. Transverse wavebreaking can occur due to the curvature of the plasma-wake phase surface (i.e., points in space sharing the same phase), which increases with the distance behind the driver~\cite{Bulanov1997}. When the radius of curvature of the phase surface is comparable to the plasma-electron displacement, the transverse mixing of electron trajectories breaks the wave with the generation of fast electrons near the axis. For stronger drivers in the blowout regime, transverse wavebreaking or self-injection [as observed in LWFA \cite{Corde2013b}] directly results from plasma-sheath electrons circulating around the ion cavity and becoming trapped at the rear of the cavity. Wavebreaking and particle trapping can be detrimental by producing ``dark current" in the plasma wake, reducing the accelerating field and the acceleration efficiency~\cite{VafaeiNajafabadi2014,Manahan2016}. It can also be desired as a means to inject a new beam of electrons into the wake~\cite{Modena1995, Corde2013b}, however injection mechanisms that are better controlled than wavebreaking and provide higher quality beams are generally preferred (see Sec.~\ref{sec:internal-injection}).

In a 1D plane-wave model, the cold relativistic wavebreaking field is $E_\mathrm{wb}=\sqrt{2(\gamma_\phi-1)}E_0$~\cite{Akhiezer1956}, where $\gamma_\phi=(1-v_\phi^2/c^2)^{-1/2}$ is the Lorentz factor associated with the phase velocity $v_\phi$ of the plasma wave. For particle-beam drivers, $\gamma_\phi$ ($=\gamma_d$ of the driver for non-evolving wakefields) is very large and thus $E_\mathrm{wb}$ can greatly exceed the cold nonrelativistic wavebreaking field $E_0$. However, when accounting for the plasma temperature, some fast electrons in the tail of the thermal velocity distribution may be trapped in the wake, thus leading to wavebreaking. In the limit $v_\phi\rightarrow c$ relevant for PWFA [see~\textcite{Esarey2009} for a discussion of warm wavebreaking relevant to LWFA], the warm relativistic wavebreaking field is given by $E_\mathrm{wb} \sim E_0(m_ec^2/k_BT)^{1/4}$~\cite{Rosenzweig1988b, Katsouleas1988}, where $T$ is the initial electron plasma temperature and $k_B$ is the Boltzmann constant.

Beyond the role of plasma temperature on the wavebreaking field, the thermal velocity spread of plasma electrons can also smooth out the field structure of the wake, for example reducing the electric-field spike at the rear of the ion cavity in the blowout regime~\cite{Lotov2003,Jain2015b}. It was also suggested by \textcite{Cao2023} that the electron plasma temperature can impact beam-breakup instability and ion motion (discussed in Secs.~\ref{sec:bbu} and~\ref{sec:ion-motion}), while an ion temperature in the few hundred keV range can completely suppress ion motion~\cite{Gholizadeh2011}. 

Plasma temperature is particularly important in the context of positron acceleration (Sec.~\ref{sec:positron-acceleration}), where smoothing the field structure and broadening the plasma electron filament used for positron focusing can be essential to linearize the transverse wakefields and to improve the uniformity of longitudinal wakefields~\cite{Silva2021,Wang2021,Diederichs2023,Cao2024}. Plasma-temperature effects are also expected to play a key role in high-repetition-rate and high-average-power plasma accelerators due to the inherent inefficiencies in the plasma-wakefield process resulting in large amounts of energy being deposited in the plasma. This stored energy is expected to manifest in the form of hot plasma electrons and ions, modifying the wakefield properties from the cold case and thus presenting the challenge of generating consistent acceleration for each consecutive bunch (see Sec.~\ref{sec:long-term-evolution}).


\subsection{Driver propagation}
\label{sec:driver-propagation}

Understanding and optimizing the propagation of the drive beam in the plasma is key to achieving efficient transfer of energy from the driver to the plasma wake as well as high-quality acceleration of a trailing bunch (covered in Sec.~\ref{sec:evolution-trailing-bunch}). Indeed, although a driver can initially excite a plasma wakefield with ideal characteristics for accelerating a trailing bunch, driver evolution during propagation in the plasma can lead to a changing and eventually sub-optimal plasma wakefield. Stable driver propagation and, more specifically, a stable plasma wakefield are thus generally desired. This requires keeping the driver focused over long distances, achievable through self-guiding (see Sec.~\ref{sec:envelope-matching}), avoiding so-called ``head erosion" or loss of charge from the front of the driver (see Sec.~\ref{sec:head_erosion}), and avoiding transverse instabilities (Secs.~\ref{sec:hosing-instability} and \ref{sec:drive_instabilities}). Having mitigated all the above-mentioned effects, the aim is to minimize the energy left in the driver at the end of a plasma-accelerator stage---so-called \textit{driver-energy depletion} (Sec.~\ref{sec:drive_plasma_efficiency}).

\subsubsection{Guiding, matching and the envelope equation}
\label{sec:envelope-matching}

For high-energy particle beams propagating in vacuum, space-charge forces are generally negligible and the beam particles move ballistically. When the drive beam is focused in vacuum to a beam size $\sigma_0$, it can maintain this beam size (to within a factor $\sqrt{2}$) over a distance given by the beta function $\beta^*=\sigma_0^2/\varepsilon_g$ [with $\varepsilon_g$ the geometric emittance, see the definition in Eq.~(\ref{eq:twiss_emittance}) below], analogous to the Rayleigh length $z_R$ for a laser pulse. $\beta^*$ is typically much larger for particle beams (meter scale) than $z_R$ for a laser pulse (mm--cm scale), which is generally considered as an advantage of particle beam drivers (PWFA) over laser drivers (LWFA). 

In the plasma, beam particles experience the transverse focusing force of the plasma wakefield, which provides guiding to the particle beam driver over distances much larger than $\beta^*$. To understand this, consider the simplest case where beam particles are subjected to a linear focusing force, with $x$-component $F_x = -g x$ and $g$ the gradient of the focusing force. The $x$-component of the equation of motion for a relativistic beam particle can then be rewritten as
\begin{align}
    \frac{\dif^2x}{\dif z^2} = - k_\beta^2 x,
    \label{eq:single-particle}
\end{align}
with $k_\beta=\sqrt{g/\gamma mc^2}$ the betatron wavenumber, $\gamma$ the Lorentz factor of the beam particle (here assumed to be constant) and $m$ its mass. The solution is a simple harmonic oscillation around the propagation axis, with a period given by $\lambda_\beta=2\pi/k_\beta$. To go from the single-particle dynamics to the dynamics of the whole beam, it is useful to introduce the Twiss or Courant-Snyder parameters~\cite{Courant1958}:
\begin{align} 
    \label{eq:twiss_parameters}
    \alpha &= -\langle xx^\prime\rangle/\varepsilon_g, \\ 
    \beta &= \langle x^2\rangle/\varepsilon_g,\\ 
    \gamma_\mathrm{twiss} &= \langle x^\prime{^2}\rangle/\varepsilon_g = 
    (1+\alpha^2)/\beta.
\end{align}
The geometric emittance \cite{Floettmann2003}, given by
\begin{equation}
    \label{eq:twiss_emittance}
    \varepsilon_g = \sqrt{\langle x^2\rangle \langle x^\prime{^2}\rangle-\langle x x^\prime \rangle^2},
\end{equation}
is a key beam parameter, as it represents the area occupied by the beam in the \textit{trace space} ($x,x^\prime$) and quantifies the transverse quality of the beam; that is, its focusability---how small it can be focused for a given numerical aperture---or parallelism---how long it can maintain its size for a given beam size \cite{Humphries1990}. The normalized emittance, representing the area occupied by the beam in the \textit{normalized phase space} ($x,p_x/mc$), can be a better representation of the transverse quality when the beam accelerates (because it is a conserved quantity during acceleration). It is given by
\begin{align}
    \label{eq:normalized_emittance}
    \varepsilon_n &= \frac{1}{mc}\sqrt{\langle x^2\rangle \langle p_x^2\rangle-\langle x p_x \rangle^2}.
\end{align}
For a mono-energetic beam, $\varepsilon_n = \varepsilon_gp/mc$, but the relationship between $\varepsilon_n$ and $\varepsilon_g$ for finite energy spread can be more complicated and depends on the correlation between $\gamma$ and the transverse trace/phase space~\cite{Floettmann2003, Antici2012, Migliorati2013, Li2019}.

From the single-particle equation [Eq.~(\ref{eq:single-particle})], the differential equations for the first two Twiss parameters are
\begin{align}
    \frac{\dif \alpha}{\dif z} = -\frac{1+\alpha^2}{\beta}+k_\beta^2\beta, \qquad 
    \frac{\dif \beta}{\dif z} = -2\alpha,
\end{align}
and the second-order equation for the evolution of the beam size $\sigma_x=\sqrt{\langle x^2\rangle}=\sqrt{\beta\varepsilon_g}$ is
\begin{align}
    \frac{\dif^2\sigma_x}{\dif z^2} = -k_\beta^2\sigma_x + \frac{\varepsilon_g^2}{\sigma_x^3},
    \label{eq:enveloppe_equation}
\end{align}
the so-called \textit{envelope equation}. In general, the solution to Eq.~(\ref{eq:enveloppe_equation}) is a beam-envelope oscillation between $\sigma_\mathrm{max}$ and $\sigma_\mathrm{min}$ at twice the betatron frequency (i.e., with a period of $\lambda_{\beta}/2=\pi/k_{\beta}$). Assuming that at the entrance of the plasma, $\alpha=0$ and $k_{\beta}^2\sigma_0\geq \varepsilon_g^2/\sigma_x^3$ and thus $\sigma_\mathrm{max}=\sigma_0$, the beam size can oscillate but always remains smaller than $\sigma_0$, and the beam can thus be effectively guided inside the plasma over very long distances, much longer than the initial beta function $\beta^*=\sigma_0^2/\varepsilon_g$, as long as the plasma provides a focusing force to the beam. 

\begin{figure}[t]
    \centering
    \includegraphics[width=\linewidth]{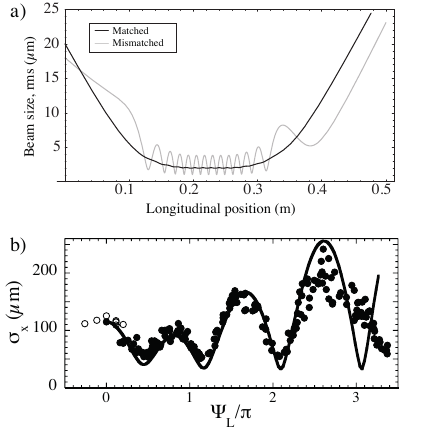}
    \caption{(a) A beam traveling in vacuum is focused at the start of a plasma cell with plasma-density ramps. If the beam is mismatched (gray line) there is an oscillation of the beam size, whereas if the beam is matched (black line) the oscillation is minimal. (b) Experimental measurement (black circles) of beam-envelope oscillations for a \SI{28.5}{GeV} mismatched beam in a plasma. The horizontal beam size varies as a function of the phase advance $\Psi_L = \int_0^L \dif s/\beta(s)\sim n_0^{1/2}L$, where the plasma density $n_0$ was scanned from 0 to \SI{2e14}{cm^{-3}} and the plasma length  stayed constant at $L=\SI{1.4}{m}$. The envelope-equation prediction (solid line) and the beam size for vacuum propagation (empty circles; all at $\Psi_L=0$) are overlaid. Adapted from \textcite{Marsh2005} and \textcite{Clayton2002}.}
    \label{fig:envelope_oscillation}
\end{figure}

The beam is said to be ``matched'' to the plasma focusing force if there is no beam-envelope oscillation [see Fig.~\ref{fig:envelope_oscillation}(a)], which corresponds to $\sigma_x(z)=\mathrm{constant}$ when $\gamma$ and $k_{\beta}$ are independent of $z$. This matching condition is satisfied if $\dif\sigma_x/\dif z=0$ and $k_{\beta}^2\sigma_x=\varepsilon_g^2/\sigma_x^3$, that is for~\cite{Barov1994, Assmann1998, Marsh2005, Mehrling2012}
\begin{align}
    \label{eq:matched_beta_function}
    \alpha_\mathrm{matched}=0, \qquad \beta_\mathrm{matched}=1/k_{\beta}.
\end{align}
Experimentally, transverse beam dynamics were first characterized in electron-driven PWFA in the blowout regime using a single drive bunch, and the beam was observed to undergo multiple envelope oscillations when mismatched~\cite{Clayton2002}; see Fig.~\ref{fig:envelope_oscillation}(b). By properly choosing the beam and plasma parameters, matching was achieved and the beam propagated through the plasma over a distance of more than 12 beta functions without envelope oscillations~\cite{Muggli2004}.

The above discussion illustrates the principle of how a beam can be guided in a plasma and how it can be matched to the plasma focusing force to avoid beam-envelope oscillations. In a realistic scenario, the driver loses energy by exciting the plasma wakefield, and thus $\gamma$ and $k_\beta$ vary during propagation. At the entrance and exit of the plasma, the driver can also propagate in a plasma-density ramp, leading to $z$-dependent $k_\beta$. Moreover, the plasma focusing force and $k_\beta$ can be different for different slices of the drive beam, thus being $\xi$-dependent, and finally, the focusing force can be nonlinear in the transverse coordinates. Note that Eq.~(\ref{eq:single-particle}) assumes a linear focusing force. In the blowout regime (see Sec.~\ref{sec:nonlin_wake}), most of the drive particles are within the blowout ion cavity and thus experience a $\xi$-independent linear focusing force with $k_\beta=k_p/\sqrt{2\gamma}$. An important refinement is to add the term $-\frac{1}{\gamma}\frac{\dif \gamma}{\dif z}\frac{\dif x}{\dif z}$ to the RHS of Eq.~(\ref{eq:single-particle}) [and $-\frac{1}{\gamma}\frac{\dif \gamma}{\dif z}\frac{\dif \sigma_x}{\dif z}$ to the RHS of Eq.~(\ref{eq:enveloppe_equation})] to account for the evolution in $\gamma$ during acceleration~\cite{Krall1995, Marsh2005}. The presence of plasma-density ramps must also be accounted for when matching the beam to the plateau of the plasma density profile~\cite{Marsh2005, Floettmann2014, Dornmair2015, Xu2016, Ariniello2019}. Furthermore, when plasma electrons are blown out, it takes a finite distance in $\xi$ for the blowout cavity to form. Thus, the particles at the head of the drive bunch experience a $\xi$-dependent and transversely nonlinear focusing force, making the beam dynamics at the head more complicated and resulting in so-called \textit{head erosion} (see Sec.~\ref{sec:head_erosion}).

In addition to the self-guiding provided by the wakefield itself, it is possible to employ external guiding. In LWFAs, this is done via a transverse plasma-density gradient (quadratic increase with radius) to guide the laser driver as in an optical fiber~\cite{Butler2002,Leemans2006}, whereas in PWFAs the opposite density gradient is required: the density must be highest on axis~\cite{Muggli2001,Adli2016a} since the local focusing force scales with density. An external magnetic field can also be applied, either via quadrupole magnets~\cite{Qian2025} or active plasma lensing~\cite{vanTilborg2015}; the latter involves conducting a longitudinal discharge current through the plasma during the acceleration process.

\subsubsection{Head erosion}
\label{sec:head_erosion}

\begin{figure}[t]
    \centering
    \includegraphics[width=\linewidth]{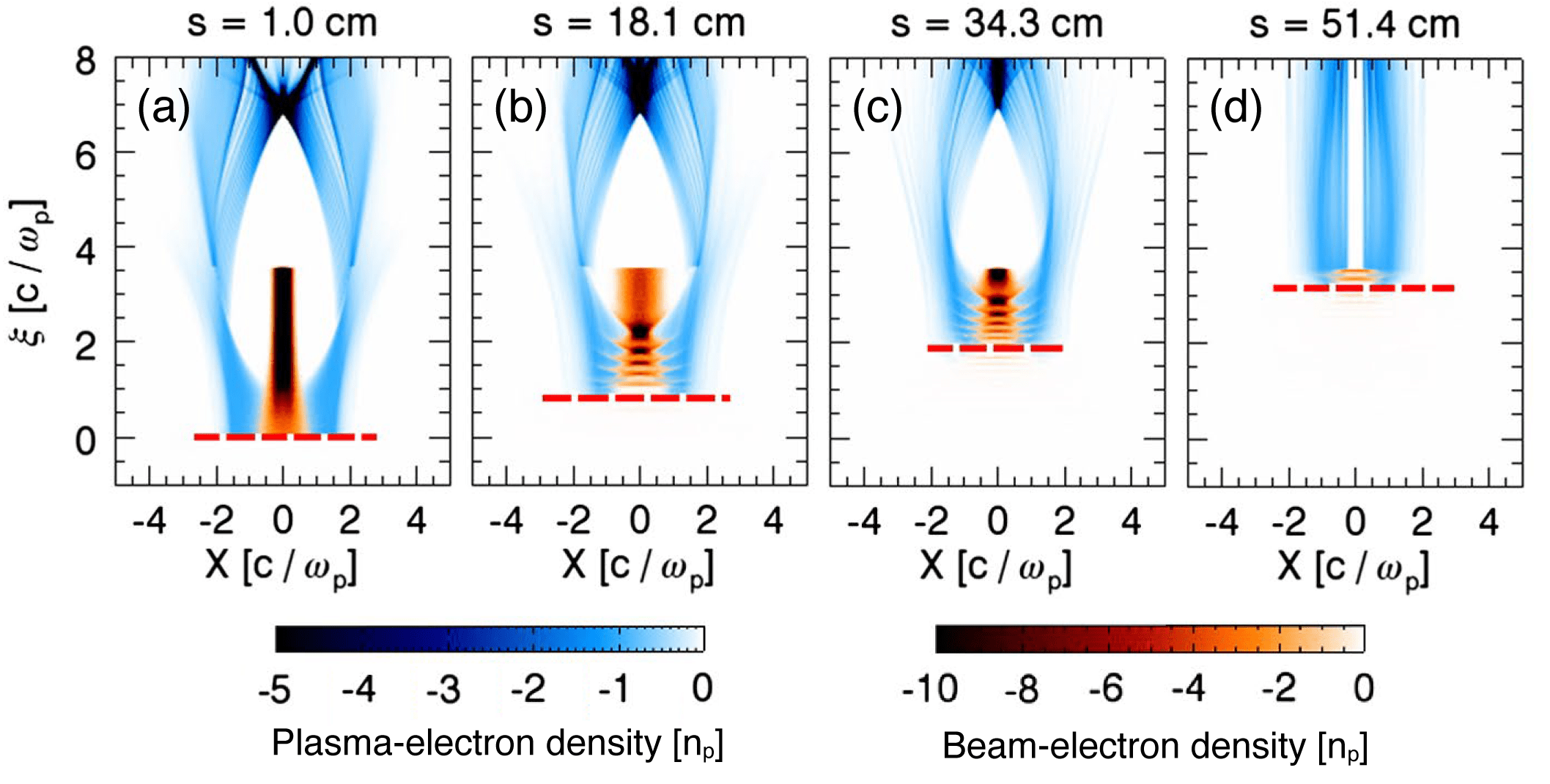}
    \caption{Simulated head erosion of a drive beam (orange color map) in a beam-ionized plasma (blue color map), showing that the leading particles (bottom) are gradually lost during propagation due to insufficient focusing (a$\to$d), which shifts the ionization front (red line) backwards. Here, the beam's initial beta function is $\beta^*=\SI{3}{cm}$. Adapted from~\textcite{An2013} (CC-BY 3.0).}
    \label{fig:head-erosion}
\end{figure}

The physics of head erosion can be understood as follows: while the particles in the bulk of the drive beam are guided by the plasma focusing force, the particles at the very head of the beam do not experience such a force because the plasma wakefield has not yet been established. As a result, the beam head expands due to its finite emittance, and is effectively lost when propagating for distances larger than $\beta^*$. The loss of particles at the head of the beam shifts the start of the wakefield backwards, causing the next-to-leading particles to propagate ahead of the guiding fields of the wake. This leads to a continuous erosion of the beam, as illustrated in Fig.~\ref{fig:head-erosion}.

Emittance-driven head erosion~\cite{Buchanan1987, Krall1989} was considered as an important limitation for PWFA in early developments~\cite{Barov1994}, in particular because it sets stringent constraints on the drive-beam emittance~\cite{Rosenzweig1998}, it reduces the wakefield amplitude and gives rise to a phase slippage of the plasma wakefield, moving it backward in the beam frame~\cite{Barov2000}. The effect of head erosion on the beam shape was experimentally evidenced through slice-resolved beam-size measurements downstream of the plasma, revealing a constant size for the rear half of the beam and a trumpet shape for the front half of the beam, with a size increasing towards the head of the bunch~\cite{Barov1998}. 

Early experimental results from Argonne (Sec.~\ref{sec:experimental:facilities:argonne}) were based on discharge plasmas. Simulation studies showed that in such pre-ionized plasmas, head erosion is important during an initial relaxation distance, after which a quasi-equilibrium is established for most slices (even near the beam head) and erosion occurs at a much slower rate~\cite{Krall1989, Barov1994, Blumenfeld2009}. The quasi-equilibrium forms because as particles are lost from the head of the bunch (where the focusing is weak or absent), the new leading particles (which initially saw strong focusing) evolve adiabatically from a matched state with strong focusing inside the blowout to a regime with weak focusing at the front of the wake. In this process, the $\beta$ function at the head increases with respect to the initial vacuum $\beta^*$, thus strongly reducing the radial expansion of the head and the accompanying erosion rate. 

The situation is different when the plasma is produced by field ionization~\cite{OConnell2006} from the space-charge field of the beam. In this beam-ionized plasma case, head erosion is exacerbated by the sudden transition from plasma to no plasma as a beam segment crosses the ionization front, which does not allow it to adiabatically evolve to an increased beta function. Importantly, head erosion in beam-ionized plasma was interpreted as limiting the maximum energy gain in the energy doubling experiment of \textcite{Blumenfeld2007} (Fig.~\ref{fig:large-energy-gain}), and has motivated the use of pre-ionized plasma~\cite{Hogan2010, An2013, Green2014}.

Head erosion in beam-ionized plasma can be modeled when the beam beta function is set to match the body of the beam inside the blowout~\cite{Zhou2007, Blumenfeld2009}. The erosion rate is estimated as $V_\mathrm{erosion}=\Delta \xi/\Delta z$ by determining the length $\Delta \xi$ of the head segment being lost, and the propagation distance $\Delta z$ over which it is lost. Combined with simulation results to determine numerical coefficients, the model leads to the following engineering formula~\cite{Blumenfeld2009}:
\begin{align}
    V_\mathrm{erosion} [\SI{}{\micro m/m}] = \SI{3.7e4}{} \frac{\varepsilon_n [\SI{}{mm.mrad}] \epsilon_i^{1.73} [\SI{}{eV}]}{\gamma I^{3/2}[\SI{}{kA}]},
    \label{eq:erosion}
\end{align}
with $\epsilon_i$ the ionization energy of the atoms or molecules to be ionized, $\gamma$ the Lorentz factor of the beam and $I$ the beam current at the ionization front. This formula highlights the need for high-peak-current and low-emittance drivers to be able to drive a beam-ionized plasma wakefield accelerator with negligible head erosion. While such drivers were not necessarily available during the previous decades, FACET-II (Sec.~\ref{sec:experimental:facilities:slac}) now offers extremely high peak-current beams~\cite{Emma2025}, which may mitigate head-erosion effects associated with beam-ionized plasma-wakefield acceleration. 

Head-erosion effects can also be mitigated with the use of mismatched drive beams, for which the model above and Eq.~(\ref{eq:erosion}) no longer apply. Indeed, a mismatched beam can have a much larger $\beta$ function than a matched case, allowing the head to expand at a smaller rate. Head erosion becomes negligible if the $\beta$ function is set to be of the same order or larger than the acceleration length. The potential of mismatched beams to mitigate head erosion was highlighted in an experiment~\cite{Corde2016} using a high-ionization-potential gas (argon; $\epsilon_i = \SI{15.8}{eV}$). For that study, Eq.~(\ref{eq:erosion}) naively (if assuming matched beams) predicts a high erosion rate and therefore a short interaction length, but the experiment demonstrated very high energy gains of \SI{27}{GeV} (from \SI{20}{GeV} to \SI{47}{GeV}) over a distance of \SI{\sim20}{cm}.

\subsubsection{Hosing instability}
\label{sec:hosing-instability}

The extremely large accelerating and focusing gradients inherent in PWFA allow for rapid acceleration of particle beams while maintaining \SI{}{\micro\m}-scale transverse beam sizes over long propagation distances. However, the small aperture (i.e., wake radius) and strong transverse focusing also makes plasma accelerators more susceptible to transverse beam offsets or asymmetries that would go unnoticed in large-aperture rf-based accelerators. This is because transverse instabilities, driven by transverse wakefields, scale rapidly with the inverse of the aperture $a$, typically as $1/a^4$ (as discussed in Secs.~\ref{sec:hc_linear} and \ref{sec:bbu}). Transverse asymmetries, in the form of beam-centroid offsets and slice-momentum deviations, arise through collective effects such as coherent synchrotron radiation or dispersion in misaligned magnets. While these can be mitigated~\cite{Guetg2015}, they can never be completely removed from the beam.

\begin{figure}[t]
    \centering\includegraphics[width=\linewidth]{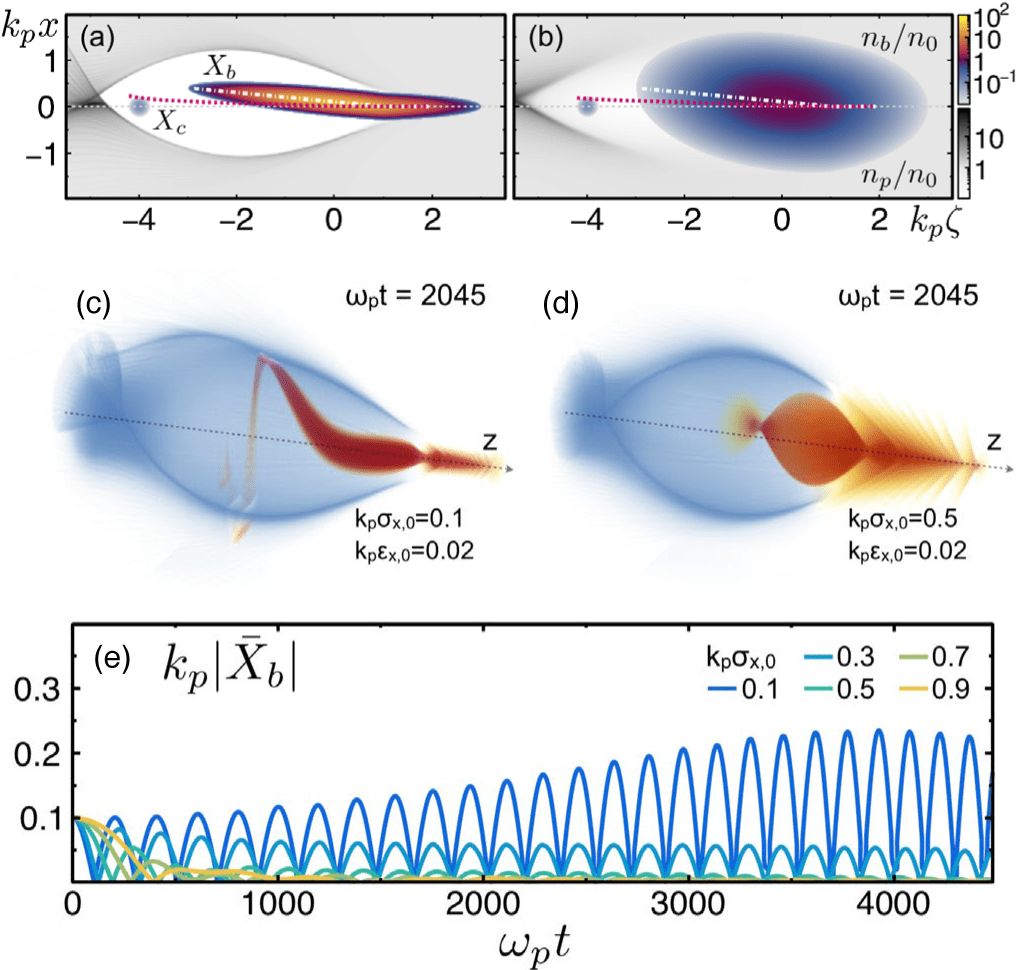}
    \caption{(a--b) Schematic showing the beam centroid offset $X_b$ acting on the plasma wake, generating an ion-channel offset $X_c$, for (a) thin and (b) wide drive beams. The hosing instability, often severe for (c) thin drive beams, can be suppressed by using (d) beam sizes comparable to the plasma wake. (e) This is demonstrated by comparing how the centroid offset grows with time $t$ for five initial transverse drive-beam sizes. Adapted from \textcite{MartinezdelaOssa2018}.}
    \label{fig:hosing_beam_size}
\end{figure}

An underlying formalism to describe the effect of these offsets during the plasma-acceleration process was originally developed by \textcite{Whittum1991}. This model describes a blowout regime for a long, adiabatically formed ion channel with a channel radius near the charge neutralization radius, $r_c = \sigma_r\sqrt{2n_b/n_0} = \sqrt{2\Lambda}/k_p$, where $\Lambda$ is the normalized peak current [Eq.~(\ref{eq:normalized-current})]. As such, the sheath-layer electrons are assumed to be at rest (i.e., at the nonrelativistic limit) resulting in a linear plasma-sheath response, whereby the sheath does not generate magnetic fields or feel the effect of them. The beam and channel centroids are described by the relations
\begin{align}
    \frac{\partial^2 X_b}{\partial t^2} + \omega_{\beta}^2(X_b - X_c) = 0 , \label{eq:hosing_beam} \\
    \frac{\partial^2 X_c}{\partial \xi^2} + \frac{k_p^2}{2}(X_c - X_b) = 0 , \label{eq:hosing_channel}
\end{align}
where $\omega_\beta=k_\beta c$ and $k_\beta = k_p/\sqrt{2\gamma}$ are the betatron frequency and wavenumber, respectively (see Sec.~\ref{sec:envelope-matching}). Any beam-centroid deviation, $X_b$, will couple to the ion channel, driving a deviation in the channel centroid, $X_c$, along the length of the beam [see Fig.~\ref{fig:hosing_beam_size}(a)]. As the beam undergoes betatron oscillations throughout the propagation, this offset will act as the ``seed" for a wiggling of the rear of the bunch, more commonly known as ``hosing". Due to the electrostatic coupling with the ion channel, each betatron oscillation of the beam will result in a movement of the ion-channel centroid, which in turn feeds back on the temporal evolution of the beam centroid. This spatio-temporal coupling will therefore result in a resonant feedback loop, leading to an ever-increasing hosing effect. If left unchecked this resonant oscillation may exponentially increase, leading to beam breakup and significant loss of beam quality. This is the so-called \textit{hosing instability} (also referred to as the \textit{hose instability} or \textit{electron-hose instability}).

\begin{figure}[t]
    \centering\includegraphics[width=\linewidth]{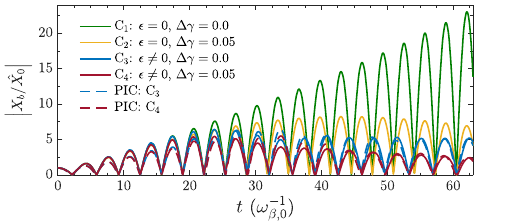}
    \caption{Hosing is largely mitigated for realistic beams in a decelerating field, for which both the energy $\epsilon$ evolves and the energy spread $\Delta\gamma$ is non-zero. From \textcite{Mehrling2017}.}
    \label{fig:hosing_parameters}
\end{figure}

Hosing has been observed experimentally \cite{Nechaeva2023} with long proton bunches in self-modulated wakefields (Sec.~\ref{sec:self-modulated-wakefields}), but so far not with electron drivers. Motivated by this lack of experimental observation, the above formalism [Eqs.~(\ref{eq:hosing_beam}) and (\ref{eq:hosing_channel})] was expanded to better represent the short bunches typically used in electron-driven PWFA experiments \cite{Huang2007} (i.e., $\omega_{\beta}t \gg k_p \xi$). Under the new conditions of a non-adiabatically-formed channel and relativistic mass corrections ($k_p\sigma_z<1$, $\Lambda \gtrsim 1$, and $k_pr_c \gtrsim 1$), the blowout radius may change along the beam and the motion in the electron sheath can become relativistic. The channel-centroid description in Eq.~(\ref{eq:hosing_channel}) becomes
\begin{equation}
    \label{eq:hosing_channel_modified}
    \frac{\partial^2 X_c}{\partial \xi^2} + \frac{k_p^2 c_r(\xi)c_{\psi}(\xi)}{2}(X_c - X_b) = 0,
\end{equation}
where the coefficients $c_r(\xi)$ and $c_\psi(\xi)$ are defined in~\textcite{Huang2007} and account for a relativistic electron sheath and for the $\xi$ dependence of the blowout radius $r_b(\xi)$ and the beam current $\lambda(\xi)$. The result provides a generalized hosing theory based on a perturbation method to the zeroth-order trajectory for the ion-channel/electron-sheath boundary, slowing the exponential growth of hosing. Simulations demonstrate an order-of-magnitude decrease in hosing growth ultimately stemming from a decrease in bunch length and from relativistic effects with $c_rc_\psi=\mathcal{O}(0.1)$.

The next evolutionary step in the underlying hosing model was made with the introduction of energy effects~\cite{Mehrling2017}, specifically the energy evolution of the drive beam as it loses energy to the wake and any correlated or uncorrelated energy spread the drive beam may have. In the model, the Lorentz factor (representing the energy) of an electron is given by $\gamma(t) = \overline{\gamma_0} + \dot{\gamma}t + \delta \gamma$, where $\overline{\gamma_0} = \overline{\gamma_0}(\xi)$ represents the initial mean energy of each longitudinal slice (i.e., the correlated energy spread along the bunch length), $\dot{\gamma}(\xi) = -eE_z(\xi)/mc$, which accounts for the slice-dependent change in energy during propagation, and $\delta \gamma = \gamma - \overline{\gamma}$ represents the uncorrelated energy spread from the mean slice energy. Due to these energy effects, electrons with different energies acquire a different phase advance, resulting in phase mixing of betatron oscillations within a slice and in the decoupling of different slices, which damps the hosing growth, as shown in Fig.~\ref{fig:hosing_parameters}.
The beam-centroid equation described in Eq.~(\ref{eq:hosing_beam}) then becomes~\cite{Mehrling2017}
\begin{equation}
\begin{aligned}
    \label{eq:hosing_beam_energydep}
    \frac{\partial^2 X_b}{\partial t^2} & + \frac{\overline{\omega_{\beta}}^2}{\overline{\omega_{\beta,0}}}\left(\epsilon + \kappa_1 \Delta \gamma^2\right)\frac{\partial X_b}{\partial t} \\
    & + \overline{\omega_{\beta}}^2\left(1 + \kappa_2 \Delta \gamma^2\right)(X_b - X_c) = 0, 
\end{aligned}
\end{equation}
where $\overline{\omega_{\beta,0}} = \omega_p / \sqrt{2 \overline{\gamma_0}}$, $\overline{\omega_{\beta}} = \overline{\omega_{\beta,0}} / \sqrt{1 + \epsilon \overline{\omega_{\beta,0}} t}$, $\kappa_1 = (\overline{\omega_{\beta}}/\overline{\omega_{\beta,0}} - (\overline{\omega_{\beta}}/\overline{\omega_{\beta,0}})^2)/\epsilon$, $\kappa_2 = (\overline{\omega_{\beta}}/\overline{\omega_{\beta,0}})^4/2 - (\overline{\omega_{\beta}}/\overline{\omega_{\beta,0}})^3/4$, $\Delta \gamma$ is the uncorrelated relative energy spread and $\epsilon=\dot{\gamma}/(\overline{\gamma_0}\:\overline{\omega_{\beta,0}})$ is the relative energy change per betatron cycle. Equation~(\ref{eq:hosing_beam_energydep}) holds for any beam in a blowout wakefield. Using this formalism, the growth rate can be calculated for variable drive-beam correlated energy spreads and energy losses, with the results demonstrating a truncation in the exponential growth of hosing for non-zero energy losses and correlated spreads; the latter analogous to BNS damping, named after \textcite*{Balakin1983}. A finite uncorrelated energy spread can also damp beam-centroid oscillations via decoherence of betatron oscillations of individual beam electrons within a slice, thus suppressing hosing.

In addition, tailoring of other experimental parameters can also reduce the growth rate of hosing. For example, a precise tapering of the plasma upramp length and profile is predicted to reduce beam-centroid oscillations and thus the hosing seed before it can significantly resonate with the plasma-electron sheath \cite{Mehrling2017}. Plasma-density steps can also be introduced to suppress the resonance \cite{Moreira2023}. Another strategy is to increase the transverse size of the driver such that the transverse wakefields are modified from uniform to nonuniform along the length of the bunch \cite{MartinezdelaOssa2018}. The nonuniform transverse fields and the head-to-tail variation of the betatron frequency are predicted to suppress hosing strongly [see Fig.~\ref{fig:hosing_beam_size}(c--e)], similarly to hosing saturation in the linear regime~\cite{Lehe2017}. Going from narrow drive beams, susceptible to hosing, to the large beams, where hosing is mitigated, can affect the acceleration performance but, for the example expounded by \textcite{MartinezdelaOssa2018}, the energy efficiency is only reduced by 10--15$\%$.

Finally, hosing is not limited to just the drive beam: the same effect can occur also for the trailing bunch, for which it is sometimes known as the \textit{beam-breakup instability}. This topic is discussed further in Sec.~\ref{sec:bbu}.

\subsubsection{Other beam--plasma instabilities}
\label{sec:drive_instabilities}

Beyond hosing, other instabilities can also develop in the interaction between particle beams and plasmas, both transversely and longitudinally within the bunch. Similarly to hosing, these instabilities can compromise the stability of the drive bunch~\cite{Su1987} thereby limiting the useful plasma accelerator length and thus the driver-energy depletion efficiency (Sec.~\ref{sec:drive_plasma_efficiency}). 

\begin{figure}[t]
    \centering\includegraphics[width=\linewidth]{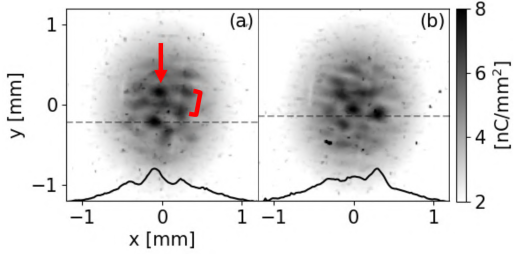}
    \caption{Two single-event transverse profiles of a \SI{400}{GeV} proton bunch containing multiple filaments (e.g., red arrow) separated by a distance (red bracket) consistent with the oblique instability in a \SI{10}{m} long plasma of density \SI{9.4e14}{\per\cubic\cm}. Adapted from \textcite{Verra2024} (CC-BY 4.0).}
    \label{fig:cfi}
\end{figure}

For ultrarelativistic beams, the two most important unstable transverse modes in beam--plasma systems are the \textit{current filamentation instability} (CFI), which is purely transverse and electromagnetic, and the \textit{oblique two-stream instability} (OTSI), which is mostly electrostatic and with a wavevector oriented obliquely with respect to the beam velocity~\cite{Bret2005, Bret2010}. The latter exhibits spatiotemporal dynamics (i.e., growing both along the bunch and during propagation) for realistic experimental conditions~\cite{SanMiguelClaveria2022}, departing substantially from the purely temporal evolution of infinite systems (only growing with propagation, but invariant along the bunch). These instabilities break up the beam into beamlets (see Fig.~\ref{fig:cfi}) and typically occur for large beams ($k_p\sigma_r\gg1$), and experiments have shown that they do not occur when $k_p\sigma_r\lesssim1$~\cite{Allen2012,Verra2024}. As a result, they are easily avoided and do not pose serious limitations to PWFA.

For narrow beams relevant to PWFA, an unstable longitudinal mode can occur, called the \textit{self-modulation instability} (SMI) \cite{Lotov1998,Kumar2010}; see Sec.~\ref{sec:self-modulated-wakefields}. If the driver is longer than a plasma wavelength ($k_p\sigma_z\gg1$), parts of the bunch are focused while others are defocused, weakly at first but with exponentially increasing strength as the focused parts get more dense and the defocused parts disappear. This leaves a self-modulated train of bunches spaced by the plasma wavelength. While this instability is avoided by using short bunches ($k_p\sigma_z\lesssim1$), it can also be useful in forming a drive bunch train that resonantly drives a wakefield.

A complete electromagnetic theory of ultrarelativistic beam-plasma instabilities based on the quasistatic formalism used in PWFA~\cite{SanMiguelClaveria2026a, SanMiguelClaveria2026b} provides a unified description of these unstable modes, including CFI, OTSI and SMI, and highlights the fundamental relationship between the OTSI for large beams $k_p\sigma_r\gg1$ and the SMI for small beams $k_p\sigma_r\ll1$.

\subsubsection{Driver-energy depletion efficiency}
\label{sec:drive_plasma_efficiency}

A plasma wakefield can be excited until the drive beam is fully depleted of its energy, which is a requirement for a high-efficiency accelerator. It is also critical to the concept of plasma beam dumps~\cite{Wu2010, Bonatto2015, Hanahoe2017, Jakobsson2019} that have been proposed as compact alternatives for the disposal of high-energy beams, with the advantage of negligible radioactivation compared to conventional beam dumps based on electromagnetic cascades in solids. Importantly, driver-energy depletion should not be considered as a limitation but as a goal when aiming for efficient acceleration. The depletion length and driver-to-plasma efficiency can be determined by the condition that some particles of the drive beam are decelerated to nonrelativistic energy. When this occurs, the energy-depleted particles recede with respect to the wakefield and are then re-accelerated when they reach the accelerating phase---changes that modify the wakefield. This modification is detrimental for both a plasma beam dump~\cite{Jakobsson2019} and high-efficiency PWFAs~\cite{Hue2023, Pena2024}, and thus justifies that acceleration should be stopped before any driver particles have decelerated to nonrelativistic energy. 

Assuming an initially monoenergetic driver and a stable wakefield, the depletion length and driver-to-plasma efficiency at depletion is~\cite{Hue2023}
\begin{align}
    L_d &= \frac{\gamma_0 mc^2}{eE_\mathrm{dec}},\\
    \eta_\mathrm{d \to p} &= \frac{\gamma_0-\langle \gamma \rangle_f}{\gamma_0} = \frac{\langle E_z \rangle_d}{E_\mathrm{dec}},
\end{align}
where $\gamma_0$ is the initial Lorentz factor of the beam, $\langle \gamma\rangle_f$ is its final average Lorentz factor at depletion, $E_\mathrm{dec}$ is the peak decelerating field experienced by the driver, and $\langle E_z \rangle_d$ is the decelerating field averaged over the particles of the driver. In particular, improving the uniformity of the decelerating field \cite{Roussel2020} can increase the driver-to-plasma efficiency at depletion (see Sec.~\ref{sec:high-transformer-ratio}). In this case, the average field $\langle E_z \rangle_d$ is closer to the peak field $E_\mathrm{dec}$. This can be done by shaping the current profile of the driver, in which case the driver-to-plasma efficiency can exceed 90\%~\cite{Lotov2005,Su2023}. Yet non-shaped Gaussian drive beams can already achieve driver-to-plasma efficiencies exceeding 75\% at depletion according to simulations~\cite{Adli2013, Hue2023}. Most of the challenge to reach high driver-to-plasma efficiencies is thus to demonstrate that the PWFA can reach the depletion length $L_d$ while mitigating other limiting phenomena such as head erosion (Sec.~\ref{sec:head_erosion}) and hosing (Sec.~\ref{sec:hosing-instability}), and ensuring beam-quality preservation for the trailing bunch over this propagation distance (Sec.~\ref{sec:evolution-trailing-bunch}).

\begin{figure}[t]
    \centering
    \includegraphics[width=\linewidth]{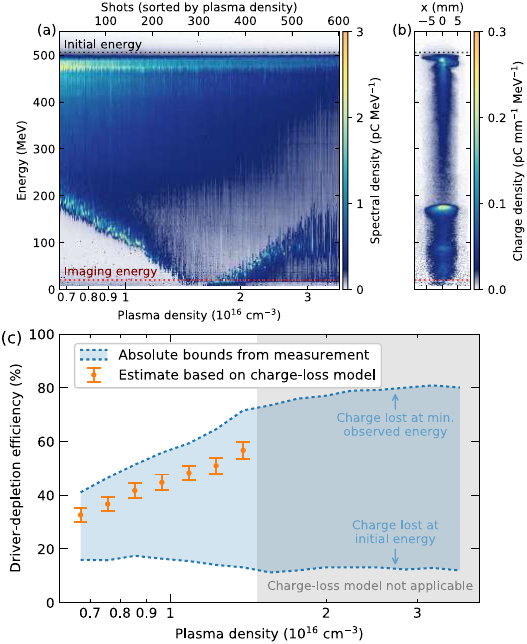}
    \caption{Driver depletion and re-acceleration: (a) driver energy spectra  are shown for a scan of plasma density and (b) a single-shot example with re-accelerated electrons (i.e., a second population below \SI{180}{MeV}). (c) The driver-depletion efficiency was found, using a model to account for charge loss during transport, to reach $(57\pm3)$\% while avoiding re-acceleration. Adapted from \textcite{Pena2024} (CC-BY 4.0).}
  \label{fig:depletion}
\end{figure}

Experimentally, energy depletion has been demonstrated using a single drive beam (see Fig.~\ref{fig:depletion}), with driver-energy depletion efficiency exceeding 50\%~\cite{Pena2024, Zhang2024}. \textcite{Pena2024} adjusted the depletion length to make it equal to the plasma length by varying the plasma density. As can be seen in Fig.~\ref{fig:depletion}(a), when the plasma density is increased, the maximum energy loss increases until the point where driver electrons reach a negligible energy. Increasing the plasma density further, re-accelerated driver electrons are observed with increasing energy, and the depletion length, which could be determined directly from the plasma light emission \cite{Oz2004,Boulton2026}, was observed to decrease with the plasma density. The optimal plasma density \SI{\sim1.5e16}{cm^{-3}} corresponds to the highest density before re-acceleration is observed [see Fig.~\ref{fig:depletion}(c)], i.e.~when depletion occurs at the very end of the plasma accelerator. Reaching higher depletion efficiency in experiments will require more careful shaping of the longitudinal and transverse beam profile.


\subsection{Evolution of the trailing bunch}
\label{sec:evolution-trailing-bunch}

The goal of a plasma accelerator is typically to add energy to a particle bunch, and to do so compactly. However, energy gain and gradient are not the only requirements; high beam quality and energy efficiency may also be important, although which combination of these properties to optimize for will depend on the application (Sec.~\ref{sec:applications}). For instance, a key metric in an FEL is the charge density in 6D phase space, the \textit{6D brightness},
\begin{equation}
    \label{eq:brightness}
    \mathcal{B}_{\mathrm{6D}} = \frac{Q}{\varepsilon_{nx}\varepsilon_{ny}\varepsilon_{nz}},
\end{equation}
where $Q$ is the bunch charge, $\varepsilon_{nz}$ is the normalized emittance in longitudinal phase space $(z,E)$, while $\varepsilon_{nx}$ and $\varepsilon_{ny}$ are the emittances in the transverse phase spaces $(x,p_x)$ and $(y,p_y)$, respectively. Another key metric, relevant to colliders, is the collision rate per energy spent, or more accurately the \textit{luminosity per wall-plug power}
\begin{equation}
\label{eq:luminosity-per-power}
    \frac{\mathcal{L}}{P_\mathrm{wp}} = \frac{1}{8\pi m_e e c^2}\frac{\eta_\mathrm{wp \to t} Q}{\sqrt{\beta^*_x \varepsilon_{nx} \beta^*_y \varepsilon_{ny}}},
\end{equation}
where $\eta_\mathrm{wp \to t}$ is the wall-plug-to-trailing-bunch energy-transfer efficiency---itself the product of the efficiencies from wall-plug to driver $(\eta_\mathrm{wp \to d})$, driver to plasma $(\eta_\mathrm{d \to p})$, and plasma to trailing bunch $(\eta_\mathrm{p \to t})$---and $\beta^*_{x/y}$ are the collision-point beta functions, which again set limits on $\varepsilon_{nz}$ through the bunch length and energy spread~\cite{Raimondi2001}. Delivering a high 6D brightness or luminosity per power at the end of the accelerator therefore requires understanding how the beam evolves in both the longitudinal (Sec.~\ref{sec:evolution-longitudinal}) and transverse (Sec.~\ref{sec:evolution-transverse}) phase spaces during acceleration. Finally, the spin polarization of the bunch can also be of relevance (Sec.~\ref{sec:spin-polarization}), especially for particle physics.

\subsubsection{Longitudinal phase space}
\label{sec:evolution-longitudinal}

\paragraph{Beam loading and energy-transfer efficiency}
\label{sec:beam-loading}

\begin{figure}[t]
    \centering\includegraphics[width=0.95\linewidth]{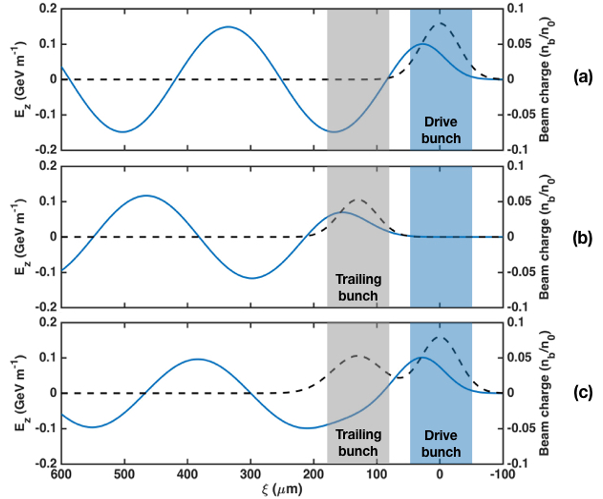}
    \caption{Beam loading in the linear regime. On-axis longitudinal electric field $E_z(\xi)$ for driver only (a), trailing bunch only (b) and for both (c), obtained as the sum of (a) and (b). The plasma is uniform with a density $n_0=\SI{e16}{cm^{-3}}$. The driver is located at $\xi=0$, the trailing bunch at $\xi=\SI{130}{\micro\m}$, $N_d=3\times 10^8$ and $N_t=2\times 10^8$. The beam density profile is shown as a dashed line. Here, 57\% of the wake energy is transferred to the trailing bunch. Adapted from \textcite{Doche2018}.}
  \label{fig:beam_loading_linear}
\end{figure}

When a trailing bunch is injected into the plasma wakefield excited by a driver in order to be accelerated, it modifies the wakefield by reducing its amplitude---a process referred to as \textit{beam loading}. This process is most simply understood in the linear regime, where the principle of superposition applies: the total wakefield is the sum of the driver wakefield and the trailing-bunch wakefield~\cite{Katsouleas1987}. Figure~\ref{fig:beam_loading_linear} shows an example of beam loading in the linear regime, with an electron driver and trailing bunch in a uniform plasma. The trailing bunch is located at the accelerating phase of the driver wakefield, adding its own wakefield contribution to it. Comparing Figs.~\ref{fig:beam_loading_linear}(a) and (c), the wakefield is modified, or loaded, by the presence of the trailing bunch. Both within and behind the trailing bunch the amplitude of the plasma wakefield is smaller than that of the original driver wakefield: part of the plasma-wave energy has been extracted by the trailing bunch and converted to beam energy.

In the language of wave interference, the trailing-bunch acceleration can be understood as the destructive interference between the driver and trailing-bunch plasma waves. The case of perfectly destructive interference, where the total wave cancels in the wake of both beams, corresponds to an energy-transfer efficiency of 100\%: all the energy in the plasma wave is being extracted by the trailing bunch, and no energy is left behind (with the caveat that the energy spread is also 100\%). In contrast, if a trailing bunch is located at a position where the driver wakefield is decelerating, the constructive interference of plasma waves is in this case associated to beam deceleration---a phenomenon that can be leveraged to generate large amplitude plasma waves using multiple drivers~\cite{Ruth1985,Manwani2021}, as demonstrated with electrons \cite{Muggli2011,Verra2025}, protons \cite{Turner2019} and multiple laser pulses~\cite{Hooker2014,Cowley2017}.

Destructive interference and beam acceleration provide two complementary pictures to understand energy transfer in this beam--plasma system. In the particle picture, the energy transferred by the driver to the plasma per unit propagation distance is simply proportional to the number of drive particles $N_d$ and the average electric field experienced by the driver $\langle E_z\rangle_d$. Similarly, the energy transferred from the plasma to the trailing bunch is proportional to its particle number $N_t$ and average field $\langle E_z\rangle_t$. The energy-transfer efficiency is defined as the ratio of the energy gained by the trailing bunch $W_\textrm{gain}$ to the energy lost by the driver $W_\textrm{loss}$:
\begin{equation}
    \eta_\mathrm{p \to t} = \frac{W_\textrm{gain}}{W_\textrm{loss}}= \left | \frac{N_t\langle E_z\rangle_t}{N_d\langle E_z\rangle_d} \right |.
    \label{eq:efficiency_particle}
\end{equation}
This can also be understood as the energy-transfer efficiency from the plasma to the trailing bunch, as $W_\textrm{loss}$ is indeed the total energy that went into the plasma. It does not include the efficiency $\eta_\mathrm{d \to p}$ from the drive beam to the plasma (discussed in Sec.~\ref{sec:drive_plasma_efficiency}), which is given by the ratio of the energy transferred to the plasma ($W_\textrm{loss}$) to the total initial drive-beam energy. 

In the wave picture, destructive interference reduces energy in the wave. The energy-transfer efficiency can thus be expressed as a ratio of wave energies, which in the linear regime can be expressed as~\cite{Hue2021}
\begin{equation}
    \eta_\mathrm{p \to t} = 1-\frac{\int E_{z0,\mathrm{tot}}^2\:\dif^2\mathbf{r}_\perp+\int E_{r0,\mathrm{tot}}^2\:\dif^2\mathbf{r}_\perp}{\int E_{z0,d}^2\:\dif^2\mathbf{r}_\perp+\int E_{r0,d}^2\:\dif^2\mathbf{r}_\perp},
    \label{eq:efficiency_wave}
\end{equation}
where $E_{z0,d}$ and $E_{r0,d}$ are the amplitudes of the longitudinal ($z$) and transverse ($r$) components of the electric field of the drive-beam plasma wave, respectively, $E_{z0,\mathrm{tot}}$ and $E_{r0,\mathrm{tot}}$ are the amplitudes for the total plasma wave behind both beams, and the integral is performed over the transverse coordinates. The $B$-field is zero behind the beams (see Fig.~\ref{fig:linear_response}), hence not contributing to $\eta_\mathrm{p \to t}$. Consider the 1D case and the short-bunch limit where the driver (resp.~trailing-bunch) density can be written as $n_d=\sigma_d \delta(\xi-\xi_d)$ [resp.~$n_t=\sigma_t \delta(\xi-\xi_t)$], with $\sigma_d$ and $\sigma_t$ the surface number density of the driver and trailing bunch and $\xi_d$ and $\xi_t$ their longitudinal position: the total electric field obtained from Eq.~(\ref{eq:Ez}) is given by
\begin{align}
\nonumber
E_z(\xi)=&-\frac{q_d\sigma_d}{\epsilon_0}\cos[k_p(\xi-\xi_d)]\Theta(\xi_d-\xi)\\
&-\frac{q_t\sigma_t}{\epsilon_0}\cos[k_p(\xi-\xi_t)]\Theta(\xi_t-\xi),
\end{align}
where $q_d$ and $q_t$ are the charges of the particles constituting the driver and trailing bunches, respectively. The efficiency can then be calculated using Eqs.~(\ref{eq:efficiency_particle}) or (\ref{eq:efficiency_wave}) as a function of the numbers of drive and trailing particles $N_d=\sigma_d A$ and $N_t=\sigma_t A$ in a cross section $A$. Importantly, $\Theta(0)$ must be equal to $1/2$ to ensure energy conservation for the beam--plasma system, so that Eqs.~(\ref{eq:efficiency_particle}) and (\ref{eq:efficiency_wave}) provide the same answer---this result is known in accelerator physics as the ``fundamental theorem of beam loading", and, simply stated, tells us that a point particle experiences half of its own wakefield~\cite{Ruth1985}. Assuming on-crest acceleration, $\xi_d-\xi_t=\lambda_p/2$ for the same particle type $q_d=q_t$ (or equivalently $\xi_d-\xi_t=\lambda_p$ for opposite particle type $q_d=-q_t$), where $\lambda_p=2\pi/k_p$ is the plasma wavelength, and short bunches compared to $\lambda_p$,  the efficiency in the 1D linear regime is given by~\cite{Katsouleas1987}
\begin{equation}
    \eta_\mathrm{p \to t,\: 1D\; linear} = \frac{N_t}{N_d}\left(2-\frac{N_t}{N_d}\right)
    \label{eq:efficiency_linear1D}
\end{equation}
and a corresponding electric-field spread $N_t/N_d$ (or energy spread if assuming acceleration from rest). Note that as $N_t$ approaches $N_d$, the efficiency approaches 100\%, but so does the energy spread. This efficiency expression [Eq.~(\ref{eq:efficiency_linear1D})] also holds true in 3D when both the driver and trailing bunch have the same density profile. The more general case with different longitudinal and transverse profiles for each beam was discussed by \textcite{Katsouleas1987} and \textcite{Hue2021}.

In the nonlinear regime, beam loading also describes how the wakefield is modified by the presence of the trailing bunch, but the principle of superposition does not hold. Yet, in the case of the blowout regime, which is particularly well suited for acceleration of electrons, the use of the electron-sheath model~\cite{Lu2006a, Lu2006b, Golovanov2023} (Sec.~\ref{sec:nonlin_wake}) provides a simple description of beam loading: the trailing bunch modifies the trajectory of the plasma-electron sheath $r_b(\xi)$ and thereby the longitudinal electric field~\cite{Tzoufras2008}. To describe beam loading using Eq.~(\ref{eq:sheath-traj}), the normalized current term needs to include both the driver and the trailing bunch (i.e., $\lambda=\lambda_d+\lambda_t$). The beam loading from the trailing electron bunch exerts a repulsive force on the electron sheath, causing it to move more slowly toward the axis, which again reduces the longitudinal electric field inside the wake. The formalism of short-range wakefields, where the jump $\Delta E_z$ in the longitudinal electric field induced by a pointlike trailing bunch inside the blowout cavity is calculated, can also be used to describe beam loading in the blowout regime~\cite{Stupakov2018}.

\begin{figure}[t]
    \centering
    \includegraphics[width=\linewidth]{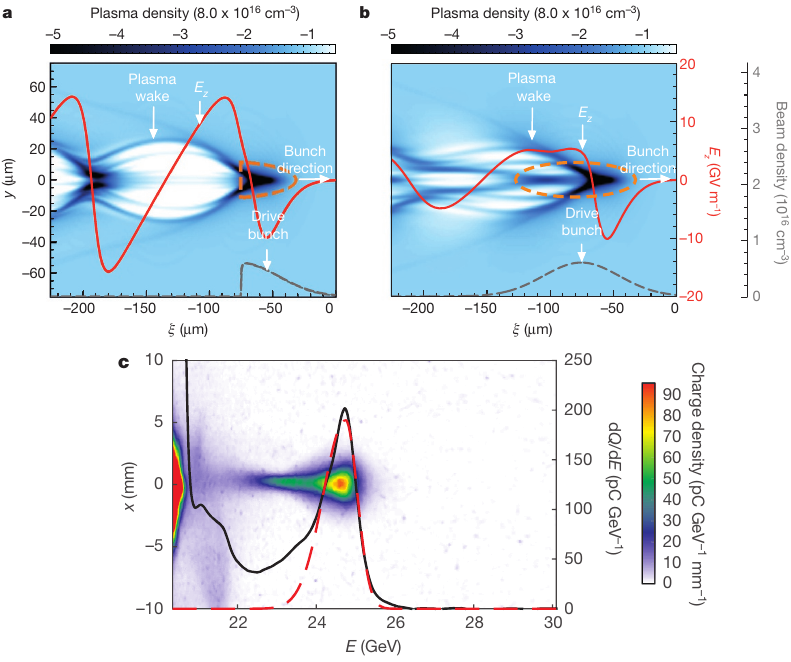}
    \caption{(a--b) Simulated beam loading with positrons, showing the plasma density (blue color map), a positron bunch (orange dashed line), the on-axis $E_z$ field (red line) and the on-axis beam density (gray dashed line). In (a), the positron bunch is cut where $E_z$ crosses zero, such that all positrons are transferring energy to the plasma and driving the plasma wakefield; the unloaded case. In (b), the positron bunch extends into the accelerating part of the wakefield, such that the rear half of the positron bunch extracts energy and loads the wakefield. (c) Experimental demonstration, showing a spectrometer image of accelerated positrons with a peaked energy spectrum. Adapted from \textcite{Corde2015}.}
  \label{fig:beam_loading_posi}
\end{figure}

In contrast to the linear and blowout regimes, nonlinear beam loading with positively charged particles like positrons cannot be simplified to a wakefield superposition or a single plasma electron-sheath differential equation. The nonlinear response of all plasma-electron trajectories to the beam load needs to be accounted for. This can be studied in simulations; an example of strong beam loading with a single positron bunch is shown in Fig.~\ref{fig:beam_loading_posi} \cite{Corde2015}. The unloaded case [Fig.~\ref{fig:beam_loading_posi}(a)], using a half-Gaussian positron bunch for which all positrons decelerate, has a plasma wake and on-axis electric field that resembles that of the blowout regime; the plasma electrons are sucked in by the positron bunch and then cross and overshoot the axis, thereby creating an ion cavity. In the loaded case, simulated in Fig.~\ref{fig:beam_loading_posi}(b) using the full positron bunch, the rear half of the bunch experiences an accelerating field, thus extracting energy and loading the plasma wakefield. In this case, beam loading comes with a more profound change of the wakefield: in the presence of the rear half of the positron bunch, there is no longer an ion cavity free of plasma electrons. Instead, an on-axis plasma-electron filament provides a focusing force to the accelerated positrons, at a position where the transverse force was defocusing in the unloaded case. Positron beam loading can thus modify the transverse wakefield---a topic discussed more in Sec.~\ref{sec:positron-acceleration}.

\paragraph{Optimal beam loading and energy spread}
\label{sec:optimal-beam-loading-energy-spread}

\begin{figure}[t]
    \centering
    \includegraphics[width=0.98\linewidth]{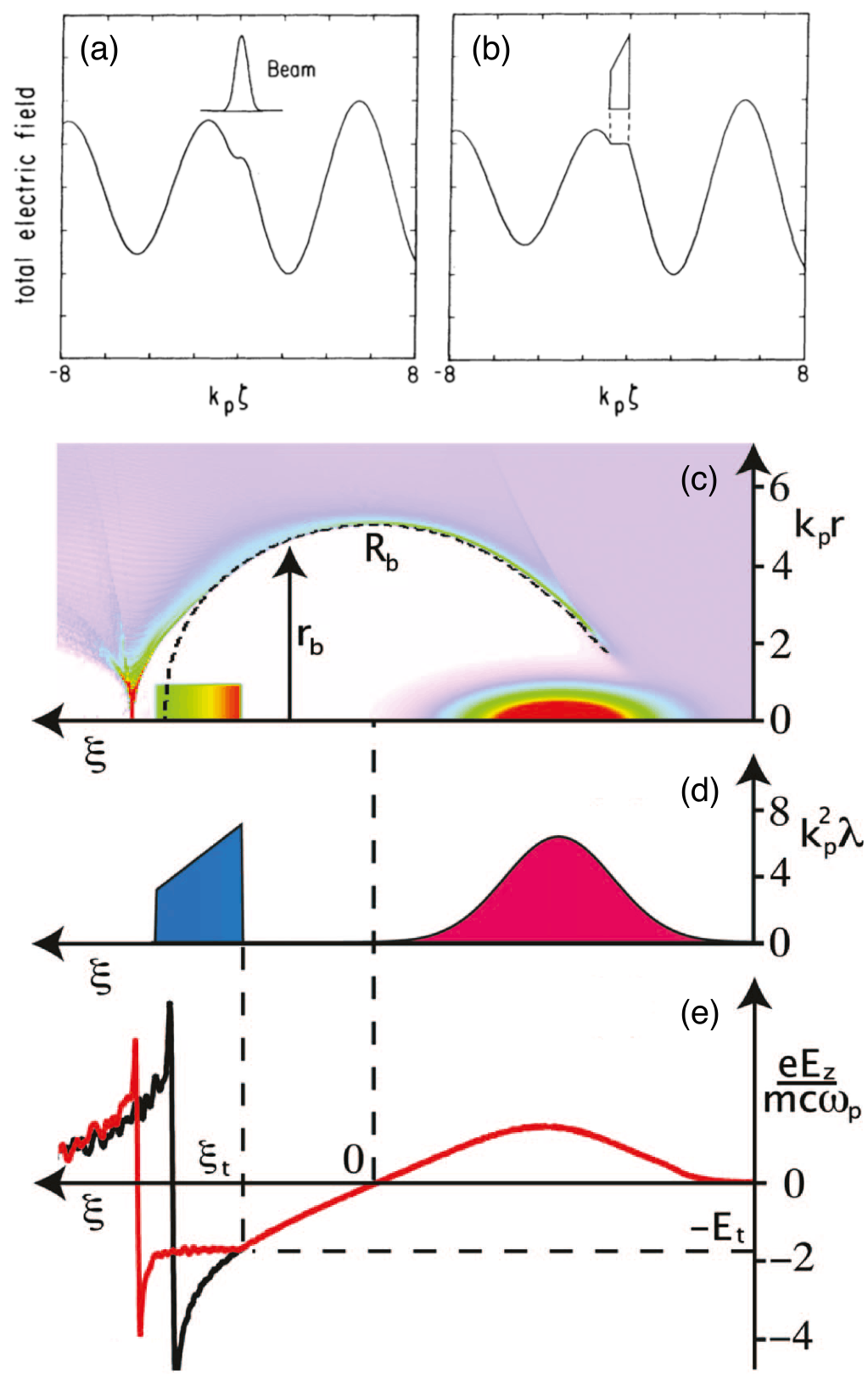}
    \caption{Flattening of $E_z$ with optimal beam loading in the linear (a,b) and nonlinear regime (c--e). (a) Loaded longitudinal electric field $E_z(\xi)$ versus $\xi$ for a trailing bunch with a Gaussian current profile and (b) a truncated triangular profile. (c) PIC simulation showing beam loading in the blowout regime, modifying the electron sheath trajectory from an unloaded wake (black dotted line) to a loaded wake (rainbow color map). (d) The driver is Gaussian whereas the trailing bunch has a truncated triangular current profile. (e) The loaded longitudinal electric field $E_z$ is flattened compared to the unloaded field  (red and black lines, respectively). Adapted from \textcite{Katsouleas1987} and \textcite{Tzoufras2009}.}
  \label{fig:beam_loading_flattening_sim}
\end{figure}

Due to the non-negligible bunch length compared to the plasma wavelength, the trailing bunch can occupy a large phase range and can thus see a commensurately large range of electric-field strengths. Figure~\ref{fig:beam_loading_linear} shows an example of beam loading in the linear regime, where the longitudinal electric field significantly varies over the length of the trailing bunch and therefore its particles do not gain the same amount of energy, resulting in a large final energy spread after acceleration in the plasma. The variation of $E_z$ within the trailing bunch is reduced due to beam loading, as observed by comparing Figs.~\ref{fig:beam_loading_linear}(a) and (c), but it remains large. However, beam loading can be optimized to induce minimal energy spread. An \textit{optimal beam loading} would be such that it results in a constant $E_z$ field along the full length of the trailing bunch. In the linear regime, optimal beam loading without any induced energy spread can be realized using a truncated triangular current profile for the trailing bunch \cite{vanderMeer1985,Katsouleas1987}, while a Gaussian bunch can provide an imperfect but reasonable optimization with some residual energy spread [see Figs.~\ref{fig:beam_loading_flattening_sim}(a) and (b)]. Interestingly, while linear and nonlinear beam-loading models are completely different, in both cases $E_z(\xi)$ can be flattened for a trailing bunch with a truncated triangular profile, as shown in Figs.~\ref{fig:beam_loading_flattening_sim}(b) and (d), respectively. Gaussian beams can also provide reasonable flattening of $E_z(\xi)$ in the nonlinear blowout regime.

\begin{figure}[t]
    \centering
    \includegraphics[width=0.98\linewidth]{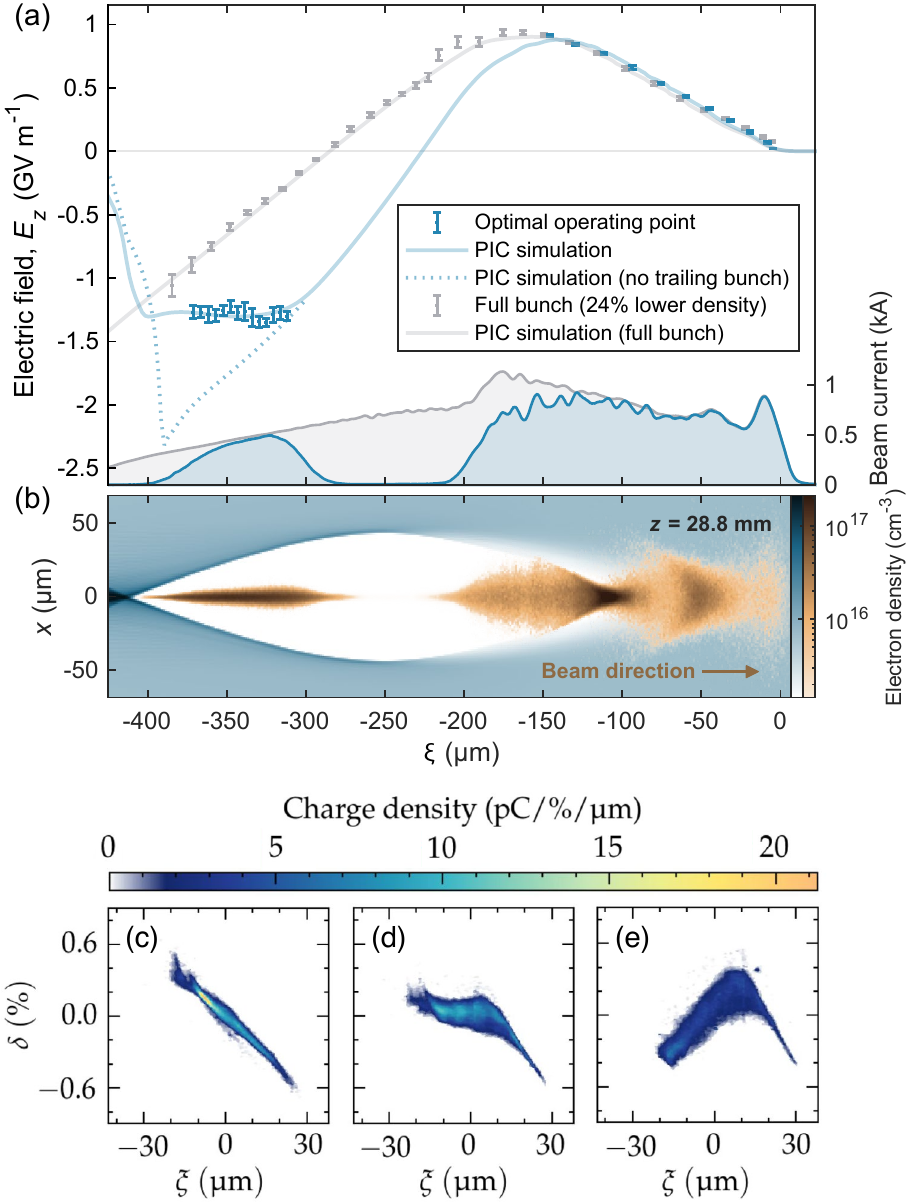}
    \caption{Experimental [points in (a)] and numerical [curves in (a) and simulated wake in (b)] results show flattening of $E_z$ with optimal beam loading in the blowout regime. (c--e) Measured longitudinal phase space of bunches from a similar experiment, showing underloaded (c), optimally loaded (d) and overloaded (e) acceleration. Adapted from \textcite{Lindstrom2021a} (CC-BY 4.0) and \textcite{Caminal2022}.}
  \label{fig:beam_loading_nonlinear_flattening_exp}
\end{figure}

Beam loading is thus a key process that plays an important role in both energy efficiency and energy spread. Achieving simultaneously high energy-transfer efficiency and low energy spread was theoretically shown to be possible with optimal beam loading in the blowout regime~\cite{Lotov2005,Tzoufras2008}. The first experimental observation of beam loading in the bubble/blowout regime was made in LWFA~\cite{Rechatin2009,Rechatin2010}, with a clear correlation between the trailing-bunch charge and the final beam energy. Further LWFA studies with high-charge bunches \cite{Couperus2017,Gotzfried2020} observed beam-loading signatures in the spectral shape and energy spread. In PWFA, evidence of beam loading in the blowout regime was reported by using distributed ionization injection~\cite{VafaeiNajafabadi2014} (Sec.~\ref{sec:ionization-injection:beam}), as well as by using a two-bunch configuration with drive and trailing bunches~\cite{Litos2014, Litos2016} (see Fig.~\ref{fig:facet-high-efficiency}) where correlations between accelerated charge, final energy and energy spread of the trailing bunch were observed. More recently, a precise optimization of beam loading was demonstrated with a driver--trailing bunch configuration, flattening $E_z$ within the trailing bunch to an unprecedented precision [see Fig.~\ref{fig:beam_loading_nonlinear_flattening_exp}(a) and Fig.~\ref{fig:emittance-preservation}(a)]. Here, a residual field variation of 2.8\% rms and an energy-transfer efficiency $\eta_\mathrm{p \to t}=42\pm 4\%$ were measured~\cite{Lindstrom2021a}, providing the first experimental evidence that beam loading in the blowout regime can simultaneously provide low energy spread and high efficiency. A similar experiment used plasma light emission \cite{Oz2004,Muggli2022} to measure the evolution of the local energy-transfer efficiency $\eta_\mathrm{p \to t}(z)$ along the plasma cell \cite{Boulton2026}; e.g.~falling from 60\% to 20\% over \SI{150}{mm} due to wake evolution or charge loss.

For positrons, the longitudinal electric field $E_z$ can also be flattened by nonlinear beam loading [see Figs.~\ref{fig:beam_loading_posi}(b) and \ref{fig:posi_nonlinear}]. Using a single bunch, \textcite{Corde2015} observed the efficient acceleration of positrons located in the rear half of the bunch, and demonstrated that positron beam loading can result in a reduced peak accelerating electric field and a smaller energy spread for the accelerated positrons [see Fig.~\ref{fig:beam_loading_posi}(c)]. Further evidence of positron beam loading in the nonlinear regime was obtained by \textcite{Doche2017} using a driver--trailing bunch configuration, observing the correlation between the trailing-bunch energy gain, energy spread and charge, as well as energy-transfer efficiencies of up to $\sim$40\%.

While optimal beam loading can mitigate \textit{correlated} energy spread (i.e., between longitudinal slices), plasma accelerators can also induce \textit{uncorrelated} energy spread (i.e., within a longitudinal slice). According to the Panofsky-Wenzel theorem [Eq.~(\ref{eq:panofsky-wenzel-theorem})], if the focusing field changes longitudinally within the trailing bunch, as in the linear regime (see Fig.~\ref{fig:linear_response}) and most positron acceleration schemes (see Fig.~\ref{fig:posi_nonlinear}), the accelerating field will not be radially constant, causing an uncorrelated energy spread. Conversely, in the blowout regime, which has a longitudinally constant focusing field, the accelerating field is radially constant and thus preserves the uncorrelated energy spread---a major benefit of the blowout regime. This was verified indirectly by \textcite{Clayton2016}, demonstrating that the focusing field was longitudinally constant, and later directly by \textcite{Caminal2022}  by measuring the longitudinal phase space of plasma-accelerated beams [see Figs.~\ref{fig:beam_loading_nonlinear_flattening_exp}(c--e)]. Lastly, note that the assumption of longitudinally constant focusing is broken in the case of ion motion (see Sec.~\ref{sec:ion-motion}).

\paragraph{Dechirping}
\label{sec:dechirping}

\begin{figure}[t]
    \centering{\includegraphics[width=0.93\linewidth]{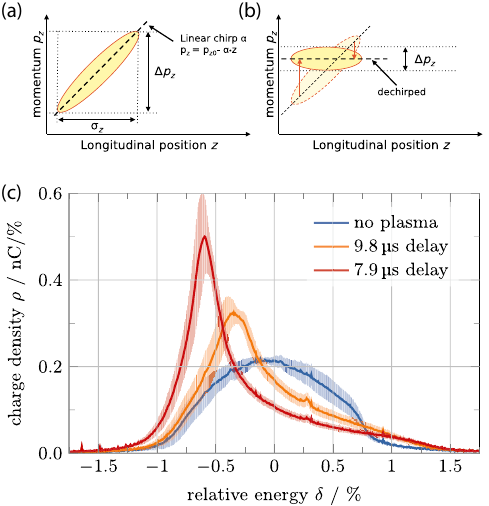}}
    \caption{Schematic showing a trailing bunch with (a) an initially correlated (chirped) particle distribution in longitudinal phase space (b) dechirped by acceleration with a longitudinally dependent field. (c) Experimental energy-spectrum measurements of a \SI{63}{\micro\meter}-long, \SI{681}{MeV} beam driver, which is dechirped by deceleration instead of acceleration, in a \SI{33}{mm}-long plasma. The initial energy spread (1.31\% FWHM; blue line) is reduced by dechirping: the effect is stronger at a shorter delay (0.33\% FWHM; red line) compared to a longer delay (orange line) after a discharge, as the plasma density is decaying. From \textcite{Dopp2018} and \textcite{DArcy2019b}.}
    \label{fig:dechirping}
\end{figure}

Flattening of the accelerating field of the wake via beam loading (Sec.~\ref{sec:optimal-beam-loading-energy-spread} above) is the most elegant solution to mitigating the energy spread growth in a plasma accelerator. However, such beam loading requires precise shaping of the current profile of the trailing bunch, often necessitating multiple rf components, photocathode lasers or collimators \cite{England2008,Loisch2018b,Ha2017,Schroeder2020b} to manipulate the longitudinal phase space of the beam sufficiently. If no remedial measures are taken, this will lead to a rapid build up of correlated energy spread, or \textit{chirp}, in the trailing bunch, thus reducing its quality. This poses a problem also for internally injected bunches (Sec.~\ref{sec:internal-injection}), as their current profiles cannot easily match that required to flatten fully the longitudinal electric field at their trapping location. Furthermore, it may in fact be beneficial to allow the accumulation of a moderate chirp during acceleration as this may help suppress the beam-breakup instability \cite{Mehrling2019} (see Sec.~\ref{sec:bbu}). In this case, a dedicated module (or additional section in the accelerating module) would be required to remove the chirp after acceleration, as shown schematically in Figs.~\ref{fig:dechirping}(a) and (b).

Modules for removing correlated energy spreads are commonplace in accelerator facilities, whereby the accelerating beam is placed in the first 90{\textdegree} of the rf phase in order to decelerate the higher-energy tail of the bunch. Such rf-based dechirping was recently demonstrated with an LWFA \cite{Winkler2025}. However, the dechirping strength of metallic cavities is limited by the same breakdown effects that also limit acceleration, which therefore dominate the length of a plasma-accelerator linac. Dechirping has been demonstrated in dielectric structures \cite{Antipov2014,Emma2014} but the dechirping strengths were still below those required for application to a plasma accelerator. Plasma dechirpers were therefore developed and demonstrated \cite{DArcy2019b, Wu2019c, Shpakov2019}, with the dechirping fields generated in plasma by an electron bunch produced and artificially chirped in an rf linac. Proof-of-principle results reached dechirping strengths of \SI{1.8}{GeV/mm/m} \cite{DArcy2019b}, far exceeding those of competing state-of-the-art techniques. In these first results, the correlated energy spread of the incoming electron bunch was reduced by a factor of 3--6 to the sub-percent level [see Fig.~\ref{fig:dechirping}(c)] required for application to photon science and high-energy physics (Sec.~\ref{sec:applications}).

In the aforementioned proof-of-principle demonstrations, the dechirped beam was the same as that used to produce the dechirping fields. This has inherent downsides, as the bunch will experience some level of head erosion (Sec.~\ref{sec:head_erosion}) and thus emittance growth. Plasma dechirpers can be modified in two ways to mitigate this effect: (1) the on-axis focusing can be removed by using a hollow-channel plasma (Sec.~\ref{sec:hc_linear}). Such a hollow-channel dechirper~\cite{Wu2019d} was recently demonstrated experimentally, reducing the full-width at half-maximum (FWHM) energy spread from 0.93\% to 0.11\%~\cite{Liu2024}. Alternatively, (2) emittance growth is avoided by using a separate leading wakefield driver. One option is to use a laser driver to generate the wake, placing the trailing bunch at the zero crossing of the longitudinal wakefield: this so-called \textit{active} plasma dechirper has been shown in simulation to dechirp the trailing bunch and reduce its energy spread by an order of magnitude \cite{FerranPousa2022}. Note that this scheme will remove a chirp of opposite sign to that of the \textit{passive} dechirper.
A similar effect is achieved by using an electron driver, as recently demonstrated in a combined plasma accelerator and dechirper~\cite{Pompili2021}, halving the energy spread from 0.2\% to 0.1\% rms while accelerating by \SI{\sim4}{MeV} in a \SI{30}{mm} long plasma.

More exotic schemes have also been proposed. This includes allowing a trailing bunch accelerating in a PWFA to accumulate chirp over a certain fraction of the acceleration length. Then, an additional \textit{escort bunch} is internally injected (see Sec.~\ref{sec:internal-injection}) from the quiescent plasma background to overload the wakefield in the location of the trailing bunch: this adds a chirp of opposite sign, effectively removing the initially accumulated chirp \cite{Manahan2017}. Finally, the chirp can also be mitigated during acceleration by applying a periodically modulated plasma density~\cite{Brinkmann2017}.

\paragraph{Dephasing and bunch length}
\label{sec:dephasing}

The longitudinal motion of the trailing bunch with respect to the plasma wave, referred to as \textit{dephasing}, is a major limitation for LWFA due to the sub-luminal laser group velocity in the plasma~\cite{Esarey2009}. In most electron- and positron-driven PWFA scenarios, dephasing is avoided, because both the driver and trailing bunch essentially propagate at the speed of light. Indeed, when the plasma wave has a phase velocity determined by that of the driver and if the Lorentz factor of the trailing bunch is much larger than that of the driver $\gamma_d$, then the trailing bunch moves forward by a distance $d\simeq L/(2\gamma_d^2)$ relative to the driver over a propagation distance $L$. For instance, it corresponds to a negligible trailing-bunch movement of the order of \SI{0.1}{\micro\m} per meter of propagation for a \SI{1}{GeV} electron driver. Dephasing in PWFA is thus only important when the plasma-wave phase velocity differs from the drive-beam velocity, which is possible in plasma-density ramps (as used in the downramp internal-injection scheme discussed in Sec.~\ref{sec:internal-injection}) and for strongly evolving drivers where the plasma wavelength effectively varies during the propagation (Sec.~\ref{sec:driver-propagation}). Dephasing also occurs when the plasma wave is associated with a growing instability, such as in self-modulated plasma wakefields (Sec.~\ref{sec:self-modulated-wakefields}). The phase velocity of self-modulated wakefields [Eq.~(\ref{eq:phase_velocity_self_modulation})] can be substantially smaller than $c$ and dephasing needs to be addressed. Moreover, dephasing can be significant for short proton drivers (i.e., not relying on self-modulation) \cite{Caldwell2009}, given their typically much smaller $\gamma_d$, but this effect can be mitigated by a gradual plasma-density upramp~\cite{Katsouleas1986,Pukhov2008}.

The bunch length is typically preserved inside a plasma accelerator, as the particles dephase negligibly with respect to each other. In practice, however, the bunch can be shortened by charge loss from the tail of the trailing bunch \cite{Lindstrom2021a}, either from defocusing by the sheath electrons or from deflection by instabilities (Sec.~\ref{sec:bbu}). Lastly, if using multiple stages separated by optics with longitudinal dispersion ($R_{56}$; i.e., the relative longitudinal displacement per relative energy offset), the bunch length and phase can evolve through a self-correction mechanism (see Sec.~\ref{sec:staging} and Fig.~\ref{fig:self-correction}).

\subsubsection{Transverse phase space and emittance}
\label{sec:evolution-transverse}

Emittance is a key parameter for applications of PWFA, directly affecting the brightness in FELs [Eq.~(\ref{eq:brightness})] and luminosity in colliders [Eq.~(\ref{eq:luminosity-per-power})]. In simple terms, it quantifies the ability of a beam to be focused to a small transverse size---the smaller the emittance, the smaller the beam size in a focus. Specifically, the \textit{normalized emittance} quantifies the area of the rms ellipse of the particle distribution in transverse phase space \cite{Floettmann2003}, as defined in Eq.~(\ref{eq:normalized_emittance}). In a vacuum or in a linear focusing field, this normalized emittance is preserved during acceleration. While the emittance cannot decrease, unless the beam is undergoing radiative damping (Sec.~\ref{sec:radiative-damping}), there are numerous ways for the emittance to increase (covered in Secs.~\ref{sec:mismatching}--\ref{sec:multiple-coulomb-scattering}). Preserving the emittance is therefore a challenge in a plasma accelerator---a topic reviewed in more detail by \textcite{Lindstrom2022}.

Emittance measurements were first performed on internally injected bunches from laser-driven plasma accelerators, using a variety of measurement techniques: the ``pepper-pot" method \cite{Fritzler2004,Brunetti2010}; the multi-shot quadrupole scan and the single-shot ``butterfly" method \cite{Weingartner2012,Barber2017}; and betatron x-ray spectroscopy \cite{Kneip2012,Plateau2012,Curcio2017}.
These methods were later applied to both internally and externally injected bunches from beam-driven plasma accelerators, including the butterfly method \cite{VafaeiNajafabadi2016} and quadrupole scans \cite{Shpakov2021}, with plans for also using betatron radiation \cite{SanMiguelClaveria2019}. The first experimental demonstration of emittance preservation, with moderate energy gain, was performed by \textcite{Lindstrom2024} using quadrupole scans [see Fig.~\ref{fig:emittance-preservation}(b)]. Simultaneous emittance preservation and large energy gain---the ultimate goal of several PWFA facilities \cite{Joshi2018,DArcy2019a,Gschwendtner2022}---has been shown in simulations \cite{Zhao2024}, but has not yet been shown in experiments.

\paragraph{Mismatching}
\label{sec:mismatching}

\begin{figure}[t]
    \centering{\includegraphics[width=\linewidth]{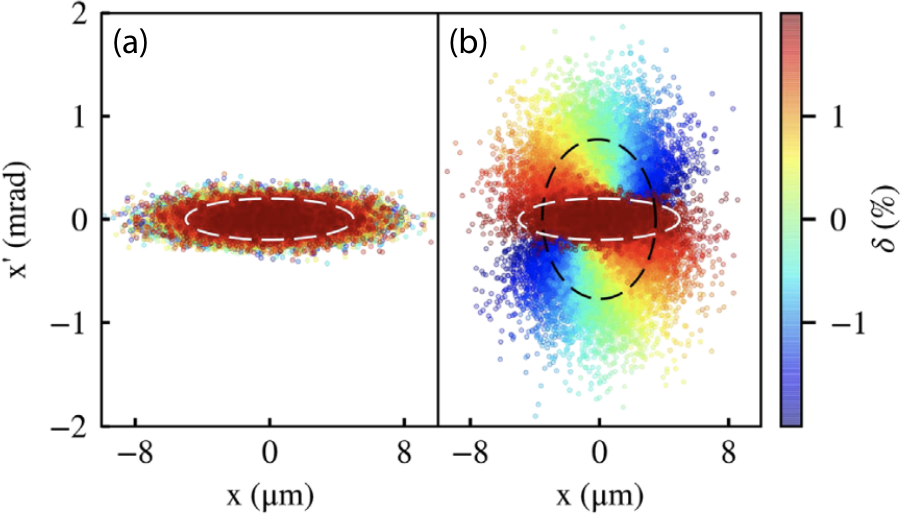}}
    \caption{A mismatched and finite-energy-spread beam, shown here after (a) zero and (b) 23.5 betatron oscillations, increases its area in phase space (i.e., emittance) as particles of different energy (rainbow colorbar) circulate the origin at different rates. This causes phase mixing, transforming the beam ellipse from unmatched (white dashed line) to matched (black dashed line). From \textcite{Ariniello2019} (CC-BY 4.0).}
    \label{fig:mismatching}
\end{figure}

Within a plasma accelerator, the ion column focuses the particles, causing them to rotate along an ellipse in phase space. If the phase-space distribution of the beam is matched, such that the beta function is $\beta_\mathrm{matched} = \sqrt{2\gamma}/k_p$ [Eq.~(\ref{eq:matched_beta_function}); assuming the blowout regime], then the beam size and divergence remain constant (if we ignore acceleration). However, if the beta function differs from $\beta_\mathrm{matched}$, they will oscillate as the beam propagates. This can be quantified with the \textit{mismatch parameter} \cite{Sands1991}
\begin{equation}
    \mathcal{M} = \frac{1}{2}\left(\tilde{\beta}+\tilde{\gamma}+\sqrt{(\tilde{\beta}+\tilde{\gamma})^2-4}\right),
\end{equation}
where $\tilde\beta = \beta/\beta_\mathrm{matched}$, $\tilde\alpha = \alpha-\alpha_\mathrm{matched}\tilde\beta$ and $\tilde\gamma = (1+\tilde\alpha^2)/\tilde\beta$ each quantify the mismatch in the individual Twiss parameters. A mismatch is problematic if the beam has a finite energy spread, as particles of different energy will oscillate at different rates, causing them to spread out over a larger area in phase space---increasing the projected emittance (see Fig.~\ref{fig:mismatching}). The emittance steadily increases until the different energy slices have fully decohered, resulting in a relative emittance growth
\begin{equation}
    \label{eq:mismatch_emittance_growth}
    \frac{\varepsilon_\mathrm{sat}}{\varepsilon_{\mathrm{in}}} = \frac{1}{2}\left(\mathcal{M}+\frac{1}{\mathcal{M}}\right),
\end{equation}
where $\varepsilon_{\mathrm{in}}$ and $\varepsilon_\mathrm{sat}$ are the initial and saturated emittances, respectively. Matching to a longitudinally flattop plasma-density profile, for which $\alpha_\mathrm{matched} = 0$, this equation simplifies to \cite{Mehrling2012}
\begin{equation}
    \frac{\varepsilon_{\mathrm{sat}}}{\varepsilon_{\mathrm{in}}} = \frac{1}{2}\left(\frac{\beta}{\beta_\mathrm{matched}}+\beta_\mathrm{matched}\frac{1+\alpha^2}{\beta}\right).
\end{equation}
Full decoherence occurs at an approximate distance
\begin{equation}
    \label{eq:decoherence-length}
    L_{dc} \approx \frac{\pi\beta_\mathrm{matched}}{\sigma_\delta},
\end{equation}
where $\sigma_\delta$ is the relative energy spread; this decoherence length corresponds to $1/\sigma_\delta$ oscillations of the beta function, or $1/(2\sigma_\delta)$ betatron wavelengths. More accurate expressions for the decoherence length and emittance evolution until saturation are given by \textcite{Ariniello2022}.

Reaching the often small beta functions required for matching can be challenging in practice. As an example, at a plasma density of \SI{e16}{\per\cm\cubed} and energy \SI{10}{GeV}, the beta function is \SI{10}{mm}, which is much smaller than that commonly seen in rf accelerators. To mitigate this, longitudinal plasma-density ramps \cite{Marsh2005,Xu2016,Ariniello2019} can be used to increase the matched beta function at the entrance and exit of the plasma accelerator, while leaving the density and accelerating gradient in the interior unchanged. Slowly varying \textit{adiabatic} ramps~\cite{Chen1990} are particularly well suited for preserving emittance. However, for positrons focused by an on-axis electron filament (Sec.~\ref{sec:positrons:with-focusing}), ramps can be detrimental \cite{Wang2025}.

The first experimental demonstration of matching was performed by \textcite{Muggli2004} with a beta function of \SI{110}{mm} (\SI{28.5}{GeV}, \SI{e14}{\per\cm\cubed}). A more recent demonstration by \textcite{Lindstrom2024} matched to a beta function of \SI{3}{mm} (\SI{1}{GeV}, \SI{e16}{\per\cm\cubed}) by using ramps that demagnified the incoming beta function by a factor 6.

\paragraph{Misalignment}
\label{sec:misalignment}

\begin{figure}[t]
    \centering{\includegraphics[width=0.9\linewidth]{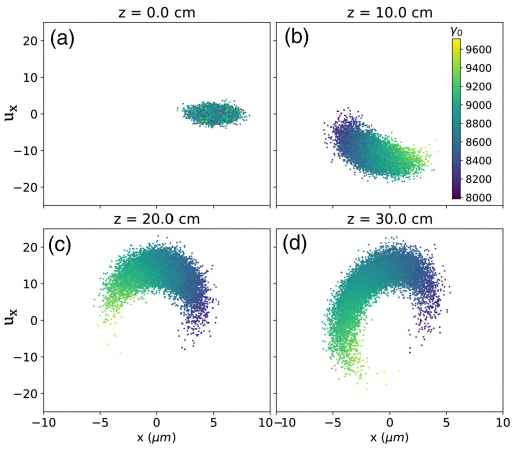}}
    \caption{An initially misaligned bunch (a), i.e.~off-center in phase space, rotating around the origin along its propagation (b--d) with particles of different energies undergoing phase mixing. This increases the area in phase space; i.e., the emittance grows. From \textcite{Thevenet2019} (CC-BY 4.0).}
    \label{fig:misalignment}
\end{figure}

Misalignment, or an offset of the bunch centroid from the origin in phase space, can also lead to emittance growth, for the same reason as when the beam is mismatched: particles of different energy rotate in phase space at different rates. As the phase mixing smears the beam into a ring in phase space (see Fig.~\ref{fig:misalignment}), the emittance grows until the decoherence length is reached [Eq.~(\ref{eq:decoherence-length})]. To quantify this, it is useful to convert the initial positional and angular misalignments, $\Delta x$ and $\Delta x'$, into normalized coordinates $\Delta x_n = \sqrt{\gamma/\beta} \Delta x$ and $\Delta x'_n = \sqrt{\gamma/\beta}(\alpha\Delta x + \beta\Delta x')$, where $\gamma$ is the Lorentz factor but $\alpha$ and $\beta$ are Twiss parameters. These can be combined into a \textit{normalized amplitude} defined by $A_n^2 = \Delta x_n^2 + \Delta {x'_n}^2$ \cite{Aksoy2011}. The resulting saturated emittance growth, which is added in quadrature with the initial emittance \cite{Lindstrom2016a,Thevenet2019}, is given by 
\begin{equation}
    \label{eq:misalignment-emit-growth}
    \Delta\varepsilon_{n,\mathrm{sat}} = \frac{A_n^2}{2} \approx \frac{\gamma}{2} \left( \frac{\Delta x^2}{\beta_\mathrm{matched}} + \beta_\mathrm{matched}\Delta x'^2\right),
\end{equation}
where we have used $\alpha = 0$ and $\beta = \beta_\mathrm{matched}$. Prior to saturation, the emittance grows quadratically with distance $s$ and relative energy spread $\sigma_\delta$: $\Delta\epsilon_{n,\mathrm{unsat}} = (s/L_\mathrm{dc})^2A_n^2/2 \sim s^2\sigma_\delta^2$ \cite{Lindstrom2016a}. More detailed expressions are required for nonuniform accelerating fields or when large energy gain occurs during decoherence \cite{Thevenet2019,Ariniello2022}.

Suppressing this emittance growth imposes strict tolerances on alignment \cite{Assmann1998,Cheshkov2000,Schulte2016}. Consider for instance a collider-like beam at \SI{100}{GeV} and a plasma density of \SI{e16}{\per\cm\cubed}; using Eq.~(\ref{eq:misalignment-emit-growth}) we find that to stay below \SI{0.1}{\milli\meter\milli\radian}, the bunch must be aligned to within \SI{0.2}{\micro\m} or \SI{6}{\micro\radian}, assuming full phase mixing. However, this may not hold for a multistage accelerator, where the misalignment is imposed repeatedly but the emittance growth in each stage typically does not saturate; this is discussed in Sec.~\ref{sec:multistage-alignment}. Use of plasma-density ramps can loosen the tolerances of the positional misalignment, but will simultaneously tighten the angular misalignment tolerance [as seen in Eq.~(\ref{eq:misalignment-emit-growth})]. Another strategy to cancel misalignment, pointed out by \textcite{Schulte2016}, is to match the stage length to an integer number of betatron oscillations. Regardless, beam alignment will pose a significant challenge to any high-beam-quality PWFA facility, likely favoring reduced plasma densities (to below \SI{e16}{\per\cm\cubed}).

\paragraph{Beam-breakup instability}
\label{sec:bbu}

The hosing instability \cite{Whittum1991} is often considered to affect the driver (see Sec.~\ref{sec:hosing-instability}), but the same effect applies to the trailing bunch. In rf accelerators, the effect is known as the \textit{beam-breakup instability} \cite{Panofsky1968}. When the trailing bunch is transversely offset, it will repel the plasma-sheath electrons closest to it a bit more strongly and those farthest from it a bit more weakly; this deforms the plasma cavity behind that point, shifting the focusing axis, and leading to a positive feedback loop with increasing oscillation amplitudes. For this reason, stronger beam loading of the wakefield (giving higher energy-transfer efficiency) will result in a stronger transverse instability. Using the formalism of the hosing theory with Eqs.~(\ref{eq:hosing_beam}) and (\ref{eq:hosing_channel_modified}), the transverse wakefield responsible for the BBU instability is directly related to the channel offset by $W_\perp/E_0 = -k_p X_c/2$, and the loading is included in the coefficient $c_r(\xi)$ involved in the equation for $X_c$.

\textcite{Stupakov2018} calculated the short-range transverse wakefield generated immediately behind a pointlike trailing bunch with a transverse offset $\Delta x$ and charge $q$ located at $\xi_0$ inside the blowout cavity. The transverse wakefield takes the form $W_\perp = w_\perp q \Delta x\propto (\xi_0-\xi)\Theta(\xi_0-\xi)$, where a good approximation for $w_\perp$, the transverse wakefield per unit charge and offset, can be obtained from
\begin{equation}
\left. \frac{\dif w_\perp}{\dif \xi} \right|_{\xi_0^-} = - \frac{2}{\pi\epsilon_0\left(r_b(\xi_0)+0.75 k_p^{-1} \right)^4},
\end{equation}
with the derivative evaluated immediately behind the trailing charge. Based on a similar approach involving short-range transverse wakefields and the short-range wake theorem, \textcite{Lebedev2017} postulated a fundamental \textit{efficiency--instability relation}:
\begin{equation}
    \label{eq:efficiency-instability-relation}
    \eta_t \approx \frac{\eta_\mathrm{p \to t}^2}{4(1-\eta_\mathrm{p \to t})},
\end{equation}
where $\eta_t$ is the ratio of the deflecting force from BBU to the focusing force from the blowout cavity experienced by a particle at the tail of the trailing bunch, and $\eta_\mathrm{p \to t}$ is the energy-transfer efficiency (plasma wake to trailing bunch). The oscillation amplitude increases exponentially with the product $\mu \eta_t$ \cite{Lebedev2017}, where $\mu$ is the integrated phase advance, leading to emittance growth via Eq.~(\ref{eq:misalignment-emit-growth}). This implies that large energy gain and high efficiency can be problematic. In detailed studies using PIC simulations, Eq.~(\ref{eq:efficiency-instability-relation}) appears to be the best-case scenario \cite{Finnerud2025}, assuming zero energy spread and immobile ions, reached only when the beam is placed such that the normalized accelerating field is
\begin{equation}
    \label{eq:efficiency-instability-relation-optimal-field}
    \frac{E_z^\mathrm{opt}}{E_0} \approx 0.23 (1-0.78\eta_\mathrm{p \to t}^{1.86})(k_pR_b)^{1.5},
\end{equation}
where $R_b$ is the maximum blowout radius---this is the optimal field that minimizes transverse instabilities.

\begin{figure}[t]
    \centering\includegraphics[width=0.9\linewidth]{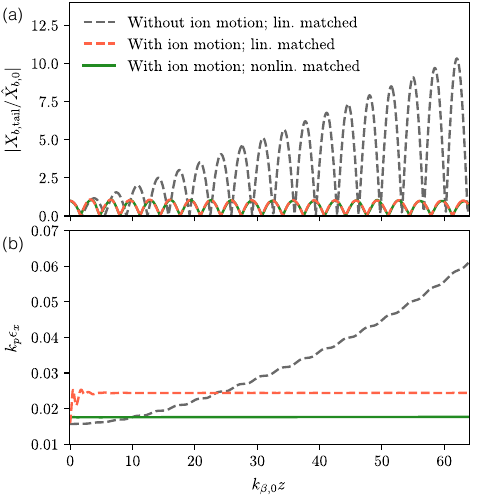}
    \caption{(a) Relative centroid displacement for the trailing-bunch tail versus propagation distance (normalized by the initial betatron wavenumber $k_{\beta,0}$), comparing a linearly matched beam with and without ion motion (orange and gray dashed lines, respectively) and a beam nonlinearly matched to the ion spike (green line), and (b) the  corresponding normalized emittance (in units of $k_p^{-1}$). From \textcite{Mehrling2018}.}
    \label{fig:hosing-ion-motion}
\end{figure}

The assumptions of no energy spread or ion motion are, however, unrealistic in plasma accelerators. The transverse oscillation of different longitudinal slices can be detuned if the focusing strength varies along the bunch, which suppresses the instability. This variation can be imposed either via an energy chirp in the trailing bunch, so-called BNS damping \cite{Balakin1983,Mehrling2017}, or by allowing ions to move, increasing their on-axis density \cite{Mehrling2018} (see Fig.~\ref{fig:hosing-ion-motion} and Sec.~\ref{sec:ion-motion}). Alternatively, a longitudinally varying focusing force can be imposed by operating in the linear or nearly-nonlinear (\textit{quasilinear}) regime \cite{Lehe2017}, but also in the nonlinear regime by employing flat drivers, which create an elliptical wake structure with a focusing force that varies both longitudinally and between the transverse planes \cite{Baturin2022,Manwani2025}.

The above considerations assume a uniform plasma density. No experiments have yet conclusively observed the beam-breakup instability in this case, although attempts have been made \cite{Adli2016b,Finnerud2026}. However, in hollow-channel plasma accelerators (Secs.~\ref{sec:hc_linear} and \ref{sec:without-focusing}), there is no on-axis focusing force to counter the instability \cite{Schroeder1999,Schroeder2001}, which will therefore rapidly lead to beam loss---this was shown experimentally for a positron beam at FACET \cite{Gessner2016b,Lindstrom2018}. Finally, in plasmas with a transverse density gradient, the wake structure becomes asymmetric, which induces a transverse wakefield even for perfectly aligned beams \cite{Doss2023,Baturin2023}.

\paragraph{Ion motion and nonlinear focusing fields}
\label{sec:ion-motion}

\begin{figure}[t]
    \centering{\includegraphics[width=\linewidth]{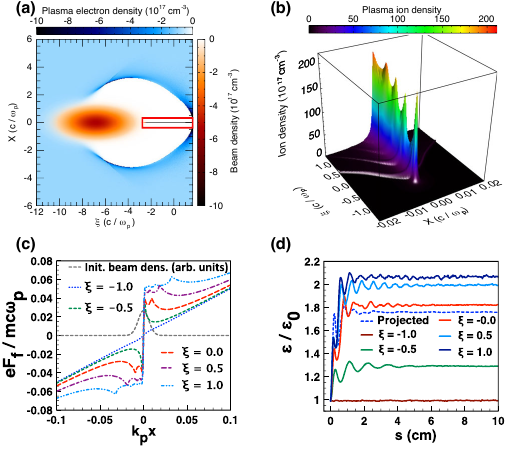}}
    \caption{Ion motion in a plasma wake (a), creating an on-axis ion density spike (b). The resulting focusing fields are highly nonlinear, and vary with longitudinal position $\xi$. (d) This results in an emittance growth until the particle distribution in transverse phase space reaches an equilibrium, where the emittance saturates. From \textcite{An2017}.}
    \label{fig:ion-motion}
\end{figure}

Just as the ions of the plasma cavity focus the accelerating electron beam, the electron beam focuses the ions. Within a long, uniform electron beam, ions will oscillate with a frequency $\omega_{i,b} = \sqrt{n_b e^2/m_i \epsilon_0} = \omega_p \sqrt{(m_e/m_i)(n_b/n_0)}$, where $n_b$ and $n_0$ are the beam and plasma densities, and $m_i$ and $m_e$ are the ion and electron masses. These oscillations typically occur on a much longer timescale [e.g., nanosecond scale \cite{DArcy2022}] than that of the plasma-electron frequency (i.e., the sub-picosecond scale). However, for a sufficiently dense beam, $n_b/n_0 \gtrsim m_i/m_e$, ion motion can occur within the beam itself \cite{Lee1999}---relevant for the high beam densities of collider beams. In this case, a density spike forms on axis with strongly nonlinear focusing forces (see Fig.~\ref{fig:ion-motion}). The spike forms when the phase advance of the ion oscillation within the bunch \cite{Rosenzweig2005}
\begin{equation}
    \label{eq:ion-motion}
    \Delta\phi_i \simeq \sqrt{\frac{\mu_0 e^2}{2} \frac{Z_i \sigma_z N}{m_i} \sqrt{\frac{r_e n_0 \gamma}{\varepsilon_{n,x}\varepsilon_{n,y}}}}
\end{equation}
exceeds a quarter oscillation; $\Delta\phi_i > \pi/2$. Here, $\sigma_z$, $N$, $\gamma$ and $\varepsilon_{n,x/y}$ are the length, number of particles, relativistic factor and emittances of the trailing bunch, respectively, $Z_i$ is the ionization state of the ions (e.g., $Z_i=1$ when singly ionized), and $r_e$ is the classical electron radius.

The resulting nonlinear focusing can cause significant emittance growth. However, the emittance only grows until a non-Gaussian equilibrium distribution is reached \cite{Lotov2017}. This implies that over-focused beams will result in less emittance growth \cite{An2017}, and that full emittance preservation can be achieved by directly matching to the equilibrium distribution \cite{Benedetti2017}. Such injection is technically challenging using an rf accelerator, but can instead be accomplished by gradually increasing the ion motion: either with an ion-mass ramp from heavy to light ions \cite{Gholizadeh2010}, or by adiabatic damping \cite{Benedetti2021}, where the beam density increases with acceleration. While ion motion is suppressed in heavier gases, this can itself cause emittance growth from increased Coulomb scattering (Sec.~\ref{sec:multiple-coulomb-scattering}). At sufficiently high ion temperatures, of order 10--\SI{100}{keV} (similar to that in a fusion reactor), ion motion is suppressed as the ion spike smears out \cite{Gholizadeh2011}. Note, however, that a controlled amount of nonlinear focusing from ion motion (e.g., limiting the emittance growth to 10--100\%) can be desirable \cite{Mehrling2018,Finnerud2025}, as this suppresses the beam-breakup instability (see Sec.~\ref{sec:bbu} and Fig.~\ref{fig:hosing-ion-motion}).

For flat beams, as employed in linear colliders (Sec.~\ref{sec:applications:colliders}), where the emittance is much larger in one transverse plane compared to the other, resonant emittance mixing can occur \cite{Diederichs2024}. As a result, the emittance increases in the low-emittance plane, and decreases in the high-emittance plane. This happens even when ion motion is suppressed (i.e., for weakly nonlinear focusing), but then only over a large number of betatron oscillations. The resonance can be suppressed if the focusing is stronger in the low-emittance plane, which can be achieved using drive bunches flat in the orthogonal direction; i.e.~a vertically flat driver when the trailing bunch is horizontally flat and vice versa.

While the above discussion applies to electron bunches, the same challenge arises for positron bunches (Sec.~\ref{sec:positron-acceleration}), where the equivalent problem of \textit{electron motion} similarly results in emittance growth, but at much lower beam densities. Lastly, ion motion can also occur within the self-modulated wakefields of a long proton bunch \cite{Vieira2012}, as recently observed at the AWAKE experiment \cite{Turner2025} (see Sec.~\ref{sec:self-modulated-wakefields}).

\paragraph{Coulomb scattering}
\label{sec:multiple-coulomb-scattering}

The presence of matter on the beam axis is unavoidable in plasma accelerators, and can lead to beam particles scattering off gas particles or ions, known as Coulomb scattering \cite{Blachman1948,Bethe1953}. While a small fraction of particles will scatter inelastically off the atomic nucleus, causing particle loss or a beam halo, most particles scatter elastically at a small angle. Multiple such scattering events will increase the rms divergence $\sqrt{\langle\theta^2\rangle}$ and, correspondingly, increase the emittance \cite{Raubenheimer1992}.

The emittance growth per length has contributions from both gas particles and ions \cite{Kirby2007}:
\begin{equation}
    \label{eq:EmitGrowthScattering}
    \frac{\dif \varepsilon_{nx}}{\dif s} = \frac{1}{2}\beta_x\gamma\left(\frac{\dif }{\dif s}\langle\theta^2\rangle_\mathrm{gas}+\frac{\dif }{\dif s}\langle\theta^2\rangle_\mathrm{ion}\right),
\end{equation}
where $\beta_x$ is the beam's beta function and $\gamma$ is the relativistic Lorentz factor. The gas-scattering term (for electrons) is approximately given by \cite{Reiser2008} 
\begin{equation}
    \label{eq:EmitGrowthScattering_gas}
    \frac{\dif }{\dif s}\langle\theta^2\rangle_\mathrm{gas} \approx 16 \pi \frac{r_e^2}{\gamma^2} n_{\mathrm{gas}} Z^2 \ln\left(\frac{204}{Z^{1/3}}\right),
\end{equation}
where $n_\mathrm{gas}$ is the gas number density, indicating that gas scattering increases quadratically with the atomic number $Z$. Note that Eq.~(\ref{eq:EmitGrowthScattering_gas}) is only an approximation of more complex forms [e.g., \textcite{Bethe1953} suggested a scaling of $Z(Z+1)$], as discussed by \textcite{Lynch1991}, that can be accurate to 11\% or better for interaction lengths between $10^{-3}$ and $100$ radiation lengths \cite{Beringer2012}, but only applies when multiple scattering events occur. Next, the ion-scattering term \cite{Montague1984}, verified with simulations \cite{Zhao2020}, is given by
\begin{equation}
    \frac{\dif }{\dif s}\langle\theta^2\rangle_\mathrm{ion} = 8\pi \frac{r_e^2 }{\gamma^2} n_\mathrm{ion} Z_i^2 \ln\left(\frac{b_{\mathrm{max}}}{b_{\mathrm{min}}}\right),
\end{equation}
where $n_\mathrm{ion}$ is the ion density, and the (Coulomb) logarithm contains the ratio of the maximum and minimum impact parameters---estimated to be $b_{\mathrm{max}} = \lambda_D$ (the Debye length) and $b_{\mathrm{min}} = r_a \approx 1.4 A^{1/3}$~fm (the atomic radius), with $A$ the atomic mass number. In a non-quasi-neutral plasma, such as in the ion cavity of the blowout regime, the maximum impact parameter increases to be approximately the blowout radius $b_{\mathrm{max}} \sim R_b$ \cite{Kirby2007}.

Integrating Eq.~(\ref{eq:EmitGrowthScattering}) over the acceleration length, and assuming that the beam is matched, the cumulative emittance growth scales as $\Delta\epsilon_n \propto \Delta\mu$ \cite{Kirby2007}, where $\Delta\mu = \sqrt{2}(\sqrt{\gamma_{f}} - \sqrt{\gamma_{i}})E_0/E_z$ is the total phase advance. Assuming a constant normalized accelerating gradient $E_z/E_0$, this emittance growth is independent of plasma density, because higher density allows smaller beta functions and shorter acceleration distance. The main strategy for suppressing Coulomb scattering is therefore the use of lighter gas species. Ultimately, Coulomb scattering can be made negligible with hydrogen, with an expected emittance growth for a TeV collider of only around \SI{1}{\nano\meter\radian} \cite{Schroeder2010}.

\paragraph{Radiative damping}
\label{sec:radiative-damping}

Particles that oscillate transversely in a focusing field will emit synchrotron radiation \cite{Esarey2002,Wang2002,Rousse2004,Corde2013}, thereby losing momentum in the direction of emission---the so-called \textit{radiation reaction}. If the particle is subsequently accelerated in the forward direction, the net effect can be a radiative damping of the transverse emittance. This is the concept behind damping rings, but is normally ignored in rf linacs due to the short acceleration length. However, in the presence of very strong focusing fields, as in a PWFA, the effect can be non-negligible \cite{Michel2006}. Each particle evolves according to a coupled set of equations for the transverse positions $x$, $y$ and the Lorentz factor $\gamma$
\begin{eqnarray}
    \ddot{x} + \left(\frac{\omega_p}{\gamma} \frac{E_z}{E_0} + \tau_R c^2K^2\right)\dot{x} + \frac{c^2K^2}{\gamma} x = 0, \\
    \ddot{y} + \left(\frac{\omega_p}{\gamma} \frac{E_z}{E_0} + \tau_R c^2K^2\right)\dot{y} + \frac{c^2K^2}{\gamma} y = 0, \\
    \label{eq:radiative-damping-longitudinal}
    \dot{\gamma} = \omega_p \frac{E_z}{E_0} -\tau_R c^2K^2\gamma^2(x^2+y^2),
\end{eqnarray}
where $\tau_R = \frac{2}{3} r_e/c$ and $K^2 = k_p^2/2$ for the blowout regime. Here, we assume no coherent radiation and no interaction between the particles and the emitted radiation. For very high plasma densities or energies (i.e., above \SI{e20}{\per\cm\cubed} and \SI{1}{TeV}), the dynamics can also be affected by strong-field quantum electrodynamics (QED) effects \cite{Zeng2021,Golovanov2022,Qian2026}.  From Eq.~(\ref{eq:radiative-damping-longitudinal}), we see that the effect of the radiation reaction is most significant in the longitudinal plane, acting as a deceleration term that scales quadratically with energy. In a bunch with finite emittance, particles oscillate with a range of different transverse amplitudes, which can therefore introduce an energy spread. Assuming no centroid offsets, radiative damping does not impose a fundamental limit on the achievable energy in plasma-based accelerators \cite{Deng2012}, as the radiative damping of emittance sufficiently reduces the oscillation amplitude during acceleration; the damping force approaches $2/3$ of the accelerating force \cite{Kostyukov2012}. However, with imperfect alignment the achievable energy will be limited---typically around the TeV range for \SI{}{\micro\meter} offsets and \SI{e16}{\per\cm\cubed} densities \cite{Saberi2023}, or alternatively placing an upper bound on the plasma density given a certain alignment tolerance.

\subsubsection{Spin polarization}
\label{sec:spin-polarization}

Spin polarization is a beam quality that is important for certain applications, in particular for nuclear and particle physics \cite{Glashausser1979,Shiltsev2021}. It is defined as
\begin{equation}
    \label{eq:spin-polarization}
    \mathbf{P} = \frac{1}{N}\sum^N_{i=1} \hat{\mathbf{s}}_i,
\end{equation}
where $\hat{\mathbf{s}}_i$ is the unit spin vector for each of the $N$ particles. Typically, a polarization of $|\mathbf{P}|=60$--80\% is required in a linear collider \cite{MoortgatPick2008}, which implies that a high polarization is required at the source, and must be maintained throughout acceleration. For a comprehensive review of plasma acceleration of spin-polarized beams, see \textcite{Reichwein2025}.

Spin-polarized electrons are conventionally produced in negative-electron-affinity GaAs photocathodes \cite{Pierce1975}, whereas polarized positrons are made with either circularly polarized photons \cite{Olsen1959,Riemann2011} or using polarized electrons \cite{Abbott2016}. High-energy storage rings will also polarize beams through the Sokolov-Ternov effect \cite{Sokolov1967}. More compact methods of producing polarized electron bunches in a plasma accelerator have been proposed, either using laser drivers \cite{Wen2019,Wu2019a,Wu2020,Fan2022,Bohlen2023} or beam drivers \cite{Wu2019b,Nie2021,Nie2022}. Here, plasma electrons are injected directly into a wake driven in a pre-polarized plasma, made e.g.~by photodissociating hydrogen halides with a UV laser \cite{Rakitzis2003}.

After generation, the particle spins can evolve during the acceleration process---a topic reviewed in detail by \textcite*{Mane2005}. Three effects must be taken into account \cite{Thomas2020}: the Stern-Gerlach force, which separates the orbits of opposite spins (relevant at low energy); the Sokolov-Ternov effect, whereby the radiation reaction depends on the spin (relevant at high energy and over long timescales; see Sec.~\ref{sec:radiative-damping}); and spin precession in magnetic fields, described by the T--BMT equation, named after \textcite{Thomas1927} and \textcite*{Bargmann1959}:
\begin{equation}
    \label{eq:t-bmt-equation}
    \frac{\dif \hat{\mathbf{s}}_i}{\dif t} = -\frac{q}{m}\left[\Omega_B \mathbf{B} - \Omega_E\frac{\mathbf{v}}{c^2}\times\mathbf{E} - \Omega_v\left(\frac{\mathbf{v}}{c}\cdot\mathbf{B}\right)\frac{\mathbf{v}}{c}\right]\times\hat{\mathbf{s}}_i,
\end{equation}
where the numerical factors for each frequency term are
\begin{equation}
    \Omega_B = a+\frac{1}{\gamma}, \hspace{1em} \Omega_E = a+\frac{1}{\gamma+1}, \hspace{1em} \Omega_v = \frac{a\gamma}{\gamma+1}.
\end{equation}
Here, $a$ is the anomalous magnetic moment of the particle ($a_e = 0.00116$ for electrons), $\mathbf{v}$ and $\gamma$ are the particle's velocity and Lorentz factor, respectively, and $\mathbf{E}$ and $\mathbf{B}$ are the electromagnetic fields it experiences. The strong fields of the plasma wake can therefore lead to spin precession, and if the fields vary sufficiently across the bunch, it can be depolarized. However, if the emittance is sufficiently small, the polarization can be preserved during acceleration \cite{Vieira2011b}, requiring that
\begin{equation}
    \label{eq:polarization-max-emittance}
    \varepsilon_n[\SI{}{\micro\meter}] \ll 188\sqrt{\alpha}\sqrt{\frac{n_0}{\SI{e16}{\per\cm\cubed}}}\sqrt{\frac{\mathcal{E}_0}{\SI{10}{GeV}}}\left(\frac{\sigma_r}{\SI{10}{\micro\meter}}\right)^2,
\end{equation}
where $n_0$ is the plasma density, $\mathcal{E}_0$ is the initial beam energy, $\sigma_r$ is the transverse beam size, and $\alpha$ quantifies the strength of the focusing force ($\alpha=1/2$ in a blowout).

No experimental demonstrations of generation or transport of spin-polarized beams in plasma have yet been performed. While simulations are promising, it is thus still unclear if polarization can truly be preserved in a plasma-accelerator stage. Further, it is not known whether spin polarization can be preserved across multiple stages---this depends on the beam optics, and especially dipole magnets, between the stages (see Sec.~\ref{sec:staging}).


\section{Advanced topics}
\label{sec:advanced-topics}

Several aspects of beam-driven plasma acceleration go beyond the simple picture of a driver, a wakefield and a trailing electron bunch. These topics include positron acceleration (Sec.~\ref{sec:positron-acceleration}), proton drivers and self-modulated wakefields (Sec.~\ref{sec:self-modulated-wakefields}), internal injection (Sec.~\ref{sec:internal-injection}), long-term evolution (Sec.~\ref{sec:long-term-evolution}) and staging (Sec.~\ref{sec:staging}).


\subsection{Positron acceleration}
\label{sec:positron-acceleration}

Acceleration of positrons in plasmas is motivated by the need for more compact electron--positron colliders (Sec.~\ref{sec:applications:colliders}). As an accelerating medium, plasmas are unique in that their nonlinear response to beams of opposite charge is asymmetric (see Sec.~\ref{sec:nonlin_wake} and Fig.~\ref{fig:nonlinear_wakefields}). This is because plasmas are composed of light, mobile electrons, and heavy, effectively immobile ions. The structure of the wakefield is determined by the motion of the plasma electrons, with the plasma ions acting as a near-stationary background. As a result, while the standard blowout regime accelerating the trailing bunch in an ion cavity is particularly favorable for electron acceleration, it is hardly suited for positron acceleration. Indeed, the transverse fields are defocusing inside the ion cavity, and the region where the fields are focusing and accelerating is where plasma electrons cross the propagation axis \cite{Lotov2007}, which is extremely narrow with very nonuniform fields unless beam loading (Sec.~\ref{sec:beam-loading}) or some specific shaping of the beam or the plasma (or both) is employed. 

The challenge for positron acceleration in plasmas, a topic reviewed in detail by \textcite{Cao2024} and \textcite{Joshi2025}, is twofold: (i) to excite a wakefield that provides an optimal region that is simultaneously accelerating and focusing (Sec.~\ref{sec:positrons:with-focusing}), or oppositely that is free of any plasma focusing (Sec.~\ref{sec:without-focusing}); and (ii) to load the wakefield with reasonable charge and energy efficiency without compromising beam quality or becoming unstable. An additional challenge is access to positrons: as an alternative, in-situ generation and injection of positron bunches has been proposed (Sec.~\ref{sec:positrons:in-situ-generation}).

\subsubsection{With plasma focusing}
\label{sec:positrons:with-focusing}

In this section, we consider positron-acceleration schemes that use on-axis plasma electrons for focusing, either unloaded (Sec.~\ref{sec:unloaded}) or loaded (Sec.~\ref{sec:loaded}), both which have been demonstrated experimentally \cite{Blue2003,Corde2015}, as well as proposed schemes that require more complex beam and/or plasma shaping (Sec.~\ref{sec:beam-plasma-shaping}).

\paragraph{Unloaded positron acceleration}
\label{sec:unloaded}

The first question to address is how to excite a wakefield with a sizeable volume that is simultaneously focusing and accelerating for positrons. One suggestion is to rely on the linear regime (Sec.~\ref{sec:lin_wake}) where the drive beam causes only a small perturbation to the plasma-electron density and for which the plasma response is symmetric to beams of opposite charge. The linear wakefield has one quarter of the plasma wave period that is focusing and accelerating for positrons, which is thus appropriate in the unloaded case where the positron beam has negligible influence on the plasma wakefield. The demonstration of positron focusing in the linear regime was in fact the first experiment involving positrons in plasmas~\cite{Ng2001}. 

In the unloaded case, the energy spread cannot be optimized with beam loading (see Sec.~\ref{sec:optimal-beam-loading-energy-spread}), but the use of a modulated plasma density---alternating between focusing and defocusing and between positive and negative slope for the accelerating field---can mitigate the accumulation of energy chirp in the plasma and thus reduce the final energy spread~\cite{Brinkmann2017}. This latter simulation result, with pC-level charge, preserved \si{\micro\m}-level emittance and 0.24\% final energy spread, was demonstrated for electrons but is also applicable to positrons by virtue of the symmetry of the linear regime. It was also found that plasma wakefields linearly driven by electron beams can be used to accelerate positron beams with a density exceeding that of the plasma without compromising beam quality~\cite{Hue2021}. In the unloaded limit, such an approach is however limited to low charge and low efficiency from plasma to positron beam, and because it is also very challenging to achieve a stable driver propagation all the way to depletion in the linear regime~\cite{Hue2020}, it is not the regime best suited for high driver-to-plasma efficiency (Sec.~\ref{sec:drive_plasma_efficiency}). 

\begin{figure}[t]
    \centering
    \includegraphics[width=\linewidth]{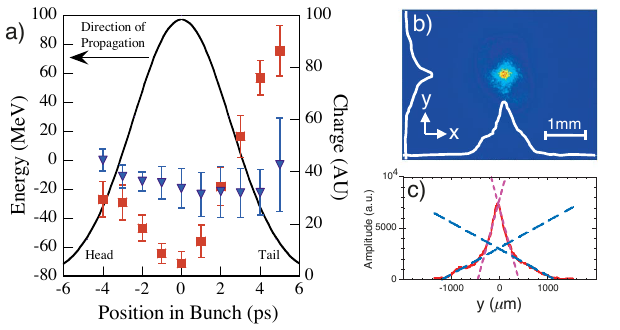}
    \caption{(a) First demonstration of positron acceleration, performed at SLAC. Positrons at the tail (right side) of a single \SI{28.5}{GeV} bunch gained energy when the \SI{1.4}{m}-long plasma of density \SI{1.8e14}{\per\cubic\cm} was on (red error bars) compared to when it was off (blue error bars). (b) A transverse profile of the positron bunch, from a separate measurement with plasma density \SI{0.7e14}{\per\cubic\cm}, shows that (c) a large fraction of the particles (the red line is the $y$-projection) is in a halo (blue dashed lines) instead of the core (purple dashed line), indicating emittance growth from nonlinear focusing. Adapted from \textcite{Blue2003} and \textcite{Muggli2008b}.}
  \label{fig:halo_formation}
\end{figure}

Experimentally, for beam densities exceeding that of the plasma, positron acceleration was observed in an unloaded, nonlinear and nonrelativistic (see Sec.~\ref{sec:nonlin_wake}) plasma wakefield using a single bunch~\cite{Blue2003}. Positrons at the bunch tail were accelerated with a broadband spectrum and with a maximum unloaded gradient of \SI{56}{MV/m}, which was limited at the time by the \SI{700}{\micro\m} rms bunch length and associated low plasma density ($\SI{1.8e14}{cm^{-3}}$) and low beam current [$\Lambda\ll1$, as defined in Eq.~(\ref{eq:normalized-current})]. Although the beam was successfully transported through the plasma~\cite{Hogan2003}, the beam--plasma interaction resulted in halo formation and a degraded beam quality~\cite{Muggli2008b} (Fig.~\ref{fig:halo_formation}).

There are several other schemes (covered in Sec.~\ref{sec:beam-plasma-shaping}) that can generate a sizable volume simultaneously focusing and accelerating for positrons and that are generally well suited for unloaded positron acceleration. For example, for the electron-driven blowout regime, the narrow focusing and accelerating region around the sheath electron crossing point can be considerably extended by the use of a plasma column whose radius is slightly smaller than the blowout radius~\cite{Diederichs2019}, or by the use of an escort electron bunch loading the back of the blowout and resulting in an elongated cavity~\cite{Wang2021}. In the unloaded limit, this elongated cavity has been shown to accelerate positrons with preserved emittance.

\paragraph{Loaded positron acceleration}
\label{sec:loaded}

Energy efficiency and charge are generally important when considering positron acceleration, as these are required for high luminosity per power [Eq.~(\ref{eq:luminosity-per-power})] in colliders (Sec.~\ref{sec:applications:colliders}). However, unloaded positron acceleration has low energy efficiency and is unable to accelerate high charge. Therefore, understanding the physics of positron-loaded plasma wakefields is crucial to model properly plasma-based positron accelerators and assess their performance. 

\begin{figure}[t]
    \centering
    \includegraphics[width=\linewidth]{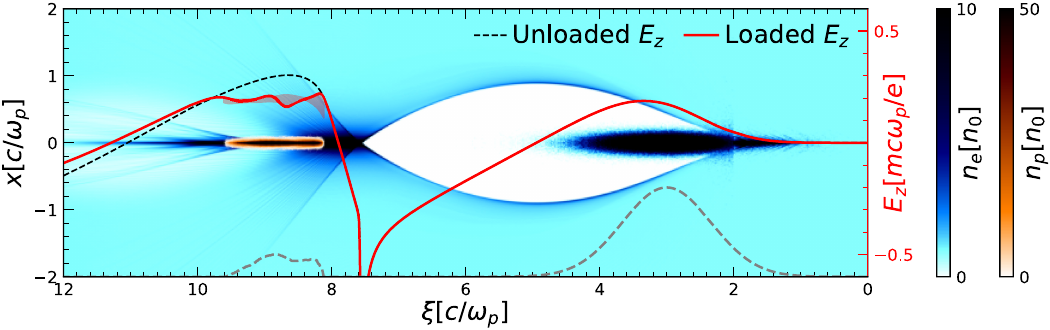}
    \caption{Simulated acceleration of positrons (orange color map; density $n_p$) in a loaded nonlinear wake driven by electrons (blue color map; density $n_e$), with electron and positron densities normalized to the plasma density $n_0$. The positron beam loading flattens the on-axis $E_z(\xi)$ (red line; the red shaded area shows the range within $\pm3$ rms beam sizes) and forms a plasma-electron filament providing focusing for the positron beam. From \textcite{Zhou2025} (CC BY 4.0).}
  \label{fig:posi_nonlinear}
\end{figure}

As for electrons, positron beam loading modifies the longitudinal profile of the accelerating field $E_z(\xi)$, as discussed in Sec.~\ref{sec:evolution-longitudinal}. Such longitudinal positron beam loading was shown experimentally by the correlation between energy gain, energy spread and positron charge in the two-bunch positron experiment of \textcite{Doche2017}, which successfully accelerated distinct positron bunches in loaded plasma wakefields spanning linear to nonlinear regimes. In the single-bunch experiment of \textcite{Corde2015}, a flattening of $E_z(\xi)$ in the second half of the bunch resulted in a low-energy-spread accelerated peak in the positron spectrum [see Figs.~\ref{fig:beam_loading_posi}(b--c)]. The nonlinear plasma wakefield, excited by the front half of the bunch, was thus self-loaded by the rear half of the bunch. With a bunch length in the range 30--\SI{50}{\micro\m}, the beam current was much higher than in previous experiments, moving from the nonrelativistic regime (for which $\Lambda\ll1$) to the relativistic regime (for which $\Lambda\sim1$) and achieving a loaded accelerating gradient of \SI{3.8}{GV/m} in a plasma of density $\SI{8e16}{cm^{-3}}$~\cite{Corde2015}. A similar flattening of $E_z(\xi)$ with a positron load located just behind the first blowout cavity (see Fig.~\ref{fig:posi_nonlinear}) was shown to be possible in an electron-driven blowout wake~\cite{Zhou2025}.

Contrary to electron beam loading in the blowout regime, positron beam loading also has a strong influence on the transverse fields because of the high mobility of the plasma electrons flowing through the positron bunch. These plasma electrons, necessary to provide plasma focusing, have a role that is analogous to the ion motion (discussed in Sec.~\ref{sec:ion-motion}) for electron acceleration, except that their detrimental effect appears at much lower beam density because $m_e\ll m_i$ [see the role of $m_i$ in the phase advance $\Delta\phi_i$ of Eq.~(\ref{eq:ion-motion})]. Electron motion induced by the positron load within the bunch itself is in fact the key challenge that can limit the performance of positron acceleration when relying on plasma focusing. Positron beam loading can also have benefits, e.g., its ability to modify also the transverse fields can turn a defocusing region (when unloaded) into a focusing region (when loaded) favorable for positron acceleration. This transverse beam-loading process, central to the experimental demonstration of positron acceleration in a self-loaded plasma wakefield~\cite{Corde2015} (Fig.~\ref{fig:beam_loading_posi}), opens the way to accelerating positrons in a loaded electron-driven blowout wake~\cite{Zhou2025}, something thought to be impossible when only considering the unloaded wake. This new path relies on an on-axis electron filament formed and maintained by the positron load, thus providing a region of the loaded wake that is simultaneously focusing and accelerating for positrons---a region that did not exist in the unloaded wake.

As we push the performance of positron acceleration towards higher charge and higher efficiency, the positron load is increased, which enhances electron motion within the bunch and can compromise the beam quality. Such electron motion typically leads to nonuniform charge and current density, and thus to a radially nonlinear and slice-dependent transverse force. By virtue of the Panofsky-Wenzel theorem [Eq.~(\ref{eq:panofsky-wenzel-theorem})]~\cite{Panofsky1956}, this also translates to a radially dependent accelerating field. This radially nonlinear and slice-dependent transverse force impacts emittance, although it is generally possible to mitigate emittance growth by quasi-matching or slice-by-slice matching the beam to be as close as possible to a self-consistent radial equilibrium~\cite{Benedetti2017, Diederichs2020, Hue2021}. The radially dependent accelerating field impacts the uncorrelated energy spread, as positrons located at different radial positions do not experience the same accelerating field (see Sec.~\ref{sec:optimal-beam-loading-energy-spread}). When increasing the positron load, the uncorrelated energy spread can easily exceed 1\%, and as a result a trade-off between energy efficiency and beam quality needs to be found~\cite{Hue2021}. It has been shown that for \si{\micro\m}-level emittances, linear, moderately nonlinear with $n_d/n_0 \approx 1$--2~\cite{Hue2021} and strongly nonlinear regimes with $n_d/n_0\gg1$~\cite{Zhou2025} can reach tens-of-percent energy-transfer efficiencies with percent-level uncorrelated energy spread and limited emittance growth. Higher nonlinearity allows for higher accelerating gradient and higher positron charge. In \textcite{Zhou2025}, a model for nonlinear positron beam loading was used to guide the optimization of the positron current profile and flatten $E_z(\xi)$ (see Fig.~\ref{fig:posi_nonlinear}), resulting in a total energy spread (correlated and uncorrelated) of about 2\% rms.

Note that while positron acceleration has been demonstrated experimentally, all have used positrons to drive the wake. No electron-driven positron acceleration scheme, including that proposed by \textcite{Zhou2025}, has yet been experimentally verified.

\paragraph{Beam and plasma shaping}
\label{sec:beam-plasma-shaping}

A natural strategy to find new schemes with conditions appropriate for positron acceleration is to shape the beam, the plasma or both. These schemes are typically designed to provide focusing and acceleration for positrons, and can be optimized to tolerate a substantial positron load. An exhaustive review of positron-acceleration schemes can be found in \textcite{Cao2024}, who identified three main directions with the potential to reach a reasonable luminosity per power [see Eq.~(\ref{eq:luminosity-per-power})]: (i) using donut-shaped drivers, (ii) finite-radius plasma columns, and (iii) asymmetric drivers in hollow channels, all discussed below. Other proposed schemes include using thin, warm, hollow plasma channels \cite{Silva2021} or using a laser-augmented blowout and ring-shaped positron beams \cite{Reichwein2022}. None of these schemes have been tested experimentally.

Donut-shaped drivers offer a straightforward way to provide plasma electrons on axis for focusing, while being operable in the strongly nonlinear regime. Indeed, whether the donut-shaped driver is a laser~\cite{Vieira2014b, Mendonca2014, Wang2020,Yu2014}, an electron beam~\cite{Jain2015, Vieira2016, Jain2019} or a non-neutral ``fireball" (i.e., electron--positron beam)~\cite{Silva2023}, the key principle is that inner plasma electrons are pushed inward or guided through the hollow core of the donut, forming a population used for the focusing of the positron bunch. Outer plasma electrons are pushed away and are then pulled back by the ions, forming a structure similar to the blowout regime, except for the presence of plasma electrons near the axis. The motion of these outer plasma electrons is responsible for the generation of the accelerating field. This donut-shaped driver regime thus benefits from the decoupling of focusing and acceleration, which are handled by two distinct populations of plasma electrons. By operating in the strongly nonlinear regime, high positron charge ($>\SI{100}{pC}$) can be accelerated at very high gradient ($>\SI{10}{GV/m}$), although with limited energy-transfer efficiency ($\eta_\mathrm{p \to t}\lesssim 5\%$)~\cite{Hue2021,Cao2024}. One important challenge lies in the propagation of the donut driver, to ensure it maintains its shape with a stable wakefield until its depletion, without being prone to filamentation instabilities or to on-axis collapse~\cite{Su1987,Pathak2016,Jain2019}.

\begin{figure}
    \centering
    \includegraphics[width=0.85\linewidth]{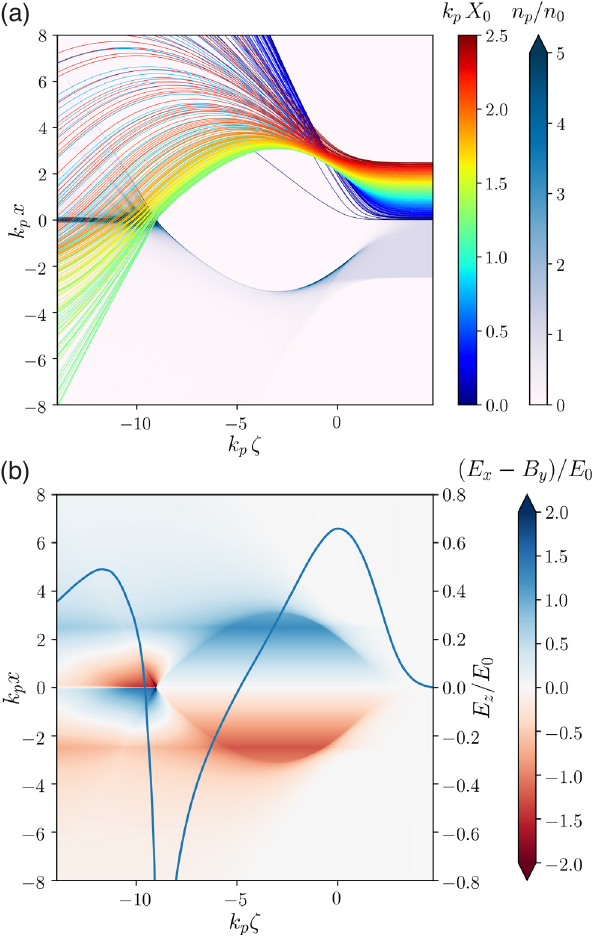}
    \caption{The finite-radius plasma column: (a) plasma-electron trajectories for different initial radii (color coded) in a plasma column (blue color map) of radius $k_pR_p=2.5$. (b) The focusing (red--blue color map) and on-axis accelerating fields (blue line) are shown, indicating a region (left) that is both accelerating and focusing for positrons. This simulation uses an electron driver with $\tilde{Q}\simeq 44$, $k_p\sigma_r=0.4$ and $k_p\sigma_z=\sqrt{2}$. From \textcite{Diederichs2019} (CC-BY 4.0).}
  \label{fig:finite_radius}
\end{figure}

Another strategy is to modify the plasma shape to a finite-radius plasma column (surrounded by neutral gas in practice), whose radius is slightly smaller than the blowout radius, $R_p\lesssim R_b$. This leads to a focusing force decreasing with $r$ outside the plasma column. Plasma electrons, which usually circulate in a thin sheath around the blowout cavity and cross the axis in a narrow region at the rear of this cavity, now follow different trajectories due to this decreasing focusing force for $r>R_p$. As a result, these plasma electrons cross the axis at different longitudinal positions $\xi$ [see Fig.~\ref{fig:finite_radius}(a)]. This decoherence of plasma electron trajectories and spreading of their crossing positions is the key principle behind the finite-radius plasma-column scheme~\cite{Diederichs2019}, making it possible to have an on-axis filament of plasma electrons providing focusing for positrons over a wide region where $E_z$ is accelerating, as shown in Fig.~\ref{fig:finite_radius}(b) for $k_p\xi\lesssim-10$. This scheme can accelerate a positron charge of about \SI{50}{pC} with high quality ($<\SI{1}{mm.mrad}$ preserved emittance, $<1\%$ energy spread) and at high gradient ($>\SI{10}{GV/m}$) \cite{Diederichs2020}, as well as with a stable propagation for both the drive and trailing bunches~\cite{Diederichs2022a,Diederichs2022b}. A non-zero plasma temperature can also help in preserving beam quality, by radially broadening the plasma electron filament and improving the radial uniformity of the accelerating field~\cite{Diederichs2023}. Importantly, because the focusing of different positron slices is provided by different plasma electrons, the limit of electron motion (analog to ion motion in electron acceleration) discussed in Sec.~\ref{sec:loaded} can be overcome. Indeed, here plasma electrons do not flow longitudinally through the entire positron bunch, but cross through it transversely [see plasma-electron trajectories crossing the axis in Fig.~\ref{fig:finite_radius}(a)]. This advantage can however pose a challenge for the energy-transfer efficiency, as the decoherence of plasma-electron trajectories limits the wakefield energy available for extraction by the positron bunch. Numerical results have demonstrated efficiencies $\eta_\mathrm{p \to t}$ of up to 5\% in this regime~\cite{Diederichs2019,Diederichs2020}.

Nonlinearly driven hollow-channel plasma wakefields can also provide focusing (and acceleration) to a positron bunch, if plasma electrons from the channel wall are made to move inward towards the channel axis. Although this regime relies on a hollow channel (Sec.~\ref{sec:hc_linear}), it starkly differs from the hollow channel that is free of any focusing and thus very susceptible to transverse instabilities. For the driver to be stable in the hollow channel, an asymmetric electron beam with $\sigma_x>\sigma_y$ can be used~\cite{Zhou2021}. This beam drives a quadrupole mode in the channel, defocusing in $x$ and focusing in $y$, and eventually splits into two beamlets propagating through the channel boundary. At this point, exposed ions from the channel can focus the beam along $x$ and balance the defocusing quadrupole mode, leading to a stable driver propagation. A positron bunch can then be accelerated, with focusing provided by inward-moving plasma electrons. This plasma focusing circumvents the transverse instability limitation of linear hollow channels. Simulations show that \SI{\sim500}{pC} of positron charge can be accelerated with high energy-transfer efficiency, $\sim$30\%, at a gradient of \SI{\sim5}{GV/m}. The challenge for this regime is related to the beam quality, with few-percent energy spreads and emittances above \SI{50}{mm.mrad} demonstrated in simulations so far~\cite{Zhou2021}.

\subsubsection{Without plasma focusing (hollow channel)}
\label{sec:without-focusing}
Although the linear hollow-channel plasma accelerator was proposed in the 1990s~\cite{Chiou1996,Chiou1998,Lee2001}, it took two decades to realize the concept experimentally due to the challenge of making a hollow plasma channel. Early attempts resulted in plasmas that had an on-axis density depression~\cite{Marsh2003}, but did not meet the definition of a hollow channel plasma. Now, there is a growing body of work on generating plasma channels with deep density depressions [see e.g.~\textcite{Kirby2009b} or \textcite{Shalloo2018}], but the only demonstrated technique for creating hollow channels with zero on-axis plasma density is with the use of high-order Bessel-shaped laser pulses that ionize plasma annuli~\cite{Fan2000}. This technique was proposed at FACET~\cite{Kimura2011} because the facility offered a high-power laser, a low-ionization-potential vapor source and short positron bunches (see Sec.~\ref{sec:experimental:facilities:slac}). The laser pulse is shaped using a diffractive optic called a \textit{kinoform}~\cite{Andreev1996}.

\begin{figure}[t]
    \centering
    \includegraphics[width=0.95\linewidth]{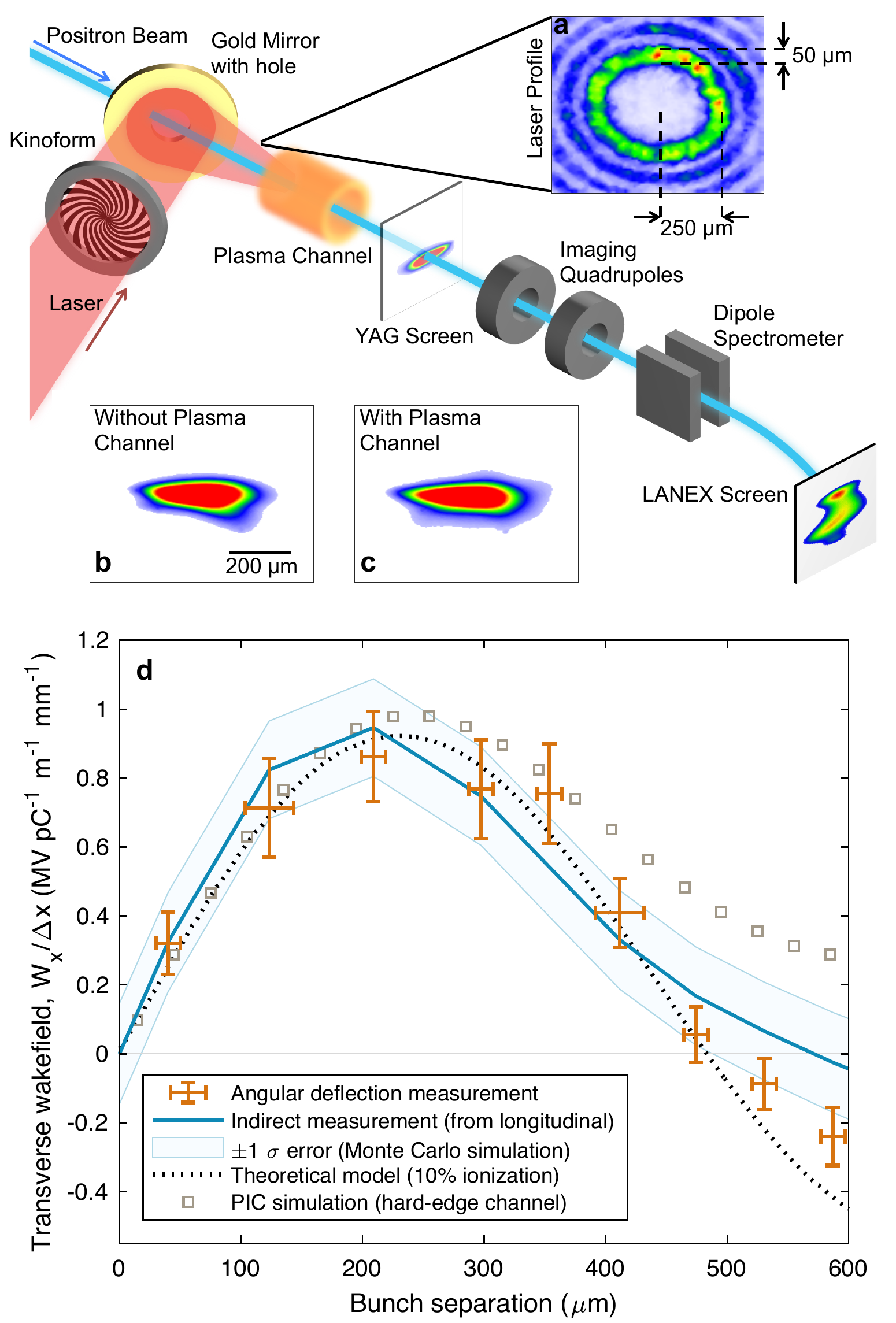}
    \caption{A laser passes through a kinoform, coupled to the beam axis by a holed mirror. Inset (a) shows the transverse laser profile upstream of the plasma source, making a hollow plasma channel of radius $\SI{250}{\micro\m}$ and density \SI{3e15}{\per\cubic\cm}. Inset (b) shows the downstream transverse profile of the positron beam  with no plasma present (laser off), whereas (c) shows the profile when the beam propagated through the hollow channel (laser on). The similarity of the profiles indicates that there are no net focusing forces in the channel. A screen downstream of a dipole measures the beam-energy spectrum. (d) Transverse wakefield from direct measurements (red crosses) and indirectly estimated via the Panofsky-Wenzel theorem (blue line) versus bunch separation measured using electro-optic sampling. These measurements are compared to theory (dotted black line) and PIC simulations (gray squares). Adapted from \textcite{Gessner2016a} (CC-BY 4.0) and \textcite{Lindstrom2018} (CC-BY 4.0).}
    \label{fig:hcpwfa}
\end{figure}

Figure~\ref{fig:hcpwfa} illustrates the experimental setup used for a demonstration performed by \textcite{Gessner2016a}. Starting from a positron beam aligned to the high-order Bessel profile, the laser offset was scanned in both transverse planes with respect to the positron beam, which induced transverse kicks to the beam~\cite{Schroeder1999}. The deflections were measured on screens and BPMs downstream and used to reconstruct the shape of the plasma channel, demonstrating that a truly hollow channel was successfully generated. The driving positron bunch induced a longitudinal wakefield in the channel [Eq.~(\ref{eq:hollow-channel-longitudinal-wakefield})], causing it to lose up to \SI{20}{MeV} in a \SI{\sim10}{cm} long channel~\cite{Gessner2016a}. In a separate experiment, a trailing positron bunch sampled this wakefield to map the accelerating and decelerating phases~\cite{Gessner2023}, finding a peak accelerating gradient of \SI{70}{MV/m}. The wavelength of the wakefield was observed to be longer than the na{\"i}ve theoretical prediction, possibly due to non-sharp channel boundaries---a slow transition from the un-ionized channel center to the fully ionized wall increases the wavelength of the on-axis longitudinal field~\cite{Shvets1996}.

These experiments also highlighted the large transverse wakefield [Eq.~(\ref{eq:hollow-channel-transverse-wakefield})], which causes a beam-breakup instability that significantly limits the applicability of the hollow channel~\cite{Schroeder1999}. Correlating the relative beam--channel offset (from a random transverse laser jitter) with the resulting angular deflection of the outgoing trailing positron bunch provided a measurement of this transverse wakefield. Using a range of different longitudinal driver-to-trailing-bunch separations, the longitudinal shape of the transverse wakefield was mapped [see Fig.~\ref{fig:hcpwfa}(d)], showing reasonable agreement between linear theory, experiment and PIC simulations. The maximum amplitude of the transverse wakefield was measured to be $0.86\pm0.13$ \si{MV.pC^{-1}.m^{-1}.mm^{-1}}~\cite{Lindstrom2018}, which is over 3000 times stronger than that in a high-frequency (\SI{12}{GHz}) rf cavity structure~\cite{Zha2016}. Traditional mitigation strategies include external focusing~\cite{Gai1997,Li2014} and BNS damping~\cite{Balakin1983}, but would necessitate large energy spreads $>10\%$. Alternative mitigation strategies are needed, but it is an open question as to whether or not mitigation can be achieved without relying on plasma-electron focusing. For example, the proposal by~\textcite{Zhou2021} relies on the focusing provided by plasma electrons from the channel wall flowing inward and into the positrons to stabilize the bunch (see Sec.~\ref{sec:beam-plasma-shaping}).

\subsubsection{In-situ positron generation and injection}
\label{sec:positrons:in-situ-generation}

Given the limited availability of conventional positron sources in accelerator test facilities,  alternative techniques to provide positrons for injection into a plasma accelerator have been considered. Using a converter foil directly within the plasma or prior to it, a pair of co-propagating electron bunches can generate positrons overlapping in space with the incident electrons~\cite{Wang2008}. In the converter foil, electrons generate bremsstrahlung gamma rays that convert into electron-positron pairs via the Bethe-Heitler process, thus providing the source of positrons. Once in the plasma, the electron and positron bunches experience either a focusing or defocusing force from the plasma wakefield, where defocusing results in loss of the bunch. Choosing an appropriate longitudinal separation between the two electron bunches, only a leading electron bunch and a trailing positron bunch remain focused, with the electron bunch losing energy to the wake and the positron bunch being accelerated by it~\cite{Wang2006,Wang2008,Wang2009,Fujii2019,Amorim2023}. Despite a trade-off between the positron charge and emittance obtained with this technique, such in-situ positron generation and injection provides a more affordable means of experimentally testing positron acceleration in plasmas. A similar method can also be employed in laser-driven plasma accelerators, where an electron bunch accelerated in a first stage is used to produce positrons in a converter foil, subsequently injected and accelerated in a second laser-driven stage~\cite{Sahai2018,Amorim2023,Terzani2023}. Going beyond positron production in converter targets, the Breit-Wheeler process has also been considered as a means to generate positrons from ultrahigh-intensity lasers~\cite{Martinez2023,Liu2022b,Sugimoto2023}.

\subsection{Proton drivers and self-modulated wakefields}
\label{sec:self-modulated-wakefields}

The vast majority of plasma-wakefield experiments utilize electron bunches as the drive beam. Recently, attention has turned towards using proton beams as drivers of plasma wakefields. This is because proton beams can be accelerated to much higher energies and with much higher bunch charges than electron beams. The total stored energy in these proton beams is in the kJ--MJ range, whereas electron beam drivers typically contain a few joules. If even a fraction of the proton beam's energy is transferred to a trailing electron bunch, it will be possible to accelerate the electron beam to TeV-scale energies in a single plasma stage~\cite{Caldwell2009}.

\begin{figure}[t]
    \centering
    \includegraphics[width=\linewidth]{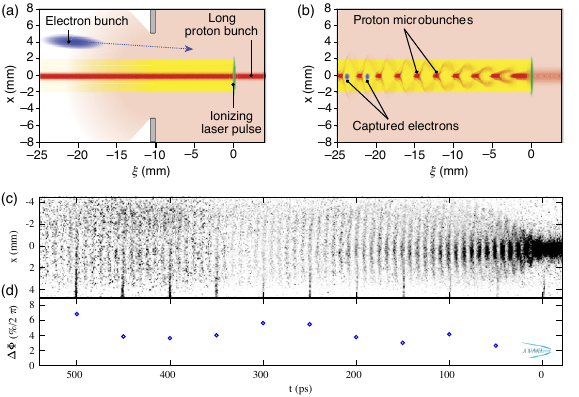}
    \caption{Schematic of electron injection into a self-modulated wakefield: (a) a laser pulse (RIF) ionizes a plasma within a long proton bunch, (b) modulating it into a train of short bunches spaced by the plasma wavelength. An electron bunch is injected off-axis, from which some electrons are captured and accelerated. (c) Experimental streak-camera image of a self-modulated proton bunch; the bunch head is to the right. This composite image, made by stitching together several images left-to-right, is made possible by (d) the few-percent phase reproducibility of the self-modulated wakefield ($\Delta\Phi$ is the rms phase variation), which is seeded by the laser. Here the plasma density is \SI{1.8e14}{\per\cm\cubed}. Adapted from \textcite{Adli2018} and \textcite{Batsch2021} (CC-BY 4.0).}
  \label{fig:stitched_ssm}
\end{figure}

The central challenge associated with the proton beam-driven approach is that the proton bunches are much longer than the plasma wavelength at relevant plasma densities. The proton bunch length $\sigma_z$ is determined by the rf frequency employed by the proton synchrotron ring, which is of order \SI{100}{MHz} [$
\mathcal{O}(\lambda_{\mathrm{rf}})\approx 1$ m]. On the other hand, even the longest plasma wavelengths $\lambda_p$ used in PWFA are less than \SI{1}{mm}: with $\sigma_z \propto \lambda_{\mathrm{rf}} \gg \lambda_p$, the proton bunch cannot drive a high-amplitude wakefield. The solution to this challenge is to modulate the proton bunch into micro-bunches spaced at the plasma frequency~\cite{Kumar2010} (covered in Sec.~\ref{sec:self-modulated-wakefields:formation}), which resonantly drive a wakefield \cite{Lotov1998} into which electrons can be injected (covered in Sec.~\ref{sec:self-modulated-wakefields:capture}), as depicted in Figs.~\ref{fig:stitched_ssm}(a--b). For a detailed review of proton-driven PWFA, see \textcite{Adli2016c}.

\subsubsection{Formation of self-modulated wakes}
\label{sec:self-modulated-wakefields:formation}

In the self-modulation process, the proton beam drives a small-amplitude wakefield that acts back on the bunch: the protons are alternately focused and defocused with a longitudinal separation determined by the plasma frequency. The focused portions of the bunch enhance the strength of the wakefield, leading to a stronger focusing--defocusing pattern. The growth rate (or $e$-folding frequency) of this exponential process is given by~\cite{Pukhov2011,Schroeder2011}
\begin{equation}
    \Gamma = \frac{3\sqrt{3}}{4}\omega_p\left(\frac{n_b m_e}{2 n_0 m_p \gamma_b}\frac{\zeta}{ct}\right)^{1/3},
\end{equation}
where $\zeta=v_bt-z$ is the longitudinal position within the bunch, $n_b$ is the bunch density, $m_p$ the proton mass, and $\gamma_b$ is the beam's Lorentz factor. Note that the growth rate is higher toward the back of the bunch (scales as $\zeta^{1/3}$), but decreases during propagation (scales as $t^{-1/3}$). The process continues until a train of micro-bunches is formed and the modulation saturates. While each micro-bunch contains only a small fraction of the total bunch charge, together they resonantly drive a large-amplitude wake~\cite{Adli2019,Turner2019}. This effect has also been observed with long ($\sigma_z \gg \lambda_p$) electron bunches propagating in plasma~\cite{Fang2014,Gross2018} as well as with laser drivers~\cite{Joshi1981,Ting1997}. 

When the self-modulation is seeded by Schottky noise from the proton drive beam, it is referred to as the \textit{self-modulation instability} (SMI)~\cite{Lotov2013}. Alternatively, the modulation may be induced by a seed pulse or structure to produce a phase-reproducible wakefield~\cite{Schroeder2013}, in which case it is referred to as \textit{self-seeded modulation} (SSM). SSM seeding can be performed using drivers with sharp features in their current profile~\cite{Fang2014,Gross2018}, a ``relativistic ionization front" (RIF) such as a laser co-propagating and ionizing the plasma within the driver ~\cite{Batsch2021,Verra2023}, or a dense electron beam propagating in front of the driver~\cite{Verra2022,vanGils2026}. Figures~\ref{fig:stitched_ssm}(c--d) show the effect of the self-modulated wakefield on the proton driver when the wake is seeded by a RIF. If the RIF is more than $2\sigma_z$ ahead of the beam centroid, the Schottky noise becomes the dominant seed and the wakefield phase is no longer reproducible~\cite{Batsch2021}. Accurate prediction of the wakefield phase is critical for successful injection of a trailing electron bunch into the self-modulated wake~\cite{Lotov2014,Olsen2018}, which is only achievable through the SSM process. Additionally, an externally-seeded wake suppresses the deleterious hosing instability~\cite{Schroeder2013,Vieira2014} (Sec.~\ref{sec:hosing-instability}). Hosing can also be seeded or suppressed by controlling the relative alignment of an electron seed bunch to the proton bunch~\cite{Nechaeva2023}.

For both seeded and unseeded wakefields, the growth rate of the wake amplitude depends on the longitudinal position within the bunch $\zeta$ and the propagation distance $ct$. Since the wakefield phase changes as the amplitude changes, and differently so at each $\zeta$ and $ct$, there is a corresponding change in the phase velocity of the wake, as given by~\cite{Pukhov2011,Schroeder2011}
\begin{equation}
\label{eq:phase_velocity_self_modulation}
    v_\phi = v_b\left[1-\frac{1}{2}\left(\frac{n_b m_e}{2 n_0 m_p \gamma_b}\frac{\zeta}{ct}\right)^{1/3}\right].
\end{equation}
The phase velocity may be considerably less than the bunch propagation speed $v_b \approx c$. This poses a challenge for accelerating electrons in the self-modulated plasma wakefield, as the electrons will reach velocity close to $c$ almost instantly and transition between accelerating and decelerating phases of the wake (see Sec.~\ref{sec:dephasing}). In addition, the wake amplitude is predicted to decay after the self-modulation process saturates~\cite{Lotov2011}.

To address issues related to the varying phase velocity and decaying amplitude of the self-modulated plasma wakefield, two approaches have been proposed. First, longitudinal plasma-density gradients have been studied~\cite{Schroeder2012,Lotov2013b} and demonstrated~\cite{Braunmller2020,MoralesGuzmn2021} to stabilize the phase velocity of the wake. Second, an abrupt step in the density gradient has been proposed to ``freeze" the wake amplitude near its maximum~\cite{Lotov2011,Lotov2015,Petrenko2016}. Steps in the plasma density have also been proposed to de-tune the hosing instability~\cite{Moreira2023}.


\subsubsection{Capture and acceleration of electrons}
\label{sec:self-modulated-wakefields:capture}

Beyond the phase reproducibility of the wake, there are challenges associated with electron beam injection into the wake due to defocusing fields in the ramps and radial edges of the plasma~\cite{Lotov2014,Gorn2018}. Off-axis injection [as shown in Fig.~\ref{fig:stitched_ssm}(a)] avoids the deleterious fields at the entrance of the plasma~\cite{Caldwell2016}, but introduces emittance growth due to the beam entering the wake with transverse momentum. This method was successfully employed in the first experimental demonstration of electron acceleration in a proton-driven wakefield~\cite{Adli2018} up to \SI{2}{GeV} (see Fig.~\ref{fig:awake-acceleration}), in the AWAKE facility at CERN (Sec.~\ref{sec:experimental:facilities:awake}). Alternatively, on-axis injection is possible if the self-modulation has fully saturated~\cite{Olsen2018}. AWAKE plans to demonstrate this concept by separating their plasma source into a self-modulation stage for the proton bunch only, followed by electron injection into an acceleration stage using the fully modulated proton beam~\cite{Gschwendtner2022,vanGils2025}. In this scenario, it is possible to preserve the emittance of a high-charge electron bunch~\cite{Olsen2018}, which superimposes a blowout wake on the self-modulated wake, but at the expense of increased energy spread due to heavy beam loading~\cite{Farmer2018} (see Sec.~\ref{sec:optimal-beam-loading-energy-spread}). Finally, it may be possible to increase the charge and energy efficiency by accelerating a train of electron bunches, spaced by a multiple of the plasma wavelength and interleaved with the proton bunch train \cite{Farmer2025}.


\subsection{Internal injection of plasma electrons}
\label{sec:internal-injection}

The injection of electrons into a plasma wakefield from inside the plasma is a common technique in LWFA experiments, but is less explored experimentally in beam-driven PWFAs. This is because self injection or wavebreaking (Sec.~\ref{sec:plasma_temperature}) cannot easily occur in beam-driven PWFAs when the plasma wave travels at the speed of the drive beam (i.e., very close to $c$). Thus electrons from the plasma, which need to reach the wake velocity within the first wakefield period to be injected, cannot achieve the required energy for wavebreaking. This can be seen as an inherent advantage of PWFAs, as unwanted trapping of charge from self-injection, or ``dark current" \cite{Manahan2016}, is typically absent. 

A variety of methods have been proposed to achieve trapping of electrons or trigger wavebreaking in a beam-driven plasma cavity. These schemes can be broken down into two categories: ionization of electrons directly within the wakefield (Sec.~\ref{sec:ionization-injection}); or by reducing the phase velocity of the wake such that streaming electrons traveling close to the speed of light can intercept the back of the elongating wake, so-called ``downramp injection" (Sec.~\ref{sec:downramp-injection}). Other schemes also exist, including using a pulsed magnetic dipole field to alter the trajectory of streaming electrons \cite{Vieira2011}. 

By promising to produce electron beams with transverse normalized emittances orders of magnitude smaller than those produced by rf photocathode guns \cite{Li2013}, such plasma-based internal injection represents an attractive prospect in ultra-low-emittance applications, e.g.~at FEL facilities or linear colliders (Sec.~\ref{sec:applications}).

\subsubsection{Ionization injection}
\label{sec:ionization-injection}

\begin{figure}[t]
    \centering\includegraphics[width=0.95\linewidth]{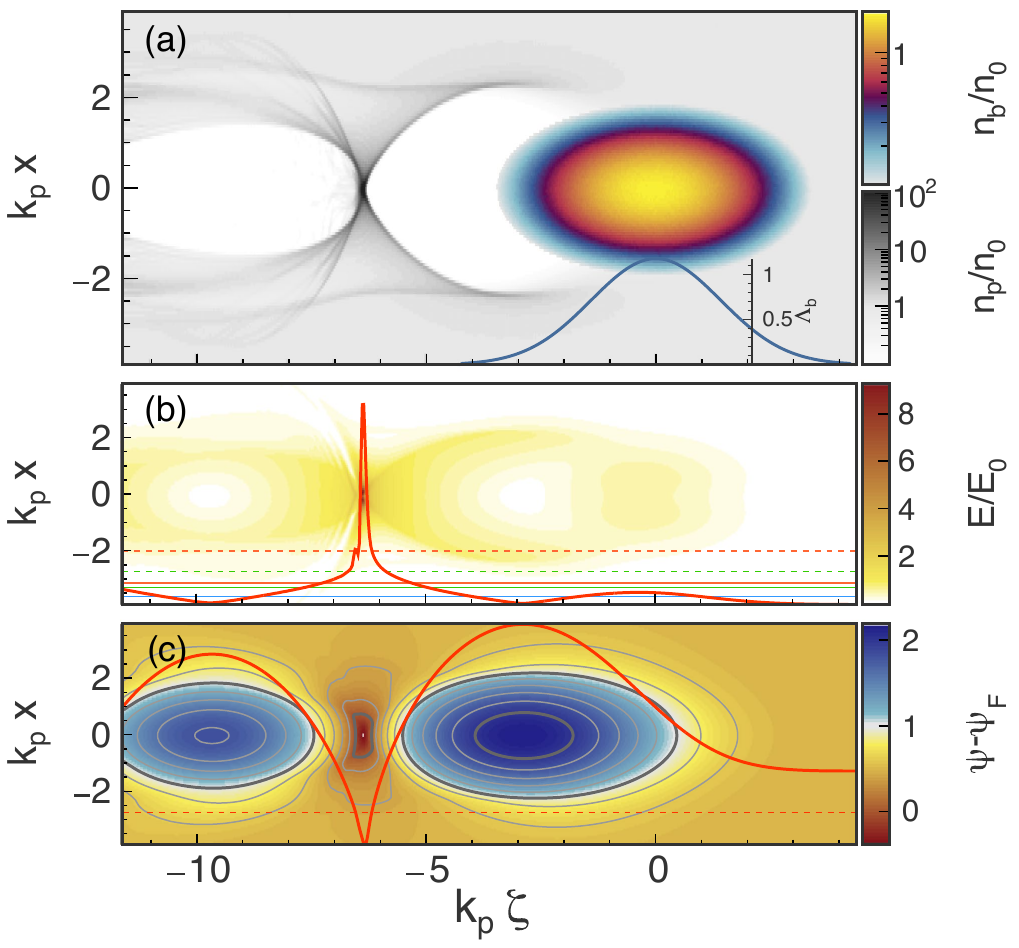}
    \caption{PIC simulation, performed with OSIRIS 3D, of a high-current (\SI{10}{kA}) electron driver traversing a plasma, showing (a) the normalized beam current (blue line) and electron density of the beam and plasma (multicolored and black--white color map, respectively). (b) The electric-field magnitude (brown color map) and corresponding value 0.1$k_p^{-1}$ off-axis (red line) can be compared to the first and second (solid and dashed line) ionization levels of He and Ne (orange and green, respectively). (c) The wake pseudo-potential (brown--blue color map) and corresponding on-axis value (red line) indicate regions from which trapping is possible (blue). Adapted from \textcite{MartinezdelaOssa2015}.}
    \label{fig:trappot}
\end{figure}

In any of the ionization-injection schemes discussed in this section, it is insufficient for electrons to simply be released inside the wakefield; they must also be trapped by the potential well of the wake. A sufficient condition for trapping is met when the velocity of the plasma electrons, $v_e$, exceeds the wake phase velocity, $v_{\phi}$. However, a more general trapping condition can be found in terms of the pseudo-potential $\psi=\phi-v_\phi a_z/c$ (introduced in Sec.~\ref{sec:nonlin_wake}), and is valid even when $v_\phi \to c$. From the Lorentz-force equation, the Hamiltonian for a test electron with Lorentz factor $\gamma$ is $H = \gamma m_ec^2 - e\Phi$, where $\Phi$ is the scalar electric potential. Electron injection can then be modeled by considering the dynamics of a test electron in a quasistatic plasma wakefield (defined in Sec.~\ref{sec:nonlin_wake}), for which $H$ depends on $z$ and $t$ only via $\xi=z-v_\phi t$. As a consequence of the quasistatic approximation ($\partial_t=-v_\phi\partial_\xi=-v_\phi\partial_z$), we have \cite{Mora1996}
\begin{align}
    &\frac{\dif H}{\dif t} = \frac{\partial H}{\partial t}=-v_\phi\frac{\partial H}{\partial \xi} = -v_\phi\frac{\partial H}{\partial z}= v_\phi \frac{\dif P_z}{\dif t},\\
    \label{Hamiltonian1}
    &\implies \frac{\dif }{\dif t}(H-v_\phi P_z)=0,
\end{align}
where $P_z = p_z - eA_z$ is the longitudinal component of the electron's canonical momentum and similarly $A_z$ of the vector potential $\mathbf{A}$. The quantity $H - v_{\phi} P_z$ is thus a constant of motion, which can be rewritten as
\begin{equation}\label{Hamiltonian2}
\gamma - v_{\phi} \frac{p_z}{m_ec^2} - \psi = \mathrm{constant}.
\end{equation}
Under the assumption that the electrons are released (or born, following ionization) from rest ($\gamma_i=1$, $p_{zi}=0$ initially) with a pseudo-potential $\psi_i$, the right-hand side of Eq.~(\ref{Hamiltonian2}) is $1-\psi_i$. For a wake traveling at the speed of light ($v_\phi=c$), Eq.~(\ref{Hamiltonian2}) and the inequality $\gamma>p_z/m_ec $ forbid the test electron to move into regions of space where $\psi \leq \psi_i-1$. For strong plasma wakes with pseudo-potential variation exceeding 1, this condition can effectively confine and trap a test electron inside regions where $\psi > \psi_i-1$ or equivalently $\Delta \psi = \psi_i - \psi < 1$. Physically, the test electron initially recedes towards the rear of the wake and is injected as $\Delta \psi \to 1$,  $p_z\to \gamma m_e c$ and $v_e\to c$, where it essentially moves at the same speed as the wake and its position asymptotically converges towards its final location $\xi_f$ defined by $\psi(\xi_f)=\psi_f=\psi_i-1$. The trapping condition can thus be stated as the existence of initial and final positions such that $\psi_i-\psi_f=1$ \cite{MartinezdelaOssa2015}. Figure~\ref{fig:trappot} illustrates the pseudo-potential well of a beam-driven wake sufficiently deep to trap electrons ionized within it.

In ionization injection, the electrons are released into the plasma wake from a previously un-ionized gas component or ionization level. This is typically achieved by generating a plasma with a low-ionization-threshold (LIT) gas species, in which a beam drives a wake. Electrons may then be born inside the wake when the ionization threshold of a high-ionization-threshold (HIT) component of the plasma is overcome. This concept was first proposed \cite{Umstadter1996} and demonstrated in LWFAs \cite{Pak2010,McGuffey2010,Clayton2010,Pollock2011}. Ionization injection can be achieved in PWFA by using the electric fields of the wake, beam, or a high-intensity laser, as discussed below.

\paragraph{Wakefield-triggered injection}
\label{sec:ionization-injection:wakefield}
The electric field of the wake can itself be strong enough to ionize and trap electrons, specifically at the rear of the ion cavity \cite{MartinezdelaOssa2013,MartinezdelaOssa2015}, as shown in Fig.~\ref{fig:trappot}(b). A high-peak-current beam ($\gtrsim$\SI{10}{kA}) is required in order to generate wakefields sufficient for this scheme. For high beam quality, the area of injection needs to be restricted to the back of the wake period such that unwanted trapping of electrons ionized by the driver fields can be avoided (discussed in Sec.~\ref{sec:ionization-injection:beam} below). This can be achieved by restricting the HIT dopant to a small region at the beginning of the plasma source before beam-pinching can occur \cite{MartinezdelaOssa2013}. So far, the mechanism has not been experimentally demonstrated but the predicted beam properties, including attosecond bunch lengths, are promising. Given that these beams could drive a blowout in a subsequent stage with a higher plasma density, ``self-similar staging" was proposed \cite{MartinezdelaOssa2015}, whereby each stage boosts the brightness of the previous stage.

\begin{figure}[t]
    \centering\includegraphics[width=\linewidth]{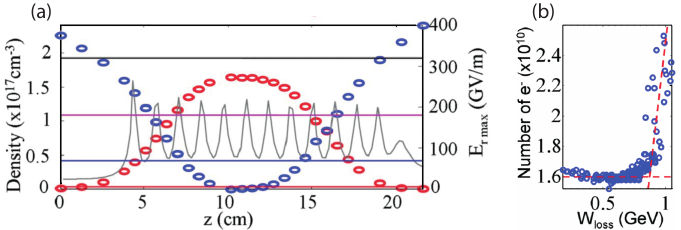}
    \caption{Ionization injection: (a) a mismatched beam enters a Li plasma (red circles) surrounded by a He buffer gas (blue circles). Simulated beam size oscillation, periodically reaching high radial electric fields (gray line) compared to the ionization thresholds of Li, He, He$^+$ and Li$^+$ (red, blue, magenta and black, respectively). (b) Increased charge, indicating injection, is observed when high energy loss (and by proxy high fields) are reached. Adapted from \textcite{Oz2007}.}
    \label{fig:ionization-injection}
\end{figure}

\paragraph{Beam-triggered injection}
\label{sec:ionization-injection:beam}
If the relativistic beam driver is focused to a very small beam size (\SI{}{\micro\m} scale), its radial electric field, often combined with the wakefield, can ionize electrons from a HIT component. Typically, this only occurs when the focusing force of the blowout ion column causes envelope oscillation and pinching of a mismatched beam driver (see Sec.~\ref{sec:envelope-matching} and Fig.~\ref{fig:mismatching}). Because of the relative simplicity of the scheme, it is one of the few with multiple experimental demonstrations, by \textcite{Oz2007} (see Fig.~\ref{fig:ionization-injection}), \textcite{Zhang2026}, as well as \textcite{VafaeiNajafabadi2014}, who also demonstrated that the injected electrons altered the wake through beam loading (Sec.~\ref{sec:beam-loading}). Further, simulations as well as experimental evidence \cite{VafaeiNajafabadi2019} indicate that this method can produce multiple distinct bunches of different energy, useful e.g.~for producing two-color FEL gain (see Sec.~\ref{sec:applications:FEL}).

\paragraph{Laser-triggered injection}
\label{sec:ionization-injection:laser}
Instead of the beam field or the wakefield itself triggering ionization, it is possible to use a high-intensity laser pulse intercepting the wake to ionize a small volume inside. This typically requires a high-peak-current beam to ensure trapping from rest, but is more relaxed than for the above schemes since the ionization may be produced in the wake center (i.e.,~at the location of maximum trapping potential) [see Fig.~\ref{fig:trappot}(c)].

\begin{figure}[t]
    \centering\includegraphics[width=0.89\linewidth]{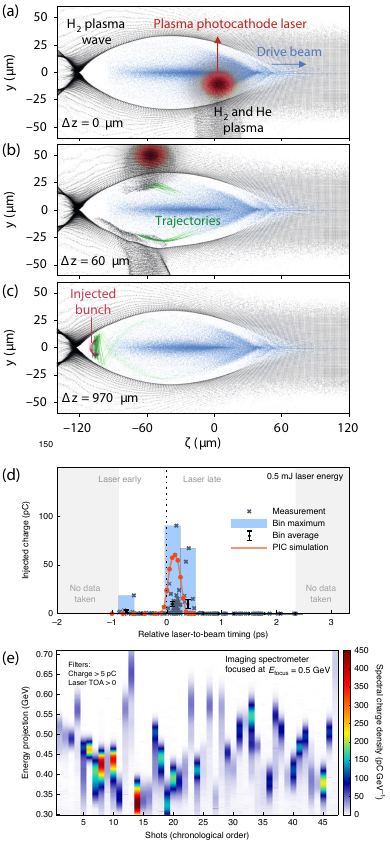}
    \caption{Plasma-photocathode injection. (a--c) A transverse laser pulse (red) traverses an electron-driven (blue) wake in a H$_2$ plasma (black). The laser ionizes He, releasing electrons inside the wake, as shown by the trajectories (green), resulting in an injected bunch (purple). (d) Experimental demonstration, showing that injection only occurs when the laser overlaps with the electron beam in time. (e) The corresponding energy spectra show mean energies around \SI{0.5}{GeV} and \%-level energy spreads. Adapted from \textcite{Deng2019}.}
    \label{fig:plasma-photocathode-injection}
\end{figure}

An example of laser-assisted injection, referred to as a ``plasma photocathode", uses a high-intensity laser co-propagating directly behind a high-peak-current beam driver with the laser focal plane situated close to the beginning of the acceleration stage \cite{Hidding2012}. The laser parameters are chosen such that it only ionizes the HIT dopant at and around its waist, so that the ionization is localized in a small volume and occurs over a short timescale---achievable with a Ti:Sapphire laser of moderate strength. This requires high spatial and temporal precision; a small fraction of the plasma wavelength and oscillation period, respectively. The resulting electron beams can have very low normalized transverse emittance (\SI{\sim 10}{nm}) when the electrons are released close to the wakefield axis, enabled by the moderate laser vector potential requirements for ionization, which mitigates ponderomotive heating. A variation on this scheme, using a transverse rather than co-propagating ionization laser [Figs.~\ref{fig:plasma-photocathode-injection}(a--c)], has been demonstrated experimentally with drivers from both an rf-accelerator \cite{Deng2019} [Figs.~\ref{fig:plasma-photocathode-injection}(d--e)] and an LWFA \cite{Ufer2026}. 

The trapping conditions on the drive-beam peak current can be further reduced by utilizing the ponderomotive force of an obliquely incident laser to ionize electrons and accelerate them into the propagation direction of the plasma wakefield to facilitate trapping \cite{Zeng2020}. However, this comes at the expense of an increase in emittance. A more complex laser-assisted injection scheme based on beat waves, whereby two counter-propagating transverse lasers interfere and trigger ionization from within the wake \cite{Li2013} (see Fig.~\ref{fig:plasma-photocathode-scheme}), may lead to ultrashort (\SI{8}{fs}) beams with moderately high peak current (\SI{0.4}{kA}) and ultralow emittance (\SI{8}{nm}) owing to a decreased injection volume.

\begin{figure}[t]\centering\includegraphics[width=0.89\linewidth]{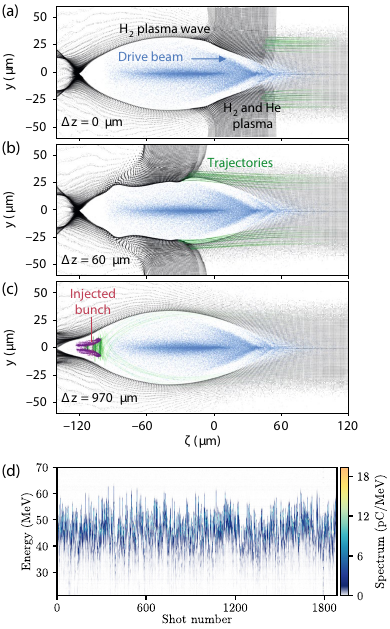}
    \caption{Density downramp (plasma torch) injection. (a--c) A beam driver (blue) traverses a laser-ionized plasma-density up- and downramp (black), injecting plasma electrons (purple) into the wake. Trajectories (green) show that trapped particles have non-negligible transverse momentum. (d) Experimental demonstration of this concept, with a stable energy spectrum. Adapted from \textcite{Deng2019} and \textcite{Knetsch2021} (CC-BY 4.0).}
    \label{fig:density-downramp-injection}
\end{figure}

\subsubsection{Phase-velocity reduction and downramp injection}
\label{sec:downramp-injection}

Streaming plasma electrons can be injected into the wakefield by temporarily reducing the phase velocity of the wake, $v_{\phi}$, to below the speed of the plasma electrons, $v_e$, via a dynamic elongation of the wake period. Defining an effective plasma frequency $\omega_{p,\mathrm{eff}} = -\partial\varphi/\partial t$ and wavenumber $k_{p,\mathrm{eff}}= \partial \varphi/\partial z$, the local phase velocity of the wake is given by $v_{\phi}= \omega_{p,\mathrm{eff}}/k_{p,\mathrm{eff}}$, i.e.
\begin{equation}\label{eq:phasevelocity}
\frac{v_{\phi}}{c} = \left(1+ \frac{\xi}{k_p}\frac{\dif k_p}{\dif z}\right)^{-1},
\end{equation}
where $\xi = z - ct$ is the co-moving coordinate and $\varphi=k_p(z)\xi$ is the local phase of the plasma wave, assuming the driver moves at $\simeq c$ and is located at $\xi=0$ \cite{Bulanov1998,Esarey2009}. According to Eq.~(\ref{eq:phasevelocity}), the phase velocity of the wake can be locally reduced by a negative wavenumber gradient in $z$ (noting that $\xi<0$ behind the driver). This is achieved in practice by introducing a density gradient in the longitudinal plasma-density profile---so-called \textit{downramp injection}. Alternatively, it can be achieved by decreasing the driver spot size in the plasma, either with strong external focusing or using a mismatched beam, such that the kinetic energy of the plasma electrons increases \cite{Dalichaouch2020}. In each case, the blowout-cavity length increases. As the wakefield driver propagates in an expanding blowout cavity, the phase velocity of the wake decreases until it becomes equal to or smaller than the maximum forward velocity of the oscillating plasma electrons, resulting in wavebreaking. A sufficient condition for trapping in an adiabatically evolving wakefield is $\Delta H^\prime<-1$ \cite{Kalmykov2009,Yi2013}, where $\Delta H^\prime$ is the variation of the  nonstationary Hamiltonian $H^\prime=H-v_\phi P_z$ of a plasma electron in the \textit{comoving} coordinate system ($\xi,x,y,t$). Here, in contrast to Eq.~(\ref{Hamiltonian1}), $H^\prime$ is not constant because the wakefield is evolving during trapping.

The concept of trapping using a density downramp was first proposed for LWFA \cite{Bulanov1998}, in which a gentle density drop much longer than the plasma skin depth facilitates injection of background plasma electrons. Later, \textcite{Suk2001} suggested that a sharp density transition with length less than a plasma-skin depth $k_p^{-1}$ be used, such that plasma-electron trapping occurs due to a localized non-laminar motion near the sharp density transition and at wake amplitudes well below the wavebreaking threshold [see Figs.~\ref{fig:density-downramp-injection}(a--c)]. 

Experimentally, density downramps are often generated hydrodynamically, including by means of shocks in a gas flow then turned into plasma, first demonstrated in LWFA \cite{Schmid2010} and later achieved in PWFA \cite{CouperusCabada2021} driven by electron bunches from an LWFA. This was extended to optically-generated hydrodynamic shocks \cite{Foerster2022}, also in a hybrid LWFA-driven PWFA (Sec.~\ref{sec:hybrid}). An alternative method for the generation of sharp density transitions is using a transverse ionization laser \cite{Wittig2015}. This method, known as ``plasma torch" injection, offers increased flexibility and was first used at FACET (Sec.~\ref{sec:experimental:facilities:slac}) to establish temporal overlap of the transverse laser with the plasma wake \cite{Scherkl2022} in the plasma-photocathode experiment \cite{Deng2019}. This was later demonstrated with improved stability by \textcite{Knetsch2021} [Fig.~\ref{fig:density-downramp-injection}(d)]. Recently, \textcite{Zhang2025} and \textcite{Wood2026} achieved beam-brightness transformation, where the downramp-injected beam had higher brightness than the driver.

The injection dynamics and the quality of the generated beams have been studied with simulations, and the results are somewhat complex. For downramp lengths of a few plasma skin depths, \textcite{MartinezdelaOssa2017,Grebenyuk2014} showed that shorter ramps result in reduced slice emittances, whereas the amount of injected charge diminishes drastically when the length of the ramp is increased. \textcite{Zhang2019} showed that the slice emittance can be reduced from 0.1 to \SI{0.03}{\milli\m\milli\radian} by using a gentler ramp and increasing its length from 5 to 25~$k_p^{-1}$. \textcite{Xu2017} uncovered a new mechanism that leads to the generation of unprecedented beam brightness in downramp injection with gentle ramps. In the high-density electron spike at the rear of the blowout cavity, electrons experience defocusing fields that can reduce their transverse momentum just as they are streaming into the wake to be trapped. These defocusing fields vanish once injected in the blowout cavity, and with proper optimization, electrons become trapped with extremely small residual transverse momenta, and thus with ultralow emittances and very high brightness. For both gentle ramps \cite{Xu2017} and evolving drivers \cite{Dalichaouch2020}, the resulting 6D brightness [Eq.~(\ref{eq:brightness})] scales linearly with the plasma density and can reach \SI{\sim e20}{A.m^{-2}.rad^{-2}/0\rm{.}1\%BW} for a plasma density of $10^{19}$ cm$^{-3}$. Driving PWFA at such plasma densities could be enabled by the intense beams of the FACET-II facility (Sec.~\ref{sec:experimental:facilities:slac}). For the slice energy spread of \SI{\sim 0.1}{\percent} reported in these studies, this also corresponds to a 5D brightness of \SI{\sim e20}{A.m^{-2}.rad^{-2}}.  Finally, \textcite{Zhang2019} showed the practical importance of cylindrical beam symmetry, with injected beam emittances up to 10 times larger for noncylindrical drivers compared to that of cylindrical drivers.


\subsection{Long-term plasma evolution}
\label{sec:long-term-evolution}

The average beam current is an important figure of merit for the performance of an accelerator as it is proportional to the integrated luminosity of a collider, i.e.~the number of collision events within a certain time, and the average brilliance of a light source (see Secs.~\ref{sec:applications:colliders} and \ref{sec:applications:FEL}). To compete with rf technology, plasma accelerators must be capable of accelerating thousands or even millions of bunches per second. Unlike with rf technology, the \textit{quality factor} of a plasma accelerator is low \cite{Ogata1998}, such that the electromagnetic wave damps after only a few to tens of oscillations. This means that energy should be maximally extracted from the first oscillation cycle, after which the plasma is left to equilibrate (or \textit{recover}) until a following, similar acceleration event can be driven. The long-term evolution of the plasma thus plays a critical role in determining the maximum number of acceleration events per second. 

The initial perturbation and subsequent motion of plasma ions is expected to be one of the main mechanisms in determining the long-term evolution of the plasma. Ions are typically considered static on the timescale of the plasma-electron wavelength due to their high mass compared to that of electrons. This is typically an adequate assumption when modeling plasma dynamics close to the driver. However, on longer timescales, and for high-density drivers, the motion of plasma ions becomes important. There are two main forces acting on the ions. Firstly, the ions may be accelerated directly by the fields of the beams. This is especially relevant for dense beams ($n_b \gg n_p$) as this can cause a collapse of the ions onto the beam axis within the first plasma-electron wavelength, resulting in a distortion of the accelerating and focusing fields~\cite{Rosenzweig2005} (see Sec.~\ref{sec:ion-motion}). Secondly, the trailing plasma wave exerts a ponderomotive force (integrated over all existing plasma-electron periods) on the ion background, which lasts until the plasma-electron wave eventually breaks down. This process was first predicted for laser drivers~\cite{Gorbunov2001,Gorbunov2003} and later adapted for long proton drivers~\cite{Vieira2012,Vieira2014c}, the latter of which has recently been demonstrated experimentally~\cite{Turner2025} at AWAKE (see Secs.~\ref{sec:self-modulated-wakefields} and \ref{sec:experimental:facilities:awake}).

\begin{figure}[t]
\includegraphics[width=\columnwidth]{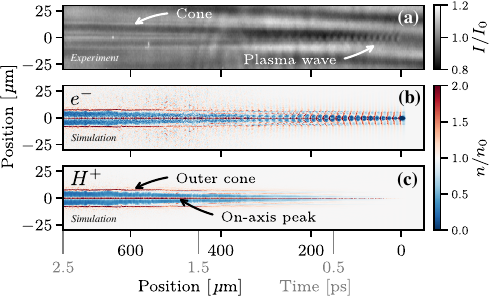}
    \caption{(a) Experimental shadowgram of the formation of a plasma wave and ion channel. Simulation showing the (b) electron and (c) ion ($\mathrm{H}^+$) density: the beam driver is at $\xi=0$ propagating to the right, followed by a plasma wave. After several plasma periods, an on-axis peak appears in the ion distribution followed by an annular channel. At this point, the plasma wave breaks down and the electrons follow the ions. Adapted from~\textcite{Gilljohann2019} (CC-BY 4.0).}
    \label{fig:long-term-evolution:transverse-shadowgrahy}
\end{figure}

These studies show that the ponderomotive force of the plasma wave leads to the formation of an annular channel with an on-axis peak in the ion density at very early timescales. A transverse shadowgraphic \cite{Buck2011,Schwab2013,Savert2015} measurement of the ion-channel formation is shown in Fig.~\ref{fig:long-term-evolution:transverse-shadowgrahy}~\cite{Gilljohann2019}. Here, a few-fs-long probe laser pulse is propagated transversely through the plasma wave, where the plasma-density perturbations lead to a local phase retardation that results in a shadowgram. The plasma wave is driven by a \SI{150}{\mega\electronvolt} and fs-long electron beam (from an LWFA; see Sec.~\ref{sec:hybrid}) with \SI{520}{\pico\coulomb} in a hydrogen plasma of density \SI{e19}{cm^{-3}}. The ion channel starts to appear after several plasma periods and lasts for at least tens of picoseconds---the diagnostic reach is limited by the maximum achievable delay between the drive- and probe-pulse. Simulations show that the formation of the ion channel causes the plasma wave to break down, after which time electrons follow the accelerated ions and the plasma becomes mostly neutral.

Going further, into the nanosecond timescale, an experiment at FACET (Sec.~\ref{sec:experimental:facilities:slac}) measured the formation of a meter-scale ion channel by using grazing-incidence shadowgraphy~\cite{Zgadzaj2020}. This technique, in which the probe and drive beams have a shallow oblique angle~\cite{Li2013b}, is suitable for probing targets not easily accessible with transverse probes, such as enclosed or long targets. This enabled the measurement of ion-channel evolution over timescales of up to \SI{1.5}{\nano\second}. Corresponding numerical simulations suggest that the majority of energy deposited in the plasma is initially transferred from the driving beam to the plasma electrons, with eventually about 90\% of the total deposited energy transferred to the plasma ions on the ns-timescale due to electron--ion and subsequent ion--ion collisions.

\begin{figure}[t]
\includegraphics[width=0.95\columnwidth]{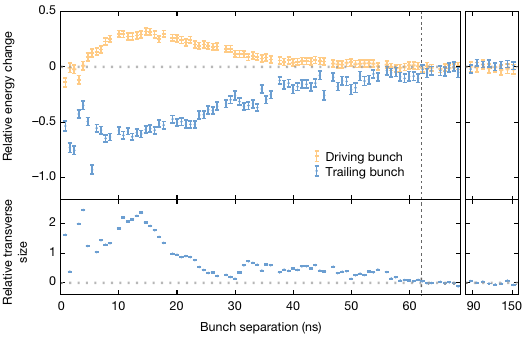}
    \caption{The residuals between the energy spectra and transverse size of the unperturbed- and perturbed-plasma events (in argon) as a function of the time separation between the leading and probing electron bunches. Extended data up to \SI{160}{\nano\second} is shown on a compressed horizontal timescale (right-hand panel). 
    The recovery time (black dashed vertical line) is reached when all three residuals are consistent with zero. Adapted from~\textcite{DArcy2022} (CC-BY 4.0).}
    \label{fig:long-term-evolution:recovery-time}
\end{figure}

Studies at FLASHForward (Sec.~\ref{sec:experimental:facilities:desy}) went even further by measuring the time it takes for the coherent motion of the plasma ions to equilibrate, thus setting a timescale for the ``recovery time" of the plasma~\cite{DArcy2022}. This was achieved by exciting the plasma with an initial GV/m wakefield event and using a probing drive--trailing bunch pair with variable delay to sample the evolving plasma properties. This showed that the plasma recovered within \SI{\sim63}{\nano\second} (see Fig.~\ref{fig:long-term-evolution:recovery-time}), after which the energy of the probe-bunch pair was no longer influenced by the presence of the previous driver. This recovery time would in principle support repetition rates on the order of \SI{10}{\mega\hertz}, making plasma accelerators compatible with the bunch-train patterns of current FELs and future linear-collider facilities. Simulations of the plasma motion show that the accelerated ions move radially outwards in the form of an ion acoustic wave, with the on-axis depletion of ions replenished by inwardly streaming cold ions from beyond the radius of the plasma wave. The equilibrated plasma profile can then be used for a subsequent acceleration event. Further studies at SPARC\_LAB (Sec.~\ref{sec:experimental:facilities:infn}) have been performed in a hydrogen plasma \cite{Pompili2024b}.

Long-term plasma motion is stimulated by residual energy transferred from the drive bunch to the plasma and not extracted by the trailing bunch (see Sec.~\ref{sec:beam-loading}). Although the average plasma profile was shown to have recovered (Fig.~\ref{fig:long-term-evolution:recovery-time}), the cumulative energy deposited in the plasma from sustained acceleration at high repetition rates will likely result in large thermal energies (i.e., $>$keV). Such thermal energies are expected to modify the wakefield properties, although limited numerical studies of non-zero plasma-temperature effects have been performed (see~Sec.~\ref{sec:plasma_temperature}). Furthermore, some proportion of the remnant energy in the plasma will conductively heat the surrounding plasma cell. High-average-power plasma stages will therefore necessitate the consideration of temperature-stabilization techniques and may require highly heat-resistant materials \cite{Crincoli2025}.


\subsection{Multistage acceleration}
\label{sec:staging}

Accelerating particle bunches of high charge to high energies is one of the main goals for PWFA. This requires efficient transfer of energy from drivers with high energy content. In linear colliders, for instance, the colliding bunches each contain of order \SI{1}{kJ} (see Sec.~\ref{sec:applications:colliders}). However, electron drivers of the required spatial dimensions typically only contain of order \SI{10}{J} or less, which means multiple drivers are required---each in their own accelerating stage. Only proton drivers are able to extend this energy-depletion limit, as proton bunches can be accelerated to very high energies in synchrotrons (see Sec.~\ref{sec:self-modulated-wakefields}); e.g., \SI{400}{GeV} per particle and \SI{6}{kJ} per bunch at Super Proton Synchrotron (SPS) \cite{Gschwendtner2016}. That said, the repetition rate and energy efficiency of proton bunch production is currently too low for linear colliders. Therefore, use of multiple stages---known as \textit{staging}---is likely required, at least for high-luminosity particle colliders. In practice, staging is challenging and has only seen rudimentary experimental exploration. For a more complete review of staging, see \textcite{Lindstrom2021b}.

The staging problem stems from the fact that plasma accelerators focus the trailing bunch strongly, which leads to small beta functions and beam sizes inside the stages and highly diverging beams outside. Since some space is required to extract the depleted driver and re-inject a new one, stages must be separated (by up to several meters, depending on the energy). This leads to issues with preservation of beam quality and overall compactness, as discussed below. In principle, the staging problem could be avoided if the beam remained focused throughout, either in a plasma as proposed for laser drivers \cite{Luo2018} or by other means such as crystal or nanostructure channeling \cite{Tajima1990,Gilljohann2023}, but no such scheme has yet been proposed for PWFA.

To date, there has been no experimental demonstration of staging of beam-driven PWFAs. This will require a dedicated medium-scale facility (larger and more complex than current single-stage experiments), concepts for which are currently being developed. However, a relevant experiment was conducted using a two-stage LWFA at Lawrence Berkeley National Lab \cite{Steinke2016}: the first stage produced \SI{120}{MeV} electron bunches with a broad energy spectrum (60\% FWHM), then refocused into the second stage, which added up to \SI{100}{MeV} of extra energy gain. While successful at accelerating electrons, this experiment highlighted the challenge of \textit{chromaticity} in the presence of finite energy spread (see Sec.~\ref{sec:multistage-chromaticity} below), as only $\sim$3.5\% of the charge was accelerated in the second stage. Nevertheless, the experiment was groundbreaking in its use of plasma lenses for refocusing, which is likely key for staging of PWFAs.

\subsubsection{Refocusing and chromaticity}
\label{sec:multistage-chromaticity}
The first issue is that the high-divergence trailing bunch expands rapidly and must be refocused into the subsequent stage. This is problematic because the beta function $\beta_{\mathrm{lens}}$ in the refocusing lens(es) is large compared to inside the stage $\beta_{\mathrm{matched}}$, which leads to strongly energy-dependent focusing, known as chromaticity \cite{Zyngier1977}. This is often quantified using the \textit{chromatic amplitude} $W$, which if unmitigated increases by approximately $\Delta W \approx \beta_{\mathrm{lens}}/f$ in a focusing element, where $f \approx L/2$ is the focal length and $L$ is the distance from the stage to the refocusing lens. Approximating $\beta_{\mathrm{lens}} \approx L^2/\beta_{\mathrm{matched}}$, the relative (squared) emittance growth from chromaticity \cite{Antici2012,Migliorati2013} can therefore be approximated as
\begin{equation}
    \frac{\Delta \varepsilon^2}{\varepsilon_0^2} = W^2 \sigma_{\delta}^2 \approx \frac{4 L^2 \sigma_{\delta}^2}{\beta_{\mathrm{matched}}^2},
\end{equation}
where $\sigma_{\delta}$ is the relative energy spread. As an example, consider a \SI{10}{GeV} beam with around 0.5\% energy spread: if the plasma density is \SI{e16}{\per\cm\cubed}, the matched beta function would be of order \SI{1}{cm}, and the stage-to-lens separation around \SI{1}{m} (assuming focusing strengths \SI{\sim100}{T/m}), resulting in an emittance growth of about 100\% per stage---clearly unacceptable if multiple stages are used. Using a more detailed model, \textcite{Thomas2021} found that a linear collider (accelerating beams to \SI{1}{TeV} over 85 stages) would require maintaining an energy spread below $10^{-5}$ in order to keep emittance growth below \SI{\sim1}{\milli\meter\milli\radian}, which would be extremely challenging in practice. Instead, we must greatly reduce the chromaticity, either by reducing the focal length of the lens or by increasing the matched beta function of the stage. Short-focal-length beam optics include plasma lenses, which come in two forms: \textit{active} plasma lenses, which use the azimuthal magnetic field from an electrical current through a plasma \cite{Panofsky1950,vanTilborg2015,Pompili2018,Lindstrom2018b}; or \textit{passive} plasma lenses, which use the strong electric fields in a plasma wake \cite{Chen1986,Ng2001,Marocchino2017,Doss2019,BjorklundSvensson2026}. Increasing the matched beta function is possible by use of plasma-density ramps \cite{Marsh2005,Dornmair2015,Xu2016,Ariniello2019} at the entrance and exit of the stage (discussed in Sec.~\ref{sec:mismatching}). Alternatively, we can introduce achromatic (and nonlinear) beam optics \cite{Montague1979,Lindstrom2016b} between the stages---a strategy successfully employed in linear-collider final-focus systems \cite{Raimondi2001,White2014}, but at the cost of significantly increasing the accelerator length. Recent work combines these solutions by using nonlinear plasma lenses, providing both compact and achromatic beam transport \cite{Drobniak2025,Lindstrom2026}.

\begin{figure*}[t]
    \centering
    \includegraphics[width=0.92\textwidth]{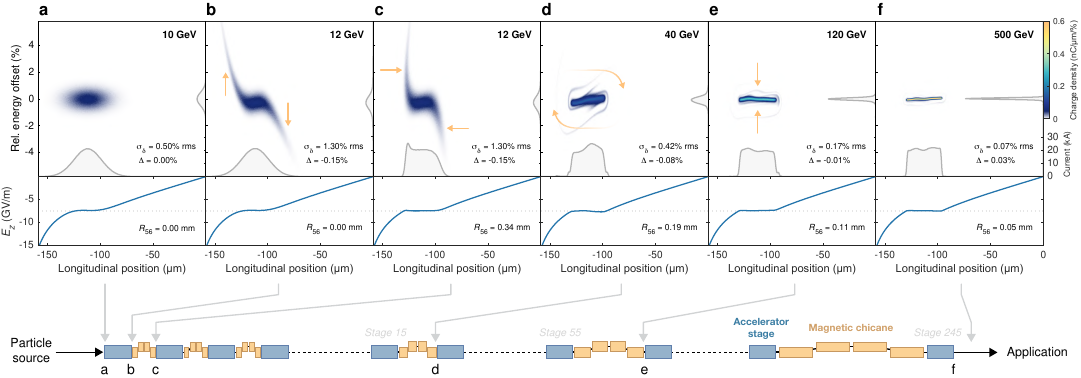}
    \caption{Multistage self-correction mechanism in longitudinal phase space. (a) A Gaussian current profile produces a non-uniform accelerating field $E_z$, which (b) results in a time--energy correlation. (c) Traversing a magnetic chicane ($R_{56} \neq 0$)  changes the current profile; (d) a process that continues with every stage. (e) With acceleration, relative energy offsets reduce, damping the energy spread and offset, and (f) creating an optimal current profile and flat wakefield. From \textcite{Lindstrom2021c}.}
  \label{fig:self-correction}
\end{figure*}

\subsubsection{Geometric accelerating gradient}
The second issue, which is related to the required optics, is that separating the stages ultimately reduces the longitudinally averaged (or \textit{geometric}) accelerating gradient of the plasma accelerator. Importantly, because refocusing is required, the interstage distance increases with energy. The length scaling can be found by considering that the focal length of the refocusing lens must be proportional to the interstage distance: $f = 1/kl \propto L$, where $l$ and $k$ are the length and strength of the refocusing lens, respectively. If the lens is operated at maximum strength $k = gc/p$, where $g$ is a constant field strength and $p$ is the particle momentum at that stage, and that the lens takes up a given fraction of the interstage distance (i.e., $l/L$ is constant), we find that $L^2 \sim 1/k \sim p/g$, or equivalently $L \sim \sqrt{p}$. As a consequence, at some energy the distance between stages will exceed the length of the stage itself, and therefore, ultimately the total length of a multistage plasma accelerator will be dominated by staging, scaling as $L_{\mathrm{total}} \sim N_{\mathrm{stages}} p^{3/2}$, where $N_{\mathrm{stages}}$ is the number of stages. For this reason, the number of stages should be minimized, although an optimization must also consider the longer rf accelerator needed to produce the drivers. Also relevant is the way drivers are distributed across multiple stages, for which concepts have been proposed \cite{Kudryavtsev1998,Adli2013,Pfingstner2016}: this will take up significant space and must be considered as part of the overall size of a facility.

\subsubsection{In- and out-coupling of drivers}
In- and out-coupling, the process of merging and separating the driver and trailing bunches, is also non-trivial. The sub-ps separation between these bunches is too short for conventional kickers, which operate with ns or longer rise times. Unless ultra-fast plasma- or laser-based kickers are introduced, the bunches must be separated by energy, which requires the use of magnetic dipoles between stages. This introduces a transverse dispersion not only in the driver but also in the trailing bunch, which must be canceled to avoid emittance growth. Moreover, incoherent and coherent synchrotron radiation may be disruptive, at high and low energies, respectively. Finally, a magnetic chicane will introduce a longitudinal dispersion, $R_{56}$, which can compress or stretch the trailing bunch. Naively, we would require $R_{56}=0$, but recent work has shown that a non-zero $R_{56}$ can in fact be beneficial, leading to energy-spread compensation, if applied across two stages \cite{FerranPousa2019}, or even a more complete self-correction mechanism in longitudinal phase space \cite{Lindstrom2021c} if applied over many stages (see Fig.~\ref{fig:self-correction}).

An alternative method is to use a train of drive bunches that traverses all the stages, and is progressively shifted backwards using a chicane between each stage \cite{Knetsch2023}. If the plasma is formed just ahead of the trailing bunch (with a laser pulse), only the rear-most driver interacts with the plasma. This gated staging scheme greatly simplifies driver in- and out-coupling and distribution, but also requires picosecond-scale driver separations---challenging to achieve in an rf accelerator, though it is in principle possible \cite{Muggli2008a}.

\subsubsection{Alignment and synchronization}
\label{sec:multistage-alignment}
Finally, staging influences the tolerances of both alignment and synchronization. The already strict misalignment tolerances for a single stage (see Sec.~\ref{sec:misalignment}) are made even stricter in a multistage accelerator, as each stage can separately contribute emittance growth. At high energy, the stages are typically short enough to not allow full decoherence, which means that the induced betatron oscillation ``leaks" into the following stage \cite{Assmann1998}. If the transverse misalignment ($\sigma_{\Delta x}$) of the driver in each stage is uncorrelated and the relative energy spread remains constant, the emittance grows approximately linearly with $N_\mathrm{stages}$ \cite{Chiu2000} [or $\sqrt{N_\mathrm{stages}\ln N_\mathrm{stages}}$ for very large $N_\mathrm{stages}$ \cite{Cheshkov2000}] assuming a constant energy gain per stage. Emittance growth from angular misalignments ($\sigma_{\Delta x'}$) scales by another factor of $\beta_\mathrm{matched}^2 \sim \gamma$ [see Eq.~(\ref{eq:misalignment-emit-growth})] and hence scales as $N_\mathrm{stages}^2$. This will dominate above a certain energy (that for which $\beta_\mathrm{matched} = \sigma_{\Delta x}/\sigma_{\Delta x'}$), which in practice implies an $N_\mathrm{stages}$-scaling between linear and quadratic; assuming the average $\beta_\mathrm{matched}$ in all stages gives a scaling of $N_\mathrm{stages}^{3/2}$ \cite{Lindstrom2016a}. If the final energy $E_\mathrm{final}$ is kept constant, then since the unsaturated emittance growth also scales as the square of the stage length $L_\mathrm{stage}^2 \sim E_\mathrm{final}^2/N_\mathrm{stages}^2$ (see Sec.~\ref{sec:misalignment}), the final emittance will scale as $E_\mathrm{final}^2/N_\mathrm{stages}$ if offset-dominated \cite{Chiu2000} or $E_\mathrm{final}^2$ if angle-dominated (or roughly $E_\mathrm{final}^2/\sqrt{N_\mathrm{stages}}$ if mixed). Simulations indicate that for a 20-stage PWFA at density \SI{e16}{\per\cm\cubed}, accelerating a beam with \SI{0.1}{\milli\meter\milli\radian} emittance and 1\% relative energy spread by \SI{25}{GeV} in each \SI{3}{m}-long stage, the misalignment tolerance is \SI{40}{nm} or \SI{1}{\micro\radian} rms \cite{Lindstrom2016a}. If the \textit{absolute} energy spread is kept constant (i.e., the relative energy spread decreases with acceleration), the final emittance saturates after a few stages \cite{Thevenet2019}. 

The situation is different for synchronization tolerances between the driver and the trailing bunch, which a priori follows a similar scaling, but can in fact be significantly improved by use of multiple stages if a non-zero $R_{56}$ and the self-correction mechanism is applied \cite{Lindstrom2021c}, as shown in Fig.~\ref{fig:self-correction}.


\section{Research methods and results}
\label{sec:research-methods-and-results}

How do researchers study beam-driven plasma wakefields and what is the experimental basis of the physics described in Secs.~\ref{sec:physics} and \ref{sec:advanced-topics}? Decades of theoretical and experimental research have yielded numerous important methods and results, covered in Secs.~\ref{sec:numerical-simulations} and \ref{sec:experiments}.


\subsection{Numerical modeling and codes}
\label{sec:numerical-simulations}

Interactions between plasma and intense beams can require the description of individual particle dynamics and nonlinear phenomena not captured by a fluid-based, or \textit{hydrodynamic}, model. In a collisionless plasma, typical for plasma accelerators, such interactions can be simulated with a \textit{kinetic} description, using the \textit{particle-in-cell} (PIC) method. Invented in the 1950s by Buneman, Hockney, Birdsall, Dawson and others [see the review by \textcite{Nishikawa2021}], the PIC method is based on charged particles moving in a grid of cells, in each of which electromagnetic fields are averaged. The method follows a basic four-step loop: (1) the charge and current density is calculated based on the particle distribution; (2) this is used to solve for the electromagnetic fields at the corners of each cell; (3) the resulting force is interpolated at the location of each particle; and finally (4) the particles are pushed forward by a time step. 

Nearly all theoretical research in plasma acceleration that goes beyond the analytical descriptions in Sec.~\ref{sec:plasma-wakefields} makes use of the PIC method. However, performing well-resolved, three-dimensional simulations based purely on the above loop requires immense computational resources. This is due to the large number of cells required to spatially resolve the often microscopic structures of beams and wakefields, as well as the correspondingly small time step \cite*{Courant1928}; for a typical \SI{}{\micro\m}-scale spatial resolution, fs-scale time resolution is required, whereas the acceleration process can last for many nanoseconds. While fully electromagnetic PIC codes, including e.g.~OOPIC \cite{Verboncoeur1995} and later OSIRIS \cite{Hemker1999,Fonseca2002}, provide an invaluable accuracy benchmark for the physics, a number of implementations have been made that speed up  calculations and lower hardware requirements by using various approximations---for a comprehensive survey of these codes, covering both laser and beam drivers, see the review by \textcite{Vay2016}. Below is a brief description of the main approximations that have been used to reduce the computational requirements of beam-driven plasma-wakefield simulations specifically.

\begin{figure}[t]
    \centering{\includegraphics[width=0.95\linewidth]{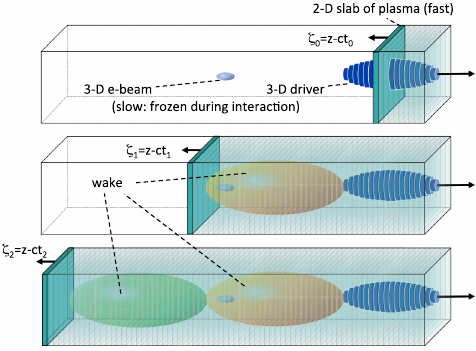}}
    \caption{Illustration of a PIC simulation using the quasistatic approximation. From \textcite{Vay2016}.}
    \label{fig:quasistatic-approximation}
\end{figure}

The \textit{quasistatic approximation} \cite{Sprangle1990,Mora1997} (see Sec.~\ref{sec:lin_wake}), which can speed up PWFA simulations by several orders of magnitude, is perhaps the most successful such approximation. It is based on the observation that a highly relativistic particle beam ($\gamma \gg 1$) evolves on a much longer timescale than that of the plasma electrons, which means that the plasma wakefield calculated from the beam distribution does not evolve (in a coordinate frame that is co-propagating with the beam) until the beam has evolved in shape (see Fig.~\ref{fig:quasistatic-approximation}). The time step used to evolve the beam is therefore not a fraction of the plasma period $\omega_p^{-1}$, but instead a fraction of the much longer betatron-oscillation period $\omega^{-1}_\beta = \sqrt{2\gamma}\omega_p^{-1}$ (see Sec.~\ref{sec:envelope-matching}). The quasistatic approximation was first implemented in the code WAKE \cite{Mora1997} and subsequently in QuickPIC \cite{Huang2006} and HiPACE \cite{Mehrling2014}. Note that quasistatic codes are not suitable for modeling internal injection (Sec.~\ref{sec:internal-injection}).

A Lorentz boost can also greatly speed up simulations, by moving from the lab frame to a \textit{boosted frame} \cite{Vay2007,Vay2011} where the (originally short) driver and the (originally long) plasma column are of similar length; in this frame, fewer time steps are needed. The presence of a relativistically streaming plasma in the simulation can, however, drive the \textit{numerical Cherenkov instability} \cite{Godfrey1974}, but this can be suppressed by using a Galilean change of coordinates \cite{Kirchen2016,Lehe2016b}, as implemented in FBPIC \cite{Lehe2016} and WarpX \cite{Vay2018}.

Reduced dimensionality is another important approximation. Early PIC codes used for the first PWFA simulations, such as WAVE \cite{Morse1971} and ZOHAR \cite{Langdon1976}, simulated only one or two spatial dimensions, in a Euclidian geometry. In order to simulate axisymmetric beams, which are more realistic, cylindrical symmetry was introduced, as in the codes LCODE \cite{Lotov1998} and INF\&RNO \cite{Benedetti2010}. A somewhat more sophisticated version of this is \textit{Fourier decomposition} \cite{Lifschitz2009}, as employed in FBPIC and QPAD \cite{Li2021}, whereby several azimuthal modes can be included to simulate the lowest-order moments of a transverse asymmetry.

If only small regions of the simulation grid require high resolution, it is possible to use \textit{adaptive mesh refinement} \cite{Vay2002}. This was first employed in the code Warp \cite{Grote2005}. The updated version of this code, WarpX, as well as FBPIC and the quasistatic code HiPACE++ \cite{Diederichs2022c}, were also early adopters of GPU acceleration, which provides a significant speedup compared to CPU operation.

Lastly, in order to study long-term evolution of the plasma after the passage of the beam (Sec.~\ref{sec:long-term-evolution}), some codes employ kinetic--fluid \textit{hybrid} methods: modeling the beam as particles, but the plasma as a hydrodynamic fluid, with the associated limitation of not capturing kinetic plasma effects such as trajectory crossing (see Sec.~\ref{sec:nonlin_wake}). Examples include the above-mentioned 2D codes LCODE and INF\&RNO, as well as the 3D code Architect \cite{Marocchino2016}. Another use of such hybrid methods is mixed dimensionality; by simplifying the description of the plasma even further, using models by \textcite{Lu2006a} (1D) or \textcite{Baxevanis2018} (2D), while maintaining a 3D beam, simulations can be very fast---the code Wake-T \cite{FerranPousa2019b}, which implements such a strategy, is currently one of the fastest codes for basic PWFA simulations. This and other codes are combined in the start-to-end simulation framework ABEL \cite{Chen2025} to simulate plasma-accelerator facilities such as a linear collider.


\subsection{Experimental facilities and results}
\label{sec:experiments}

\begin{table*}
    \centering
    \begin{tabular}{l|ccccc|cccc}
         &  &  &  & \textit{Bunch} & \textit{Norm.} &  &  & \textit{Plasma} & \textit{Plasma} \\
         \textit{PWFA facility} & \textit{Driver} & \textit{Energy} & \textit{Charge} & \textit{length} & \textit{emittance} & \textit{Plasma} &  \textit{Ionization}  & \textit{density} & \textit{length} \\
        \textit{(laboratory)} & \textit{species} & (GeV) & (nC) & (mm) & (mm mrad) & \textit{species} & \textit{method} & (\SI{}{\per\cubic\cm}) & (m) \\
        \hline
        AWAKE (CERN) & $p^+$ & 400 & 50 & 60--120 & 3.5 & Rb, He/Ar/Xe & Laser, HV & \SI{e14}{}--\SI{e15}{} & 10 \\
        FFTB (SLAC) & $e^-$, $e^+$ & 28--42 & 3.2 & 0.01--0.7 & $\sim$100$\times$10 & Li & Laser/beam & \SI{e13}{}--\SI{e17}{} & 0.85--1.4 \\
        FACET (SLAC) & $e^-$, $e^+$ & 20 & 3.2 & 0.02 & $\sim$100$\times$10 & H/He/Li/Ar & Laser/beam & \SI{e16}{}--\SI{e17}{} & 0.4--1.3 \\
        FACET-II (SLAC) & $e^-$ & 10 & 2.0 & $<$0.01 & $<$20 & H/He/Li & Laser/beam & \SI{e16}{}--\SI{e17}{} & 0.4--2.8\\
        FLASHForward (DESY) & $e^-$ & 0.5--1.2 & 1 & 0.04--0.3& 1--10 & H/He/Ne/Kr/Ar & HV/laser & \SI{e15}{}--\SI{e17}{} & 0.03--0.5 \\
        PF \& TRISTAN (KEK) & $e^-$ & 0.25--0.5 & 7\footnote{Distributed over 6 bunches spaced by \SI{350}{ps}.} & 3.3 & 3 & Ar & HV & \SI{e11}{}--\SI{e13}{} & 1 \\
        SPARC\_LAB (INFN) & $e^-$ & 0.08--0.15 & 0.2 & 0.07 & 1 & H & HV & \SI{e13}{}--\SI{e16}{} & 0.03\\
        ATF (Brookhaven) & $e^-$ & 0.06 & 0.5 & 0.5 & 1 & H & HV & \SI{e14}{}--\SI{e17}{} & 0.006--0.02\\
        AWA (Argonne) & $e^-$ & 0.04 & $\sim$15 & $\sim$2 & 150 & Ar & HV & \SI{e13}{}--\SI{e14}{} & 0.06--0.12 \\
        Tsinghua University & $e^-$ & 0.04 & 0.05 & $\sim$0.5 & 1.4 & N & Laser & \SI{e14}{}--\SI{e15}{} & 0.2 \\
        Tokyo University & $e^-$ & 0.028 & 0.5 & $\sim$3 & 180 & Ar & HV & \SI{e10}{}--\SI{e11}{} & 0.36 \\
        PITZ (DESY) & $e^-$ & 0.025 & 1 & $\sim$2.5 & $\sim$0.4 & Li, Ar & Laser, HV & \SI{e12}{}--\SI{e16}{} & 0.1 \\
        AATF (Argonne) & $e^-$ & 0.021 & 4 & 2.1 & 290 & Ar & HV & \SI{e12}{}--\SI{e14}{} & 0.2--0.35 \\
        A0 (Fermilab) & $e^-$ & 0.014 & 4--8 & 0.6 & 45 & Ar & HV & \SI{e14}{} & 0.08 \\
        SPA (Los Alamos) & $e^-$ & 0.008 & 1--2 & 0.8 & 2.5 & Xe & Beam & \SI{e17}{}--\SI{e18}{} & 0.003 \\
        \hline
    \end{tabular}
    \caption{Past and present PWFA facilities, ordered by driver energy, showing typical values or ranges for both beam and plasma parameters. Laser-driven (hybrid) PWFA facilities and planned facilities (EuPRAXIA, IHEP, SXFEL) are not shown.}
    \label{tab:facilities}
\end{table*}

Since the inception of PWFA \cite{Chen1985,Ruth1985}, experimental test facilities have been planned and operated. Early beam--plasma experiments were performed in Soviet Ukraine \cite{Kharchenko1960,Fainberg1968}, but modern investigations began with those conducted in the United States by \textcite{Rosenzweig1985}. Results from facilities across the US, Japan and Europe are covered lab-by-lab in chronological order of first commissioning. Key facility parameters are summarized in Table~\ref{tab:facilities}. For a general overview of experimental methods and diagnostics for plasma accelerators, see the review by \textcite{Downer2018}. For an overview of plasma sources, see the review by \textcite{Garland2020}.

\subsubsection{Argonne National Laboratory: AATF and AWA}
\label{sec:experimental:facilities:argonne}

\begin{figure}[t]
    \centering{\includegraphics[width=0.78\linewidth]{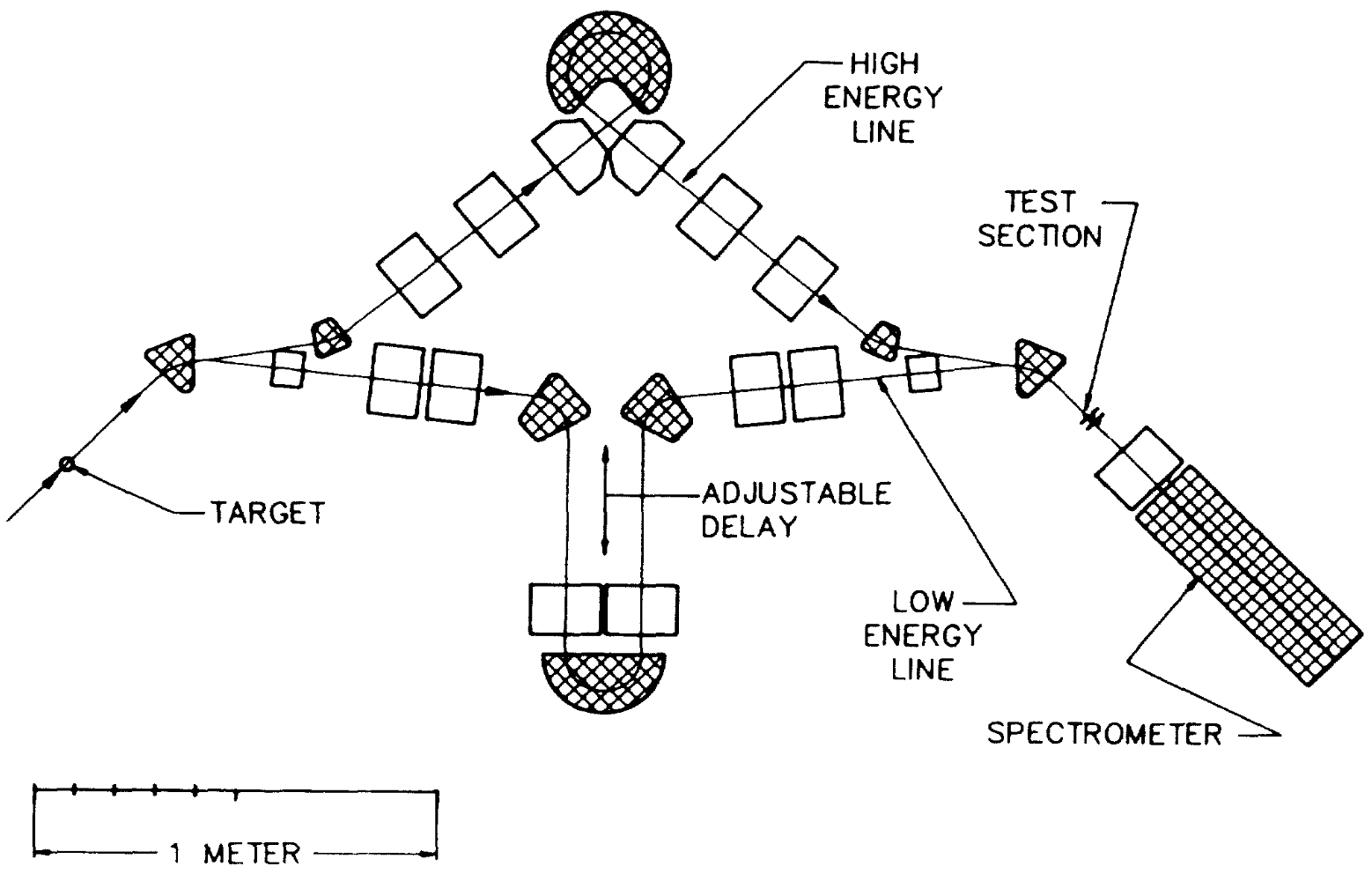}}
    \caption{Experimental setup in the AATF at Argonne National Lab, where two electron bunches with an adjustable delay interacted in a plasma before being diagnosed with a spectrometer. Adapted from \textcite{Rosenzweig1988}.}
    \label{fig:aatf-setup}
\end{figure}

The first proof-of-principle demonstration of beam-driven plasma wakefields by \textcite{Rosenzweig1988} (see Fig.~\ref{fig:first-observation}) was achieved at Argonne National Laboratory's Advanced Accelerator Test Facility (AATF). Here (see Fig.~\ref{fig:aatf-setup}), two bunches of energy 21 and \SI{15}{MeV} were created by sending a small portion (transversely) of a single bunch through a carbon target, transporting the low- and high-energy bunches through separate beam lines---one with an adjustable ``trombone" delay section---and combining them into a 20--\SI{35}{cm}-long hollow-cathode arc source filled with argon operating at a plasma density of 0.7--\SI{7e13}{\per\cm\cubed} \cite{Rosenzweig1987b}. The driver had \SI{\sim2.1}{nC} of charge and a length of \SI{2.4}{mm} rms. The observed wakefield, at around \SI{1}{MV/m}, agreed with the linear theory \cite{Chen1985}. Later experiments with higher charge (2.9--\SI{4.0}{nC}) showed nonlinear steepening of the wakefields and a density perturbation of around $\delta n/n_0 \simeq 33\%$ \cite{Rosenzweig1989}. Additionally, electron self-focusing, or passive plasma lensing, was observed \cite{Rosenzweig1990}. 

Later, the Argonne Wakefield Accelerator (AWA) test facility, which provided low-energy (\SI{\sim15.6}{MeV}) but high-charge beams (\SI{\sim18}{nC}), was the first to reach the blowout regime  \cite{Barov2000} by producing a beam density (\SI{4e13}{\per\cm\cubed}) higher than the plasma density (\SI{1.3e13}{\per\cm\cubed}). More recently, an emittance-exchange chicane was installed to shape the current profile, which was used to demonstrate a transformer ratio as high as~7.8~\cite{Roussel2020} (see Fig.~\ref{fig:high-transformer-ratio} and Sec.~\ref{sec:high-transformer-ratio}).

\subsubsection{KEK and Tokyo University}

Early PWFA experiments were also performed in Japan \cite{Ogata1992} between 1989 and the mid-1990s at KEK and Tokyo University, discussed below. Related experiments were also conducted at Utsunomiya University using low-energy ion bunches \cite{Nishida1991}.

The first experiments were performed at KEK \cite{Nakajima1990} using the positron linac of the Photon Factory (PF) and TRISTAN Accumulation Ring, which could deliver electrons and positrons at \SI{500}{MeV}. Six nC-level electron bunches spaced by \SI{350}{ps} were injected into a \SI{1}{m}-long helicon source ionizing argon to a density of \SI{e11}{}--\SI{e13}{\per\cm\cubed}. The plasma density was measured with a Langmuir probe, which also measured a temperature of 2--\SI{5}{eV}, while a streak camera and dipole spectrometer were used to measure the beam in longitudinal phase space. Accelerating fields higher than \SI{20}{MV/m} were observed, but the resonant build-up was not consistent with linear theory after the first three bunches, likely due to transverse misalignment of the later bunches.

Lower-energy experiments were performed at Tokyo University \cite{Nakanishi1993}, using beams from two independent linacs operated at different energies (14--18 and 24--\SI{28}{MeV}) combined into a similar plasma source as that described above. Early experiments, using only one \SI{14}{MeV} beam, showed resonant buildup of the wakefield from a train of multiple bunches, measured with a coaxial diode, reaching 400 times the amplitude of a single bunch \cite{Ogata1991}. Twin-linac experiments were then performed, first in the overdense regime \cite{Ogata1994} and later in the underdense regime \cite{Ogata1995,Kozawa1997}, the latter showing hints of nonlinear behavior.

\subsubsection{Fermi National Accelerator Laboratory: A0}
The A0 facility \cite{Carneiro1999} at Fermilab, USA, briefly performed PWFA experiments using a \SI{14}{MeV} electron beam with a bunch length of \SI{0.6}{mm} rms and charge 4--\SI{8}{nC}, interacting with an \SI{8}{cm}-long argon discharge plasma at density \SI{e14}{\per\cubic\cm} \cite{Barov2002}. This decelerated some electrons down to \SI{3}{MeV} (\SI{140}{MV/m}) and accelerated up to \SI{20}{MeV} (\SI{72}{MV/m}).

\subsubsection{Los Alamos National Laboratory: SPA}
Another small PWFA experiment was performed at the Sub-Picosecond Accelerator (SPA) facility \cite{Carlsten1996} at Los Alamos National Laboratory, USA. It used up to five \SI{8}{MeV}, 1--\SI{2}{nC}, \SI{0.8}{mm}-rms-long electron bunches (spaced by \SI{9.2}{ns}) that beam-ionized a \SI{3}{mm}-long xenon gas jet at density \SI{5e17}{\per\cubic\cm}, achieving decelerating fields of up to \SI{74}{MV/m} \cite{Russell2002}.

\subsubsection{Brookhaven National Laboratory: ATF}

\begin{figure}[t]
    \centering{\includegraphics[width=0.95\linewidth]{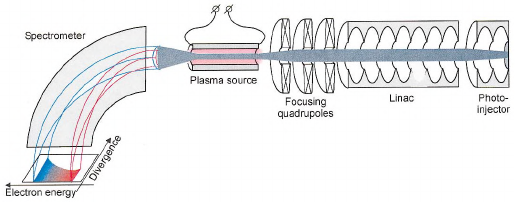}}
    \caption{Experimental setup at ATF, where the beam (moving right to left) is produced and accelerated in an rf accelerator, interacts with the plasma, and is diagnosed by a dipole spectrometer. Most PWFA experiments are variations of this basic setup. Adapted from \textcite{Yakimenko2003}.}
    \label{fig:atf-setup}
\end{figure}

The Accelerator Test Facility (ATF) at Brookhaven National Laboratory, USA, performed PWFA experiments \cite{Pogorelsky2014} using a high-quality \SI{60}{MeV} electron beam from a photocathode (see Fig.~\ref{fig:atf-setup}). Initial experiments, using a \SI{17}{mm}-long capillary-discharge plasma source \cite{Kaganovich1997}, demonstrated that the accelerating and focusing fields had a phase offset of $\pi/2$ \cite{Yakimenko2003}, as expected in the linear regime. The facility pioneered the use of collimator masks in energetically dispersive sections to generate bunches separated by picoseconds \cite{Yakimenko2006,Muggli2007} and was the first to use this technique for double-bunch plasma acceleration \cite{Kallos2008}, showing a loaded accelerating field up to \SI{150}{MV/m} (\SI{\sim315}{MV/m} unloaded) at a plasma density of \SI{e16}{\per\cubic\cm}. Resonant excitation using multiple bunches was also demonstrated \cite{Muggli2008a,Muggli2010b}. Finally, experiments were performed that demonstrated a current-filamentation instability \cite{Allen2012} as well as seeding of the self-modulation instability \cite{Fang2014} (see Secs.~\ref{sec:drive_instabilities} and \ref{sec:self-modulated-wakefields}).

\subsubsection{SLAC National Accelerator Laboratory: FFTB, FACET, and FACET-II}
\label{sec:experimental:facilities:slac}

SLAC has hosted three beam test facilities that support PWFA research. The Final Focus Test Beam (FFTB) operated from 1993--2006~\cite{Burke1991}. The Facility for Advanced Accelerator Tests (FACET) operated from 2011--2016~\cite{Clarke2011}. The successor to FACET, FACET-II, has been operating since 2020~\cite{Yakimenko2019}. Each of these facilities leverage the infrastructure that was built for the original SLAC complex, including electron and positron sources and the high energy linac. Additionally, PWFA research at SLAC greatly benefited from the construction of the Sub-Picosecond Photon Source (SPPS) chicane that was installed in the main linac to test bunch compression concepts for LCLS~\cite{Bentson2003}.

The FFTB PWFA program consisted of four experiments: E157, E162, E164 and E167. The experiments used meter-scale lithium vapor heat-pipe ovens~\cite{Vidal1969} that were photo-ionized or beam-ionized to produce a plasma~\cite{Muggli1999,OConnell2006} and electron/positron bunches with an energy up to \SI{42}{GeV} (but typically \SI{28.5}{GeV}) and charge \SI{3.2}{nC}. The E157 experiment explored PWFA with the mm-long electron bunch~\cite{Katsouleas1998,Assmann1998a,Hogan2000,Clayton2002}. Results from E157 included the refraction and guiding of electron beams inside a plasma column \cite{Muggli2001a,Muggli2001}, matched beam propagation in plasma~\cite{Muggli2004}, and observation of x-ray betatron radiation from the plasma~\cite{Wang2002}. The E162 experiment replaced electron beams with positron beams. Results included positron transport through meter-scale plasmas~\cite{Hogan2003}, acceleration of positrons in plasma~\cite{Blue2003}, and halo formation due to nonlinear focusing of positron beams in plasma~\cite{Muggli2008b}; see Fig.~\ref{fig:halo_formation}.

The E164 and E167 experiments operated with shorter electron bunches which enabled the use of higher plasma densities and gradients. This led to the observation of multi-GeV energy gain~\cite{Hogan2005} and eventually the \SI{42}{GeV} energy-doubling result~\cite{Blumenfeld2007} (see Fig.~\ref{fig:large-energy-gain}). Additional results include positron generation from betatron radiation~\cite{Johnson2006} and ionization injection of bright bunches~\cite{Oz2007,Kirby2009}. 

The FFTB program ended with the construction of LCLS, which only required the final kilometer of the SLAC linac for operation. Based on the success of the FFTB PWFA program, FACET was proposed as a user facility to host accelerator R\&D with a specific emphasis on PWFA research~\cite{Clarke2011} using \SI{20.35}{GeV}, \SI{3.2}{nC} electron or positron bunches. FACET added a variable $R_{56}$ chicane at the end of the linac to provide additional compression and control of the longitudinal bunch profile. The final chicane included a notch collimator in a dispersive region that was used to convert a single bunch to two bunches with \SI{100}{\micro\m}-scale separation---much smaller than the rf bucket spacing. FACET was also capable of running with compressed positron bunches and included a \SI{20}{TW} Ti:Sapphire laser used to pre-ionize the plasma source~\cite{Green2014}.

\begin{figure}[t]
    \centering{\includegraphics[width=0.96\linewidth]{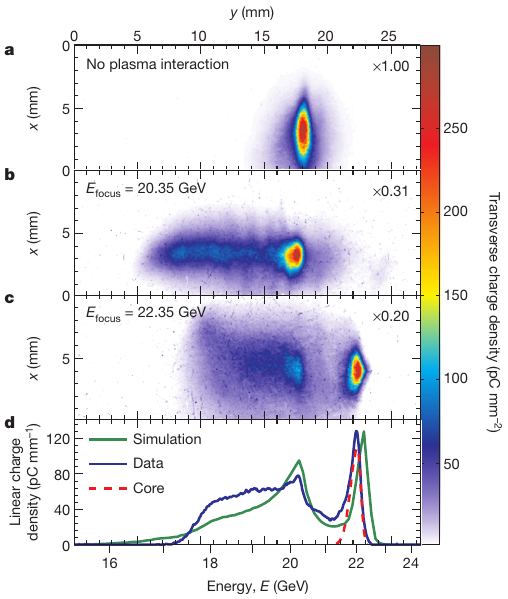}}
    \caption{Acceleration of a distinct trailing bunch with multi-GeV energy gain, low energy spread, and high energy-transfer efficiency (up to 30\%) in an experiment performed at FACET, SLAC National Accelerator Lab. From \textcite{Litos2014}.}
    \label{fig:facet-high-efficiency}
\end{figure}

The E200 experiment~\cite{Hogan2010} demonstrated high energy-transfer efficiency PWFA~\cite{Litos2014}, with efficiencies as high as 30\% with \SI{2}{GeV} energy gain (see Fig.~\ref{fig:facet-high-efficiency}). With a longer oven, the energy gain was raised to \SI{9}{GeV}~\cite{Litos2016}. FACET also demonstrated the acceleration of low-energy spread positron bunches~\cite{Corde2015} [see Fig.~\ref{fig:beam_loading_posi}(c)], two-bunch positron acceleration~\cite{Doche2017}, and a positron beam-driven hollow-channel plasma accelerator~\cite{Gessner2016a} (see Sec.~\ref{sec:without-focusing} and Fig.~\ref{fig:hcpwfa}). Further results on hollow-channel positron acceleration included the measurement of transverse wakefields~\cite{Lindstrom2018} and acceleration of a trailing positron bunch in the channel~\cite{Gessner2023}.

The wider E200 program included beam physics~\cite{Clayton2016,Adli2016a} and high-field acceleration studies~\cite{Corde2016}. A number of ionization-injection experiments were pursued at FACET~\cite{VafaeiNajafabadi2014,VafaeiNajafabadi2016,VafaeiNajafabadi2019}, including the plasma photocathode~\cite{Deng2019} (see Sec.~\ref{sec:internal-injection} and Fig.~\ref{fig:plasma-photocathode-injection}). Novel techniques to measure the long-term evolution of the plasma using an oblique-angle laser probe helped to determine the lifetime and energy deposited in the plasma~\cite{Zgadzaj2020} (see Sec.~\ref{sec:long-term-evolution}).

The FACET program ended in 2016 to allow the re-purposing of the first third of the SLAC linac for the LCLS-II superconducting rf FEL. The facility concept for FACET-II~\cite{Yakimenko2019} replicates the LCLS normal-conducting rf design~\cite{FACET2016}.
The normal-conducting rf photocathode gun provides short, low-emittance electron beams. The cathode laser may be double pulsed to provide two electron bunches in the same rf bucket. This level of control was not possible with damping-ring sources at FACET. FACET-II is capable of producing extremely high peak-current electron beams with energy \SI{10}{GeV} and charge \SI{2}{nC}, which can be diagnosed in part through their interactions with the plasma~\cite{Zhang2024,Emma2025}. Experiments include E300 \cite{Joshi2018,Storey2024}, which aims to demonstrate controlled two-bunch acceleration with high overall efficiency (Sec.~\ref{sec:beam-loading}) and beam-quality preservation (Sec.~\ref{sec:evolution-transverse}), and E304 \cite{Zhang2025}, boosting beam brightness through high-quality density-downramp injection (Sec.~\ref{sec:downramp-injection}).

\subsubsection{CERN: AWAKE}
\label{sec:experimental:facilities:awake}

The AWAKE facility at CERN is the world's only proton-beam-driven PWFA facility~\cite{Assmann2014,Caldwell2016,Gschwendtner2016,Muggli2017}, using \SI{400}{GeV} proton bunches with charge up to \SI{50}{nC} and length 6--\SI{12}{cm} delivered by the SPS 2--4 times per minute. It occupies the tunnel formerly used by the CNGS neutrino experiment \cite{Gschwendtner2006}. AWAKE has an electron beamline producing \SI{20}{MeV} bunches with \SI{0.7}{nC} charge and $\sim$mm bunch length~\cite{Pepitone2018,Kim2020} used for external injection into a self-modulated wakefield (Sec.~\ref{sec:self-modulated-wakefields}). The experiment employs a unique, laser-ionized~\cite{Demeter2021} \SI{10}{m}-long rubidium plasma source \cite{Oz2014,Plyushchev2017} with the ability to impose density gradients and density steps \cite{Muggli2022}. Diagnostic systems include streak cameras~\cite{Rieger2017}, beam-halo monitors~\cite{Turner2017}, and an electron-energy spectrometer~\cite{Bauche2019}.

\begin{figure}[t]
    \centering{\includegraphics[width=\linewidth]{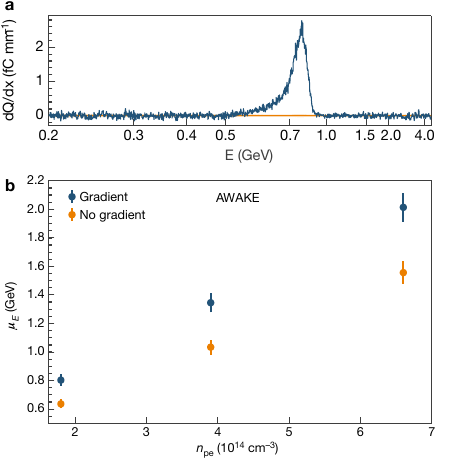}}
    \caption{(a) Acceleration of an electron bunch in a self-modulated proton-driven wakefield, achieved at the AWAKE experiment at CERN, showing a peak in the spectrum around \SI{0.8}{GeV}. (b) Energies up to \SI{2}{GeV} were achieved at higher gradients and when introducing a small plasma-density gradient. Adapted from \textcite{Adli2018} (CC-BY 4.0).}
    \label{fig:awake-acceleration}
\end{figure}

AWAKE demonstrated acceleration of electrons in a proton beam-driven plasma wakefield up to \SI{2}{GeV}~\cite{Adli2018,Gschwendtner2019} (see Fig.~\ref{fig:awake-acceleration} and Sec.~\ref{sec:self-modulated-wakefields:capture}). Subsequent studies mapped the strength of the self-modulated wakefield by scanning both the timing of the laser seed pulse that creates the relativistic ionization front and the timing of the injected electron bunch~\cite{Turner2020,Chappell2021}. The facility is also well suited to measuring fundamental plasma-physics processes, using its long plasma source and unique suite of diagnostics. Measurements include observation of proton-bunch hosing \cite{Nechaeva2023} (Sec.~\ref{sec:hosing-instability}), observation of transverse filamentation~\cite{Verra2024} (Sec.~\ref{sec:drive_instabilities}), and the effect of ion motion on wake formation~\cite{Turner2025}. 

An upgrade of the AWAKE facility has been planned~\cite{Gschwendtner2022}, making space for longer plasma cells. The plasma cell would be split into a modulation stage and an acceleration stage, between which \SI{150}{MeV} electrons are injected \cite{vanGils2025}---higher electron energy is used in order to increase the charge-capture efficiency~\cite{Ramjiawan2023}. The upgraded facility would feature novel, extendable plasma sources such as helicon~\cite{Buttenschn2018} and discharge plasma cells~\cite{Torrado2023}, the latter which was already used in the above-mentioned ion-motion result.

\subsubsection{INFN: SPARC\_LAB and EuPRAXIA}
\label{sec:experimental:facilities:infn}

The SPARC\_LAB facility \cite{Ferrario2013}---merging the capabilities of the former SPARC and PLASMONX projects  \cite{Alesini2005}---is located at INFN in Frascati, Italy, and builds on the existing SPARC accelerator and FEL facility \cite{Alesini2003}. It uses a \SI{150}{MeV} electron rf accelerator with velocity bunching to perform PWFA experiments. Multi-bunch generation is based on comb-like laser pulses directly on the photocathode \cite{Ferrario2011,Cianchi2015}. A set of permanent magnetic quadrupoles provides compact final focusing into the plasma \cite{Pompili2018b}. The plasma source, used for PWFA as well as for active-plasma-lens experiments \cite{Marocchino2017,Pompili2018b}, is a \SI{33}{mm}-long 3D-printed discharge capillary \cite{Filippi2018}, characterized using optical spectroscopy \cite{Costa2022}.

\begin{figure}[t]
    \centering{\includegraphics[width=\linewidth]{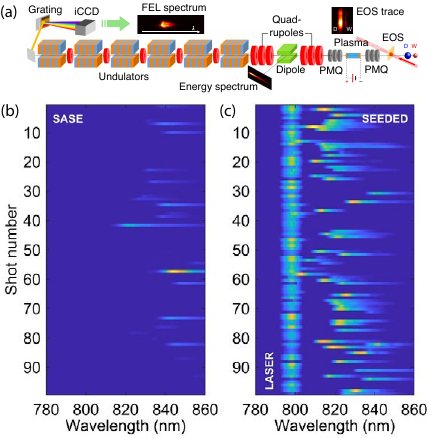}}
    \caption{(a) Schematic of the PWFA-boosted FEL beamline at SPARC\_LAB, INFN. Photon spectra show (b) self-amplified and (c) seeded FEL gain at infrared wavelengths. Adapted from \textcite{Pompili2022} and \textcite{Galletti2022}.}
    \label{fig:sparc_lab-fel-gain}
\end{figure}

Early experiments included reduction of energy spread, either through dechirping of a single bunch \cite{Shpakov2019} or using a double-bunch setup where the dechirped bunch was also accelerated \cite{Pompili2021}. First emittance measurements were also performed, using a multi-shot quadrupole scan \cite{Shpakov2021}. A seminal result was the use of the PWFA module to boost the energy of a high-quality bunch (1.4$\times$\SI{1.2}{\milli\m\milli\radian} emittance, 0.4\% rms energy spread and \SI{270}{A} peak current), from 88 to \SI{94}{MeV}, before using it in an FEL application; the beam quality was sufficiently preserved to demonstrate FEL gain \cite{Pompili2022}. This experiment included seeding of the FEL process with an infrared laser \cite{Galletti2022} (see Fig.~\ref{fig:sparc_lab-fel-gain}). Recent experiments include the use of an integrated plasma source with upstream and downstream plasma lenses for improved matching \cite{Pompili2024a}, measurements of ns-scale recovery of the plasma density in hydrogen \cite{Pompili2024b}, observation of plasma screening of relativistic beam fields~\cite{Verra2024b}, detailed studies of a laser-induced plasma filament \cite{Galletti2025} and resonant excitation of wakefields using a train of electron bunches~\cite{Verra2025}.

The EuPRAXIA project \cite{Assmann2020} aims to demonstrate a plasma-based FEL user facility with high-brightness 1--\SI{5}{GeV} electron beams. EuPRAXIA@SPARC\_LAB \cite{Ferrario2018} has been selected as the beam-driven arm of this project, and is currently being commissioned \cite{Villa2023}, aiming to deliver 1--\SI{1.2}{GeV}, \SI{50}{pC} electron bunches with high quality (\SI{0.5}{\milli\m\milli\radian} emittance and 0.05\% energy spread per slice) at \SI{100}{Hz} repetition rate \cite{DelDotto2025,Romeo2024}.

\subsubsection{DESY: PITZ and FLASHForward}
\label{sec:experimental:facilities:desy}

Two separate facilities have been performing PWFA experiments at the Deutsches Elektronen-Synchrotron (DESY) in Germany: PITZ at the Zeuthen site and FLASHForward at the Hamburg site.

The PITZ facility \cite{Gross2014}, primarily dedicated to photo-injector research, delivers \SI{\sim20}{ps}-long electron bunches with low emittance (\SI{\sim 0.4}{\milli\m\milli\radian}) at energies up to \SI{25}{MeV}. Two plasma sources were used: a heat-pipe oven and a UV laser that created an \SI{8}{cm}-long lithium plasma at densities up to \SI{e15}{\per\cm\cubed}, or a \SI{10}{cm}-long high-voltage discharge that ionized argon at densities up to \SI{e16}{\per\cm\cubed} \cite{Lishilin2019}. The key experimental results achieved were the first demonstration of the self-modulation instability, measured in longitudinal phase space using a combination of an rf deflector and a magnetic dipole \cite{Gross2018}, and the first demonstration of high transformer ratio ($\mathcal{R}=|E_\mathrm{acc}|/|E_\mathrm{dec}|$; see Sec.~\ref{sec:high-transformer-ratio}), up to 4.6, by employing a long driver with ramped current profile \cite{Loisch2018}. Later work includes plasma-density measurements via the self-modulation instability \cite{Loisch2019b}, as well as development of low-jitter discharge sources \cite{Loisch2019} and polymer foil windows to protect the upstream vacuum \cite{Engel2020}.

The FLASHForward facility \cite{Aschikhin2016,DArcy2019a} makes use of \SI{1.2}{GeV}, \SI{1}{nC} electron bunches accelerated in the FLASH linac, a free-electron laser facility based on superconducting rf cavities and a high-quality photocathode electron source. The experimental area includes a collimator system for double-bunch generation \cite{Schroeder2020b}, a capillary-based plasma source, ionized by either a \SI{25}{TW} Ti:sapphire laser or a high-voltage discharge, an imaging-spectrometer setup with both broadband and high-resolution screens, and a polarizable transverse-deflection structure~\cite{Caminal2024} for measuring the longitudinal phase space. The facility is unique in its ability to deliver bunches at up to \SI{3}{MHz} repetition rate and, in principle, up to \SI{30}{kW} of beam power. Three major directions have been pursued: downramp injection for beam-brightness transformation (Sec.~\ref{sec:downramp-injection}), external injection with beam-quality preservation and high energy-transfer efficiency (Sec.~\ref{sec:evolution-trailing-bunch}), and experiments investigating the limits of repetition rate (Sec.~\ref{sec:long-term-evolution}).

\begin{figure}[t]
    \centering{\includegraphics[width=0.95\linewidth]{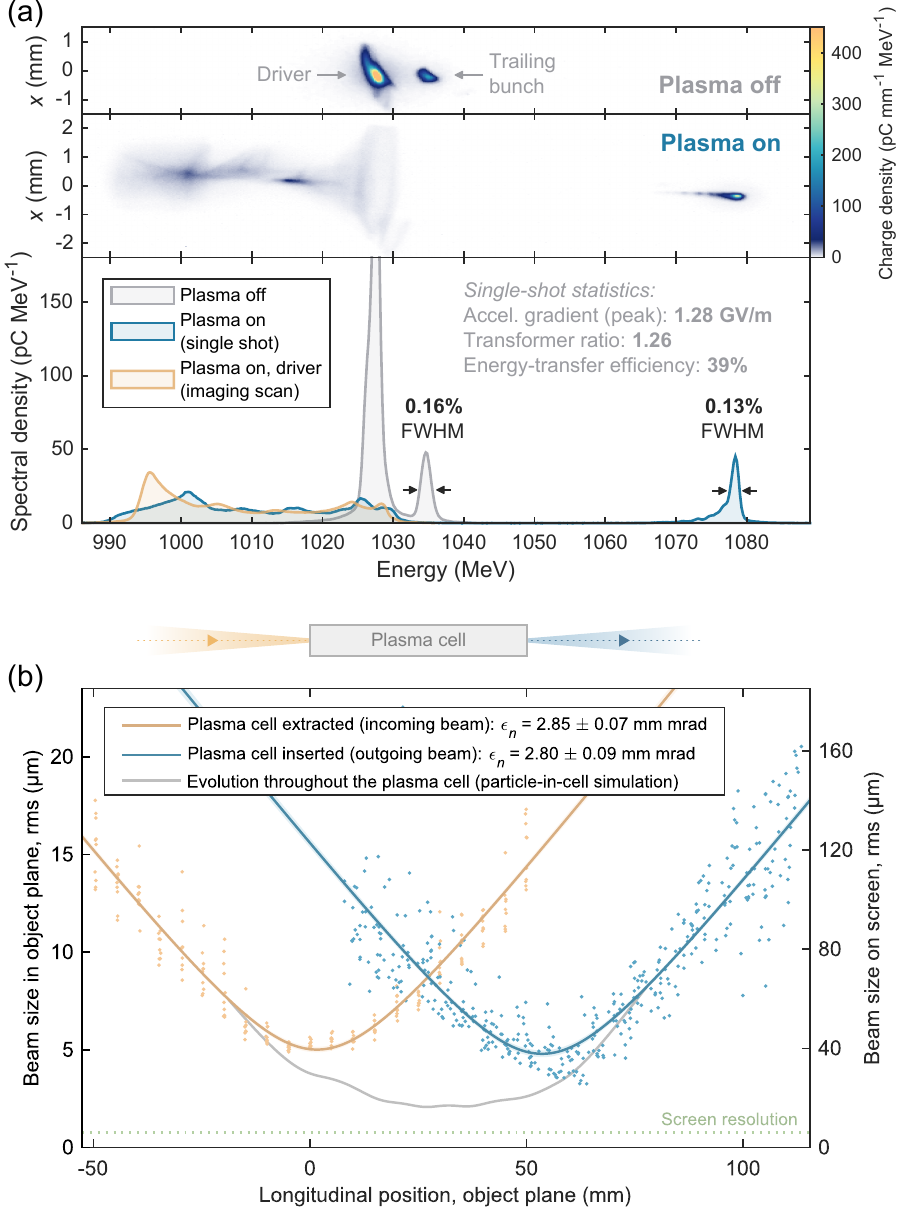}}
    \caption{Beam-quality preservation: first demonstrations of energy-spread preservation (a) and emittance preservation (b), performed in two separate experiments at FLASHForward. Energy-spread preservation required optimal beam loading of the wakefield and high-resolution energy spectrum measurements. Emittance preservation required precise alignment and beta-function matching as well as a high-resolution quadrupole scan, measuring the virtual beam size (i.e., as it would be with no plasma) in various planes throughout the plasma cell. From \textcite{Lindstrom2021a} and \textcite{Lindstrom2024} (CC-BY 4.0).}
    \label{fig:emittance-preservation}
\end{figure}

Initial experiments at FLASHForward included a demonstration of dechirping \cite{DArcy2019b} (see Fig.~\ref{fig:dechirping}), in which the energy spread of a bunch was reduced from 1.31\% to 0.33\% FWHM in a \SI{33}{mm}-long plasma; high-resolution measurements of plasma wakefields \cite{Schroeder2020} based on collimation; and a novel BPM-based diagnostics technique for measuring small beta functions \cite{Lindstrom2020a}. Stable density-downramp injection was achieved \cite{Knetsch2021} [see Fig.~\ref{fig:density-downramp-injection}(d)], with a 95\% injection probability and a normalized emittance of \SI{9.3}{mm.mrad} for the injected electron bunch. This was later improved down to \SI{1.2}{\milli\meter\milli\radian}, demonstrating a beam-brightness boost between the driver and the injected bunch \cite{Wood2026}. 
Optimal beam loading was demonstrated \cite{Lindstrom2021a} [see Fig.~\ref{fig:beam_loading_nonlinear_flattening_exp}(a)], showing simultaneous high energy-transfer efficiency of $\eta_\mathrm{p \to t}=42\%$ and preservation of 0.16\% FWHM energy spread and \SI{100}{pC} of charge, followed by a demonstration of emittance preservation at \SI{2.8}{\milli\m\milli\radian} with \SI{40}{MeV} energy gain \cite{Lindstrom2024}, shown in Fig.~\ref{fig:emittance-preservation}(a) and (b), respectively. Later experiments with sixfold higher gain (\SI{252}{MeV}) lowered the energy spread \textit{per gain} from a few percent down to 0.4\% rms \cite{Wood2024}. High driver-energy depletion was also demonstrated with $\eta_\mathrm{d \to p}=57\%$ \cite{Pena2024} (Fig.~\ref{fig:depletion} and Sec.~\ref{sec:drive_plasma_efficiency}). The longitudinal phase space of beam-loaded bunches was measured for the first time [see Figs.~\ref{fig:beam_loading_nonlinear_flattening_exp}(c--e)], which was used to set a new upper bound on the transverse uniformity of the accelerating field observed in the plasma wake: 0.6\% rms at 68\% confidence level \cite{Caminal2022}. Limits on repetition rate in PWFA were also established \cite{DArcy2022}, indicating that long-term ion motion requires at least 10--\SI{100}{ns} (in argon) before repeatable acceleration is recovered (see Fig.~\ref{fig:long-term-evolution:recovery-time}).

\subsubsection{Tsinghua University, IHEP and SXFEL}
PWFA research is ramping up in several locations across China. Small-scale experiments on dechirping with hollow plasma channels \cite{Liu2024} (see Secs.~\ref{sec:without-focusing} and \ref{sec:dechirping}) have already been performed using a sixth-order Bessel laser beam and a \SI{50}{pC}, \SI{39}{MeV} electron beam from TTX platform at Tsinghua University, Beijing \cite{Tang2009}. The construction of a larger-scale PWFA research facility was initiated in 2023 at IHEP, Beijing \cite{Li2025}, using the \SI{2}{GeV} BEPC-II electron and positron linacs and a new \SI{150}{MeV} electron linac. It aims to study beam quality, efficiency, staging as well as positron acceleration. Finally, plans also exist to upgrade the SXFEL facility in Shanghai using PWFA to double the energy of the \SI{\sim 1}{GeV} beam \cite{Peng2026,Li2026}.

\subsubsection{Hybrid LWFA-driven PWFA}

\label{sec:hybrid}

LWFA can generate electron bunches that are suitable for driving the PWFA process. A transition from laser-driven to beam-driven plasma waves can naturally happen in LWFA when the laser driver depletes and the accelerated electron bunch takes the role of a driver. This has been observed in early numerical studies~\cite{Tsung2004,Pae2010}, and several experimental studies found indications for this transition~\cite{Corde2011,MassonLaborde2014,Heigoldt2015,Guillaume2015}.

\begin{figure}[t]
    \centering{\includegraphics[width=0.95\linewidth]{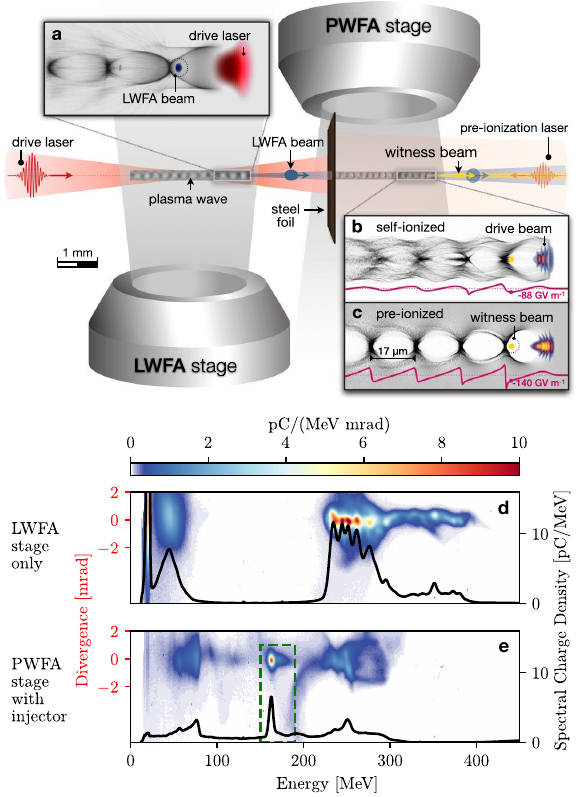}}
    \caption{Schematic of a LWFA-driven PWFA: the first stage (a) is driven by a laser pulse, after which laser-injected electrons drive a self-ionized (b) or pre-ionized (c) second stage. The beam quality of the laser-injected beams (d) can be surpassed by high-brightness beams injected in the PWFA stage (e). Adapted from \textcite{Kurz2021} and \textcite{Foerster2022} (CC-BY 4.0).}
    \label{fig:hybrid-schematic}
\end{figure}

A hybrid acceleration scheme, where LWFA is used specifically to generate PWFA drivers, is depicted in Figs.~\ref{fig:hybrid-schematic}(a--c). The concept was proposed as an energy booster for LWFA~\cite{Hidding2010}, but it soon became apparent that its prospects are much broader [see e.g.~\textcite{Hidding2019a,Hidding2019b}]. The hybrid scheme allows to utilize the strengths of both LWFA and PWFA, i.e., the availability, compactness and comparatively low cost of LWFAs, and the prospects of high beam qualities from PWFAs. Such hybrid accelerators may then be used, e.g., as compact and low-cost drivers for FELs (Sec.~\ref{sec:applications:FEL}). Hybrid LWFA--PWFA can also open new opportunities for ultrabright gamma-ray sources, which can be obtained from betatron radiation~\cite{Wang2002} in a high-density PWFA driven by an electron beam generated in a low-density LWFA~\cite{Ferri2018}.

Experimental progress in hybrid LWFA-driven PWFA, reviewed by \textcite{Hidding2023}, has mainly been driven by three facilities: (i) the Laboratory for Extreme Photonics (LEX) at the Ludwig-Maximilians-Universität München with the
\SI{300}{\tera\watt} ATLAS 300 laser (until 2016); (ii) the Helmholtz-Zentrum Dresden-Rossendorf (HZDR) with the \si{\peta\watt}-class DRACO laser; and (iii) the Centre for Advanced Laser Applications (CALA) at the Ludwig-Maximilians-Universität München with the upgraded \SI{3}{\peta\watt} ATLAS 3000 laser (from 2017).

Proof-of-principle experiments demonstrated that PWFA can be powered by LWFA-generated electron bunches, by observing plasma lensing~\cite{Kuschel2016} and deceleration~\cite{Chou2016} of the drive electron bunch and the acceleration of a trailing electron bunch~\cite{Kurz2021} in the PWFA stage, and by taking the first snapshots of PWFA plasma waves~\cite{Gilljohann2019} (see Fig.~\ref{fig:long-term-evolution:transverse-shadowgrahy}). Shortly after, internal injection based on density-downramp injection (Sec.~\ref{sec:downramp-injection}) in the PWFA stage and subsequent acceleration have been achieved in two different laboratories~\cite{CouperusCabada2021,Foerster2022}. \textcite{Foerster2022} showed that the hybrid scheme can act as a quality transformer, where lower-quality LWFA beams can generate higher-quality electron beams in the PWFA process [see Figs.~\ref{fig:hybrid-schematic}(d--e)], and later achieved a record-high driver-to-trailing-bunch energy-transfer efficiency of 20\% \cite{Foerster2026}.

The hybrid scheme also enables fundamental PWFA research in widely available laser laboratories, which can support the research towards large-scale high-energy accelerators and colliders (Sec.~\ref{sec:applications:colliders}). \textcite{Gotzfried2020} used a pair of electron bunches from LWFA to study PWFA with a two-bunch configuration. \textcite{Schoebel2022} observed the damping of PWFA plasma waves in a self-ionized plasma (i.e.,~an initially neutral gas ionized by the fields of the drive electron beam) and the elongation of the first PWFA plasma period with higher PWFA drive charge. The inherently synchronized laser that drives the LWFA can easily be used for optical diagnostics like few-cycle shadowgraphy~\cite{Buck2011,Schwab2013,Savert2015}. This allowed, e.g., to study the long-term plasma response with high spatial and temporal resolution~\cite{Gilljohann2019} (see Sec.~\ref{sec:long-term-evolution} and Fig.~\ref{fig:long-term-evolution:transverse-shadowgrahy}). This inherent synchronization facilitates advanced concepts of internal injection that require an additional laser to trigger injection, e.g.~the plasma-photocathode scheme \cite{Ufer2026}.


\section{Applications}
\label{sec:applications}

Beam-driven PWFA has historically been motivated by the potential to shrink the footprint and cost of large-scale and high-power particle accelerators, such as those used for high-energy physics and photon science. This most prominently includes particle colliders (Sec.~\ref{sec:applications:colliders}) and free-electron lasers (Sec.~\ref{sec:applications:FEL}), as detailed below. 

Other potential applications include fixed-target experiments \cite{Wing2019}, synchrotron injectors \cite{Wang2022}, laboratory astrophysics including strong-field QED experiments \cite{Cole2018,Poder2018,Mirzaie2024,Matheron2024}, medical applications such as ultra-high-energy electron (UHEE) radiotherapy \cite{Kokurewicz2019}, and extreme ultraviolet (EUV) semiconductor lithography \cite{Goldstein2005}.

\subsection{Particle colliders}
\label{sec:applications:colliders}

The concept of PWFA emerged from basic research on laser--plasma \cite{Tajima1979} and beam--plasma interactions \cite{Chen1985,Ruth1985}, and it was immediately recognized by the high-energy physics community as a promising technology for future colliders at the TeV scale~\cite{Richter1988,Joshi2012}. However, building a collider requires tackling a number of challenges related to luminosity and power, as laid out by \textcite{Schulte2016}, which have been progressively addressed over the several generations of concepts discussed below.

\begin{figure}[t]
    \centering
    \includegraphics[width=\linewidth]{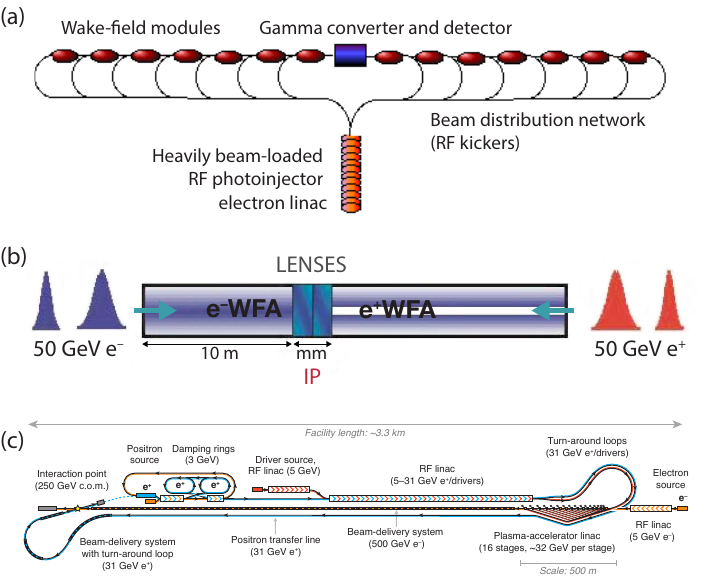}
    \caption{Various electron-driven PWFA collider concepts, including: (a) a multistage gamma--gamma collider; (b) the plasma afterburner or energy doubler; and (c) HALHF. Adapted from \textcite{Rosenzweig1996}, \textcite{Lee2002} (CC-BY 3.0) and \textcite{Foster2023} (CC-BY 4.0).}
  \label{fig:various-pwfa-colliders}
\end{figure}

The first conceptual design of a collider based on beam-driven plasma accelerators was introduced by \textcite{Rosenzweig1996,Rosenzweig1998}, as shown in Fig.~\ref{fig:various-pwfa-colliders}(a). It was based on one low-gradient (\SI{6}{MV/m}) and high-power (\SI{104}{MW}) rf linac providing \SI{3}{GeV} electron drivers for two PWFA linacs, each with multiple \SI{5.7}{m}-long stages accelerating at \SI{1}{GV/m}. This collider concept was envisioned as a gamma--gamma collider \cite{Ginzburg1983}, where both arms accelerate electron beams converted to photon beams shortly before the interaction point.

The next concept to be introduced was the ``plasma afterburner" or ``energy doubler" \cite{Lee2002} [see Fig.~\ref{fig:various-pwfa-colliders}(b)]. This uses beams from an existing linear collider, splits them longitudinally into two bunches, doubling the energy of the trailing bunch just prior to the interaction point. It was also proposed to use hollow channels for positrons (Sec.~\ref{sec:without-focusing}) and plasma lenses for final focusing \cite{Chen1987}. The original study used \SI{50}{GeV} beams from the Stanford Linear Collider, but was later extended to \SI{250}{GeV} beams from ILC \cite{Raubenheimer2004}, inspiring studies into the detrimental effect of ion motion \cite{Rosenzweig2005} (Sec.~\ref{sec:ion-motion}). A multi-bunch version of the afterburner concept was proposed by \textcite{Maeda2004}, using up to five unequally spaced drivers to give higher energy gain (2.8$\times$ the initial energy) and lower energy spread (4\%).

A new iteration of the multistage collider, colliding electrons and positrons, was proposed by \textcite{Seryi2009,Pei2009}, as shown in Fig.~\ref{fig:PWFA-collider-concept}. Inspired by CLIC \cite{Aicheler2012}, it fleshed out the rf driver linac design, beam loading (Sec.~\ref{sec:beam-loading}), and the driver-distribution system using kickers and 180-degree turn-arounds. An updated version was later introduced by \textcite{Adli2013,Delahaye2014}, who simplified the driver-distribution system with an undulating chicane design instead of turn-arounds. This version was estimated to reach a wall-plug energy efficiency of approximately 10\% and a similar luminosity-per-power [Eq.~(\ref{eq:luminosity-per-power})] to that of CLIC. However, it also focused a spotlight on the challenge of high-quality positron acceleration \cite{Cao2024} (Sec.~\ref{sec:positron-acceleration}) as well as compact, quality-preserving staging (Sec.~\ref{sec:staging}), which remained unsolved in these designs \cite{Adli2019}.

The hybrid, asymmetric, linear Higgs factory (HALHF) concept by \textcite{Foster2023} sidesteps the positron problem by only accelerating electrons with PWFA (to high energy), colliding them with positrons accelerated by an rf accelerator (to lower energy), as illustrated in Fig.~\ref{fig:various-pwfa-colliders}(c). HALHF initially proposed to collide $E_e=\SI{500}{GeV}$ electrons with $E_p=\SI{31}{GeV}$ positrons, resulting in a \SI{250}{GeV} center-of-mass energy; an updated baseline \cite{Foster2025,Adli2025a} uses $E_e=\SI{375}{GeV}$ and $E_p=\SI{42}{GeV}$. Asymmetric collisions require more energy overall, as the collision products have a forward boost, but this can be mostly re-couped by using asymmetric bunch charges: more charge in the low-energy positrons ($Q_p=\SI{4.8}{nC}$) and less in the high-energy electrons ($Q_e=\SI{1.6}{nC}$). This is because the luminosity scales as $Q_eQ_p$ but the power scales as $Q_eE_e+Q_pE_p$. Moreover, the higher electron energy allows for a higher normalized emittance (but similar geometric emittance) in the PWFA linac. This, along with an order-of-magnitude reduction in plasma density as well as the introduction of a controlled amount of ion motion \cite{An2017,Benedetti2017}, mitigates the issue of beam-breakup instability and improves the transverse tolerances \cite{Adli2025b}. High-charge (\SI{8}{nC}), medium-energy (\SI{4}{GeV}) drive bunches from a CLIC-like rf driver linac \cite{Aicheler2012} ensures a high wall-plug-to-driver power efficiency and compact driver distribution. Achromatic staging is performed with the use of nonlinear plasma lenses \cite{Drobniak2025,Lindstrom2026}, which through self-correction significantly improves the timing and current-shaping tolerances \cite{Lindstrom2021c}. Cost estimates obtained by scaling from ILC and CLIC showed significant cost savings compared to rf-only-based colliders \cite{Lindstrom2025b}. The HALHF concept therefore addresses all the key challenges laid out by \textcite{Schulte2016}: positron acceleration, efficiency, staging, beam quality, transverse tolerances and instability, as well as cost effectiveness. However, as with previous concepts, more realism in design leads to new specific challenges, including plasma heating and cooling (Sec.~\ref{sec:long-term-evolution}), radiation shielding and cross-plane emittance mixing of flat beams \cite{Diederichs2024}.

\begin{table}[b]
    \centering
    \begin{tabular}{l|cc|cc}
        & \multicolumn{2}{c}{\textit{Requirements}} & \multicolumn{2}{c}{\textit{State-of-the-art}} \\
        \textit{PWFA parameter} & \textit{Collider} & \textit{FEL} & \textit{Separat.} & \textit{Simult.} \\
        \hline
        Accelerated particle type & $e^\pm$ & $e^-$ & $e^\pm$ & $e^-$ \\
        Energy gain (GeV/stage) & 1--100 & 1--10 & 29\footnote{\textcite{Blumenfeld2007} showed \SI{42}{GeV} gain, but not for a distinct trailing bunch.} & 0.04 \\
        Energy spread (per gain) & $\sim$0.3\% & $\sim$0.1\% & 0.4\% & 1.6\%  \\
        Charge (nC) & $\sim$1 & $\sim$0.1 & 0.1 & 0.04 \\
        Emittance (mm mrad) & $\sim$0.1 & $\sim$1 & 1.2 & 2.8 \\
        Driver-to-beam efficiency & 10--50\% & 1--10\% & 20\% & 0.4\%\\
        Repetition rate (Hz) & $\sim$\SI{e4}{} & 1--\SI{e6}{} & 10 & 10 \\
        Number of stages & 10--100\footnote{If using proton drivers, a single long stage may be sufficient.} & 1--10 & 1 & 1 \\
        \hline
    \end{tabular}
    \caption{Parameter requirements for collider and FEL applications, compared with numbers achieved in state-of-the-art two-bunch PWFA experiments, both separately and simultaneously. The former are quoted from \textcite{Zhang2025}, \textcite{Wood2024}, \textcite{Lindstrom2021a}, \textcite{Wood2026} and \textcite{Foerster2026} (in hybrid LWFA/PWFA), while the latter are quoted from \textcite{Lindstrom2024}, where the beam quality was preserved.}
    \label{tab:parameter-requirements}
\end{table}

While all the above concepts used electron drivers, proton-driven collider concepts have also been proposed. One concept is the very high energy electron--proton collider (VHEeP) proposed by \textcite{Caldwell2016b}, using \SI{7}{TeV} protons from the LHC to accelerate electrons up to \SI{3}{TeV} in a self-modulated wake (Sec.~\ref{sec:self-modulated-wakefields}) colliding with another LHC proton beam. Specifically, this could reach parton momentum fractions, $x$, down to \SI{e-8}{} for photon virtualities, $Q^2$, of \SI{1}{GeV^2}. Another concept is the plasma electron--proton/ion collider (PEPIC) \cite{Gschwendtner2018}, siphoning off LHC beams to a proton-driven PWFA to produce and re-inject \SI{70}{GeV} electrons into the LHC for $e$--$p$ collisions. Lastly, an electron--positron collider based on PWFA driven by short proton bunches was recently proposed by \textcite{Farmer2024}, utilizing fast-ramping proton rings.

A collider application ultimately imposes stringent parameter requirements on PWFAs. Table \ref{tab:parameter-requirements} compares these requirements with state-of-the-art experimental results, indicating clear progress but also that significant experimental effort is still required, and highlighting the importance of medium-term applications including FELs.

\subsection{Free-electron lasers}
\label{sec:applications:FEL}

In photon science, which makes use of bright photon beams for research in e.g.~biology, chemistry, and material science \cite{Bostedt2016}, PWFAs can open new possibilities with the application of a plasma-based free-electron laser. Recently, FELs powered by plasma accelerators saw their first real-world demonstrations, for both LWFA \cite{Wang2021b,Labat2023} and PWFA \cite{Pompili2022, Galletti2022}, successfully operating in either self-amplified spontaneous emission (SASE) or seeded configurations (see Fig.~\ref{fig:sparc_lab-fel-gain}). These experimental breakthroughs represent a cornerstone in the development of photon-science applications of plasma accelerators, opening prospects for both more compact FELs and higher photon brightness \cite{Emma2021c,Galletti2024}, as discussed below.

PWFA offers several avenues for reducing the size and cost of FELs. Table \ref{tab:parameter-requirements} summarizes the typical beam parameters required in FELs. A standard quality-preserving PWFA module can be used as a cost-effective energy booster or upgrade for the FEL electron beam driver, which can be of interest to existing FEL facilities \cite{Israeli2016,Schroeder2025} as well as new facilities, e.g.~EuPRAXIA \cite{Assmann2020}. In the latter case, the PWFA ``pillar" of EuPRAXIA uses a 0.5--\SI{1}{GeV} driver and trailing bunch from an X-band linac \cite{Vaccarezza2018} to reach 1--\SI{2}{GeV} after the PWFA stage, lasing in the undulators at a wavelength of a few nm \cite{Ferrario2018, Petrillo2018}. Looking ahead, PWFA R\&D may allow use of high-transformer ratio (Sec.~\ref{sec:high-transformer-ratio}) or multistage acceleration (Sec.~\ref{sec:staging}) to extend the energy reach beyond what a standard PWFA module can provide. Ultimately, this could enable x-ray FELs (XFELs) \cite{Pellegrini2016} with electron drivers in the 10--\SI{20}{GeV} range with a total accelerator length (i.e., rf linac and PWFA) of less than \SI{100}{meters}, comparable or shorter than the undulator section. 
Finally, plasmas or lasers can themselves be used as compact undulators, replacing magnetic undulators by a ``plasma wiggler" in the so-called \textit{ion-channel laser} \cite{Whittum1990,Ersfeld2014,davoine2018} or by an optical undulator \cite{Xu2024}.

\begin{figure}[t]
    \centering
    \includegraphics[width=0.92\linewidth]{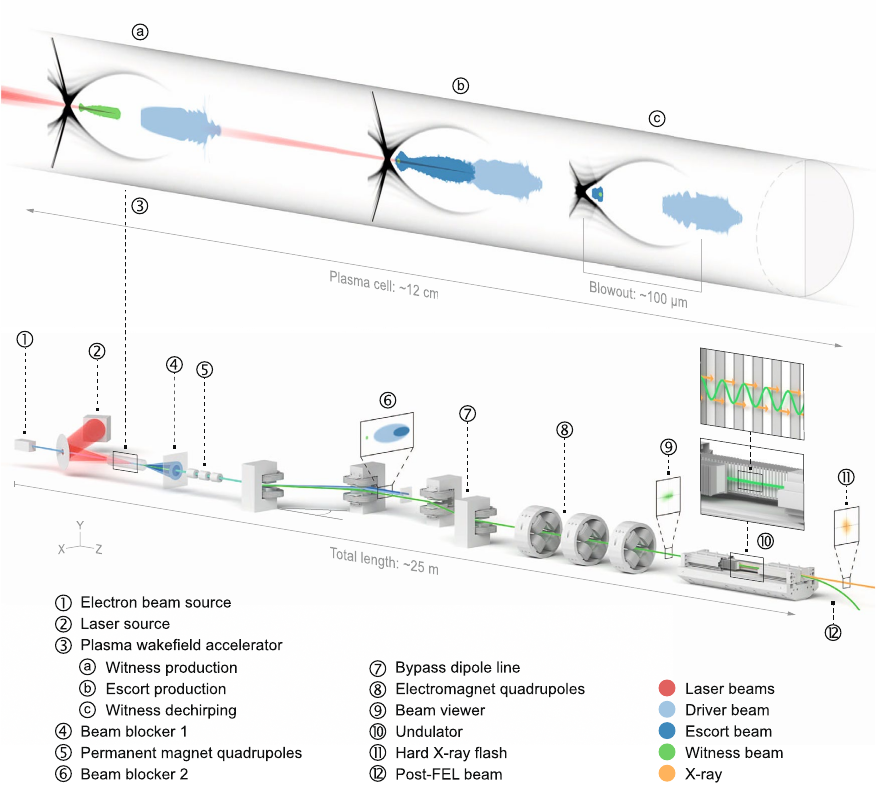}
    \caption{PWFA-based FEL concept based on a plasma photocathode. From \textcite{Habib2023} (CC-BY 4.0).}
  \label{fig:fel}
\end{figure}

The other key motivation, with a potentially strong physics impact, comes from the new capabilities opened by ultrabright and ultrashort beams from PWFA. Internal injection (Sec.~\ref{sec:internal-injection}) in PWFA offers new means to generate beams of superior brightness, e.g.~using the plasma photocathode scheme~\cite{Hidding2012} that can deliver 6D brightness [Eq.~(\ref{eq:brightness})] of \SI{\sim e18}{A.m^{-2}.rad^{-2}/0\rm{.}1\%BW} \cite{Manahan2017,Habib2023}, as well as pre-bunched beams \cite{Xu2022}. Beams internally injected and accelerated in a PWFA are generally chirped (see Sec.~\ref{sec:dechirping}) and can be further compressed to attosecond duration and near-MA currents in a magnetic chicane \cite{Emma2021a, Hessami2024}. Alternatively, an externally injected trailing bunch from the linac can be chirped in a PWFA stage for further compression to near-MA currents \cite{Emma2021b}. These unique beam properties enable novel possibilities with near-single-spike temporally coherent FEL radiation pulses with duration of the order of \SI{100}{as} FWHM or below and GW-to-TW peak power. In addition, higher brightness and beam current lead to much reduced FEL gain length, resulting in a more compact undulator section \cite{Baxevanis2017}. For example, in the conceptual design depicted in Fig.~\ref{fig:fel}, the plasma photocathode, beam transport and undulators fit in a total length of \SI{25}{m} while delivering attosecond and angstrom FEL light \cite{Habib2023}. 

Combining the above advantages of PWFA, a very competitive PWFA-based FEL facility design was proposed by \textcite{Hogan2016} (see Fig.~17 therein), leveraging high-repetition-rate beam drivers to drive multiple FELs, high-transformer-ratio PWFA for compact acceleration, and plasma photocathode for compact undulators and for single-spike attosecond FEL light whose photon brightness exceeds the current state-of-the-art. An alternative design for a compact FEL relies on hybrid LWFA-driven PWFA (Sec.~\ref{sec:hybrid}), which benefits from the beam quality offered by PWFA, the compactness and comparatively low cost of LWFA, and the inherently precise synchronization between lasers and beams that originate from a single laser system.


\section{Conclusion and outlook}
\label{sec:conclusion}

The field of beam-driven plasma acceleration has witnessed extraordinary progress since the 1980s. Collaborative efforts among national laboratories and university research groups have successfully merged diverse expertise in accelerators, lasers, plasmas, and computational methodologies, resulting in pioneering experiments at state-of-the-art accelerator test facilities. Major experimental results include demonstrations of large energy gain, positron acceleration, high energy-transfer efficiency, high transformer ratios, energy-spread and emittance preservation as well as internal injection of high-quality beams.

In PWFA, the power source is a high-energy particle beam, and the accelerating medium is a plasma. Ongoing developments in rf accelerators, at facilities like the European XFEL and LCLS-II, are producing electron beams with higher energy, quality and power. Coupled with increasingly well-controlled and -diagnosed plasmas, able to sustain GV/m accelerating fields and MT/m focusing gradients, PWFA is well positioned to deliver both more cost-effective and higher-brightness electron beams, adding GeVs of energy while preserving beam quality. This promises to transform high-energy particle-accelerator facilities, such as future particle colliders. The most ambitious concepts envisage employing PWFAs to drastically reduce the size, cost and carbon footprint of particle colliders examining the Higgs boson and subsequently the energy frontier of particle physics. Furthermore, the extremely short and bright bunches from PWFAs could unlock enhanced photon generation, opening new avenues for applications in photon science.

The concepts and results presented in this Review illustrate the potential of PWFA, but also highlight critical challenges that need to be tackled. In particular, for a collider application, the particle-physics community requires not only high energy but also high luminosity-per-power \cite{Roser2023}. While a variety of beam-parameter combinations can be used to satisfy this, it typically implies using several beam-quality-preserving stages with \SI{10}{GeV}-level energy gain, transferring up to 50\% of the driver energy to beams with $\sim$\SI{1}{nC} charge, $\sim$0.3\% energy spread and normalized emittances as low as $\sim$\SI{0.1}{\milli\m\milli\radian} at a repetition rate of $\sim$\SI{10}{kHz}. FELs require nearly as high-quality beams (\SI{0.1}{nC}, \SI{1}{\milli\m\milli\radian} and 0.1\% energy spread), but at lower repetition rate and efficiency. Simulations indicate that this is achievable, and some of these numbers have been individually achieved in experiments (see Table \ref{tab:parameter-requirements}): \SI{29}{GeV} energy gain, 20\% driver-to-beam efficiency, \SI{0.1}{nC} charge, \SI{1.2}{\milli\m\milli\radian} emittance and 0.4\% rms energy spread per gain. State-of-the-art experiments attempting all at once have so far demonstrated beam-quality preservation for beams with \SI{2.8}{\milli\m\milli\radian} emittance, 0.06\% rms energy spread (1.6\% rms per gain) and \SI{40}{pC} charge accelerated by \SI{40}{MeV} with 0.4\% driver-to-beam efficiency at \SI{10}{Hz}. This simultaneously motivates and highlights the need for continued experimental and theoretical effort. 

Recent strategy workshops \cite{2016AACreport, 2022AACforESPP} have culminated in comprehensive road maps that prioritize research areas such as: emittance preservation for large energy gain, in particular related to tight misalignment tolerances, instability suppression and ion motion for intense bunches; demonstrating compact and achromatic staging; the development of systems capable of delivering high average beam power; as well as positron acceleration with both high beam quality and efficiency, though this may be circumvented using gamma--gamma or hybrid (plasma--rf) colliders. Design initiatives, such as HALHF \cite{Foster2023,Foster2025}, EuPRAXIA \cite{Assmann2020} and the ongoing study of a \SI{10}{TeV} parton center-of-momentum wakefield-based energy frontier collider \cite{Gessner2025}, will provide further guidance and focus for research directions and will identify needs for future R\&D and demonstrator facilities. In the near-to-medium term this likely involves the implementation of a plasma-based FEL user facility (EuPRAXIA) and a multistage plasma accelerator facility that can demonstrate staging and achieve energies of order 50--\SI{100}{GeV}. Moreover, the development of advanced simulation tools, integrating reduced models, artificial intelligence and machine-learning surrogates, and exascale computing, will be vital for exploring the vast available parameter space, optimizing performance, and determining tolerances. Finally, it is critical to identify and apply PWFA to new medium-term applications, such as strong-field QED~\cite{Lindstrom2025} or high-energy-physics studies including precision quantum chromodynamics and ``Beyond the Standard Model" physics measurements \cite{Bulanov2024}, where achieving \SI{100}{GeV}-level energy will enable forefront studies but the requirements for beam quality and power are somewhat relaxed.

In conclusion, this is an exhilarating era for plasma-wakefield accelerator research, marked by the promise of groundbreaking applications and theoretical refinement. The coming years are poised to yield further discoveries and the first realization of practical applications, leading to refined concepts for light sources and energy-frontier colliders. As the field continues to evolve, it is set to significantly impact high-energy physics and beyond.


\begin{acknowledgments}
    The authors would like to thank B.~Foster for valuable feedback. This work was funded by the Research Council of Norway (NFR Grant No.~313770) and the European Research Council (ERC Grant Agreement No.~715807 and No.~101116161). This work was supported in part by the Director, Office of Science, Office of High Energy Physics, of the U.S.~Department of Energy, under contract numbers DE-AC02-76SF00515 and DE-AC02-05CH11231.
\end{acknowledgments}


%


\begin{thebibliography}{642}%
\makeatletter
\providecommand \@ifxundefined [1]{%
 \@ifx{#1\undefined}
}%
\providecommand \@ifnum [1]{%
 \ifnum #1\expandafter \@firstoftwo
 \else \expandafter \@secondoftwo
 \fi
}%
\providecommand \@ifx [1]{%
 \ifx #1\expandafter \@firstoftwo
 \else \expandafter \@secondoftwo
 \fi
}%
\providecommand \natexlab [1]{#1}%
\providecommand \enquote  [1]{``#1''}%
\providecommand \bibnamefont  [1]{#1}%
\providecommand \bibfnamefont [1]{#1}%
\providecommand \citenamefont [1]{#1}%
\providecommand \href@noop [0]{\@secondoftwo}%
\providecommand \href [0]{\begingroup \@sanitize@url \@href}%
\providecommand \@href[1]{\@@startlink{#1}\@@href}%
\providecommand \@@href[1]{\endgroup#1\@@endlink}%
\providecommand \@sanitize@url [0]{\catcode `\\12\catcode `\$12\catcode
  `\&12\catcode `\#12\catcode `\^12\catcode `\_12\catcode `\%12\relax}%
\providecommand \@@startlink[1]{}%
\providecommand \@@endlink[0]{}%
\providecommand \url  [0]{\begingroup\@sanitize@url \@url }%
\providecommand \@url [1]{\endgroup\@href {#1}{\urlprefix }}%
\providecommand \urlprefix  [0]{URL }%
\providecommand \Eprint [0]{\href }%
\providecommand \doibase [0]{http://dx.doi.org/}%
\providecommand \selectlanguage [0]{\@gobble}%
\providecommand \bibinfo  [0]{\@secondoftwo}%
\providecommand \bibfield  [0]{\@secondoftwo}%
\providecommand \translation [1]{[#1]}%
\providecommand \BibitemOpen [0]{}%
\providecommand \bibitemStop [0]{}%
\providecommand \bibitemNoStop [0]{.\EOS\space}%
\providecommand \EOS [0]{\spacefactor3000\relax}%
\providecommand \BibitemShut  [1]{\csname bibitem#1\endcsname}%
\let\auto@bib@innerbib\@empty
\bibitem [{\citenamefont {Abada}\ \emph
  {et~al.}(2019{\natexlab{a}})\citenamefont {Abada} \emph {et~al.}}]{FCC2019a}%
  \BibitemOpen
  \bibfield  {author} {\bibinfo {author} {\bibnamefont {Abada}, \bibfnamefont
  {A}},  \emph {et~al.} (\bibinfo {collaboration} {FCC Collaboration})}
  (\bibinfo {year} {2019}{\natexlab{a}}),\ \href {\doibase
  10.1140/epjst/e2019-900045-4} {\bibfield  {journal} {\bibinfo  {journal}
  {Eur. Phys. J. Spec. Top.}\ }\textbf {\bibinfo {volume} {228}},\ \bibinfo
  {pages} {261–623}}\BibitemShut {NoStop}%
\bibitem [{\citenamefont {Abada}\ \emph
  {et~al.}(2019{\natexlab{b}})\citenamefont {Abada} \emph {et~al.}}]{FCC2019b}%
  \BibitemOpen
  \bibfield  {author} {\bibinfo {author} {\bibnamefont {Abada}, \bibfnamefont
  {A}},  \emph {et~al.} (\bibinfo {collaboration} {FCC Collaboration})}
  (\bibinfo {year} {2019}{\natexlab{b}}),\ \href {\doibase
  10.1140/epjst/e2019-900087-0} {\bibfield  {journal} {\bibinfo  {journal}
  {Eur. Phys. J. Spec. Top.}\ }\textbf {\bibinfo {volume} {228}},\ \bibinfo
  {pages} {755--1107}}\BibitemShut {NoStop}%
\bibitem [{\citenamefont {Abbott}\ \emph {et~al.}(2016)\citenamefont {Abbott}
  \emph {et~al.}}]{Abbott2016}%
  \BibitemOpen
  \bibfield  {author} {\bibinfo {author} {\bibnamefont {Abbott}, \bibfnamefont
  {D}},  \emph {et~al.} (\bibinfo {collaboration} {PEPPo Collaboration})}
  (\bibinfo {year} {2016}),\ \href
  {https://doi.org/10.1103/PhysRevLett.116.214801} {\bibfield  {journal}
  {\bibinfo  {journal} {Phys. Rev. Lett.}\ }\textbf {\bibinfo {volume} {116}},\
  \bibinfo {pages} {214801}}\BibitemShut {NoStop}%
\bibitem [{\citenamefont {Adli}\ \emph
  {et~al.}(2025{\natexlab{a}})\citenamefont {Adli}, \citenamefont {Chen},
  \citenamefont {Drobniak}, \citenamefont {Finnerud}, \citenamefont {Kalvik},
  \citenamefont {Lindstrøm},\ and\ \citenamefont {Sjobak}}]{Adli2025b}%
  \BibitemOpen
  \bibfield  {author} {\bibinfo {author} {\bibnamefont {Adli}, \bibfnamefont
  {E}}, \bibinfo {author} {\bibfnamefont {J.~B.~B.}\ \bibnamefont {Chen}},
  \bibinfo {author} {\bibfnamefont {P.}~\bibnamefont {Drobniak}}, \bibinfo
  {author} {\bibfnamefont {O.~G.}\ \bibnamefont {Finnerud}}, \bibinfo {author}
  {\bibfnamefont {D.}~\bibnamefont {Kalvik}}, \bibinfo {author} {\bibfnamefont
  {C.~A.}\ \bibnamefont {Lindstrøm}}, \ and\ \bibinfo {author} {\bibfnamefont
  {K.}~\bibnamefont {Sjobak}}} (\bibinfo {year} {2025}{\natexlab{a}}),\ in\
  \href {\doibase 10.18429/JACOW-IPAC2025-TUPS011} {\emph {\bibinfo {booktitle}
  {Proceedings of the 16th Int. Particle Accelerator Conf.}}}\ (\bibinfo
  {publisher} {JACoW},\ \bibinfo {address} {Geneva, Switzerland})\ pp.\
  \bibinfo {pages} {1434--1437}\BibitemShut {NoStop}%
\bibitem [{\citenamefont {Adli}\ \emph {et~al.}(2013)\citenamefont {Adli},
  \citenamefont {Delahaye}, \citenamefont {Gessner}, \citenamefont {Hogan},
  \citenamefont {Raubenheimer}, \citenamefont {An}, \citenamefont {Joshi},\
  and\ \citenamefont {Mori}}]{Adli2013}%
  \BibitemOpen
  \bibfield  {author} {\bibinfo {author} {\bibnamefont {Adli}, \bibfnamefont
  {E}}, \bibinfo {author} {\bibfnamefont {J.-P.}\ \bibnamefont {Delahaye}},
  \bibinfo {author} {\bibfnamefont {S.~J.}\ \bibnamefont {Gessner}}, \bibinfo
  {author} {\bibfnamefont {M.~J.}\ \bibnamefont {Hogan}}, \bibinfo {author}
  {\bibfnamefont {T.}~\bibnamefont {Raubenheimer}}, \bibinfo {author}
  {\bibfnamefont {W.}~\bibnamefont {An}}, \bibinfo {author} {\bibfnamefont
  {C.}~\bibnamefont {Joshi}}, \ and\ \bibinfo {author} {\bibfnamefont
  {W.}~\bibnamefont {Mori}}} (\bibinfo {year} {2013}),\ \href@noop {} {}\Eprint
  {http://arxiv.org/abs/1308.1145} {arXiv:1308.1145} \BibitemShut {NoStop}%
\bibitem [{\citenamefont {Adli}\ and\ \citenamefont
  {Muggli}(2016)}]{Adli2016c}%
  \BibitemOpen
  \bibfield  {author} {\bibinfo {author} {\bibnamefont {Adli}, \bibfnamefont
  {E}}, \ and\ \bibinfo {author} {\bibfnamefont {P.}~\bibnamefont {Muggli}}}
  (\bibinfo {year} {2016}),\ \href {https://doi.org/10.1142/S1793626816300048}
  {\bibfield  {journal} {\bibinfo  {journal} {Rev. Accel. Sci. Tech.}\ }\textbf
  {\bibinfo {volume} {9}},\ \bibinfo {pages} {85--104}}\BibitemShut {NoStop}%
\bibitem [{\citenamefont {Adli}\ \emph
  {et~al.}(2016{\natexlab{a}})\citenamefont {Adli} \emph {et~al.}}]{Adli2016a}%
  \BibitemOpen
  \bibfield  {author} {\bibinfo {author} {\bibnamefont {Adli}, \bibfnamefont
  {E}},  \emph {et~al.}} (\bibinfo {year} {2016}{\natexlab{a}}),\ \href
  {https://doi.org/10.1088/1367-2630/18/10/103013} {\bibfield  {journal}
  {\bibinfo  {journal} {New J. Phys.}\ }\textbf {\bibinfo {volume} {18}},\
  \bibinfo {pages} {103013}}\BibitemShut {NoStop}%
\bibitem [{\citenamefont {Adli}\ \emph
  {et~al.}(2016{\natexlab{b}})\citenamefont {Adli} \emph {et~al.}}]{Adli2016b}%
  \BibitemOpen
  \bibfield  {author} {\bibinfo {author} {\bibnamefont {Adli}, \bibfnamefont
  {E}},  \emph {et~al.}} (\bibinfo {year} {2016}{\natexlab{b}}),\ \href
  {https://doi.org/10.1016/j.nima.2016.02.054} {\bibfield  {journal} {\bibinfo
  {journal} {Nucl. Instrum. Methods Phys. Res. A}\ }\textbf {\bibinfo {volume}
  {829}},\ \bibinfo {pages} {94--98}}\BibitemShut {NoStop}%
\bibitem [{\citenamefont {Adli}\ \emph {et~al.}(2018)\citenamefont {Adli} \emph
  {et~al.}}]{Adli2018}%
  \BibitemOpen
  \bibfield  {author} {\bibinfo {author} {\bibnamefont {Adli}, \bibfnamefont
  {E}},  \emph {et~al.} (\bibinfo {collaboration} {AWAKE Collaboration})}
  (\bibinfo {year} {2018}),\ \href {https://doi.org/10.1038/s41586-018-0485-4}
  {\bibfield  {journal} {\bibinfo  {journal} {Nature (London)}\ }\textbf
  {\bibinfo {volume} {561}},\ \bibinfo {pages} {363--367}}\BibitemShut
  {NoStop}%
\bibitem [{\citenamefont {Adli}\ \emph {et~al.}(2019)\citenamefont {Adli} \emph
  {et~al.}}]{Adli2019}%
  \BibitemOpen
  \bibfield  {author} {\bibinfo {author} {\bibnamefont {Adli}, \bibfnamefont
  {E}},  \emph {et~al.} (\bibinfo {collaboration} {AWAKE Collaboration})}
  (\bibinfo {year} {2019}),\ \href
  {https://doi.org/10.1103/physrevlett.122.054802} {\bibfield  {journal}
  {\bibinfo  {journal} {Phys. Rev. Lett.}\ }\textbf {\bibinfo {volume} {122}},\
  \bibinfo {pages} {054802}}\BibitemShut {NoStop}%
\bibitem [{\citenamefont {Adli}\ \emph
  {et~al.}(2025{\natexlab{b}})\citenamefont {Adli} \emph {et~al.}}]{Adli2025a}%
  \BibitemOpen
  \bibfield  {author} {\bibinfo {author} {\bibnamefont {Adli}, \bibfnamefont
  {E}},  \emph {et~al.} (\bibinfo {collaboration} {HALHF Collaboration})}
  (\bibinfo {year} {2025}{\natexlab{b}}),\ \href
  {https://doi.org/10.48550/arxiv.2503.19880} {}\Eprint
  {http://arxiv.org/abs/2503.19880} {arXiv:2503.19880} \BibitemShut {NoStop}%
\bibitem [{\citenamefont {Ahmed}\ \emph {et~al.}(2015)\citenamefont {Ahmed},
  \citenamefont {Kaplan}, \citenamefont {Roberts},\ and\ \citenamefont
  {Spentzouris}}]{Ahmed2015}%
  \BibitemOpen
  \bibfield  {author} {\bibinfo {author} {\bibnamefont {Ahmed}, \bibfnamefont
  {S}}, \bibinfo {author} {\bibfnamefont {D.~M.}\ \bibnamefont {Kaplan}},
  \bibinfo {author} {\bibfnamefont {T.~J.}\ \bibnamefont {Roberts}}, \ and\
  \bibinfo {author} {\bibfnamefont {L.~K.}\ \bibnamefont {Spentzouris}}}
  (\bibinfo {year} {2015}),\ \href {\doibase 10.1016/j.nima.2015.05.063}
  {\bibfield  {journal} {\bibinfo  {journal} {Nucl. Instrum. Methods Phys. Res.
  A}\ }\textbf {\bibinfo {volume} {797}},\ \bibinfo {pages}
  {8--12}}\BibitemShut {NoStop}%
\bibitem [{\citenamefont {Aicheler}\ \emph {et~al.}(2012)\citenamefont
  {Aicheler} \emph {et~al.}}]{Aicheler2012}%
  \BibitemOpen
  \bibfield  {author} {\bibinfo {author} {\bibnamefont {Aicheler},
  \bibfnamefont {M}},  \emph {et~al.}} (\bibinfo {year} {2012}),\ \href
  {https://cds.cern.ch/record/1500095} {\emph {\bibinfo {title} {{A Multi-TeV
  Linear Collider Based on CLIC Technology: CLIC Conceptual Design Report}}}},\
  \bibinfo {series} {CERN Yellow Reports: Monographs}\ No.\ \bibinfo {number}
  {CERN-2012-007}\ (\bibinfo  {publisher} {CERN},\ \bibinfo {address} {Geneva,
  Switzerland})\BibitemShut {NoStop}%
\bibitem [{\citenamefont {Akhiezer}\ and\ \citenamefont
  {Polovin}(1956)}]{Akhiezer1956}%
  \BibitemOpen
  \bibfield  {author} {\bibinfo {author} {\bibnamefont {Akhiezer},
  \bibfnamefont {A~I}}, \ and\ \bibinfo {author} {\bibfnamefont {R.~V.}\
  \bibnamefont {Polovin}}} (\bibinfo {year} {1956}),\ \href
  {http://www.jetp.ras.ru/cgi-bin/dn/e_003_05_0696.pdf} {\bibfield  {journal}
  {\bibinfo  {journal} {J. Exp. Theor. Phys. (U.S.S.R.)}\ }\textbf {\bibinfo
  {volume} {3}},\ \bibinfo {pages} {696--705}}\BibitemShut {NoStop}%
\bibitem [{\citenamefont {Aksoy}\ \emph {et~al.}(2011)\citenamefont {Aksoy},
  \citenamefont {Schulte},\ and\ \citenamefont {Yavaş}}]{Aksoy2011}%
  \BibitemOpen
  \bibfield  {author} {\bibinfo {author} {\bibnamefont {Aksoy}, \bibfnamefont
  {A}}, \bibinfo {author} {\bibfnamefont {D.}~\bibnamefont {Schulte}}, \ and\
  \bibinfo {author} {\bibfnamefont {\"{O}.}\ \bibnamefont {Yavaş}}} (\bibinfo
  {year} {2011}),\ \href {\doibase 10.1103/physrevstab.14.084402} {\bibfield
  {journal} {\bibinfo  {journal} {Phys. Rev. ST Accel. Beams}\ }\textbf
  {\bibinfo {volume} {14}},\ \bibinfo {pages} {084402}}\BibitemShut {NoStop}%
\bibitem [{\citenamefont {Alesini}\ \emph {et~al.}(2003)\citenamefont {Alesini}
  \emph {et~al.}}]{Alesini2003}%
  \BibitemOpen
  \bibfield  {author} {\bibinfo {author} {\bibnamefont {Alesini}, \bibfnamefont
  {D}},  \emph {et~al.}} (\bibinfo {year} {2003}),\ \href
  {https://doi.org/10.1016/S0168-9002(03)00943-4} {\bibfield  {journal}
  {\bibinfo  {journal} {Nucl. Instrum. Methods Phys. Res. A}\ }\textbf
  {\bibinfo {volume} {507}},\ \bibinfo {pages} {345--349}}\BibitemShut
  {NoStop}%
\bibitem [{\citenamefont {Alesini}\ \emph {et~al.}(2005)\citenamefont {Alesini}
  \emph {et~al.}}]{Alesini2005}%
  \BibitemOpen
  \bibfield  {author} {\bibinfo {author} {\bibnamefont {Alesini}, \bibfnamefont
  {D}},  \emph {et~al.}} (\bibinfo {year} {2005}),\ in\ \href {\doibase
  10.1109/pac.2005.1590575} {\emph {\bibinfo {booktitle} {Proceedings of the
  2005 Particle Accelerator Conf.}}}\ (\bibinfo  {publisher} {IEEE},\ \bibinfo
  {address} {New York, NY, United States})\ pp.\ \bibinfo {pages}
  {820--822}\BibitemShut {NoStop}%
\bibitem [{\citenamefont {Allen}\ \emph {et~al.}(2012)\citenamefont {Allen},
  \citenamefont {Yakimenko}, \citenamefont {Babzien}, \citenamefont {Fedurin},
  \citenamefont {Kusche},\ and\ \citenamefont {Muggli}}]{Allen2012}%
  \BibitemOpen
  \bibfield  {author} {\bibinfo {author} {\bibnamefont {Allen}, \bibfnamefont
  {B}}, \bibinfo {author} {\bibfnamefont {V.}~\bibnamefont {Yakimenko}},
  \bibinfo {author} {\bibfnamefont {M.}~\bibnamefont {Babzien}}, \bibinfo
  {author} {\bibfnamefont {M.}~\bibnamefont {Fedurin}}, \bibinfo {author}
  {\bibfnamefont {K.}~\bibnamefont {Kusche}}, \ and\ \bibinfo {author}
  {\bibfnamefont {P.}~\bibnamefont {Muggli}}} (\bibinfo {year} {2012}),\ \href
  {https://link.aps.org/doi/10.1103/PhysRevLett.109.185007} {\bibfield
  {journal} {\bibinfo  {journal} {Phys. Rev. Lett.}\ }\textbf {\bibinfo
  {volume} {109}},\ \bibinfo {pages} {185007}}\BibitemShut {NoStop}%
\bibitem [{\citenamefont {Amorim}\ \emph {et~al.}(2023)\citenamefont {Amorim},
  \citenamefont {Benedetti}, \citenamefont {Bulanov}, \citenamefont {Terzani},
  \citenamefont {Huebl}, \citenamefont {Schroeder}, \citenamefont {Vay},\ and\
  \citenamefont {Esarey}}]{Amorim2023}%
  \BibitemOpen
  \bibfield  {author} {\bibinfo {author} {\bibnamefont {Amorim}, \bibfnamefont
  {L~D}}, \bibinfo {author} {\bibfnamefont {C.}~\bibnamefont {Benedetti}},
  \bibinfo {author} {\bibfnamefont {S.~S.}\ \bibnamefont {Bulanov}}, \bibinfo
  {author} {\bibfnamefont {D.}~\bibnamefont {Terzani}}, \bibinfo {author}
  {\bibfnamefont {A.}~\bibnamefont {Huebl}}, \bibinfo {author} {\bibfnamefont
  {C.~B.}\ \bibnamefont {Schroeder}}, \bibinfo {author} {\bibfnamefont {J.-L.}\
  \bibnamefont {Vay}}, \ and\ \bibinfo {author} {\bibfnamefont
  {E.}~\bibnamefont {Esarey}}} (\bibinfo {year} {2023}),\ \href
  {https://doi.org/10.1088/1361-6587/ace3f1} {\bibfield  {journal} {\bibinfo
  {journal} {Plasma Phys. Control. Fusion}\ }\textbf {\bibinfo {volume} {65}},\
  \bibinfo {pages} {085016}}\BibitemShut {NoStop}%
\bibitem [{\citenamefont {An}\ \emph {et~al.}(2017)\citenamefont {An},
  \citenamefont {Lu}, \citenamefont {Huang}, \citenamefont {Xu}, \citenamefont
  {Hogan}, \citenamefont {Joshi},\ and\ \citenamefont {Mori}}]{An2017}%
  \BibitemOpen
  \bibfield  {author} {\bibinfo {author} {\bibnamefont {An}, \bibfnamefont
  {W}}, \bibinfo {author} {\bibfnamefont {W.}~\bibnamefont {Lu}}, \bibinfo
  {author} {\bibfnamefont {C.}~\bibnamefont {Huang}}, \bibinfo {author}
  {\bibfnamefont {X.}~\bibnamefont {Xu}}, \bibinfo {author} {\bibfnamefont
  {M.~J.}\ \bibnamefont {Hogan}}, \bibinfo {author} {\bibfnamefont
  {C.}~\bibnamefont {Joshi}}, \ and\ \bibinfo {author} {\bibfnamefont {W.~B.}\
  \bibnamefont {Mori}}} (\bibinfo {year} {2017}),\ \href
  {https://doi.org/10.1103/physrevlett.118.244801} {\bibfield  {journal}
  {\bibinfo  {journal} {Phys. Rev. Lett.}\ }\textbf {\bibinfo {volume} {118}},\
  \bibinfo {pages} {244801}}\BibitemShut {NoStop}%
\bibitem [{\citenamefont {An}\ \emph {et~al.}(2013)\citenamefont {An} \emph
  {et~al.}}]{An2013}%
  \BibitemOpen
  \bibfield  {author} {\bibinfo {author} {\bibnamefont {An}, \bibfnamefont
  {W}},  \emph {et~al.}} (\bibinfo {year} {2013}),\ \href
  {https://doi.org/10.1103/PhysRevSTAB.16.101301} {\bibfield  {journal}
  {\bibinfo  {journal} {Phys. Rev. ST Accel. Beams}\ }\textbf {\bibinfo
  {volume} {16}},\ \bibinfo {pages} {101301}}\BibitemShut {NoStop}%
\bibitem [{\citenamefont {Andreev}\ \emph {et~al.}(1996)\citenamefont
  {Andreev}, \citenamefont {Bychkov}, \citenamefont {Kotlyar}, \citenamefont
  {Margolin}, \citenamefont {Pyatnitskii},\ and\ \citenamefont
  {Serafimovich}}]{Andreev1996}%
  \BibitemOpen
  \bibfield  {author} {\bibinfo {author} {\bibnamefont {Andreev}, \bibfnamefont
  {N~E}}, \bibinfo {author} {\bibfnamefont {S.~S.}\ \bibnamefont {Bychkov}},
  \bibinfo {author} {\bibfnamefont {V.~V.}\ \bibnamefont {Kotlyar}}, \bibinfo
  {author} {\bibfnamefont {L.~Ya.}\ \bibnamefont {Margolin}}, \bibinfo {author}
  {\bibfnamefont {L.~N.}\ \bibnamefont {Pyatnitskii}}, \ and\ \bibinfo {author}
  {\bibfnamefont {P.~G.}\ \bibnamefont {Serafimovich}}} (\bibinfo {year}
  {1996}),\ \href {https://doi.org/10.1070/QE1996v026n02ABEH000607} {\bibfield
  {journal} {\bibinfo  {journal} {Quantum Electron.}\ }\textbf {\bibinfo
  {volume} {26}},\ \bibinfo {pages} {126--130}}\BibitemShut {NoStop}%
\bibitem [{\citenamefont {Antici}\ \emph {et~al.}(2012)\citenamefont {Antici}
  \emph {et~al.}}]{Antici2012}%
  \BibitemOpen
  \bibfield  {author} {\bibinfo {author} {\bibnamefont {Antici}, \bibfnamefont
  {P}},  \emph {et~al.}} (\bibinfo {year} {2012}),\ \href
  {https://doi.org/10.1063/1.4740456} {\bibfield  {journal} {\bibinfo
  {journal} {J. Appl. Phys.}\ }\textbf {\bibinfo {volume} {112}},\ \bibinfo
  {pages} {044902}}\BibitemShut {NoStop}%
\bibitem [{\citenamefont {Antipov}\ \emph {et~al.}(2014)\citenamefont
  {Antipov}, \citenamefont {Baturin}, \citenamefont {Jing}, \citenamefont
  {Fedurin}, \citenamefont {Kanareykin}, \citenamefont {Swinson}, \citenamefont
  {Schoessow}, \citenamefont {Gai},\ and\ \citenamefont
  {Zholents}}]{Antipov2014}%
  \BibitemOpen
  \bibfield  {author} {\bibinfo {author} {\bibnamefont {Antipov}, \bibfnamefont
  {S}}, \bibinfo {author} {\bibfnamefont {S.}~\bibnamefont {Baturin}}, \bibinfo
  {author} {\bibfnamefont {C.}~\bibnamefont {Jing}}, \bibinfo {author}
  {\bibfnamefont {M.}~\bibnamefont {Fedurin}}, \bibinfo {author} {\bibfnamefont
  {A.}~\bibnamefont {Kanareykin}}, \bibinfo {author} {\bibfnamefont
  {C.}~\bibnamefont {Swinson}}, \bibinfo {author} {\bibfnamefont
  {P.}~\bibnamefont {Schoessow}}, \bibinfo {author} {\bibfnamefont
  {W.}~\bibnamefont {Gai}}, \ and\ \bibinfo {author} {\bibfnamefont
  {A.}~\bibnamefont {Zholents}}} (\bibinfo {year} {2014}),\ \href
  {https://doi.org/10.1103/PhysRevLett.112.114801} {\bibfield  {journal}
  {\bibinfo  {journal} {Phys. Rev. Lett.}\ }\textbf {\bibinfo {volume} {112}},\
  \bibinfo {pages} {114801}}\BibitemShut {NoStop}%
\bibitem [{\citenamefont {Ariniello}\ \emph {et~al.}(2019)\citenamefont
  {Ariniello}, \citenamefont {Doss}, \citenamefont {Hunt-Stone}, \citenamefont
  {Cary},\ and\ \citenamefont {Litos}}]{Ariniello2019}%
  \BibitemOpen
  \bibfield  {author} {\bibinfo {author} {\bibnamefont {Ariniello},
  \bibfnamefont {R}}, \bibinfo {author} {\bibfnamefont {C.~E.}\ \bibnamefont
  {Doss}}, \bibinfo {author} {\bibfnamefont {K.}~\bibnamefont {Hunt-Stone}},
  \bibinfo {author} {\bibfnamefont {J.~R.}\ \bibnamefont {Cary}}, \ and\
  \bibinfo {author} {\bibfnamefont {M.~D.}\ \bibnamefont {Litos}}} (\bibinfo
  {year} {2019}),\ \href {https://doi.org/10.1103/physrevaccelbeams.22.041304}
  {\bibfield  {journal} {\bibinfo  {journal} {Phys. Rev. Accel. Beams}\
  }\textbf {\bibinfo {volume} {22}},\ \bibinfo {pages} {041304}}\BibitemShut
  {NoStop}%
\bibitem [{\citenamefont {Ariniello}\ \emph {et~al.}(2022)\citenamefont
  {Ariniello}, \citenamefont {Doss}, \citenamefont {Lee}, \citenamefont
  {Hansel}, \citenamefont {Cary},\ and\ \citenamefont {Litos}}]{Ariniello2022}%
  \BibitemOpen
  \bibfield  {author} {\bibinfo {author} {\bibnamefont {Ariniello},
  \bibfnamefont {R}}, \bibinfo {author} {\bibfnamefont {C.~E.}\ \bibnamefont
  {Doss}}, \bibinfo {author} {\bibfnamefont {V.}~\bibnamefont {Lee}}, \bibinfo
  {author} {\bibfnamefont {C.}~\bibnamefont {Hansel}}, \bibinfo {author}
  {\bibfnamefont {J.~R.}\ \bibnamefont {Cary}}, \ and\ \bibinfo {author}
  {\bibfnamefont {M.~D.}\ \bibnamefont {Litos}}} (\bibinfo {year} {2022}),\
  \href {\doibase 10.1103/physrevresearch.4.043120} {\bibfield  {journal}
  {\bibinfo  {journal} {Phys. Rev. Res.}\ }\textbf {\bibinfo {volume} {4}},\
  \bibinfo {pages} {043120}}\BibitemShut {NoStop}%
\bibitem [{\citenamefont {Aschikhin}\ \emph {et~al.}(2016)\citenamefont
  {Aschikhin} \emph {et~al.}}]{Aschikhin2016}%
  \BibitemOpen
  \bibfield  {author} {\bibinfo {author} {\bibnamefont {Aschikhin},
  \bibfnamefont {A}},  \emph {et~al.}} (\bibinfo {year} {2016}),\ \href
  {https://doi.org/10.1016/j.nima.2015.10.005} {\bibfield  {journal} {\bibinfo
  {journal} {Nucl. Instrum. Methods Phys. Res. A}\ }\textbf {\bibinfo {volume}
  {806}},\ \bibinfo {pages} {175--183}}\BibitemShut {NoStop}%
\bibitem [{\citenamefont {Assmann}\ and\ \citenamefont
  {Yokoya}(1998)}]{Assmann1998}%
  \BibitemOpen
  \bibfield  {author} {\bibinfo {author} {\bibnamefont {Assmann}, \bibfnamefont
  {R}}, \ and\ \bibinfo {author} {\bibfnamefont {K.}~\bibnamefont {Yokoya}}}
  (\bibinfo {year} {1998}),\ \href
  {https://doi.org/10.1016/s0168-9002(98)00187-9} {\bibfield  {journal}
  {\bibinfo  {journal} {Nucl. Instrum. Methods Phys. Res. A}\ }\textbf
  {\bibinfo {volume} {410}},\ \bibinfo {pages} {544--548}}\BibitemShut
  {NoStop}%
\bibitem [{\citenamefont {Assmann}\ \emph {et~al.}(1998)\citenamefont {Assmann}
  \emph {et~al.}}]{Assmann1998a}%
  \BibitemOpen
  \bibfield  {author} {\bibinfo {author} {\bibnamefont {Assmann}, \bibfnamefont
  {R}},  \emph {et~al.}} (\bibinfo {year} {1998}),\ \href
  {https://doi.org/10.1016/S0168-9002(98)00169-7} {\bibfield  {journal}
  {\bibinfo  {journal} {Nucl. Instrum. Methods Phys. Res. A}\ }\textbf
  {\bibinfo {volume} {410}},\ \bibinfo {pages} {396--406}}\BibitemShut
  {NoStop}%
\bibitem [{\citenamefont {Assmann}\ \emph {et~al.}(2014)\citenamefont {Assmann}
  \emph {et~al.}}]{Assmann2014}%
  \BibitemOpen
  \bibfield  {author} {\bibinfo {author} {\bibnamefont {Assmann}, \bibfnamefont
  {R}},  \emph {et~al.}} (\bibinfo {year} {2014}),\ \href
  {https://doi.org/10.1088/0741-3335/56/8/084013} {\bibfield  {journal}
  {\bibinfo  {journal} {Plasma Phys. Control. Fusion}\ }\textbf {\bibinfo
  {volume} {56}},\ \bibinfo {pages} {084013}}\BibitemShut {NoStop}%
\bibitem [{\citenamefont {Assmann}\ \emph {et~al.}(2020)\citenamefont {Assmann}
  \emph {et~al.}}]{Assmann2020}%
  \BibitemOpen
  \bibfield  {author} {\bibinfo {author} {\bibnamefont {Assmann}, \bibfnamefont
  {R}},  \emph {et~al.}} (\bibinfo {year} {2020}),\ \href
  {https://doi.org/10.1140/epjst/e2020-000127-8} {\bibfield  {journal}
  {\bibinfo  {journal} {Eur. Phys. J. Spec. Top.}\ }\textbf {\bibinfo {volume}
  {229}},\ \bibinfo {pages} {3675--4284}}\BibitemShut {NoStop}%
\bibitem [{\citenamefont {Balakin}\ \emph {et~al.}(1983)\citenamefont
  {Balakin}, \citenamefont {Novokhatsky},\ and\ \citenamefont
  {Smirnov}}]{Balakin1983}%
  \BibitemOpen
  \bibfield  {author} {\bibinfo {author} {\bibnamefont {Balakin}, \bibfnamefont
  {V~E}}, \bibinfo {author} {\bibfnamefont {A.~V.}\ \bibnamefont
  {Novokhatsky}}, \ and\ \bibinfo {author} {\bibfnamefont {V.~P.}\ \bibnamefont
  {Smirnov}}} (\bibinfo {year} {1983}),\ in\ \href
  {https://inspirehep.net/literature/198113} {\emph {\bibinfo {booktitle} {12th
  Int. Conf. on High-Energy Accelerators, HEACC}}},\ Vol.\ \bibinfo {volume}
  {830811}\ (\bibinfo  {publisher} {Fermilab},\ \bibinfo {address} {Batavia,
  IL, United States})\ pp.\ \bibinfo {pages} {119--120}\BibitemShut {NoStop}%
\bibitem [{\citenamefont {Bane}\ \emph
  {et~al.}(1985{\natexlab{a}})\citenamefont {Bane}, \citenamefont {Chen},\ and\
  \citenamefont {Wilson}}]{Bane1985a}%
  \BibitemOpen
  \bibfield  {author} {\bibinfo {author} {\bibnamefont {Bane}, \bibfnamefont
  {K~L~F}}, \bibinfo {author} {\bibfnamefont {Pisin}\ \bibnamefont {Chen}}, \
  and\ \bibinfo {author} {\bibfnamefont {P.~B.}\ \bibnamefont {Wilson}}}
  (\bibinfo {year} {1985}{\natexlab{a}}),\ \href
  {https://doi.org/10.1109/TNS.1985.4334416} {\bibfield  {journal} {\bibinfo
  {journal} {IEEE T. Nucl. Sci.}\ }\textbf {\bibinfo {volume} {32}},\ \bibinfo
  {pages} {3524--3526}}\BibitemShut {NoStop}%
\bibitem [{\citenamefont {Bane}\ \emph
  {et~al.}(1985{\natexlab{b}})\citenamefont {Bane}, \citenamefont {Wilson},\
  and\ \citenamefont {Weiland}}]{Bane1985b}%
  \BibitemOpen
  \bibfield  {author} {\bibinfo {author} {\bibnamefont {Bane}, \bibfnamefont
  {K~L~F}}, \bibinfo {author} {\bibfnamefont {P.~B.}\ \bibnamefont {Wilson}}, \
  and\ \bibinfo {author} {\bibfnamefont {T.}~\bibnamefont {Weiland}}} (\bibinfo
  {year} {1985}{\natexlab{b}}),\ in\ \href {https://doi.org/10.1063/1.35182}
  {\emph {\bibinfo {booktitle} {{AIP} Conf. Proc.}}},\ Vol.\ \bibinfo {volume}
  {127}\ (\bibinfo  {publisher} {{AIP}})\ pp.\ \bibinfo {pages}
  {875--928}\BibitemShut {NoStop}%
\bibitem [{\citenamefont {Barber}\ \emph {et~al.}(2017)\citenamefont {Barber}
  \emph {et~al.}}]{Barber2017}%
  \BibitemOpen
  \bibfield  {author} {\bibinfo {author} {\bibnamefont {Barber}, \bibfnamefont
  {S~K}},  \emph {et~al.}} (\bibinfo {year} {2017}),\ \href
  {https://doi.org/10.1103/physrevlett.119.104801} {\bibfield  {journal}
  {\bibinfo  {journal} {Phys. Rev. Lett.}\ }\textbf {\bibinfo {volume} {119}},\
  \bibinfo {pages} {104801}}\BibitemShut {NoStop}%
\bibitem [{\citenamefont {Bargmann}\ \emph {et~al.}(1959)\citenamefont
  {Bargmann}, \citenamefont {Michel},\ and\ \citenamefont
  {Telegdi}}]{Bargmann1959}%
  \BibitemOpen
  \bibfield  {author} {\bibinfo {author} {\bibnamefont {Bargmann},
  \bibfnamefont {V}}, \bibinfo {author} {\bibfnamefont {L.}~\bibnamefont
  {Michel}}, \ and\ \bibinfo {author} {\bibfnamefont {V.~L.}\ \bibnamefont
  {Telegdi}}} (\bibinfo {year} {1959}),\ \href
  {https://doi.org/10.1103/PhysRevLett.2.435} {\bibfield  {journal} {\bibinfo
  {journal} {Phys. Rev. Lett.}\ }\textbf {\bibinfo {volume} {2}},\ \bibinfo
  {pages} {435--436}}\BibitemShut {NoStop}%
\bibitem [{\citenamefont {Barov}\ \emph {et~al.}(1998)\citenamefont {Barov},
  \citenamefont {Conde}, \citenamefont {Gai},\ and\ \citenamefont
  {Rosenzweig}}]{Barov1998}%
  \BibitemOpen
  \bibfield  {author} {\bibinfo {author} {\bibnamefont {Barov}, \bibfnamefont
  {N}}, \bibinfo {author} {\bibfnamefont {M.~E.}\ \bibnamefont {Conde}},
  \bibinfo {author} {\bibfnamefont {W.}~\bibnamefont {Gai}}, \ and\ \bibinfo
  {author} {\bibfnamefont {J.~B.}\ \bibnamefont {Rosenzweig}}} (\bibinfo {year}
  {1998}),\ \href {https://doi.org/10.1103/physrevlett.80.81} {\bibfield
  {journal} {\bibinfo  {journal} {Phys. Rev. Lett.}\ }\textbf {\bibinfo
  {volume} {80}},\ \bibinfo {pages} {81--84}}\BibitemShut {NoStop}%
\bibitem [{\citenamefont {Barov}\ and\ \citenamefont
  {Rosenzweig}(1994)}]{Barov1994}%
  \BibitemOpen
  \bibfield  {author} {\bibinfo {author} {\bibnamefont {Barov}, \bibfnamefont
  {N}}, \ and\ \bibinfo {author} {\bibfnamefont {J.~B.}\ \bibnamefont
  {Rosenzweig}}} (\bibinfo {year} {1994}),\ \href
  {https://doi.org/10.1103/PhysRevE.49.4407} {\bibfield  {journal} {\bibinfo
  {journal} {Phys. Rev. E}\ }\textbf {\bibinfo {volume} {49}},\ \bibinfo
  {pages} {4407--4416}}\BibitemShut {NoStop}%
\bibitem [{\citenamefont {Barov}\ \emph {et~al.}(2000)\citenamefont {Barov},
  \citenamefont {Rosenzweig}, \citenamefont {Conde}, \citenamefont {Gai},\ and\
  \citenamefont {Power}}]{Barov2000}%
  \BibitemOpen
  \bibfield  {author} {\bibinfo {author} {\bibnamefont {Barov}, \bibfnamefont
  {N}}, \bibinfo {author} {\bibfnamefont {J.~B.}\ \bibnamefont {Rosenzweig}},
  \bibinfo {author} {\bibfnamefont {M.~E.}\ \bibnamefont {Conde}}, \bibinfo
  {author} {\bibfnamefont {W.}~\bibnamefont {Gai}}, \ and\ \bibinfo {author}
  {\bibfnamefont {J.~G.}\ \bibnamefont {Power}}} (\bibinfo {year} {2000}),\
  \href {https://doi.org/10.1103/physrevstab.3.011301} {\bibfield  {journal}
  {\bibinfo  {journal} {Phys. Rev. ST Accel. Beams}\ }\textbf {\bibinfo
  {volume} {3}},\ \bibinfo {pages} {011301}}\BibitemShut {NoStop}%
\bibitem [{\citenamefont {Barov}\ \emph {et~al.}(2004)\citenamefont {Barov},
  \citenamefont {Rosenzweig}, \citenamefont {Thompson},\ and\ \citenamefont
  {Yoder}}]{Barov2004}%
  \BibitemOpen
  \bibfield  {author} {\bibinfo {author} {\bibnamefont {Barov}, \bibfnamefont
  {N}}, \bibinfo {author} {\bibfnamefont {J.~B.}\ \bibnamefont {Rosenzweig}},
  \bibinfo {author} {\bibfnamefont {M.~C.}\ \bibnamefont {Thompson}}, \ and\
  \bibinfo {author} {\bibfnamefont {R.~B.}\ \bibnamefont {Yoder}}} (\bibinfo
  {year} {2004}),\ \href {https://doi.org/10.1103/physrevstab.7.061301}
  {\bibfield  {journal} {\bibinfo  {journal} {Phys. Rev. ST Accel. Beams}\
  }\textbf {\bibinfo {volume} {7}},\ \bibinfo {pages} {061301}}\BibitemShut
  {NoStop}%
\bibitem [{\citenamefont {Barov}\ \emph {et~al.}(2002)\citenamefont {Barov}
  \emph {et~al.}}]{Barov2002}%
  \BibitemOpen
  \bibfield  {author} {\bibinfo {author} {\bibnamefont {Barov}, \bibfnamefont
  {N}},  \emph {et~al.}} (\bibinfo {year} {2002}),\ in\ \href {\doibase
  10.1109/pac.2001.988316} {\emph {\bibinfo {booktitle} {Proceedings of the
  2001 Particle Accelerator Conf.}}}\ (\bibinfo  {publisher} {IEEE},\ \bibinfo
  {address} {New York, NY, United States})\ pp.\ \bibinfo {pages}
  {126--128}\BibitemShut {NoStop}%
\bibitem [{\citenamefont {Batsch}\ \emph {et~al.}(2021)\citenamefont {Batsch}
  \emph {et~al.}}]{Batsch2021}%
  \BibitemOpen
  \bibfield  {author} {\bibinfo {author} {\bibnamefont {Batsch}, \bibfnamefont
  {F}},  \emph {et~al.} (\bibinfo {collaboration} {AWAKE Collaboration})}
  (\bibinfo {year} {2021}),\ \href
  {https://doi.org/10.1103/physrevlett.126.164802} {\bibfield  {journal}
  {\bibinfo  {journal} {Phys. Rev. Lett.}\ }\textbf {\bibinfo {volume} {126}},\
  \bibinfo {pages} {164802}}\BibitemShut {NoStop}%
\bibitem [{\citenamefont {Baturin}(2022)}]{Baturin2022}%
  \BibitemOpen
  \bibfield  {author} {\bibinfo {author} {\bibnamefont {Baturin}, \bibfnamefont
  {S~S}}} (\bibinfo {year} {2022}),\ \href {\doibase
  10.1103/physrevaccelbeams.25.081301} {\bibfield  {journal} {\bibinfo
  {journal} {Phys. Rev. Accel. Beams}\ }\textbf {\bibinfo {volume} {25}},\
  \bibinfo {pages} {081301}}\BibitemShut {NoStop}%
\bibitem [{\citenamefont {Baturin}(2023)}]{Baturin2023}%
  \BibitemOpen
  \bibfield  {author} {\bibinfo {author} {\bibnamefont {Baturin}, \bibfnamefont
  {S~S}}} (\bibinfo {year} {2023}),\ \href {\doibase
  10.1103/physreve.108.015202} {\bibfield  {journal} {\bibinfo  {journal}
  {Phys. Rev. E}\ }\textbf {\bibinfo {volume} {108}},\ \bibinfo {pages}
  {015202}}\BibitemShut {NoStop}%
\bibitem [{\citenamefont {Bauche}\ \emph {et~al.}(2019)\citenamefont {Bauche}
  \emph {et~al.}}]{Bauche2019}%
  \BibitemOpen
  \bibfield  {author} {\bibinfo {author} {\bibnamefont {Bauche}, \bibfnamefont
  {J}},  \emph {et~al.}} (\bibinfo {year} {2019}),\ \href
  {https://doi.org/10.1016/j.nima.2019.05.067} {\bibfield  {journal} {\bibinfo
  {journal} {Nucl. Instrum. Methods Phys. Res. A}\ }\textbf {\bibinfo {volume}
  {940}},\ \bibinfo {pages} {103--108}}\BibitemShut {NoStop}%
\bibitem [{\citenamefont {Baxevanis}\ \emph {et~al.}(2017)\citenamefont
  {Baxevanis}, \citenamefont {Hogan}, \citenamefont {Huang}, \citenamefont
  {Litos}, \citenamefont {O'Shea}, \citenamefont {Raubenheimer}, \citenamefont
  {Frisch}, \citenamefont {White}, \citenamefont {Xu},\ and\ \citenamefont
  {Mori}}]{Baxevanis2017}%
  \BibitemOpen
  \bibfield  {author} {\bibinfo {author} {\bibnamefont {Baxevanis},
  \bibfnamefont {P}}, \bibinfo {author} {\bibfnamefont {M.~J.}\ \bibnamefont
  {Hogan}}, \bibinfo {author} {\bibfnamefont {Z.}~\bibnamefont {Huang}},
  \bibinfo {author} {\bibfnamefont {M.}~\bibnamefont {Litos}}, \bibinfo
  {author} {\bibfnamefont {B.}~\bibnamefont {O'Shea}}, \bibinfo {author}
  {\bibfnamefont {T.~O.}\ \bibnamefont {Raubenheimer}}, \bibinfo {author}
  {\bibfnamefont {J.~C.}\ \bibnamefont {Frisch}}, \bibinfo {author}
  {\bibfnamefont {G.}~\bibnamefont {White}}, \bibinfo {author} {\bibfnamefont
  {X.~L.}\ \bibnamefont {Xu}}, \ and\ \bibinfo {author} {\bibfnamefont
  {W.}~\bibnamefont {Mori}}} (\bibinfo {year} {2017}),\ in\ \href
  {https://doi.org/10.1063/1.4975911} {\emph {\bibinfo {booktitle} {{AIP} Conf.
  Proc.}}},\ Vol.\ \bibinfo {volume} {1812}\ (\bibinfo  {publisher} {{AIP}})\
  p.\ \bibinfo {pages} {100013}\BibitemShut {NoStop}%
\bibitem [{\citenamefont {Baxevanis}\ and\ \citenamefont
  {Stupakov}(2018)}]{Baxevanis2018}%
  \BibitemOpen
  \bibfield  {author} {\bibinfo {author} {\bibnamefont {Baxevanis},
  \bibfnamefont {P}}, \ and\ \bibinfo {author} {\bibfnamefont {G.}~\bibnamefont
  {Stupakov}}} (\bibinfo {year} {2018}),\ \href
  {https://doi.org/10.1103/physrevaccelbeams.21.071301} {\bibfield  {journal}
  {\bibinfo  {journal} {Phys. Rev. Accel. Beams}\ }\textbf {\bibinfo {volume}
  {21}},\ \bibinfo {pages} {071301}}\BibitemShut {NoStop}%
\bibitem [{\citenamefont {Benedetti}\ \emph {et~al.}(2021)\citenamefont
  {Benedetti}, \citenamefont {Mehrling}, \citenamefont {Schroeder},
  \citenamefont {Geddes},\ and\ \citenamefont {Esarey}}]{Benedetti2021}%
  \BibitemOpen
  \bibfield  {author} {\bibinfo {author} {\bibnamefont {Benedetti},
  \bibfnamefont {C}}, \bibinfo {author} {\bibfnamefont {T.~J.}\ \bibnamefont
  {Mehrling}}, \bibinfo {author} {\bibfnamefont {C.~B.}\ \bibnamefont
  {Schroeder}}, \bibinfo {author} {\bibfnamefont {C.~G.~R.}\ \bibnamefont
  {Geddes}}, \ and\ \bibinfo {author} {\bibfnamefont {E.}~\bibnamefont
  {Esarey}}} (\bibinfo {year} {2021}),\ \href
  {https://doi.org/10.1063/5.0043847} {\bibfield  {journal} {\bibinfo
  {journal} {Phys. Plasmas}\ }\textbf {\bibinfo {volume} {28}},\ \bibinfo
  {pages} {053102}}\BibitemShut {NoStop}%
\bibitem [{\citenamefont {Benedetti}\ \emph {et~al.}(2010)\citenamefont
  {Benedetti}, \citenamefont {Schroeder}, \citenamefont {Esarey}, \citenamefont
  {Geddes},\ and\ \citenamefont {Leemans}}]{Benedetti2010}%
  \BibitemOpen
  \bibfield  {author} {\bibinfo {author} {\bibnamefont {Benedetti},
  \bibfnamefont {C}}, \bibinfo {author} {\bibfnamefont {C.~B.}\ \bibnamefont
  {Schroeder}}, \bibinfo {author} {\bibfnamefont {E.}~\bibnamefont {Esarey}},
  \bibinfo {author} {\bibfnamefont {C.~G.~R.}\ \bibnamefont {Geddes}}, \ and\
  \bibinfo {author} {\bibfnamefont {W.~P.}\ \bibnamefont {Leemans}}} (\bibinfo
  {year} {2010}),\ in\ \href {https://doi.org/10.1063/1.3520323} {\emph
  {\bibinfo {booktitle} {{AIP} Conf. Proc.}}},\ Vol.\ \bibinfo {volume} {1299}\
  (\bibinfo  {publisher} {{AIP}})\ pp.\ \bibinfo {pages} {250--255}\BibitemShut
  {NoStop}%
\bibitem [{\citenamefont {Benedetti}\ \emph {et~al.}(2017)\citenamefont
  {Benedetti}, \citenamefont {Schroeder}, \citenamefont {Esarey},\ and\
  \citenamefont {Leemans}}]{Benedetti2017}%
  \BibitemOpen
  \bibfield  {author} {\bibinfo {author} {\bibnamefont {Benedetti},
  \bibfnamefont {C}}, \bibinfo {author} {\bibfnamefont {C.~B.}\ \bibnamefont
  {Schroeder}}, \bibinfo {author} {\bibfnamefont {E.}~\bibnamefont {Esarey}}, \
  and\ \bibinfo {author} {\bibfnamefont {W.~P.}\ \bibnamefont {Leemans}}}
  (\bibinfo {year} {2017}),\ \href
  {https://doi.org/10.1103/physrevaccelbeams.20.111301} {\bibfield  {journal}
  {\bibinfo  {journal} {Phys. Rev. Accel. Beams}\ }\textbf {\bibinfo {volume}
  {20}},\ \bibinfo {pages} {111301}}\BibitemShut {NoStop}%
\bibitem [{\citenamefont {Bentson}\ \emph {et~al.}(2003)\citenamefont
  {Bentson}, \citenamefont {Bolton}, \citenamefont {Bong}, \citenamefont
  {Emma}, \citenamefont {Galayda}, \citenamefont {Hastings}, \citenamefont
  {Krejcik}, \citenamefont {Rago}, \citenamefont {Rifkin},\ and\ \citenamefont
  {Spencer}}]{Bentson2003}%
  \BibitemOpen
  \bibfield  {author} {\bibinfo {author} {\bibnamefont {Bentson}, \bibfnamefont
  {L}}, \bibinfo {author} {\bibfnamefont {P.}~\bibnamefont {Bolton}}, \bibinfo
  {author} {\bibfnamefont {E.}~\bibnamefont {Bong}}, \bibinfo {author}
  {\bibfnamefont {P.}~\bibnamefont {Emma}}, \bibinfo {author} {\bibfnamefont
  {J.}~\bibnamefont {Galayda}}, \bibinfo {author} {\bibfnamefont
  {J.}~\bibnamefont {Hastings}}, \bibinfo {author} {\bibfnamefont
  {P.}~\bibnamefont {Krejcik}}, \bibinfo {author} {\bibfnamefont
  {C.}~\bibnamefont {Rago}}, \bibinfo {author} {\bibfnamefont {J.}~\bibnamefont
  {Rifkin}}, \ and\ \bibinfo {author} {\bibfnamefont {C.~M.}\ \bibnamefont
  {Spencer}}} (\bibinfo {year} {2003}),\ \href
  {https://doi.org/10.1016/S0168-9002(03)00875-1} {\bibfield  {journal}
  {\bibinfo  {journal} {Nucl. Instrum. Methods Phys. Res. A}\ }\textbf
  {\bibinfo {volume} {507}},\ \bibinfo {pages} {205--209}}\BibitemShut
  {NoStop}%
\bibitem [{\citenamefont {Beringer}\ \emph {et~al.}(2012)\citenamefont
  {Beringer} \emph {et~al.}}]{Beringer2012}%
  \BibitemOpen
  \bibfield  {author} {\bibinfo {author} {\bibnamefont {Beringer},
  \bibfnamefont {J}},  \emph {et~al.}} (\bibinfo {year} {2012}),\ \href
  {https://doi.org/10.1103/physrevd.86.010001} {\bibfield  {journal} {\bibinfo
  {journal} {Phys. Rev. D}\ }\textbf {\bibinfo {volume} {86}},\ \bibinfo
  {pages} {328}}\BibitemShut {NoStop}%
\bibitem [{\citenamefont {Bethe}(1953)}]{Bethe1953}%
  \BibitemOpen
  \bibfield  {author} {\bibinfo {author} {\bibnamefont {Bethe}, \bibfnamefont
  {H~A}}} (\bibinfo {year} {1953}),\ \href
  {https://doi.org/10.1103/physrev.89.1256} {\bibfield  {journal} {\bibinfo
  {journal} {Phys. Rev.}\ }\textbf {\bibinfo {volume} {89}},\ \bibinfo {pages}
  {1256--1266}}\BibitemShut {NoStop}%
\bibitem [{\citenamefont {Bj\"{o}rklund~Svensson}\ \emph
  {et~al.}(2026)\citenamefont {Bj\"{o}rklund~Svensson} \emph
  {et~al.}}]{BjorklundSvensson2026}%
  \BibitemOpen
  \bibfield  {author} {\bibinfo {author} {\bibnamefont
  {Bj\"{o}rklund~Svensson}, \bibfnamefont {J}},  \emph {et~al.}} (\bibinfo
  {year} {2026}),\ \href {\doibase 10.1103/jjxz-3fr4} {\bibfield  {journal}
  {\bibinfo  {journal} {Phys. Rev. Res.}\ }\textbf {\bibinfo {volume} {8}},\
  \bibinfo {pages} {033242}}\BibitemShut {NoStop}%
\bibitem [{\citenamefont {Blachman}\ and\ \citenamefont
  {Courant}(1948)}]{Blachman1948}%
  \BibitemOpen
  \bibfield  {author} {\bibinfo {author} {\bibnamefont {Blachman},
  \bibfnamefont {N~M}}, \ and\ \bibinfo {author} {\bibfnamefont {E.~D.}\
  \bibnamefont {Courant}}} (\bibinfo {year} {1948}),\ \href
  {https://doi.org/10.1103/physrev.74.140} {\bibfield  {journal} {\bibinfo
  {journal} {Phys. Rev.}\ }\textbf {\bibinfo {volume} {74}},\ \bibinfo {pages}
  {140--144}}\BibitemShut {NoStop}%
\bibitem [{\citenamefont {Blue}\ \emph {et~al.}(2003)\citenamefont {Blue} \emph
  {et~al.}}]{Blue2003}%
  \BibitemOpen
  \bibfield  {author} {\bibinfo {author} {\bibnamefont {Blue}, \bibfnamefont
  {B~E}},  \emph {et~al.}} (\bibinfo {year} {2003}),\ \href
  {https://doi.org/10.1103/physrevlett.90.214801} {\bibfield  {journal}
  {\bibinfo  {journal} {Phys. Rev. Lett.}\ }\textbf {\bibinfo {volume} {90}},\
  \bibinfo {pages} {214801}}\BibitemShut {NoStop}%
\bibitem [{\citenamefont {Blumenfeld}(2009)}]{Blumenfeld2009}%
  \BibitemOpen
  \bibfield  {author} {\bibinfo {author} {\bibnamefont {Blumenfeld},
  \bibfnamefont {I}}} (\bibinfo {year} {2009}),\ \href@noop {} {\bibinfo {type}
  {Ph{D} {T}hesis}}\ (\bibinfo  {school} {Stanford University})\BibitemShut
  {NoStop}%
\bibitem [{\citenamefont {Blumenfeld}\ \emph {et~al.}(2007)\citenamefont
  {Blumenfeld} \emph {et~al.}}]{Blumenfeld2007}%
  \BibitemOpen
  \bibfield  {author} {\bibinfo {author} {\bibnamefont {Blumenfeld},
  \bibfnamefont {I}},  \emph {et~al.}} (\bibinfo {year} {2007}),\ \href
  {https://doi.org/10.1038/nature05538} {\bibfield  {journal} {\bibinfo
  {journal} {Nature (London)}\ }\textbf {\bibinfo {volume} {445}},\ \bibinfo
  {pages} {741--744}}\BibitemShut {NoStop}%
\bibitem [{\citenamefont {Bohlen}\ \emph {et~al.}(2023)\citenamefont {Bohlen},
  \citenamefont {Gong}, \citenamefont {Quin}, \citenamefont {Tamburini},\ and\
  \citenamefont {P\~oder}}]{Bohlen2023}%
  \BibitemOpen
  \bibfield  {author} {\bibinfo {author} {\bibnamefont {Bohlen}, \bibfnamefont
  {S}}, \bibinfo {author} {\bibfnamefont {Z.}~\bibnamefont {Gong}}, \bibinfo
  {author} {\bibfnamefont {M.~J.}\ \bibnamefont {Quin}}, \bibinfo {author}
  {\bibfnamefont {M.}~\bibnamefont {Tamburini}}, \ and\ \bibinfo {author}
  {\bibfnamefont {K.}~\bibnamefont {P\~oder}}} (\bibinfo {year} {2023}),\ \href
  {https://doi.org/10.1103/PhysRevResearch.5.033205} {\bibfield  {journal}
  {\bibinfo  {journal} {Phys. Rev. Res.}\ }\textbf {\bibinfo {volume} {5}},\
  \bibinfo {pages} {033205}}\BibitemShut {NoStop}%
\bibitem [{\citenamefont {Bonatto}\ \emph {et~al.}(2015)\citenamefont
  {Bonatto}, \citenamefont {Schroeder}, \citenamefont {Vay}, \citenamefont
  {Geddes}, \citenamefont {Benedetti}, \citenamefont {Esarey},\ and\
  \citenamefont {Leemans}}]{Bonatto2015}%
  \BibitemOpen
  \bibfield  {author} {\bibinfo {author} {\bibnamefont {Bonatto}, \bibfnamefont
  {A}}, \bibinfo {author} {\bibfnamefont {C.~B.}\ \bibnamefont {Schroeder}},
  \bibinfo {author} {\bibfnamefont {J.-L.}\ \bibnamefont {Vay}}, \bibinfo
  {author} {\bibfnamefont {C.~G.~R.}\ \bibnamefont {Geddes}}, \bibinfo {author}
  {\bibfnamefont {C.}~\bibnamefont {Benedetti}}, \bibinfo {author}
  {\bibfnamefont {E.}~\bibnamefont {Esarey}}, \ and\ \bibinfo {author}
  {\bibfnamefont {W.~P.}\ \bibnamefont {Leemans}}} (\bibinfo {year} {2015}),\
  \href {http://aip.scitation.org/doi/10.1063/1.4928379} {\bibfield  {journal}
  {\bibinfo  {journal} {Phys. Plasmas}\ }\textbf {\bibinfo {volume} {22}},\
  \bibinfo {pages} {083106}}\BibitemShut {NoStop}%
\bibitem [{\citenamefont {Bostedt}\ \emph {et~al.}(2016)\citenamefont {Bostedt}
  \emph {et~al.}}]{Bostedt2016}%
  \BibitemOpen
  \bibfield  {author} {\bibinfo {author} {\bibnamefont {Bostedt}, \bibfnamefont
  {C}},  \emph {et~al.}} (\bibinfo {year} {2016}),\ \href
  {https://doi.org/10.1103/RevModPhys.88.015007} {\bibfield  {journal}
  {\bibinfo  {journal} {Rev. Mod. Phys.}\ }\textbf {\bibinfo {volume} {88}},\
  \bibinfo {pages} {015007}}\BibitemShut {NoStop}%
\bibitem [{\citenamefont {Boulton}\ \emph {et~al.}(2026)\citenamefont {Boulton}
  \emph {et~al.}}]{Boulton2026}%
  \BibitemOpen
  \bibfield  {author} {\bibinfo {author} {\bibnamefont {Boulton}, \bibfnamefont
  {L}},  \emph {et~al.}} (\bibinfo {year} {2026}),\ \href {\doibase
  10.1103/b8hj-shvc} {\bibfield  {journal} {\bibinfo  {journal} {Phys. Rev.
  Res.}\ }\textbf {\bibinfo {volume} {8}},\ \bibinfo {pages}
  {023231}}\BibitemShut {NoStop}%
\bibitem [{\citenamefont {Braunm\"{u}ller}\ \emph {et~al.}(2020)\citenamefont
  {Braunm\"{u}ller} \emph {et~al.}}]{Braunmller2020}%
  \BibitemOpen
  \bibfield  {author} {\bibinfo {author} {\bibnamefont {Braunm\"{u}ller},
  \bibfnamefont {F}},  \emph {et~al.} (\bibinfo {collaboration} {AWAKE
  Collaboration})} (\bibinfo {year} {2020}),\ \href
  {https://doi.org/10.1103/physrevlett.125.264801} {\bibfield  {journal}
  {\bibinfo  {journal} {Phys. Rev. Lett.}\ }\textbf {\bibinfo {volume} {125}},\
  \bibinfo {pages} {264801}}\BibitemShut {NoStop}%
\bibitem [{\citenamefont {Bret}\ \emph {et~al.}(2005)\citenamefont {Bret},
  \citenamefont {Firpo},\ and\ \citenamefont {Deutsch}}]{Bret2005}%
  \BibitemOpen
  \bibfield  {author} {\bibinfo {author} {\bibnamefont {Bret}, \bibfnamefont
  {A}}, \bibinfo {author} {\bibfnamefont {M.-C.}\ \bibnamefont {Firpo}}, \ and\
  \bibinfo {author} {\bibfnamefont {C.}~\bibnamefont {Deutsch}}} (\bibinfo
  {year} {2005}),\ \href
  {https://link.aps.org/doi/10.1103/PhysRevLett.94.115002} {\bibfield
  {journal} {\bibinfo  {journal} {Phys. Rev. Lett.}\ }\textbf {\bibinfo
  {volume} {94}},\ \bibinfo {pages} {115002}}\BibitemShut {NoStop}%
\bibitem [{\citenamefont {Bret}\ \emph {et~al.}(2010)\citenamefont {Bret},
  \citenamefont {Gremillet},\ and\ \citenamefont {Dieckmann}}]{Bret2010}%
  \BibitemOpen
  \bibfield  {author} {\bibinfo {author} {\bibnamefont {Bret}, \bibfnamefont
  {A}}, \bibinfo {author} {\bibfnamefont {L.}~\bibnamefont {Gremillet}}, \ and\
  \bibinfo {author} {\bibfnamefont {M.~E.}\ \bibnamefont {Dieckmann}}}
  (\bibinfo {year} {2010}),\ \href {https://doi.org/10.1063/1.3514586}
  {\bibfield  {journal} {\bibinfo  {journal} {Phys. Plasmas}\ }\textbf
  {\bibinfo {volume} {17}},\ \bibinfo {pages} {120501}}\BibitemShut {NoStop}%
\bibitem [{\citenamefont {Brinkmann}\ \emph {et~al.}(2017)\citenamefont
  {Brinkmann} \emph {et~al.}}]{Brinkmann2017}%
  \BibitemOpen
  \bibfield  {author} {\bibinfo {author} {\bibnamefont {Brinkmann},
  \bibfnamefont {R}},  \emph {et~al.}} (\bibinfo {year} {2017}),\ \href
  {http://link.aps.org/doi/10.1103/PhysRevLett.118.214801} {\bibfield
  {journal} {\bibinfo  {journal} {Phys. Rev. Lett.}\ }\textbf {\bibinfo
  {volume} {118}},\ \bibinfo {pages} {214801}}\BibitemShut {NoStop}%
\bibitem [{\citenamefont {Brunetti}\ \emph {et~al.}(2010)\citenamefont
  {Brunetti} \emph {et~al.}}]{Brunetti2010}%
  \BibitemOpen
  \bibfield  {author} {\bibinfo {author} {\bibnamefont {Brunetti},
  \bibfnamefont {E}},  \emph {et~al.}} (\bibinfo {year} {2010}),\ \href
  {https://doi.org/10.1103/physrevlett.105.215007} {\bibfield  {journal}
  {\bibinfo  {journal} {Phys. Rev. Lett.}\ }\textbf {\bibinfo {volume} {105}},\
  \bibinfo {pages} {215007}}\BibitemShut {NoStop}%
\bibitem [{\citenamefont {Buchanan}(1987)}]{Buchanan1987}%
  \BibitemOpen
  \bibfield  {author} {\bibinfo {author} {\bibnamefont {Buchanan},
  \bibfnamefont {H~Lee}}} (\bibinfo {year} {1987}),\ \href
  {https://aip.scitation.org/doi/abs/10.1063/1.866173} {\bibfield  {journal}
  {\bibinfo  {journal} {Phys. Fluids}\ }\textbf {\bibinfo {volume} {30}},\
  \bibinfo {pages} {221--231}}\BibitemShut {NoStop}%
\bibitem [{\citenamefont {Buck}\ \emph {et~al.}(2011)\citenamefont {Buck},
  \citenamefont {Nicolai}, \citenamefont {Schmid}, \citenamefont {Sears},
  \citenamefont {S\"{a}vert}, \citenamefont {Mikhailova}, \citenamefont
  {Krausz}, \citenamefont {Kaluza},\ and\ \citenamefont {Veisz}}]{Buck2011}%
  \BibitemOpen
  \bibfield  {author} {\bibinfo {author} {\bibnamefont {Buck}, \bibfnamefont
  {A}}, \bibinfo {author} {\bibfnamefont {M.}~\bibnamefont {Nicolai}}, \bibinfo
  {author} {\bibfnamefont {K.}~\bibnamefont {Schmid}}, \bibinfo {author}
  {\bibfnamefont {C.~M.~S.}\ \bibnamefont {Sears}}, \bibinfo {author}
  {\bibfnamefont {A.}~\bibnamefont {S\"{a}vert}}, \bibinfo {author}
  {\bibfnamefont {J.~M.}\ \bibnamefont {Mikhailova}}, \bibinfo {author}
  {\bibfnamefont {F.}~\bibnamefont {Krausz}}, \bibinfo {author} {\bibfnamefont
  {M.~C.}\ \bibnamefont {Kaluza}}, \ and\ \bibinfo {author} {\bibfnamefont
  {L.}~\bibnamefont {Veisz}}} (\bibinfo {year} {2011}),\ \href
  {https://doi.org/10.1038/nphys1942} {\bibfield  {journal} {\bibinfo
  {journal} {Nat. Phys.}\ }\textbf {\bibinfo {volume} {7}},\ \bibinfo {pages}
  {543--548}}\BibitemShut {NoStop}%
\bibitem [{\citenamefont {Bulanov}\ \emph {et~al.}(1998)\citenamefont
  {Bulanov}, \citenamefont {Naumova}, \citenamefont {Pegoraro},\ and\
  \citenamefont {Sakai}}]{Bulanov1998}%
  \BibitemOpen
  \bibfield  {author} {\bibinfo {author} {\bibnamefont {Bulanov}, \bibfnamefont
  {S}}, \bibinfo {author} {\bibfnamefont {N.}~\bibnamefont {Naumova}}, \bibinfo
  {author} {\bibfnamefont {F.}~\bibnamefont {Pegoraro}}, \ and\ \bibinfo
  {author} {\bibfnamefont {J.}~\bibnamefont {Sakai}}} (\bibinfo {year}
  {1998}),\ \href {https://doi.org/10.1103/PhysRevE.58.R5257} {\bibfield
  {journal} {\bibinfo  {journal} {Phys. Rev. E}\ }\textbf {\bibinfo {volume}
  {58}},\ \bibinfo {pages} {R5257--R5260}}\BibitemShut {NoStop}%
\bibitem [{\citenamefont {Bulanov}\ \emph {et~al.}(2024)\citenamefont {Bulanov}
  \emph {et~al.}}]{Bulanov2024}%
  \BibitemOpen
  \bibfield  {author} {\bibinfo {author} {\bibnamefont {Bulanov}, \bibfnamefont
  {S~S}},  \emph {et~al.}} (\bibinfo {year} {2024}),\ \href
  {https://doi.org/10.1088/1748-0221/19/01/T01010} {\bibfield  {journal}
  {\bibinfo  {journal} {J. Instrum.}\ }\textbf {\bibinfo {volume} {19}},\
  \bibinfo {pages} {T01010}}\BibitemShut {NoStop}%
\bibitem [{\citenamefont {Bulanov}\ \emph {et~al.}(1997)\citenamefont
  {Bulanov}, \citenamefont {Pegoraro}, \citenamefont {Pukhov},\ and\
  \citenamefont {Sakharov}}]{Bulanov1997}%
  \BibitemOpen
  \bibfield  {author} {\bibinfo {author} {\bibnamefont {Bulanov}, \bibfnamefont
  {S~V}}, \bibinfo {author} {\bibfnamefont {F.}~\bibnamefont {Pegoraro}},
  \bibinfo {author} {\bibfnamefont {A.~M.}\ \bibnamefont {Pukhov}}, \ and\
  \bibinfo {author} {\bibfnamefont {A.~S.}\ \bibnamefont {Sakharov}}} (\bibinfo
  {year} {1997}),\ \href {https://doi.org/10.1103/PhysRevLett.78.4205}
  {\bibfield  {journal} {\bibinfo  {journal} {Phys. Rev. Lett.}\ }\textbf
  {\bibinfo {volume} {78}},\ \bibinfo {pages} {4205--4208}}\BibitemShut
  {NoStop}%
\bibitem [{\citenamefont {Burke}(1991)}]{Burke1991}%
  \BibitemOpen
  \bibfield  {author} {\bibinfo {author} {\bibnamefont {Burke}, \bibfnamefont
  {D}}} (\bibinfo {year} {1991}),\ in\ \href
  {https://doi.org/10.1109/PAC.1991.164866} {\emph {\bibinfo {booktitle}
  {Proceedings of the 1991 Particle Accelerator Conf.}}}\ (\bibinfo
  {publisher} {IEEE},\ \bibinfo {address} {New York, NY, United States})\ p.\
  \bibinfo {pages} {2055–2057}\BibitemShut {NoStop}%
\bibitem [{\citenamefont {Butler}\ \emph {et~al.}(2002)\citenamefont {Butler},
  \citenamefont {Spence},\ and\ \citenamefont {Hooker}}]{Butler2002}%
  \BibitemOpen
  \bibfield  {author} {\bibinfo {author} {\bibnamefont {Butler}, \bibfnamefont
  {A}}, \bibinfo {author} {\bibfnamefont {D.~J.}\ \bibnamefont {Spence}}, \
  and\ \bibinfo {author} {\bibfnamefont {S.~M.}\ \bibnamefont {Hooker}}}
  (\bibinfo {year} {2002}),\ \href
  {https://doi.org/10.1103/PhysRevLett.89.185003} {\bibfield  {journal}
  {\bibinfo  {journal} {Phys. Rev. Lett.}\ }\textbf {\bibinfo {volume} {89}},\
  \bibinfo {pages} {185003}}\BibitemShut {NoStop}%
\bibitem [{\citenamefont {Buttensch\"{o}n}\ \emph {et~al.}(2018)\citenamefont
  {Buttensch\"{o}n}, \citenamefont {Fahrenkamp},\ and\ \citenamefont
  {Grulke}}]{Buttenschn2018}%
  \BibitemOpen
  \bibfield  {author} {\bibinfo {author} {\bibnamefont {Buttensch\"{o}n},
  \bibfnamefont {B}}, \bibinfo {author} {\bibfnamefont {N.}~\bibnamefont
  {Fahrenkamp}}, \ and\ \bibinfo {author} {\bibfnamefont {O.}~\bibnamefont
  {Grulke}}} (\bibinfo {year} {2018}),\ \href
  {https://doi.org/10.1088/1361-6587/aac13a} {\bibfield  {journal} {\bibinfo
  {journal} {Plasma Phys. Control. Fusion}\ }\textbf {\bibinfo {volume} {60}},\
  \bibinfo {pages} {075005}}\BibitemShut {NoStop}%
\bibitem [{\citenamefont {Caldwell}\ \emph {et~al.}(2009)\citenamefont
  {Caldwell}, \citenamefont {Lotov}, \citenamefont {Pukhov},\ and\
  \citenamefont {Simon}}]{Caldwell2009}%
  \BibitemOpen
  \bibfield  {author} {\bibinfo {author} {\bibnamefont {Caldwell},
  \bibfnamefont {A}}, \bibinfo {author} {\bibfnamefont {K.}~\bibnamefont
  {Lotov}}, \bibinfo {author} {\bibfnamefont {A.}~\bibnamefont {Pukhov}}, \
  and\ \bibinfo {author} {\bibfnamefont {F.}~\bibnamefont {Simon}}} (\bibinfo
  {year} {2009}),\ \href {https://doi.org/10.1038/nphys1248} {\bibfield
  {journal} {\bibinfo  {journal} {Nat. Phys.}\ }\textbf {\bibinfo {volume}
  {5}},\ \bibinfo {pages} {363--367}}\BibitemShut {NoStop}%
\bibitem [{\citenamefont {Caldwell}\ and\ \citenamefont
  {Wing}(2016)}]{Caldwell2016b}%
  \BibitemOpen
  \bibfield  {author} {\bibinfo {author} {\bibnamefont {Caldwell},
  \bibfnamefont {A}}, \ and\ \bibinfo {author} {\bibfnamefont {M.}~\bibnamefont
  {Wing}}} (\bibinfo {year} {2016}),\ \href
  {https://doi.org/10.1140/epjc/s10052-016-4316-1} {\bibfield  {journal}
  {\bibinfo  {journal} {Eur. Phys. J. C}\ }\textbf {\bibinfo {volume} {76}},\
  \bibinfo {pages} {463}}\BibitemShut {NoStop}%
\bibitem [{\citenamefont {Caldwell}\ \emph {et~al.}(2016)\citenamefont
  {Caldwell} \emph {et~al.}}]{Caldwell2016}%
  \BibitemOpen
  \bibfield  {author} {\bibinfo {author} {\bibnamefont {Caldwell},
  \bibfnamefont {A}},  \emph {et~al.} (\bibinfo {collaboration} {AWAKE
  Collaboration})} (\bibinfo {year} {2016}),\ \href
  {https://doi.org/10.1016/j.nima.2015.12.050} {\bibfield  {journal} {\bibinfo
  {journal} {Nucl. Instrum. Methods Phys. Res. A}\ }\textbf {\bibinfo {volume}
  {829}},\ \bibinfo {pages} {3--16}}\BibitemShut {NoStop}%
\bibitem [{\citenamefont {Cao}(2023)}]{Cao2023}%
  \BibitemOpen
  \bibfield  {author} {\bibinfo {author} {\bibnamefont {Cao}, \bibfnamefont
  {G~J}}} (\bibinfo {year} {2023}),\ \href
  {https://doi.org/10.3390/instruments7040037} {\bibfield  {journal} {\bibinfo
  {journal} {Instruments}\ }\textbf {\bibinfo {volume} {7}},\ \bibinfo {pages}
  {37}}\BibitemShut {NoStop}%
\bibitem [{\citenamefont {Cao}\ \emph {et~al.}(2024)\citenamefont {Cao},
  \citenamefont {Lindstr{\o}m}, \citenamefont {Adli}, \citenamefont {Corde},\
  and\ \citenamefont {Gessner}}]{Cao2024}%
  \BibitemOpen
  \bibfield  {author} {\bibinfo {author} {\bibnamefont {Cao}, \bibfnamefont
  {G~J}}, \bibinfo {author} {\bibfnamefont {C.~A.}\ \bibnamefont
  {Lindstr{\o}m}}, \bibinfo {author} {\bibfnamefont {E.}~\bibnamefont {Adli}},
  \bibinfo {author} {\bibfnamefont {S.}~\bibnamefont {Corde}}, \ and\ \bibinfo
  {author} {\bibfnamefont {S.}~\bibnamefont {Gessner}}} (\bibinfo {year}
  {2024}),\ \href
  {https://link.aps.org/doi/10.1103/PhysRevAccelBeams.27.034801} {\bibfield
  {journal} {\bibinfo  {journal} {Phys. Rev. Accel. Beams}\ }\textbf {\bibinfo
  {volume} {27}},\ \bibinfo {pages} {034801}}\BibitemShut {NoStop}%
\bibitem [{\citenamefont {Carlsten}\ \emph {et~al.}(1996)\citenamefont
  {Carlsten} \emph {et~al.}}]{Carlsten1996}%
  \BibitemOpen
  \bibfield  {author} {\bibinfo {author} {\bibnamefont {Carlsten},
  \bibfnamefont {B~E}},  \emph {et~al.}} (\bibinfo {year} {1996}),\ in\ \href
  {\doibase 10.1063/1.50307} {\emph {\bibinfo {booktitle} {AIP Conf. Proc.}}},\
  Vol.\ \bibinfo {volume} {367}\ (\bibinfo  {publisher} {AIP})\ pp.\ \bibinfo
  {pages} {21--35}\BibitemShut {NoStop}%
\bibitem [{\citenamefont {Carneiro}\ \emph {et~al.}(1999)\citenamefont
  {Carneiro} \emph {et~al.}}]{Carneiro1999}%
  \BibitemOpen
  \bibfield  {author} {\bibinfo {author} {\bibnamefont {Carneiro},
  \bibfnamefont {J-P}},  \emph {et~al.}} (\bibinfo {year} {1999}),\ in\ \href
  {https://jacow.org/p99/papers/WEA60.pdf} {\emph {\bibinfo {booktitle}
  {Proceedings of the 1999 Particle Accelerator Conf.}}}\ (\bibinfo
  {publisher} {JACoW},\ \bibinfo {address} {Geneva, Switzerland})\ pp.\
  \bibinfo {pages} {2027--2029}\BibitemShut {NoStop}%
\bibitem [{\citenamefont {Chao}(1993)}]{Chao1993}%
  \BibitemOpen
  \bibfield  {author} {\bibinfo {author} {\bibnamefont {Chao}, \bibfnamefont
  {A~W}}} (\bibinfo {year} {1993}),\ \href@noop {} {\emph {\bibinfo {title}
  {Physics of Collective Beam Instabilities in High Energy Accelerators}}}\
  (\bibinfo  {publisher} {John Wiley \& Sons},\ \bibinfo {address} {New York,
  NY, United States})\BibitemShut {NoStop}%
\bibitem [{\citenamefont {Chappell}\ \emph {et~al.}(2021)\citenamefont
  {Chappell} \emph {et~al.}}]{Chappell2021}%
  \BibitemOpen
  \bibfield  {author} {\bibinfo {author} {\bibnamefont {Chappell},
  \bibfnamefont {J}},  \emph {et~al.} (\bibinfo {collaboration} {AWAKE
  Collaboration})} (\bibinfo {year} {2021}),\ \href
  {https://doi.org/10.1103/physrevaccelbeams.24.011301} {\bibfield  {journal}
  {\bibinfo  {journal} {Phys. Rev. Accel. Beams}\ }\textbf {\bibinfo {volume}
  {24}},\ \bibinfo {pages} {011301}}\BibitemShut {NoStop}%
\bibitem [{\citenamefont {Chattopadhyay}\ \emph {et~al.}(2020)\citenamefont
  {Chattopadhyay}, \citenamefont {Mourou}, \citenamefont {Shiltsev},\ and\
  \citenamefont {Tajima}}]{Chattopadhyay2020}%
  \BibitemOpen
  \bibfield  {author} {\bibinfo {author} {\bibnamefont {Chattopadhyay},
  \bibfnamefont {S}}, \bibinfo {author} {\bibfnamefont {G.}~\bibnamefont
  {Mourou}}, \bibinfo {author} {\bibfnamefont {V.~D.}\ \bibnamefont
  {Shiltsev}}, \ and\ \bibinfo {author} {\bibfnamefont {T.}~\bibnamefont
  {Tajima}}} (\bibinfo {year} {2020}),\ \href {\doibase 10.1142/11742} {\emph
  {\bibinfo {title} {Beam Acceleration in Crystals and Nanostructures}}}\
  (\bibinfo  {publisher} {World Scientific},\ \bibinfo {address}
  {Singapore})\BibitemShut {NoStop}%
\bibitem [{\citenamefont {Chen}\ \emph {et~al.}(2025)\citenamefont {Chen} \emph
  {et~al.}}]{Chen2025}%
  \BibitemOpen
  \bibfield  {author} {\bibinfo {author} {\bibnamefont {Chen}, \bibfnamefont
  {J~B~B}},  \emph {et~al.}} (\bibinfo {year} {2025}),\ in\ \href {\doibase
  10.18429/JACOW-IPAC2025-TUPS012} {\emph {\bibinfo {booktitle} {Proceedings of
  the 16th Int. Particle Accelerator Conf.}}}\ (\bibinfo  {publisher} {JACoW},\
  \bibinfo {address} {Geneva, Switzerland})\ pp.\ \bibinfo {pages}
  {1438--1441}\BibitemShut {NoStop}%
\bibitem [{\citenamefont {Chen}(1987)}]{Chen1987}%
  \BibitemOpen
  \bibfield  {author} {\bibinfo {author} {\bibnamefont {Chen}, \bibfnamefont
  {P}}} (\bibinfo {year} {1987}),\ \href
  {https://cds.cern.ch/record/166083/files/p171.pdf} {\bibfield  {journal}
  {\bibinfo  {journal} {Part. Accel.}\ }\textbf {\bibinfo {volume} {20}},\
  \bibinfo {pages} {171--182}}\BibitemShut {NoStop}%
\bibitem [{\citenamefont {Chen}\ \emph {et~al.}(1985)\citenamefont {Chen},
  \citenamefont {Dawson}, \citenamefont {Huff},\ and\ \citenamefont
  {Katsouleas}}]{Chen1985}%
  \BibitemOpen
  \bibfield  {author} {\bibinfo {author} {\bibnamefont {Chen}, \bibfnamefont
  {P}}, \bibinfo {author} {\bibfnamefont {J.~M.}\ \bibnamefont {Dawson}},
  \bibinfo {author} {\bibfnamefont {R.~W.}\ \bibnamefont {Huff}}, \ and\
  \bibinfo {author} {\bibfnamefont {T.}~\bibnamefont {Katsouleas}}} (\bibinfo
  {year} {1985}),\ \href {https://doi.org/10.1103/physrevlett.54.693}
  {\bibfield  {journal} {\bibinfo  {journal} {Phys. Rev. Lett.}\ }\textbf
  {\bibinfo {volume} {54}},\ \bibinfo {pages} {693--696}}\BibitemShut {NoStop}%
\bibitem [{\citenamefont {Chen}\ and\ \citenamefont {Liu}(2026)}]{Chen2026}%
  \BibitemOpen
  \bibfield  {author} {\bibinfo {author} {\bibnamefont {Chen}, \bibfnamefont
  {P}}, \ and\ \bibinfo {author} {\bibfnamefont {Y.-K.}\ \bibnamefont {Liu}}}
  (\bibinfo {year} {2026}),\ \href {\doibase 10.1007/s41614-026-00215-z}
  {\bibfield  {journal} {\bibinfo  {journal} {Rev. Mod. Plasma Phys.}\ }\textbf
  {\bibinfo {volume} {10}},\ \bibinfo {pages} {7}}\BibitemShut {NoStop}%
\bibitem [{\citenamefont {Chen}\ \emph {et~al.}(1990)\citenamefont {Chen},
  \citenamefont {Oide}, \citenamefont {Sessler},\ and\ \citenamefont
  {Yu}}]{Chen1990}%
  \BibitemOpen
  \bibfield  {author} {\bibinfo {author} {\bibnamefont {Chen}, \bibfnamefont
  {P}}, \bibinfo {author} {\bibfnamefont {K.}~\bibnamefont {Oide}}, \bibinfo
  {author} {\bibfnamefont {A.~M.}\ \bibnamefont {Sessler}}, \ and\ \bibinfo
  {author} {\bibfnamefont {S.~S.}\ \bibnamefont {Yu}}} (\bibinfo {year}
  {1990}),\ \href {https://doi.org/10.1103/PhysRevLett.64.1231} {\bibfield
  {journal} {\bibinfo  {journal} {Phys. Rev. Lett.}\ }\textbf {\bibinfo
  {volume} {64}},\ \bibinfo {pages} {1231--1234}}\BibitemShut {NoStop}%
\bibitem [{\citenamefont {Chen}\ \emph {et~al.}(1986)\citenamefont {Chen},
  \citenamefont {Su}, \citenamefont {Dawson}, \citenamefont {Bane},\ and\
  \citenamefont {Wilson}}]{Chen1986}%
  \BibitemOpen
  \bibfield  {author} {\bibinfo {author} {\bibnamefont {Chen}, \bibfnamefont
  {P}}, \bibinfo {author} {\bibfnamefont {J.~J.}\ \bibnamefont {Su}}, \bibinfo
  {author} {\bibfnamefont {J.~M.}\ \bibnamefont {Dawson}}, \bibinfo {author}
  {\bibfnamefont {K.~L.~F.}\ \bibnamefont {Bane}}, \ and\ \bibinfo {author}
  {\bibfnamefont {P.~B.}\ \bibnamefont {Wilson}}} (\bibinfo {year} {1986}),\
  \href {https://doi.org/10.1103/physrevlett.56.1252} {\bibfield  {journal}
  {\bibinfo  {journal} {Phys. Rev. Lett.}\ }\textbf {\bibinfo {volume} {56}},\
  \bibinfo {pages} {1252--1255}}\BibitemShut {NoStop}%
\bibitem [{\citenamefont {Cheshkov}\ \emph {et~al.}(2000)\citenamefont
  {Cheshkov}, \citenamefont {Tajima}, \citenamefont {Horton},\ and\
  \citenamefont {Yokoya}}]{Cheshkov2000}%
  \BibitemOpen
  \bibfield  {author} {\bibinfo {author} {\bibnamefont {Cheshkov},
  \bibfnamefont {S}}, \bibinfo {author} {\bibfnamefont {T.}~\bibnamefont
  {Tajima}}, \bibinfo {author} {\bibfnamefont {W.}~\bibnamefont {Horton}}, \
  and\ \bibinfo {author} {\bibfnamefont {K.}~\bibnamefont {Yokoya}}} (\bibinfo
  {year} {2000}),\ \href {https://doi.org/10.1103/physrevstab.3.071301}
  {\bibfield  {journal} {\bibinfo  {journal} {Phys. Rev. ST Accel. Beams}\
  }\textbf {\bibinfo {volume} {3}},\ \bibinfo {pages} {071301}}\BibitemShut
  {NoStop}%
\bibitem [{\citenamefont {Chiou}\ and\ \citenamefont
  {Katsouleas}(1998)}]{Chiou1998}%
  \BibitemOpen
  \bibfield  {author} {\bibinfo {author} {\bibnamefont {Chiou}, \bibfnamefont
  {T~C}}, \ and\ \bibinfo {author} {\bibfnamefont {T.}~\bibnamefont
  {Katsouleas}}} (\bibinfo {year} {1998}),\ \href
  {https://doi.org/10.1103/physrevlett.81.3411} {\bibfield  {journal} {\bibinfo
   {journal} {Phys. Rev. Lett.}\ }\textbf {\bibinfo {volume} {81}},\ \bibinfo
  {pages} {3411--3414}}\BibitemShut {NoStop}%
\bibitem [{\citenamefont {Chiou}\ \emph {et~al.}(1996)\citenamefont {Chiou},
  \citenamefont {Katsouleas},\ and\ \citenamefont {Mori}}]{Chiou1996}%
  \BibitemOpen
  \bibfield  {author} {\bibinfo {author} {\bibnamefont {Chiou}, \bibfnamefont
  {T~C}}, \bibinfo {author} {\bibfnamefont {T.}~\bibnamefont {Katsouleas}}, \
  and\ \bibinfo {author} {\bibfnamefont {W.~B.}\ \bibnamefont {Mori}}}
  (\bibinfo {year} {1996}),\ \href {https://doi.org/10.1063/1.871689}
  {\bibfield  {journal} {\bibinfo  {journal} {Phys. Plasmas}\ }\textbf
  {\bibinfo {volume} {3}},\ \bibinfo {pages} {1700--1708}}\BibitemShut
  {NoStop}%
\bibitem [{\citenamefont {Chiu}\ \emph {et~al.}(2000)\citenamefont {Chiu},
  \citenamefont {Cheshkov},\ and\ \citenamefont {Tajima}}]{Chiu2000}%
  \BibitemOpen
  \bibfield  {author} {\bibinfo {author} {\bibnamefont {Chiu}, \bibfnamefont
  {C}}, \bibinfo {author} {\bibfnamefont {S.}~\bibnamefont {Cheshkov}}, \ and\
  \bibinfo {author} {\bibfnamefont {T.}~\bibnamefont {Tajima}}} (\bibinfo
  {year} {2000}),\ \href {\doibase 10.1103/physrevstab.3.101301} {\bibfield
  {journal} {\bibinfo  {journal} {Phys. Rev. ST Accel. Beams}\ }\textbf
  {\bibinfo {volume} {3}},\ \bibinfo {pages} {101301}}\BibitemShut {NoStop}%
\bibitem [{\citenamefont {Chou}\ \emph {et~al.}(2016)\citenamefont {Chou},
  \citenamefont {Xu}, \citenamefont {Khrennikov}, \citenamefont {Cardenas},
  \citenamefont {Wenz}, \citenamefont {Heigoldt}, \citenamefont {Hofmann},
  \citenamefont {Veisz},\ and\ \citenamefont {Karsch}}]{Chou2016}%
  \BibitemOpen
  \bibfield  {author} {\bibinfo {author} {\bibnamefont {Chou}, \bibfnamefont
  {S}}, \bibinfo {author} {\bibfnamefont {J.}~\bibnamefont {Xu}}, \bibinfo
  {author} {\bibfnamefont {K.}~\bibnamefont {Khrennikov}}, \bibinfo {author}
  {\bibfnamefont {D.~E.}\ \bibnamefont {Cardenas}}, \bibinfo {author}
  {\bibfnamefont {J.}~\bibnamefont {Wenz}}, \bibinfo {author} {\bibfnamefont
  {M.}~\bibnamefont {Heigoldt}}, \bibinfo {author} {\bibfnamefont
  {L.}~\bibnamefont {Hofmann}}, \bibinfo {author} {\bibfnamefont
  {L.}~\bibnamefont {Veisz}}, \ and\ \bibinfo {author} {\bibfnamefont
  {S.}~\bibnamefont {Karsch}}} (\bibinfo {year} {2016}),\ \href
  {https://doi.org/10.1103/PhysRevLett.117.144801} {\bibfield  {journal}
  {\bibinfo  {journal} {Phys. Rev. Lett.}\ }\textbf {\bibinfo {volume} {117}},\
  \bibinfo {pages} {144801}}\BibitemShut {NoStop}%
\bibitem [{\citenamefont {Cianchi}\ \emph {et~al.}(2015)\citenamefont {Cianchi}
  \emph {et~al.}}]{Cianchi2015}%
  \BibitemOpen
  \bibfield  {author} {\bibinfo {author} {\bibnamefont {Cianchi}, \bibfnamefont
  {A}},  \emph {et~al.}} (\bibinfo {year} {2015}),\ \href
  {https://doi.org/10.1103/PhysRevSTAB.18.082804} {\bibfield  {journal}
  {\bibinfo  {journal} {Phys. Rev. ST Accel. Beams}\ }\textbf {\bibinfo
  {volume} {18}},\ \bibinfo {pages} {082804}}\BibitemShut {NoStop}%
\bibitem [{\citenamefont {Clarke}\ \emph {et~al.}(2011)\citenamefont {Clarke}
  \emph {et~al.}}]{Clarke2011}%
  \BibitemOpen
  \bibfield  {author} {\bibinfo {author} {\bibnamefont {Clarke}, \bibfnamefont
  {C~I}},  \emph {et~al.}} (\bibinfo {year} {2011}),\ in\ \href
  {https://jacow.org/IPAC2011/papers/WEOAB02.pdf} {\emph {\bibinfo {booktitle}
  {Proceedings of the 2nd Int. Particle Accelerator Conf.}}}\ (\bibinfo
  {publisher} {JACoW},\ \bibinfo {address} {Geneva, Switzerland})\ pp.\
  \bibinfo {pages} {1953--1955}\BibitemShut {NoStop}%
\bibitem [{\citenamefont {Clayton}\ \emph {et~al.}(2002)\citenamefont {Clayton}
  \emph {et~al.}}]{Clayton2002}%
  \BibitemOpen
  \bibfield  {author} {\bibinfo {author} {\bibnamefont {Clayton}, \bibfnamefont
  {C~E}},  \emph {et~al.}} (\bibinfo {year} {2002}),\ \href
  {https://doi.org/10.1103/PhysRevLett.88.154801} {\bibfield  {journal}
  {\bibinfo  {journal} {Phys. Rev. Lett.}\ }\textbf {\bibinfo {volume} {88}},\
  \bibinfo {pages} {154801}}\BibitemShut {NoStop}%
\bibitem [{\citenamefont {Clayton}\ \emph {et~al.}(2010)\citenamefont {Clayton}
  \emph {et~al.}}]{Clayton2010}%
  \BibitemOpen
  \bibfield  {author} {\bibinfo {author} {\bibnamefont {Clayton}, \bibfnamefont
  {C~E}},  \emph {et~al.}} (\bibinfo {year} {2010}),\ \href
  {https://doi.org/10.1103/PhysRevLett.105.105003} {\bibfield  {journal}
  {\bibinfo  {journal} {Phys. Rev. Lett.}\ }\textbf {\bibinfo {volume} {105}},\
  \bibinfo {pages} {105003}}\BibitemShut {NoStop}%
\bibitem [{\citenamefont {Clayton}\ \emph {et~al.}(2016)\citenamefont {Clayton}
  \emph {et~al.}}]{Clayton2016}%
  \BibitemOpen
  \bibfield  {author} {\bibinfo {author} {\bibnamefont {Clayton}, \bibfnamefont
  {C~E}},  \emph {et~al.}} (\bibinfo {year} {2016}),\ \href
  {https://doi.org/10.1038/ncomms12483} {\bibfield  {journal} {\bibinfo
  {journal} {Nat. Commun.}\ }\textbf {\bibinfo {volume} {7}},\ \bibinfo {pages}
  {12483}}\BibitemShut {NoStop}%
\bibitem [{\citenamefont {Cole}\ \emph {et~al.}(2018)\citenamefont {Cole} \emph
  {et~al.}}]{Cole2018}%
  \BibitemOpen
  \bibfield  {author} {\bibinfo {author} {\bibnamefont {Cole}, \bibfnamefont
  {J~M}},  \emph {et~al.}} (\bibinfo {year} {2018}),\ \href
  {https://link.aps.org/doi/10.1103/PhysRevX.8.011020} {\bibfield  {journal}
  {\bibinfo  {journal} {Phys. Rev. X}\ }\textbf {\bibinfo {volume} {8}},\
  \bibinfo {pages} {011020}}\BibitemShut {NoStop}%
\bibitem [{\citenamefont {Corde}\ \emph
  {et~al.}(2013{\natexlab{a}})\citenamefont {Corde}, \citenamefont {{Ta
  Phuoc}}, \citenamefont {Lambert}, \citenamefont {Fitour}, \citenamefont
  {Malka}, \citenamefont {Rousse}, \citenamefont {Beck},\ and\ \citenamefont
  {Lefebvre}}]{Corde2013}%
  \BibitemOpen
  \bibfield  {author} {\bibinfo {author} {\bibnamefont {Corde}, \bibfnamefont
  {S}}, \bibinfo {author} {\bibfnamefont {K.}~\bibnamefont {{Ta Phuoc}}},
  \bibinfo {author} {\bibfnamefont {G.}~\bibnamefont {Lambert}}, \bibinfo
  {author} {\bibfnamefont {R.}~\bibnamefont {Fitour}}, \bibinfo {author}
  {\bibfnamefont {V.}~\bibnamefont {Malka}}, \bibinfo {author} {\bibfnamefont
  {A.}~\bibnamefont {Rousse}}, \bibinfo {author} {\bibfnamefont
  {A.}~\bibnamefont {Beck}}, \ and\ \bibinfo {author} {\bibfnamefont
  {E.}~\bibnamefont {Lefebvre}}} (\bibinfo {year} {2013}{\natexlab{a}}),\ \href
  {https://doi.org/10.1103/revmodphys.85.1} {\bibfield  {journal} {\bibinfo
  {journal} {Rev. Mod. Phys.}\ }\textbf {\bibinfo {volume} {85}},\ \bibinfo
  {pages} {1--48}}\BibitemShut {NoStop}%
\bibitem [{\citenamefont {Corde}\ \emph
  {et~al.}(2013{\natexlab{b}})\citenamefont {Corde}, \citenamefont {Thaury},
  \citenamefont {Lifschitz}, \citenamefont {Lambert}, \citenamefont {Ta~Phuoc},
  \citenamefont {Davoine}, \citenamefont {Lehe}, \citenamefont {Douillet},
  \citenamefont {Rousse},\ and\ \citenamefont {Malka}}]{Corde2013b}%
  \BibitemOpen
  \bibfield  {author} {\bibinfo {author} {\bibnamefont {Corde}, \bibfnamefont
  {S}}, \bibinfo {author} {\bibfnamefont {C.}~\bibnamefont {Thaury}}, \bibinfo
  {author} {\bibfnamefont {A.}~\bibnamefont {Lifschitz}}, \bibinfo {author}
  {\bibfnamefont {G.}~\bibnamefont {Lambert}}, \bibinfo {author} {\bibfnamefont
  {K.}~\bibnamefont {Ta~Phuoc}}, \bibinfo {author} {\bibfnamefont
  {X.}~\bibnamefont {Davoine}}, \bibinfo {author} {\bibfnamefont
  {R.}~\bibnamefont {Lehe}}, \bibinfo {author} {\bibfnamefont {D.}~\bibnamefont
  {Douillet}}, \bibinfo {author} {\bibfnamefont {A.}~\bibnamefont {Rousse}}, \
  and\ \bibinfo {author} {\bibfnamefont {V.}~\bibnamefont {Malka}}} (\bibinfo
  {year} {2013}{\natexlab{b}}),\ \href {https://doi.org/10.1038/ncomms2528}
  {\bibfield  {journal} {\bibinfo  {journal} {Nat. Commun.}\ }\textbf {\bibinfo
  {volume} {4}},\ \bibinfo {pages} {1501}}\BibitemShut {NoStop}%
\bibitem [{\citenamefont {Corde}\ \emph {et~al.}(2011)\citenamefont {Corde}
  \emph {et~al.}}]{Corde2011}%
  \BibitemOpen
  \bibfield  {author} {\bibinfo {author} {\bibnamefont {Corde}, \bibfnamefont
  {S}},  \emph {et~al.}} (\bibinfo {year} {2011}),\ \href
  {https://doi.org/10.1103/PhysRevLett.107.215004} {\bibfield  {journal}
  {\bibinfo  {journal} {Phys. Rev. Lett.}\ }\textbf {\bibinfo {volume} {107}},\
  \bibinfo {pages} {215004}}\BibitemShut {NoStop}%
\bibitem [{\citenamefont {Corde}\ \emph {et~al.}(2015)\citenamefont {Corde}
  \emph {et~al.}}]{Corde2015}%
  \BibitemOpen
  \bibfield  {author} {\bibinfo {author} {\bibnamefont {Corde}, \bibfnamefont
  {S}},  \emph {et~al.}} (\bibinfo {year} {2015}),\ \href
  {https://doi.org/10.1038/nature14890} {\bibfield  {journal} {\bibinfo
  {journal} {Nature (London)}\ }\textbf {\bibinfo {volume} {524}},\ \bibinfo
  {pages} {442--445}}\BibitemShut {NoStop}%
\bibitem [{\citenamefont {Corde}\ \emph {et~al.}(2016)\citenamefont {Corde}
  \emph {et~al.}}]{Corde2016}%
  \BibitemOpen
  \bibfield  {author} {\bibinfo {author} {\bibnamefont {Corde}, \bibfnamefont
  {S}},  \emph {et~al.}} (\bibinfo {year} {2016}),\ \href
  {https://doi.org/10.1038/ncomms11898} {\bibfield  {journal} {\bibinfo
  {journal} {Nat. Commun.}\ }\textbf {\bibinfo {volume} {7}},\ \bibinfo {pages}
  {11898}}\BibitemShut {NoStop}%
\bibitem [{\citenamefont {Costa}\ \emph {et~al.}(2022)\citenamefont {Costa}
  \emph {et~al.}}]{Costa2022}%
  \BibitemOpen
  \bibfield  {author} {\bibinfo {author} {\bibnamefont {Costa}, \bibfnamefont
  {G}},  \emph {et~al.}} (\bibinfo {year} {2022}),\ \href
  {https://doi.org/10.1088/1361-6587/ac5477} {\bibfield  {journal} {\bibinfo
  {journal} {Plasma Phys. Control. Fusion}\ }\textbf {\bibinfo {volume} {64}},\
  \bibinfo {pages} {044012}}\BibitemShut {NoStop}%
\bibitem [{\citenamefont {Couperus}\ \emph {et~al.}(2017)\citenamefont
  {Couperus} \emph {et~al.}}]{Couperus2017}%
  \BibitemOpen
  \bibfield  {author} {\bibinfo {author} {\bibnamefont {Couperus},
  \bibfnamefont {J~P}},  \emph {et~al.}} (\bibinfo {year} {2017}),\ \href
  {https://doi.org/10.1038/s41467-017-00592-7} {\bibfield  {journal} {\bibinfo
  {journal} {Nat. Commun.}\ }\textbf {\bibinfo {volume} {8}},\ \bibinfo {pages}
  {487}}\BibitemShut {NoStop}%
\bibitem [{\citenamefont {{Couperus Cabada{\u{g}}}}\ \emph
  {et~al.}(2021)\citenamefont {{Couperus Cabada{\u{g}}}} \emph
  {et~al.}}]{CouperusCabada2021}%
  \BibitemOpen
  \bibfield  {author} {\bibinfo {author} {\bibnamefont {{Couperus
  Cabada{\u{g}}}}, \bibfnamefont {J~P}},  \emph {et~al.}} (\bibinfo {year}
  {2021}),\ \href {https://doi.org/10.1103/physrevresearch.3.l042005}
  {\bibfield  {journal} {\bibinfo  {journal} {Phys. Rev. Res.}\ }\textbf
  {\bibinfo {volume} {3}},\ \bibinfo {pages} {L042005}}\BibitemShut {NoStop}%
\bibitem [{\citenamefont {Courant}\ and\ \citenamefont
  {Snyder}(1958)}]{Courant1958}%
  \BibitemOpen
  \bibfield  {author} {\bibinfo {author} {\bibnamefont {Courant}, \bibfnamefont
  {E~D}}, \ and\ \bibinfo {author} {\bibfnamefont {H.~S.}\ \bibnamefont
  {Snyder}}} (\bibinfo {year} {1958}),\ \href
  {https://doi.org/10.1016/0003-4916(58)90012-5} {\bibfield  {journal}
  {\bibinfo  {journal} {Ann. Phys.}\ }\textbf {\bibinfo {volume} {3}},\
  \bibinfo {pages} {1--48}}\BibitemShut {NoStop}%
\bibitem [{\citenamefont {Courant}\ \emph {et~al.}(1928)\citenamefont
  {Courant}, \citenamefont {Friedrichs},\ and\ \citenamefont
  {Lewy}}]{Courant1928}%
  \BibitemOpen
  \bibfield  {author} {\bibinfo {author} {\bibnamefont {Courant}, \bibfnamefont
  {R}}, \bibinfo {author} {\bibfnamefont {K.}~\bibnamefont {Friedrichs}}, \
  and\ \bibinfo {author} {\bibfnamefont {H.}~\bibnamefont {Lewy}}} (\bibinfo
  {year} {1928}),\ \href {\doibase 10.1007/bf01448839} {\bibfield  {journal}
  {\bibinfo  {journal} {Math. Ann.}\ }\textbf {\bibinfo {volume} {100}},\
  \bibinfo {pages} {32--74}}\BibitemShut {NoStop}%
\bibitem [{\citenamefont {Cowley}\ \emph {et~al.}(2017)\citenamefont {Cowley}
  \emph {et~al.}}]{Cowley2017}%
  \BibitemOpen
  \bibfield  {author} {\bibinfo {author} {\bibnamefont {Cowley}, \bibfnamefont
  {J}},  \emph {et~al.}} (\bibinfo {year} {2017}),\ \href
  {https://doi.org/10.1103/physrevlett.119.044802} {\bibfield  {journal}
  {\bibinfo  {journal} {Phys. Rev. Lett.}\ }\textbf {\bibinfo {volume} {119}},\
  \bibinfo {pages} {044802}}\BibitemShut {NoStop}%
\bibitem [{\citenamefont {Crincoli}\ \emph {et~al.}(2025)\citenamefont
  {Crincoli}, \citenamefont {Demitra}, \citenamefont {Lollo}, \citenamefont
  {Pellegrini}, \citenamefont {Pitti}, \citenamefont {Pronti}, \citenamefont
  {Romani}, \citenamefont {Ferrario},\ and\ \citenamefont
  {Biagioni}}]{Crincoli2025}%
  \BibitemOpen
  \bibfield  {author} {\bibinfo {author} {\bibnamefont {Crincoli},
  \bibfnamefont {L}}, \bibinfo {author} {\bibfnamefont {R.}~\bibnamefont
  {Demitra}}, \bibinfo {author} {\bibfnamefont {V.}~\bibnamefont {Lollo}},
  \bibinfo {author} {\bibfnamefont {D.}~\bibnamefont {Pellegrini}}, \bibinfo
  {author} {\bibfnamefont {M.}~\bibnamefont {Pitti}}, \bibinfo {author}
  {\bibfnamefont {L.}~\bibnamefont {Pronti}}, \bibinfo {author} {\bibfnamefont
  {M.}~\bibnamefont {Romani}}, \bibinfo {author} {\bibfnamefont
  {M.}~\bibnamefont {Ferrario}}, \ and\ \bibinfo {author} {\bibfnamefont
  {A.}~\bibnamefont {Biagioni}}} (\bibinfo {year} {2025}),\ \href {\doibase
  10.1038/s41598-025-96882-y} {\bibfield  {journal} {\bibinfo  {journal} {Sci.
  Rep.}\ }\textbf {\bibinfo {volume} {15}},\ \bibinfo {pages}
  {12456}}\BibitemShut {NoStop}%
\bibitem [{\citenamefont {Curcio}\ \emph {et~al.}(2017)\citenamefont {Curcio}
  \emph {et~al.}}]{Curcio2017}%
  \BibitemOpen
  \bibfield  {author} {\bibinfo {author} {\bibnamefont {Curcio}, \bibfnamefont
  {A}},  \emph {et~al.}} (\bibinfo {year} {2017}),\ \href
  {https://doi.org/10.1103/physrevaccelbeams.20.012801} {\bibfield  {journal}
  {\bibinfo  {journal} {Phys. Rev. Accel. Beams}\ }\textbf {\bibinfo {volume}
  {20}},\ \bibinfo {pages} {012801}}\BibitemShut {NoStop}%
\bibitem [{\citenamefont {Dalichaouch}\ \emph {et~al.}(2020)\citenamefont
  {Dalichaouch}, \citenamefont {Xu}, \citenamefont {Li}, \citenamefont
  {Tableman}, \citenamefont {Tsung}, \citenamefont {An},\ and\ \citenamefont
  {Mori}}]{Dalichaouch2020}%
  \BibitemOpen
  \bibfield  {author} {\bibinfo {author} {\bibnamefont {Dalichaouch},
  \bibfnamefont {T~N}}, \bibinfo {author} {\bibfnamefont {X.~L.}\ \bibnamefont
  {Xu}}, \bibinfo {author} {\bibfnamefont {F.}~\bibnamefont {Li}}, \bibinfo
  {author} {\bibfnamefont {A.}~\bibnamefont {Tableman}}, \bibinfo {author}
  {\bibfnamefont {F.~S.}\ \bibnamefont {Tsung}}, \bibinfo {author}
  {\bibfnamefont {W.}~\bibnamefont {An}}, \ and\ \bibinfo {author}
  {\bibfnamefont {W.~B.}\ \bibnamefont {Mori}}} (\bibinfo {year} {2020}),\
  \href {https://doi.org/10.1103/physrevaccelbeams.23.021304} {\bibfield
  {journal} {\bibinfo  {journal} {Phys. Rev. Accel. Beams}\ }\textbf {\bibinfo
  {volume} {23}},\ \bibinfo {pages} {021304}}\BibitemShut {NoStop}%
\bibitem [{\citenamefont {Dalichaouch}\ \emph {et~al.}(2021)\citenamefont
  {Dalichaouch}, \citenamefont {Xu}, \citenamefont {Tableman}, \citenamefont
  {Li}, \citenamefont {Tsung},\ and\ \citenamefont {Mori}}]{Dalichaouch2021}%
  \BibitemOpen
  \bibfield  {author} {\bibinfo {author} {\bibnamefont {Dalichaouch},
  \bibfnamefont {T~N}}, \bibinfo {author} {\bibfnamefont {X.~L.}\ \bibnamefont
  {Xu}}, \bibinfo {author} {\bibfnamefont {A.}~\bibnamefont {Tableman}},
  \bibinfo {author} {\bibfnamefont {F.}~\bibnamefont {Li}}, \bibinfo {author}
  {\bibfnamefont {F.~S.}\ \bibnamefont {Tsung}}, \ and\ \bibinfo {author}
  {\bibfnamefont {W.~B.}\ \bibnamefont {Mori}}} (\bibinfo {year} {2021}),\
  \href {https://doi.org/10.1063/5.0051282} {\bibfield  {journal} {\bibinfo
  {journal} {Phys. Plasmas}\ }\textbf {\bibinfo {volume} {28}},\ \bibinfo
  {pages} {063103}}\BibitemShut {NoStop}%
\bibitem [{\citenamefont {D'Arcy}\ \emph
  {et~al.}(2019{\natexlab{a}})\citenamefont {D'Arcy} \emph
  {et~al.}}]{DArcy2019a}%
  \BibitemOpen
  \bibfield  {author} {\bibinfo {author} {\bibnamefont {D'Arcy}, \bibfnamefont
  {R}},  \emph {et~al.}} (\bibinfo {year} {2019}{\natexlab{a}}),\ \href
  {https://doi.org/10.1098/rsta.2018.0392} {\bibfield  {journal} {\bibinfo
  {journal} {Philos. Trans. R. Soc. A}\ }\textbf {\bibinfo {volume} {377}},\
  \bibinfo {pages} {20180392}}\BibitemShut {NoStop}%
\bibitem [{\citenamefont {D'Arcy}\ \emph
  {et~al.}(2019{\natexlab{b}})\citenamefont {D'Arcy} \emph
  {et~al.}}]{DArcy2019b}%
  \BibitemOpen
  \bibfield  {author} {\bibinfo {author} {\bibnamefont {D'Arcy}, \bibfnamefont
  {R}},  \emph {et~al.}} (\bibinfo {year} {2019}{\natexlab{b}}),\ \href
  {https://doi.org/10.1103/physrevlett.122.034801} {\bibfield  {journal}
  {\bibinfo  {journal} {Phys. Rev. Lett.}\ }\textbf {\bibinfo {volume} {122}},\
  \bibinfo {pages} {034801}}\BibitemShut {NoStop}%
\bibitem [{\citenamefont {D'Arcy}\ \emph {et~al.}(2022)\citenamefont {D'Arcy}
  \emph {et~al.}}]{DArcy2022}%
  \BibitemOpen
  \bibfield  {author} {\bibinfo {author} {\bibnamefont {D'Arcy}, \bibfnamefont
  {R}},  \emph {et~al.}} (\bibinfo {year} {2022}),\ \href
  {https://doi.org/10.1038/s41586-021-04348-8} {\bibfield  {journal} {\bibinfo
  {journal} {Nature (London)}\ }\textbf {\bibinfo {volume} {603}},\ \bibinfo
  {pages} {58--62}}\BibitemShut {NoStop}%
\bibitem [{\citenamefont {Davoine}\ \emph {et~al.}(2018)\citenamefont
  {Davoine}, \citenamefont {Fiúza}, \citenamefont {Fonseca}, \citenamefont
  {Mori},\ and\ \citenamefont {Silva}}]{davoine2018}%
  \BibitemOpen
  \bibfield  {author} {\bibinfo {author} {\bibnamefont {Davoine}, \bibfnamefont
  {X}}, \bibinfo {author} {\bibfnamefont {F.}~\bibnamefont {Fiúza}}, \bibinfo
  {author} {\bibfnamefont {R.~A.}\ \bibnamefont {Fonseca}}, \bibinfo {author}
  {\bibfnamefont {W.~B.}\ \bibnamefont {Mori}}, \ and\ \bibinfo {author}
  {\bibfnamefont {L.~O.}\ \bibnamefont {Silva}}} (\bibinfo {year} {2018}),\
  \href {https://doi.org/10.1017/S0022377818000429} {\bibfield  {journal}
  {\bibinfo  {journal} {J. Plasma Phys.}\ }\textbf {\bibinfo {volume} {84}},\
  \bibinfo {pages} {905840304}}\BibitemShut {NoStop}%
\bibitem [{\citenamefont {Dawson}(1959)}]{Dawson1959}%
  \BibitemOpen
  \bibfield  {author} {\bibinfo {author} {\bibnamefont {Dawson}, \bibfnamefont
  {J~M}}} (\bibinfo {year} {1959}),\ \href
  {https://doi.org/10.1103/physrev.113.383} {\bibfield  {journal} {\bibinfo
  {journal} {Phys. Rev.}\ }\textbf {\bibinfo {volume} {113}},\ \bibinfo {pages}
  {383--387}}\BibitemShut {NoStop}%
\bibitem [{\citenamefont {Dawson}(2001)}]{Dawson2001}%
  \BibitemOpen
  \bibfield  {author} {\bibinfo {author} {\bibnamefont {Dawson}, \bibfnamefont
  {J~M}}} (\bibinfo {year} {2001}),\ in\ \href
  {https://doi.org/10.1063/1.1384328} {\emph {\bibinfo {booktitle} {{AIP} Conf.
  Proc.}}},\ Vol.\ \bibinfo {volume} {569}\ (\bibinfo  {publisher} {{AIP}})\
  pp.\ \bibinfo {pages} {3--22}\BibitemShut {NoStop}%
\bibitem [{\citenamefont {Del~Dotto}\ \emph {et~al.}(2025)\citenamefont
  {Del~Dotto} \emph {et~al.}}]{DelDotto2025}%
  \BibitemOpen
  \bibfield  {author} {\bibinfo {author} {\bibnamefont {Del~Dotto},
  \bibfnamefont {A}},  \emph {et~al.}} (\bibinfo {year} {2025}),\ \href
  {\doibase 10.1088/1742-6596/3124/1/012003} {\bibfield  {journal} {\bibinfo
  {journal} {J. Phys.: Conf. Ser.}\ }\textbf {\bibinfo {volume} {3124}},\
  \bibinfo {pages} {012003}}\BibitemShut {NoStop}%
\bibitem [{\citenamefont {Delahaye}\ \emph {et~al.}(2014)\citenamefont
  {Delahaye}, \citenamefont {Adli}, \citenamefont {An}, \citenamefont
  {Gessner}, \citenamefont {Hogan}, \citenamefont {Joshi}, \citenamefont
  {Mori},\ and\ \citenamefont {Raubenheimer}}]{Delahaye2014}%
  \BibitemOpen
  \bibfield  {author} {\bibinfo {author} {\bibnamefont {Delahaye},
  \bibfnamefont {J-P}}, \bibinfo {author} {\bibfnamefont {E.}~\bibnamefont
  {Adli}}, \bibinfo {author} {\bibfnamefont {W.}~\bibnamefont {An}}, \bibinfo
  {author} {\bibfnamefont {S.}~\bibnamefont {Gessner}}, \bibinfo {author}
  {\bibfnamefont {M.}~\bibnamefont {Hogan}}, \bibinfo {author} {\bibfnamefont
  {C.}~\bibnamefont {Joshi}}, \bibinfo {author} {\bibfnamefont
  {W.}~\bibnamefont {Mori}}, \ and\ \bibinfo {author} {\bibfnamefont
  {T.}~\bibnamefont {Raubenheimer}}} (\bibinfo {year} {2014}),\ in\ \href
  {https://doi.org/10.18429/JACOW-IPAC2014-THPRI013} {\emph {\bibinfo
  {booktitle} {Proceedings of the 5th Int. Particle Accelerator Conf.}}}\
  (\bibinfo  {publisher} {JACoW},\ \bibinfo {address} {Geneva, Switzerland})\
  pp.\ \bibinfo {pages} {2755--2758}\BibitemShut {NoStop}%
\bibitem [{\citenamefont {Demeter}\ \emph {et~al.}(2021)\citenamefont {Demeter}
  \emph {et~al.}}]{Demeter2021}%
  \BibitemOpen
  \bibfield  {author} {\bibinfo {author} {\bibnamefont {Demeter}, \bibfnamefont
  {G}},  \emph {et~al.}} (\bibinfo {year} {2021}),\ \href
  {https://doi.org/10.1103/physreva.104.033506} {\bibfield  {journal} {\bibinfo
   {journal} {Phys. Rev. A}\ }\textbf {\bibinfo {volume} {104}},\ \bibinfo
  {pages} {033506}}\BibitemShut {NoStop}%
\bibitem [{\citenamefont {Deng}\ \emph {et~al.}(2012)\citenamefont {Deng},
  \citenamefont {Nakajima}, \citenamefont {Liu}, \citenamefont {Shen},
  \citenamefont {Zhang}, \citenamefont {Yu}, \citenamefont {Li}, \citenamefont
  {Li},\ and\ \citenamefont {Xu}}]{Deng2012}%
  \BibitemOpen
  \bibfield  {author} {\bibinfo {author} {\bibnamefont {Deng}, \bibfnamefont
  {A}}, \bibinfo {author} {\bibfnamefont {K.}~\bibnamefont {Nakajima}},
  \bibinfo {author} {\bibfnamefont {J.}~\bibnamefont {Liu}}, \bibinfo {author}
  {\bibfnamefont {B.}~\bibnamefont {Shen}}, \bibinfo {author} {\bibfnamefont
  {X.}~\bibnamefont {Zhang}}, \bibinfo {author} {\bibfnamefont
  {Y.}~\bibnamefont {Yu}}, \bibinfo {author} {\bibfnamefont {W.}~\bibnamefont
  {Li}}, \bibinfo {author} {\bibfnamefont {R.}~\bibnamefont {Li}}, \ and\
  \bibinfo {author} {\bibfnamefont {Z.}~\bibnamefont {Xu}}} (\bibinfo {year}
  {2012}),\ \href {https://doi.org/10.1103/physrevstab.15.081303} {\bibfield
  {journal} {\bibinfo  {journal} {Phys. Rev. ST Accel. Beams}\ }\textbf
  {\bibinfo {volume} {15}},\ \bibinfo {pages} {081303}}\BibitemShut {NoStop}%
\bibitem [{\citenamefont {Deng}\ \emph {et~al.}(2019)\citenamefont {Deng} \emph
  {et~al.}}]{Deng2019}%
  \BibitemOpen
  \bibfield  {author} {\bibinfo {author} {\bibnamefont {Deng}, \bibfnamefont
  {A}},  \emph {et~al.}} (\bibinfo {year} {2019}),\ \href
  {https://doi.org/10.1038/s41567-019-0610-9} {\bibfield  {journal} {\bibinfo
  {journal} {Nat. Phys.}\ }\textbf {\bibinfo {volume} {15}},\ \bibinfo {pages}
  {1156--1160}}\BibitemShut {NoStop}%
\bibitem [{\citenamefont {Diederichs}\ \emph {et~al.}(2020)\citenamefont
  {Diederichs}, \citenamefont {Benedetti}, \citenamefont {Esarey},
  \citenamefont {Osterhoff},\ and\ \citenamefont {Schroeder}}]{Diederichs2020}%
  \BibitemOpen
  \bibfield  {author} {\bibinfo {author} {\bibnamefont {Diederichs},
  \bibfnamefont {S}}, \bibinfo {author} {\bibfnamefont {C.}~\bibnamefont
  {Benedetti}}, \bibinfo {author} {\bibfnamefont {E.}~\bibnamefont {Esarey}},
  \bibinfo {author} {\bibfnamefont {J.}~\bibnamefont {Osterhoff}}, \ and\
  \bibinfo {author} {\bibfnamefont {C.~B.}\ \bibnamefont {Schroeder}}}
  (\bibinfo {year} {2020}),\ \href
  {https://link.aps.org/doi/10.1103/PhysRevAccelBeams.23.121301} {\bibfield
  {journal} {\bibinfo  {journal} {Phys. Rev. Accel. Beams}\ }\textbf {\bibinfo
  {volume} {23}},\ \bibinfo {pages} {121301}}\BibitemShut {NoStop}%
\bibitem [{\citenamefont {Diederichs}\ \emph
  {et~al.}(2022{\natexlab{a}})\citenamefont {Diederichs}, \citenamefont
  {Benedetti}, \citenamefont {Esarey}, \citenamefont {Th\'evenet},
  \citenamefont {Osterhoff},\ and\ \citenamefont
  {Schroeder}}]{Diederichs2022b}%
  \BibitemOpen
  \bibfield  {author} {\bibinfo {author} {\bibnamefont {Diederichs},
  \bibfnamefont {S}}, \bibinfo {author} {\bibfnamefont {C.}~\bibnamefont
  {Benedetti}}, \bibinfo {author} {\bibfnamefont {E.}~\bibnamefont {Esarey}},
  \bibinfo {author} {\bibfnamefont {M.}~\bibnamefont {Th\'evenet}}, \bibinfo
  {author} {\bibfnamefont {J.}~\bibnamefont {Osterhoff}}, \ and\ \bibinfo
  {author} {\bibfnamefont {C.~B.}\ \bibnamefont {Schroeder}}} (\bibinfo {year}
  {2022}{\natexlab{a}}),\ \href {https://doi.org/10.1063/5.0087807} {\bibfield
  {journal} {\bibinfo  {journal} {Phys. Plasmas}\ }\textbf {\bibinfo {volume}
  {29}},\ \bibinfo {pages} {043101}}\BibitemShut {NoStop}%
\bibitem [{\citenamefont {Diederichs}\ \emph {et~al.}(2023)\citenamefont
  {Diederichs}, \citenamefont {Benedetti}, \citenamefont {Esarey},
  \citenamefont {Th\'evenet}, \citenamefont {Sinn}, \citenamefont {Osterhoff},\
  and\ \citenamefont {Schroeder}}]{Diederichs2023}%
  \BibitemOpen
  \bibfield  {author} {\bibinfo {author} {\bibnamefont {Diederichs},
  \bibfnamefont {S}}, \bibinfo {author} {\bibfnamefont {C.}~\bibnamefont
  {Benedetti}}, \bibinfo {author} {\bibfnamefont {E.}~\bibnamefont {Esarey}},
  \bibinfo {author} {\bibfnamefont {M.}~\bibnamefont {Th\'evenet}}, \bibinfo
  {author} {\bibfnamefont {A.}~\bibnamefont {Sinn}}, \bibinfo {author}
  {\bibfnamefont {J.}~\bibnamefont {Osterhoff}}, \ and\ \bibinfo {author}
  {\bibfnamefont {C.~B.}\ \bibnamefont {Schroeder}}} (\bibinfo {year} {2023}),\
  \href {https://doi.org/10.1063/5.0155489} {\bibfield  {journal} {\bibinfo
  {journal} {Phys. Plasmas}\ }\textbf {\bibinfo {volume} {30}},\ \bibinfo
  {pages} {073104}}\BibitemShut {NoStop}%
\bibitem [{\citenamefont {Diederichs}\ \emph {et~al.}(2024)\citenamefont
  {Diederichs}, \citenamefont {Benedetti}, \citenamefont {Ferran~Pousa},
  \citenamefont {Sinn}, \citenamefont {Osterhoff}, \citenamefont {Schroeder},\
  and\ \citenamefont {Thévenet}}]{Diederichs2024}%
  \BibitemOpen
  \bibfield  {author} {\bibinfo {author} {\bibnamefont {Diederichs},
  \bibfnamefont {S}}, \bibinfo {author} {\bibfnamefont {C.}~\bibnamefont
  {Benedetti}}, \bibinfo {author} {\bibfnamefont {A.}~\bibnamefont
  {Ferran~Pousa}}, \bibinfo {author} {\bibfnamefont {A.}~\bibnamefont {Sinn}},
  \bibinfo {author} {\bibfnamefont {J.}~\bibnamefont {Osterhoff}}, \bibinfo
  {author} {\bibfnamefont {C.~B.}\ \bibnamefont {Schroeder}}, \ and\ \bibinfo
  {author} {\bibfnamefont {M.}~\bibnamefont {Thévenet}}} (\bibinfo {year}
  {2024}),\ \href {https://doi.org/10.1103/PhysRevLett.133.265003} {\bibfield
  {journal} {\bibinfo  {journal} {Phys. Rev. Lett.}\ }\textbf {\bibinfo
  {volume} {133}},\ \bibinfo {pages} {265003}}\BibitemShut {NoStop}%
\bibitem [{\citenamefont {Diederichs}\ \emph
  {et~al.}(2022{\natexlab{b}})\citenamefont {Diederichs}, \citenamefont
  {Benedetti}, \citenamefont {Huebl}, \citenamefont {Lehe}, \citenamefont
  {Myers}, \citenamefont {Sinn}, \citenamefont {Vay}, \citenamefont {Zhang},\
  and\ \citenamefont {Thévenet}}]{Diederichs2022c}%
  \BibitemOpen
  \bibfield  {author} {\bibinfo {author} {\bibnamefont {Diederichs},
  \bibfnamefont {S}}, \bibinfo {author} {\bibfnamefont {C.}~\bibnamefont
  {Benedetti}}, \bibinfo {author} {\bibfnamefont {A.}~\bibnamefont {Huebl}},
  \bibinfo {author} {\bibfnamefont {R.}~\bibnamefont {Lehe}}, \bibinfo {author}
  {\bibfnamefont {A.}~\bibnamefont {Myers}}, \bibinfo {author} {\bibfnamefont
  {A.}~\bibnamefont {Sinn}}, \bibinfo {author} {\bibfnamefont {J.-L.}\
  \bibnamefont {Vay}}, \bibinfo {author} {\bibfnamefont {W.}~\bibnamefont
  {Zhang}}, \ and\ \bibinfo {author} {\bibfnamefont {M.}~\bibnamefont
  {Thévenet}}} (\bibinfo {year} {2022}{\natexlab{b}}),\ \href
  {https://doi.org/10.1016/j.cpc.2022.108421} {\bibfield  {journal} {\bibinfo
  {journal} {Comput. Phys. Commun.}\ }\textbf {\bibinfo {volume} {278}},\
  \bibinfo {pages} {108421}}\BibitemShut {NoStop}%
\bibitem [{\citenamefont {Diederichs}\ \emph
  {et~al.}(2022{\natexlab{c}})\citenamefont {Diederichs}, \citenamefont
  {Benedetti}, \citenamefont {Th\'evenet}, \citenamefont {Esarey},
  \citenamefont {Osterhoff},\ and\ \citenamefont
  {Schroeder}}]{Diederichs2022a}%
  \BibitemOpen
  \bibfield  {author} {\bibinfo {author} {\bibnamefont {Diederichs},
  \bibfnamefont {S}}, \bibinfo {author} {\bibfnamefont {C.}~\bibnamefont
  {Benedetti}}, \bibinfo {author} {\bibfnamefont {M.}~\bibnamefont
  {Th\'evenet}}, \bibinfo {author} {\bibfnamefont {E.}~\bibnamefont {Esarey}},
  \bibinfo {author} {\bibfnamefont {J.}~\bibnamefont {Osterhoff}}, \ and\
  \bibinfo {author} {\bibfnamefont {C.~B.}\ \bibnamefont {Schroeder}}}
  (\bibinfo {year} {2022}{\natexlab{c}}),\ \href
  {https://doi.org/10.1103/PhysRevAccelBeams.25.091304} {\bibfield  {journal}
  {\bibinfo  {journal} {Phys. Rev. Accel. Beams}\ }\textbf {\bibinfo {volume}
  {25}},\ \bibinfo {pages} {091304}}\BibitemShut {NoStop}%
\bibitem [{\citenamefont {Diederichs}\ \emph {et~al.}(2019)\citenamefont
  {Diederichs}, \citenamefont {Mehrling}, \citenamefont {Benedetti},
  \citenamefont {Schroeder}, \citenamefont {Knetsch}, \citenamefont {Esarey},\
  and\ \citenamefont {Osterhoff}}]{Diederichs2019}%
  \BibitemOpen
  \bibfield  {author} {\bibinfo {author} {\bibnamefont {Diederichs},
  \bibfnamefont {S}}, \bibinfo {author} {\bibfnamefont {T.~J.}\ \bibnamefont
  {Mehrling}}, \bibinfo {author} {\bibfnamefont {C.}~\bibnamefont {Benedetti}},
  \bibinfo {author} {\bibfnamefont {C.~B.}\ \bibnamefont {Schroeder}}, \bibinfo
  {author} {\bibfnamefont {A.}~\bibnamefont {Knetsch}}, \bibinfo {author}
  {\bibfnamefont {E.}~\bibnamefont {Esarey}}, \ and\ \bibinfo {author}
  {\bibfnamefont {J.}~\bibnamefont {Osterhoff}}} (\bibinfo {year} {2019}),\
  \href {https://doi.org/10.1103/physrevaccelbeams.22.081301} {\bibfield
  {journal} {\bibinfo  {journal} {Phys. Rev. Accel. Beams}\ }\textbf {\bibinfo
  {volume} {22}},\ \bibinfo {pages} {081301}}\BibitemShut {NoStop}%
\bibitem [{\citenamefont {Doche}(2018)}]{Doche2018}%
  \BibitemOpen
  \bibfield  {author} {\bibinfo {author} {\bibnamefont {Doche}, \bibfnamefont
  {A}}} (\bibinfo {year} {2018}),\ \href
  {https://pastel.archives-ouvertes.fr/tel-01767745} {\bibinfo {type} {Ph{D}
  {T}hesis}}\ (\bibinfo  {school} {{Universit{\'e} Paris Saclay}})\BibitemShut
  {NoStop}%
\bibitem [{\citenamefont {Doche}\ \emph {et~al.}(2017)\citenamefont {Doche}
  \emph {et~al.}}]{Doche2017}%
  \BibitemOpen
  \bibfield  {author} {\bibinfo {author} {\bibnamefont {Doche}, \bibfnamefont
  {A}},  \emph {et~al.}} (\bibinfo {year} {2017}),\ \href
  {https://doi.org/10.1038/s41598-017-14524-4} {\bibfield  {journal} {\bibinfo
  {journal} {Sci. Rep.}\ }\textbf {\bibinfo {volume} {7}},\ \bibinfo {pages}
  {14180}}\BibitemShut {NoStop}%
\bibitem [{\citenamefont {D\"{o}pp}\ \emph {et~al.}(2018)\citenamefont
  {D\"{o}pp} \emph {et~al.}}]{Dopp2018}%
  \BibitemOpen
  \bibfield  {author} {\bibinfo {author} {\bibnamefont {D\"{o}pp},
  \bibfnamefont {A}},  \emph {et~al.}} (\bibinfo {year} {2018}),\ \href
  {https://doi.org/10.1103/PhysRevLett.121.074802} {\bibfield  {journal}
  {\bibinfo  {journal} {Phys. Rev. Lett.}\ }\textbf {\bibinfo {volume} {121}},\
  \bibinfo {pages} {074802}}\BibitemShut {NoStop}%
\bibitem [{\citenamefont {Dornmair}\ \emph {et~al.}(2015)\citenamefont
  {Dornmair}, \citenamefont {Floettmann},\ and\ \citenamefont
  {Maier}}]{Dornmair2015}%
  \BibitemOpen
  \bibfield  {author} {\bibinfo {author} {\bibnamefont {Dornmair},
  \bibfnamefont {I}}, \bibinfo {author} {\bibfnamefont {K.}~\bibnamefont
  {Floettmann}}, \ and\ \bibinfo {author} {\bibfnamefont {A.~R.}\ \bibnamefont
  {Maier}}} (\bibinfo {year} {2015}),\ \href
  {https://doi.org/10.1103/physrevstab.18.041302} {\bibfield  {journal}
  {\bibinfo  {journal} {Phys. Rev. ST Accel. Beams}\ }\textbf {\bibinfo
  {volume} {18}},\ \bibinfo {pages} {041302}}\BibitemShut {NoStop}%
\bibitem [{\citenamefont {Doss}\ \emph {et~al.}(2019)\citenamefont {Doss} \emph
  {et~al.}}]{Doss2019}%
  \BibitemOpen
  \bibfield  {author} {\bibinfo {author} {\bibnamefont {Doss}, \bibfnamefont
  {C~E}},  \emph {et~al.}} (\bibinfo {year} {2019}),\ \href
  {https://doi.org/10.1103/PhysRevAccelBeams.22.111001} {\bibfield  {journal}
  {\bibinfo  {journal} {Phys. Rev. Accel. Beams}\ }\textbf {\bibinfo {volume}
  {22}},\ \bibinfo {pages} {111001}}\BibitemShut {NoStop}%
\bibitem [{\citenamefont {Doss}\ \emph {et~al.}(2023)\citenamefont {Doss} \emph
  {et~al.}}]{Doss2023}%
  \BibitemOpen
  \bibfield  {author} {\bibinfo {author} {\bibnamefont {Doss}, \bibfnamefont
  {C~E}},  \emph {et~al.}} (\bibinfo {year} {2023}),\ \href {\doibase
  10.1103/physrevaccelbeams.26.031302} {\bibfield  {journal} {\bibinfo
  {journal} {Phys. Rev. Accel. Beams}\ }\textbf {\bibinfo {volume} {26}},\
  \bibinfo {pages} {031302}}\BibitemShut {NoStop}%
\bibitem [{\citenamefont {Downer}\ \emph {et~al.}(2018)\citenamefont {Downer},
  \citenamefont {Zgadzaj}, \citenamefont {Debus}, \citenamefont {Schramm},\
  and\ \citenamefont {Kaluza}}]{Downer2018}%
  \BibitemOpen
  \bibfield  {author} {\bibinfo {author} {\bibnamefont {Downer}, \bibfnamefont
  {M~C}}, \bibinfo {author} {\bibfnamefont {R.}~\bibnamefont {Zgadzaj}},
  \bibinfo {author} {\bibfnamefont {A.}~\bibnamefont {Debus}}, \bibinfo
  {author} {\bibfnamefont {U.}~\bibnamefont {Schramm}}, \ and\ \bibinfo
  {author} {\bibfnamefont {M.~C.}\ \bibnamefont {Kaluza}}} (\bibinfo {year}
  {2018}),\ \href {https://doi.org/10.1103/revmodphys.90.035002} {\bibfield
  {journal} {\bibinfo  {journal} {Rev. Mod. Phys.}\ }\textbf {\bibinfo {volume}
  {90}},\ \bibinfo {pages} {035002}}\BibitemShut {NoStop}%
\bibitem [{\citenamefont {Drobniak}\ \emph {et~al.}(2025)\citenamefont
  {Drobniak}, \citenamefont {Adli}, \citenamefont {Anderson}, \citenamefont
  {Dyson}, \citenamefont {Mewes}, \citenamefont {Sjobak}, \citenamefont
  {Th{\'e}venet},\ and\ \citenamefont {Lindstr{\o}m}}]{Drobniak2025}%
  \BibitemOpen
  \bibfield  {author} {\bibinfo {author} {\bibnamefont {Drobniak},
  \bibfnamefont {P}}, \bibinfo {author} {\bibfnamefont {E.}~\bibnamefont
  {Adli}}, \bibinfo {author} {\bibfnamefont {H.~Bergravf}\ \bibnamefont
  {Anderson}}, \bibinfo {author} {\bibfnamefont {A.}~\bibnamefont {Dyson}},
  \bibinfo {author} {\bibfnamefont {S.~M.}\ \bibnamefont {Mewes}}, \bibinfo
  {author} {\bibfnamefont {K.~N.}\ \bibnamefont {Sjobak}}, \bibinfo {author}
  {\bibfnamefont {M.}~\bibnamefont {Th{\'e}venet}}, \ and\ \bibinfo {author}
  {\bibfnamefont {C.~A.}\ \bibnamefont {Lindstr{\o}m}}} (\bibinfo {year}
  {2025}),\ \href {https://doi.org/10.1016/j.nima.2025.170223} {\bibfield
  {journal} {\bibinfo  {journal} {Nucl. Instrum. Methods Phys. Res. A}\
  }\textbf {\bibinfo {volume} {1072}},\ \bibinfo {pages} {170223}}\BibitemShut
  {NoStop}%
\bibitem [{\citenamefont {Emma}\ \emph
  {et~al.}(2021{\natexlab{a}})\citenamefont {Emma}, \citenamefont {Xu},
  \citenamefont {Fisher}, \citenamefont {Robles}, \citenamefont {MacArthur},
  \citenamefont {Cryan}, \citenamefont {Hogan}, \citenamefont {Musumeci},
  \citenamefont {White},\ and\ \citenamefont {Marinelli}}]{Emma2021a}%
  \BibitemOpen
  \bibfield  {author} {\bibinfo {author} {\bibnamefont {Emma}, \bibfnamefont
  {C}}, \bibinfo {author} {\bibfnamefont {X.}~\bibnamefont {Xu}}, \bibinfo
  {author} {\bibfnamefont {A.}~\bibnamefont {Fisher}}, \bibinfo {author}
  {\bibfnamefont {R.}~\bibnamefont {Robles}}, \bibinfo {author} {\bibfnamefont
  {J.~P.}\ \bibnamefont {MacArthur}}, \bibinfo {author} {\bibfnamefont
  {J.}~\bibnamefont {Cryan}}, \bibinfo {author} {\bibfnamefont {M.~J.}\
  \bibnamefont {Hogan}}, \bibinfo {author} {\bibfnamefont {P.}~\bibnamefont
  {Musumeci}}, \bibinfo {author} {\bibfnamefont {G.}~\bibnamefont {White}}, \
  and\ \bibinfo {author} {\bibfnamefont {A.}~\bibnamefont {Marinelli}}}
  (\bibinfo {year} {2021}{\natexlab{a}}),\ \href
  {https://doi.org/10.1063/5.0050693} {\bibfield  {journal} {\bibinfo
  {journal} {APL Photonics}\ }\textbf {\bibinfo {volume} {6}},\ \bibinfo
  {pages} {076107}}\BibitemShut {NoStop}%
\bibitem [{\citenamefont {Emma}\ \emph
  {et~al.}(2021{\natexlab{b}})\citenamefont {Emma} \emph {et~al.}}]{Emma2021c}%
  \BibitemOpen
  \bibfield  {author} {\bibinfo {author} {\bibnamefont {Emma}, \bibfnamefont
  {C}},  \emph {et~al.}} (\bibinfo {year} {2021}{\natexlab{b}}),\ \href
  {https://doi.org/10.1017/hpl.2021.39} {\bibfield  {journal} {\bibinfo
  {journal} {High Pow. Laser Sci. Eng.}\ }\textbf {\bibinfo {volume} {9}},\
  \bibinfo {pages} {e57}}\BibitemShut {NoStop}%
\bibitem [{\citenamefont {Emma}\ \emph
  {et~al.}(2021{\natexlab{c}})\citenamefont {Emma} \emph {et~al.}}]{Emma2021b}%
  \BibitemOpen
  \bibfield  {author} {\bibinfo {author} {\bibnamefont {Emma}, \bibfnamefont
  {C}},  \emph {et~al.}} (\bibinfo {year} {2021}{\natexlab{c}}),\ in\ \href
  {https://doi.org/10.18429/JACOW-IPAC2021-WEPAB072} {\emph {\bibinfo
  {booktitle} {Proceedings of the 12th Int. Particle Accelerator Conf.}}}\
  (\bibinfo  {publisher} {JACoW},\ \bibinfo {address} {Geneva, Switzerland})\
  pp.\ \bibinfo {pages} {2755--2758}\BibitemShut {NoStop}%
\bibitem [{\citenamefont {Emma}\ \emph {et~al.}(2025)\citenamefont {Emma} \emph
  {et~al.}}]{Emma2025}%
  \BibitemOpen
  \bibfield  {author} {\bibinfo {author} {\bibnamefont {Emma}, \bibfnamefont
  {C}},  \emph {et~al.}} (\bibinfo {year} {2025}),\ \href
  {https://doi.org/10.1103/PhysRevLett.134.085001} {\bibfield  {journal}
  {\bibinfo  {journal} {Phys. Rev. Lett.}\ }\textbf {\bibinfo {volume} {134}},\
  \bibinfo {pages} {085001}}\BibitemShut {NoStop}%
\bibitem [{\citenamefont {Emma}\ \emph {et~al.}(2014)\citenamefont {Emma} \emph
  {et~al.}}]{Emma2014}%
  \BibitemOpen
  \bibfield  {author} {\bibinfo {author} {\bibnamefont {Emma}, \bibfnamefont
  {P}},  \emph {et~al.}} (\bibinfo {year} {2014}),\ \href
  {https://doi.org/10.1103/PhysRevLett.112.034801} {\bibfield  {journal}
  {\bibinfo  {journal} {Phys. Rev. Lett.}\ }\textbf {\bibinfo {volume} {112}},\
  \bibinfo {pages} {034801}}\BibitemShut {NoStop}%
\bibitem [{\citenamefont {US~Department~of Energy}(2016)}]{2016AACreport}%
  \BibitemOpen
  \bibfield  {author} {\bibinfo {author} {\bibnamefont {US~Department~of
  Energy}, \bibfnamefont {Office of~Science}}} (\bibinfo {year} {2016}),\ \href
  {https://doi.org/10.2172/1358081} {\emph {\bibinfo {title} {Advanced
  Accelerator Development Strategy Report: DOE Advanced Accelerator Concepts
  Research Roadmap Workshop}}},\ \bibinfo {type} {Tech. Rep.}\BibitemShut
  {Stop}%
\bibitem [{\citenamefont {Engel}\ \emph {et~al.}(2020)\citenamefont {Engel},
  \citenamefont {Gross}, \citenamefont {Koss}, \citenamefont {Lishilin},
  \citenamefont {Loisch}, \citenamefont {Philipp}, \citenamefont {Richter},\
  and\ \citenamefont {Stephan}}]{Engel2020}%
  \BibitemOpen
  \bibfield  {author} {\bibinfo {author} {\bibnamefont {Engel}, \bibfnamefont
  {J}}, \bibinfo {author} {\bibfnamefont {M.}~\bibnamefont {Gross}}, \bibinfo
  {author} {\bibfnamefont {G.}~\bibnamefont {Koss}}, \bibinfo {author}
  {\bibfnamefont {O.}~\bibnamefont {Lishilin}}, \bibinfo {author}
  {\bibfnamefont {G.}~\bibnamefont {Loisch}}, \bibinfo {author} {\bibfnamefont
  {S.}~\bibnamefont {Philipp}}, \bibinfo {author} {\bibfnamefont
  {D.}~\bibnamefont {Richter}}, \ and\ \bibinfo {author} {\bibfnamefont
  {F.}~\bibnamefont {Stephan}}} (\bibinfo {year} {2020}),\ \href {\doibase
  10.1063/1.5143258} {\bibfield  {journal} {\bibinfo  {journal} {AIP Adv.}\
  }\textbf {\bibinfo {volume} {10}},\ \bibinfo {pages} {025224}}\BibitemShut
  {NoStop}%
\bibitem [{\citenamefont {England}\ \emph {et~al.}(2008)\citenamefont
  {England}, \citenamefont {Rosenzweig},\ and\ \citenamefont
  {Travish}}]{England2008}%
  \BibitemOpen
  \bibfield  {author} {\bibinfo {author} {\bibnamefont {England}, \bibfnamefont
  {R~J}}, \bibinfo {author} {\bibfnamefont {J.~B.}\ \bibnamefont {Rosenzweig}},
  \ and\ \bibinfo {author} {\bibfnamefont {G.}~\bibnamefont {Travish}}}
  (\bibinfo {year} {2008}),\ \href
  {https://doi.org/10.1103/PhysRevLett.100.214802} {\bibfield  {journal}
  {\bibinfo  {journal} {Phys. Rev. Lett.}\ }\textbf {\bibinfo {volume} {100}},\
  \bibinfo {pages} {214802}}\BibitemShut {NoStop}%
\bibitem [{\citenamefont {Ersfeld}\ \emph {et~al.}(2014)\citenamefont
  {Ersfeld}, \citenamefont {Bonifacio}, \citenamefont {Chen}, \citenamefont
  {Islam}, \citenamefont {Smorenburg},\ and\ \citenamefont
  {Jaroszynski}}]{Ersfeld2014}%
  \BibitemOpen
  \bibfield  {author} {\bibinfo {author} {\bibnamefont {Ersfeld}, \bibfnamefont
  {B}}, \bibinfo {author} {\bibfnamefont {R}~\bibnamefont {Bonifacio}},
  \bibinfo {author} {\bibfnamefont {S}~\bibnamefont {Chen}}, \bibinfo {author}
  {\bibfnamefont {M~R}\ \bibnamefont {Islam}}, \bibinfo {author} {\bibfnamefont
  {P~W}\ \bibnamefont {Smorenburg}}, \ and\ \bibinfo {author} {\bibfnamefont
  {D~A}\ \bibnamefont {Jaroszynski}}} (\bibinfo {year} {2014}),\ \href
  {https://doi.org/10.1088/1367-2630/16/9/093025} {\bibfield  {journal}
  {\bibinfo  {journal} {New J. Phys.}\ }\textbf {\bibinfo {volume} {16}},\
  \bibinfo {pages} {093025}}\BibitemShut {NoStop}%
\bibitem [{\citenamefont {Esarey}\ \emph {et~al.}(2009)\citenamefont {Esarey},
  \citenamefont {Schroeder},\ and\ \citenamefont {Leemans}}]{Esarey2009}%
  \BibitemOpen
  \bibfield  {author} {\bibinfo {author} {\bibnamefont {Esarey}, \bibfnamefont
  {E}}, \bibinfo {author} {\bibfnamefont {C.~B.}\ \bibnamefont {Schroeder}}, \
  and\ \bibinfo {author} {\bibfnamefont {W.~P.}\ \bibnamefont {Leemans}}}
  (\bibinfo {year} {2009}),\ \href {https://doi.org/10.1103/revmodphys.81.1229}
  {\bibfield  {journal} {\bibinfo  {journal} {Rev. Mod. Phys.}\ }\textbf
  {\bibinfo {volume} {81}},\ \bibinfo {pages} {1229--1285}}\BibitemShut
  {NoStop}%
\bibitem [{\citenamefont {Esarey}\ \emph {et~al.}(2002)\citenamefont {Esarey},
  \citenamefont {Shadwick}, \citenamefont {Catravas},\ and\ \citenamefont
  {Leemans}}]{Esarey2002}%
  \BibitemOpen
  \bibfield  {author} {\bibinfo {author} {\bibnamefont {Esarey}, \bibfnamefont
  {E}}, \bibinfo {author} {\bibfnamefont {B.~A.}\ \bibnamefont {Shadwick}},
  \bibinfo {author} {\bibfnamefont {P.}~\bibnamefont {Catravas}}, \ and\
  \bibinfo {author} {\bibfnamefont {W.~P.}\ \bibnamefont {Leemans}}} (\bibinfo
  {year} {2002}),\ \href {https://doi.org/10.1103/physreve.65.056505}
  {\bibfield  {journal} {\bibinfo  {journal} {Phys. Rev. E}\ }\textbf {\bibinfo
  {volume} {65}},\ \bibinfo {pages} {056505}}\BibitemShut {NoStop}%
\bibitem [{\citenamefont {FACET}(2016)}]{FACET2016}%
  \BibitemOpen
  \bibfield  {author} {\bibinfo {author} {\bibnamefont {FACET},}} (\bibinfo
  {year} {2016}),\ \href {\doibase 10.2172/1340171} {\emph {\bibinfo {title}
  {Technical Design Report for the FACET-II Project at SLAC National
  Accelerator Laboratory}}}\BibitemShut {NoStop}%
\bibitem [{\citenamefont {Fa{\u{\i}}nberg}(1956)}]{Fainberg1956}%
  \BibitemOpen
  \bibfield  {author} {\bibinfo {author} {\bibnamefont {Fa{\u{\i}}nberg},
  \bibfnamefont {Ya~B}}} (\bibinfo {year} {1956}),\ in\ \href
  {http://cds.cern.ch/record/1241564} {\emph {\bibinfo {booktitle} {CERN
  Symposium on High Energy Accelerators and Pion Physics}}}\ (\bibinfo
  {publisher} {CERN})\ pp.\ \bibinfo {pages} {84--90}\BibitemShut {NoStop}%
\bibitem [{\citenamefont {Fa{\u{\i}}nberg}(1960)}]{Fainberg1960}%
  \BibitemOpen
  \bibfield  {author} {\bibinfo {author} {\bibnamefont {Fa{\u{\i}}nberg},
  \bibfnamefont {Ya~B}}} (\bibinfo {year} {1960}),\ \href
  {https://doi.org/10.1007/BF01479735} {\bibfield  {journal} {\bibinfo
  {journal} {Sov. At. Energy.}\ }\textbf {\bibinfo {volume} {6}},\ \bibinfo
  {pages} {297--309}}\BibitemShut {NoStop}%
\bibitem [{\citenamefont {Fa{\u{\i}}nberg}(1968)}]{Fainberg1968}%
  \BibitemOpen
  \bibfield  {author} {\bibinfo {author} {\bibnamefont {Fa{\u{\i}}nberg},
  \bibfnamefont {Ya~B}}} (\bibinfo {year} {1968}),\ \href
  {https://doi.org/10.1070/pu1968v010n06abeh003715} {\bibfield  {journal}
  {\bibinfo  {journal} {Sov. Phys. Uspekhi}\ }\textbf {\bibinfo {volume}
  {10}},\ \bibinfo {pages} {750--758}}\BibitemShut {NoStop}%
\bibitem [{\citenamefont {Fan}\ \emph {et~al.}(2022)\citenamefont {Fan} \emph
  {et~al.}}]{Fan2022}%
  \BibitemOpen
  \bibfield  {author} {\bibinfo {author} {\bibnamefont {Fan}, \bibfnamefont
  {H~C}},  \emph {et~al.}} (\bibinfo {year} {2022}),\ \href
  {https://doi.org/10.1088/1367-2630/ac8951} {\bibfield  {journal} {\bibinfo
  {journal} {New J. Phys.}\ }\textbf {\bibinfo {volume} {24}},\ \bibinfo
  {pages} {083047}}\BibitemShut {NoStop}%
\bibitem [{\citenamefont {Fan}\ \emph {et~al.}(2000)\citenamefont {Fan},
  \citenamefont {Parra}, \citenamefont {Alexeev}, \citenamefont {Kim},
  \citenamefont {Milchberg}, \citenamefont {Margolin},\ and\ \citenamefont
  {Pyatnitskii}}]{Fan2000}%
  \BibitemOpen
  \bibfield  {author} {\bibinfo {author} {\bibnamefont {Fan}, \bibfnamefont
  {J}}, \bibinfo {author} {\bibfnamefont {E.}~\bibnamefont {Parra}}, \bibinfo
  {author} {\bibfnamefont {I.}~\bibnamefont {Alexeev}}, \bibinfo {author}
  {\bibfnamefont {K.~Y.}\ \bibnamefont {Kim}}, \bibinfo {author} {\bibfnamefont
  {H.~M.}\ \bibnamefont {Milchberg}}, \bibinfo {author} {\bibfnamefont
  {L.~Ya.}\ \bibnamefont {Margolin}}, \ and\ \bibinfo {author} {\bibfnamefont
  {L.~N.}\ \bibnamefont {Pyatnitskii}}} (\bibinfo {year} {2000}),\ \href
  {https://doi.org/10.1103/physreve.62.r7603} {\bibfield  {journal} {\bibinfo
  {journal} {Phys. Rev. E}\ }\textbf {\bibinfo {volume} {62}},\ \bibinfo
  {pages} {R7603--R7606}}\BibitemShut {NoStop}%
\bibitem [{\citenamefont {Fang}\ \emph {et~al.}(2011)\citenamefont {Fang},
  \citenamefont {Muggli}, \citenamefont {Huang},\ and\ \citenamefont
  {Mori}}]{Fang2011}%
  \BibitemOpen
  \bibfield  {author} {\bibinfo {author} {\bibnamefont {Fang}, \bibfnamefont
  {Y}}, \bibinfo {author} {\bibfnamefont {P.}~\bibnamefont {Muggli}}, \bibinfo
  {author} {\bibfnamefont {C.}~\bibnamefont {Huang}}, \ and\ \bibinfo {author}
  {\bibfnamefont {W.}~\bibnamefont {Mori}}} (\bibinfo {year} {2011}),\ in\
  \href {https://proceedings.jacow.org/PAC2011/papers/mop158.pdf} {\emph
  {\bibinfo {booktitle} {Proceedings of the 2011 Particle Accelerator Conf.}}}\
  (\bibinfo  {publisher} {IEEE},\ \bibinfo {address} {New York, NY, United
  States})\ pp.\ \bibinfo {pages} {244--246}\BibitemShut {NoStop}%
\bibitem [{\citenamefont {Fang}\ \emph {et~al.}(2014)\citenamefont {Fang},
  \citenamefont {Yakimenko}, \citenamefont {Babzien}, \citenamefont {Fedurin},
  \citenamefont {Kusche}, \citenamefont {Malone}, \citenamefont {Vieira},
  \citenamefont {Mori},\ and\ \citenamefont {Muggli}}]{Fang2014}%
  \BibitemOpen
  \bibfield  {author} {\bibinfo {author} {\bibnamefont {Fang}, \bibfnamefont
  {Y}}, \bibinfo {author} {\bibfnamefont {V.~E.}\ \bibnamefont {Yakimenko}},
  \bibinfo {author} {\bibfnamefont {M.}~\bibnamefont {Babzien}}, \bibinfo
  {author} {\bibfnamefont {M.}~\bibnamefont {Fedurin}}, \bibinfo {author}
  {\bibfnamefont {K.~P.}\ \bibnamefont {Kusche}}, \bibinfo {author}
  {\bibfnamefont {R.}~\bibnamefont {Malone}}, \bibinfo {author} {\bibfnamefont
  {J.}~\bibnamefont {Vieira}}, \bibinfo {author} {\bibfnamefont {W.~B.}\
  \bibnamefont {Mori}}, \ and\ \bibinfo {author} {\bibfnamefont
  {P.}~\bibnamefont {Muggli}}} (\bibinfo {year} {2014}),\ \href
  {https://doi.org/10.1103/physrevlett.112.045001} {\bibfield  {journal}
  {\bibinfo  {journal} {Phys. Rev. Lett.}\ }\textbf {\bibinfo {volume} {112}},\
  \bibinfo {pages} {045001}}\BibitemShut {NoStop}%
\bibitem [{\citenamefont {Farmer}\ \emph {et~al.}(2024)\citenamefont {Farmer},
  \citenamefont {Caldwell},\ and\ \citenamefont {Pukhov}}]{Farmer2024}%
  \BibitemOpen
  \bibfield  {author} {\bibinfo {author} {\bibnamefont {Farmer}, \bibfnamefont
  {J}}, \bibinfo {author} {\bibfnamefont {A.}~\bibnamefont {Caldwell}}, \ and\
  \bibinfo {author} {\bibfnamefont {A.}~\bibnamefont {Pukhov}}} (\bibinfo
  {year} {2024}),\ \href {https://doi.org/10.1088/1367-2630/ad8fc5} {\bibfield
  {journal} {\bibinfo  {journal} {New J. Phys.}\ }\textbf {\bibinfo {volume}
  {26}},\ \bibinfo {pages} {113011}}\BibitemShut {NoStop}%
\bibitem [{\citenamefont {Farmer}\ \emph {et~al.}(2021)\citenamefont {Farmer},
  \citenamefont {Gschwendtner}, \citenamefont {Liang}, \citenamefont {Muggli},\
  and\ \citenamefont {Weidl}}]{Farmer2018}%
  \BibitemOpen
  \bibfield  {author} {\bibinfo {author} {\bibnamefont {Farmer}, \bibfnamefont
  {J}}, \bibinfo {author} {\bibfnamefont {E.}~\bibnamefont {Gschwendtner}},
  \bibinfo {author} {\bibfnamefont {L.}~\bibnamefont {Liang}}, \bibinfo
  {author} {\bibfnamefont {P.}~\bibnamefont {Muggli}}, \ and\ \bibinfo {author}
  {\bibfnamefont {M.}~\bibnamefont {Weidl}}} (\bibinfo {year} {2021}),\ in\
  \href {https://jacow.org/ipac2021/doi/JACoW-IPAC2021-TUPAB158.html} {\emph
  {\bibinfo {booktitle} {Proceedings of the 12th Int. Particle Accelerator
  Conf.}}}\ (\bibinfo  {publisher} {JACoW},\ \bibinfo {address} {Geneva,
  Switzerland})\ pp.\ \bibinfo {pages} {1753--1756}\BibitemShut {NoStop}%
\bibitem [{\citenamefont {Farmer}\ \emph {et~al.}(2015)\citenamefont {Farmer},
  \citenamefont {Martorelli},\ and\ \citenamefont {Pukhov}}]{Farmer2015}%
  \BibitemOpen
  \bibfield  {author} {\bibinfo {author} {\bibnamefont {Farmer}, \bibfnamefont
  {J~P}}, \bibinfo {author} {\bibfnamefont {R.}~\bibnamefont {Martorelli}}, \
  and\ \bibinfo {author} {\bibfnamefont {A.}~\bibnamefont {Pukhov}}} (\bibinfo
  {year} {2015}),\ \href {https://doi.org/10.1063/1.4938038} {\bibfield
  {journal} {\bibinfo  {journal} {Phys. Plasmas}\ }\textbf {\bibinfo {volume}
  {22}},\ \bibinfo {pages} {123113}}\BibitemShut {NoStop}%
\bibitem [{\citenamefont {Farmer}\ and\ \citenamefont {Zevi
  Della~Porta}(2025)}]{Farmer2025}%
  \BibitemOpen
  \bibfield  {author} {\bibinfo {author} {\bibnamefont {Farmer}, \bibfnamefont
  {J~P}}, \ and\ \bibinfo {author} {\bibfnamefont {G.}~\bibnamefont {Zevi
  Della~Porta}}} (\bibinfo {year} {2025}),\ \href {\doibase
  10.1103/physrevresearch.7.l012055} {\bibfield  {journal} {\bibinfo  {journal}
  {Phys. Rev. Res.}\ }\textbf {\bibinfo {volume} {7}},\ \bibinfo {pages}
  {L012055}}\BibitemShut {NoStop}%
\bibitem [{\citenamefont {Ferran~Pousa}\ \emph {et~al.}(2019)\citenamefont
  {Ferran~Pousa}, \citenamefont {Assmann},\ and\ \citenamefont {{Martinez de la
  Ossa}}}]{FerranPousa2019b}%
  \BibitemOpen
  \bibfield  {author} {\bibinfo {author} {\bibnamefont {Ferran~Pousa},
  \bibfnamefont {\'{A}}}, \bibinfo {author} {\bibfnamefont {R.}~\bibnamefont
  {Assmann}}, \ and\ \bibinfo {author} {\bibfnamefont {A.}~\bibnamefont
  {{Martinez de la Ossa}}}} (\bibinfo {year} {2019}),\ in\ \href
  {http://jacow.org/ipac2019/doi/JACoW-IPAC2019-THPGW012.html} {\emph {\bibinfo
  {booktitle} {Proceedings of the 10th Int. Particle Accelerator Conf.}}}\
  (\bibinfo  {publisher} {JACoW},\ \bibinfo {address} {Geneva, Switzerland})\
  pp.\ \bibinfo {pages} {3601--3604}\BibitemShut {NoStop}%
\bibitem [{\citenamefont {{Ferran Pousa}}\ \emph {et~al.}(2019)\citenamefont
  {{Ferran Pousa}}, \citenamefont {{Martinez de la Ossa}}, \citenamefont
  {Brinkmann},\ and\ \citenamefont {Assmann}}]{FerranPousa2019}%
  \BibitemOpen
  \bibfield  {author} {\bibinfo {author} {\bibnamefont {{Ferran Pousa}},
  \bibfnamefont {A}}, \bibinfo {author} {\bibfnamefont {A.}~\bibnamefont
  {{Martinez de la Ossa}}}, \bibinfo {author} {\bibfnamefont {R.}~\bibnamefont
  {Brinkmann}}, \ and\ \bibinfo {author} {\bibfnamefont {R.~W.}\ \bibnamefont
  {Assmann}}} (\bibinfo {year} {2019}),\ \href
  {https://doi.org/10.1103/physrevlett.123.054801} {\bibfield  {journal}
  {\bibinfo  {journal} {Phys. Rev. Lett.}\ }\textbf {\bibinfo {volume} {123}},\
  \bibinfo {pages} {054801}}\BibitemShut {NoStop}%
\bibitem [{\citenamefont {Ferran~Pousa}\ \emph {et~al.}(2022)\citenamefont
  {Ferran~Pousa} \emph {et~al.}}]{FerranPousa2022}%
  \BibitemOpen
  \bibfield  {author} {\bibinfo {author} {\bibnamefont {Ferran~Pousa},
  \bibfnamefont {\'{A}}},  \emph {et~al.}} (\bibinfo {year} {2022}),\ \href
  {https://link.aps.org/doi/10.1103/PhysRevLett.129.094801} {\bibfield
  {journal} {\bibinfo  {journal} {Phys. Rev. Lett.}\ }\textbf {\bibinfo
  {volume} {129}},\ \bibinfo {pages} {094801}}\BibitemShut {NoStop}%
\bibitem [{\citenamefont {Ferrario}\ \emph {et~al.}(2011)\citenamefont
  {Ferrario} \emph {et~al.}}]{Ferrario2011}%
  \BibitemOpen
  \bibfield  {author} {\bibinfo {author} {\bibnamefont {Ferrario},
  \bibfnamefont {M}},  \emph {et~al.}} (\bibinfo {year} {2011}),\ \href
  {https://doi.org/10.1016/j.nima.2010.02.018} {\bibfield  {journal} {\bibinfo
  {journal} {Nucl. Instrum. Methods Phys. Res. A}\ }\textbf {\bibinfo {volume}
  {637}},\ \bibinfo {pages} {S43--S46}}\BibitemShut {NoStop}%
\bibitem [{\citenamefont {Ferrario}\ \emph {et~al.}(2013)\citenamefont
  {Ferrario} \emph {et~al.}}]{Ferrario2013}%
  \BibitemOpen
  \bibfield  {author} {\bibinfo {author} {\bibnamefont {Ferrario},
  \bibfnamefont {M}},  \emph {et~al.}} (\bibinfo {year} {2013}),\ \href
  {https://doi.org/10.1016/j.nimb.2013.03.049} {\bibfield  {journal} {\bibinfo
  {journal} {Nucl. Instrum. Methods Phys. Res. B}\ }\textbf {\bibinfo {volume}
  {309}},\ \bibinfo {pages} {183--188}}\BibitemShut {NoStop}%
\bibitem [{\citenamefont {Ferrario}\ \emph {et~al.}(2018)\citenamefont
  {Ferrario} \emph {et~al.}}]{Ferrario2018}%
  \BibitemOpen
  \bibfield  {author} {\bibinfo {author} {\bibnamefont {Ferrario},
  \bibfnamefont {M}},  \emph {et~al.}} (\bibinfo {year} {2018}),\ \href
  {https://doi.org/10.1016/j.nima.2018.01.094} {\bibfield  {journal} {\bibinfo
  {journal} {Nucl. Instrum. Methods Phys. Res. A}\ }\textbf {\bibinfo {volume}
  {909}},\ \bibinfo {pages} {134--138}}\BibitemShut {NoStop}%
\bibitem [{\citenamefont {Ferri}\ \emph {et~al.}(2018)\citenamefont {Ferri}
  \emph {et~al.}}]{Ferri2018}%
  \BibitemOpen
  \bibfield  {author} {\bibinfo {author} {\bibnamefont {Ferri}, \bibfnamefont
  {J}},  \emph {et~al.}} (\bibinfo {year} {2018}),\ \href
  {https://doi.org/10.1103/physrevlett.120.254802} {\bibfield  {journal}
  {\bibinfo  {journal} {Phys. Rev. Lett.}\ }\textbf {\bibinfo {volume} {120}},\
  \bibinfo {pages} {254802}}\BibitemShut {NoStop}%
\bibitem [{\citenamefont {Filippi}\ \emph {et~al.}(2018)\citenamefont
  {Filippi}, \citenamefont {Anania}, \citenamefont {Biagioni}, \citenamefont
  {Chiadroni}, \citenamefont {Cianchi}, \citenamefont {Ferber}, \citenamefont
  {Ferrario},\ and\ \citenamefont {Zigler}}]{Filippi2018}%
  \BibitemOpen
  \bibfield  {author} {\bibinfo {author} {\bibnamefont {Filippi}, \bibfnamefont
  {F}}, \bibinfo {author} {\bibfnamefont {M.~P.}\ \bibnamefont {Anania}},
  \bibinfo {author} {\bibfnamefont {A.}~\bibnamefont {Biagioni}}, \bibinfo
  {author} {\bibfnamefont {E.}~\bibnamefont {Chiadroni}}, \bibinfo {author}
  {\bibfnamefont {A.}~\bibnamefont {Cianchi}}, \bibinfo {author} {\bibfnamefont
  {Y.}~\bibnamefont {Ferber}}, \bibinfo {author} {\bibfnamefont
  {M.}~\bibnamefont {Ferrario}}, \ and\ \bibinfo {author} {\bibfnamefont
  {A.}~\bibnamefont {Zigler}}} (\bibinfo {year} {2018}),\ \href
  {https://doi.org/10.1063/1.5010264} {\bibfield  {journal} {\bibinfo
  {journal} {Rev. Sci. Instrum.}\ }\textbf {\bibinfo {volume} {89}},\ \bibinfo
  {pages} {083502}}\BibitemShut {NoStop}%
\bibitem [{\citenamefont {Finnerud}\ \emph {et~al.}(2025)\citenamefont
  {Finnerud}, \citenamefont {Lindstrøm},\ and\ \citenamefont
  {Adli}}]{Finnerud2025}%
  \BibitemOpen
  \bibfield  {author} {\bibinfo {author} {\bibnamefont {Finnerud},
  \bibfnamefont {O~G}}, \bibinfo {author} {\bibfnamefont {C.~A.}\ \bibnamefont
  {Lindstrøm}}, \ and\ \bibinfo {author} {\bibfnamefont {E.}~\bibnamefont
  {Adli}}} (\bibinfo {year} {2025}),\ \href {\doibase 10.1103/72qb-517d}
  {\bibfield  {journal} {\bibinfo  {journal} {Phys. Rev. Accel. Beams}\
  }\textbf {\bibinfo {volume} {28}},\ \bibinfo {pages} {071301}}\BibitemShut
  {NoStop}%
\bibitem [{\citenamefont {Finnerud}\ \emph {et~al.}(2026)\citenamefont
  {Finnerud} \emph {et~al.}}]{Finnerud2026}%
  \BibitemOpen
  \bibfield  {author} {\bibinfo {author} {\bibnamefont {Finnerud},
  \bibfnamefont {O~G}},  \emph {et~al.}} (\bibinfo {year} {2026}),\ \href
  {\doibase 10.1088/1742-6596/3266/1/012001} {\bibfield  {journal} {\bibinfo
  {journal} {J. Phys.: Conf. Ser.}\ }\textbf {\bibinfo {volume} {3266}},\
  \bibinfo {pages} {012001}}\BibitemShut {NoStop}%
\bibitem [{\citenamefont {Floettmann}(2003)}]{Floettmann2003}%
  \BibitemOpen
  \bibfield  {author} {\bibinfo {author} {\bibnamefont {Floettmann},
  \bibfnamefont {K}}} (\bibinfo {year} {2003}),\ \href
  {https://doi.org/10.1103/PhysRevSTAB.6.034202} {\bibfield  {journal}
  {\bibinfo  {journal} {Phys. Rev. ST Accel. Beams}\ }\textbf {\bibinfo
  {volume} {6}},\ \bibinfo {pages} {034202}}\BibitemShut {NoStop}%
\bibitem [{\citenamefont {Floettmann}(2014)}]{Floettmann2014}%
  \BibitemOpen
  \bibfield  {author} {\bibinfo {author} {\bibnamefont {Floettmann},
  \bibfnamefont {K}}} (\bibinfo {year} {2014}),\ \href
  {https://doi.org/10.1103/PhysRevSTAB.17.054402} {\bibfield  {journal}
  {\bibinfo  {journal} {Phys. Rev. ST Accel. Beams}\ }\textbf {\bibinfo
  {volume} {17}},\ \bibinfo {pages} {054402}}\BibitemShut {NoStop}%
\bibitem [{\citenamefont {Foerster}\ \emph {et~al.}(2022)\citenamefont
  {Foerster} \emph {et~al.}}]{Foerster2022}%
  \BibitemOpen
  \bibfield  {author} {\bibinfo {author} {\bibnamefont {Foerster},
  \bibfnamefont {F~M}},  \emph {et~al.}} (\bibinfo {year} {2022}),\ \href
  {https://doi.org/10.1103/physrevx.12.041016} {\bibfield  {journal} {\bibinfo
  {journal} {Phys. Rev. X}\ }\textbf {\bibinfo {volume} {12}},\ \bibinfo
  {pages} {041016}}\BibitemShut {NoStop}%
\bibitem [{\citenamefont {Foerster}\ \emph {et~al.}(2026)\citenamefont
  {Foerster} \emph {et~al.}}]{Foerster2026}%
  \BibitemOpen
  \bibfield  {author} {\bibinfo {author} {\bibnamefont {Foerster},
  \bibfnamefont {F~M}},  \emph {et~al.}} (\bibinfo {year} {2026}),\ \href
  {\doibase 10.48550/arxiv.2602.24107} {}\Eprint
  {http://arxiv.org/abs/2602.24107} {arXiv:2602.24107} \BibitemShut {NoStop}%
\bibitem [{\citenamefont {Fonseca}\ \emph {et~al.}(2002)\citenamefont {Fonseca}
  \emph {et~al.}}]{Fonseca2002}%
  \BibitemOpen
  \bibfield  {author} {\bibinfo {author} {\bibnamefont {Fonseca}, \bibfnamefont
  {R~A}},  \emph {et~al.}} (\bibinfo {year} {2002}),\ in\ \href
  {https://doi.org/10.1007/3-540-47789-6_36} {\emph {\bibinfo {booktitle}
  {Lect. Notes Comput. Sci.}}}\ (\bibinfo  {publisher} {Springer, Berlin,
  Heidelberg})\ pp.\ \bibinfo {pages} {342--351}\BibitemShut {NoStop}%
\bibitem [{\citenamefont {Foster}\ \emph {et~al.}(2023)\citenamefont {Foster},
  \citenamefont {D’Arcy},\ and\ \citenamefont {Lindstr{\o}m}}]{Foster2023}%
  \BibitemOpen
  \bibfield  {author} {\bibinfo {author} {\bibnamefont {Foster}, \bibfnamefont
  {B}}, \bibinfo {author} {\bibfnamefont {R.}~\bibnamefont {D’Arcy}}, \ and\
  \bibinfo {author} {\bibfnamefont {C.~A.}\ \bibnamefont {Lindstr{\o}m}}}
  (\bibinfo {year} {2023}),\ \href {https://doi.org/10.1088/1367-2630/acf395}
  {\bibfield  {journal} {\bibinfo  {journal} {New J. Phys.}\ }\textbf {\bibinfo
  {volume} {25}},\ \bibinfo {pages} {093037}}\BibitemShut {NoStop}%
\bibitem [{\citenamefont {Foster}\ \emph {et~al.}(2025)\citenamefont {Foster}
  \emph {et~al.}}]{Foster2025}%
  \BibitemOpen
  \bibfield  {author} {\bibinfo {author} {\bibnamefont {Foster}, \bibfnamefont
  {B}},  \emph {et~al.}} (\bibinfo {year} {2025}),\ \href
  {https://doi.org/10.1016/j.physo.2025.100261} {\bibfield  {journal} {\bibinfo
   {journal} {Phys. Open}\ }\textbf {\bibinfo {volume} {23}},\ \bibinfo {pages}
  {100261}}\BibitemShut {NoStop}%
\bibitem [{\citenamefont {Fritzler}\ \emph {et~al.}(2004)\citenamefont
  {Fritzler}, \citenamefont {Lefebvre}, \citenamefont {Malka}, \citenamefont
  {Burgy}, \citenamefont {Dangor}, \citenamefont {Krushelnick}, \citenamefont
  {Mangles}, \citenamefont {Najmudin}, \citenamefont {Rousseau},\ and\
  \citenamefont {Walton}}]{Fritzler2004}%
  \BibitemOpen
  \bibfield  {author} {\bibinfo {author} {\bibnamefont {Fritzler},
  \bibfnamefont {S}}, \bibinfo {author} {\bibfnamefont {E.}~\bibnamefont
  {Lefebvre}}, \bibinfo {author} {\bibfnamefont {V.}~\bibnamefont {Malka}},
  \bibinfo {author} {\bibfnamefont {F.}~\bibnamefont {Burgy}}, \bibinfo
  {author} {\bibfnamefont {A.~E.}\ \bibnamefont {Dangor}}, \bibinfo {author}
  {\bibfnamefont {K.}~\bibnamefont {Krushelnick}}, \bibinfo {author}
  {\bibfnamefont {S.~P.~D.}\ \bibnamefont {Mangles}}, \bibinfo {author}
  {\bibfnamefont {Z.}~\bibnamefont {Najmudin}}, \bibinfo {author}
  {\bibfnamefont {J.-P.}\ \bibnamefont {Rousseau}}, \ and\ \bibinfo {author}
  {\bibfnamefont {B.}~\bibnamefont {Walton}}} (\bibinfo {year} {2004}),\ \href
  {https://doi.org/10.1103/physrevlett.92.165006} {\bibfield  {journal}
  {\bibinfo  {journal} {Phys. Rev. Lett.}\ }\textbf {\bibinfo {volume} {92}},\
  \bibinfo {pages} {165006}}\BibitemShut {NoStop}%
\bibitem [{\citenamefont {Fujii}\ \emph {et~al.}(2019)\citenamefont {Fujii},
  \citenamefont {Marsh}, \citenamefont {An}, \citenamefont {Corde},
  \citenamefont {Hogan}, \citenamefont {Yakimenko},\ and\ \citenamefont
  {Joshi}}]{Fujii2019}%
  \BibitemOpen
  \bibfield  {author} {\bibinfo {author} {\bibnamefont {Fujii}, \bibfnamefont
  {H}}, \bibinfo {author} {\bibfnamefont {K.~A.}\ \bibnamefont {Marsh}},
  \bibinfo {author} {\bibfnamefont {W.}~\bibnamefont {An}}, \bibinfo {author}
  {\bibfnamefont {S.}~\bibnamefont {Corde}}, \bibinfo {author} {\bibfnamefont
  {M.~J.}\ \bibnamefont {Hogan}}, \bibinfo {author} {\bibfnamefont
  {V.}~\bibnamefont {Yakimenko}}, \ and\ \bibinfo {author} {\bibfnamefont
  {C.}~\bibnamefont {Joshi}}} (\bibinfo {year} {2019}),\ \href
  {https://doi.org/10.1103/PhysRevAccelBeams.22.091301} {\bibfield  {journal}
  {\bibinfo  {journal} {Phys. Rev. Accel. Beams}\ }\textbf {\bibinfo {volume}
  {22}},\ \bibinfo {pages} {091301}}\BibitemShut {NoStop}%
\bibitem [{\citenamefont {Gai}\ \emph {et~al.}(1997)\citenamefont {Gai},
  \citenamefont {Kanareykin}, \citenamefont {Kustov},\ and\ \citenamefont
  {Simpson}}]{Gai1997}%
  \BibitemOpen
  \bibfield  {author} {\bibinfo {author} {\bibnamefont {Gai}, \bibfnamefont
  {W}}, \bibinfo {author} {\bibfnamefont {A.~D.}\ \bibnamefont {Kanareykin}},
  \bibinfo {author} {\bibfnamefont {A.~L.}\ \bibnamefont {Kustov}}, \ and\
  \bibinfo {author} {\bibfnamefont {J.}~\bibnamefont {Simpson}}} (\bibinfo
  {year} {1997}),\ \href {https://doi.org/10.1103/PhysRevE.55.3481} {\bibfield
  {journal} {\bibinfo  {journal} {Phys. Rev. E}\ }\textbf {\bibinfo {volume}
  {55}},\ \bibinfo {pages} {3481--3488}}\BibitemShut {NoStop}%
\bibitem [{\citenamefont {Galletti}\ \emph {et~al.}(2024)\citenamefont
  {Galletti}, \citenamefont {Assmann}, \citenamefont {Couprie}, \citenamefont
  {Ferrario}, \citenamefont {Giannessi}, \citenamefont {Irman}, \citenamefont
  {Pompili},\ and\ \citenamefont {Wang}}]{Galletti2024}%
  \BibitemOpen
  \bibfield  {author} {\bibinfo {author} {\bibnamefont {Galletti},
  \bibfnamefont {M}}, \bibinfo {author} {\bibfnamefont {R.}~\bibnamefont
  {Assmann}}, \bibinfo {author} {\bibfnamefont {M.~E.}\ \bibnamefont
  {Couprie}}, \bibinfo {author} {\bibfnamefont {M.}~\bibnamefont {Ferrario}},
  \bibinfo {author} {\bibfnamefont {L.}~\bibnamefont {Giannessi}}, \bibinfo
  {author} {\bibfnamefont {A.}~\bibnamefont {Irman}}, \bibinfo {author}
  {\bibfnamefont {R.}~\bibnamefont {Pompili}}, \ and\ \bibinfo {author}
  {\bibfnamefont {W.}~\bibnamefont {Wang}}} (\bibinfo {year} {2024}),\ \href
  {https://doi.org/10.1038/s41566-024-01474-3} {\bibfield  {journal} {\bibinfo
  {journal} {Nat. Photon.}\ }\textbf {\bibinfo {volume} {18}},\ \bibinfo
  {pages} {780–791}}\BibitemShut {NoStop}%
\bibitem [{\citenamefont {Galletti}\ \emph {et~al.}(2025)\citenamefont
  {Galletti}, \citenamefont {Crincoli}, \citenamefont {Pompili}, \citenamefont
  {Verra}, \citenamefont {Villa}, \citenamefont {Demitra}, \citenamefont
  {Biagioni}, \citenamefont {Zigler},\ and\ \citenamefont
  {Ferrario}}]{Galletti2025}%
  \BibitemOpen
  \bibfield  {author} {\bibinfo {author} {\bibnamefont {Galletti},
  \bibfnamefont {M}}, \bibinfo {author} {\bibfnamefont {L.}~\bibnamefont
  {Crincoli}}, \bibinfo {author} {\bibfnamefont {R.}~\bibnamefont {Pompili}},
  \bibinfo {author} {\bibfnamefont {L.}~\bibnamefont {Verra}}, \bibinfo
  {author} {\bibfnamefont {F.}~\bibnamefont {Villa}}, \bibinfo {author}
  {\bibfnamefont {R.}~\bibnamefont {Demitra}}, \bibinfo {author} {\bibfnamefont
  {A.}~\bibnamefont {Biagioni}}, \bibinfo {author} {\bibfnamefont
  {A.}~\bibnamefont {Zigler}}, \ and\ \bibinfo {author} {\bibfnamefont
  {M.}~\bibnamefont {Ferrario}}} (\bibinfo {year} {2025}),\ \href {\doibase
  10.1103/physreve.111.025202} {\bibfield  {journal} {\bibinfo  {journal}
  {Phys. Rev. E}\ }\textbf {\bibinfo {volume} {111}},\ \bibinfo {pages}
  {025202}}\BibitemShut {NoStop}%
\bibitem [{\citenamefont {Galletti}\ \emph {et~al.}(2022)\citenamefont
  {Galletti} \emph {et~al.}}]{Galletti2022}%
  \BibitemOpen
  \bibfield  {author} {\bibinfo {author} {\bibnamefont {Galletti},
  \bibfnamefont {M}},  \emph {et~al.}} (\bibinfo {year} {2022}),\ \href
  {https://doi.org/10.1103/physrevlett.129.234801} {\bibfield  {journal}
  {\bibinfo  {journal} {Phys. Rev. Lett.}\ }\textbf {\bibinfo {volume} {129}},\
  \bibinfo {pages} {234801}}\BibitemShut {NoStop}%
\bibitem [{\citenamefont {Gao}\ \emph {et~al.}(2018)\citenamefont {Gao} \emph
  {et~al.}}]{Gao2018}%
  \BibitemOpen
  \bibfield  {author} {\bibinfo {author} {\bibnamefont {Gao}, \bibfnamefont
  {Q}},  \emph {et~al.}} (\bibinfo {year} {2018}),\ \href
  {https://link.aps.org/doi/10.1103/PhysRevLett.120.114801} {\bibfield
  {journal} {\bibinfo  {journal} {Phys. Rev. Lett.}\ }\textbf {\bibinfo
  {volume} {120}},\ \bibinfo {pages} {114801}}\BibitemShut {NoStop}%
\bibitem [{\citenamefont {Garland}\ \emph {et~al.}(2020)\citenamefont
  {Garland}, \citenamefont {Wood}, \citenamefont {Boyle},\ and\ \citenamefont
  {Osterhoff}}]{Garland2020}%
  \BibitemOpen
  \bibfield  {author} {\bibinfo {author} {\bibnamefont {Garland}, \bibfnamefont
  {M~J}}, \bibinfo {author} {\bibfnamefont {J.~C.}\ \bibnamefont {Wood}},
  \bibinfo {author} {\bibfnamefont {G.}~\bibnamefont {Boyle}}, \ and\ \bibinfo
  {author} {\bibfnamefont {J.}~\bibnamefont {Osterhoff}}} (\bibinfo {year}
  {2020}),\ in\ \href {\doibase 10.17181/CAS2019SESIMBRA} {\emph {\bibinfo
  {booktitle} {Proceedings of the 2019 CERN--Accelerator--School course on High
  Gradient Wakefield Accelerators, Sesimbra, (Portugal)}}}\ (\bibinfo
  {publisher} {CERN},\ \bibinfo {address} {Geneva, Switzerland})\ p.\ \bibinfo
  {pages} {158}\BibitemShut {NoStop}%
\bibitem [{\citenamefont {Gessner}\ \emph {et~al.}(2025)\citenamefont
  {Gessner}, \citenamefont {Osterhoff} \emph {et~al.}}]{Gessner2025}%
  \BibitemOpen
  \bibfield  {author} {\bibinfo {author} {\bibnamefont {Gessner}, \bibfnamefont
  {S}}, \bibinfo {author} {\bibfnamefont {J.}~\bibnamefont {Osterhoff}},  \emph
  {et~al.}} (\bibinfo {year} {2025}),\ \href
  {https://doi.org/10.48550/arXiv.2503.20214} {}\Eprint
  {http://arxiv.org/abs/2503.20214} {arXiv:2503.20214} \BibitemShut {NoStop}%
\bibitem [{\citenamefont {Gessner}\ \emph {et~al.}(2016)\citenamefont {Gessner}
  \emph {et~al.}}]{Gessner2016a}%
  \BibitemOpen
  \bibfield  {author} {\bibinfo {author} {\bibnamefont {Gessner}, \bibfnamefont
  {S}},  \emph {et~al.}} (\bibinfo {year} {2016}),\ \href
  {https://doi.org/10.1038/ncomms11785} {\bibfield  {journal} {\bibinfo
  {journal} {Nat. Commun.}\ }\textbf {\bibinfo {volume} {7}},\ \bibinfo {pages}
  {11785}}\BibitemShut {NoStop}%
\bibitem [{\citenamefont {Gessner}\ \emph {et~al.}(2023)\citenamefont {Gessner}
  \emph {et~al.}}]{Gessner2023}%
  \BibitemOpen
  \bibfield  {author} {\bibinfo {author} {\bibnamefont {Gessner}, \bibfnamefont
  {S}},  \emph {et~al.}} (\bibinfo {year} {2023}),\ \href
  {https://doi.org/10.48550/arXiv.2304.01700} {}\Eprint
  {http://arxiv.org/abs/2304.01700} {arXiv:2304.01700} \BibitemShut {NoStop}%
\bibitem [{\citenamefont {Gessner}(2016)}]{Gessner2016b}%
  \BibitemOpen
  \bibfield  {author} {\bibinfo {author} {\bibnamefont {Gessner}, \bibfnamefont
  {S~J}}} (\bibinfo {year} {2016}),\ \href {https://doi.org/10.2172/1340170}
  {\bibinfo {type} {Ph{D} {T}hesis}}\ (\bibinfo  {school} {Stanford
  University})\BibitemShut {NoStop}%
\bibitem [{\citenamefont {Gholizadeh}\ \emph {et~al.}(2011)\citenamefont
  {Gholizadeh}, \citenamefont {Katsouleas}, \citenamefont {Huang},
  \citenamefont {Mori},\ and\ \citenamefont {Muggli}}]{Gholizadeh2011}%
  \BibitemOpen
  \bibfield  {author} {\bibinfo {author} {\bibnamefont {Gholizadeh},
  \bibfnamefont {R}}, \bibinfo {author} {\bibfnamefont {T.}~\bibnamefont
  {Katsouleas}}, \bibinfo {author} {\bibfnamefont {C.}~\bibnamefont {Huang}},
  \bibinfo {author} {\bibfnamefont {W.~B.}\ \bibnamefont {Mori}}, \ and\
  \bibinfo {author} {\bibfnamefont {P.}~\bibnamefont {Muggli}}} (\bibinfo
  {year} {2011}),\ \href {https://doi.org/10.1103/PhysRevSTAB.14.021303}
  {\bibfield  {journal} {\bibinfo  {journal} {Phys. Rev. ST Accel. Beams}\
  }\textbf {\bibinfo {volume} {14}},\ \bibinfo {pages} {021303}}\BibitemShut
  {NoStop}%
\bibitem [{\citenamefont {Gholizadeh}\ \emph {et~al.}(2010)\citenamefont
  {Gholizadeh}, \citenamefont {Katsouleas}, \citenamefont {Muggli},
  \citenamefont {Huang},\ and\ \citenamefont {Mori}}]{Gholizadeh2010}%
  \BibitemOpen
  \bibfield  {author} {\bibinfo {author} {\bibnamefont {Gholizadeh},
  \bibfnamefont {R}}, \bibinfo {author} {\bibfnamefont {T.}~\bibnamefont
  {Katsouleas}}, \bibinfo {author} {\bibfnamefont {P.}~\bibnamefont {Muggli}},
  \bibinfo {author} {\bibfnamefont {C.}~\bibnamefont {Huang}}, \ and\ \bibinfo
  {author} {\bibfnamefont {W.}~\bibnamefont {Mori}}} (\bibinfo {year} {2010}),\
  \href {https://doi.org/10.1103/PhysRevLett.104.155001} {\bibfield  {journal}
  {\bibinfo  {journal} {Phys. Rev. Lett.}\ }\textbf {\bibinfo {volume} {104}},\
  \bibinfo {pages} {155001}}\BibitemShut {NoStop}%
\bibitem [{\citenamefont {Gilljohann}\ \emph {et~al.}(2019)\citenamefont
  {Gilljohann} \emph {et~al.}}]{Gilljohann2019}%
  \BibitemOpen
  \bibfield  {author} {\bibinfo {author} {\bibnamefont {Gilljohann},
  \bibfnamefont {M~F}},  \emph {et~al.}} (\bibinfo {year} {2019}),\ \href
  {https://doi.org/10.1103/physrevx.9.011046} {\bibfield  {journal} {\bibinfo
  {journal} {Phys. Rev. X}\ }\textbf {\bibinfo {volume} {9}},\ \bibinfo {pages}
  {011046}}\BibitemShut {NoStop}%
\bibitem [{\citenamefont {Gilljohann}\ \emph {et~al.}(2023)\citenamefont
  {Gilljohann} \emph {et~al.}}]{Gilljohann2023}%
  \BibitemOpen
  \bibfield  {author} {\bibinfo {author} {\bibnamefont {Gilljohann},
  \bibfnamefont {M~F}},  \emph {et~al.}} (\bibinfo {year} {2023}),\ \href
  {\doibase 10.1088/1748-0221/18/11/p11008} {\bibfield  {journal} {\bibinfo
  {journal} {J. Instrum.}\ }\textbf {\bibinfo {volume} {18}},\ \bibinfo {pages}
  {P11008}}\BibitemShut {NoStop}%
\bibitem [{\citenamefont {van Gils}\ \emph {et~al.}(2025)\citenamefont {van
  Gils} \emph {et~al.}}]{vanGils2025}%
  \BibitemOpen
  \bibfield  {author} {\bibinfo {author} {\bibnamefont {van Gils},
  \bibfnamefont {N~Z}},  \emph {et~al.} (\bibinfo {collaboration} {AWAKE
  Collaboration})} (\bibinfo {year} {2025}),\ \href {\doibase
  10.1088/1742-6596/3124/1/012006} {\bibfield  {journal} {\bibinfo  {journal}
  {J. Phys.: Conf. Ser.}\ }\textbf {\bibinfo {volume} {3124}},\ \bibinfo
  {pages} {012006}}\BibitemShut {NoStop}%
\bibitem [{\citenamefont {van Gils}\ \emph {et~al.}(2026)\citenamefont {van
  Gils} \emph {et~al.}}]{vanGils2026}%
  \BibitemOpen
  \bibfield  {author} {\bibinfo {author} {\bibnamefont {van Gils},
  \bibfnamefont {N~Z}},  \emph {et~al.} (\bibinfo {collaboration} {AWAKE
  Collaboration})} (\bibinfo {year} {2026}),\ \href {\doibase
  10.1103/c7jm-hf4q} {\bibfield  {journal} {\bibinfo  {journal} {Phys. Rev.
  Accel. Beams}\ }\textbf {\bibinfo {volume} {29}},\ \bibinfo {pages}
  {083601}}\BibitemShut {NoStop}%
\bibitem [{\citenamefont {Ginzburg}\ \emph {et~al.}(1983)\citenamefont
  {Ginzburg}, \citenamefont {Kotkin}, \citenamefont {Serbo},\ and\
  \citenamefont {Telnov}}]{Ginzburg1983}%
  \BibitemOpen
  \bibfield  {author} {\bibinfo {author} {\bibnamefont {Ginzburg},
  \bibfnamefont {I~F}}, \bibinfo {author} {\bibfnamefont {G.~L.}\ \bibnamefont
  {Kotkin}}, \bibinfo {author} {\bibfnamefont {V.~G.}\ \bibnamefont {Serbo}}, \
  and\ \bibinfo {author} {\bibfnamefont {V.~I.}\ \bibnamefont {Telnov}}}
  (\bibinfo {year} {1983}),\ \href
  {https://doi.org/10.1016/0167-5087(83)90173-4} {\bibfield  {journal}
  {\bibinfo  {journal} {Nucl. Instrum. Methods Phys. Res.}\ }\textbf {\bibinfo
  {volume} {205}},\ \bibinfo {pages} {47--68}}\BibitemShut {NoStop}%
\bibitem [{\citenamefont {Glashausser}(1979)}]{Glashausser1979}%
  \BibitemOpen
  \bibfield  {author} {\bibinfo {author} {\bibnamefont {Glashausser},
  \bibfnamefont {C}}} (\bibinfo {year} {1979}),\ \href
  {https://doi.org/10.1146/annurev.ns.29.120179.000341} {\bibfield  {journal}
  {\bibinfo  {journal} {Annu. Rev. Nucl. Part. Sci.}\ }\textbf {\bibinfo
  {volume} {29}},\ \bibinfo {pages} {33--68}}\BibitemShut {NoStop}%
\bibitem [{\citenamefont {Godfrey}(1974)}]{Godfrey1974}%
  \BibitemOpen
  \bibfield  {author} {\bibinfo {author} {\bibnamefont {Godfrey}, \bibfnamefont
  {B~B}}} (\bibinfo {year} {1974}),\ \href {\doibase
  10.1016/0021-9991(74)90076-x} {\bibfield  {journal} {\bibinfo  {journal} {J.
  Comput. Phys.}\ }\textbf {\bibinfo {volume} {15}},\ \bibinfo {pages}
  {504--521}}\BibitemShut {NoStop}%
\bibitem [{\citenamefont {Goldstein}\ \emph {et~al.}(2005)\citenamefont
  {Goldstein} \emph {et~al.}}]{Goldstein2005}%
  \BibitemOpen
  \bibfield  {author} {\bibinfo {author} {\bibnamefont {Goldstein},
  \bibfnamefont {M}},  \emph {et~al.}} (\bibinfo {year} {2005}),\ in\ \href
  {https://proceedings.jacow.org/f05/PAPERS/THOA002.PDF} {\emph {\bibinfo
  {booktitle} {Proceedings of the 27th Int. Free Electron Laser Conf.}}}\
  (\bibinfo  {publisher} {JACoW},\ \bibinfo {address} {Geneva, Switzerland})\
  pp.\ \bibinfo {pages} {422--425}\BibitemShut {NoStop}%
\bibitem [{\citenamefont {Goldston}\ and\ \citenamefont
  {Rutherford}(1995)}]{Goldston1995}%
  \BibitemOpen
  \bibfield  {author} {\bibinfo {author} {\bibnamefont {Goldston},
  \bibfnamefont {R~J}}, \ and\ \bibinfo {author} {\bibfnamefont {P.~H.}\
  \bibnamefont {Rutherford}}} (\bibinfo {year} {1995}),\ \href@noop {} {\emph
  {\bibinfo {title} {Introduction to Plasma Physics}}}\ (\bibinfo  {publisher}
  {Taylor \& Francis Group},\ \bibinfo {address} {New York, NY, United
  States})\BibitemShut {NoStop}%
\bibitem [{\citenamefont {Golovanov}\ \emph {et~al.}(2023)\citenamefont
  {Golovanov}, \citenamefont {Kostyukov}, \citenamefont {Pukhov},\ and\
  \citenamefont {Malka}}]{Golovanov2023}%
  \BibitemOpen
  \bibfield  {author} {\bibinfo {author} {\bibnamefont {Golovanov},
  \bibfnamefont {A}}, \bibinfo {author} {\bibfnamefont {I.~Yu.}\ \bibnamefont
  {Kostyukov}}, \bibinfo {author} {\bibfnamefont {A.}~\bibnamefont {Pukhov}}, \
  and\ \bibinfo {author} {\bibfnamefont {V.}~\bibnamefont {Malka}}} (\bibinfo
  {year} {2023}),\ \href
  {https://link.aps.org/doi/10.1103/PhysRevLett.130.105001} {\bibfield
  {journal} {\bibinfo  {journal} {Phys. Rev. Lett.}\ }\textbf {\bibinfo
  {volume} {130}},\ \bibinfo {pages} {105001}}\BibitemShut {NoStop}%
\bibitem [{\citenamefont {Golovanov}\ \emph {et~al.}(2017)\citenamefont
  {Golovanov}, \citenamefont {Kostyukov}, \citenamefont {Thomas},\ and\
  \citenamefont {Pukhov}}]{Golovanov2017}%
  \BibitemOpen
  \bibfield  {author} {\bibinfo {author} {\bibnamefont {Golovanov},
  \bibfnamefont {A~A}}, \bibinfo {author} {\bibfnamefont {I.~Yu.}\ \bibnamefont
  {Kostyukov}}, \bibinfo {author} {\bibfnamefont {J.}~\bibnamefont {Thomas}}, \
  and\ \bibinfo {author} {\bibfnamefont {A.}~\bibnamefont {Pukhov}}} (\bibinfo
  {year} {2017}),\ \href {https://doi.org/10.1063/1.4996856} {\bibfield
  {journal} {\bibinfo  {journal} {Phys. Plasmas}\ }\textbf {\bibinfo {volume}
  {24}},\ \bibinfo {pages} {103104}}\BibitemShut {NoStop}%
\bibitem [{\citenamefont {Golovanov}\ \emph {et~al.}(2022)\citenamefont
  {Golovanov}, \citenamefont {Nerush},\ and\ \citenamefont
  {Kostyukov}}]{Golovanov2022}%
  \BibitemOpen
  \bibfield  {author} {\bibinfo {author} {\bibnamefont {Golovanov},
  \bibfnamefont {A~A}}, \bibinfo {author} {\bibfnamefont {E.~N.}\ \bibnamefont
  {Nerush}}, \ and\ \bibinfo {author} {\bibfnamefont {I.~Yu.}\ \bibnamefont
  {Kostyukov}}} (\bibinfo {year} {2022}),\ \href
  {https://doi.org/10.1088/1367-2630/ac53b9} {\bibfield  {journal} {\bibinfo
  {journal} {New J. Phys.}\ }\textbf {\bibinfo {volume} {24}},\ \bibinfo
  {pages} {033011}}\BibitemShut {NoStop}%
\bibitem [{\citenamefont {Gonz{\'a}lez~Caminal}(2022)}]{Caminal2022}%
  \BibitemOpen
  \bibfield  {author} {\bibinfo {author} {\bibnamefont {Gonz{\'a}lez~Caminal},
  \bibfnamefont {P}}} (\bibinfo {year} {2022}),\ \href
  {https://bib-pubdb1.desy.de/record/480259} {\bibinfo {type} {Ph{D}
  {T}hesis}}\ (\bibinfo  {school} {Hamburg University})\BibitemShut {NoStop}%
\bibitem [{\citenamefont {Gonz{\'a}lez~Caminal}\ \emph
  {et~al.}(2024)\citenamefont {Gonz{\'a}lez~Caminal} \emph
  {et~al.}}]{Caminal2024}%
  \BibitemOpen
  \bibfield  {author} {\bibinfo {author} {\bibnamefont {Gonz{\'a}lez~Caminal},
  \bibfnamefont {P}},  \emph {et~al.}} (\bibinfo {year} {2024}),\ \href
  {https://doi.org/10.1103/PhysRevAccelBeams.27.032801} {\bibfield  {journal}
  {\bibinfo  {journal} {Phys. Rev. Accel. Beams}\ }\textbf {\bibinfo {volume}
  {27}},\ \bibinfo {pages} {032801}}\BibitemShut {NoStop}%
\bibitem [{\citenamefont {Gorbunov}\ \emph {et~al.}(2001)\citenamefont
  {Gorbunov}, \citenamefont {Mora},\ and\ \citenamefont
  {Solodov}}]{Gorbunov2001}%
  \BibitemOpen
  \bibfield  {author} {\bibinfo {author} {\bibnamefont {Gorbunov},
  \bibfnamefont {L~M}}, \bibinfo {author} {\bibfnamefont {P.}~\bibnamefont
  {Mora}}, \ and\ \bibinfo {author} {\bibfnamefont {A.~A.}\ \bibnamefont
  {Solodov}}} (\bibinfo {year} {2001}),\ \href
  {https://doi.org/10.1103/PhysRevLett.86.3332} {\bibfield  {journal} {\bibinfo
   {journal} {Phys. Rev. Lett.}\ }\textbf {\bibinfo {volume} {86}},\ \bibinfo
  {pages} {3332–3335}}\BibitemShut {NoStop}%
\bibitem [{\citenamefont {Gorbunov}\ \emph {et~al.}(2003)\citenamefont
  {Gorbunov}, \citenamefont {Mora},\ and\ \citenamefont
  {Solodov}}]{Gorbunov2003}%
  \BibitemOpen
  \bibfield  {author} {\bibinfo {author} {\bibnamefont {Gorbunov},
  \bibfnamefont {L~M}}, \bibinfo {author} {\bibfnamefont {P.}~\bibnamefont
  {Mora}}, \ and\ \bibinfo {author} {\bibfnamefont {A.~A.}\ \bibnamefont
  {Solodov}}} (\bibinfo {year} {2003}),\ \href
  {https://doi.org/10.1063/1.1559011} {\bibfield  {journal} {\bibinfo
  {journal} {Phys. Plasmas}\ }\textbf {\bibinfo {volume} {10}},\ \bibinfo
  {pages} {1124--1134}}\BibitemShut {NoStop}%
\bibitem [{\citenamefont {Gorn}\ \emph {et~al.}(2018)\citenamefont {Gorn},
  \citenamefont {Tuev}, \citenamefont {Petrenko}, \citenamefont {Sosedkin},\
  and\ \citenamefont {Lotov}}]{Gorn2018}%
  \BibitemOpen
  \bibfield  {author} {\bibinfo {author} {\bibnamefont {Gorn}, \bibfnamefont
  {A~A}}, \bibinfo {author} {\bibfnamefont {P.~V.}\ \bibnamefont {Tuev}},
  \bibinfo {author} {\bibfnamefont {A.~V.}\ \bibnamefont {Petrenko}}, \bibinfo
  {author} {\bibfnamefont {A.~P.}\ \bibnamefont {Sosedkin}}, \ and\ \bibinfo
  {author} {\bibfnamefont {K.~V.}\ \bibnamefont {Lotov}}} (\bibinfo {year}
  {2018}),\ \href {https://doi.org/10.1063/1.5039803} {\bibfield  {journal}
  {\bibinfo  {journal} {Phys. Plasmas}\ }\textbf {\bibinfo {volume} {25}},\
  \bibinfo {pages} {063108}}\BibitemShut {NoStop}%
\bibitem [{\citenamefont {G\"otzfried}\ \emph {et~al.}(2020)\citenamefont
  {G\"otzfried}, \citenamefont {D\"opp}, \citenamefont {Gilljohann},
  \citenamefont {Foerster}, \citenamefont {Ding}, \citenamefont {Schindler},
  \citenamefont {Schilling}, \citenamefont {Buck}, \citenamefont {Veisz},\ and\
  \citenamefont {Karsch}}]{Gotzfried2020}%
  \BibitemOpen
  \bibfield  {author} {\bibinfo {author} {\bibnamefont {G\"otzfried},
  \bibfnamefont {J}}, \bibinfo {author} {\bibfnamefont {A.}~\bibnamefont
  {D\"opp}}, \bibinfo {author} {\bibfnamefont {M.~F.}\ \bibnamefont
  {Gilljohann}}, \bibinfo {author} {\bibfnamefont {F.~M.}\ \bibnamefont
  {Foerster}}, \bibinfo {author} {\bibfnamefont {H.}~\bibnamefont {Ding}},
  \bibinfo {author} {\bibfnamefont {S.}~\bibnamefont {Schindler}}, \bibinfo
  {author} {\bibfnamefont {G.}~\bibnamefont {Schilling}}, \bibinfo {author}
  {\bibfnamefont {A.}~\bibnamefont {Buck}}, \bibinfo {author} {\bibfnamefont
  {L.}~\bibnamefont {Veisz}}, \ and\ \bibinfo {author} {\bibfnamefont
  {S.}~\bibnamefont {Karsch}}} (\bibinfo {year} {2020}),\ \href
  {https://doi.org/10.1103/PhysRevX.10.041015} {\bibfield  {journal} {\bibinfo
  {journal} {Phys. Rev. X}\ }\textbf {\bibinfo {volume} {10}},\ \bibinfo
  {pages} {041015}}\BibitemShut {NoStop}%
\bibitem [{\citenamefont {Grebenyuk}\ \emph {et~al.}(2014)\citenamefont
  {Grebenyuk}, \citenamefont {Martinez de~la Ossa}, \citenamefont {Mehrling},\
  and\ \citenamefont {Osterhoff}}]{Grebenyuk2014}%
  \BibitemOpen
  \bibfield  {author} {\bibinfo {author} {\bibnamefont {Grebenyuk},
  \bibfnamefont {J}}, \bibinfo {author} {\bibfnamefont {A.}~\bibnamefont
  {Martinez de~la Ossa}}, \bibinfo {author} {\bibfnamefont {T.}~\bibnamefont
  {Mehrling}}, \ and\ \bibinfo {author} {\bibfnamefont {J.}~\bibnamefont
  {Osterhoff}}} (\bibinfo {year} {2014}),\ \href
  {https://doi.org/10.1016/j.nima.2013.10.054} {\bibfield  {journal} {\bibinfo
  {journal} {Nucl. Instrum. Methods Phys. Res. A}\ }\textbf {\bibinfo {volume}
  {740}},\ \bibinfo {pages} {246--249}}\BibitemShut {NoStop}%
\bibitem [{\citenamefont {Green}\ \emph {et~al.}(2014)\citenamefont {Green}
  \emph {et~al.}}]{Green2014}%
  \BibitemOpen
  \bibfield  {author} {\bibinfo {author} {\bibnamefont {Green}, \bibfnamefont
  {S~Z}},  \emph {et~al.}} (\bibinfo {year} {2014}),\ \href
  {https://doi.org/10.1088/0741-3335/56/8/084011} {\bibfield  {journal}
  {\bibinfo  {journal} {Plasma Phys. Control. Fusion}\ }\textbf {\bibinfo
  {volume} {56}},\ \bibinfo {pages} {084011}}\BibitemShut {NoStop}%
\bibitem [{\citenamefont {Gross}\ \emph {et~al.}(2014)\citenamefont {Gross},
  \citenamefont {Brinkmann}, \citenamefont {Good}, \citenamefont {Gr\"{u}ner},
  \citenamefont {Khojoyan}, \citenamefont {Martinez de~la Ossa}, \citenamefont
  {Osterhoff}, \citenamefont {Pathak}, \citenamefont {Schroeder},\ and\
  \citenamefont {Stephan}}]{Gross2014}%
  \BibitemOpen
  \bibfield  {author} {\bibinfo {author} {\bibnamefont {Gross}, \bibfnamefont
  {M}}, \bibinfo {author} {\bibfnamefont {R.}~\bibnamefont {Brinkmann}},
  \bibinfo {author} {\bibfnamefont {J.~D.}\ \bibnamefont {Good}}, \bibinfo
  {author} {\bibfnamefont {F.}~\bibnamefont {Gr\"{u}ner}}, \bibinfo {author}
  {\bibfnamefont {M.}~\bibnamefont {Khojoyan}}, \bibinfo {author}
  {\bibfnamefont {A.}~\bibnamefont {Martinez de~la Ossa}}, \bibinfo {author}
  {\bibfnamefont {J.}~\bibnamefont {Osterhoff}}, \bibinfo {author}
  {\bibfnamefont {G.}~\bibnamefont {Pathak}}, \bibinfo {author} {\bibfnamefont
  {C.}~\bibnamefont {Schroeder}}, \ and\ \bibinfo {author} {\bibfnamefont
  {F.}~\bibnamefont {Stephan}}} (\bibinfo {year} {2014}),\ \href
  {https://doi.org/10.1016/j.nima.2013.11.042} {\bibfield  {journal} {\bibinfo
  {journal} {Nucl. Instrum. Methods Phys. Res. A}\ }\textbf {\bibinfo {volume}
  {740}},\ \bibinfo {pages} {74--80}}\BibitemShut {NoStop}%
\bibitem [{\citenamefont {Gross}\ \emph {et~al.}(2018)\citenamefont {Gross}
  \emph {et~al.}}]{Gross2018}%
  \BibitemOpen
  \bibfield  {author} {\bibinfo {author} {\bibnamefont {Gross}, \bibfnamefont
  {M}},  \emph {et~al.}} (\bibinfo {year} {2018}),\ \href
  {https://doi.org/10.1103/physrevlett.120.144802} {\bibfield  {journal}
  {\bibinfo  {journal} {Phys. Rev. Lett.}\ }\textbf {\bibinfo {volume} {120}},\
  \bibinfo {pages} {144802}}\BibitemShut {NoStop}%
\bibitem [{\citenamefont {Grote}\ \emph {et~al.}(2005)\citenamefont {Grote},
  \citenamefont {Friedman}, \citenamefont {Vay},\ and\ \citenamefont
  {Haber}}]{Grote2005}%
  \BibitemOpen
  \bibfield  {author} {\bibinfo {author} {\bibnamefont {Grote}, \bibfnamefont
  {D~P}}, \bibinfo {author} {\bibfnamefont {A.}~\bibnamefont {Friedman}},
  \bibinfo {author} {\bibfnamefont {J.-L.}\ \bibnamefont {Vay}}, \ and\
  \bibinfo {author} {\bibfnamefont {I.}~\bibnamefont {Haber}}} (\bibinfo {year}
  {2005}),\ in\ \href {https://doi.org/10.1063/1.1893366} {\emph {\bibinfo
  {booktitle} {{AIP} Conf. Proc.}}},\ Vol.\ \bibinfo {volume} {749}\ (\bibinfo
  {publisher} {{AIP}})\ pp.\ \bibinfo {pages} {55--58}\BibitemShut {NoStop}%
\bibitem [{\citenamefont {Grudiev}\ \emph {et~al.}(2009)\citenamefont
  {Grudiev}, \citenamefont {Calatroni},\ and\ \citenamefont
  {Wuensch}}]{Grudiev2009}%
  \BibitemOpen
  \bibfield  {author} {\bibinfo {author} {\bibnamefont {Grudiev}, \bibfnamefont
  {A}}, \bibinfo {author} {\bibfnamefont {S.}~\bibnamefont {Calatroni}}, \ and\
  \bibinfo {author} {\bibfnamefont {W.}~\bibnamefont {Wuensch}}} (\bibinfo
  {year} {2009}),\ \href {\doibase 10.1103/physrevstab.12.102001} {\bibfield
  {journal} {\bibinfo  {journal} {Phys. Rev. ST Accel. Beams}\ }\textbf
  {\bibinfo {volume} {12}},\ \bibinfo {pages} {102001}}\BibitemShut {NoStop}%
\bibitem [{\citenamefont {Gschwendtner}(2006)}]{Gschwendtner2006}%
  \BibitemOpen
  \bibfield  {author} {\bibinfo {author} {\bibnamefont {Gschwendtner},
  \bibfnamefont {E}}} (\bibinfo {year} {2006}),\ in\ \href
  {https://doi.org/10.1109/NSSMIC.2006.354181} {\emph {\bibinfo {booktitle}
  {2006 IEEE Nuclear Science Symposium Conference Record}}}\ (\bibinfo
  {publisher} {IEEE},\ \bibinfo {address} {New York, NY, United States})\ pp.\
  \bibinfo {pages} {1489--1492}\BibitemShut {NoStop}%
\bibitem [{\citenamefont {Gschwendtner}\ \emph {et~al.}(2016)\citenamefont
  {Gschwendtner} \emph {et~al.}}]{Gschwendtner2016}%
  \BibitemOpen
  \bibfield  {author} {\bibinfo {author} {\bibnamefont {Gschwendtner},
  \bibfnamefont {E}},  \emph {et~al.} (\bibinfo {collaboration} {AWAKE
  Collaboration})} (\bibinfo {year} {2016}),\ \href
  {https://doi.org/10.1016/j.nima.2016.02.026} {\bibfield  {journal} {\bibinfo
  {journal} {Nucl. Instrum. Methods Phys. Res. A}\ }\textbf {\bibinfo {volume}
  {829}},\ \bibinfo {pages} {76--82}}\BibitemShut {NoStop}%
\bibitem [{\citenamefont {Gschwendtner}\ \emph {et~al.}(2018)\citenamefont
  {Gschwendtner} \emph {et~al.}}]{Gschwendtner2018}%
  \BibitemOpen
  \bibfield  {author} {\bibinfo {author} {\bibnamefont {Gschwendtner},
  \bibfnamefont {E}},  \emph {et~al.}} (\bibinfo {year} {2018}),\ \href
  {http://cds.cern.ch/record/2651319} {}\bibinfo {type} {Tech. Rep.}\ \bibinfo
  {number} {CERN-PBC-REPORT-2018-005}\ (\bibinfo  {institution} {CERN},\
  \bibinfo {address} {Geneva, Switzerland})\BibitemShut {NoStop}%
\bibitem [{\citenamefont {Gschwendtner}\ \emph {et~al.}(2019)\citenamefont
  {Gschwendtner} \emph {et~al.}}]{Gschwendtner2019}%
  \BibitemOpen
  \bibfield  {author} {\bibinfo {author} {\bibnamefont {Gschwendtner},
  \bibfnamefont {E}},  \emph {et~al.} (\bibinfo {collaboration} {AWAKE
  Collaboration})} (\bibinfo {year} {2019}),\ \href
  {https://doi.org/10.1098/rsta.2018.0418} {\bibfield  {journal} {\bibinfo
  {journal} {Philos. Trans. R. Soc. A}\ }\textbf {\bibinfo {volume} {377}},\
  \bibinfo {pages} {20180418}}\BibitemShut {NoStop}%
\bibitem [{\citenamefont {Gschwendtner}\ \emph {et~al.}(2022)\citenamefont
  {Gschwendtner} \emph {et~al.}}]{Gschwendtner2022}%
  \BibitemOpen
  \bibfield  {author} {\bibinfo {author} {\bibnamefont {Gschwendtner},
  \bibfnamefont {E}},  \emph {et~al.} (\bibinfo {collaboration} {AWAKE
  Collaboration})} (\bibinfo {year} {2022}),\ \href
  {https://doi.org/10.3390/sym14081680} {\bibfield  {journal} {\bibinfo
  {journal} {Symmetry}\ }\textbf {\bibinfo {volume} {14}},\ \bibinfo {pages}
  {1680}}\BibitemShut {NoStop}%
\bibitem [{\citenamefont {Guetg}\ \emph {et~al.}(2015)\citenamefont {Guetg},
  \citenamefont {Beutner}, \citenamefont {Prat},\ and\ \citenamefont
  {Reiche}}]{Guetg2015}%
  \BibitemOpen
  \bibfield  {author} {\bibinfo {author} {\bibnamefont {Guetg}, \bibfnamefont
  {M~W}}, \bibinfo {author} {\bibfnamefont {B.}~\bibnamefont {Beutner}},
  \bibinfo {author} {\bibfnamefont {E.}~\bibnamefont {Prat}}, \ and\ \bibinfo
  {author} {\bibfnamefont {S.}~\bibnamefont {Reiche}}} (\bibinfo {year}
  {2015}),\ \href {https://doi.org/10.1103/PhysRevSTAB.18.030701} {\bibfield
  {journal} {\bibinfo  {journal} {Phys. Rev. ST Accel. Beams}\ }\textbf
  {\bibinfo {volume} {18}},\ \bibinfo {pages} {030701}}\BibitemShut {NoStop}%
\bibitem [{\citenamefont {Guillaume}\ \emph {et~al.}(2015)\citenamefont
  {Guillaume} \emph {et~al.}}]{Guillaume2015}%
  \BibitemOpen
  \bibfield  {author} {\bibinfo {author} {\bibnamefont {Guillaume},
  \bibfnamefont {E}},  \emph {et~al.}} (\bibinfo {year} {2015}),\ \href
  {https://doi.org/10.1103/PhysRevSTAB.18.061301} {\bibfield  {journal}
  {\bibinfo  {journal} {Phys. Rev. ST Accel. Beams}\ }\textbf {\bibinfo
  {volume} {18}},\ \bibinfo {pages} {061301}}\BibitemShut {NoStop}%
\bibitem [{\citenamefont {Ha}\ \emph {et~al.}(2017)\citenamefont {Ha} \emph
  {et~al.}}]{Ha2017}%
  \BibitemOpen
  \bibfield  {author} {\bibinfo {author} {\bibnamefont {Ha}, \bibfnamefont
  {G}},  \emph {et~al.}} (\bibinfo {year} {2017}),\ \href
  {https://doi.org/10.1103/PhysRevLett.118.104801} {\bibfield  {journal}
  {\bibinfo  {journal} {Phys. Rev. Lett.}\ }\textbf {\bibinfo {volume} {118}},\
  \bibinfo {pages} {104801}}\BibitemShut {NoStop}%
\bibitem [{\citenamefont {Habib}\ \emph {et~al.}(2023)\citenamefont {Habib}
  \emph {et~al.}}]{Habib2023}%
  \BibitemOpen
  \bibfield  {author} {\bibinfo {author} {\bibnamefont {Habib}, \bibfnamefont
  {A~F}},  \emph {et~al.}} (\bibinfo {year} {2023}),\ \href
  {https://doi.org/10.1038/s41467-023-36592-z} {\bibfield  {journal} {\bibinfo
  {journal} {Nat. Commun.}\ }\textbf {\bibinfo {volume} {14}},\ \bibinfo
  {pages} {1054}}\BibitemShut {NoStop}%
\bibitem [{\citenamefont {Hanahoe}\ \emph {et~al.}(2017)\citenamefont
  {Hanahoe}, \citenamefont {Xia}, \citenamefont {Islam}, \citenamefont {Li},
  \citenamefont {Mete-Apsimon}, \citenamefont {Hidding},\ and\ \citenamefont
  {Smith}}]{Hanahoe2017}%
  \BibitemOpen
  \bibfield  {author} {\bibinfo {author} {\bibnamefont {Hanahoe}, \bibfnamefont
  {K}}, \bibinfo {author} {\bibfnamefont {G.}~\bibnamefont {Xia}}, \bibinfo
  {author} {\bibfnamefont {M.}~\bibnamefont {Islam}}, \bibinfo {author}
  {\bibfnamefont {Y.}~\bibnamefont {Li}}, \bibinfo {author} {\bibfnamefont
  {Ö.}\ \bibnamefont {Mete-Apsimon}}, \bibinfo {author} {\bibfnamefont
  {B.}~\bibnamefont {Hidding}}, \ and\ \bibinfo {author} {\bibfnamefont
  {J.}~\bibnamefont {Smith}}} (\bibinfo {year} {2017}),\ \href
  {https://doi.org/10.1063/1.4977449} {\bibfield  {journal} {\bibinfo
  {journal} {Phys. Plasmas}\ }\textbf {\bibinfo {volume} {24}},\ \bibinfo
  {pages} {023120}}\BibitemShut {NoStop}%
\bibitem [{\citenamefont {Heigoldt}\ \emph {et~al.}(2015)\citenamefont
  {Heigoldt}, \citenamefont {Popp}, \citenamefont {Khrennikov}, \citenamefont
  {Wenz}, \citenamefont {Chou}, \citenamefont {Karsch}, \citenamefont
  {Bajlekov}, \citenamefont {Hooker},\ and\ \citenamefont
  {Schmidt}}]{Heigoldt2015}%
  \BibitemOpen
  \bibfield  {author} {\bibinfo {author} {\bibnamefont {Heigoldt},
  \bibfnamefont {M}}, \bibinfo {author} {\bibfnamefont {A.}~\bibnamefont
  {Popp}}, \bibinfo {author} {\bibfnamefont {K.}~\bibnamefont {Khrennikov}},
  \bibinfo {author} {\bibfnamefont {J.}~\bibnamefont {Wenz}}, \bibinfo {author}
  {\bibfnamefont {S.~W.}\ \bibnamefont {Chou}}, \bibinfo {author}
  {\bibfnamefont {S.}~\bibnamefont {Karsch}}, \bibinfo {author} {\bibfnamefont
  {S.~I.}\ \bibnamefont {Bajlekov}}, \bibinfo {author} {\bibfnamefont {S.~M.}\
  \bibnamefont {Hooker}}, \ and\ \bibinfo {author} {\bibfnamefont
  {B.}~\bibnamefont {Schmidt}}} (\bibinfo {year} {2015}),\ \href
  {https://doi.org/10.1103/PhysRevSTAB.18.121302} {\bibfield  {journal}
  {\bibinfo  {journal} {Phys. Rev. ST Accel. Beams}\ }\textbf {\bibinfo
  {volume} {18}},\ \bibinfo {pages} {121302}}\BibitemShut {NoStop}%
\bibitem [{\citenamefont {Hemker}\ \emph {et~al.}(1999)\citenamefont {Hemker},
  \citenamefont {Tsung}, \citenamefont {Decyk}, \citenamefont {Mori},
  \citenamefont {Lee},\ and\ \citenamefont {Katsouleas}}]{Hemker1999}%
  \BibitemOpen
  \bibfield  {author} {\bibinfo {author} {\bibnamefont {Hemker}, \bibfnamefont
  {R~G}}, \bibinfo {author} {\bibfnamefont {F.~S.}\ \bibnamefont {Tsung}},
  \bibinfo {author} {\bibfnamefont {V.~K.}\ \bibnamefont {Decyk}}, \bibinfo
  {author} {\bibfnamefont {W.~B.}\ \bibnamefont {Mori}}, \bibinfo {author}
  {\bibfnamefont {S.}~\bibnamefont {Lee}}, \ and\ \bibinfo {author}
  {\bibfnamefont {T.}~\bibnamefont {Katsouleas}}} (\bibinfo {year} {1999}),\
  in\ \href {https://doi.org/10.1109/PAC.1999.792407} {\emph {\bibinfo
  {booktitle} {Proceedings of the 1999 Particle Accelerator Conf.}}}\ (\bibinfo
   {publisher} {IEEE},\ \bibinfo {address} {New York, NY, United States})\ pp.\
  \bibinfo {pages} {3672--3674}\BibitemShut {NoStop}%
\bibitem [{\citenamefont {Hessami}\ \emph {et~al.}(2024)\citenamefont
  {Hessami}, \citenamefont {Morgan}, \citenamefont {Robles}, \citenamefont
  {Larsen}, \citenamefont {Marinelli},\ and\ \citenamefont
  {Emma}}]{Hessami2024}%
  \BibitemOpen
  \bibfield  {author} {\bibinfo {author} {\bibnamefont {Hessami}, \bibfnamefont
  {R}}, \bibinfo {author} {\bibfnamefont {J.}~\bibnamefont {Morgan}}, \bibinfo
  {author} {\bibfnamefont {R.}~\bibnamefont {Robles}}, \bibinfo {author}
  {\bibfnamefont {K.~A.}\ \bibnamefont {Larsen}}, \bibinfo {author}
  {\bibfnamefont {A.}~\bibnamefont {Marinelli}}, \ and\ \bibinfo {author}
  {\bibfnamefont {C.}~\bibnamefont {Emma}}} (\bibinfo {year} {2024}),\ \href
  {https://doi.org/10.1103/PhysRevAccelBeams.27.070701} {\bibfield  {journal}
  {\bibinfo  {journal} {Phys. Rev. Accel. Beams}\ }\textbf {\bibinfo {volume}
  {27}},\ \bibinfo {pages} {070701}}\BibitemShut {NoStop}%
\bibitem [{\citenamefont {Hidding}\ \emph
  {et~al.}(2019{\natexlab{a}})\citenamefont {Hidding}, \citenamefont {Foster},
  \citenamefont {Hogan}, \citenamefont {Muggli},\ and\ \citenamefont
  {Rosenzweig}}]{Hidding2019b}%
  \BibitemOpen
  \bibfield  {author} {\bibinfo {author} {\bibnamefont {Hidding}, \bibfnamefont
  {B}}, \bibinfo {author} {\bibfnamefont {B.}~\bibnamefont {Foster}}, \bibinfo
  {author} {\bibfnamefont {M.~J.}\ \bibnamefont {Hogan}}, \bibinfo {author}
  {\bibfnamefont {P.}~\bibnamefont {Muggli}}, \ and\ \bibinfo {author}
  {\bibfnamefont {J.~B.}\ \bibnamefont {Rosenzweig}}} (\bibinfo {year}
  {2019}{\natexlab{a}}),\ \href {https://doi.org/10.1098/rsta.2019.0215}
  {\bibfield  {journal} {\bibinfo  {journal} {Philos. Trans. R. Soc. A}\
  }\textbf {\bibinfo {volume} {377}},\ \bibinfo {pages} {20190215}}\BibitemShut
  {NoStop}%
\bibitem [{\citenamefont {Hidding}\ \emph {et~al.}(2010)\citenamefont
  {Hidding}, \citenamefont {K\"{o}nigstein}, \citenamefont {Osterholz},
  \citenamefont {Karsch}, \citenamefont {Willi},\ and\ \citenamefont
  {Pretzler}}]{Hidding2010}%
  \BibitemOpen
  \bibfield  {author} {\bibinfo {author} {\bibnamefont {Hidding}, \bibfnamefont
  {B}}, \bibinfo {author} {\bibfnamefont {T.}~\bibnamefont {K\"{o}nigstein}},
  \bibinfo {author} {\bibfnamefont {J.}~\bibnamefont {Osterholz}}, \bibinfo
  {author} {\bibfnamefont {S.}~\bibnamefont {Karsch}}, \bibinfo {author}
  {\bibfnamefont {O.}~\bibnamefont {Willi}}, \ and\ \bibinfo {author}
  {\bibfnamefont {G.}~\bibnamefont {Pretzler}}} (\bibinfo {year} {2010}),\
  \href {https://doi.org/10.1103/PhysRevLett.104.195002} {\bibfield  {journal}
  {\bibinfo  {journal} {Phys. Rev. Lett.}\ }\textbf {\bibinfo {volume} {104}},\
  \bibinfo {pages} {195002}}\BibitemShut {NoStop}%
\bibitem [{\citenamefont {Hidding}\ \emph {et~al.}(2012)\citenamefont
  {Hidding}, \citenamefont {Pretzler}, \citenamefont {Rosenzweig},
  \citenamefont {K\"{o}nigstein}, \citenamefont {Schiller},\ and\ \citenamefont
  {Bruhwiler}}]{Hidding2012}%
  \BibitemOpen
  \bibfield  {author} {\bibinfo {author} {\bibnamefont {Hidding}, \bibfnamefont
  {B}}, \bibinfo {author} {\bibfnamefont {G.}~\bibnamefont {Pretzler}},
  \bibinfo {author} {\bibfnamefont {J.~B.}\ \bibnamefont {Rosenzweig}},
  \bibinfo {author} {\bibfnamefont {T.}~\bibnamefont {K\"{o}nigstein}},
  \bibinfo {author} {\bibfnamefont {D.}~\bibnamefont {Schiller}}, \ and\
  \bibinfo {author} {\bibfnamefont {D.~L.}\ \bibnamefont {Bruhwiler}}}
  (\bibinfo {year} {2012}),\ \href
  {https://doi.org/10.1103/physrevlett.108.035001} {\bibfield  {journal}
  {\bibinfo  {journal} {Phys. Rev. Lett.}\ }\textbf {\bibinfo {volume} {108}},\
  \bibinfo {pages} {035001}}\BibitemShut {NoStop}%
\bibitem [{\citenamefont {Hidding}\ \emph
  {et~al.}(2019{\natexlab{b}})\citenamefont {Hidding} \emph
  {et~al.}}]{Hidding2019a}%
  \BibitemOpen
  \bibfield  {author} {\bibinfo {author} {\bibnamefont {Hidding}, \bibfnamefont
  {B}},  \emph {et~al.}} (\bibinfo {year} {2019}{\natexlab{b}}),\ \href
  {https://doi.org/10.3390/app9132626} {\bibfield  {journal} {\bibinfo
  {journal} {Appl. Sci.}\ }\textbf {\bibinfo {volume} {9}},\ \bibinfo {pages}
  {2626}}\BibitemShut {NoStop}%
\bibitem [{\citenamefont {Hidding}\ \emph {et~al.}(2023)\citenamefont {Hidding}
  \emph {et~al.}}]{Hidding2023}%
  \BibitemOpen
  \bibfield  {author} {\bibinfo {author} {\bibnamefont {Hidding}, \bibfnamefont
  {B}},  \emph {et~al.}} (\bibinfo {year} {2023}),\ \href
  {https://doi.org/10.3390/photonics10020099} {\bibfield  {journal} {\bibinfo
  {journal} {Photonics}\ }\textbf {\bibinfo {volume} {10}},\ \bibinfo {pages}
  {99}}\BibitemShut {NoStop}%
\bibitem [{\citenamefont {Hogan}(2016)}]{Hogan2016}%
  \BibitemOpen
  \bibfield  {author} {\bibinfo {author} {\bibnamefont {Hogan}, \bibfnamefont
  {M~J}}} (\bibinfo {year} {2016}),\ \href
  {https://doi.org/10.1142/S1793626816300036} {\bibfield  {journal} {\bibinfo
  {journal} {Rev. Accel. Sci. Tech.}\ }\textbf {\bibinfo {volume} {9}},\
  \bibinfo {pages} {209--233}}\BibitemShut {NoStop}%
\bibitem [{\citenamefont {Hogan}\ \emph {et~al.}(2000)\citenamefont {Hogan}
  \emph {et~al.}}]{Hogan2000}%
  \BibitemOpen
  \bibfield  {author} {\bibinfo {author} {\bibnamefont {Hogan}, \bibfnamefont
  {M~J}},  \emph {et~al.}} (\bibinfo {year} {2000}),\ \href
  {https://doi.org/10.1063/1.874059} {\bibfield  {journal} {\bibinfo  {journal}
  {Phys. Plasmas}\ }\textbf {\bibinfo {volume} {7}},\ \bibinfo {pages}
  {2241--2248}}\BibitemShut {NoStop}%
\bibitem [{\citenamefont {Hogan}\ \emph {et~al.}(2003)\citenamefont {Hogan}
  \emph {et~al.}}]{Hogan2003}%
  \BibitemOpen
  \bibfield  {author} {\bibinfo {author} {\bibnamefont {Hogan}, \bibfnamefont
  {M~J}},  \emph {et~al.}} (\bibinfo {year} {2003}),\ \href
  {https://doi.org/10.1103/physrevlett.90.205002} {\bibfield  {journal}
  {\bibinfo  {journal} {Phys. Rev. Lett.}\ }\textbf {\bibinfo {volume} {90}},\
  \bibinfo {pages} {205002}}\BibitemShut {NoStop}%
\bibitem [{\citenamefont {Hogan}\ \emph {et~al.}(2005)\citenamefont {Hogan}
  \emph {et~al.}}]{Hogan2005}%
  \BibitemOpen
  \bibfield  {author} {\bibinfo {author} {\bibnamefont {Hogan}, \bibfnamefont
  {M~J}},  \emph {et~al.}} (\bibinfo {year} {2005}),\ \href
  {https://doi.org/10.1103/physrevlett.95.054802} {\bibfield  {journal}
  {\bibinfo  {journal} {Phys. Rev. Lett.}\ }\textbf {\bibinfo {volume} {95}},\
  \bibinfo {pages} {054802}}\BibitemShut {NoStop}%
\bibitem [{\citenamefont {Hogan}\ \emph {et~al.}(2010)\citenamefont {Hogan}
  \emph {et~al.}}]{Hogan2010}%
  \BibitemOpen
  \bibfield  {author} {\bibinfo {author} {\bibnamefont {Hogan}, \bibfnamefont
  {M~J}},  \emph {et~al.}} (\bibinfo {year} {2010}),\ \href
  {https://iopscience.iop.org/article/10.1088/1367-2630/12/5/055030} {\bibfield
   {journal} {\bibinfo  {journal} {New J. Phys.}\ }\textbf {\bibinfo {volume}
  {12}},\ \bibinfo {pages} {055030}}\BibitemShut {NoStop}%
\bibitem [{\citenamefont {Hooker}\ \emph {et~al.}(2014)\citenamefont {Hooker},
  \citenamefont {Bartolini}, \citenamefont {Mangles}, \citenamefont
  {T\"{u}nnermann}, \citenamefont {Corner}, \citenamefont {Limpert},
  \citenamefont {Seryi},\ and\ \citenamefont {Walczak}}]{Hooker2014}%
  \BibitemOpen
  \bibfield  {author} {\bibinfo {author} {\bibnamefont {Hooker}, \bibfnamefont
  {S~M}}, \bibinfo {author} {\bibfnamefont {R.}~\bibnamefont {Bartolini}},
  \bibinfo {author} {\bibfnamefont {S.~P.~D.}\ \bibnamefont {Mangles}},
  \bibinfo {author} {\bibfnamefont {A.}~\bibnamefont {T\"{u}nnermann}},
  \bibinfo {author} {\bibfnamefont {L.}~\bibnamefont {Corner}}, \bibinfo
  {author} {\bibfnamefont {J.}~\bibnamefont {Limpert}}, \bibinfo {author}
  {\bibfnamefont {A.}~\bibnamefont {Seryi}}, \ and\ \bibinfo {author}
  {\bibfnamefont {R.}~\bibnamefont {Walczak}}} (\bibinfo {year} {2014}),\ \href
  {https://doi.org/10.1088/0953-4075/47/23/234003} {\bibfield  {journal}
  {\bibinfo  {journal} {J. Phys. B}\ }\textbf {\bibinfo {volume} {47}},\
  \bibinfo {pages} {234003}}\BibitemShut {NoStop}%
\bibitem [{\citenamefont {Huang}\ \emph {et~al.}(2006)\citenamefont {Huang},
  \citenamefont {Decyk}, \citenamefont {Ren}, \citenamefont {Zhou},
  \citenamefont {Lu}, \citenamefont {Mori}, \citenamefont {Cooley},
  \citenamefont {Antonsen},\ and\ \citenamefont {Katsouleas}}]{Huang2006}%
  \BibitemOpen
  \bibfield  {author} {\bibinfo {author} {\bibnamefont {Huang}, \bibfnamefont
  {C}}, \bibinfo {author} {\bibfnamefont {V.~K.}\ \bibnamefont {Decyk}},
  \bibinfo {author} {\bibfnamefont {C.}~\bibnamefont {Ren}}, \bibinfo {author}
  {\bibfnamefont {M.}~\bibnamefont {Zhou}}, \bibinfo {author} {\bibfnamefont
  {W.}~\bibnamefont {Lu}}, \bibinfo {author} {\bibfnamefont {W.~B.}\
  \bibnamefont {Mori}}, \bibinfo {author} {\bibfnamefont {J.~H.}\ \bibnamefont
  {Cooley}}, \bibinfo {author} {\bibfnamefont {T.~M.}\ \bibnamefont
  {Antonsen}}, \ and\ \bibinfo {author} {\bibfnamefont {T.}~\bibnamefont
  {Katsouleas}}} (\bibinfo {year} {2006}),\ \href
  {https://doi.org/10.1016/j.jcp.2006.01.039} {\bibfield  {journal} {\bibinfo
  {journal} {J. Comput. Phys.}\ }\textbf {\bibinfo {volume} {217}},\ \bibinfo
  {pages} {658--679}}\BibitemShut {NoStop}%
\bibitem [{\citenamefont {Huang}\ \emph {et~al.}(2007)\citenamefont {Huang}
  \emph {et~al.}}]{Huang2007}%
  \BibitemOpen
  \bibfield  {author} {\bibinfo {author} {\bibnamefont {Huang}, \bibfnamefont
  {C}},  \emph {et~al.}} (\bibinfo {year} {2007}),\ \href
  {https://doi.org/10.1103/physrevlett.99.255001} {\bibfield  {journal}
  {\bibinfo  {journal} {Phys. Rev. Lett.}\ }\textbf {\bibinfo {volume} {99}},\
  \bibinfo {pages} {255001}}\BibitemShut {NoStop}%
\bibitem [{\citenamefont {Hue}(2020)}]{Hue2020}%
  \BibitemOpen
  \bibfield  {author} {\bibinfo {author} {\bibnamefont {Hue}, \bibfnamefont
  {C}}} (\bibinfo {year} {2020}),\ \href
  {https://theses.hal.science/tel-03592844} {\bibinfo {type} {Ph{D} {T}hesis}}\
  (\bibinfo  {school} {{Institut Polytechnique de Paris}})\BibitemShut
  {NoStop}%
\bibitem [{\citenamefont {Hue}\ \emph {et~al.}(2023)\citenamefont {Hue},
  \citenamefont {Golovanov}, \citenamefont {Tata}, \citenamefont {Corde},\ and\
  \citenamefont {Malka}}]{Hue2023}%
  \BibitemOpen
  \bibfield  {author} {\bibinfo {author} {\bibnamefont {Hue}, \bibfnamefont
  {C}}, \bibinfo {author} {\bibfnamefont {A.}~\bibnamefont {Golovanov}},
  \bibinfo {author} {\bibfnamefont {S.}~\bibnamefont {Tata}}, \bibinfo {author}
  {\bibfnamefont {S.}~\bibnamefont {Corde}}, \ and\ \bibinfo {author}
  {\bibfnamefont {V.}~\bibnamefont {Malka}}} (\bibinfo {year} {2023}),\ \href
  {https://doi.org/10.1017/S0022377823001162} {\bibfield  {journal} {\bibinfo
  {journal} {J. Plasma Phys.}\ }\textbf {\bibinfo {volume} {89}},\ \bibinfo
  {pages} {965890502}}\BibitemShut {NoStop}%
\bibitem [{\citenamefont {Hue}\ \emph {et~al.}(2021)\citenamefont {Hue},
  \citenamefont {Cao}, \citenamefont {Andriyash}, \citenamefont {Knetsch},
  \citenamefont {Hogan}, \citenamefont {Adli}, \citenamefont {Gessner},\ and\
  \citenamefont {Corde}}]{Hue2021}%
  \BibitemOpen
  \bibfield  {author} {\bibinfo {author} {\bibnamefont {Hue}, \bibfnamefont
  {C~S}}, \bibinfo {author} {\bibfnamefont {G.~J.}\ \bibnamefont {Cao}},
  \bibinfo {author} {\bibfnamefont {I.~A.}\ \bibnamefont {Andriyash}}, \bibinfo
  {author} {\bibfnamefont {A.}~\bibnamefont {Knetsch}}, \bibinfo {author}
  {\bibfnamefont {M.~J.}\ \bibnamefont {Hogan}}, \bibinfo {author}
  {\bibfnamefont {E.}~\bibnamefont {Adli}}, \bibinfo {author} {\bibfnamefont
  {S.}~\bibnamefont {Gessner}}, \ and\ \bibinfo {author} {\bibfnamefont
  {S.}~\bibnamefont {Corde}}} (\bibinfo {year} {2021}),\ \href
  {https://doi.org/10.1103/PhysRevResearch.3.043063} {\bibfield  {journal}
  {\bibinfo  {journal} {Phys. Rev. Res.}\ }\textbf {\bibinfo {volume} {3}},\
  \bibinfo {pages} {043063}}\BibitemShut {NoStop}%
\bibitem [{\citenamefont {Humphries}(1990)}]{Humphries1990}%
  \BibitemOpen
  \bibfield  {author} {\bibinfo {author} {\bibnamefont {Humphries},
  \bibfnamefont {S}}} (\bibinfo {year} {1990}),\ \href@noop {} {\emph {\bibinfo
  {title} {Charged Particle Beams}}}\ (\bibinfo  {publisher} {John Wiley \&
  Sons},\ \bibinfo {address} {New York, NY, United States})\BibitemShut
  {NoStop}%
\bibitem [{\citenamefont {Israeli}\ \emph {et~al.}(2016)\citenamefont
  {Israeli}, \citenamefont {Vieira}, \citenamefont {Reiche}, \citenamefont
  {Pedrozzi},\ and\ \citenamefont {Muggli}}]{Israeli2016}%
  \BibitemOpen
  \bibfield  {author} {\bibinfo {author} {\bibnamefont {Israeli}, \bibfnamefont
  {Y}}, \bibinfo {author} {\bibfnamefont {J.}~\bibnamefont {Vieira}}, \bibinfo
  {author} {\bibfnamefont {S.}~\bibnamefont {Reiche}}, \bibinfo {author}
  {\bibfnamefont {M.}~\bibnamefont {Pedrozzi}}, \ and\ \bibinfo {author}
  {\bibfnamefont {P.}~\bibnamefont {Muggli}}} (\bibinfo {year} {2016}),\ \href
  {https://doi.org/10.48550/arXiv.1601.00480} {}\Eprint
  {http://arxiv.org/abs/1601.00480} {arXiv:1601.00480} \BibitemShut {NoStop}%
\bibitem [{\citenamefont {Jain}(2019)}]{Jain2019}%
  \BibitemOpen
  \bibfield  {author} {\bibinfo {author} {\bibnamefont {Jain}, \bibfnamefont
  {N}}} (\bibinfo {year} {2019}),\ \href {https://doi.org/10.1063/1.5065375}
  {\bibfield  {journal} {\bibinfo  {journal} {Phys. Plasmas}\ }\textbf
  {\bibinfo {volume} {26}},\ \bibinfo {pages} {023107}}\BibitemShut {NoStop}%
\bibitem [{\citenamefont {Jain}\ \emph
  {et~al.}(2015{\natexlab{a}})\citenamefont {Jain}, \citenamefont {Antonsen},\
  and\ \citenamefont {Palastro}}]{Jain2015}%
  \BibitemOpen
  \bibfield  {author} {\bibinfo {author} {\bibnamefont {Jain}, \bibfnamefont
  {N}}, \bibinfo {author} {\bibfnamefont {T.~M.}\ \bibnamefont {Antonsen}}, \
  and\ \bibinfo {author} {\bibfnamefont {J.~P.}\ \bibnamefont {Palastro}}}
  (\bibinfo {year} {2015}{\natexlab{a}}),\ \href
  {https://link.aps.org/doi/10.1103/PhysRevLett.115.195001} {\bibfield
  {journal} {\bibinfo  {journal} {Phys. Rev. Lett.}\ }\textbf {\bibinfo
  {volume} {115}},\ \bibinfo {pages} {195001}}\BibitemShut {NoStop}%
\bibitem [{\citenamefont {Jain}\ \emph
  {et~al.}(2015{\natexlab{b}})\citenamefont {Jain}, \citenamefont {Palastro},
  \citenamefont {{Antonsen, Jr.}}, \citenamefont {Mori},\ and\ \citenamefont
  {An}}]{Jain2015b}%
  \BibitemOpen
  \bibfield  {author} {\bibinfo {author} {\bibnamefont {Jain}, \bibfnamefont
  {N}}, \bibinfo {author} {\bibfnamefont {J.}~\bibnamefont {Palastro}},
  \bibinfo {author} {\bibfnamefont {T.~M.}\ \bibnamefont {{Antonsen, Jr.}}},
  \bibinfo {author} {\bibfnamefont {W.~B.}\ \bibnamefont {Mori}}, \ and\
  \bibinfo {author} {\bibfnamefont {W.}~\bibnamefont {An}}} (\bibinfo {year}
  {2015}{\natexlab{b}}),\ \href {https://doi.org/10.1063/1.4907159} {\bibfield
  {journal} {\bibinfo  {journal} {Phys. Plasmas}\ }\textbf {\bibinfo {volume}
  {22}},\ \bibinfo {pages} {023103}}\BibitemShut {NoStop}%
\bibitem [{\citenamefont {Jakobsson}\ \emph {et~al.}(2019)\citenamefont
  {Jakobsson}, \citenamefont {Bonatto}, \citenamefont {Li}, \citenamefont
  {Zhao}, \citenamefont {Nunes}, \citenamefont {Williamson}, \citenamefont
  {Xia},\ and\ \citenamefont {Tajima}}]{Jakobsson2019}%
  \BibitemOpen
  \bibfield  {author} {\bibinfo {author} {\bibnamefont {Jakobsson},
  \bibfnamefont {O}}, \bibinfo {author} {\bibfnamefont {A.}~\bibnamefont
  {Bonatto}}, \bibinfo {author} {\bibfnamefont {Y.}~\bibnamefont {Li}},
  \bibinfo {author} {\bibfnamefont {Y.}~\bibnamefont {Zhao}}, \bibinfo {author}
  {\bibfnamefont {R.~P.}\ \bibnamefont {Nunes}}, \bibinfo {author}
  {\bibfnamefont {B.}~\bibnamefont {Williamson}}, \bibinfo {author}
  {\bibfnamefont {G.}~\bibnamefont {Xia}}, \ and\ \bibinfo {author}
  {\bibfnamefont {T.}~\bibnamefont {Tajima}}} (\bibinfo {year} {2019}),\ \href
  {https://iopscience.iop.org/article/10.1088/1361-6587/ab4cfb} {\bibfield
  {journal} {\bibinfo  {journal} {Plasma Phys. Control. Fusion}\ }\textbf
  {\bibinfo {volume} {61}},\ \bibinfo {pages} {124002}}\BibitemShut {NoStop}%
\bibitem [{\citenamefont {Jiang}\ \emph {et~al.}(2012)\citenamefont {Jiang},
  \citenamefont {Jing}, \citenamefont {Schoessow}, \citenamefont {Power},\ and\
  \citenamefont {Gai}}]{Jiang2012}%
  \BibitemOpen
  \bibfield  {author} {\bibinfo {author} {\bibnamefont {Jiang}, \bibfnamefont
  {B}}, \bibinfo {author} {\bibfnamefont {C.}~\bibnamefont {Jing}}, \bibinfo
  {author} {\bibfnamefont {P.}~\bibnamefont {Schoessow}}, \bibinfo {author}
  {\bibfnamefont {J.}~\bibnamefont {Power}}, \ and\ \bibinfo {author}
  {\bibfnamefont {W.}~\bibnamefont {Gai}}} (\bibinfo {year} {2012}),\ \href
  {https://link.aps.org/doi/10.1103/PhysRevSTAB.15.011301} {\bibfield
  {journal} {\bibinfo  {journal} {Phys. Rev. ST Accel. Beams}\ }\textbf
  {\bibinfo {volume} {15}},\ \bibinfo {pages} {011301}}\BibitemShut {NoStop}%
\bibitem [{\citenamefont {Jing}\ \emph {et~al.}(2007)\citenamefont {Jing},
  \citenamefont {Kanareykin}, \citenamefont {Power}, \citenamefont {Conde},
  \citenamefont {Yusof}, \citenamefont {Schoessow},\ and\ \citenamefont
  {Gai}}]{Jing2007}%
  \BibitemOpen
  \bibfield  {author} {\bibinfo {author} {\bibnamefont {Jing}, \bibfnamefont
  {C}}, \bibinfo {author} {\bibfnamefont {A.}~\bibnamefont {Kanareykin}},
  \bibinfo {author} {\bibfnamefont {J.~G.}\ \bibnamefont {Power}}, \bibinfo
  {author} {\bibfnamefont {M.}~\bibnamefont {Conde}}, \bibinfo {author}
  {\bibfnamefont {Z.}~\bibnamefont {Yusof}}, \bibinfo {author} {\bibfnamefont
  {P.}~\bibnamefont {Schoessow}}, \ and\ \bibinfo {author} {\bibfnamefont
  {W.}~\bibnamefont {Gai}}} (\bibinfo {year} {2007}),\ \href
  {https://link.aps.org/doi/10.1103/PhysRevLett.98.144801} {\bibfield
  {journal} {\bibinfo  {journal} {Phys. Rev. Lett.}\ }\textbf {\bibinfo
  {volume} {98}},\ \bibinfo {pages} {144801}}\BibitemShut {NoStop}%
\bibitem [{\citenamefont {Jing}\ \emph {et~al.}(2011)\citenamefont {Jing},
  \citenamefont {Power}, \citenamefont {Conde}, \citenamefont {Liu},
  \citenamefont {Yusof}, \citenamefont {Kanareykin},\ and\ \citenamefont
  {Gai}}]{Jing2011}%
  \BibitemOpen
  \bibfield  {author} {\bibinfo {author} {\bibnamefont {Jing}, \bibfnamefont
  {C}}, \bibinfo {author} {\bibfnamefont {J.~G.}\ \bibnamefont {Power}},
  \bibinfo {author} {\bibfnamefont {M.}~\bibnamefont {Conde}}, \bibinfo
  {author} {\bibfnamefont {W.}~\bibnamefont {Liu}}, \bibinfo {author}
  {\bibfnamefont {Z.}~\bibnamefont {Yusof}}, \bibinfo {author} {\bibfnamefont
  {A.}~\bibnamefont {Kanareykin}}, \ and\ \bibinfo {author} {\bibfnamefont
  {W.}~\bibnamefont {Gai}}} (\bibinfo {year} {2011}),\ \href
  {https://link.aps.org/doi/10.1103/PhysRevSTAB.14.021302} {\bibfield
  {journal} {\bibinfo  {journal} {Phys. Rev. ST Accel. Beams}\ }\textbf
  {\bibinfo {volume} {14}},\ \bibinfo {pages} {021302}}\BibitemShut {NoStop}%
\bibitem [{\citenamefont {Johnson}\ \emph {et~al.}(2006)\citenamefont {Johnson}
  \emph {et~al.}}]{Johnson2006}%
  \BibitemOpen
  \bibfield  {author} {\bibinfo {author} {\bibnamefont {Johnson}, \bibfnamefont
  {D~K}},  \emph {et~al.}} (\bibinfo {year} {2006}),\ \href
  {https://doi.org/10.1103/physrevlett.97.175003} {\bibfield  {journal}
  {\bibinfo  {journal} {Phys. Rev. Lett.}\ }\textbf {\bibinfo {volume} {97}},\
  \bibinfo {pages} {175003}}\BibitemShut {NoStop}%
\bibitem [{\citenamefont {{Joint Commission on Atomic
  Energy}}(1972)}]{AEC1972}%
  \BibitemOpen
  \bibfield  {author} {\bibinfo {author} {\bibnamefont {{Joint Commission on
  Atomic Energy}},}} (\bibinfo {year} {1972}),\ in\ \href
  {https://babel.hathitrust.org/cgi/pt?id=umn.31951d034406374&view=1up&seq=601}
  {\emph {\bibinfo {booktitle} {AEC authorizing legislation, fiscal year 1973:
  Hearings, Ninety-second Congress, second session}}}\ (\bibinfo  {publisher}
  {{U.S. Government Printing Office}})\ pp.\ \bibinfo {pages}
  {2005--2046}\BibitemShut {NoStop}%
\bibitem [{\citenamefont {Joshi}(2001)}]{Joshi2001}%
  \BibitemOpen
  \bibfield  {author} {\bibinfo {author} {\bibnamefont {Joshi}, \bibfnamefont
  {C}}} (\bibinfo {year} {2001}),\ in\ \href
  {https://doi.org/10.1063/1.1384330} {\emph {\bibinfo {booktitle} {{AIP} Conf.
  Proc.}}},\ Vol.\ \bibinfo {volume} {569}\ (\bibinfo  {publisher} {{AIP}})\
  pp.\ \bibinfo {pages} {26--30}\BibitemShut {NoStop}%
\bibitem [{\citenamefont {Joshi}(2007)}]{Joshi2007}%
  \BibitemOpen
  \bibfield  {author} {\bibinfo {author} {\bibnamefont {Joshi}, \bibfnamefont
  {C}}} (\bibinfo {year} {2007}),\ \href {\doibase 10.1063/1.2721965}
  {\bibfield  {journal} {\bibinfo  {journal} {Phys. Plasmas}\ }\textbf
  {\bibinfo {volume} {14}},\ \bibinfo {pages} {055501}}\BibitemShut {NoStop}%
\bibitem [{\citenamefont {Joshi}(2012)}]{Joshi2012}%
  \BibitemOpen
  \bibfield  {author} {\bibinfo {author} {\bibnamefont {Joshi}, \bibfnamefont
  {C}}} (\bibinfo {year} {2012}),\ in\ \href
  {https://doi.org/10.1063/1.4773677} {\emph {\bibinfo {booktitle} {{AIP} Conf.
  Proc.}}},\ Vol.~\bibinfo {volume} {91}\ (\bibinfo  {publisher} {{AIP}})\ pp.\
  \bibinfo {pages} {61--66}\BibitemShut {NoStop}%
\bibitem [{\citenamefont {Joshi}\ and\ \citenamefont
  {Katsouleas}(2003)}]{Joshi2003}%
  \BibitemOpen
  \bibfield  {author} {\bibinfo {author} {\bibnamefont {Joshi}, \bibfnamefont
  {C}}, \ and\ \bibinfo {author} {\bibfnamefont {T.}~\bibnamefont
  {Katsouleas}}} (\bibinfo {year} {2003}),\ \href {\doibase 10.1063/1.1595054}
  {\bibfield  {journal} {\bibinfo  {journal} {Phys. Today}\ }\textbf {\bibinfo
  {volume} {56}},\ \bibinfo {pages} {47--53}}\BibitemShut {NoStop}%
\bibitem [{\citenamefont {Joshi}\ \emph {et~al.}(2025)\citenamefont {Joshi},
  \citenamefont {Mori},\ and\ \citenamefont {Hogan}}]{Joshi2025}%
  \BibitemOpen
  \bibfield  {author} {\bibinfo {author} {\bibnamefont {Joshi}, \bibfnamefont
  {C}}, \bibinfo {author} {\bibfnamefont {W.~B.}\ \bibnamefont {Mori}}, \ and\
  \bibinfo {author} {\bibfnamefont {M.~J.}\ \bibnamefont {Hogan}}} (\bibinfo
  {year} {2025}),\ \href {\doibase 10.1038/s41567-025-02910-z} {\bibfield
  {journal} {\bibinfo  {journal} {Nat. Physics}\ }\textbf {\bibinfo {volume}
  {21}},\ \bibinfo {pages} {885--894}}\BibitemShut {NoStop}%
\bibitem [{\citenamefont {Joshi}\ \emph {et~al.}(1981)\citenamefont {Joshi},
  \citenamefont {Tajima}, \citenamefont {Dawson}, \citenamefont {Baldis},\ and\
  \citenamefont {Ebrahim}}]{Joshi1981}%
  \BibitemOpen
  \bibfield  {author} {\bibinfo {author} {\bibnamefont {Joshi}, \bibfnamefont
  {C}}, \bibinfo {author} {\bibfnamefont {T.}~\bibnamefont {Tajima}}, \bibinfo
  {author} {\bibfnamefont {J.~M.}\ \bibnamefont {Dawson}}, \bibinfo {author}
  {\bibfnamefont {H.~A.}\ \bibnamefont {Baldis}}, \ and\ \bibinfo {author}
  {\bibfnamefont {N.~A.}\ \bibnamefont {Ebrahim}}} (\bibinfo {year} {1981}),\
  \href {https://doi.org/10.1103/physrevlett.47.1285} {\bibfield  {journal}
  {\bibinfo  {journal} {Phys. Rev. Lett.}\ }\textbf {\bibinfo {volume} {47}},\
  \bibinfo {pages} {1285--1288}}\BibitemShut {NoStop}%
\bibitem [{\citenamefont {Joshi}\ \emph {et~al.}(2018)\citenamefont {Joshi}
  \emph {et~al.}}]{Joshi2018}%
  \BibitemOpen
  \bibfield  {author} {\bibinfo {author} {\bibnamefont {Joshi}, \bibfnamefont
  {C}},  \emph {et~al.}} (\bibinfo {year} {2018}),\ \href
  {https://iopscience.iop.org/article/10.1088/1361-6587/aaa2e3} {\bibfield
  {journal} {\bibinfo  {journal} {Plasma Phys. Control. Fusion}\ }\textbf
  {\bibinfo {volume} {60}},\ \bibinfo {pages} {034001}}\BibitemShut {NoStop}%
\bibitem [{\citenamefont {Kaganovich}\ \emph {et~al.}(1997)\citenamefont
  {Kaganovich}, \citenamefont {Sasorov}, \citenamefont {Ehrlich}, \citenamefont
  {Cohen},\ and\ \citenamefont {Zigler}}]{Kaganovich1997}%
  \BibitemOpen
  \bibfield  {author} {\bibinfo {author} {\bibnamefont {Kaganovich},
  \bibfnamefont {D}}, \bibinfo {author} {\bibfnamefont {P.~V.}\ \bibnamefont
  {Sasorov}}, \bibinfo {author} {\bibfnamefont {Y.}~\bibnamefont {Ehrlich}},
  \bibinfo {author} {\bibfnamefont {C.}~\bibnamefont {Cohen}}, \ and\ \bibinfo
  {author} {\bibfnamefont {A.}~\bibnamefont {Zigler}}} (\bibinfo {year}
  {1997}),\ \href {https://doi.org/10.1063/1.120217} {\bibfield  {journal}
  {\bibinfo  {journal} {Appl. Phys. Lett.}\ }\textbf {\bibinfo {volume} {71}},\
  \bibinfo {pages} {2925--2927}}\BibitemShut {NoStop}%
\bibitem [{\citenamefont {Kallos}\ \emph {et~al.}(2008)\citenamefont {Kallos},
  \citenamefont {Katsouleas}, \citenamefont {Kimura}, \citenamefont {Kusche},
  \citenamefont {Muggli}, \citenamefont {Pavlishin}, \citenamefont
  {Pogorelsky}, \citenamefont {Stolyarov},\ and\ \citenamefont
  {Yakimenko}}]{Kallos2008}%
  \BibitemOpen
  \bibfield  {author} {\bibinfo {author} {\bibnamefont {Kallos}, \bibfnamefont
  {E}}, \bibinfo {author} {\bibfnamefont {T.}~\bibnamefont {Katsouleas}},
  \bibinfo {author} {\bibfnamefont {W.~D.}\ \bibnamefont {Kimura}}, \bibinfo
  {author} {\bibfnamefont {K.}~\bibnamefont {Kusche}}, \bibinfo {author}
  {\bibfnamefont {P.}~\bibnamefont {Muggli}}, \bibinfo {author} {\bibfnamefont
  {I.}~\bibnamefont {Pavlishin}}, \bibinfo {author} {\bibfnamefont
  {I.}~\bibnamefont {Pogorelsky}}, \bibinfo {author} {\bibfnamefont
  {D.}~\bibnamefont {Stolyarov}}, \ and\ \bibinfo {author} {\bibfnamefont
  {V.}~\bibnamefont {Yakimenko}}} (\bibinfo {year} {2008}),\ \href
  {https://doi.org/10.1103/physrevlett.100.074802} {\bibfield  {journal}
  {\bibinfo  {journal} {Phys. Rev. Lett.}\ }\textbf {\bibinfo {volume} {100}},\
  \bibinfo {pages} {074802}}\BibitemShut {NoStop}%
\bibitem [{\citenamefont {Kallos}\ \emph {et~al.}(2007)\citenamefont {Kallos},
  \citenamefont {Katsouleas}, \citenamefont {Muggli}, \citenamefont
  {Pavlishin}, \citenamefont {Pogorelsky}, \citenamefont {Stolyarov},
  \citenamefont {Yakimenko},\ and\ \citenamefont {Kimura}}]{Kallos2007}%
  \BibitemOpen
  \bibfield  {author} {\bibinfo {author} {\bibnamefont {Kallos}, \bibfnamefont
  {E}}, \bibinfo {author} {\bibfnamefont {T.}~\bibnamefont {Katsouleas}},
  \bibinfo {author} {\bibfnamefont {P.}~\bibnamefont {Muggli}}, \bibinfo
  {author} {\bibfnamefont {I.}~\bibnamefont {Pavlishin}}, \bibinfo {author}
  {\bibfnamefont {I.}~\bibnamefont {Pogorelsky}}, \bibinfo {author}
  {\bibfnamefont {D.}~\bibnamefont {Stolyarov}}, \bibinfo {author}
  {\bibfnamefont {V.}~\bibnamefont {Yakimenko}}, \ and\ \bibinfo {author}
  {\bibfnamefont {W.~D.}\ \bibnamefont {Kimura}}} (\bibinfo {year} {2007}),\
  in\ \href {\doibase 10.1109/pac.2007.4440671} {\emph {\bibinfo {booktitle}
  {Proceedings of the 2007 Particle Accelerator Conf.}}}\ (\bibinfo
  {publisher} {IEEE},\ \bibinfo {address} {New York, NY, United States})\ pp.\
  \bibinfo {pages} {3070--3072}\BibitemShut {NoStop}%
\bibitem [{\citenamefont {Kalmykov}\ \emph {et~al.}(2009)\citenamefont
  {Kalmykov}, \citenamefont {Yi}, \citenamefont {Khudik},\ and\ \citenamefont
  {Shvets}}]{Kalmykov2009}%
  \BibitemOpen
  \bibfield  {author} {\bibinfo {author} {\bibnamefont {Kalmykov},
  \bibfnamefont {S}}, \bibinfo {author} {\bibfnamefont {S.~A.}\ \bibnamefont
  {Yi}}, \bibinfo {author} {\bibfnamefont {V.}~\bibnamefont {Khudik}}, \ and\
  \bibinfo {author} {\bibfnamefont {G.}~\bibnamefont {Shvets}}} (\bibinfo
  {year} {2009}),\ \href {https://doi.org/10.1103/PhysRevLett.103.135004}
  {\bibfield  {journal} {\bibinfo  {journal} {Phys. Rev. Lett.}\ }\textbf
  {\bibinfo {volume} {103}},\ \bibinfo {pages} {135004}}\BibitemShut {NoStop}%
\bibitem [{\citenamefont {Katsouleas}(1986)}]{Katsouleas1986}%
  \BibitemOpen
  \bibfield  {author} {\bibinfo {author} {\bibnamefont {Katsouleas},
  \bibfnamefont {T}}} (\bibinfo {year} {1986}),\ \href
  {https://link.aps.org/doi/10.1103/PhysRevA.33.2056} {\bibfield  {journal}
  {\bibinfo  {journal} {Phys. Rev. A}\ }\textbf {\bibinfo {volume} {33}},\
  \bibinfo {pages} {2056--2064}}\BibitemShut {NoStop}%
\bibitem [{\citenamefont {Katsouleas}\ and\ \citenamefont
  {Mori}(1988)}]{Katsouleas1988}%
  \BibitemOpen
  \bibfield  {author} {\bibinfo {author} {\bibnamefont {Katsouleas},
  \bibfnamefont {T}}, \ and\ \bibinfo {author} {\bibfnamefont {W.~B.}\
  \bibnamefont {Mori}}} (\bibinfo {year} {1988}),\ \href
  {https://doi.org/10.1103/physrevlett.61.90} {\bibfield  {journal} {\bibinfo
  {journal} {Phys. Rev. Lett.}\ }\textbf {\bibinfo {volume} {61}},\ \bibinfo
  {pages} {90--93}}\BibitemShut {NoStop}%
\bibitem [{\citenamefont {Katsouleas}\ \emph {et~al.}(1998)\citenamefont
  {Katsouleas} \emph {et~al.}}]{Katsouleas1998}%
  \BibitemOpen
  \bibfield  {author} {\bibinfo {author} {\bibnamefont {Katsouleas},
  \bibfnamefont {T}},  \emph {et~al.}} (\bibinfo {year} {1998}),\ in\ \href
  {https://doi.org/10.1109/pac.1997.749806} {\emph {\bibinfo {booktitle}
  {Proceedings of the 1997 Particle Accelerator Conf.}}}\ (\bibinfo
  {publisher} {{IEEE}},\ \bibinfo {address} {New York, NY, United States})\
  pp.\ \bibinfo {pages} {687--689}\BibitemShut {NoStop}%
\bibitem [{\citenamefont {Katsouleas}\ \emph {et~al.}(1987)\citenamefont
  {Katsouleas}, \citenamefont {Wilks}, \citenamefont {Chen}, \citenamefont
  {Dawson},\ and\ \citenamefont {Su}}]{Katsouleas1987}%
  \BibitemOpen
  \bibfield  {author} {\bibinfo {author} {\bibnamefont {Katsouleas},
  \bibfnamefont {T~C}}, \bibinfo {author} {\bibfnamefont {S.}~\bibnamefont
  {Wilks}}, \bibinfo {author} {\bibfnamefont {P.}~\bibnamefont {Chen}},
  \bibinfo {author} {\bibfnamefont {J.~M.}\ \bibnamefont {Dawson}}, \ and\
  \bibinfo {author} {\bibfnamefont {J.~J.}\ \bibnamefont {Su}}} (\bibinfo
  {year} {1987}),\ \href {https://inspirehep.net/literature/253298} {\bibfield
  {journal} {\bibinfo  {journal} {Part. Accel.}\ }\textbf {\bibinfo {volume}
  {22}},\ \bibinfo {pages} {81--99}}\BibitemShut {NoStop}%
\bibitem [{\citenamefont {Keinigs}\ and\ \citenamefont
  {Jones}(1987)}]{Keinigs1987}%
  \BibitemOpen
  \bibfield  {author} {\bibinfo {author} {\bibnamefont {Keinigs}, \bibfnamefont
  {R}}, \ and\ \bibinfo {author} {\bibfnamefont {M.~E.}\ \bibnamefont {Jones}}}
  (\bibinfo {year} {1987}),\ \href {https://doi.org/10.1063/1.866183}
  {\bibfield  {journal} {\bibinfo  {journal} {Phys. Fluids}\ }\textbf {\bibinfo
  {volume} {30}},\ \bibinfo {pages} {252--263}}\BibitemShut {NoStop}%
\bibitem [{\citenamefont {Kharchenko}\ \emph {et~al.}(1960)\citenamefont
  {Kharchenko}, \citenamefont {Fa{\u{\i}}nberg}, \citenamefont {Nikolaev},
  \citenamefont {Kornilov}, \citenamefont {Lutsenko},\ and\ \citenamefont
  {Pedenko}}]{Kharchenko1960}%
  \BibitemOpen
  \bibfield  {author} {\bibinfo {author} {\bibnamefont {Kharchenko},
  \bibfnamefont {I~F}}, \bibinfo {author} {\bibfnamefont {Ya.~B.}\ \bibnamefont
  {Fa{\u{\i}}nberg}}, \bibinfo {author} {\bibfnamefont {R.~M.}\ \bibnamefont
  {Nikolaev}}, \bibinfo {author} {\bibfnamefont {E.~A.}\ \bibnamefont
  {Kornilov}}, \bibinfo {author} {\bibfnamefont {E.~A.}\ \bibnamefont
  {Lutsenko}}, \ and\ \bibinfo {author} {\bibfnamefont {N.~S.}\ \bibnamefont
  {Pedenko}}} (\bibinfo {year} {1960}),\ \href
  {http://www.jetp.ac.ru/cgi-bin/dn/e_011_03_0493.pdf} {\bibfield  {journal}
  {\bibinfo  {journal} {J. Exp. Theor. Phys. (U.S.S.R.)}\ }\textbf {\bibinfo
  {volume} {38}},\ \bibinfo {pages} {685--692}}\BibitemShut {NoStop}%
\bibitem [{\citenamefont {Kim}\ \emph {et~al.}(2020)\citenamefont {Kim} \emph
  {et~al.}}]{Kim2020}%
  \BibitemOpen
  \bibfield  {author} {\bibinfo {author} {\bibnamefont {Kim}, \bibfnamefont
  {S-Y}},  \emph {et~al.}} (\bibinfo {year} {2020}),\ \href
  {https://doi.org/10.1016/j.nima.2019.163194} {\bibfield  {journal} {\bibinfo
  {journal} {Nucl. Instrum. Methods Phys. Res. A}\ }\textbf {\bibinfo {volume}
  {953}},\ \bibinfo {pages} {163194}}\BibitemShut {NoStop}%
\bibitem [{\citenamefont {Kimura}\ \emph {et~al.}(2011)\citenamefont {Kimura},
  \citenamefont {Milchberg}, \citenamefont {Muggli}, \citenamefont {Li},\ and\
  \citenamefont {Mori}}]{Kimura2011}%
  \BibitemOpen
  \bibfield  {author} {\bibinfo {author} {\bibnamefont {Kimura}, \bibfnamefont
  {W~D}}, \bibinfo {author} {\bibfnamefont {H.~M.}\ \bibnamefont {Milchberg}},
  \bibinfo {author} {\bibfnamefont {P.}~\bibnamefont {Muggli}}, \bibinfo
  {author} {\bibfnamefont {X.}~\bibnamefont {Li}}, \ and\ \bibinfo {author}
  {\bibfnamefont {W.~B.}\ \bibnamefont {Mori}}} (\bibinfo {year} {2011}),\
  \href {https://doi.org/10.1103/physrevstab.14.041301} {\bibfield  {journal}
  {\bibinfo  {journal} {Phys. Rev. ST Accel. Beams}\ }\textbf {\bibinfo
  {volume} {14}},\ \bibinfo {pages} {041301}}\BibitemShut {NoStop}%
\bibitem [{\citenamefont {Kirby}\ \emph {et~al.}(2007)\citenamefont {Kirby},
  \citenamefont {Berry}, \citenamefont {Blumenfeld}, \citenamefont {Hogan},
  \citenamefont {Ischebeck},\ and\ \citenamefont {Siemann}}]{Kirby2007}%
  \BibitemOpen
  \bibfield  {author} {\bibinfo {author} {\bibnamefont {Kirby}, \bibfnamefont
  {N}}, \bibinfo {author} {\bibfnamefont {M.}~\bibnamefont {Berry}}, \bibinfo
  {author} {\bibfnamefont {I.}~\bibnamefont {Blumenfeld}}, \bibinfo {author}
  {\bibfnamefont {M.~J.}\ \bibnamefont {Hogan}}, \bibinfo {author}
  {\bibfnamefont {R.}~\bibnamefont {Ischebeck}}, \ and\ \bibinfo {author}
  {\bibfnamefont {R.}~\bibnamefont {Siemann}}} (\bibinfo {year} {2007}),\ in\
  \href {https://doi.org/10.1109/pac.2007.4440680} {\emph {\bibinfo {booktitle}
  {Proceedings of the 2007 Particle Accelerator Conf.}}}\ (\bibinfo
  {publisher} {{IEEE}},\ \bibinfo {address} {New York, NY, United States})\
  pp.\ \bibinfo {pages} {3097--3099}\BibitemShut {NoStop}%
\bibitem [{\citenamefont {Kirby}\ \emph
  {et~al.}(2009{\natexlab{a}})\citenamefont {Kirby}, \citenamefont
  {Blumenfeld}, \citenamefont {Hogan}, \citenamefont {Siemann}, \citenamefont
  {Walz}, \citenamefont {Davidson},\ and\ \citenamefont {Huang}}]{Kirby2009b}%
  \BibitemOpen
  \bibfield  {author} {\bibinfo {author} {\bibnamefont {Kirby}, \bibfnamefont
  {N}}, \bibinfo {author} {\bibfnamefont {I.}~\bibnamefont {Blumenfeld}},
  \bibinfo {author} {\bibfnamefont {M.~J.}\ \bibnamefont {Hogan}}, \bibinfo
  {author} {\bibfnamefont {R.~H.}\ \bibnamefont {Siemann}}, \bibinfo {author}
  {\bibfnamefont {D.~R.}\ \bibnamefont {Walz}}, \bibinfo {author}
  {\bibfnamefont {A.~W.}\ \bibnamefont {Davidson}}, \ and\ \bibinfo {author}
  {\bibfnamefont {C.}~\bibnamefont {Huang}}} (\bibinfo {year}
  {2009}{\natexlab{a}}),\ in\ \href
  {https://accelconf.web.cern.ch/PAC2009/papers/fr5rfp017.pdf} {\emph {\bibinfo
  {booktitle} {Proceedings of the 2009 Particle Accelerator Conf.}}}\ (\bibinfo
   {publisher} {{JACoW}},\ \bibinfo {address} {Geneva, Switzerland})\ pp.\
  \bibinfo {pages} {4566--4568}\BibitemShut {NoStop}%
\bibitem [{\citenamefont {Kirby}\ \emph
  {et~al.}(2009{\natexlab{b}})\citenamefont {Kirby} \emph
  {et~al.}}]{Kirby2009}%
  \BibitemOpen
  \bibfield  {author} {\bibinfo {author} {\bibnamefont {Kirby}, \bibfnamefont
  {N}},  \emph {et~al.}} (\bibinfo {year} {2009}{\natexlab{b}}),\ \href
  {https://doi.org/10.1103/physrevstab.12.051302} {\bibfield  {journal}
  {\bibinfo  {journal} {Phys. Rev. ST Accel. Beams}\ }\textbf {\bibinfo
  {volume} {12}},\ \bibinfo {pages} {051302}}\BibitemShut {NoStop}%
\bibitem [{\citenamefont {Kirchen}\ \emph {et~al.}(2016)\citenamefont
  {Kirchen}, \citenamefont {Lehe}, \citenamefont {Godfrey}, \citenamefont
  {Dornmair}, \citenamefont {Jalas}, \citenamefont {Peters}, \citenamefont
  {Vay},\ and\ \citenamefont {Maier}}]{Kirchen2016}%
  \BibitemOpen
  \bibfield  {author} {\bibinfo {author} {\bibnamefont {Kirchen}, \bibfnamefont
  {M}}, \bibinfo {author} {\bibfnamefont {R.}~\bibnamefont {Lehe}}, \bibinfo
  {author} {\bibfnamefont {B.~B.}\ \bibnamefont {Godfrey}}, \bibinfo {author}
  {\bibfnamefont {I.}~\bibnamefont {Dornmair}}, \bibinfo {author}
  {\bibfnamefont {S.}~\bibnamefont {Jalas}}, \bibinfo {author} {\bibfnamefont
  {K.}~\bibnamefont {Peters}}, \bibinfo {author} {\bibfnamefont {J.-L.}\
  \bibnamefont {Vay}}, \ and\ \bibinfo {author} {\bibfnamefont {A.~R.}\
  \bibnamefont {Maier}}} (\bibinfo {year} {2016}),\ \href {\doibase
  10.1063/1.4964770} {\bibfield  {journal} {\bibinfo  {journal} {Phys.
  Plasmas}\ }\textbf {\bibinfo {volume} {23}},\ \bibinfo {pages}
  {100704}}\BibitemShut {NoStop}%
\bibitem [{\citenamefont {Kiselev}\ \emph {et~al.}(1976)\citenamefont
  {Kiselev}, \citenamefont {Berezin},\ and\ \citenamefont
  {Fa{\u{\i}}nberg}}]{Kiselev1976}%
  \BibitemOpen
  \bibfield  {author} {\bibinfo {author} {\bibnamefont {Kiselev}, \bibfnamefont
  {V~A}}, \bibinfo {author} {\bibfnamefont {A.~K.}\ \bibnamefont {Berezin}}, \
  and\ \bibinfo {author} {\bibfnamefont {Ya.~B.}\ \bibnamefont
  {Fa{\u{\i}}nberg}}} (\bibinfo {year} {1976}),\ \href
  {http://jetp.ac.ru/cgi-bin/dn/e_044_01_0101.pdf} {\bibfield  {journal}
  {\bibinfo  {journal} {J. Exp. Theor. Phys. (U.S.S.R.)}\ }\textbf {\bibinfo
  {volume} {44}},\ \bibinfo {pages} {101--105}}\BibitemShut {NoStop}%
\bibitem [{\citenamefont {Kneip}\ \emph {et~al.}(2012)\citenamefont {Kneip}
  \emph {et~al.}}]{Kneip2012}%
  \BibitemOpen
  \bibfield  {author} {\bibinfo {author} {\bibnamefont {Kneip}, \bibfnamefont
  {S}},  \emph {et~al.}} (\bibinfo {year} {2012}),\ \href
  {https://doi.org/10.1103/PhysRevSTAB.15.021302} {\bibfield  {journal}
  {\bibinfo  {journal} {Phys. Rev. ST Accel. Beams}\ }\textbf {\bibinfo
  {volume} {15}},\ \bibinfo {pages} {021302}}\BibitemShut {NoStop}%
\bibitem [{\citenamefont {Knetsch}\ \emph {et~al.}(2023)\citenamefont
  {Knetsch}, \citenamefont {Andriyash}, \citenamefont {Gilljohann},
  \citenamefont {Kononenko}, \citenamefont {Matheron}, \citenamefont
  {Mankovska}, \citenamefont {San Miguel~Claveria}, \citenamefont {Zakharova},
  \citenamefont {Adli},\ and\ \citenamefont {Corde}}]{Knetsch2023}%
  \BibitemOpen
  \bibfield  {author} {\bibinfo {author} {\bibnamefont {Knetsch}, \bibfnamefont
  {A}}, \bibinfo {author} {\bibfnamefont {I.~A.}\ \bibnamefont {Andriyash}},
  \bibinfo {author} {\bibfnamefont {M.}~\bibnamefont {Gilljohann}}, \bibinfo
  {author} {\bibfnamefont {O.}~\bibnamefont {Kononenko}}, \bibinfo {author}
  {\bibfnamefont {A.}~\bibnamefont {Matheron}}, \bibinfo {author}
  {\bibfnamefont {Y.}~\bibnamefont {Mankovska}}, \bibinfo {author}
  {\bibfnamefont {P.}~\bibnamefont {San Miguel~Claveria}}, \bibinfo {author}
  {\bibfnamefont {V.}~\bibnamefont {Zakharova}}, \bibinfo {author}
  {\bibfnamefont {E.}~\bibnamefont {Adli}}, \ and\ \bibinfo {author}
  {\bibfnamefont {S.}~\bibnamefont {Corde}}} (\bibinfo {year} {2023}),\ \href
  {https://doi.org/10.1103/PhysRevLett.131.135001} {\bibfield  {journal}
  {\bibinfo  {journal} {Phys. Rev. Lett.}\ }\textbf {\bibinfo {volume} {131}},\
  \bibinfo {pages} {135001}}\BibitemShut {NoStop}%
\bibitem [{\citenamefont {Knetsch}\ \emph {et~al.}(2021)\citenamefont {Knetsch}
  \emph {et~al.}}]{Knetsch2021}%
  \BibitemOpen
  \bibfield  {author} {\bibinfo {author} {\bibnamefont {Knetsch}, \bibfnamefont
  {A}},  \emph {et~al.}} (\bibinfo {year} {2021}),\ \href
  {https://doi.org/10.1103/physrevaccelbeams.24.101302} {\bibfield  {journal}
  {\bibinfo  {journal} {Phys. Rev. Accel. Beams}\ }\textbf {\bibinfo {volume}
  {24}},\ \bibinfo {pages} {101302}}\BibitemShut {NoStop}%
\bibitem [{\citenamefont {Kokurewicz}\ \emph {et~al.}(2019)\citenamefont
  {Kokurewicz} \emph {et~al.}}]{Kokurewicz2019}%
  \BibitemOpen
  \bibfield  {author} {\bibinfo {author} {\bibnamefont {Kokurewicz},
  \bibfnamefont {K}},  \emph {et~al.}} (\bibinfo {year} {2019}),\ \href
  {\doibase 10.1038/s41598-019-46630-w} {\bibfield  {journal} {\bibinfo
  {journal} {Sci. Rep.}\ }\textbf {\bibinfo {volume} {9}},\ \bibinfo {pages}
  {10837}}\BibitemShut {NoStop}%
\bibitem [{\citenamefont {Kostyukov}\ \emph {et~al.}(2004)\citenamefont
  {Kostyukov}, \citenamefont {Pukhov},\ and\ \citenamefont
  {Kiselev}}]{Kostyukov2004}%
  \BibitemOpen
  \bibfield  {author} {\bibinfo {author} {\bibnamefont {Kostyukov},
  \bibfnamefont {I}}, \bibinfo {author} {\bibfnamefont {A.}~\bibnamefont
  {Pukhov}}, \ and\ \bibinfo {author} {\bibfnamefont {S.}~\bibnamefont
  {Kiselev}}} (\bibinfo {year} {2004}),\ \href
  {https://doi.org/10.1063/1.1799371} {\bibfield  {journal} {\bibinfo
  {journal} {Phys. Plasmas}\ }\textbf {\bibinfo {volume} {11}},\ \bibinfo
  {pages} {5256--5264}}\BibitemShut {NoStop}%
\bibitem [{\citenamefont {Kostyukov}\ \emph {et~al.}(2012)\citenamefont
  {Kostyukov}, \citenamefont {Nerush},\ and\ \citenamefont
  {Litvak}}]{Kostyukov2012}%
  \BibitemOpen
  \bibfield  {author} {\bibinfo {author} {\bibnamefont {Kostyukov},
  \bibfnamefont {I~Yu}}, \bibinfo {author} {\bibfnamefont {E.~N.}\ \bibnamefont
  {Nerush}}, \ and\ \bibinfo {author} {\bibfnamefont {A.~G.}\ \bibnamefont
  {Litvak}}} (\bibinfo {year} {2012}),\ \href
  {https://doi.org/10.1103/physrevstab.15.111001} {\bibfield  {journal}
  {\bibinfo  {journal} {Phys. Rev. ST Accel. Beams}\ }\textbf {\bibinfo
  {volume} {15}},\ \bibinfo {pages} {111001}}\BibitemShut {NoStop}%
\bibitem [{\citenamefont {Kozawa}\ \emph {et~al.}(1997)\citenamefont {Kozawa}
  \emph {et~al.}}]{Kozawa1997}%
  \BibitemOpen
  \bibfield  {author} {\bibinfo {author} {\bibnamefont {Kozawa}, \bibfnamefont
  {T}},  \emph {et~al.}} (\bibinfo {year} {1997}),\ in\ \href
  {https://doi.org/10.1063/1.52939} {\emph {\bibinfo {booktitle} {AIP Conf.
  Proc.}}},\ Vol.\ \bibinfo {volume} {395}\ (\bibinfo  {publisher} {AIP})\ pp.\
  \bibinfo {pages} {343--353}\BibitemShut {NoStop}%
\bibitem [{\citenamefont {Krall}\ and\ \citenamefont
  {Joyce}(1995)}]{Krall1995}%
  \BibitemOpen
  \bibfield  {author} {\bibinfo {author} {\bibnamefont {Krall}, \bibfnamefont
  {J}}, \ and\ \bibinfo {author} {\bibfnamefont {G.}~\bibnamefont {Joyce}}}
  (\bibinfo {year} {1995}),\ \href {https://doi.org/10.1063/1.871344}
  {\bibfield  {journal} {\bibinfo  {journal} {Phys. Plasmas}\ }\textbf
  {\bibinfo {volume} {2}},\ \bibinfo {pages} {1326--1331}}\BibitemShut
  {NoStop}%
\bibitem [{\citenamefont {Krall}\ \emph {et~al.}(1991)\citenamefont {Krall},
  \citenamefont {Joyce},\ and\ \citenamefont {Esarey}}]{Krall1991}%
  \BibitemOpen
  \bibfield  {author} {\bibinfo {author} {\bibnamefont {Krall}, \bibfnamefont
  {J}}, \bibinfo {author} {\bibfnamefont {G.}~\bibnamefont {Joyce}}, \ and\
  \bibinfo {author} {\bibfnamefont {E.}~\bibnamefont {Esarey}}} (\bibinfo
  {year} {1991}),\ \href {https://doi.org/10.1103/physreva.44.6854} {\bibfield
  {journal} {\bibinfo  {journal} {Phys. Rev. A}\ }\textbf {\bibinfo {volume}
  {44}},\ \bibinfo {pages} {6854--6861}}\BibitemShut {NoStop}%
\bibitem [{\citenamefont {Krall}\ \emph {et~al.}(1989)\citenamefont {Krall},
  \citenamefont {Nguyen},\ and\ \citenamefont {Joyce}}]{Krall1989}%
  \BibitemOpen
  \bibfield  {author} {\bibinfo {author} {\bibnamefont {Krall}, \bibfnamefont
  {J}}, \bibinfo {author} {\bibfnamefont {K.}~\bibnamefont {Nguyen}}, \ and\
  \bibinfo {author} {\bibfnamefont {G.}~\bibnamefont {Joyce}}} (\bibinfo {year}
  {1989}),\ \href {https://doi.org/10.1063/1.859074} {\bibfield  {journal}
  {\bibinfo  {journal} {Phys. Fluids B}\ }\textbf {\bibinfo {volume} {1}},\
  \bibinfo {pages} {2099--2105}}\BibitemShut {NoStop}%
\bibitem [{\citenamefont {Kruer}(2003)}]{Kruer2003}%
  \BibitemOpen
  \bibfield  {author} {\bibinfo {author} {\bibnamefont {Kruer}, \bibfnamefont
  {W~L}}} (\bibinfo {year} {2003}),\ \href
  {https://doi.org/10.1201/9781003003243} {\emph {\bibinfo {title} {The Physics
  of Laser Plasma Interactions}}}\ (\bibinfo  {publisher} {CRC Press},\
  \bibinfo {address} {Boca Raton, FL, United States})\BibitemShut {NoStop}%
\bibitem [{\citenamefont {Kudryavtsev}\ \emph {et~al.}(1998)\citenamefont
  {Kudryavtsev}, \citenamefont {Lotov},\ and\ \citenamefont
  {Skrinsky}}]{Kudryavtsev1998}%
  \BibitemOpen
  \bibfield  {author} {\bibinfo {author} {\bibnamefont {Kudryavtsev},
  \bibfnamefont {A~M}}, \bibinfo {author} {\bibfnamefont {K.~V.}\ \bibnamefont
  {Lotov}}, \ and\ \bibinfo {author} {\bibfnamefont {A.~N.}\ \bibnamefont
  {Skrinsky}}} (\bibinfo {year} {1998}),\ \href
  {https://doi.org/10.1016/S0168-9002(98)00168-5} {\bibfield  {journal}
  {\bibinfo  {journal} {Nucl. Instrum. Methods Phys. Res. A}\ }\textbf
  {\bibinfo {volume} {410}},\ \bibinfo {pages} {388--395}}\BibitemShut
  {NoStop}%
\bibitem [{\citenamefont {Kumar}\ \emph {et~al.}(2010)\citenamefont {Kumar},
  \citenamefont {Pukhov},\ and\ \citenamefont {Lotov}}]{Kumar2010}%
  \BibitemOpen
  \bibfield  {author} {\bibinfo {author} {\bibnamefont {Kumar}, \bibfnamefont
  {N}}, \bibinfo {author} {\bibfnamefont {A.}~\bibnamefont {Pukhov}}, \ and\
  \bibinfo {author} {\bibfnamefont {K.}~\bibnamefont {Lotov}}} (\bibinfo {year}
  {2010}),\ \href {https://doi.org/10.1103/physrevlett.104.255003} {\bibfield
  {journal} {\bibinfo  {journal} {Phys. Rev. Lett.}\ }\textbf {\bibinfo
  {volume} {104}},\ \bibinfo {pages} {255003}}\BibitemShut {NoStop}%
\bibitem [{\citenamefont {Kurz}\ \emph {et~al.}(2021)\citenamefont {Kurz} \emph
  {et~al.}}]{Kurz2021}%
  \BibitemOpen
  \bibfield  {author} {\bibinfo {author} {\bibnamefont {Kurz}, \bibfnamefont
  {T}},  \emph {et~al.}} (\bibinfo {year} {2021}),\ \href
  {https://doi.org/10.1038/s41467-021-23000-7} {\bibfield  {journal} {\bibinfo
  {journal} {Nat. Commun.}\ }\textbf {\bibinfo {volume} {12}},\ \bibinfo
  {pages} {2895}}\BibitemShut {NoStop}%
\bibitem [{\citenamefont {Kuschel}\ \emph {et~al.}(2016)\citenamefont {Kuschel}
  \emph {et~al.}}]{Kuschel2016}%
  \BibitemOpen
  \bibfield  {author} {\bibinfo {author} {\bibnamefont {Kuschel}, \bibfnamefont
  {S}},  \emph {et~al.}} (\bibinfo {year} {2016}),\ \href
  {https://doi.org/10.1103/PhysRevAccelBeams.19.071301} {\bibfield  {journal}
  {\bibinfo  {journal} {Phys. Rev. Accel. Beams}\ }\textbf {\bibinfo {volume}
  {19}},\ \bibinfo {pages} {071301}}\BibitemShut {NoStop}%
\bibitem [{\citenamefont {Labat}\ \emph {et~al.}(2023)\citenamefont {Labat}
  \emph {et~al.}}]{Labat2023}%
  \BibitemOpen
  \bibfield  {author} {\bibinfo {author} {\bibnamefont {Labat}, \bibfnamefont
  {M}},  \emph {et~al.}} (\bibinfo {year} {2023}),\ \href
  {https://doi.org/10.1038/s41566-022-01104-w} {\bibfield  {journal} {\bibinfo
  {journal} {Nat. Photon.}\ }\textbf {\bibinfo {volume} {17}},\ \bibinfo
  {pages} {150–156}}\BibitemShut {NoStop}%
\bibitem [{\citenamefont {Langdon}\ and\ \citenamefont
  {Lasinski}(1976)}]{Langdon1976}%
  \BibitemOpen
  \bibfield  {author} {\bibinfo {author} {\bibnamefont {Langdon}, \bibfnamefont
  {A~B}}, \ and\ \bibinfo {author} {\bibfnamefont {B.~F.}\ \bibnamefont
  {Lasinski}}} (\bibinfo {year} {1976}),\ in\ \href
  {https://doi.org/10.1016/B978-0-12-460816-0.50014-2} {\emph {\bibinfo
  {booktitle} {Controlled Fusion}}}\ (\bibinfo  {publisher} {Elsevier})\ pp.\
  \bibinfo {pages} {327--366}\BibitemShut {NoStop}%
\bibitem [{\citenamefont {Lawson}(1972)}]{Lawson1972}%
  \BibitemOpen
  \bibfield  {author} {\bibinfo {author} {\bibnamefont {Lawson}, \bibfnamefont
  {J~D}}} (\bibinfo {year} {1972}),\ \href
  {https://cds.cern.ch/record/1107899/files/p21.pdf} {\bibfield  {journal}
  {\bibinfo  {journal} {Part. Accel.}\ }\textbf {\bibinfo {volume} {3}},\
  \bibinfo {pages} {21--33}}\BibitemShut {NoStop}%
\bibitem [{\citenamefont {Lebedev}\ \emph {et~al.}(2017)\citenamefont
  {Lebedev}, \citenamefont {Burov},\ and\ \citenamefont
  {Nagaitsev}}]{Lebedev2017}%
  \BibitemOpen
  \bibfield  {author} {\bibinfo {author} {\bibnamefont {Lebedev}, \bibfnamefont
  {V}}, \bibinfo {author} {\bibfnamefont {A.}~\bibnamefont {Burov}}, \ and\
  \bibinfo {author} {\bibfnamefont {S.}~\bibnamefont {Nagaitsev}}} (\bibinfo
  {year} {2017}),\ \href {https://doi.org/10.1103/physrevaccelbeams.20.121301}
  {\bibfield  {journal} {\bibinfo  {journal} {Phys. Rev. Accel. Beams}\
  }\textbf {\bibinfo {volume} {20}},\ \bibinfo {pages} {121301}}\BibitemShut
  {NoStop}%
\bibitem [{\citenamefont {Lee}\ and\ \citenamefont
  {Katsouleas}(1999)}]{Lee1999}%
  \BibitemOpen
  \bibfield  {author} {\bibinfo {author} {\bibnamefont {Lee}, \bibfnamefont
  {S}}, \ and\ \bibinfo {author} {\bibfnamefont {T.}~\bibnamefont
  {Katsouleas}}} (\bibinfo {year} {1999}),\ in\ \href
  {https://doi.org/10.1063/1.58913} {\emph {\bibinfo {booktitle} {{AIP} Conf.
  Proc.}}},\ Vol.\ \bibinfo {volume} {472}\ (\bibinfo  {publisher} {{AIP}})\
  pp.\ \bibinfo {pages} {524--533}\BibitemShut {NoStop}%
\bibitem [{\citenamefont {Lee}\ \emph {et~al.}(2000)\citenamefont {Lee},
  \citenamefont {Katsouleas}, \citenamefont {Hemker},\ and\ \citenamefont
  {Mori}}]{Lee2000}%
  \BibitemOpen
  \bibfield  {author} {\bibinfo {author} {\bibnamefont {Lee}, \bibfnamefont
  {S}}, \bibinfo {author} {\bibfnamefont {T.}~\bibnamefont {Katsouleas}},
  \bibinfo {author} {\bibfnamefont {R.}~\bibnamefont {Hemker}}, \ and\ \bibinfo
  {author} {\bibfnamefont {W.~B.}\ \bibnamefont {Mori}}} (\bibinfo {year}
  {2000}),\ \href {https://doi.org/10.1103/physreve.61.7014} {\bibfield
  {journal} {\bibinfo  {journal} {Phys. Rev. E}\ }\textbf {\bibinfo {volume}
  {61}},\ \bibinfo {pages} {7014--7021}}\BibitemShut {NoStop}%
\bibitem [{\citenamefont {Lee}\ \emph {et~al.}(2001)\citenamefont {Lee},
  \citenamefont {Katsouleas}, \citenamefont {Hemker}, \citenamefont {Dodd},\
  and\ \citenamefont {Mori}}]{Lee2001}%
  \BibitemOpen
  \bibfield  {author} {\bibinfo {author} {\bibnamefont {Lee}, \bibfnamefont
  {S}}, \bibinfo {author} {\bibfnamefont {T.}~\bibnamefont {Katsouleas}},
  \bibinfo {author} {\bibfnamefont {R.~G.}\ \bibnamefont {Hemker}}, \bibinfo
  {author} {\bibfnamefont {E.~S.}\ \bibnamefont {Dodd}}, \ and\ \bibinfo
  {author} {\bibfnamefont {W.~B.}\ \bibnamefont {Mori}}} (\bibinfo {year}
  {2001}),\ \href {https://doi.org/10.1103/physreve.64.045501} {\bibfield
  {journal} {\bibinfo  {journal} {Phys. Rev. E}\ }\textbf {\bibinfo {volume}
  {64}},\ \bibinfo {pages} {045501}}\BibitemShut {NoStop}%
\bibitem [{\citenamefont {Lee}\ \emph {et~al.}(2002)\citenamefont {Lee} \emph
  {et~al.}}]{Lee2002}%
  \BibitemOpen
  \bibfield  {author} {\bibinfo {author} {\bibnamefont {Lee}, \bibfnamefont
  {S}},  \emph {et~al.}} (\bibinfo {year} {2002}),\ \href
  {https://doi.org/10.1103/physrevstab.5.011001} {\bibfield  {journal}
  {\bibinfo  {journal} {Phys. Rev. ST Accel. Beams}\ }\textbf {\bibinfo
  {volume} {5}},\ \bibinfo {pages} {011001}}\BibitemShut {NoStop}%
\bibitem [{\citenamefont {Leemans}\ \emph {et~al.}(2006)\citenamefont
  {Leemans}, \citenamefont {Nagler}, \citenamefont {Gonsalves}, \citenamefont
  {Tóth}, \citenamefont {Nakamura}, \citenamefont {Geddes}, \citenamefont
  {Esarey}, \citenamefont {Schroeder},\ and\ \citenamefont
  {Hooker}}]{Leemans2006}%
  \BibitemOpen
  \bibfield  {author} {\bibinfo {author} {\bibnamefont {Leemans}, \bibfnamefont
  {W~P}}, \bibinfo {author} {\bibfnamefont {B.}~\bibnamefont {Nagler}},
  \bibinfo {author} {\bibfnamefont {A.~J.}\ \bibnamefont {Gonsalves}}, \bibinfo
  {author} {\bibfnamefont {Cs.}\ \bibnamefont {Tóth}}, \bibinfo {author}
  {\bibfnamefont {K.}~\bibnamefont {Nakamura}}, \bibinfo {author}
  {\bibfnamefont {C.~G.~R.}\ \bibnamefont {Geddes}}, \bibinfo {author}
  {\bibfnamefont {E.}~\bibnamefont {Esarey}}, \bibinfo {author} {\bibfnamefont
  {C.~B.}\ \bibnamefont {Schroeder}}, \ and\ \bibinfo {author} {\bibfnamefont
  {S.~M.}\ \bibnamefont {Hooker}}} (\bibinfo {year} {2006}),\ \href
  {https://doi.org/10.1038/nphys418} {\bibfield  {journal} {\bibinfo  {journal}
  {Nat. Phys.}\ }\textbf {\bibinfo {volume} {2}},\ \bibinfo {pages}
  {696--699}}\BibitemShut {NoStop}%
\bibitem [{\citenamefont {Lehe}\ \emph
  {et~al.}(2016{\natexlab{a}})\citenamefont {Lehe}, \citenamefont {Kirchen},
  \citenamefont {Andriyash}, \citenamefont {Godfrey},\ and\ \citenamefont
  {Vay}}]{Lehe2016}%
  \BibitemOpen
  \bibfield  {author} {\bibinfo {author} {\bibnamefont {Lehe}, \bibfnamefont
  {R}}, \bibinfo {author} {\bibfnamefont {M.}~\bibnamefont {Kirchen}}, \bibinfo
  {author} {\bibfnamefont {I.~A.}\ \bibnamefont {Andriyash}}, \bibinfo {author}
  {\bibfnamefont {B.~B.}\ \bibnamefont {Godfrey}}, \ and\ \bibinfo {author}
  {\bibfnamefont {J.-L.}\ \bibnamefont {Vay}}} (\bibinfo {year}
  {2016}{\natexlab{a}}),\ \href {https://doi.org/10.1016/j.cpc.2016.02.007}
  {\bibfield  {journal} {\bibinfo  {journal} {Comput. Phys. Commun.}\ }\textbf
  {\bibinfo {volume} {203}},\ \bibinfo {pages} {66–82}}\BibitemShut {NoStop}%
\bibitem [{\citenamefont {Lehe}\ \emph
  {et~al.}(2016{\natexlab{b}})\citenamefont {Lehe}, \citenamefont {Kirchen},
  \citenamefont {Godfrey}, \citenamefont {Maier},\ and\ \citenamefont
  {Vay}}]{Lehe2016b}%
  \BibitemOpen
  \bibfield  {author} {\bibinfo {author} {\bibnamefont {Lehe}, \bibfnamefont
  {R}}, \bibinfo {author} {\bibfnamefont {M.}~\bibnamefont {Kirchen}}, \bibinfo
  {author} {\bibfnamefont {B.~B.}\ \bibnamefont {Godfrey}}, \bibinfo {author}
  {\bibfnamefont {A.~R.}\ \bibnamefont {Maier}}, \ and\ \bibinfo {author}
  {\bibfnamefont {J.-L.}\ \bibnamefont {Vay}}} (\bibinfo {year}
  {2016}{\natexlab{b}}),\ \href {\doibase 10.1103/physreve.94.053305}
  {\bibfield  {journal} {\bibinfo  {journal} {Phys. Rev. E}\ }\textbf {\bibinfo
  {volume} {94}},\ \bibinfo {pages} {053305}}\BibitemShut {NoStop}%
\bibitem [{\citenamefont {Lehe}\ \emph {et~al.}(2017)\citenamefont {Lehe},
  \citenamefont {Schroeder}, \citenamefont {Vay}, \citenamefont {Esarey},\ and\
  \citenamefont {Leemans}}]{Lehe2017}%
  \BibitemOpen
  \bibfield  {author} {\bibinfo {author} {\bibnamefont {Lehe}, \bibfnamefont
  {R}}, \bibinfo {author} {\bibfnamefont {C.~B.}\ \bibnamefont {Schroeder}},
  \bibinfo {author} {\bibfnamefont {J.-L.}\ \bibnamefont {Vay}}, \bibinfo
  {author} {\bibfnamefont {E.}~\bibnamefont {Esarey}}, \ and\ \bibinfo {author}
  {\bibfnamefont {W.~P.}\ \bibnamefont {Leemans}}} (\bibinfo {year} {2017}),\
  \href {https://doi.org/10.1103/physrevlett.119.244801} {\bibfield  {journal}
  {\bibinfo  {journal} {Phys. Rev. Lett.}\ }\textbf {\bibinfo {volume} {119}},\
  \bibinfo {pages} {244801}}\BibitemShut {NoStop}%
\bibitem [{\citenamefont {Lemery}\ and\ \citenamefont
  {Piot}(2015)}]{Lemery2015}%
  \BibitemOpen
  \bibfield  {author} {\bibinfo {author} {\bibnamefont {Lemery}, \bibfnamefont
  {F}}, \ and\ \bibinfo {author} {\bibfnamefont {P.}~\bibnamefont {Piot}}}
  (\bibinfo {year} {2015}),\ \href
  {https://link.aps.org/doi/10.1103/PhysRevSTAB.18.081301} {\bibfield
  {journal} {\bibinfo  {journal} {Phys. Rev. ST Accel. Beams}\ }\textbf
  {\bibinfo {volume} {18}},\ \bibinfo {pages} {081301}}\BibitemShut {NoStop}%
\bibitem [{\citenamefont {Li}\ \emph {et~al.}(2014)\citenamefont {Li},
  \citenamefont {Gai}, \citenamefont {Jing}, \citenamefont {Power},
  \citenamefont {Tang},\ and\ \citenamefont {Zholents}}]{Li2014}%
  \BibitemOpen
  \bibfield  {author} {\bibinfo {author} {\bibnamefont {Li}, \bibfnamefont
  {C}}, \bibinfo {author} {\bibfnamefont {W.}~\bibnamefont {Gai}}, \bibinfo
  {author} {\bibfnamefont {C.}~\bibnamefont {Jing}}, \bibinfo {author}
  {\bibfnamefont {J.~G.}\ \bibnamefont {Power}}, \bibinfo {author}
  {\bibfnamefont {C.~X.}\ \bibnamefont {Tang}}, \ and\ \bibinfo {author}
  {\bibfnamefont {A.}~\bibnamefont {Zholents}}} (\bibinfo {year} {2014}),\
  \href {https://doi.org/10.1103/PhysRevSTAB.17.091302} {\bibfield  {journal}
  {\bibinfo  {journal} {Phys. Rev. ST Accel. Beams}\ }\textbf {\bibinfo
  {volume} {17}},\ \bibinfo {pages} {091302}}\BibitemShut {NoStop}%
\bibitem [{\citenamefont {Li}\ \emph {et~al.}(2021)\citenamefont {Li},
  \citenamefont {An}, \citenamefont {Decyk}, \citenamefont {Xu}, \citenamefont
  {Hogan},\ and\ \citenamefont {Mori}}]{Li2021}%
  \BibitemOpen
  \bibfield  {author} {\bibinfo {author} {\bibnamefont {Li}, \bibfnamefont
  {F}}, \bibinfo {author} {\bibfnamefont {W.}~\bibnamefont {An}}, \bibinfo
  {author} {\bibfnamefont {V.~K.}\ \bibnamefont {Decyk}}, \bibinfo {author}
  {\bibfnamefont {X.}~\bibnamefont {Xu}}, \bibinfo {author} {\bibfnamefont
  {M.~J.}\ \bibnamefont {Hogan}}, \ and\ \bibinfo {author} {\bibfnamefont
  {W.~B.}\ \bibnamefont {Mori}}} (\bibinfo {year} {2021}),\ \href
  {https://doi.org/10.1016/j.cpc.2020.107784} {\bibfield  {journal} {\bibinfo
  {journal} {Comput. Phys. Commun.}\ }\textbf {\bibinfo {volume} {261}},\
  \bibinfo {pages} {107784}}\BibitemShut {NoStop}%
\bibitem [{\citenamefont {Li}\ \emph {et~al.}(2013)\citenamefont {Li} \emph
  {et~al.}}]{Li2013}%
  \BibitemOpen
  \bibfield  {author} {\bibinfo {author} {\bibnamefont {Li}, \bibfnamefont
  {F}},  \emph {et~al.}} (\bibinfo {year} {2013}),\ \href
  {https://doi.org/10.1103/physrevlett.111.015003} {\bibfield  {journal}
  {\bibinfo  {journal} {Phys. Rev. Lett.}\ }\textbf {\bibinfo {volume} {111}},\
  \bibinfo {pages} {015003}}\BibitemShut {NoStop}%
\bibitem [{\citenamefont {Li}\ \emph {et~al.}(2026)\citenamefont {Li},
  \citenamefont {Liu}, \citenamefont {Liu},\ and\ \citenamefont
  {Liu}}]{Li2026}%
  \BibitemOpen
  \bibfield  {author} {\bibinfo {author} {\bibnamefont {Li}, \bibfnamefont
  {H}}, \bibinfo {author} {\bibfnamefont {Y.}~\bibnamefont {Liu}}, \bibinfo
  {author} {\bibfnamefont {T.}~\bibnamefont {Liu}}, \ and\ \bibinfo {author}
  {\bibfnamefont {B.}~\bibnamefont {Liu}}} (\bibinfo {year} {2026}),\ \href
  {\doibase 10.1088/1402-4896/ae9141} {\bibfield  {journal} {\bibinfo
  {journal} {Phys. Scr.}\ }\textbf {\bibinfo {volume} {101}},\ \bibinfo {pages}
  {315601}}\BibitemShut {NoStop}%
\bibitem [{\citenamefont {Li}\ \emph {et~al.}(2019)\citenamefont {Li},
  \citenamefont {Chanc\'e},\ and\ \citenamefont {Nghiem}}]{Li2019}%
  \BibitemOpen
  \bibfield  {author} {\bibinfo {author} {\bibnamefont {Li}, \bibfnamefont
  {X}}, \bibinfo {author} {\bibfnamefont {A.}~\bibnamefont {Chanc\'e}}, \ and\
  \bibinfo {author} {\bibfnamefont {P.~A.~P.}\ \bibnamefont {Nghiem}}}
  (\bibinfo {year} {2019}),\ \href
  {https://doi.org/10.1103/PhysRevAccelBeams.22.021304} {\bibfield  {journal}
  {\bibinfo  {journal} {Phys. Rev. Accel. Beams}\ }\textbf {\bibinfo {volume}
  {22}},\ \bibinfo {pages} {021304}}\BibitemShut {NoStop}%
\bibitem [{\citenamefont {Li}\ \emph {et~al.}(2025)\citenamefont {Li} \emph
  {et~al.}}]{Li2025}%
  \BibitemOpen
  \bibfield  {author} {\bibinfo {author} {\bibnamefont {Li}, \bibfnamefont
  {Y}},  \emph {et~al.}} (\bibinfo {year} {2025}),\ \href {\doibase
  10.1017/hpl.2024.69} {\bibfield  {journal} {\bibinfo  {journal} {High Pow.
  Laser Sci. Eng.}\ }\textbf {\bibinfo {volume} {13}},\ \bibinfo {pages}
  {e12}}\BibitemShut {NoStop}%
\bibitem [{\citenamefont {Li}\ \emph {et~al.}(2012)\citenamefont {Li},
  \citenamefont {Zgadzaj}, \citenamefont {Wang}, \citenamefont {Chang},\ and\
  \citenamefont {Downer}}]{Li2013b}%
  \BibitemOpen
  \bibfield  {author} {\bibinfo {author} {\bibnamefont {Li}, \bibfnamefont
  {Z}}, \bibinfo {author} {\bibfnamefont {R.}~\bibnamefont {Zgadzaj}}, \bibinfo
  {author} {\bibfnamefont {X.}~\bibnamefont {Wang}}, \bibinfo {author}
  {\bibfnamefont {Y.-Y.}\ \bibnamefont {Chang}}, \ and\ \bibinfo {author}
  {\bibfnamefont {M.~C.}\ \bibnamefont {Downer}}} (\bibinfo {year} {2012}),\
  in\ \href {https://doi.org/10.1063/1.4773689} {\emph {\bibinfo {booktitle}
  {{AIP} Conf. Proc.}}}\ (\bibinfo  {publisher} {{AIP}})\ pp.\ \bibinfo {pages}
  {160--168}\BibitemShut {NoStop}%
\bibitem [{\citenamefont {Lien}(1970)}]{Lien1970}%
  \BibitemOpen
  \bibfield  {author} {\bibinfo {author} {\bibnamefont {Lien}, \bibfnamefont
  {E~L}}} (\bibinfo {year} {1970}),\ in\ \href
  {https://doi.org/10.1109/IEDM.1969.188154} {\emph {\bibinfo {booktitle} {1969
  International Electron Devices Meeting}}},\ p.~\bibinfo {pages}
  {98}\BibitemShut {NoStop}%
\bibitem [{\citenamefont {Lifschitz}\ \emph {et~al.}(2009)\citenamefont
  {Lifschitz}, \citenamefont {Davoine}, \citenamefont {Lefebvre}, \citenamefont
  {Faure}, \citenamefont {Rechatin},\ and\ \citenamefont
  {Malka}}]{Lifschitz2009}%
  \BibitemOpen
  \bibfield  {author} {\bibinfo {author} {\bibnamefont {Lifschitz},
  \bibfnamefont {A~F}}, \bibinfo {author} {\bibfnamefont {X.}~\bibnamefont
  {Davoine}}, \bibinfo {author} {\bibfnamefont {E.}~\bibnamefont {Lefebvre}},
  \bibinfo {author} {\bibfnamefont {J.}~\bibnamefont {Faure}}, \bibinfo
  {author} {\bibfnamefont {C.}~\bibnamefont {Rechatin}}, \ and\ \bibinfo
  {author} {\bibfnamefont {V.}~\bibnamefont {Malka}}} (\bibinfo {year}
  {2009}),\ \href {https://doi.org/10.1016/j.jcp.2008.11.017} {\bibfield
  {journal} {\bibinfo  {journal} {J. Comput. Phys.}\ }\textbf {\bibinfo
  {volume} {228}},\ \bibinfo {pages} {1803--1814}}\BibitemShut {NoStop}%
\bibitem [{\citenamefont {Lindstr{\o}m}(2021{\natexlab{a}})}]{Lindstrom2021b}%
  \BibitemOpen
  \bibfield  {author} {\bibinfo {author} {\bibnamefont {Lindstr{\o}m},
  \bibfnamefont {C~A}}} (\bibinfo {year} {2021}{\natexlab{a}}),\ \href
  {https://doi.org/10.1103/physrevaccelbeams.24.014801} {\bibfield  {journal}
  {\bibinfo  {journal} {Phys. Rev. Accel. Beams}\ }\textbf {\bibinfo {volume}
  {24}},\ \bibinfo {pages} {014801}}\BibitemShut {NoStop}%
\bibitem [{\citenamefont {Lindstr{\o}m}(2021{\natexlab{b}})}]{Lindstrom2021c}%
  \BibitemOpen
  \bibfield  {author} {\bibinfo {author} {\bibnamefont {Lindstr{\o}m},
  \bibfnamefont {C~A}}} (\bibinfo {year} {2021}{\natexlab{b}}),\ \href
  {https://arxiv.org/abs/2104.14460} {}\Eprint
  {http://arxiv.org/abs/2104.14460} {arXiv:2104.14460} \BibitemShut {NoStop}%
\bibitem [{\citenamefont {Lindstr{\o}m}\ and\ \citenamefont
  {Adli}(2016)}]{Lindstrom2016b}%
  \BibitemOpen
  \bibfield  {author} {\bibinfo {author} {\bibnamefont {Lindstr{\o}m},
  \bibfnamefont {C~A}}, \ and\ \bibinfo {author} {\bibfnamefont
  {E.}~\bibnamefont {Adli}}} (\bibinfo {year} {2016}),\ \href
  {https://doi.org/10.1103/physrevaccelbeams.19.071002} {\bibfield  {journal}
  {\bibinfo  {journal} {Phys. Rev. Accel. Beams}\ }\textbf {\bibinfo {volume}
  {19}},\ \bibinfo {pages} {071002}}\BibitemShut {NoStop}%
\bibitem [{\citenamefont {Lindstr{\o}m}\ \emph {et~al.}(2016)\citenamefont
  {Lindstr{\o}m}, \citenamefont {Adli}, \citenamefont {Marín}, \citenamefont
  {Pfingstner},\ and\ \citenamefont {Schulte}}]{Lindstrom2016a}%
  \BibitemOpen
  \bibfield  {author} {\bibinfo {author} {\bibnamefont {Lindstr{\o}m},
  \bibfnamefont {C~A}}, \bibinfo {author} {\bibfnamefont {E.}~\bibnamefont
  {Adli}}, \bibinfo {author} {\bibfnamefont {E.}~\bibnamefont {Marín}},
  \bibinfo {author} {\bibfnamefont {J.}~\bibnamefont {Pfingstner}}, \ and\
  \bibinfo {author} {\bibfnamefont {D.}~\bibnamefont {Schulte}}} (\bibinfo
  {year} {2016}),\ in\ \href {http://jacow.org/ipac2016/papers/wepmy009.pdf}
  {\emph {\bibinfo {booktitle} {Proceedings of the 7th Int. Particle
  Accelerator Conf.}}}\ (\bibinfo  {publisher} {JACoW},\ \bibinfo {address}
  {Geneva, Switzerland})\ pp.\ \bibinfo {pages} {2561--2564}\BibitemShut
  {NoStop}%
\bibitem [{\citenamefont {Lindstr{\o}m}\ \emph {et~al.}(2020)\citenamefont
  {Lindstr{\o}m}, \citenamefont {D'Arcy}, \citenamefont {Garland},
  \citenamefont {Gonzalez}, \citenamefont {Schmidt}, \citenamefont
  {Schr\"{o}der}, \citenamefont {Wesch},\ and\ \citenamefont
  {Osterhoff}}]{Lindstrom2020a}%
  \BibitemOpen
  \bibfield  {author} {\bibinfo {author} {\bibnamefont {Lindstr{\o}m},
  \bibfnamefont {C~A}}, \bibinfo {author} {\bibfnamefont {R.}~\bibnamefont
  {D'Arcy}}, \bibinfo {author} {\bibfnamefont {M.~J.}\ \bibnamefont {Garland}},
  \bibinfo {author} {\bibfnamefont {P.}~\bibnamefont {Gonzalez}}, \bibinfo
  {author} {\bibfnamefont {B.}~\bibnamefont {Schmidt}}, \bibinfo {author}
  {\bibfnamefont {S.}~\bibnamefont {Schr\"{o}der}}, \bibinfo {author}
  {\bibfnamefont {S.}~\bibnamefont {Wesch}}, \ and\ \bibinfo {author}
  {\bibfnamefont {J.}~\bibnamefont {Osterhoff}}} (\bibinfo {year} {2020}),\
  \href {https://doi.org/10.1103/physrevaccelbeams.23.052802} {\bibfield
  {journal} {\bibinfo  {journal} {Phys. Rev. Accel. Beams}\ }\textbf {\bibinfo
  {volume} {23}},\ \bibinfo {pages} {052802}}\BibitemShut {NoStop}%
\bibitem [{\citenamefont {Lindstr{\o}m}\ and\ \citenamefont
  {Th{\'{e}}venet}(2022)}]{Lindstrom2022}%
  \BibitemOpen
  \bibfield  {author} {\bibinfo {author} {\bibnamefont {Lindstr{\o}m},
  \bibfnamefont {C~A}}, \ and\ \bibinfo {author} {\bibfnamefont
  {M.}~\bibnamefont {Th{\'{e}}venet}}} (\bibinfo {year} {2022}),\ \href
  {https://doi.org/10.1088/1748-0221/17/05/p05016} {\bibfield  {journal}
  {\bibinfo  {journal} {J. Instrum.}\ }\textbf {\bibinfo {volume} {17}},\
  \bibinfo {pages} {P05016}}\BibitemShut {NoStop}%
\bibitem [{\citenamefont {Lindstr{\o}m}\ \emph
  {et~al.}(2018{\natexlab{a}})\citenamefont {Lindstr{\o}m} \emph
  {et~al.}}]{Lindstrom2018}%
  \BibitemOpen
  \bibfield  {author} {\bibinfo {author} {\bibnamefont {Lindstr{\o}m},
  \bibfnamefont {C~A}},  \emph {et~al.}} (\bibinfo {year}
  {2018}{\natexlab{a}}),\ \href
  {https://doi.org/10.1103/physrevlett.120.124802} {\bibfield  {journal}
  {\bibinfo  {journal} {Phys. Rev. Lett.}\ }\textbf {\bibinfo {volume} {120}},\
  \bibinfo {pages} {124802}}\BibitemShut {NoStop}%
\bibitem [{\citenamefont {Lindstr{\o}m}\ \emph
  {et~al.}(2018{\natexlab{b}})\citenamefont {Lindstr{\o}m} \emph
  {et~al.}}]{Lindstrom2018b}%
  \BibitemOpen
  \bibfield  {author} {\bibinfo {author} {\bibnamefont {Lindstr{\o}m},
  \bibfnamefont {C~A}},  \emph {et~al.}} (\bibinfo {year}
  {2018}{\natexlab{b}}),\ \href
  {https://doi.org/10.1103/physrevlett.121.194801} {\bibfield  {journal}
  {\bibinfo  {journal} {Phys. Rev. Lett.}\ }\textbf {\bibinfo {volume} {121}},\
  \bibinfo {pages} {194801}}\BibitemShut {NoStop}%
\bibitem [{\citenamefont {Lindstr{\o}m}\ \emph {et~al.}(2021)\citenamefont
  {Lindstr{\o}m} \emph {et~al.}}]{Lindstrom2021a}%
  \BibitemOpen
  \bibfield  {author} {\bibinfo {author} {\bibnamefont {Lindstr{\o}m},
  \bibfnamefont {C~A}},  \emph {et~al.}} (\bibinfo {year} {2021}),\ \href
  {https://doi.org/10.1103/PhysRevLett.126.014801} {\bibfield  {journal}
  {\bibinfo  {journal} {Phys. Rev. Lett.}\ }\textbf {\bibinfo {volume} {126}},\
  \bibinfo {pages} {014801}}\BibitemShut {NoStop}%
\bibitem [{\citenamefont {Lindstr{\o}m}\ \emph {et~al.}(2024)\citenamefont
  {Lindstr{\o}m} \emph {et~al.}}]{Lindstrom2024}%
  \BibitemOpen
  \bibfield  {author} {\bibinfo {author} {\bibnamefont {Lindstr{\o}m},
  \bibfnamefont {C~A}},  \emph {et~al.}} (\bibinfo {year} {2024}),\ \href
  {https://doi.org/10.1038/s41467-024-50320-1} {\bibfield  {journal} {\bibinfo
  {journal} {Nat. Commun.}\ }\textbf {\bibinfo {volume} {15}},\ \bibinfo
  {pages} {6097}}\BibitemShut {NoStop}%
\bibitem [{\citenamefont {Lindstr{\o}m}\ \emph
  {et~al.}(2025{\natexlab{a}})\citenamefont {Lindstr{\o}m} \emph
  {et~al.}}]{Lindstrom2025b}%
  \BibitemOpen
  \bibfield  {author} {\bibinfo {author} {\bibnamefont {Lindstr{\o}m},
  \bibfnamefont {C~A}},  \emph {et~al.}} (\bibinfo {year}
  {2025}{\natexlab{a}}),\ in\ \href {\doibase 10.18429/JACOW-IPAC2025-MOCD1}
  {\emph {\bibinfo {booktitle} {Proceedings of the 16th Int. Particle
  Accelerator Conf.}}}\ (\bibinfo  {publisher} {JACoW},\ \bibinfo {address}
  {Geneva, Switzerland})\ pp.\ \bibinfo {pages} {53--56}\BibitemShut {NoStop}%
\bibitem [{\citenamefont {Lindstr{\o}m}\ \emph
  {et~al.}(2025{\natexlab{b}})\citenamefont {Lindstr{\o}m} \emph
  {et~al.}}]{Lindstrom2025}%
  \BibitemOpen
  \bibfield  {author} {\bibinfo {author} {\bibnamefont {Lindstr{\o}m},
  \bibfnamefont {C~A}},  \emph {et~al.}} (\bibinfo {year}
  {2025}{\natexlab{b}}),\ in\ \href {\doibase 10.18429/JACOW-IPAC2025-TUPS120}
  {\emph {\bibinfo {booktitle} {Proceedings of the 16th Int. Particle
  Accelerator Conf.}}}\ (\bibinfo  {publisher} {JACoW},\ \bibinfo {address}
  {Geneva, Switzerland})\ pp.\ \bibinfo {pages} {1615--1618}\BibitemShut
  {NoStop}%
\bibitem [{\citenamefont {Lindstrøm}\ \emph {et~al.}(2026)\citenamefont
  {Lindstrøm}, \citenamefont {Adli}, \citenamefont {Chen}, \citenamefont
  {Drobniak}, \citenamefont {Huebl}, \citenamefont {Kalvik}, \citenamefont
  {Mitchell}, \citenamefont {Peña},\ and\ \citenamefont
  {Sjobak}}]{Lindstrom2026}%
  \BibitemOpen
  \bibfield  {author} {\bibinfo {author} {\bibnamefont {Lindstrøm},
  \bibfnamefont {C~A}}, \bibinfo {author} {\bibfnamefont {E.}~\bibnamefont
  {Adli}}, \bibinfo {author} {\bibfnamefont {J.~B.~B.}\ \bibnamefont {Chen}},
  \bibinfo {author} {\bibfnamefont {P.}~\bibnamefont {Drobniak}}, \bibinfo
  {author} {\bibfnamefont {A.}~\bibnamefont {Huebl}}, \bibinfo {author}
  {\bibfnamefont {D.}~\bibnamefont {Kalvik}}, \bibinfo {author} {\bibfnamefont
  {C.~E.}\ \bibnamefont {Mitchell}}, \bibinfo {author} {\bibfnamefont
  {F.}~\bibnamefont {Peña}}, \ and\ \bibinfo {author} {\bibfnamefont {K.~N.}\
  \bibnamefont {Sjobak}}} (\bibinfo {year} {2026}),\ \href {\doibase
  10.48550/arxiv.2604.17605} {}\Eprint {http://arxiv.org/abs/2604.17605}
  {arXiv:2604.17605} \BibitemShut {NoStop}%
\bibitem [{\citenamefont {Lishilin}\ \emph {et~al.}(2019)\citenamefont
  {Lishilin} \emph {et~al.}}]{Lishilin2019}%
  \BibitemOpen
  \bibfield  {author} {\bibinfo {author} {\bibnamefont {Lishilin},
  \bibfnamefont {O}},  \emph {et~al.}} (\bibinfo {year} {2019}),\ \href
  {\doibase 10.1088/1742-6596/1350/1/012057} {\bibfield  {journal} {\bibinfo
  {journal} {J. Phys.: Conf. Ser.}\ }\textbf {\bibinfo {volume} {1350}},\
  \bibinfo {pages} {012057}}\BibitemShut {NoStop}%
\bibitem [{\citenamefont {Litos}\ \emph {et~al.}(2014)\citenamefont {Litos}
  \emph {et~al.}}]{Litos2014}%
  \BibitemOpen
  \bibfield  {author} {\bibinfo {author} {\bibnamefont {Litos}, \bibfnamefont
  {M}},  \emph {et~al.}} (\bibinfo {year} {2014}),\ \href
  {https://doi.org/10.1038/nature13882} {\bibfield  {journal} {\bibinfo
  {journal} {Nature (London)}\ }\textbf {\bibinfo {volume} {515}},\ \bibinfo
  {pages} {92--95}}\BibitemShut {NoStop}%
\bibitem [{\citenamefont {Litos}\ \emph {et~al.}(2016)\citenamefont {Litos}
  \emph {et~al.}}]{Litos2016}%
  \BibitemOpen
  \bibfield  {author} {\bibinfo {author} {\bibnamefont {Litos}, \bibfnamefont
  {M}},  \emph {et~al.}} (\bibinfo {year} {2016}),\ \href
  {https://doi.org/10.1088/0741-3335/58/3/034017} {\bibfield  {journal}
  {\bibinfo  {journal} {Plasma Phys. Control. Fusion}\ }\textbf {\bibinfo
  {volume} {58}},\ \bibinfo {pages} {034017}}\BibitemShut {NoStop}%
\bibitem [{\citenamefont {Liu}\ \emph {et~al.}(2024)\citenamefont {Liu} \emph
  {et~al.}}]{Liu2024}%
  \BibitemOpen
  \bibfield  {author} {\bibinfo {author} {\bibnamefont {Liu}, \bibfnamefont
  {S}},  \emph {et~al.}} (\bibinfo {year} {2024}),\ \href
  {https://doi.org/10.1103/PhysRevLett.133.175001} {\bibfield  {journal}
  {\bibinfo  {journal} {Phys. Rev. Lett.}\ }\textbf {\bibinfo {volume} {133}},\
  \bibinfo {pages} {175001}}\BibitemShut {NoStop}%
\bibitem [{\citenamefont {Liu}\ \emph {et~al.}(2022)\citenamefont {Liu},
  \citenamefont {Xue}, \citenamefont {Wan}, \citenamefont {Chen}, \citenamefont
  {Li}, \citenamefont {Liu}, \citenamefont {Weng}, \citenamefont {Sheng},\ and\
  \citenamefont {Zhang}}]{Liu2022b}%
  \BibitemOpen
  \bibfield  {author} {\bibinfo {author} {\bibnamefont {Liu}, \bibfnamefont
  {W-Y}}, \bibinfo {author} {\bibfnamefont {K.}~\bibnamefont {Xue}}, \bibinfo
  {author} {\bibfnamefont {F.}~\bibnamefont {Wan}}, \bibinfo {author}
  {\bibfnamefont {M.}~\bibnamefont {Chen}}, \bibinfo {author} {\bibfnamefont
  {J.-X.}\ \bibnamefont {Li}}, \bibinfo {author} {\bibfnamefont
  {F.}~\bibnamefont {Liu}}, \bibinfo {author} {\bibfnamefont {S.-M.}\
  \bibnamefont {Weng}}, \bibinfo {author} {\bibfnamefont {Z.-M.}\ \bibnamefont
  {Sheng}}, \ and\ \bibinfo {author} {\bibfnamefont {J.}~\bibnamefont {Zhang}}}
  (\bibinfo {year} {2022}),\ \href
  {https://link.aps.org/doi/10.1103/PhysRevResearch.4.L022028} {\bibfield
  {journal} {\bibinfo  {journal} {Phys. Rev. Res.}\ }\textbf {\bibinfo {volume}
  {4}},\ \bibinfo {pages} {L022028}}\BibitemShut {NoStop}%
\bibitem [{\citenamefont {Liu}\ \emph {et~al.}(2025)\citenamefont {Liu},
  \citenamefont {Zeng}, \citenamefont {Reichwein},\ and\ \citenamefont
  {Pukhov}}]{Liu2025}%
  \BibitemOpen
  \bibfield  {author} {\bibinfo {author} {\bibnamefont {Liu}, \bibfnamefont
  {Yulong}}, \bibinfo {author} {\bibfnamefont {Ming}\ \bibnamefont {Zeng}},
  \bibinfo {author} {\bibfnamefont {Lars}\ \bibnamefont {Reichwein}}, \ and\
  \bibinfo {author} {\bibfnamefont {Alexander}\ \bibnamefont {Pukhov}}}
  (\bibinfo {year} {2025}),\ \href {\doibase 10.1103/physrevresearch.7.023101}
  {\bibfield  {journal} {\bibinfo  {journal} {Phys. Rev. Res.}\ }\textbf
  {\bibinfo {volume} {7}},\ \bibinfo {pages} {023101}}\BibitemShut {NoStop}%
\bibitem [{\citenamefont {Loisch}\ \emph
  {et~al.}(2018{\natexlab{a}})\citenamefont {Loisch} \emph
  {et~al.}}]{Loisch2018}%
  \BibitemOpen
  \bibfield  {author} {\bibinfo {author} {\bibnamefont {Loisch}, \bibfnamefont
  {G}},  \emph {et~al.}} (\bibinfo {year} {2018}{\natexlab{a}}),\ \href
  {https://doi.org/10.1103/physrevlett.121.064801} {\bibfield  {journal}
  {\bibinfo  {journal} {Phys. Rev. Lett.}\ }\textbf {\bibinfo {volume} {121}},\
  \bibinfo {pages} {064801}}\BibitemShut {NoStop}%
\bibitem [{\citenamefont {Loisch}\ \emph
  {et~al.}(2018{\natexlab{b}})\citenamefont {Loisch} \emph
  {et~al.}}]{Loisch2018b}%
  \BibitemOpen
  \bibfield  {author} {\bibinfo {author} {\bibnamefont {Loisch}, \bibfnamefont
  {G}},  \emph {et~al.}} (\bibinfo {year} {2018}{\natexlab{b}}),\ \href
  {https://doi.org/10.1016/j.nima.2018.02.043} {\bibfield  {journal} {\bibinfo
  {journal} {Nucl. Instrum. Methods Phys. Res. A}\ }\textbf {\bibinfo {volume}
  {909}},\ \bibinfo {pages} {107--110}}\BibitemShut {NoStop}%
\bibitem [{\citenamefont {Loisch}\ \emph
  {et~al.}(2019{\natexlab{a}})\citenamefont {Loisch} \emph
  {et~al.}}]{Loisch2019b}%
  \BibitemOpen
  \bibfield  {author} {\bibinfo {author} {\bibnamefont {Loisch}, \bibfnamefont
  {G}},  \emph {et~al.}} (\bibinfo {year} {2019}{\natexlab{a}}),\ \href
  {\doibase 10.1088/1361-6587/ab04b9} {\bibfield  {journal} {\bibinfo
  {journal} {Plasma Phys. Control. Fusion}\ }\textbf {\bibinfo {volume} {61}},\
  \bibinfo {pages} {045012}}\BibitemShut {NoStop}%
\bibitem [{\citenamefont {Loisch}\ \emph
  {et~al.}(2019{\natexlab{b}})\citenamefont {Loisch} \emph
  {et~al.}}]{Loisch2019}%
  \BibitemOpen
  \bibfield  {author} {\bibinfo {author} {\bibnamefont {Loisch}, \bibfnamefont
  {G}},  \emph {et~al.}} (\bibinfo {year} {2019}{\natexlab{b}}),\ \href
  {https://doi.org/10.1063/1.5068753} {\bibfield  {journal} {\bibinfo
  {journal} {J. Appl. Phys.}\ }\textbf {\bibinfo {volume} {125}},\ \bibinfo
  {pages} {063301}}\BibitemShut {NoStop}%
\bibitem [{\citenamefont {Lotov}(1998)}]{Lotov1998}%
  \BibitemOpen
  \bibfield  {author} {\bibinfo {author} {\bibnamefont {Lotov}, \bibfnamefont
  {K~V}}} (\bibinfo {year} {1998}),\ \href {https://doi.org/10.1063/1.872765}
  {\bibfield  {journal} {\bibinfo  {journal} {Phys. Plasmas}\ }\textbf
  {\bibinfo {volume} {5}},\ \bibinfo {pages} {785--791}}\BibitemShut {NoStop}%
\bibitem [{\citenamefont {Lotov}(2003)}]{Lotov2003}%
  \BibitemOpen
  \bibfield  {author} {\bibinfo {author} {\bibnamefont {Lotov}, \bibfnamefont
  {K~V}}} (\bibinfo {year} {2003}),\ \href
  {https://doi.org/10.1103/PhysRevSTAB.6.061301} {\bibfield  {journal}
  {\bibinfo  {journal} {Phys. Rev. ST Accel. Beams}\ }\textbf {\bibinfo
  {volume} {6}},\ \bibinfo {pages} {061301}}\BibitemShut {NoStop}%
\bibitem [{\citenamefont {Lotov}(2004)}]{Lotov2004}%
  \BibitemOpen
  \bibfield  {author} {\bibinfo {author} {\bibnamefont {Lotov}, \bibfnamefont
  {K~V}}} (\bibinfo {year} {2004}),\ \href
  {https://doi.org/10.1103/physreve.69.046405} {\bibfield  {journal} {\bibinfo
  {journal} {Phys. Rev. E}\ }\textbf {\bibinfo {volume} {69}},\ \bibinfo
  {pages} {046405}}\BibitemShut {NoStop}%
\bibitem [{\citenamefont {Lotov}(2005)}]{Lotov2005}%
  \BibitemOpen
  \bibfield  {author} {\bibinfo {author} {\bibnamefont {Lotov}, \bibfnamefont
  {K~V}}} (\bibinfo {year} {2005}),\ \href {https://doi.org/10.1063/1.1889444}
  {\bibfield  {journal} {\bibinfo  {journal} {Phys. Plasmas}\ }\textbf
  {\bibinfo {volume} {12}},\ \bibinfo {pages} {053105}}\BibitemShut {NoStop}%
\bibitem [{\citenamefont {Lotov}(2007)}]{Lotov2007}%
  \BibitemOpen
  \bibfield  {author} {\bibinfo {author} {\bibnamefont {Lotov}, \bibfnamefont
  {K~V}}} (\bibinfo {year} {2007}),\ \href {https://doi.org/10.1063/1.2434793}
  {\bibfield  {journal} {\bibinfo  {journal} {Phys. Plasmas}\ }\textbf
  {\bibinfo {volume} {14}},\ \bibinfo {pages} {023101}}\BibitemShut {NoStop}%
\bibitem [{\citenamefont {Lotov}(2011)}]{Lotov2011}%
  \BibitemOpen
  \bibfield  {author} {\bibinfo {author} {\bibnamefont {Lotov}, \bibfnamefont
  {K~V}}} (\bibinfo {year} {2011}),\ \href {https://doi.org/10.1063/1.3558697}
  {\bibfield  {journal} {\bibinfo  {journal} {Phys. Plasmas}\ }\textbf
  {\bibinfo {volume} {18}},\ \bibinfo {pages} {024501}}\BibitemShut {NoStop}%
\bibitem [{\citenamefont {Lotov}(2015)}]{Lotov2015}%
  \BibitemOpen
  \bibfield  {author} {\bibinfo {author} {\bibnamefont {Lotov}, \bibfnamefont
  {K~V}}} (\bibinfo {year} {2015}),\ \href {https://doi.org/10.1063/1.4933129}
  {\bibfield  {journal} {\bibinfo  {journal} {Phys. Plasmas}\ }\textbf
  {\bibinfo {volume} {22}},\ \bibinfo {pages} {103110}}\BibitemShut {NoStop}%
\bibitem [{\citenamefont {Lotov}(2017)}]{Lotov2017}%
  \BibitemOpen
  \bibfield  {author} {\bibinfo {author} {\bibnamefont {Lotov}, \bibfnamefont
  {K~V}}} (\bibinfo {year} {2017}),\ \href {https://doi.org/10.1063/1.4977058}
  {\bibfield  {journal} {\bibinfo  {journal} {Phys. Plasmas}\ }\textbf
  {\bibinfo {volume} {24}},\ \bibinfo {pages} {023119}}\BibitemShut {NoStop}%
\bibitem [{\citenamefont {Lotov}\ \emph
  {et~al.}(2013{\natexlab{a}})\citenamefont {Lotov}, \citenamefont {Lotova},
  \citenamefont {Lotov}, \citenamefont {Upadhyay}, \citenamefont
  {T\"{u}ckmantel}, \citenamefont {Pukhov},\ and\ \citenamefont
  {Caldwell}}]{Lotov2013}%
  \BibitemOpen
  \bibfield  {author} {\bibinfo {author} {\bibnamefont {Lotov}, \bibfnamefont
  {K~V}}, \bibinfo {author} {\bibfnamefont {G.~Z.}\ \bibnamefont {Lotova}},
  \bibinfo {author} {\bibfnamefont {V.~I.}\ \bibnamefont {Lotov}}, \bibinfo
  {author} {\bibfnamefont {A.}~\bibnamefont {Upadhyay}}, \bibinfo {author}
  {\bibfnamefont {T.}~\bibnamefont {T\"{u}ckmantel}}, \bibinfo {author}
  {\bibfnamefont {A.}~\bibnamefont {Pukhov}}, \ and\ \bibinfo {author}
  {\bibfnamefont {A.}~\bibnamefont {Caldwell}}} (\bibinfo {year}
  {2013}{\natexlab{a}}),\ \href {https://doi.org/10.1103/physrevstab.16.041301}
  {\bibfield  {journal} {\bibinfo  {journal} {Phys. Rev. ST Accel. Beams}\
  }\textbf {\bibinfo {volume} {16}},\ \bibinfo {pages} {041301}}\BibitemShut
  {NoStop}%
\bibitem [{\citenamefont {Lotov}\ \emph
  {et~al.}(2013{\natexlab{b}})\citenamefont {Lotov}, \citenamefont {Pukhov},\
  and\ \citenamefont {Caldwell}}]{Lotov2013b}%
  \BibitemOpen
  \bibfield  {author} {\bibinfo {author} {\bibnamefont {Lotov}, \bibfnamefont
  {K~V}}, \bibinfo {author} {\bibfnamefont {A.}~\bibnamefont {Pukhov}}, \ and\
  \bibinfo {author} {\bibfnamefont {A.}~\bibnamefont {Caldwell}}} (\bibinfo
  {year} {2013}{\natexlab{b}}),\ \href {https://doi.org/10.1063/1.4773905}
  {\bibfield  {journal} {\bibinfo  {journal} {Phys. Plasmas}\ }\textbf
  {\bibinfo {volume} {20}},\ \bibinfo {pages} {013102}}\BibitemShut {NoStop}%
\bibitem [{\citenamefont {Lotov}\ \emph {et~al.}(2014)\citenamefont {Lotov},
  \citenamefont {Sosedkin}, \citenamefont {Petrenko}, \citenamefont {Amorim},
  \citenamefont {Vieira}, \citenamefont {Fonseca}, \citenamefont {Silva},
  \citenamefont {Gschwendtner},\ and\ \citenamefont {Muggli}}]{Lotov2014}%
  \BibitemOpen
  \bibfield  {author} {\bibinfo {author} {\bibnamefont {Lotov}, \bibfnamefont
  {K~V}}, \bibinfo {author} {\bibfnamefont {A.~P.}\ \bibnamefont {Sosedkin}},
  \bibinfo {author} {\bibfnamefont {A.~V.}\ \bibnamefont {Petrenko}}, \bibinfo
  {author} {\bibfnamefont {L.~D.}\ \bibnamefont {Amorim}}, \bibinfo {author}
  {\bibfnamefont {J.}~\bibnamefont {Vieira}}, \bibinfo {author} {\bibfnamefont
  {R.~A.}\ \bibnamefont {Fonseca}}, \bibinfo {author} {\bibfnamefont {L.~O.}\
  \bibnamefont {Silva}}, \bibinfo {author} {\bibfnamefont {E.}~\bibnamefont
  {Gschwendtner}}, \ and\ \bibinfo {author} {\bibfnamefont {P.}~\bibnamefont
  {Muggli}}} (\bibinfo {year} {2014}),\ \href
  {https://doi.org/10.1063/1.4904365} {\bibfield  {journal} {\bibinfo
  {journal} {Phys. Plasmas}\ }\textbf {\bibinfo {volume} {21}},\ \bibinfo
  {pages} {123116}}\BibitemShut {NoStop}%
\bibitem [{\citenamefont {Lu}(2006)}]{Lu2006c}%
  \BibitemOpen
  \bibfield  {author} {\bibinfo {author} {\bibnamefont {Lu}, \bibfnamefont
  {W}}} (\bibinfo {year} {2006}),\ \href
  {https://picksc.physics.ucla.edu/reports-notes/weilu.pdf} {\bibinfo {type}
  {Ph{D} {T}hesis}}\ (\bibinfo  {school} {University of California, Los
  Angeles})\BibitemShut {NoStop}%
\bibitem [{\citenamefont {Lu}\ \emph {et~al.}(2010)\citenamefont {Lu},
  \citenamefont {An}, \citenamefont {Zhou}, \citenamefont {Joshi},
  \citenamefont {Huang},\ and\ \citenamefont {Mori}}]{Lu2010}%
  \BibitemOpen
  \bibfield  {author} {\bibinfo {author} {\bibnamefont {Lu}, \bibfnamefont
  {W}}, \bibinfo {author} {\bibfnamefont {W.}~\bibnamefont {An}}, \bibinfo
  {author} {\bibfnamefont {M.}~\bibnamefont {Zhou}}, \bibinfo {author}
  {\bibfnamefont {C.}~\bibnamefont {Joshi}}, \bibinfo {author} {\bibfnamefont
  {C.}~\bibnamefont {Huang}}, \ and\ \bibinfo {author} {\bibfnamefont {W.~B.}\
  \bibnamefont {Mori}}} (\bibinfo {year} {2010}),\ \href
  {https://doi.org/10.1088/1367-2630/12/8/085002} {\bibfield  {journal}
  {\bibinfo  {journal} {New J. Phys.}\ }\textbf {\bibinfo {volume} {12}},\
  \bibinfo {pages} {085002}}\BibitemShut {NoStop}%
\bibitem [{\citenamefont {Lu}\ \emph {et~al.}(2006{\natexlab{a}})\citenamefont
  {Lu}, \citenamefont {Huang}, \citenamefont {Zhou}, \citenamefont {Mori},\
  and\ \citenamefont {Katsouleas}}]{Lu2006a}%
  \BibitemOpen
  \bibfield  {author} {\bibinfo {author} {\bibnamefont {Lu}, \bibfnamefont
  {W}}, \bibinfo {author} {\bibfnamefont {C.}~\bibnamefont {Huang}}, \bibinfo
  {author} {\bibfnamefont {M.}~\bibnamefont {Zhou}}, \bibinfo {author}
  {\bibfnamefont {W.~B.}\ \bibnamefont {Mori}}, \ and\ \bibinfo {author}
  {\bibfnamefont {T.}~\bibnamefont {Katsouleas}}} (\bibinfo {year}
  {2006}{\natexlab{a}}),\ \href {https://doi.org/10.1103/physrevlett.96.165002}
  {\bibfield  {journal} {\bibinfo  {journal} {Phys. Rev. Lett.}\ }\textbf
  {\bibinfo {volume} {96}},\ \bibinfo {pages} {165002}}\BibitemShut {NoStop}%
\bibitem [{\citenamefont {Lu}\ \emph {et~al.}(2006{\natexlab{b}})\citenamefont
  {Lu}, \citenamefont {Huang}, \citenamefont {Zhou}, \citenamefont {Tzoufras},
  \citenamefont {Tsung}, \citenamefont {Mori},\ and\ \citenamefont
  {Katsouleas}}]{Lu2006b}%
  \BibitemOpen
  \bibfield  {author} {\bibinfo {author} {\bibnamefont {Lu}, \bibfnamefont
  {W}}, \bibinfo {author} {\bibfnamefont {C.}~\bibnamefont {Huang}}, \bibinfo
  {author} {\bibfnamefont {M.}~\bibnamefont {Zhou}}, \bibinfo {author}
  {\bibfnamefont {M.}~\bibnamefont {Tzoufras}}, \bibinfo {author}
  {\bibfnamefont {F.~S.}\ \bibnamefont {Tsung}}, \bibinfo {author}
  {\bibfnamefont {W.~B.}\ \bibnamefont {Mori}}, \ and\ \bibinfo {author}
  {\bibfnamefont {T.}~\bibnamefont {Katsouleas}}} (\bibinfo {year}
  {2006}{\natexlab{b}}),\ \href {https://doi.org/10.1063/1.2203364} {\bibfield
  {journal} {\bibinfo  {journal} {Phys. Plasmas}\ }\textbf {\bibinfo {volume}
  {13}},\ \bibinfo {pages} {056709}}\BibitemShut {NoStop}%
\bibitem [{\citenamefont {Lu}\ \emph {et~al.}(2005)\citenamefont {Lu},
  \citenamefont {Huang}, \citenamefont {Zhou}, \citenamefont {Mori},\ and\
  \citenamefont {Katsouleas}}]{Lu2005}%
  \BibitemOpen
  \bibfield  {author} {\bibinfo {author} {\bibnamefont {Lu}, \bibfnamefont
  {W}}, \bibinfo {author} {\bibfnamefont {C.}~\bibnamefont {Huang}}, \bibinfo
  {author} {\bibfnamefont {M.~M.}\ \bibnamefont {Zhou}}, \bibinfo {author}
  {\bibfnamefont {W.~B.}\ \bibnamefont {Mori}}, \ and\ \bibinfo {author}
  {\bibfnamefont {T.}~\bibnamefont {Katsouleas}}} (\bibinfo {year} {2005}),\
  \href {\doibase 10.1063/1.1905587} {\bibfield  {journal} {\bibinfo  {journal}
  {Phys. Plasmas}\ }\textbf {\bibinfo {volume} {12}},\ \bibinfo {pages}
  {063101}}\BibitemShut {NoStop}%
\bibitem [{\citenamefont {Luo}\ \emph {et~al.}(2018)\citenamefont {Luo} \emph
  {et~al.}}]{Luo2018}%
  \BibitemOpen
  \bibfield  {author} {\bibinfo {author} {\bibnamefont {Luo}, \bibfnamefont
  {J}},  \emph {et~al.}} (\bibinfo {year} {2018}),\ \href {\doibase
  10.1103/physrevlett.120.154801} {\bibfield  {journal} {\bibinfo  {journal}
  {Phys. Rev. Lett.}\ }\textbf {\bibinfo {volume} {120}},\ \bibinfo {pages}
  {154801}}\BibitemShut {NoStop}%
\bibitem [{\citenamefont {Lynch}\ and\ \citenamefont {Dahl}(1991)}]{Lynch1991}%
  \BibitemOpen
  \bibfield  {author} {\bibinfo {author} {\bibnamefont {Lynch}, \bibfnamefont
  {G~R}}, \ and\ \bibinfo {author} {\bibfnamefont {O.~I.}\ \bibnamefont
  {Dahl}}} (\bibinfo {year} {1991}),\ \href {\doibase
  10.1016/0168-583X(91)95671-Y} {\bibfield  {journal} {\bibinfo  {journal}
  {Nucl. Instrum. Methods Phys. Res. B}\ }\textbf {\bibinfo {volume} {58}},\
  \bibinfo {pages} {6--10}}\BibitemShut {NoStop}%
\bibitem [{\citenamefont {Maeda}\ \emph {et~al.}(2004)\citenamefont {Maeda},
  \citenamefont {Katsouleas}, \citenamefont {Muggli}, \citenamefont {Joshi},
  \citenamefont {Mori},\ and\ \citenamefont {Quillinan}}]{Maeda2004}%
  \BibitemOpen
  \bibfield  {author} {\bibinfo {author} {\bibnamefont {Maeda}, \bibfnamefont
  {R}}, \bibinfo {author} {\bibfnamefont {T.}~\bibnamefont {Katsouleas}},
  \bibinfo {author} {\bibfnamefont {P.}~\bibnamefont {Muggli}}, \bibinfo
  {author} {\bibfnamefont {C.}~\bibnamefont {Joshi}}, \bibinfo {author}
  {\bibfnamefont {W.}~\bibnamefont {Mori}}, \ and\ \bibinfo {author}
  {\bibfnamefont {W.}~\bibnamefont {Quillinan}}} (\bibinfo {year} {2004}),\
  \href {https://doi.org/10.1103/physrevstab.7.111301} {\bibfield  {journal}
  {\bibinfo  {journal} {Phys. Rev. ST Accel. Beams}\ }\textbf {\bibinfo
  {volume} {7}},\ \bibinfo {pages} {111301}}\BibitemShut {NoStop}%
\bibitem [{\citenamefont {Manahan}\ \emph {et~al.}(2016)\citenamefont {Manahan}
  \emph {et~al.}}]{Manahan2016}%
  \BibitemOpen
  \bibfield  {author} {\bibinfo {author} {\bibnamefont {Manahan}, \bibfnamefont
  {G~G}},  \emph {et~al.}} (\bibinfo {year} {2016}),\ \href
  {https://doi.org/10.1103/PhysRevAccelBeams.19.011303} {\bibfield  {journal}
  {\bibinfo  {journal} {Phys. Rev. Accel. Beams}\ }\textbf {\bibinfo {volume}
  {19}},\ \bibinfo {pages} {011303}}\BibitemShut {NoStop}%
\bibitem [{\citenamefont {Manahan}\ \emph {et~al.}(2017)\citenamefont {Manahan}
  \emph {et~al.}}]{Manahan2017}%
  \BibitemOpen
  \bibfield  {author} {\bibinfo {author} {\bibnamefont {Manahan}, \bibfnamefont
  {G~G}},  \emph {et~al.}} (\bibinfo {year} {2017}),\ \href
  {https://doi.org/10.1038/ncomms15705} {\bibfield  {journal} {\bibinfo
  {journal} {Nat. Commun.}\ }\textbf {\bibinfo {volume} {8}},\ \bibinfo {pages}
  {15705}}\BibitemShut {NoStop}%
\bibitem [{\citenamefont {Mane}\ \emph {et~al.}(2005)\citenamefont {Mane},
  \citenamefont {Shatunov},\ and\ \citenamefont {Yokoya}}]{Mane2005}%
  \BibitemOpen
  \bibfield  {author} {\bibinfo {author} {\bibnamefont {Mane}, \bibfnamefont
  {S~R}}, \bibinfo {author} {\bibfnamefont {Yu.~M.}\ \bibnamefont {Shatunov}},
  \ and\ \bibinfo {author} {\bibfnamefont {K.}~\bibnamefont {Yokoya}}}
  (\bibinfo {year} {2005}),\ \href {https://doi.org/10.1088/0034-4885/68/9/R01}
  {\bibfield  {journal} {\bibinfo  {journal} {Rep. Prog. Phys.}\ }\textbf
  {\bibinfo {volume} {68}},\ \bibinfo {pages} {1997--2265}}\BibitemShut
  {NoStop}%
\bibitem [{\citenamefont {Manwani}\ \emph {et~al.}(2025)\citenamefont
  {Manwani}, \citenamefont {Kang}, \citenamefont {Mann}, \citenamefont
  {Naranjo}, \citenamefont {Andonian},\ and\ \citenamefont
  {Rosenzweig}}]{Manwani2025}%
  \BibitemOpen
  \bibfield  {author} {\bibinfo {author} {\bibnamefont {Manwani}, \bibfnamefont
  {P}}, \bibinfo {author} {\bibfnamefont {Y.}~\bibnamefont {Kang}}, \bibinfo
  {author} {\bibfnamefont {J.}~\bibnamefont {Mann}}, \bibinfo {author}
  {\bibfnamefont {B.}~\bibnamefont {Naranjo}}, \bibinfo {author} {\bibfnamefont
  {G.}~\bibnamefont {Andonian}}, \ and\ \bibinfo {author} {\bibfnamefont
  {J.~B.}\ \bibnamefont {Rosenzweig}}} (\bibinfo {year} {2025}),\ \href
  {\doibase 10.1103/28c4-blhg} {\bibfield  {journal} {\bibinfo  {journal}
  {Phys. Rev. Lett.}\ }\textbf {\bibinfo {volume} {135}},\ \bibinfo {pages}
  {095001}}\BibitemShut {NoStop}%
\bibitem [{\citenamefont {Manwani}\ \emph {et~al.}(2021)\citenamefont
  {Manwani}, \citenamefont {Majernik}, \citenamefont {Yadav}, \citenamefont
  {Hansel},\ and\ \citenamefont {Rosenzweig}}]{Manwani2021}%
  \BibitemOpen
  \bibfield  {author} {\bibinfo {author} {\bibnamefont {Manwani}, \bibfnamefont
  {P}}, \bibinfo {author} {\bibfnamefont {N.}~\bibnamefont {Majernik}},
  \bibinfo {author} {\bibfnamefont {M.}~\bibnamefont {Yadav}}, \bibinfo
  {author} {\bibfnamefont {C.}~\bibnamefont {Hansel}}, \ and\ \bibinfo {author}
  {\bibfnamefont {J.~B.}\ \bibnamefont {Rosenzweig}}} (\bibinfo {year}
  {2021}),\ \href {\doibase 10.1103/physrevaccelbeams.24.051302} {\bibfield
  {journal} {\bibinfo  {journal} {Phys. Rev. Accel. Beams}\ }\textbf {\bibinfo
  {volume} {24}},\ \bibinfo {pages} {051302}}\BibitemShut {NoStop}%
\bibitem [{\citenamefont {Marocchino}\ \emph {et~al.}(2016)\citenamefont
  {Marocchino}, \citenamefont {Massimo}, \citenamefont {Rossi}, \citenamefont
  {Chiadroni},\ and\ \citenamefont {Ferrario}}]{Marocchino2016}%
  \BibitemOpen
  \bibfield  {author} {\bibinfo {author} {\bibnamefont {Marocchino},
  \bibfnamefont {A}}, \bibinfo {author} {\bibfnamefont {F.}~\bibnamefont
  {Massimo}}, \bibinfo {author} {\bibfnamefont {A.~R.}\ \bibnamefont {Rossi}},
  \bibinfo {author} {\bibfnamefont {E.}~\bibnamefont {Chiadroni}}, \ and\
  \bibinfo {author} {\bibfnamefont {M.}~\bibnamefont {Ferrario}}} (\bibinfo
  {year} {2016}),\ \href {https://doi.org/10.1016/j.nima.2016.03.005}
  {\bibfield  {journal} {\bibinfo  {journal} {Nucl. Instrum. Methods Phys. Res.
  A}\ }\textbf {\bibinfo {volume} {829}},\ \bibinfo {pages}
  {386--391}}\BibitemShut {NoStop}%
\bibitem [{\citenamefont {Marocchino}\ \emph {et~al.}(2017)\citenamefont
  {Marocchino} \emph {et~al.}}]{Marocchino2017}%
  \BibitemOpen
  \bibfield  {author} {\bibinfo {author} {\bibnamefont {Marocchino},
  \bibfnamefont {A}},  \emph {et~al.}} (\bibinfo {year} {2017}),\ \href
  {https://doi.org/10.1063/1.4999010} {\bibfield  {journal} {\bibinfo
  {journal} {Appl. Phys. Lett.}\ }\textbf {\bibinfo {volume} {111}},\ \bibinfo
  {pages} {184101}}\BibitemShut {NoStop}%
\bibitem [{\citenamefont {Marsh}\ \emph {et~al.}(2003)\citenamefont {Marsh}
  \emph {et~al.}}]{Marsh2003}%
  \BibitemOpen
  \bibfield  {author} {\bibinfo {author} {\bibnamefont {Marsh}, \bibfnamefont
  {K~A}},  \emph {et~al.}} (\bibinfo {year} {2003}),\ in\ \href
  {https://doi.org/10.1109/pac.2003.1289023} {\emph {\bibinfo {booktitle}
  {Proceedings of the 2003 Particle Accelerator Conf.}}}\ (\bibinfo
  {publisher} {{IEEE}},\ \bibinfo {address} {New York, NY, United States})\
  pp.\ \bibinfo {pages} {731--733}\BibitemShut {NoStop}%
\bibitem [{\citenamefont {Marsh}\ \emph {et~al.}(2005)\citenamefont {Marsh}
  \emph {et~al.}}]{Marsh2005}%
  \BibitemOpen
  \bibfield  {author} {\bibinfo {author} {\bibnamefont {Marsh}, \bibfnamefont
  {K~A}},  \emph {et~al.}} (\bibinfo {year} {2005}),\ in\ \href
  {https://doi.org/10.1109/pac.2005.1591234} {\emph {\bibinfo {booktitle}
  {Proceedings of the 2005 Particle Accelerator Conf.}}}\ (\bibinfo
  {publisher} {{IEEE}},\ \bibinfo {address} {New York, NY, United States})\
  pp.\ \bibinfo {pages} {2702--2704}\BibitemShut {NoStop}%
\bibitem [{\citenamefont {Martinez}\ \emph {et~al.}(2023)\citenamefont
  {Martinez}, \citenamefont {Barbosa},\ and\ \citenamefont
  {Vranic}}]{Martinez2023}%
  \BibitemOpen
  \bibfield  {author} {\bibinfo {author} {\bibnamefont {Martinez},
  \bibfnamefont {B}}, \bibinfo {author} {\bibfnamefont {B.}~\bibnamefont
  {Barbosa}}, \ and\ \bibinfo {author} {\bibfnamefont {M.}~\bibnamefont
  {Vranic}}} (\bibinfo {year} {2023}),\ \href
  {https://link.aps.org/doi/10.1103/PhysRevAccelBeams.26.011301} {\bibfield
  {journal} {\bibinfo  {journal} {Phys. Rev. Accel. Beams}\ }\textbf {\bibinfo
  {volume} {26}},\ \bibinfo {pages} {011301}}\BibitemShut {NoStop}%
\bibitem [{\citenamefont {{Martinez de la Ossa}}\ \emph
  {et~al.}(2013)\citenamefont {{Martinez de la Ossa}}, \citenamefont
  {Grebenyuk}, \citenamefont {Mehrling}, \citenamefont {Schaper},\ and\
  \citenamefont {Osterhoff}}]{MartinezdelaOssa2013}%
  \BibitemOpen
  \bibfield  {author} {\bibinfo {author} {\bibnamefont {{Martinez de la Ossa}},
  \bibfnamefont {A}}, \bibinfo {author} {\bibfnamefont {J.}~\bibnamefont
  {Grebenyuk}}, \bibinfo {author} {\bibfnamefont {T.}~\bibnamefont {Mehrling}},
  \bibinfo {author} {\bibfnamefont {L.}~\bibnamefont {Schaper}}, \ and\
  \bibinfo {author} {\bibfnamefont {J.}~\bibnamefont {Osterhoff}}} (\bibinfo
  {year} {2013}),\ \href {https://doi.org/10.1103/physrevlett.111.245003}
  {\bibfield  {journal} {\bibinfo  {journal} {Phys. Rev. Lett.}\ }\textbf
  {\bibinfo {volume} {111}},\ \bibinfo {pages} {245003}}\BibitemShut {NoStop}%
\bibitem [{\citenamefont {{Martinez de la Ossa}}\ \emph
  {et~al.}(2017)\citenamefont {{Martinez de la Ossa}}, \citenamefont {Hu},
  \citenamefont {Streeter}, \citenamefont {Mehrling}, \citenamefont
  {Kononenko}, \citenamefont {Sheeran},\ and\ \citenamefont
  {Osterhoff}}]{MartinezdelaOssa2017}%
  \BibitemOpen
  \bibfield  {author} {\bibinfo {author} {\bibnamefont {{Martinez de la Ossa}},
  \bibfnamefont {A}}, \bibinfo {author} {\bibfnamefont {Z.}~\bibnamefont {Hu}},
  \bibinfo {author} {\bibfnamefont {M.~J.~V.}\ \bibnamefont {Streeter}},
  \bibinfo {author} {\bibfnamefont {T.~J.}\ \bibnamefont {Mehrling}}, \bibinfo
  {author} {\bibfnamefont {O.}~\bibnamefont {Kononenko}}, \bibinfo {author}
  {\bibfnamefont {B.}~\bibnamefont {Sheeran}}, \ and\ \bibinfo {author}
  {\bibfnamefont {J.}~\bibnamefont {Osterhoff}}} (\bibinfo {year} {2017}),\
  \href {https://doi.org/10.1103/PhysRevAccelBeams.20.091301} {\bibfield
  {journal} {\bibinfo  {journal} {Phys. Rev. Accel. Beams}\ }\textbf {\bibinfo
  {volume} {20}},\ \bibinfo {pages} {091301}}\BibitemShut {NoStop}%
\bibitem [{\citenamefont {{Martinez de la Ossa}}\ \emph
  {et~al.}(2018)\citenamefont {{Martinez de la Ossa}}, \citenamefont
  {Mehrling},\ and\ \citenamefont {Osterhoff}}]{MartinezdelaOssa2018}%
  \BibitemOpen
  \bibfield  {author} {\bibinfo {author} {\bibnamefont {{Martinez de la Ossa}},
  \bibfnamefont {A}}, \bibinfo {author} {\bibfnamefont {T.~J.}\ \bibnamefont
  {Mehrling}}, \ and\ \bibinfo {author} {\bibfnamefont {J.}~\bibnamefont
  {Osterhoff}}} (\bibinfo {year} {2018}),\ \href
  {https://doi.org/10.1103/physrevlett.121.064803} {\bibfield  {journal}
  {\bibinfo  {journal} {Phys. Rev. Lett.}\ }\textbf {\bibinfo {volume} {121}},\
  \bibinfo {pages} {064803}}\BibitemShut {NoStop}%
\bibitem [{\citenamefont {{Martinez de la Ossa}}\ \emph
  {et~al.}(2015)\citenamefont {{Martinez de la Ossa}}, \citenamefont
  {Mehrling}, \citenamefont {Schaper}, \citenamefont {Streeter},\ and\
  \citenamefont {Osterhoff}}]{MartinezdelaOssa2015}%
  \BibitemOpen
  \bibfield  {author} {\bibinfo {author} {\bibnamefont {{Martinez de la Ossa}},
  \bibfnamefont {A}}, \bibinfo {author} {\bibfnamefont {T.~J.}\ \bibnamefont
  {Mehrling}}, \bibinfo {author} {\bibfnamefont {L.}~\bibnamefont {Schaper}},
  \bibinfo {author} {\bibfnamefont {M.~J.~V.}\ \bibnamefont {Streeter}}, \ and\
  \bibinfo {author} {\bibfnamefont {J.}~\bibnamefont {Osterhoff}}} (\bibinfo
  {year} {2015}),\ \href {https://doi.org/10.1063/1.4929921} {\bibfield
  {journal} {\bibinfo  {journal} {Phys. Plasmas}\ }\textbf {\bibinfo {volume}
  {22}},\ \bibinfo {pages} {093107}}\BibitemShut {NoStop}%
\bibitem [{\citenamefont {Massimo}\ \emph {et~al.}(2014)\citenamefont
  {Massimo}, \citenamefont {Marocchino}, \citenamefont {Chiadroni},
  \citenamefont {Ferrario}, \citenamefont {Mostacci}, \citenamefont
  {Musumeci},\ and\ \citenamefont {Palumbo}}]{Massimo2014}%
  \BibitemOpen
  \bibfield  {author} {\bibinfo {author} {\bibnamefont {Massimo}, \bibfnamefont
  {F}}, \bibinfo {author} {\bibfnamefont {A.}~\bibnamefont {Marocchino}},
  \bibinfo {author} {\bibfnamefont {E.}~\bibnamefont {Chiadroni}}, \bibinfo
  {author} {\bibfnamefont {M.}~\bibnamefont {Ferrario}}, \bibinfo {author}
  {\bibfnamefont {A.}~\bibnamefont {Mostacci}}, \bibinfo {author}
  {\bibfnamefont {P.}~\bibnamefont {Musumeci}}, \ and\ \bibinfo {author}
  {\bibfnamefont {L.}~\bibnamefont {Palumbo}}} (\bibinfo {year} {2014}),\ \href
  {https://doi.org/10.1016/j.nima.2013.10.046} {\bibfield  {journal} {\bibinfo
  {journal} {Nucl. Instrum. Methods Phys. Res. A}\ }\textbf {\bibinfo {volume}
  {740}},\ \bibinfo {pages} {242--245}}\BibitemShut {NoStop}%
\bibitem [{\citenamefont {Masson-Laborde}\ \emph {et~al.}(2014)\citenamefont
  {Masson-Laborde}, \citenamefont {Mo}, \citenamefont {Ali}, \citenamefont
  {Fourmaux}, \citenamefont {Lassonde}, \citenamefont {Kieffer}, \citenamefont
  {Rozmus}, \citenamefont {Teychenn{\'{e}}},\ and\ \citenamefont
  {Fedosejevs}}]{MassonLaborde2014}%
  \BibitemOpen
  \bibfield  {author} {\bibinfo {author} {\bibnamefont {Masson-Laborde},
  \bibfnamefont {P~E}}, \bibinfo {author} {\bibfnamefont {M.~Z.}\ \bibnamefont
  {Mo}}, \bibinfo {author} {\bibfnamefont {A.}~\bibnamefont {Ali}}, \bibinfo
  {author} {\bibfnamefont {S.}~\bibnamefont {Fourmaux}}, \bibinfo {author}
  {\bibfnamefont {P.}~\bibnamefont {Lassonde}}, \bibinfo {author}
  {\bibfnamefont {J.~C.}\ \bibnamefont {Kieffer}}, \bibinfo {author}
  {\bibfnamefont {W.}~\bibnamefont {Rozmus}}, \bibinfo {author} {\bibfnamefont
  {D.}~\bibnamefont {Teychenn{\'{e}}}}, \ and\ \bibinfo {author} {\bibfnamefont
  {R.}~\bibnamefont {Fedosejevs}}} (\bibinfo {year} {2014}),\ \href
  {https://doi.org/10.1063/1.4903851} {\bibfield  {journal} {\bibinfo
  {journal} {Phys. Plasmas}\ }\textbf {\bibinfo {volume} {21}},\ \bibinfo
  {pages} {123113}}\BibitemShut {NoStop}%
\bibitem [{\citenamefont {Matheron}\ \emph {et~al.}(2024)\citenamefont
  {Matheron} \emph {et~al.}}]{Matheron2024}%
  \BibitemOpen
  \bibfield  {author} {\bibinfo {author} {\bibnamefont {Matheron},
  \bibfnamefont {A}},  \emph {et~al.}} (\bibinfo {year} {2024}),\ \href
  {\doibase 10.48550/arxiv.2412.19337} {}\Eprint
  {http://arxiv.org/abs/2412.19337} {arXiv:2412.19337} \BibitemShut {NoStop}%
\bibitem [{\citenamefont {McGuffey}\ \emph {et~al.}(2010)\citenamefont
  {McGuffey} \emph {et~al.}}]{McGuffey2010}%
  \BibitemOpen
  \bibfield  {author} {\bibinfo {author} {\bibnamefont {McGuffey},
  \bibfnamefont {C}},  \emph {et~al.}} (\bibinfo {year} {2010}),\ \href
  {https://doi.org/10.1103/PhysRevLett.104.025004} {\bibfield  {journal}
  {\bibinfo  {journal} {Phys. Rev. Lett.}\ }\textbf {\bibinfo {volume} {104}},\
  \bibinfo {pages} {025004}}\BibitemShut {NoStop}%
\bibitem [{\citenamefont {van~der Meer}(1985)}]{vanderMeer1985}%
  \BibitemOpen
  \bibfield  {author} {\bibinfo {author} {\bibnamefont {van~der Meer},
  \bibfnamefont {S}}} (\bibinfo {year} {1985}),\ \href
  {https://cds.cern.ch/record/163918/files/CM-P00058040.pdf} {}\bibinfo {type}
  {Tech. Rep.}\ \bibinfo {number} {CERN/PS/85-65}\ (\bibinfo  {institution}
  {CERN},\ \bibinfo {address} {Geneva, Switzerland})\BibitemShut {NoStop}%
\bibitem [{\citenamefont {Mehrling}\ \emph {et~al.}(2014)\citenamefont
  {Mehrling}, \citenamefont {Benedetti}, \citenamefont {Schroeder},\ and\
  \citenamefont {Osterhoff}}]{Mehrling2014}%
  \BibitemOpen
  \bibfield  {author} {\bibinfo {author} {\bibnamefont {Mehrling},
  \bibfnamefont {T}}, \bibinfo {author} {\bibfnamefont {C.}~\bibnamefont
  {Benedetti}}, \bibinfo {author} {\bibfnamefont {C.~B.}\ \bibnamefont
  {Schroeder}}, \ and\ \bibinfo {author} {\bibfnamefont {J.}~\bibnamefont
  {Osterhoff}}} (\bibinfo {year} {2014}),\ \href
  {https://doi.org/10.1088/0741-3335/56/8/084012} {\bibfield  {journal}
  {\bibinfo  {journal} {Plasma Phys. Control. Fusion}\ }\textbf {\bibinfo
  {volume} {56}},\ \bibinfo {pages} {084012}}\BibitemShut {NoStop}%
\bibitem [{\citenamefont {Mehrling}\ \emph {et~al.}(2012)\citenamefont
  {Mehrling}, \citenamefont {Grebenyuk}, \citenamefont {Tsung}, \citenamefont
  {Floettmann},\ and\ \citenamefont {Osterhoff}}]{Mehrling2012}%
  \BibitemOpen
  \bibfield  {author} {\bibinfo {author} {\bibnamefont {Mehrling},
  \bibfnamefont {T}}, \bibinfo {author} {\bibfnamefont {J.}~\bibnamefont
  {Grebenyuk}}, \bibinfo {author} {\bibfnamefont {F.~S.}\ \bibnamefont
  {Tsung}}, \bibinfo {author} {\bibfnamefont {K.}~\bibnamefont {Floettmann}}, \
  and\ \bibinfo {author} {\bibfnamefont {J.}~\bibnamefont {Osterhoff}}}
  (\bibinfo {year} {2012}),\ \href
  {https://doi.org/10.1103/physrevstab.15.111303} {\bibfield  {journal}
  {\bibinfo  {journal} {Phys. Rev. ST Accel. Beams}\ }\textbf {\bibinfo
  {volume} {15}},\ \bibinfo {pages} {111303}}\BibitemShut {NoStop}%
\bibitem [{\citenamefont {Mehrling}\ \emph {et~al.}(2018)\citenamefont
  {Mehrling}, \citenamefont {Benedetti}, \citenamefont {Schroeder},
  \citenamefont {Esarey},\ and\ \citenamefont {Leemans}}]{Mehrling2018}%
  \BibitemOpen
  \bibfield  {author} {\bibinfo {author} {\bibnamefont {Mehrling},
  \bibfnamefont {T~J}}, \bibinfo {author} {\bibfnamefont {C.}~\bibnamefont
  {Benedetti}}, \bibinfo {author} {\bibfnamefont {C.~B.}\ \bibnamefont
  {Schroeder}}, \bibinfo {author} {\bibfnamefont {E.}~\bibnamefont {Esarey}}, \
  and\ \bibinfo {author} {\bibfnamefont {W.~P.}\ \bibnamefont {Leemans}}}
  (\bibinfo {year} {2018}),\ \href
  {https://doi.org/10.1103/physrevlett.121.264802} {\bibfield  {journal}
  {\bibinfo  {journal} {Phys. Rev. Lett.}\ }\textbf {\bibinfo {volume} {121}},\
  \bibinfo {pages} {264802}}\BibitemShut {NoStop}%
\bibitem [{\citenamefont {Mehrling}\ \emph {et~al.}(2017)\citenamefont
  {Mehrling}, \citenamefont {Fonseca}, \citenamefont {{Martinez de la Ossa}},\
  and\ \citenamefont {Vieira}}]{Mehrling2017}%
  \BibitemOpen
  \bibfield  {author} {\bibinfo {author} {\bibnamefont {Mehrling},
  \bibfnamefont {T~J}}, \bibinfo {author} {\bibfnamefont {R.~A.}\ \bibnamefont
  {Fonseca}}, \bibinfo {author} {\bibfnamefont {A.}~\bibnamefont {{Martinez de
  la Ossa}}}, \ and\ \bibinfo {author} {\bibfnamefont {J.}~\bibnamefont
  {Vieira}}} (\bibinfo {year} {2017}),\ \href
  {https://doi.org/10.1103/physrevlett.118.174801} {\bibfield  {journal}
  {\bibinfo  {journal} {Phys. Rev. Lett.}\ }\textbf {\bibinfo {volume} {118}},\
  \bibinfo {pages} {174801}}\BibitemShut {NoStop}%
\bibitem [{\citenamefont {Mehrling}\ \emph {et~al.}(2019)\citenamefont
  {Mehrling}, \citenamefont {Fonseca}, \citenamefont {Martinez de~la Ossa},\
  and\ \citenamefont {Vieira}}]{Mehrling2019}%
  \BibitemOpen
  \bibfield  {author} {\bibinfo {author} {\bibnamefont {Mehrling},
  \bibfnamefont {T~J}}, \bibinfo {author} {\bibfnamefont {R.~A.}\ \bibnamefont
  {Fonseca}}, \bibinfo {author} {\bibfnamefont {A.}~\bibnamefont {Martinez
  de~la Ossa}}, \ and\ \bibinfo {author} {\bibfnamefont {J.}~\bibnamefont
  {Vieira}}} (\bibinfo {year} {2019}),\ \href
  {https://doi.org/10.1103/PhysRevAccelBeams.22.031302} {\bibfield  {journal}
  {\bibinfo  {journal} {Phys. Rev. Accel. Beams}\ }\textbf {\bibinfo {volume}
  {22}},\ \bibinfo {pages} {031302}}\BibitemShut {NoStop}%
\bibitem [{\citenamefont {Mendonça}\ and\ \citenamefont
  {Vieira}(2014)}]{Mendonca2014}%
  \BibitemOpen
  \bibfield  {author} {\bibinfo {author} {\bibnamefont {Mendonça},
  \bibfnamefont {J~T}}, \ and\ \bibinfo {author} {\bibfnamefont
  {J.}~\bibnamefont {Vieira}}} (\bibinfo {year} {2014}),\ \href
  {https://doi.org/10.1063/1.4868967} {\bibfield  {journal} {\bibinfo
  {journal} {Phys. Plasmas}\ }\textbf {\bibinfo {volume} {21}},\ \bibinfo
  {pages} {033107}}\BibitemShut {NoStop}%
\bibitem [{\citenamefont {Michel}\ \emph {et~al.}(2006)\citenamefont {Michel},
  \citenamefont {Schroeder}, \citenamefont {Shadwick}, \citenamefont {Esarey},\
  and\ \citenamefont {Leemans}}]{Michel2006}%
  \BibitemOpen
  \bibfield  {author} {\bibinfo {author} {\bibnamefont {Michel}, \bibfnamefont
  {P}}, \bibinfo {author} {\bibfnamefont {C.~B.}\ \bibnamefont {Schroeder}},
  \bibinfo {author} {\bibfnamefont {B.~A.}\ \bibnamefont {Shadwick}}, \bibinfo
  {author} {\bibfnamefont {E.}~\bibnamefont {Esarey}}, \ and\ \bibinfo {author}
  {\bibfnamefont {W.~P.}\ \bibnamefont {Leemans}}} (\bibinfo {year} {2006}),\
  \href {https://doi.org/10.1103/physreve.74.026501} {\bibfield  {journal}
  {\bibinfo  {journal} {Phys. Rev. E}\ }\textbf {\bibinfo {volume} {74}},\
  \bibinfo {pages} {026501}}\BibitemShut {NoStop}%
\bibitem [{\citenamefont {Migliorati}\ \emph {et~al.}(2013)\citenamefont
  {Migliorati}, \citenamefont {Bacci}, \citenamefont {Benedetti}, \citenamefont
  {Chiadroni}, \citenamefont {Ferrario}, \citenamefont {Mostacci},
  \citenamefont {Palumbo}, \citenamefont {Rossi}, \citenamefont {Serafini},\
  and\ \citenamefont {Antici}}]{Migliorati2013}%
  \BibitemOpen
  \bibfield  {author} {\bibinfo {author} {\bibnamefont {Migliorati},
  \bibfnamefont {M}}, \bibinfo {author} {\bibfnamefont {A.}~\bibnamefont
  {Bacci}}, \bibinfo {author} {\bibfnamefont {C.}~\bibnamefont {Benedetti}},
  \bibinfo {author} {\bibfnamefont {E.}~\bibnamefont {Chiadroni}}, \bibinfo
  {author} {\bibfnamefont {M.}~\bibnamefont {Ferrario}}, \bibinfo {author}
  {\bibfnamefont {A.}~\bibnamefont {Mostacci}}, \bibinfo {author}
  {\bibfnamefont {L.}~\bibnamefont {Palumbo}}, \bibinfo {author} {\bibfnamefont
  {A.~R.}\ \bibnamefont {Rossi}}, \bibinfo {author} {\bibfnamefont
  {L.}~\bibnamefont {Serafini}}, \ and\ \bibinfo {author} {\bibfnamefont
  {P.}~\bibnamefont {Antici}}} (\bibinfo {year} {2013}),\ \href
  {https://doi.org/10.1103/physrevstab.16.011302} {\bibfield  {journal}
  {\bibinfo  {journal} {Phys. Rev. ST Accel. Beams}\ }\textbf {\bibinfo
  {volume} {16}},\ \bibinfo {pages} {011302}}\BibitemShut {NoStop}%
\bibitem [{\citenamefont {Mirzaie}\ \emph {et~al.}(2024)\citenamefont {Mirzaie}
  \emph {et~al.}}]{Mirzaie2024}%
  \BibitemOpen
  \bibfield  {author} {\bibinfo {author} {\bibnamefont {Mirzaie}, \bibfnamefont
  {M}},  \emph {et~al.}} (\bibinfo {year} {2024}),\ \href
  {https://doi.org/10.1038/s41566-024-01550-8} {\bibfield  {journal} {\bibinfo
  {journal} {Nat. Photon.}\ }\textbf {\bibinfo {volume} {18}},\ \bibinfo
  {pages} {1212--1217}}\BibitemShut {NoStop}%
\bibitem [{\citenamefont {Modena}\ \emph {et~al.}(1995)\citenamefont {Modena}
  \emph {et~al.}}]{Modena1995}%
  \BibitemOpen
  \bibfield  {author} {\bibinfo {author} {\bibnamefont {Modena}, \bibfnamefont
  {A}},  \emph {et~al.}} (\bibinfo {year} {1995}),\ \href
  {https://doi.org/10.1038/377606a0} {\bibfield  {journal} {\bibinfo  {journal}
  {Nature (London)}\ }\textbf {\bibinfo {volume} {377}},\ \bibinfo {pages}
  {606--608}}\BibitemShut {NoStop}%
\bibitem [{\citenamefont {Montague}(1979)}]{Montague1979}%
  \BibitemOpen
  \bibfield  {author} {\bibinfo {author} {\bibnamefont {Montague},
  \bibfnamefont {B~W}}} (\bibinfo {year} {1979}),\ \href
  {http://cds.cern.ch/record/67243} {}\bibinfo {type} {Tech. Rep.}\ \bibinfo
  {number} {LEP Note 165}\ (\bibinfo  {institution} {CERN},\ \bibinfo {address}
  {Geneva, Switzerland})\BibitemShut {NoStop}%
\bibitem [{\citenamefont {Montague}(1984)}]{Montague1984}%
  \BibitemOpen
  \bibfield  {author} {\bibinfo {author} {\bibnamefont {Montague},
  \bibfnamefont {B~W}}} (\bibinfo {year} {1984}),\ \href@noop {} {}\bibinfo
  {type} {Tech. Rep.}\ \bibinfo {number} {CERN-LEP-TH-84-24}\ (\bibinfo
  {institution} {CERN},\ \bibinfo {address} {Geneva, Switzerland})\BibitemShut
  {NoStop}%
\bibitem [{\citenamefont {Moortgat-Pick}\ \emph {et~al.}(2008)\citenamefont
  {Moortgat-Pick} \emph {et~al.}}]{MoortgatPick2008}%
  \BibitemOpen
  \bibfield  {author} {\bibinfo {author} {\bibnamefont {Moortgat-Pick},
  \bibfnamefont {G}},  \emph {et~al.}} (\bibinfo {year} {2008}),\ \href
  {https://doi.org/10.1016/j.physrep.2007.12.003} {\bibfield  {journal}
  {\bibinfo  {journal} {Phys. Rep.}\ }\textbf {\bibinfo {volume} {460}},\
  \bibinfo {pages} {131–243}}\BibitemShut {NoStop}%
\bibitem [{\citenamefont {Mora}\ and\ \citenamefont
  {Antonsen}(1996)}]{Mora1996}%
  \BibitemOpen
  \bibfield  {author} {\bibinfo {author} {\bibnamefont {Mora}, \bibfnamefont
  {P}}, \ and\ \bibinfo {author} {\bibfnamefont {T.~M.}\ \bibnamefont
  {Antonsen}}} (\bibinfo {year} {1996}),\ \href
  {https://doi.org/10.1103/PhysRevE.53.R2068} {\bibfield  {journal} {\bibinfo
  {journal} {Phys. Rev. E}\ }\textbf {\bibinfo {volume} {53}},\ \bibinfo
  {pages} {R2068--R2071}}\BibitemShut {NoStop}%
\bibitem [{\citenamefont {Mora}\ and\ \citenamefont
  {Antonsen}(1997)}]{Mora1997}%
  \BibitemOpen
  \bibfield  {author} {\bibinfo {author} {\bibnamefont {Mora}, \bibfnamefont
  {P}}, \ and\ \bibinfo {author} {\bibfnamefont {T.~M.}\ \bibnamefont
  {Antonsen}}} (\bibinfo {year} {1997}),\ \href
  {https://doi.org/10.1063/1.872134} {\bibfield  {journal} {\bibinfo  {journal}
  {Phys. Plasmas}\ }\textbf {\bibinfo {volume} {4}},\ \bibinfo {pages}
  {217--229}}\BibitemShut {NoStop}%
\bibitem [{\citenamefont {{Morales Guzm{\'{a}}n}}\ \emph
  {et~al.}(2021)\citenamefont {{Morales Guzm{\'{a}}n}} \emph
  {et~al.}}]{MoralesGuzmn2021}%
  \BibitemOpen
  \bibfield  {author} {\bibinfo {author} {\bibnamefont {{Morales
  Guzm{\'{a}}n}}, \bibfnamefont {P~I}},  \emph {et~al.} (\bibinfo
  {collaboration} {AWAKE Collaboration})} (\bibinfo {year} {2021}),\ \href
  {https://doi.org/10.1103/physrevaccelbeams.24.101301} {\bibfield  {journal}
  {\bibinfo  {journal} {Phys. Rev. Accel. Beams}\ }\textbf {\bibinfo {volume}
  {24}},\ \bibinfo {pages} {101301}}\BibitemShut {NoStop}%
\bibitem [{\citenamefont {Moreira}\ \emph {et~al.}(2023)\citenamefont
  {Moreira}, \citenamefont {Muggli},\ and\ \citenamefont
  {Vieira}}]{Moreira2023}%
  \BibitemOpen
  \bibfield  {author} {\bibinfo {author} {\bibnamefont {Moreira}, \bibfnamefont
  {M}}, \bibinfo {author} {\bibfnamefont {P.}~\bibnamefont {Muggli}}, \ and\
  \bibinfo {author} {\bibfnamefont {J.}~\bibnamefont {Vieira}}} (\bibinfo
  {year} {2023}),\ \href {https://doi.org/10.1103/physrevlett.130.115001}
  {\bibfield  {journal} {\bibinfo  {journal} {Phys. Rev. Lett.}\ }\textbf
  {\bibinfo {volume} {130}},\ \bibinfo {pages} {115001}}\BibitemShut {NoStop}%
\bibitem [{\citenamefont {Morse}\ and\ \citenamefont
  {Nielson}(1971)}]{Morse1971}%
  \BibitemOpen
  \bibfield  {author} {\bibinfo {author} {\bibnamefont {Morse}, \bibfnamefont
  {R~L}}, \ and\ \bibinfo {author} {\bibfnamefont {C.~W.}\ \bibnamefont
  {Nielson}}} (\bibinfo {year} {1971}),\ \href
  {https://doi.org/10.1063/1.1693518} {\bibfield  {journal} {\bibinfo
  {journal} {Phys. Fluids}\ }\textbf {\bibinfo {volume} {14}},\ \bibinfo
  {pages} {830--840}}\BibitemShut {NoStop}%
\bibitem [{\citenamefont {Mounet}(2022)}]{2022AACforESPP}%
  \BibitemOpen
  \bibinfo {editor} {\bibnamefont {Mounet}, \bibfnamefont {N}},\ Ed. (\bibinfo
  {year} {2022}),\ \href {https://doi.org/10.23731/CYRM-2022-001} {\emph
  {\bibinfo {title} {European Strategy for Particle Physics - Accelerator R\&D
  Roadmap}}}\ (\bibinfo  {publisher} {CERN},\ \bibinfo {address} {Geneva,
  Switzerland})\BibitemShut {NoStop}%
\bibitem [{\citenamefont {Muggli}(2016)}]{Muggli2016}%
  \BibitemOpen
  \bibfield  {author} {\bibinfo {author} {\bibnamefont {Muggli}, \bibfnamefont
  {P}}} (\bibinfo {year} {2016}),\ in\ \href {\doibase
  10.5170/CERN-2016-001.119} {\emph {\bibinfo {booktitle} {Proceedings of the
  CAS-CERN Accelerator School: Plasma Wake Acceleration}}}\ (\bibinfo
  {publisher} {CERN, Geneva})\ pp.\ \bibinfo {pages} {119--142}\BibitemShut
  {NoStop}%
\bibitem [{\citenamefont {Muggli}\ \emph {et~al.}(2010)\citenamefont {Muggli},
  \citenamefont {Allen}, \citenamefont {Yakimenko}, \citenamefont {Fedurin},
  \citenamefont {Kusche},\ and\ \citenamefont {Babzien}}]{Muggli2010b}%
  \BibitemOpen
  \bibfield  {author} {\bibinfo {author} {\bibnamefont {Muggli}, \bibfnamefont
  {P}}, \bibinfo {author} {\bibfnamefont {B.}~\bibnamefont {Allen}}, \bibinfo
  {author} {\bibfnamefont {V.}~\bibnamefont {Yakimenko}}, \bibinfo {author}
  {\bibfnamefont {M.}~\bibnamefont {Fedurin}}, \bibinfo {author} {\bibfnamefont
  {K.}~\bibnamefont {Kusche}}, \ and\ \bibinfo {author} {\bibfnamefont
  {M.}~\bibnamefont {Babzien}}} (\bibinfo {year} {2010}),\ in\ \href
  {https://doi.org/10.1063/1.3520372} {\emph {\bibinfo {booktitle} {{AIP} Conf.
  Proc.}}},\ Vol.\ \bibinfo {volume} {1299}\ (\bibinfo  {publisher} {{AIP}})\
  pp.\ \bibinfo {pages} {495--499}\BibitemShut {NoStop}%
\bibitem [{\citenamefont {Muggli}\ \emph {et~al.}(2022)\citenamefont {Muggli},
  \citenamefont {Bergamaschi}, \citenamefont {Pucek}, \citenamefont {Easton},
  \citenamefont {Pisani},\ and\ \citenamefont {Uncles}}]{Muggli2022}%
  \BibitemOpen
  \bibfield  {author} {\bibinfo {author} {\bibnamefont {Muggli}, \bibfnamefont
  {P}}, \bibinfo {author} {\bibfnamefont {M.}~\bibnamefont {Bergamaschi}},
  \bibinfo {author} {\bibfnamefont {J.}~\bibnamefont {Pucek}}, \bibinfo
  {author} {\bibfnamefont {D.}~\bibnamefont {Easton}}, \bibinfo {author}
  {\bibfnamefont {J.}~\bibnamefont {Pisani}}, \ and\ \bibinfo {author}
  {\bibfnamefont {J.}~\bibnamefont {Uncles}}} (\bibinfo {year} {2022}),\ in\
  \href {\doibase 10.1109/aac55212.2022.10822987} {\emph {\bibinfo {booktitle}
  {2022 IEEE Advanced Accelerator Concepts Workshop (AAC)}}}\ (\bibinfo
  {publisher} {IEEE},\ \bibinfo {address} {New York, NY, United States})\ pp.\
  \bibinfo {pages} {1--6}\BibitemShut {NoStop}%
\bibitem [{\citenamefont {Muggli}\ \emph {et~al.}(2007)\citenamefont {Muggli},
  \citenamefont {Kimura}, \citenamefont {Kallos}, \citenamefont {Katsouleas},
  \citenamefont {Kusche}, \citenamefont {Pavlishin}, \citenamefont
  {Stolyarov},\ and\ \citenamefont {Yakimenko}}]{Muggli2007}%
  \BibitemOpen
  \bibfield  {author} {\bibinfo {author} {\bibnamefont {Muggli}, \bibfnamefont
  {P}}, \bibinfo {author} {\bibfnamefont {W.~D.}\ \bibnamefont {Kimura}},
  \bibinfo {author} {\bibfnamefont {E.}~\bibnamefont {Kallos}}, \bibinfo
  {author} {\bibfnamefont {T.~C.}\ \bibnamefont {Katsouleas}}, \bibinfo
  {author} {\bibfnamefont {K.~P.}\ \bibnamefont {Kusche}}, \bibinfo {author}
  {\bibfnamefont {I.~V.}\ \bibnamefont {Pavlishin}}, \bibinfo {author}
  {\bibfnamefont {D.}~\bibnamefont {Stolyarov}}, \ and\ \bibinfo {author}
  {\bibfnamefont {V.~E.}\ \bibnamefont {Yakimenko}}} (\bibinfo {year} {2007}),\
  in\ \href {\doibase 10.1109/pac.2007.4440672} {\emph {\bibinfo {booktitle}
  {Proceedings of the 2007 Particle Accelerator Conf.}}}\ (\bibinfo
  {publisher} {IEEE},\ \bibinfo {address} {New York, NY, United States})\ pp.\
  \bibinfo {pages} {3073--3075}\BibitemShut {NoStop}%
\bibitem [{\citenamefont {Muggli}\ \emph {et~al.}(1999)\citenamefont {Muggli},
  \citenamefont {Marsh}, \citenamefont {Wang}, \citenamefont {Clayton},
  \citenamefont {Lee}, \citenamefont {Katsouleas},\ and\ \citenamefont
  {Joshi}}]{Muggli1999}%
  \BibitemOpen
  \bibfield  {author} {\bibinfo {author} {\bibnamefont {Muggli}, \bibfnamefont
  {P}}, \bibinfo {author} {\bibfnamefont {K.~A.}\ \bibnamefont {Marsh}},
  \bibinfo {author} {\bibfnamefont {S.}~\bibnamefont {Wang}}, \bibinfo {author}
  {\bibfnamefont {C.~E.}\ \bibnamefont {Clayton}}, \bibinfo {author}
  {\bibfnamefont {S.}~\bibnamefont {Lee}}, \bibinfo {author} {\bibfnamefont
  {T.~C.}\ \bibnamefont {Katsouleas}}, \ and\ \bibinfo {author} {\bibfnamefont
  {C.}~\bibnamefont {Joshi}}} (\bibinfo {year} {1999}),\ \href
  {https://doi.org/10.1109/27.774685} {\bibfield  {journal} {\bibinfo
  {journal} {{IEEE} T. Plasma Sci.}\ }\textbf {\bibinfo {volume} {27}},\
  \bibinfo {pages} {791--799}}\BibitemShut {NoStop}%
\bibitem [{\citenamefont {Muggli}\ \emph
  {et~al.}(2008{\natexlab{a}})\citenamefont {Muggli}, \citenamefont
  {Yakimenko}, \citenamefont {Babzien}, \citenamefont {Kallos},\ and\
  \citenamefont {Kusche}}]{Muggli2008a}%
  \BibitemOpen
  \bibfield  {author} {\bibinfo {author} {\bibnamefont {Muggli}, \bibfnamefont
  {P}}, \bibinfo {author} {\bibfnamefont {V.}~\bibnamefont {Yakimenko}},
  \bibinfo {author} {\bibfnamefont {M.}~\bibnamefont {Babzien}}, \bibinfo
  {author} {\bibfnamefont {E.}~\bibnamefont {Kallos}}, \ and\ \bibinfo {author}
  {\bibfnamefont {K.~P.}\ \bibnamefont {Kusche}}} (\bibinfo {year}
  {2008}{\natexlab{a}}),\ \href
  {https://doi.org/10.1103/physrevlett.101.054801} {\bibfield  {journal}
  {\bibinfo  {journal} {Phys. Rev. Lett.}\ }\textbf {\bibinfo {volume} {101}},\
  \bibinfo {pages} {054801}}\BibitemShut {NoStop}%
\bibitem [{\citenamefont {Muggli}\ \emph
  {et~al.}(2001{\natexlab{a}})\citenamefont {Muggli} \emph
  {et~al.}}]{Muggli2001}%
  \BibitemOpen
  \bibfield  {author} {\bibinfo {author} {\bibnamefont {Muggli}, \bibfnamefont
  {P}},  \emph {et~al.}} (\bibinfo {year} {2001}{\natexlab{a}}),\ \href
  {https://doi.org/10.1103/physrevstab.4.091301} {\bibfield  {journal}
  {\bibinfo  {journal} {Phys. Rev. ST Accel. Beams}\ }\textbf {\bibinfo
  {volume} {4}},\ \bibinfo {pages} {091301}}\BibitemShut {NoStop}%
\bibitem [{\citenamefont {Muggli}\ \emph
  {et~al.}(2001{\natexlab{b}})\citenamefont {Muggli} \emph
  {et~al.}}]{Muggli2001a}%
  \BibitemOpen
  \bibfield  {author} {\bibinfo {author} {\bibnamefont {Muggli}, \bibfnamefont
  {P}},  \emph {et~al.}} (\bibinfo {year} {2001}{\natexlab{b}}),\ \href
  {https://doi.org/10.1038/35075144} {\bibfield  {journal} {\bibinfo  {journal}
  {Nature (London)}\ }\textbf {\bibinfo {volume} {411}},\ \bibinfo {pages}
  {43}}\BibitemShut {NoStop}%
\bibitem [{\citenamefont {Muggli}\ \emph {et~al.}(2004)\citenamefont {Muggli}
  \emph {et~al.}}]{Muggli2004}%
  \BibitemOpen
  \bibfield  {author} {\bibinfo {author} {\bibnamefont {Muggli}, \bibfnamefont
  {P}},  \emph {et~al.}} (\bibinfo {year} {2004}),\ \href
  {https://doi.org/10.1103/physrevlett.93.014802} {\bibfield  {journal}
  {\bibinfo  {journal} {Phys. Rev. Lett.}\ }\textbf {\bibinfo {volume} {93}},\
  \bibinfo {pages} {014802}}\BibitemShut {NoStop}%
\bibitem [{\citenamefont {Muggli}\ \emph
  {et~al.}(2008{\natexlab{b}})\citenamefont {Muggli} \emph
  {et~al.}}]{Muggli2008b}%
  \BibitemOpen
  \bibfield  {author} {\bibinfo {author} {\bibnamefont {Muggli}, \bibfnamefont
  {P}},  \emph {et~al.}} (\bibinfo {year} {2008}{\natexlab{b}}),\ \href
  {https://doi.org/10.1103/physrevlett.101.055001} {\bibfield  {journal}
  {\bibinfo  {journal} {Phys. Rev. Lett.}\ }\textbf {\bibinfo {volume} {101}},\
  \bibinfo {pages} {055001}}\BibitemShut {NoStop}%
\bibitem [{\citenamefont {Muggli}\ \emph {et~al.}(2011)\citenamefont {Muggli}
  \emph {et~al.}}]{Muggli2011}%
  \BibitemOpen
  \bibfield  {author} {\bibinfo {author} {\bibnamefont {Muggli}, \bibfnamefont
  {P}},  \emph {et~al.}} (\bibinfo {year} {2011}),\ in\ \href
  {https://proceedings.jacow.org/PAC2011/papers/tuobn3.pdf} {\emph {\bibinfo
  {booktitle} {Proceedings of the 2011 Particle Accelerator Conf.}}}\ (\bibinfo
   {publisher} {IEEE},\ \bibinfo {address} {New York, NY, United States})\ pp.\
  \bibinfo {pages} {712--714}\BibitemShut {NoStop}%
\bibitem [{\citenamefont {Muggli}\ \emph {et~al.}(2017)\citenamefont {Muggli}
  \emph {et~al.}}]{Muggli2017}%
  \BibitemOpen
  \bibfield  {author} {\bibinfo {author} {\bibnamefont {Muggli}, \bibfnamefont
  {P}},  \emph {et~al.}} (\bibinfo {year} {2017}),\ \href
  {https://doi.org/10.1088/1361-6587/aa941c} {\bibfield  {journal} {\bibinfo
  {journal} {Plasma Phys. Control. Fusion}\ }\textbf {\bibinfo {volume} {60}},\
  \bibinfo {pages} {014046}}\BibitemShut {NoStop}%
\bibitem [{\citenamefont {Nakajima}\ \emph {et~al.}(1990)\citenamefont
  {Nakajima}, \citenamefont {Enomoto}, \citenamefont {Kobayashi}, \citenamefont
  {Nakanishi}, \citenamefont {Nishida}, \citenamefont {Ogata}, \citenamefont
  {Ohsawa}, \citenamefont {Oogoe}, \citenamefont {Shoji},\ and\ \citenamefont
  {Urano}}]{Nakajima1990}%
  \BibitemOpen
  \bibfield  {author} {\bibinfo {author} {\bibnamefont {Nakajima},
  \bibfnamefont {K}}, \bibinfo {author} {\bibfnamefont {A.}~\bibnamefont
  {Enomoto}}, \bibinfo {author} {\bibfnamefont {H.}~\bibnamefont {Kobayashi}},
  \bibinfo {author} {\bibfnamefont {H.}~\bibnamefont {Nakanishi}}, \bibinfo
  {author} {\bibfnamefont {Y.}~\bibnamefont {Nishida}}, \bibinfo {author}
  {\bibfnamefont {A.}~\bibnamefont {Ogata}}, \bibinfo {author} {\bibfnamefont
  {S.}~\bibnamefont {Ohsawa}}, \bibinfo {author} {\bibfnamefont
  {T.}~\bibnamefont {Oogoe}}, \bibinfo {author} {\bibfnamefont
  {T.}~\bibnamefont {Shoji}}, \ and\ \bibinfo {author} {\bibfnamefont
  {T.}~\bibnamefont {Urano}}} (\bibinfo {year} {1990}),\ \href
  {https://doi.org/10.1016/0168-9002(90)91729-u} {\bibfield  {journal}
  {\bibinfo  {journal} {Nucl. Instrum. Methods Phys. Res. A}\ }\textbf
  {\bibinfo {volume} {292}},\ \bibinfo {pages} {12--20}}\BibitemShut {NoStop}%
\bibitem [{\citenamefont {Nakanishi}\ \emph {et~al.}(1993)\citenamefont
  {Nakanishi} \emph {et~al.}}]{Nakanishi1993}%
  \BibitemOpen
  \bibfield  {author} {\bibinfo {author} {\bibnamefont {Nakanishi},
  \bibfnamefont {H}},  \emph {et~al.}} (\bibinfo {year} {1993}),\ \href
  {https://doi.org/10.1016/0168-9002(93)90681-7} {\bibfield  {journal}
  {\bibinfo  {journal} {Nucl. Instrum. Methods Phys. Res. A}\ }\textbf
  {\bibinfo {volume} {328}},\ \bibinfo {pages} {596--598}}\BibitemShut
  {NoStop}%
\bibitem [{\citenamefont {Nechaeva}\ \emph {et~al.}(2024)\citenamefont
  {Nechaeva} \emph {et~al.}}]{Nechaeva2023}%
  \BibitemOpen
  \bibfield  {author} {\bibinfo {author} {\bibnamefont {Nechaeva},
  \bibfnamefont {T}},  \emph {et~al.} (\bibinfo {collaboration} {AWAKE
  Collaboration})} (\bibinfo {year} {2024}),\ \href
  {https://link.aps.org/doi/10.1103/PhysRevLett.132.075001} {\bibfield
  {journal} {\bibinfo  {journal} {Phys. Rev. Lett.}\ }\textbf {\bibinfo
  {volume} {132}},\ \bibinfo {pages} {075001}}\BibitemShut {NoStop}%
\bibitem [{\citenamefont {Ng}\ \emph {et~al.}(2001)\citenamefont {Ng} \emph
  {et~al.}}]{Ng2001}%
  \BibitemOpen
  \bibfield  {author} {\bibinfo {author} {\bibnamefont {Ng}, \bibfnamefont
  {J~S~T}},  \emph {et~al.}} (\bibinfo {year} {2001}),\ \href
  {https://doi.org/10.1103/physrevlett.87.244801} {\bibfield  {journal}
  {\bibinfo  {journal} {Phys. Rev. Lett.}\ }\textbf {\bibinfo {volume} {87}},\
  \bibinfo {pages} {244801}}\BibitemShut {NoStop}%
\bibitem [{\citenamefont {Nie}\ \emph {et~al.}(2021)\citenamefont {Nie} \emph
  {et~al.}}]{Nie2021}%
  \BibitemOpen
  \bibfield  {author} {\bibinfo {author} {\bibnamefont {Nie}, \bibfnamefont
  {Z}},  \emph {et~al.}} (\bibinfo {year} {2021}),\ \href
  {https://doi.org/10.1103/physrevlett.126.054801} {\bibfield  {journal}
  {\bibinfo  {journal} {Phys. Rev. Lett.}\ }\textbf {\bibinfo {volume} {126}},\
  \bibinfo {pages} {054801}}\BibitemShut {NoStop}%
\bibitem [{\citenamefont {Nie}\ \emph {et~al.}(2022)\citenamefont {Nie} \emph
  {et~al.}}]{Nie2022}%
  \BibitemOpen
  \bibfield  {author} {\bibinfo {author} {\bibnamefont {Nie}, \bibfnamefont
  {Z}},  \emph {et~al.}} (\bibinfo {year} {2022}),\ \href
  {https://doi.org/10.1103/physrevresearch.4.033015} {\bibfield  {journal}
  {\bibinfo  {journal} {Phys. Rev. Res.}\ }\textbf {\bibinfo {volume} {4}},\
  \bibinfo {pages} {033015}}\BibitemShut {NoStop}%
\bibitem [{\citenamefont {Nishida}\ \emph {et~al.}(1991)\citenamefont
  {Nishida}, \citenamefont {Okazaki}, \citenamefont {Yugami},\ and\
  \citenamefont {Nagasawa}}]{Nishida1991}%
  \BibitemOpen
  \bibfield  {author} {\bibinfo {author} {\bibnamefont {Nishida}, \bibfnamefont
  {Y}}, \bibinfo {author} {\bibfnamefont {T.}~\bibnamefont {Okazaki}}, \bibinfo
  {author} {\bibfnamefont {N.}~\bibnamefont {Yugami}}, \ and\ \bibinfo {author}
  {\bibfnamefont {T.}~\bibnamefont {Nagasawa}}} (\bibinfo {year} {1991}),\
  \href {https://doi.org/10.1103/physrevlett.66.2328} {\bibfield  {journal}
  {\bibinfo  {journal} {Phys. Rev. Lett.}\ }\textbf {\bibinfo {volume} {66}},\
  \bibinfo {pages} {2328--2331}}\BibitemShut {NoStop}%
\bibitem [{\citenamefont {Nishikawa}\ \emph {et~al.}(2021)\citenamefont
  {Nishikawa}, \citenamefont {Du\c{t}an}, \citenamefont {K\"{o}hn},\ and\
  \citenamefont {Mizuno}}]{Nishikawa2021}%
  \BibitemOpen
  \bibfield  {author} {\bibinfo {author} {\bibnamefont {Nishikawa},
  \bibfnamefont {K}}, \bibinfo {author} {\bibfnamefont {I.}~\bibnamefont
  {Du\c{t}an}}, \bibinfo {author} {\bibfnamefont {C.}~\bibnamefont {K\"{o}hn}},
  \ and\ \bibinfo {author} {\bibfnamefont {Y.}~\bibnamefont {Mizuno}}}
  (\bibinfo {year} {2021}),\ \href {https://doi.org/10.1007/s41115-021-00012-0}
  {\bibfield  {journal} {\bibinfo  {journal} {Living Rev. Comput. Astrophys.}\
  }\textbf {\bibinfo {volume} {7}},\ \bibinfo {pages} {1}}\BibitemShut
  {NoStop}%
\bibitem [{\citenamefont {O'Connell}\ \emph {et~al.}(2006)\citenamefont
  {O'Connell} \emph {et~al.}}]{OConnell2006}%
  \BibitemOpen
  \bibfield  {author} {\bibinfo {author} {\bibnamefont {O'Connell},
  \bibfnamefont {C~L}},  \emph {et~al.}} (\bibinfo {year} {2006}),\ \href
  {https://doi.org/10.1103/physrevstab.9.101301} {\bibfield  {journal}
  {\bibinfo  {journal} {Phys. Rev. ST Accel. Beams}\ }\textbf {\bibinfo
  {volume} {9}},\ \bibinfo {pages} {101301}}\BibitemShut {NoStop}%
\bibitem [{\citenamefont {Ogata}(1992)}]{Ogata1992}%
  \BibitemOpen
  \bibfield  {author} {\bibinfo {author} {\bibnamefont {Ogata}, \bibfnamefont
  {A}}} (\bibinfo {year} {1992}),\ in\ \href {https://doi.org/10.1063/1.44061}
  {\emph {\bibinfo {booktitle} {{AIP} Conf. Proc.}}},\ Vol.\ \bibinfo {volume}
  {279}\ (\bibinfo  {publisher} {{AIP}})\ pp.\ \bibinfo {pages}
  {420--449}\BibitemShut {NoStop}%
\bibitem [{\citenamefont {Ogata}\ \emph {et~al.}(1998)\citenamefont {Ogata},
  \citenamefont {Rosenzweig},\ and\ \citenamefont {Ferrario}}]{Ogata1998}%
  \BibitemOpen
  \bibfield  {author} {\bibinfo {author} {\bibnamefont {Ogata}, \bibfnamefont
  {A}}, \bibinfo {author} {\bibfnamefont {J.~B.}\ \bibnamefont {Rosenzweig}}, \
  and\ \bibinfo {author} {\bibfnamefont {M.}~\bibnamefont {Ferrario}}}
  (\bibinfo {year} {1998}),\ \href
  {https://doi.org/10.1016/S0168-9002(98)00188-0} {\bibfield  {journal}
  {\bibinfo  {journal} {Nucl. Instrum. Methods Phys. Res. A}\ }\textbf
  {\bibinfo {volume} {410}},\ \bibinfo {pages} {549--561}}\BibitemShut
  {NoStop}%
\bibitem [{\citenamefont {Ogata}\ \emph {et~al.}(1991)\citenamefont {Ogata},
  \citenamefont {Yoshida}, \citenamefont {Yugami}, \citenamefont {Nishida},
  \citenamefont {Nakanishi}, \citenamefont {Nakajima}, \citenamefont {Shibata},
  \citenamefont {Kozawa}, \citenamefont {Kobayashi},\ and\ \citenamefont
  {Ueda}}]{Ogata1991}%
  \BibitemOpen
  \bibfield  {author} {\bibinfo {author} {\bibnamefont {Ogata}, \bibfnamefont
  {A}}, \bibinfo {author} {\bibfnamefont {Y.}~\bibnamefont {Yoshida}}, \bibinfo
  {author} {\bibfnamefont {N.}~\bibnamefont {Yugami}}, \bibinfo {author}
  {\bibfnamefont {Y.}~\bibnamefont {Nishida}}, \bibinfo {author} {\bibfnamefont
  {H.}~\bibnamefont {Nakanishi}}, \bibinfo {author} {\bibfnamefont
  {K.}~\bibnamefont {Nakajima}}, \bibinfo {author} {\bibfnamefont
  {H.}~\bibnamefont {Shibata}}, \bibinfo {author} {\bibfnamefont
  {T.}~\bibnamefont {Kozawa}}, \bibinfo {author} {\bibfnamefont
  {T.}~\bibnamefont {Kobayashi}}, \ and\ \bibinfo {author} {\bibfnamefont
  {T.}~\bibnamefont {Ueda}}} (\bibinfo {year} {1991}),\ in\ \href
  {https://doi.org/10.1109/PAC.1991.164383} {\emph {\bibinfo {booktitle}
  {Proceedings of the 1991 Particle Accelerator Conf.}}}\ (\bibinfo
  {publisher} {IEEE},\ \bibinfo {address} {New York, NY, United States})\ pp.\
  \bibinfo {pages} {622--624}\BibitemShut {NoStop}%
\bibitem [{\citenamefont {Ogata}\ \emph {et~al.}(1994)\citenamefont {Ogata}
  \emph {et~al.}}]{Ogata1994}%
  \BibitemOpen
  \bibfield  {author} {\bibinfo {author} {\bibnamefont {Ogata}, \bibfnamefont
  {A}},  \emph {et~al.}} (\bibinfo {year} {1994}),\ \href
  {https://doi.org/10.1088/0031-8949/1994/t52/011} {\bibfield  {journal}
  {\bibinfo  {journal} {Phys. Scr.}\ }\textbf {\bibinfo {volume} {T52}},\
  \bibinfo {pages} {69--72}}\BibitemShut {NoStop}%
\bibitem [{\citenamefont {Ogata}\ \emph {et~al.}(1995)\citenamefont {Ogata}
  \emph {et~al.}}]{Ogata1995}%
  \BibitemOpen
  \bibfield  {author} {\bibinfo {author} {\bibnamefont {Ogata}, \bibfnamefont
  {A}},  \emph {et~al.}} (\bibinfo {year} {1995}),\ in\ \href
  {https://doi.org/10.1063/1.48289} {\emph {\bibinfo {booktitle} {{AIP} Conf.
  Proc.}}},\ Vol.\ \bibinfo {volume} {335}\ (\bibinfo  {publisher} {{AIP}})\
  pp.\ \bibinfo {pages} {501--504}\BibitemShut {NoStop}%
\bibitem [{\citenamefont {Olsen}\ and\ \citenamefont
  {Maximon}(1959)}]{Olsen1959}%
  \BibitemOpen
  \bibfield  {author} {\bibinfo {author} {\bibnamefont {Olsen}, \bibfnamefont
  {H}}, \ and\ \bibinfo {author} {\bibfnamefont {L.~C.}\ \bibnamefont
  {Maximon}}} (\bibinfo {year} {1959}),\ \href
  {https://doi.org/10.1103/PhysRev.114.887} {\bibfield  {journal} {\bibinfo
  {journal} {Phys. Rev.}\ }\textbf {\bibinfo {volume} {114}},\ \bibinfo {pages}
  {887--904}}\BibitemShut {NoStop}%
\bibitem [{\citenamefont {Olsen}\ \emph {et~al.}(2018)\citenamefont {Olsen},
  \citenamefont {Adli},\ and\ \citenamefont {Muggli}}]{Olsen2018}%
  \BibitemOpen
  \bibfield  {author} {\bibinfo {author} {\bibnamefont {Olsen}, \bibfnamefont
  {V~K~B}}, \bibinfo {author} {\bibfnamefont {E.}~\bibnamefont {Adli}}, \ and\
  \bibinfo {author} {\bibfnamefont {P.}~\bibnamefont {Muggli}}} (\bibinfo
  {year} {2018}),\ \href {https://doi.org/10.1103/physrevaccelbeams.21.011301}
  {\bibfield  {journal} {\bibinfo  {journal} {Phys. Rev. Accel. Beams}\
  }\textbf {\bibinfo {volume} {21}},\ \bibinfo {pages} {011301}}\BibitemShut
  {NoStop}%
\bibitem [{\citenamefont {O'Shea}\ \emph {et~al.}(2012)\citenamefont {O'Shea},
  \citenamefont {Rosenzweig}, \citenamefont {Barber}, \citenamefont {Fukasawa},
  \citenamefont {Williams}, \citenamefont {Muggli}, \citenamefont {Yakimenko},\
  and\ \citenamefont {Kusche}}]{OShea2012}%
  \BibitemOpen
  \bibfield  {author} {\bibinfo {author} {\bibnamefont {O'Shea}, \bibfnamefont
  {B}}, \bibinfo {author} {\bibfnamefont {J.}~\bibnamefont {Rosenzweig}},
  \bibinfo {author} {\bibfnamefont {S.}~\bibnamefont {Barber}}, \bibinfo
  {author} {\bibfnamefont {A.}~\bibnamefont {Fukasawa}}, \bibinfo {author}
  {\bibfnamefont {O.}~\bibnamefont {Williams}}, \bibinfo {author}
  {\bibfnamefont {P.}~\bibnamefont {Muggli}}, \bibinfo {author} {\bibfnamefont
  {V.}~\bibnamefont {Yakimenko}}, \ and\ \bibinfo {author} {\bibfnamefont
  {K.}~\bibnamefont {Kusche}}} (\bibinfo {year} {2012}),\ in\ \href
  {https://doi.org/10.1063/1.4773766} {\emph {\bibinfo {booktitle} {{AIP} Conf.
  Proc.}}},\ Vol.\ \bibinfo {volume} {1507}\ (\bibinfo  {publisher} {{AIP}})\
  pp.\ \bibinfo {pages} {606--611}\BibitemShut {NoStop}%
\bibitem [{\citenamefont {O'Shea}\ \emph {et~al.}(2016)\citenamefont {O'Shea}
  \emph {et~al.}}]{OShea2016}%
  \BibitemOpen
  \bibfield  {author} {\bibinfo {author} {\bibnamefont {O'Shea}, \bibfnamefont
  {B~D}},  \emph {et~al.}} (\bibinfo {year} {2016}),\ \href
  {https://doi.org/10.1038/ncomms12763} {\bibfield  {journal} {\bibinfo
  {journal} {Nat. Commun.}\ }\textbf {\bibinfo {volume} {7}},\ \bibinfo {pages}
  {12763}}\BibitemShut {NoStop}%
\bibitem [{\citenamefont {\"{O}z}\ and\ \citenamefont {Muggli}(2014)}]{Oz2014}%
  \BibitemOpen
  \bibfield  {author} {\bibinfo {author} {\bibnamefont {\"{O}z}, \bibfnamefont
  {E}}, \ and\ \bibinfo {author} {\bibfnamefont {P.}~\bibnamefont {Muggli}}}
  (\bibinfo {year} {2014}),\ \href {https://doi.org/10.1016/j.nima.2013.10.093}
  {\bibfield  {journal} {\bibinfo  {journal} {Nucl. Instrum. Methods Phys. Res.
  A}\ }\textbf {\bibinfo {volume} {740}},\ \bibinfo {pages}
  {197--202}}\BibitemShut {NoStop}%
\bibitem [{\citenamefont {{\"O}z}\ \emph {et~al.}(2004)\citenamefont {{\"O}z}
  \emph {et~al.}}]{Oz2004}%
  \BibitemOpen
  \bibfield  {author} {\bibinfo {author} {\bibnamefont {{\"O}z}, \bibfnamefont
  {E}},  \emph {et~al.}} (\bibinfo {year} {2004}),\ in\ \href {\doibase
  10.1063/1.1842612} {\emph {\bibinfo {booktitle} {AIP Conf. Proc.}}},\ Vol.\
  \bibinfo {volume} {737}\ (\bibinfo  {publisher} {AIP})\ pp.\ \bibinfo {pages}
  {708--714}\BibitemShut {NoStop}%
\bibitem [{\citenamefont {{\"O}z}\ \emph {et~al.}(2007)\citenamefont {{\"O}z}
  \emph {et~al.}}]{Oz2007}%
  \BibitemOpen
  \bibfield  {author} {\bibinfo {author} {\bibnamefont {{\"O}z}, \bibfnamefont
  {E}},  \emph {et~al.}} (\bibinfo {year} {2007}),\ \href
  {https://doi.org/10.1103/physrevlett.98.084801} {\bibfield  {journal}
  {\bibinfo  {journal} {Phys. Rev. Lett.}\ }\textbf {\bibinfo {volume} {98}},\
  \bibinfo {pages} {084801}}\BibitemShut {NoStop}%
\bibitem [{\citenamefont {P\~{o}der}\ \emph {et~al.}(2018)\citenamefont
  {P\~{o}der} \emph {et~al.}}]{Poder2018}%
  \BibitemOpen
  \bibfield  {author} {\bibinfo {author} {\bibnamefont {P\~{o}der},
  \bibfnamefont {K}},  \emph {et~al.}} (\bibinfo {year} {2018}),\ \href
  {https://link.aps.org/doi/10.1103/PhysRevX.8.031004} {\bibfield  {journal}
  {\bibinfo  {journal} {Phys. Rev. X}\ }\textbf {\bibinfo {volume} {8}},\
  \bibinfo {pages} {031004}}\BibitemShut {NoStop}%
\bibitem [{\citenamefont {Pae}\ \emph {et~al.}(2010)\citenamefont {Pae},
  \citenamefont {Choi},\ and\ \citenamefont {Lee}}]{Pae2010}%
  \BibitemOpen
  \bibfield  {author} {\bibinfo {author} {\bibnamefont {Pae}, \bibfnamefont
  {K~H}}, \bibinfo {author} {\bibfnamefont {I.~W.}\ \bibnamefont {Choi}}, \
  and\ \bibinfo {author} {\bibfnamefont {J.}~\bibnamefont {Lee}}} (\bibinfo
  {year} {2010}),\ \href {\doibase 10.1063/1.3522757} {\bibfield  {journal}
  {\bibinfo  {journal} {Phys. Plasmas}\ }\textbf {\bibinfo {volume} {17}},\
  \bibinfo {pages} {123104}}\BibitemShut {NoStop}%
\bibitem [{\citenamefont {Pak}\ \emph {et~al.}(2010)\citenamefont {Pak},
  \citenamefont {Marsh}, \citenamefont {Martins}, \citenamefont {Lu},
  \citenamefont {Mori},\ and\ \citenamefont {Joshi}}]{Pak2010}%
  \BibitemOpen
  \bibfield  {author} {\bibinfo {author} {\bibnamefont {Pak}, \bibfnamefont
  {A}}, \bibinfo {author} {\bibfnamefont {K.~A.}\ \bibnamefont {Marsh}},
  \bibinfo {author} {\bibfnamefont {S.~F.}\ \bibnamefont {Martins}}, \bibinfo
  {author} {\bibfnamefont {W.}~\bibnamefont {Lu}}, \bibinfo {author}
  {\bibfnamefont {W.~B.}\ \bibnamefont {Mori}}, \ and\ \bibinfo {author}
  {\bibfnamefont {C.}~\bibnamefont {Joshi}}} (\bibinfo {year} {2010}),\ \href
  {https://doi.org/10.1103/PhysRevLett.104.025003} {\bibfield  {journal}
  {\bibinfo  {journal} {Phys. Rev. Lett.}\ }\textbf {\bibinfo {volume} {104}},\
  \bibinfo {pages} {025003}}\BibitemShut {NoStop}%
\bibitem [{\citenamefont {Panofsky}\ and\ \citenamefont
  {Baker}(1950)}]{Panofsky1950}%
  \BibitemOpen
  \bibfield  {author} {\bibinfo {author} {\bibnamefont {Panofsky},
  \bibfnamefont {W~K~H}}, \ and\ \bibinfo {author} {\bibfnamefont {W.~R.}\
  \bibnamefont {Baker}}} (\bibinfo {year} {1950}),\ \href
  {https://doi.org/10.1063/1.1745611} {\bibfield  {journal} {\bibinfo
  {journal} {Rev. Sci. Instrum.}\ }\textbf {\bibinfo {volume} {21}},\ \bibinfo
  {pages} {445--447}}\BibitemShut {NoStop}%
\bibitem [{\citenamefont {Panofsky}\ and\ \citenamefont
  {Bander}(1968)}]{Panofsky1968}%
  \BibitemOpen
  \bibfield  {author} {\bibinfo {author} {\bibnamefont {Panofsky},
  \bibfnamefont {W~K~H}}, \ and\ \bibinfo {author} {\bibfnamefont
  {M.}~\bibnamefont {Bander}}} (\bibinfo {year} {1968}),\ \href
  {https://doi.org/10.1063/1.1683315} {\bibfield  {journal} {\bibinfo
  {journal} {Rev. Sci. Instrum.}\ }\textbf {\bibinfo {volume} {39}},\ \bibinfo
  {pages} {206--212}}\BibitemShut {NoStop}%
\bibitem [{\citenamefont {Panofsky}\ and\ \citenamefont
  {Wenzel}(1956)}]{Panofsky1956}%
  \BibitemOpen
  \bibfield  {author} {\bibinfo {author} {\bibnamefont {Panofsky},
  \bibfnamefont {W~K~H}}, \ and\ \bibinfo {author} {\bibfnamefont {W.~A.}\
  \bibnamefont {Wenzel}}} (\bibinfo {year} {1956}),\ \href
  {https://doi.org/10.1063/1.1715427} {\bibfield  {journal} {\bibinfo
  {journal} {Rev. Sci. Instrum.}\ }\textbf {\bibinfo {volume} {27}},\ \bibinfo
  {pages} {967}}\BibitemShut {NoStop}%
\bibitem [{\citenamefont {Pathak}\ \emph {et~al.}(2016)\citenamefont {Pathak},
  \citenamefont {Zhidkov}, \citenamefont {Nakanii}, \citenamefont {Masuda},
  \citenamefont {Hosokai},\ and\ \citenamefont {Kodama}}]{Pathak2016}%
  \BibitemOpen
  \bibfield  {author} {\bibinfo {author} {\bibnamefont {Pathak}, \bibfnamefont
  {N}}, \bibinfo {author} {\bibfnamefont {A.}~\bibnamefont {Zhidkov}}, \bibinfo
  {author} {\bibfnamefont {N.}~\bibnamefont {Nakanii}}, \bibinfo {author}
  {\bibfnamefont {S.}~\bibnamefont {Masuda}}, \bibinfo {author} {\bibfnamefont
  {T.}~\bibnamefont {Hosokai}}, \ and\ \bibinfo {author} {\bibfnamefont
  {R.}~\bibnamefont {Kodama}}} (\bibinfo {year} {2016}),\ \href
  {https://doi.org/10.1063/1.4942942} {\bibfield  {journal} {\bibinfo
  {journal} {Phys. Plasmas}\ }\textbf {\bibinfo {volume} {23}},\ \bibinfo
  {pages} {033102}}\BibitemShut {NoStop}%
\bibitem [{\citenamefont {Pei}\ \emph {et~al.}(2009)\citenamefont {Pei},
  \citenamefont {Hogan}, \citenamefont {Raubenheimer}, \citenamefont {Seryi},
  \citenamefont {Braun}, \citenamefont {Corsini},\ and\ \citenamefont
  {Delahaye}}]{Pei2009}%
  \BibitemOpen
  \bibfield  {author} {\bibinfo {author} {\bibnamefont {Pei}, \bibfnamefont
  {S}}, \bibinfo {author} {\bibfnamefont {M.~J.}\ \bibnamefont {Hogan}},
  \bibinfo {author} {\bibfnamefont {T.~O.}\ \bibnamefont {Raubenheimer}},
  \bibinfo {author} {\bibfnamefont {A.}~\bibnamefont {Seryi}}, \bibinfo
  {author} {\bibfnamefont {H.~H.}\ \bibnamefont {Braun}}, \bibinfo {author}
  {\bibfnamefont {R.}~\bibnamefont {Corsini}}, \ and\ \bibinfo {author}
  {\bibfnamefont {J.~P.}\ \bibnamefont {Delahaye}}} (\bibinfo {year} {2009}),\
  in\ \href {https://accelconf.web.cern.ch/PAC2009/papers/we6pfp079.pdf} {\emph
  {\bibinfo {booktitle} {Proceedings of the 2009 Particle Accelerator
  Conf.}}},\ pp.\ \bibinfo {pages} {2682--2684}\BibitemShut {NoStop}%
\bibitem [{\citenamefont {Pellegrini}\ \emph {et~al.}(2016)\citenamefont
  {Pellegrini}, \citenamefont {Marinelli},\ and\ \citenamefont
  {Reiche}}]{Pellegrini2016}%
  \BibitemOpen
  \bibfield  {author} {\bibinfo {author} {\bibnamefont {Pellegrini},
  \bibfnamefont {C}}, \bibinfo {author} {\bibfnamefont {A.}~\bibnamefont
  {Marinelli}}, \ and\ \bibinfo {author} {\bibfnamefont {S.}~\bibnamefont
  {Reiche}}} (\bibinfo {year} {2016}),\ \href
  {https://doi.org/10.1103/RevModPhys.88.015006} {\bibfield  {journal}
  {\bibinfo  {journal} {Rev. Mod. Phys.}\ }\textbf {\bibinfo {volume} {88}},\
  \bibinfo {pages} {015006}}\BibitemShut {NoStop}%
\bibitem [{\citenamefont {Pe{\~n}a}\ \emph {et~al.}(2024)\citenamefont
  {Pe{\~n}a} \emph {et~al.}}]{Pena2024}%
  \BibitemOpen
  \bibfield  {author} {\bibinfo {author} {\bibnamefont {Pe{\~n}a},
  \bibfnamefont {F}},  \emph {et~al.}} (\bibinfo {year} {2024}),\ \href
  {https://link.aps.org/doi/10.1103/PhysRevResearch.6.043090} {\bibfield
  {journal} {\bibinfo  {journal} {Phys. Rev. Res.}\ }\textbf {\bibinfo {volume}
  {6}},\ \bibinfo {pages} {043090}}\BibitemShut {NoStop}%
\bibitem [{\citenamefont {Peng}\ \emph {et~al.}(2026)\citenamefont {Peng},
  \citenamefont {Wan}, \citenamefont {Wang}, \citenamefont {Hua},\ and\
  \citenamefont {Lu}}]{Peng2026}%
  \BibitemOpen
  \bibfield  {author} {\bibinfo {author} {\bibnamefont {Peng}, \bibfnamefont
  {B}}, \bibinfo {author} {\bibfnamefont {Y.}~\bibnamefont {Wan}}, \bibinfo
  {author} {\bibfnamefont {Z.}~\bibnamefont {Wang}}, \bibinfo {author}
  {\bibfnamefont {J.}~\bibnamefont {Hua}}, \ and\ \bibinfo {author}
  {\bibfnamefont {W.}~\bibnamefont {Lu}}} (\bibinfo {year} {2026}),\ \href
  {\doibase 10.1103/ksms-1x6l} {\bibfield  {journal} {\bibinfo  {journal}
  {Phys. Rev. Accel. Beams}\ }\textbf {\bibinfo {volume} {29}},\ \bibinfo
  {pages} {050702}}\BibitemShut {NoStop}%
\bibitem [{\citenamefont {Pepitone}\ \emph {et~al.}(2018)\citenamefont
  {Pepitone} \emph {et~al.}}]{Pepitone2018}%
  \BibitemOpen
  \bibfield  {author} {\bibinfo {author} {\bibnamefont {Pepitone},
  \bibfnamefont {K}},  \emph {et~al.} (\bibinfo {collaboration} {AWAKE
  Collaboration})} (\bibinfo {year} {2018}),\ \href
  {https://doi.org/10.1016/j.nima.2018.02.044} {\bibfield  {journal} {\bibinfo
  {journal} {Nucl. Instrum. Methods Phys. Res. A}\ }\textbf {\bibinfo {volume}
  {909}},\ \bibinfo {pages} {102--106}}\BibitemShut {NoStop}%
\bibitem [{\citenamefont {Petrenko}\ \emph {et~al.}(2016)\citenamefont
  {Petrenko}, \citenamefont {Lotov},\ and\ \citenamefont
  {Sosedkin}}]{Petrenko2016}%
  \BibitemOpen
  \bibfield  {author} {\bibinfo {author} {\bibnamefont {Petrenko},
  \bibfnamefont {A}}, \bibinfo {author} {\bibfnamefont {K.}~\bibnamefont
  {Lotov}}, \ and\ \bibinfo {author} {\bibfnamefont {A.}~\bibnamefont
  {Sosedkin}}} (\bibinfo {year} {2016}),\ \href
  {https://doi.org/10.1016/j.nima.2016.01.063} {\bibfield  {journal} {\bibinfo
  {journal} {Nucl. Instrum. Methods Phys. Res. A}\ }\textbf {\bibinfo {volume}
  {829}},\ \bibinfo {pages} {63--66}}\BibitemShut {NoStop}%
\bibitem [{\citenamefont {Petrillo}\ \emph {et~al.}(2018)\citenamefont
  {Petrillo} \emph {et~al.}}]{Petrillo2018}%
  \BibitemOpen
  \bibfield  {author} {\bibinfo {author} {\bibnamefont {Petrillo},
  \bibfnamefont {V}},  \emph {et~al.}} (\bibinfo {year} {2018}),\ \href
  {https://doi.org/10.1016/j.nima.2018.02.036} {\bibfield  {journal} {\bibinfo
  {journal} {Nucl. Instrum. Methods Phys. Res. A}\ }\textbf {\bibinfo {volume}
  {909}},\ \bibinfo {pages} {303--308}}\BibitemShut {NoStop}%
\bibitem [{\citenamefont {Pfingstner}\ \emph {et~al.}(2016)\citenamefont
  {Pfingstner}, \citenamefont {Adli}, \citenamefont {Lindstr{\o}m},
  \citenamefont {Marín},\ and\ \citenamefont {Schulte}}]{Pfingstner2016}%
  \BibitemOpen
  \bibfield  {author} {\bibinfo {author} {\bibnamefont {Pfingstner},
  \bibfnamefont {J}}, \bibinfo {author} {\bibfnamefont {E.}~\bibnamefont
  {Adli}}, \bibinfo {author} {\bibfnamefont {C.}~\bibnamefont {Lindstr{\o}m}},
  \bibinfo {author} {\bibfnamefont {E.}~\bibnamefont {Marín}}, \ and\ \bibinfo
  {author} {\bibfnamefont {D.}~\bibnamefont {Schulte}}} (\bibinfo {year}
  {2016}),\ in\ \href
  {http://jacow.org/ipac2016/doi/JACoW-IPAC2016-WEPMY010.html} {\emph {\bibinfo
  {booktitle} {Proceedings of the 7th Int. Particle Accelerator Conf.}}}\
  (\bibinfo  {publisher} {JACoW},\ \bibinfo {address} {Geneva, Switzerland})\
  pp.\ \bibinfo {pages} {2565--2568}\BibitemShut {NoStop}%
\bibitem [{\citenamefont {Pierce}\ \emph {et~al.}(1975)\citenamefont {Pierce},
  \citenamefont {Meier},\ and\ \citenamefont {Z\"{u}rcher}}]{Pierce1975}%
  \BibitemOpen
  \bibfield  {author} {\bibinfo {author} {\bibnamefont {Pierce}, \bibfnamefont
  {D~T}}, \bibinfo {author} {\bibfnamefont {F.}~\bibnamefont {Meier}}, \ and\
  \bibinfo {author} {\bibfnamefont {P.}~\bibnamefont {Z\"{u}rcher}}} (\bibinfo
  {year} {1975}),\ \href {https://doi.org/10.1063/1.88030} {\bibfield
  {journal} {\bibinfo  {journal} {Appl. Phys. Lett.}\ }\textbf {\bibinfo
  {volume} {26}},\ \bibinfo {pages} {670--672}}\BibitemShut {NoStop}%
\bibitem [{\citenamefont {Plateau}\ \emph {et~al.}(2012)\citenamefont {Plateau}
  \emph {et~al.}}]{Plateau2012}%
  \BibitemOpen
  \bibfield  {author} {\bibinfo {author} {\bibnamefont {Plateau}, \bibfnamefont
  {G~R}},  \emph {et~al.}} (\bibinfo {year} {2012}),\ \href
  {https://doi.org/10.1103/physrevlett.109.064802} {\bibfield  {journal}
  {\bibinfo  {journal} {Phys. Rev. Lett.}\ }\textbf {\bibinfo {volume} {109}},\
  \bibinfo {pages} {064802}}\BibitemShut {NoStop}%
\bibitem [{\citenamefont {Plyushchev}\ \emph {et~al.}(2017)\citenamefont
  {Plyushchev}, \citenamefont {Kersevan}, \citenamefont {Petrenko},\ and\
  \citenamefont {Muggli}}]{Plyushchev2017}%
  \BibitemOpen
  \bibfield  {author} {\bibinfo {author} {\bibnamefont {Plyushchev},
  \bibfnamefont {G}}, \bibinfo {author} {\bibfnamefont {R.}~\bibnamefont
  {Kersevan}}, \bibinfo {author} {\bibfnamefont {A.}~\bibnamefont {Petrenko}},
  \ and\ \bibinfo {author} {\bibfnamefont {P.}~\bibnamefont {Muggli}}}
  (\bibinfo {year} {2017}),\ \href {https://doi.org/10.1088/1361-6463/aa9dd7}
  {\bibfield  {journal} {\bibinfo  {journal} {J. Phys. D: Appl. Phys.}\
  }\textbf {\bibinfo {volume} {51}},\ \bibinfo {pages} {025203}}\BibitemShut
  {NoStop}%
\bibitem [{\citenamefont {Pogorelsky}\ and\ \citenamefont
  {Ben-Zvi}(2014)}]{Pogorelsky2014}%
  \BibitemOpen
  \bibfield  {author} {\bibinfo {author} {\bibnamefont {Pogorelsky},
  \bibfnamefont {I~V}}, \ and\ \bibinfo {author} {\bibfnamefont
  {I.}~\bibnamefont {Ben-Zvi}}} (\bibinfo {year} {2014}),\ \href
  {https://doi.org/10.1088/0741-3335/56/8/084017} {\bibfield  {journal}
  {\bibinfo  {journal} {Plasma Phys. Control. Fusion}\ }\textbf {\bibinfo
  {volume} {56}},\ \bibinfo {pages} {084017}}\BibitemShut {NoStop}%
\bibitem [{\citenamefont {Pollock}\ \emph {et~al.}(2011)\citenamefont {Pollock}
  \emph {et~al.}}]{Pollock2011}%
  \BibitemOpen
  \bibfield  {author} {\bibinfo {author} {\bibnamefont {Pollock}, \bibfnamefont
  {B~B}},  \emph {et~al.}} (\bibinfo {year} {2011}),\ \href
  {https://doi.org/10.1103/PhysRevLett.107.045001} {\bibfield  {journal}
  {\bibinfo  {journal} {Phys. Rev. Lett.}\ }\textbf {\bibinfo {volume} {107}},\
  \bibinfo {pages} {045001}}\BibitemShut {NoStop}%
\bibitem [{\citenamefont {Pompili}\ \emph
  {et~al.}(2018{\natexlab{a}})\citenamefont {Pompili} \emph
  {et~al.}}]{Pompili2018}%
  \BibitemOpen
  \bibfield  {author} {\bibinfo {author} {\bibnamefont {Pompili}, \bibfnamefont
  {R}},  \emph {et~al.}} (\bibinfo {year} {2018}{\natexlab{a}}),\ \href
  {https://doi.org/10.1103/physrevlett.121.174801} {\bibfield  {journal}
  {\bibinfo  {journal} {Phys. Rev. Lett.}\ }\textbf {\bibinfo {volume} {121}},\
  \bibinfo {pages} {174801}}\BibitemShut {NoStop}%
\bibitem [{\citenamefont {Pompili}\ \emph
  {et~al.}(2018{\natexlab{b}})\citenamefont {Pompili} \emph
  {et~al.}}]{Pompili2018b}%
  \BibitemOpen
  \bibfield  {author} {\bibinfo {author} {\bibnamefont {Pompili}, \bibfnamefont
  {R}},  \emph {et~al.}} (\bibinfo {year} {2018}{\natexlab{b}}),\ \href
  {https://doi.org/10.1063/1.5006134} {\bibfield  {journal} {\bibinfo
  {journal} {Rev. Sci. Instrum.}\ }\textbf {\bibinfo {volume} {89}},\ \bibinfo
  {pages} {033302}}\BibitemShut {NoStop}%
\bibitem [{\citenamefont {Pompili}\ \emph {et~al.}(2021)\citenamefont {Pompili}
  \emph {et~al.}}]{Pompili2021}%
  \BibitemOpen
  \bibfield  {author} {\bibinfo {author} {\bibnamefont {Pompili}, \bibfnamefont
  {R}},  \emph {et~al.}} (\bibinfo {year} {2021}),\ \href
  {https://doi.org/10.1038/s41567-020-01116-9} {\bibfield  {journal} {\bibinfo
  {journal} {Nat. Phys.}\ }\textbf {\bibinfo {volume} {17}},\ \bibinfo {pages}
  {499--503}}\BibitemShut {NoStop}%
\bibitem [{\citenamefont {Pompili}\ \emph {et~al.}(2022)\citenamefont {Pompili}
  \emph {et~al.}}]{Pompili2022}%
  \BibitemOpen
  \bibfield  {author} {\bibinfo {author} {\bibnamefont {Pompili}, \bibfnamefont
  {R}},  \emph {et~al.}} (\bibinfo {year} {2022}),\ \href
  {https://doi.org/10.1038/s41586-022-04589-1} {\bibfield  {journal} {\bibinfo
  {journal} {Nature (London)}\ }\textbf {\bibinfo {volume} {605}},\ \bibinfo
  {pages} {659--662}}\BibitemShut {NoStop}%
\bibitem [{\citenamefont {Pompili}\ \emph
  {et~al.}(2024{\natexlab{a}})\citenamefont {Pompili} \emph
  {et~al.}}]{Pompili2024b}%
  \BibitemOpen
  \bibfield  {author} {\bibinfo {author} {\bibnamefont {Pompili}, \bibfnamefont
  {R}},  \emph {et~al.}} (\bibinfo {year} {2024}{\natexlab{a}}),\ \href
  {https://doi.org/10.1038/s42005-024-01739-x} {\bibfield  {journal} {\bibinfo
  {journal} {Commun. Phys.}\ }\textbf {\bibinfo {volume} {7}},\ \bibinfo
  {pages} {241}}\BibitemShut {NoStop}%
\bibitem [{\citenamefont {Pompili}\ \emph
  {et~al.}(2024{\natexlab{b}})\citenamefont {Pompili} \emph
  {et~al.}}]{Pompili2024a}%
  \BibitemOpen
  \bibfield  {author} {\bibinfo {author} {\bibnamefont {Pompili}, \bibfnamefont
  {R}},  \emph {et~al.}} (\bibinfo {year} {2024}{\natexlab{b}}),\ \href
  {https://doi.org/10.1103/PhysRevE.109.055202} {\bibfield  {journal} {\bibinfo
   {journal} {Phys. Rev. E}\ }\textbf {\bibinfo {volume} {109}},\ \bibinfo
  {pages} {055202}}\BibitemShut {NoStop}%
\bibitem [{\citenamefont {Pukhov}\ \emph {et~al.}(2004)\citenamefont {Pukhov},
  \citenamefont {Gordienko}, \citenamefont {Kiselev},\ and\ \citenamefont
  {Kostyukov}}]{Pukhov2004}%
  \BibitemOpen
  \bibfield  {author} {\bibinfo {author} {\bibnamefont {Pukhov}, \bibfnamefont
  {A}}, \bibinfo {author} {\bibfnamefont {S.}~\bibnamefont {Gordienko}},
  \bibinfo {author} {\bibfnamefont {S.}~\bibnamefont {Kiselev}}, \ and\
  \bibinfo {author} {\bibfnamefont {I.}~\bibnamefont {Kostyukov}}} (\bibinfo
  {year} {2004}),\ \href {https://doi.org/10.1088/0741-3335/46/12b/016}
  {\bibfield  {journal} {\bibinfo  {journal} {Plasma Phys. Control. Fusion}\
  }\textbf {\bibinfo {volume} {46}},\ \bibinfo {pages}
  {B179--B186}}\BibitemShut {NoStop}%
\bibitem [{\citenamefont {Pukhov}\ and\ \citenamefont
  {Kostyukov}(2008)}]{Pukhov2008}%
  \BibitemOpen
  \bibfield  {author} {\bibinfo {author} {\bibnamefont {Pukhov}, \bibfnamefont
  {A}}, \ and\ \bibinfo {author} {\bibfnamefont {I.}~\bibnamefont {Kostyukov}}}
  (\bibinfo {year} {2008}),\ \href {https://doi.org/10.1103/PhysRevE.77.025401}
  {\bibfield  {journal} {\bibinfo  {journal} {Phys. Rev. E}\ }\textbf {\bibinfo
  {volume} {77}},\ \bibinfo {pages} {025401}}\BibitemShut {NoStop}%
\bibitem [{\citenamefont {Pukhov}\ \emph {et~al.}(2011)\citenamefont {Pukhov},
  \citenamefont {Kumar}, \citenamefont {T\"{u}ckmantel}, \citenamefont
  {Upadhyay}, \citenamefont {Lotov}, \citenamefont {Muggli}, \citenamefont
  {Khudik}, \citenamefont {Siemon},\ and\ \citenamefont {Shvets}}]{Pukhov2011}%
  \BibitemOpen
  \bibfield  {author} {\bibinfo {author} {\bibnamefont {Pukhov}, \bibfnamefont
  {A}}, \bibinfo {author} {\bibfnamefont {N.}~\bibnamefont {Kumar}}, \bibinfo
  {author} {\bibfnamefont {T.}~\bibnamefont {T\"{u}ckmantel}}, \bibinfo
  {author} {\bibfnamefont {A.}~\bibnamefont {Upadhyay}}, \bibinfo {author}
  {\bibfnamefont {K.}~\bibnamefont {Lotov}}, \bibinfo {author} {\bibfnamefont
  {P.}~\bibnamefont {Muggli}}, \bibinfo {author} {\bibfnamefont
  {V.}~\bibnamefont {Khudik}}, \bibinfo {author} {\bibfnamefont
  {C.}~\bibnamefont {Siemon}}, \ and\ \bibinfo {author} {\bibfnamefont
  {G.}~\bibnamefont {Shvets}}} (\bibinfo {year} {2011}),\ \href
  {https://doi.org/10.1103/physrevlett.107.145003} {\bibfield  {journal}
  {\bibinfo  {journal} {Phys. Rev. Lett.}\ }\textbf {\bibinfo {volume} {107}},\
  \bibinfo {pages} {145003}}\BibitemShut {NoStop}%
\bibitem [{\citenamefont {Qian}\ \emph {et~al.}(2025)\citenamefont {Qian},
  \citenamefont {Kasa}, \citenamefont {Xu}, \citenamefont {Doran},
  \citenamefont {Lee}, \citenamefont {Sorsher}, \citenamefont {Strelnikov},
  \citenamefont {Trakhtenberg},\ and\ \citenamefont {Zholents}}]{Qian2025}%
  \BibitemOpen
  \bibfield  {author} {\bibinfo {author} {\bibnamefont {Qian}, \bibfnamefont
  {M}}, \bibinfo {author} {\bibfnamefont {M.}~\bibnamefont {Kasa}}, \bibinfo
  {author} {\bibfnamefont {J.}~\bibnamefont {Xu}}, \bibinfo {author}
  {\bibfnamefont {S.}~\bibnamefont {Doran}}, \bibinfo {author} {\bibfnamefont
  {S.}~\bibnamefont {Lee}}, \bibinfo {author} {\bibfnamefont {S.}~\bibnamefont
  {Sorsher}}, \bibinfo {author} {\bibfnamefont {N.}~\bibnamefont {Strelnikov}},
  \bibinfo {author} {\bibfnamefont {E.}~\bibnamefont {Trakhtenberg}}, \ and\
  \bibinfo {author} {\bibfnamefont {A.}~\bibnamefont {Zholents}}} (\bibinfo
  {year} {2025}),\ \href {https://doi.org/10.1103/PhysRevAccelBeams.28.012401}
  {\bibfield  {journal} {\bibinfo  {journal} {Phys. Rev. Accel. Beams}\
  }\textbf {\bibinfo {volume} {28}},\ \bibinfo {pages} {012401}}\BibitemShut
  {NoStop}%
\bibitem [{\citenamefont {Qian}\ \emph {et~al.}(2026)\citenamefont {Qian},
  \citenamefont {Seipt}, \citenamefont {Ma}, \citenamefont {Bulanov},
  \citenamefont {Schroeder},\ and\ \citenamefont {Thomas}}]{Qian2026}%
  \BibitemOpen
  \bibfield  {author} {\bibinfo {author} {\bibnamefont {Qian}, \bibfnamefont
  {Q}}, \bibinfo {author} {\bibfnamefont {D.}~\bibnamefont {Seipt}}, \bibinfo
  {author} {\bibfnamefont {Y.}~\bibnamefont {Ma}}, \bibinfo {author}
  {\bibfnamefont {S.~S.}\ \bibnamefont {Bulanov}}, \bibinfo {author}
  {\bibfnamefont {C.~B.}\ \bibnamefont {Schroeder}}, \ and\ \bibinfo {author}
  {\bibfnamefont {A.~G.~R.}\ \bibnamefont {Thomas}}} (\bibinfo {year} {2026}),\
  \href {\doibase 10.1088/1361-6587/ae60a2} {\bibfield  {journal} {\bibinfo
  {journal} {Plasma Phys. Control. Fusion}\ }\textbf {\bibinfo {volume} {68}},\
  \bibinfo {pages} {055011}}\BibitemShut {NoStop}%
\bibitem [{\citenamefont {Raimondi}\ and\ \citenamefont
  {Seryi}(2001)}]{Raimondi2001}%
  \BibitemOpen
  \bibfield  {author} {\bibinfo {author} {\bibnamefont {Raimondi},
  \bibfnamefont {P}}, \ and\ \bibinfo {author} {\bibfnamefont {A.}~\bibnamefont
  {Seryi}}} (\bibinfo {year} {2001}),\ \href
  {https://doi.org/10.1103/physrevlett.86.3779} {\bibfield  {journal} {\bibinfo
   {journal} {Phys. Rev. Lett.}\ }\textbf {\bibinfo {volume} {86}},\ \bibinfo
  {pages} {3779--3782}}\BibitemShut {NoStop}%
\bibitem [{\citenamefont {Rakitzis}\ \emph {et~al.}(2003)\citenamefont
  {Rakitzis}, \citenamefont {Samartzis}, \citenamefont {Toomes}, \citenamefont
  {Kitsopoulos}, \citenamefont {Brown}, \citenamefont {Balint-Kurti},
  \citenamefont {Vasyutinskii},\ and\ \citenamefont {Beswick}}]{Rakitzis2003}%
  \BibitemOpen
  \bibfield  {author} {\bibinfo {author} {\bibnamefont {Rakitzis},
  \bibfnamefont {T~P}}, \bibinfo {author} {\bibfnamefont {P.~C.}\ \bibnamefont
  {Samartzis}}, \bibinfo {author} {\bibfnamefont {R.~L.}\ \bibnamefont
  {Toomes}}, \bibinfo {author} {\bibfnamefont {T.~N.}\ \bibnamefont
  {Kitsopoulos}}, \bibinfo {author} {\bibfnamefont {A.}~\bibnamefont {Brown}},
  \bibinfo {author} {\bibfnamefont {G.~G.}\ \bibnamefont {Balint-Kurti}},
  \bibinfo {author} {\bibfnamefont {O.~S.}\ \bibnamefont {Vasyutinskii}}, \
  and\ \bibinfo {author} {\bibfnamefont {J.~A.}\ \bibnamefont {Beswick}}}
  (\bibinfo {year} {2003}),\ \href {https://doi.org/10.1126/SCIENCE.1084809}
  {\bibfield  {journal} {\bibinfo  {journal} {Science}\ }\textbf {\bibinfo
  {volume} {300}},\ \bibinfo {pages} {1936--1938}}\BibitemShut {NoStop}%
\bibitem [{\citenamefont {Ramjiawan}\ \emph {et~al.}(2023)\citenamefont
  {Ramjiawan}, \citenamefont {Bencini}, \citenamefont {Burrows},\ and\
  \citenamefont {Velotti}}]{Ramjiawan2023}%
  \BibitemOpen
  \bibfield  {author} {\bibinfo {author} {\bibnamefont {Ramjiawan},
  \bibfnamefont {R}}, \bibinfo {author} {\bibfnamefont {V.}~\bibnamefont
  {Bencini}}, \bibinfo {author} {\bibfnamefont {P.~N.}\ \bibnamefont
  {Burrows}}, \ and\ \bibinfo {author} {\bibfnamefont {F.~M.}\ \bibnamefont
  {Velotti}}} (\bibinfo {year} {2023}),\ \href
  {https://doi.org/10.1016/j.nima.2023.168094} {\bibfield  {journal} {\bibinfo
  {journal} {Nucl. Instrum. Methods Phys. Res. A}\ }\textbf {\bibinfo {volume}
  {1049}},\ \bibinfo {pages} {168094}}\BibitemShut {NoStop}%
\bibitem [{\citenamefont {Raubenheimer}(1992)}]{Raubenheimer1992}%
  \BibitemOpen
  \bibfield  {author} {\bibinfo {author} {\bibnamefont {Raubenheimer},
  \bibfnamefont {T~O}}} (\bibinfo {year} {1992}),\ \href
  {https://lib-extopc.kek.jp/preprints/PDF/1992/9224/9224007.pdf} {}\bibinfo
  {type} {Tech. Rep.}\ \bibinfo {number} {KEK-92-7}\ (\bibinfo  {institution}
  {KEK},\ \bibinfo {address} {Tsukuba, Japan})\BibitemShut {NoStop}%
\bibitem [{\citenamefont {Raubenheimer}(2004)}]{Raubenheimer2004}%
  \BibitemOpen
  \bibfield  {author} {\bibinfo {author} {\bibnamefont {Raubenheimer},
  \bibfnamefont {T~O}}} (\bibinfo {year} {2004}),\ in\ \href
  {https://doi.org/10.1063/1.1842536} {\emph {\bibinfo {booktitle} {{AIP} Conf.
  Proc.}}},\ Vol.\ \bibinfo {volume} {737}\ (\bibinfo  {publisher} {{AIP}})\
  pp.\ \bibinfo {pages} {86--94}\BibitemShut {NoStop}%
\bibitem [{\citenamefont {Rechatin}\ \emph {et~al.}(2009)\citenamefont
  {Rechatin}, \citenamefont {Davoine}, \citenamefont {Lifschitz}, \citenamefont
  {Ismail}, \citenamefont {Lim}, \citenamefont {Lefebvre}, \citenamefont
  {Faure},\ and\ \citenamefont {Malka}}]{Rechatin2009}%
  \BibitemOpen
  \bibfield  {author} {\bibinfo {author} {\bibnamefont {Rechatin},
  \bibfnamefont {C}}, \bibinfo {author} {\bibfnamefont {X.}~\bibnamefont
  {Davoine}}, \bibinfo {author} {\bibfnamefont {A.}~\bibnamefont {Lifschitz}},
  \bibinfo {author} {\bibfnamefont {A.~Ben}\ \bibnamefont {Ismail}}, \bibinfo
  {author} {\bibfnamefont {J.}~\bibnamefont {Lim}}, \bibinfo {author}
  {\bibfnamefont {E.}~\bibnamefont {Lefebvre}}, \bibinfo {author}
  {\bibfnamefont {J.}~\bibnamefont {Faure}}, \ and\ \bibinfo {author}
  {\bibfnamefont {V.}~\bibnamefont {Malka}}} (\bibinfo {year} {2009}),\ \href
  {https://doi.org/10.1103/PhysRevLett.103.194804} {\bibfield  {journal}
  {\bibinfo  {journal} {Phys. Rev. Lett.}\ }\textbf {\bibinfo {volume} {103}},\
  \bibinfo {pages} {194804}}\BibitemShut {NoStop}%
\bibitem [{\citenamefont {Rechatin}\ \emph {et~al.}(2010)\citenamefont
  {Rechatin} \emph {et~al.}}]{Rechatin2010}%
  \BibitemOpen
  \bibfield  {author} {\bibinfo {author} {\bibnamefont {Rechatin},
  \bibfnamefont {C}},  \emph {et~al.}} (\bibinfo {year} {2010}),\ \href
  {https://doi.org/10.1088/1367-2630/12/4/045023} {\bibfield  {journal}
  {\bibinfo  {journal} {New J. Phys.}\ }\textbf {\bibinfo {volume} {12}},\
  \bibinfo {pages} {045023}}\BibitemShut {NoStop}%
\bibitem [{\citenamefont {Reichwein}\ \emph {et~al.}(2025)\citenamefont
  {Reichwein}, \citenamefont {Gong}, \citenamefont {Zheng}, \citenamefont {Ji},
  \citenamefont {Pukhov},\ and\ \citenamefont {B\"{u}scher}}]{Reichwein2025}%
  \BibitemOpen
  \bibfield  {author} {\bibinfo {author} {\bibnamefont {Reichwein},
  \bibfnamefont {L}}, \bibinfo {author} {\bibfnamefont {Z}~\bibnamefont
  {Gong}}, \bibinfo {author} {\bibfnamefont {C}~\bibnamefont {Zheng}}, \bibinfo
  {author} {\bibfnamefont {L~L}\ \bibnamefont {Ji}}, \bibinfo {author}
  {\bibfnamefont {A}~\bibnamefont {Pukhov}}, \ and\ \bibinfo {author}
  {\bibfnamefont {M}~\bibnamefont {B\"{u}scher}}} (\bibinfo {year} {2025}),\
  \href {\doibase 10.1088/1361-6633/ae1988} {\bibfield  {journal} {\bibinfo
  {journal} {Rep. Prog. Phys.}\ }\textbf {\bibinfo {volume} {88}},\ \bibinfo
  {pages} {117001}}\BibitemShut {NoStop}%
\bibitem [{\citenamefont {Reichwein}\ \emph {et~al.}(2022)\citenamefont
  {Reichwein}, \citenamefont {Pukhov}, \citenamefont {Golovanov},\ and\
  \citenamefont {Kostyukov}}]{Reichwein2022}%
  \BibitemOpen
  \bibfield  {author} {\bibinfo {author} {\bibnamefont {Reichwein},
  \bibfnamefont {L}}, \bibinfo {author} {\bibfnamefont {A.}~\bibnamefont
  {Pukhov}}, \bibinfo {author} {\bibfnamefont {A.}~\bibnamefont {Golovanov}}, \
  and\ \bibinfo {author} {\bibfnamefont {I.~Yu.}\ \bibnamefont {Kostyukov}}}
  (\bibinfo {year} {2022}),\ \href {https://10.1103/PhysRevE.105.055207}
  {\bibfield  {journal} {\bibinfo  {journal} {Phys. Rev. E}\ }\textbf {\bibinfo
  {volume} {105}},\ \bibinfo {eid} {055207}}\BibitemShut {NoStop}%
\bibitem [{\citenamefont {Reiser}(2008)}]{Reiser2008}%
  \BibitemOpen
  \bibfield  {author} {\bibinfo {author} {\bibnamefont {Reiser}, \bibfnamefont
  {M}}} (\bibinfo {year} {2008}),\ \href
  {https://doi.org/10.1002/9783527622047} {\emph {\bibinfo {title} {Theory and
  Design of Charged Particle Beams}}}\ (\bibinfo  {publisher} {Wiley-VCH},\
  \bibinfo {address} {Weinheim, Germany})\BibitemShut {NoStop}%
\bibitem [{\citenamefont {Richter}(1988)}]{Richter1988}%
  \BibitemOpen
  \bibfield  {author} {\bibinfo {author} {\bibnamefont {Richter}, \bibfnamefont
  {B}}} (\bibinfo {year} {1988}),\ \href@noop {} {}\bibinfo {type} {Tech.
  Rep.}\ \bibinfo {number} {SLAC-PUB-4614}\ (\bibinfo  {institution} {SLAC},\
  \bibinfo {address} {Menlo Park, CA, United States})\BibitemShut {NoStop}%
\bibitem [{\citenamefont {Rieger}\ \emph {et~al.}(2017)\citenamefont {Rieger},
  \citenamefont {Caldwell}, \citenamefont {Reimann},\ and\ \citenamefont
  {Muggli}}]{Rieger2017}%
  \BibitemOpen
  \bibfield  {author} {\bibinfo {author} {\bibnamefont {Rieger}, \bibfnamefont
  {K}}, \bibinfo {author} {\bibfnamefont {A.}~\bibnamefont {Caldwell}},
  \bibinfo {author} {\bibfnamefont {O.}~\bibnamefont {Reimann}}, \ and\
  \bibinfo {author} {\bibfnamefont {P.}~\bibnamefont {Muggli}}} (\bibinfo
  {year} {2017}),\ \href {https://doi.org/10.1063/1.4975380} {\bibfield
  {journal} {\bibinfo  {journal} {Rev. Sci. Instrum.}\ }\textbf {\bibinfo
  {volume} {88}},\ \bibinfo {pages} {025110}}\BibitemShut {NoStop}%
\bibitem [{\citenamefont {Riemann}\ \emph {et~al.}(2011)\citenamefont
  {Riemann}, \citenamefont {Schalicke},\ and\ \citenamefont
  {Ushakov}}]{Riemann2011}%
  \BibitemOpen
  \bibfield  {author} {\bibinfo {author} {\bibnamefont {Riemann}, \bibfnamefont
  {S}}, \bibinfo {author} {\bibfnamefont {A.}~\bibnamefont {Schalicke}}, \ and\
  \bibinfo {author} {\bibfnamefont {A.}~\bibnamefont {Ushakov}}} (\bibinfo
  {year} {2011}),\ \href {https://doi.org/10.1088/1742-6596/298/1/012020}
  {\bibfield  {journal} {\bibinfo  {journal} {J. Phys.: Conf. Ser.}\ }\textbf
  {\bibinfo {volume} {298}},\ \bibinfo {pages} {012020}}\BibitemShut {NoStop}%
\bibitem [{\citenamefont {Romeo}\ \emph {et~al.}(2024)\citenamefont {Romeo},
  \citenamefont {Del~Dotto}, \citenamefont {Ferrario}, \citenamefont
  {Giribono}, \citenamefont {Rossi}, \citenamefont {Silvi},\ and\ \citenamefont
  {Vaccarezza}}]{Romeo2024}%
  \BibitemOpen
  \bibfield  {author} {\bibinfo {author} {\bibnamefont {Romeo}, \bibfnamefont
  {S}}, \bibinfo {author} {\bibfnamefont {A.}~\bibnamefont {Del~Dotto}},
  \bibinfo {author} {\bibfnamefont {M.}~\bibnamefont {Ferrario}}, \bibinfo
  {author} {\bibfnamefont {A.}~\bibnamefont {Giribono}}, \bibinfo {author}
  {\bibfnamefont {A.~R.}\ \bibnamefont {Rossi}}, \bibinfo {author}
  {\bibfnamefont {G.~J.}\ \bibnamefont {Silvi}}, \ and\ \bibinfo {author}
  {\bibfnamefont {C.}~\bibnamefont {Vaccarezza}}} (\bibinfo {year} {2024}),\
  \href {\doibase 10.1088/1742-6596/2687/4/042008} {\bibfield  {journal}
  {\bibinfo  {journal} {J. Phys.: Conf. Ser.}\ }\textbf {\bibinfo {volume}
  {2687}},\ \bibinfo {pages} {042008}}\BibitemShut {NoStop}%
\bibitem [{\citenamefont {Rosenzweig}\ \emph {et~al.}(1996)\citenamefont
  {Rosenzweig}, \citenamefont {Barov}, \citenamefont {Colby},\ and\
  \citenamefont {Colestock}}]{Rosenzweig1996}%
  \BibitemOpen
  \bibfield  {author} {\bibinfo {author} {\bibnamefont {Rosenzweig},
  \bibfnamefont {J}}, \bibinfo {author} {\bibfnamefont {N.}~\bibnamefont
  {Barov}}, \bibinfo {author} {\bibfnamefont {E.}~\bibnamefont {Colby}}, \ and\
  \bibinfo {author} {\bibfnamefont {P.}~\bibnamefont {Colestock}}} (\bibinfo
  {year} {1996}),\ in\ \href
  {https://www.slac.stanford.edu/pubs/snowmass96/PDF/ACC067.PDF} {\emph
  {\bibinfo {booktitle} {Proceedings of Snowmass 1996}}}\ (\bibinfo
  {publisher} {SLAC},\ \bibinfo {address} {Menlo Park, CA, United States})\
  pp.\ \bibinfo {pages} {394--398}\BibitemShut {NoStop}%
\bibitem [{\citenamefont {Rosenzweig}\ \emph {et~al.}(1998)\citenamefont
  {Rosenzweig}, \citenamefont {Barov}, \citenamefont {Murokh}, \citenamefont
  {Colby},\ and\ \citenamefont {Colestock}}]{Rosenzweig1998}%
  \BibitemOpen
  \bibfield  {author} {\bibinfo {author} {\bibnamefont {Rosenzweig},
  \bibfnamefont {J}}, \bibinfo {author} {\bibfnamefont {N.}~\bibnamefont
  {Barov}}, \bibinfo {author} {\bibfnamefont {A.}~\bibnamefont {Murokh}},
  \bibinfo {author} {\bibfnamefont {E.}~\bibnamefont {Colby}}, \ and\ \bibinfo
  {author} {\bibfnamefont {P.}~\bibnamefont {Colestock}}} (\bibinfo {year}
  {1998}),\ \href
  {https://www.sciencedirect.com/science/article/pii/S0168900298001867}
  {\bibfield  {journal} {\bibinfo  {journal} {Nucl. Instrum. Methods Phys. Res.
  A}\ }\textbf {\bibinfo {volume} {410}},\ \bibinfo {pages}
  {532--543}}\BibitemShut {NoStop}%
\bibitem [{\citenamefont {Rosenzweig}\ \emph {et~al.}(1987)\citenamefont
  {Rosenzweig}, \citenamefont {Cline}, \citenamefont {Cole}, \citenamefont
  {Detra},\ and\ \citenamefont {Sealy}}]{Rosenzweig1987b}%
  \BibitemOpen
  \bibfield  {author} {\bibinfo {author} {\bibnamefont {Rosenzweig},
  \bibfnamefont {J}}, \bibinfo {author} {\bibfnamefont {D.}~\bibnamefont
  {Cline}}, \bibinfo {author} {\bibfnamefont {B.}~\bibnamefont {Cole}},
  \bibinfo {author} {\bibfnamefont {J.}~\bibnamefont {Detra}}, \ and\ \bibinfo
  {author} {\bibfnamefont {P.}~\bibnamefont {Sealy}}} (\bibinfo {year}
  {1987}),\ in\ \href {https://doi.org/10.1063/1.36459} {\emph {\bibinfo
  {booktitle} {{AIP} Conf. Proc.}}},\ Vol.\ \bibinfo {volume} {156}\ (\bibinfo
  {publisher} {{AIP}})\ pp.\ \bibinfo {pages} {231--246}\BibitemShut {NoStop}%
\bibitem [{\citenamefont {Rosenzweig}(1987)}]{Rosenzweig1987}%
  \BibitemOpen
  \bibfield  {author} {\bibinfo {author} {\bibnamefont {Rosenzweig},
  \bibfnamefont {J~B}}} (\bibinfo {year} {1987}),\ \href
  {https://doi.org/10.1103/physrevlett.58.555} {\bibfield  {journal} {\bibinfo
  {journal} {Phys. Rev. Lett.}\ }\textbf {\bibinfo {volume} {58}},\ \bibinfo
  {pages} {555--558}}\BibitemShut {NoStop}%
\bibitem [{\citenamefont {Rosenzweig}(1988)}]{Rosenzweig1988b}%
  \BibitemOpen
  \bibfield  {author} {\bibinfo {author} {\bibnamefont {Rosenzweig},
  \bibfnamefont {J~B}}} (\bibinfo {year} {1988}),\ \href
  {https://doi.org/10.1103/PhysRevA.38.3634} {\bibfield  {journal} {\bibinfo
  {journal} {Phys. Rev. A}\ }\textbf {\bibinfo {volume} {38}},\ \bibinfo
  {pages} {3634--3642}}\BibitemShut {NoStop}%
\bibitem [{\citenamefont {Rosenzweig}\ \emph {et~al.}(2004)\citenamefont
  {Rosenzweig}, \citenamefont {Barov}, \citenamefont {Thompson},\ and\
  \citenamefont {Yoder}}]{Rosenzweig2004}%
  \BibitemOpen
  \bibfield  {author} {\bibinfo {author} {\bibnamefont {Rosenzweig},
  \bibfnamefont {J~B}}, \bibinfo {author} {\bibfnamefont {N.}~\bibnamefont
  {Barov}}, \bibinfo {author} {\bibfnamefont {M.~C.}\ \bibnamefont {Thompson}},
  \ and\ \bibinfo {author} {\bibfnamefont {R.~B.}\ \bibnamefont {Yoder}}}
  (\bibinfo {year} {2004}),\ \href
  {https://doi.org/10.1103/physrevstab.7.061302} {\bibfield  {journal}
  {\bibinfo  {journal} {Phys. Rev. ST Accel. Beams}\ }\textbf {\bibinfo
  {volume} {7}},\ \bibinfo {pages} {061302}}\BibitemShut {NoStop}%
\bibitem [{\citenamefont {Rosenzweig}\ \emph {et~al.}(1991)\citenamefont
  {Rosenzweig}, \citenamefont {Breizman}, \citenamefont {Katsouleas},\ and\
  \citenamefont {Su}}]{Rosenzweig1991}%
  \BibitemOpen
  \bibfield  {author} {\bibinfo {author} {\bibnamefont {Rosenzweig},
  \bibfnamefont {J~B}}, \bibinfo {author} {\bibfnamefont {B.}~\bibnamefont
  {Breizman}}, \bibinfo {author} {\bibfnamefont {T.}~\bibnamefont
  {Katsouleas}}, \ and\ \bibinfo {author} {\bibfnamefont {J.~J.}\ \bibnamefont
  {Su}}} (\bibinfo {year} {1991}),\ \href
  {https://doi.org/10.1103/physreva.44.r6189} {\bibfield  {journal} {\bibinfo
  {journal} {Phys. Rev. A}\ }\textbf {\bibinfo {volume} {44}},\ \bibinfo
  {pages} {R6189--R6192}}\BibitemShut {NoStop}%
\bibitem [{\citenamefont {Rosenzweig}\ \emph {et~al.}(1988)\citenamefont
  {Rosenzweig}, \citenamefont {Cline}, \citenamefont {Cole}, \citenamefont
  {Figueroa}, \citenamefont {Gai}, \citenamefont {Konecny}, \citenamefont
  {Norem}, \citenamefont {Schoessow},\ and\ \citenamefont
  {Simpson}}]{Rosenzweig1988}%
  \BibitemOpen
  \bibfield  {author} {\bibinfo {author} {\bibnamefont {Rosenzweig},
  \bibfnamefont {J~B}}, \bibinfo {author} {\bibfnamefont {D.~B.}\ \bibnamefont
  {Cline}}, \bibinfo {author} {\bibfnamefont {B.}~\bibnamefont {Cole}},
  \bibinfo {author} {\bibfnamefont {H.}~\bibnamefont {Figueroa}}, \bibinfo
  {author} {\bibfnamefont {W.}~\bibnamefont {Gai}}, \bibinfo {author}
  {\bibfnamefont {R.}~\bibnamefont {Konecny}}, \bibinfo {author} {\bibfnamefont
  {J.}~\bibnamefont {Norem}}, \bibinfo {author} {\bibfnamefont
  {P.}~\bibnamefont {Schoessow}}, \ and\ \bibinfo {author} {\bibfnamefont
  {J.}~\bibnamefont {Simpson}}} (\bibinfo {year} {1988}),\ \href
  {https://doi.org/10.1103/physrevlett.61.98} {\bibfield  {journal} {\bibinfo
  {journal} {Phys. Rev. Lett.}\ }\textbf {\bibinfo {volume} {61}},\ \bibinfo
  {pages} {98--101}}\BibitemShut {NoStop}%
\bibitem [{\citenamefont {Rosenzweig}\ \emph {et~al.}(1985)\citenamefont
  {Rosenzweig}, \citenamefont {Cline}, \citenamefont {Dexter}, \citenamefont
  {Larson}, \citenamefont {Leonard}, \citenamefont {Mengelt}, \citenamefont
  {Sprott}, \citenamefont {Mills},\ and\ \citenamefont
  {Cole}}]{Rosenzweig1985}%
  \BibitemOpen
  \bibfield  {author} {\bibinfo {author} {\bibnamefont {Rosenzweig},
  \bibfnamefont {J~B}}, \bibinfo {author} {\bibfnamefont {D.~B.}\ \bibnamefont
  {Cline}}, \bibinfo {author} {\bibfnamefont {R.~N.}\ \bibnamefont {Dexter}},
  \bibinfo {author} {\bibfnamefont {D.~J.}\ \bibnamefont {Larson}}, \bibinfo
  {author} {\bibfnamefont {A.~W.}\ \bibnamefont {Leonard}}, \bibinfo {author}
  {\bibfnamefont {K.~R.}\ \bibnamefont {Mengelt}}, \bibinfo {author}
  {\bibfnamefont {J.~C.}\ \bibnamefont {Sprott}}, \bibinfo {author}
  {\bibfnamefont {F.~E.}\ \bibnamefont {Mills}}, \ and\ \bibinfo {author}
  {\bibfnamefont {F.~T.}\ \bibnamefont {Cole}}} (\bibinfo {year} {1985}),\ in\
  \href {https://doi.org/10.1063/1.35323} {\emph {\bibinfo {booktitle} {{AIP}
  Conf. Proc.}}},\ Vol.\ \bibinfo {volume} {130}\ (\bibinfo  {publisher}
  {{AIP}})\ pp.\ \bibinfo {pages} {226--233}\BibitemShut {NoStop}%
\bibitem [{\citenamefont {Rosenzweig}\ \emph {et~al.}(2005)\citenamefont
  {Rosenzweig}, \citenamefont {Cook}, \citenamefont {Scott}, \citenamefont
  {Thompson},\ and\ \citenamefont {Yoder}}]{Rosenzweig2005}%
  \BibitemOpen
  \bibfield  {author} {\bibinfo {author} {\bibnamefont {Rosenzweig},
  \bibfnamefont {J~B}}, \bibinfo {author} {\bibfnamefont {A.~M.}\ \bibnamefont
  {Cook}}, \bibinfo {author} {\bibfnamefont {A.}~\bibnamefont {Scott}},
  \bibinfo {author} {\bibfnamefont {M.~C.}\ \bibnamefont {Thompson}}, \ and\
  \bibinfo {author} {\bibfnamefont {R.~B.}\ \bibnamefont {Yoder}}} (\bibinfo
  {year} {2005}),\ \href {https://doi.org/10.1103/physrevlett.95.195002}
  {\bibfield  {journal} {\bibinfo  {journal} {Phys. Rev. Lett.}\ }\textbf
  {\bibinfo {volume} {95}},\ \bibinfo {pages} {195002}}\BibitemShut {NoStop}%
\bibitem [{\citenamefont {Rosenzweig}\ \emph {et~al.}(1989)\citenamefont
  {Rosenzweig}, \citenamefont {Schoessow}, \citenamefont {Cole}, \citenamefont
  {Gai}, \citenamefont {Konecny}, \citenamefont {Norem},\ and\ \citenamefont
  {Simpson}}]{Rosenzweig1989}%
  \BibitemOpen
  \bibfield  {author} {\bibinfo {author} {\bibnamefont {Rosenzweig},
  \bibfnamefont {J~B}}, \bibinfo {author} {\bibfnamefont {P.}~\bibnamefont
  {Schoessow}}, \bibinfo {author} {\bibfnamefont {B.}~\bibnamefont {Cole}},
  \bibinfo {author} {\bibfnamefont {W.}~\bibnamefont {Gai}}, \bibinfo {author}
  {\bibfnamefont {R.}~\bibnamefont {Konecny}}, \bibinfo {author} {\bibfnamefont
  {J.}~\bibnamefont {Norem}}, \ and\ \bibinfo {author} {\bibfnamefont
  {J.}~\bibnamefont {Simpson}}} (\bibinfo {year} {1989}),\ \href
  {https://doi.org/10.1103/physreva.39.1586} {\bibfield  {journal} {\bibinfo
  {journal} {Phys. Rev. A}\ }\textbf {\bibinfo {volume} {39}},\ \bibinfo
  {pages} {1586--1589}}\BibitemShut {NoStop}%
\bibitem [{\citenamefont {Rosenzweig}\ \emph {et~al.}(1990)\citenamefont
  {Rosenzweig}, \citenamefont {Schoessow}, \citenamefont {Cole}, \citenamefont
  {Ho}, \citenamefont {Gai}, \citenamefont {Konecny}, \citenamefont {Mtingwa},
  \citenamefont {Norem}, \citenamefont {Rosing},\ and\ \citenamefont
  {Simpson}}]{Rosenzweig1990}%
  \BibitemOpen
  \bibfield  {author} {\bibinfo {author} {\bibnamefont {Rosenzweig},
  \bibfnamefont {J~B}}, \bibinfo {author} {\bibfnamefont {P.}~\bibnamefont
  {Schoessow}}, \bibinfo {author} {\bibfnamefont {B.}~\bibnamefont {Cole}},
  \bibinfo {author} {\bibfnamefont {C.}~\bibnamefont {Ho}}, \bibinfo {author}
  {\bibfnamefont {W.}~\bibnamefont {Gai}}, \bibinfo {author} {\bibfnamefont
  {R.}~\bibnamefont {Konecny}}, \bibinfo {author} {\bibfnamefont
  {S.}~\bibnamefont {Mtingwa}}, \bibinfo {author} {\bibfnamefont
  {J.}~\bibnamefont {Norem}}, \bibinfo {author} {\bibfnamefont
  {M.}~\bibnamefont {Rosing}}, \ and\ \bibinfo {author} {\bibfnamefont
  {J.}~\bibnamefont {Simpson}}} (\bibinfo {year} {1990}),\ \href
  {https://doi.org/10.1063/1.859559} {\bibfield  {journal} {\bibinfo  {journal}
  {Phys. Fluids B}\ }\textbf {\bibinfo {volume} {2}},\ \bibinfo {pages}
  {1376--1383}}\BibitemShut {NoStop}%
\bibitem [{\citenamefont {Roser}\ \emph {et~al.}(2023)\citenamefont {Roser}
  \emph {et~al.}}]{Roser2023}%
  \BibitemOpen
  \bibfield  {author} {\bibinfo {author} {\bibnamefont {Roser}, \bibfnamefont
  {T}},  \emph {et~al.}} (\bibinfo {year} {2023}),\ \href {\doibase
  10.1088/1748-0221/18/05/p05018} {\bibfield  {journal} {\bibinfo  {journal}
  {J. Instrum.}\ }\textbf {\bibinfo {volume} {18}},\ \bibinfo {pages}
  {P05018}}\BibitemShut {NoStop}%
\bibitem [{\citenamefont {Rousse}\ \emph {et~al.}(2004)\citenamefont {Rousse}
  \emph {et~al.}}]{Rousse2004}%
  \BibitemOpen
  \bibfield  {author} {\bibinfo {author} {\bibnamefont {Rousse}, \bibfnamefont
  {A}},  \emph {et~al.}} (\bibinfo {year} {2004}),\ \href
  {https://doi.org/10.1103/physrevlett.93.135005} {\bibfield  {journal}
  {\bibinfo  {journal} {Phys. Rev. Lett.}\ }\textbf {\bibinfo {volume} {93}},\
  \bibinfo {pages} {135005}}\BibitemShut {NoStop}%
\bibitem [{\citenamefont {Roussel}\ \emph {et~al.}(2020)\citenamefont {Roussel}
  \emph {et~al.}}]{Roussel2020}%
  \BibitemOpen
  \bibfield  {author} {\bibinfo {author} {\bibnamefont {Roussel}, \bibfnamefont
  {R}},  \emph {et~al.}} (\bibinfo {year} {2020}),\ \href
  {https://link.aps.org/doi/10.1103/PhysRevLett.124.044802} {\bibfield
  {journal} {\bibinfo  {journal} {Phys. Rev. Lett.}\ }\textbf {\bibinfo
  {volume} {124}},\ \bibinfo {pages} {044802}}\BibitemShut {NoStop}%
\bibitem [{\citenamefont {Russell}\ \emph {et~al.}(2002)\citenamefont
  {Russell}, \citenamefont {Goettee},\ and\ \citenamefont
  {Carlsten}}]{Russell2002}%
  \BibitemOpen
  \bibfield  {author} {\bibinfo {author} {\bibnamefont {Russell}, \bibfnamefont
  {S~J}}, \bibinfo {author} {\bibfnamefont {J.~D.}\ \bibnamefont {Goettee}}, \
  and\ \bibinfo {author} {\bibfnamefont {B.~E.}\ \bibnamefont {Carlsten}}}
  (\bibinfo {year} {2002}),\ in\ \href {\doibase 10.1109/pac.2001.988316}
  {\emph {\bibinfo {booktitle} {Proceedings of the 2001 Particle Accelerator
  Conf.}}}\ (\bibinfo  {publisher} {IEEE},\ \bibinfo {address} {New York, NY,
  United States})\ pp.\ \bibinfo {pages} {3975--3977}\BibitemShut {NoStop}%
\bibitem [{\citenamefont {Ruth}\ \emph {et~al.}(1985)\citenamefont {Ruth},
  \citenamefont {Chao}, \citenamefont {Morton},\ and\ \citenamefont
  {Wilson}}]{Ruth1985}%
  \BibitemOpen
  \bibfield  {author} {\bibinfo {author} {\bibnamefont {Ruth}, \bibfnamefont
  {R~D}}, \bibinfo {author} {\bibfnamefont {A.~W.}\ \bibnamefont {Chao}},
  \bibinfo {author} {\bibfnamefont {P.~L.}\ \bibnamefont {Morton}}, \ and\
  \bibinfo {author} {\bibfnamefont {P.~B.}\ \bibnamefont {Wilson}}} (\bibinfo
  {year} {1985}),\ \href {http://inspirehep.net/record/206328/} {\bibfield
  {journal} {\bibinfo  {journal} {Part. Accel.}\ }\textbf {\bibinfo {volume}
  {17}},\ \bibinfo {pages} {171--189}}\BibitemShut {NoStop}%
\bibitem [{\citenamefont {Saberi}\ \emph {et~al.}(2023)\citenamefont {Saberi},
  \citenamefont {Xia}, \citenamefont {Islam}, \citenamefont {Liang},\ and\
  \citenamefont {Davut}}]{Saberi2023}%
  \BibitemOpen
  \bibfield  {author} {\bibinfo {author} {\bibnamefont {Saberi}, \bibfnamefont
  {H}}, \bibinfo {author} {\bibfnamefont {G.}~\bibnamefont {Xia}}, \bibinfo
  {author} {\bibfnamefont {M.~R.}\ \bibnamefont {Islam}}, \bibinfo {author}
  {\bibfnamefont {L.}~\bibnamefont {Liang}}, \ and\ \bibinfo {author}
  {\bibfnamefont {C.}~\bibnamefont {Davut}}} (\bibinfo {year} {2023}),\ \href
  {https://doi.org/10.1063/5.0140525} {\bibfield  {journal} {\bibinfo
  {journal} {Phys. Plasmas}\ }\textbf {\bibinfo {volume} {30}},\ \bibinfo
  {pages} {043104}}\BibitemShut {NoStop}%
\bibitem [{\citenamefont {Saberi}\ \emph {et~al.}(2026)\citenamefont {Saberi},
  \citenamefont {Xia}, \citenamefont {Lei}, \citenamefont {Si},\ and\
  \citenamefont {Huang}}]{Saberi2026}%
  \BibitemOpen
  \bibfield  {author} {\bibinfo {author} {\bibnamefont {Saberi}, \bibfnamefont
  {H}}, \bibinfo {author} {\bibfnamefont {G.}~\bibnamefont {Xia}}, \bibinfo
  {author} {\bibfnamefont {B.}~\bibnamefont {Lei}}, \bibinfo {author}
  {\bibfnamefont {M.}~\bibnamefont {Si}}, \ and\ \bibinfo {author}
  {\bibfnamefont {Y.}~\bibnamefont {Huang}}} (\bibinfo {year} {2026}),\ \href
  {\doibase 10.1016/j.physrep.2026.07.008} {\bibfield  {journal} {\bibinfo
  {journal} {Phys. Rep.}\ }\textbf {\bibinfo {volume} {1199}},\ \bibinfo
  {pages} {1--50}}\BibitemShut {NoStop}%
\bibitem [{\citenamefont {Sahai}(2018)}]{Sahai2018}%
  \BibitemOpen
  \bibfield  {author} {\bibinfo {author} {\bibnamefont {Sahai}, \bibfnamefont
  {A~A}}} (\bibinfo {year} {2018}),\ \href
  {https://link.aps.org/doi/10.1103/PhysRevAccelBeams.21.081301} {\bibfield
  {journal} {\bibinfo  {journal} {Phys. Rev. Accel. Beams}\ }\textbf {\bibinfo
  {volume} {21}},\ \bibinfo {pages} {081301}}\BibitemShut {NoStop}%
\bibitem [{\citenamefont {San Miguel~Claveria}\ \emph
  {et~al.}(2026{\natexlab{a}})\citenamefont {San Miguel~Claveria},
  \citenamefont {Gremillet}, \citenamefont {Davoine}, \citenamefont {Labro},
  \citenamefont {Matheron}, \citenamefont {Tamburini}, \citenamefont {Fiuza},\
  and\ \citenamefont {Corde}}]{SanMiguelClaveria2026b}%
  \BibitemOpen
  \bibfield  {author} {\bibinfo {author} {\bibnamefont {San Miguel~Claveria},
  \bibfnamefont {P}}, \bibinfo {author} {\bibfnamefont {L.}~\bibnamefont
  {Gremillet}}, \bibinfo {author} {\bibfnamefont {X.}~\bibnamefont {Davoine}},
  \bibinfo {author} {\bibfnamefont {Q.}~\bibnamefont {Labro}}, \bibinfo
  {author} {\bibfnamefont {A.}~\bibnamefont {Matheron}}, \bibinfo {author}
  {\bibfnamefont {M.}~\bibnamefont {Tamburini}}, \bibinfo {author}
  {\bibfnamefont {F.}~\bibnamefont {Fiuza}}, \ and\ \bibinfo {author}
  {\bibfnamefont {S.}~\bibnamefont {Corde}}} (\bibinfo {year}
  {2026}{\natexlab{a}}),\ \href {https://arxiv.org/abs/2607.24234} {}\Eprint
  {http://arxiv.org/abs/2607.24234} {arXiv:2607.24234} \BibitemShut {NoStop}%
\bibitem [{\citenamefont {San Miguel~Claveria}\ \emph
  {et~al.}(2026{\natexlab{b}})\citenamefont {San Miguel~Claveria},
  \citenamefont {Labro}, \citenamefont {Davoine}, \citenamefont {Matheron},
  \citenamefont {Tamburini}, \citenamefont {Corde}, \citenamefont {Gremillet},\
  and\ \citenamefont {Fiuza}}]{SanMiguelClaveria2026a}%
  \BibitemOpen
  \bibfield  {author} {\bibinfo {author} {\bibnamefont {San Miguel~Claveria},
  \bibfnamefont {P}}, \bibinfo {author} {\bibfnamefont {Q.}~\bibnamefont
  {Labro}}, \bibinfo {author} {\bibfnamefont {X.}~\bibnamefont {Davoine}},
  \bibinfo {author} {\bibfnamefont {A.}~\bibnamefont {Matheron}}, \bibinfo
  {author} {\bibfnamefont {M.}~\bibnamefont {Tamburini}}, \bibinfo {author}
  {\bibfnamefont {S.}~\bibnamefont {Corde}}, \bibinfo {author} {\bibfnamefont
  {L.}~\bibnamefont {Gremillet}}, \ and\ \bibinfo {author} {\bibfnamefont
  {F.}~\bibnamefont {Fiuza}}} (\bibinfo {year} {2026}{\natexlab{b}}),\ \href
  {https://arxiv.org/abs/2607.24214} {}\Eprint
  {http://arxiv.org/abs/2607.24214} {arXiv:2607.24214} \BibitemShut {NoStop}%
\bibitem [{\citenamefont {San Miguel~Claveria}\ \emph
  {et~al.}(2019)\citenamefont {San Miguel~Claveria} \emph
  {et~al.}}]{SanMiguelClaveria2019}%
  \BibitemOpen
  \bibfield  {author} {\bibinfo {author} {\bibnamefont {San Miguel~Claveria},
  \bibfnamefont {P}},  \emph {et~al.}} (\bibinfo {year} {2019}),\ \href
  {https://doi.org/10.1098/rsta.2018.0173} {\bibfield  {journal} {\bibinfo
  {journal} {Philos. Trans. R. Soc. A}\ }\textbf {\bibinfo {volume} {377}},\
  \bibinfo {pages} {20180173}}\BibitemShut {NoStop}%
\bibitem [{\citenamefont {San Miguel~Claveria}\ \emph
  {et~al.}(2022)\citenamefont {San Miguel~Claveria} \emph
  {et~al.}}]{SanMiguelClaveria2022}%
  \BibitemOpen
  \bibfield  {author} {\bibinfo {author} {\bibnamefont {San Miguel~Claveria},
  \bibfnamefont {P}},  \emph {et~al.}} (\bibinfo {year} {2022}),\ \href
  {https://link.aps.org/doi/10.1103/PhysRevResearch.4.023085} {\bibfield
  {journal} {\bibinfo  {journal} {Phys. Rev. Res.}\ }\textbf {\bibinfo {volume}
  {4}},\ \bibinfo {pages} {023085}}\BibitemShut {NoStop}%
\bibitem [{\citenamefont {Sands}(1991)}]{Sands1991}%
  \BibitemOpen
  \bibfield  {author} {\bibinfo {author} {\bibnamefont {Sands}, \bibfnamefont
  {M}}} (\bibinfo {year} {1991}),\ \href@noop {} {}\bibinfo {type} {Tech.
  Rep.}\ \bibinfo {number} {SLAC-AP-85}\ (\bibinfo  {institution} {SLAC},\
  \bibinfo {address} {Menlo Park, CA, United States})\BibitemShut {NoStop}%
\bibitem [{\citenamefont {S\"avert}\ \emph {et~al.}(2015)\citenamefont
  {S\"avert} \emph {et~al.}}]{Savert2015}%
  \BibitemOpen
  \bibfield  {author} {\bibinfo {author} {\bibnamefont {S\"avert},
  \bibfnamefont {A}},  \emph {et~al.}} (\bibinfo {year} {2015}),\ \href
  {https://doi.org/10.1103/PhysRevLett.115.055002} {\bibfield  {journal}
  {\bibinfo  {journal} {Phys. Rev. Lett.}\ }\textbf {\bibinfo {volume} {115}},\
  \bibinfo {pages} {055002}}\BibitemShut {NoStop}%
\bibitem [{\citenamefont {Scherkl}\ \emph {et~al.}(2022)\citenamefont {Scherkl}
  \emph {et~al.}}]{Scherkl2022}%
  \BibitemOpen
  \bibfield  {author} {\bibinfo {author} {\bibnamefont {Scherkl}, \bibfnamefont
  {P}},  \emph {et~al.}} (\bibinfo {year} {2022}),\ \href
  {https://doi.org/10.1103/PhysRevAccelBeams.25.052803} {\bibfield  {journal}
  {\bibinfo  {journal} {Phys. Rev. Accel. Beams}\ }\textbf {\bibinfo {volume}
  {25}},\ \bibinfo {pages} {052803}}\BibitemShut {NoStop}%
\bibitem [{\citenamefont {Schmid}\ \emph {et~al.}(2010)\citenamefont {Schmid},
  \citenamefont {Buck}, \citenamefont {Sears}, \citenamefont {Mikhailova},
  \citenamefont {Tautz}, \citenamefont {Herrmann}, \citenamefont {Geissler},
  \citenamefont {Krausz},\ and\ \citenamefont {Veisz}}]{Schmid2010}%
  \BibitemOpen
  \bibfield  {author} {\bibinfo {author} {\bibnamefont {Schmid}, \bibfnamefont
  {K}}, \bibinfo {author} {\bibfnamefont {A.}~\bibnamefont {Buck}}, \bibinfo
  {author} {\bibfnamefont {C.~M.~S.}\ \bibnamefont {Sears}}, \bibinfo {author}
  {\bibfnamefont {J.~M.}\ \bibnamefont {Mikhailova}}, \bibinfo {author}
  {\bibfnamefont {R.}~\bibnamefont {Tautz}}, \bibinfo {author} {\bibfnamefont
  {D.}~\bibnamefont {Herrmann}}, \bibinfo {author} {\bibfnamefont
  {M.}~\bibnamefont {Geissler}}, \bibinfo {author} {\bibfnamefont
  {F.}~\bibnamefont {Krausz}}, \ and\ \bibinfo {author} {\bibfnamefont
  {L.}~\bibnamefont {Veisz}}} (\bibinfo {year} {2010}),\ \href
  {https://doi.org/10.1103/PhysRevSTAB.13.091301} {\bibfield  {journal}
  {\bibinfo  {journal} {Phys. Rev. ST Accel. Beams}\ }\textbf {\bibinfo
  {volume} {13}},\ \bibinfo {pages} {091301}}\BibitemShut {NoStop}%
\bibitem [{\citenamefont {Schr\"{o}der}\ \emph {et~al.}(2025)\citenamefont
  {Schr\"{o}der}, \citenamefont {Osterhoff}, \citenamefont {Wesch},\ and\
  \citenamefont {Schroeder}}]{Schroeder2025}%
  \BibitemOpen
  \bibfield  {author} {\bibinfo {author} {\bibnamefont {Schr\"{o}der},
  \bibfnamefont {S}}, \bibinfo {author} {\bibfnamefont {J.}~\bibnamefont
  {Osterhoff}}, \bibinfo {author} {\bibfnamefont {S.}~\bibnamefont {Wesch}}, \
  and\ \bibinfo {author} {\bibfnamefont {C.~B.}\ \bibnamefont {Schroeder}}}
  (\bibinfo {year} {2025}),\ \href {\doibase 10.1088/1742-6596/3124/1/012018}
  {\bibfield  {journal} {\bibinfo  {journal} {J. Phys.: Conf. Ser.}\ }\textbf
  {\bibinfo {volume} {3124}},\ \bibinfo {pages} {012018}}\BibitemShut {NoStop}%
\bibitem [{\citenamefont {Schr\"{o}der}\ \emph
  {et~al.}(2020{\natexlab{a}})\citenamefont {Schr\"{o}der} \emph
  {et~al.}}]{Schroeder2020b}%
  \BibitemOpen
  \bibfield  {author} {\bibinfo {author} {\bibnamefont {Schr\"{o}der},
  \bibfnamefont {S}},  \emph {et~al.}} (\bibinfo {year} {2020}{\natexlab{a}}),\
  \href {https://doi.org/10.1088/1742-6596/1596/1/012002} {\bibfield  {journal}
  {\bibinfo  {journal} {J. Phys.: Conf. Ser.}\ }\textbf {\bibinfo {volume}
  {1596}},\ \bibinfo {pages} {012002}}\BibitemShut {NoStop}%
\bibitem [{\citenamefont {Schr\"{o}der}\ \emph
  {et~al.}(2020{\natexlab{b}})\citenamefont {Schr\"{o}der} \emph
  {et~al.}}]{Schroeder2020}%
  \BibitemOpen
  \bibfield  {author} {\bibinfo {author} {\bibnamefont {Schr\"{o}der},
  \bibfnamefont {S}},  \emph {et~al.}} (\bibinfo {year} {2020}{\natexlab{b}}),\
  \href {https://doi.org/10.1038/s41467-020-19811-9} {\bibfield  {journal}
  {\bibinfo  {journal} {Nat. Commun.}\ }\textbf {\bibinfo {volume} {11}},\
  \bibinfo {pages} {5984}}\BibitemShut {NoStop}%
\bibitem [{\citenamefont {Schroeder}\ \emph {et~al.}(2011)\citenamefont
  {Schroeder}, \citenamefont {Benedetti}, \citenamefont {Esarey}, \citenamefont
  {Gr\"{u}ner},\ and\ \citenamefont {Leemans}}]{Schroeder2011}%
  \BibitemOpen
  \bibfield  {author} {\bibinfo {author} {\bibnamefont {Schroeder},
  \bibfnamefont {C~B}}, \bibinfo {author} {\bibfnamefont {C.}~\bibnamefont
  {Benedetti}}, \bibinfo {author} {\bibfnamefont {E.}~\bibnamefont {Esarey}},
  \bibinfo {author} {\bibfnamefont {F.~J.}\ \bibnamefont {Gr\"{u}ner}}, \ and\
  \bibinfo {author} {\bibfnamefont {W.~P.}\ \bibnamefont {Leemans}}} (\bibinfo
  {year} {2011}),\ \href {https://doi.org/10.1103/physrevlett.107.145002}
  {\bibfield  {journal} {\bibinfo  {journal} {Phys. Rev. Lett.}\ }\textbf
  {\bibinfo {volume} {107}},\ \bibinfo {pages} {145002}}\BibitemShut {NoStop}%
\bibitem [{\citenamefont {Schroeder}\ \emph {et~al.}(2012)\citenamefont
  {Schroeder}, \citenamefont {Benedetti}, \citenamefont {Esarey}, \citenamefont
  {Gr\"{u}ner},\ and\ \citenamefont {Leemans}}]{Schroeder2012}%
  \BibitemOpen
  \bibfield  {author} {\bibinfo {author} {\bibnamefont {Schroeder},
  \bibfnamefont {C~B}}, \bibinfo {author} {\bibfnamefont {C.}~\bibnamefont
  {Benedetti}}, \bibinfo {author} {\bibfnamefont {E.}~\bibnamefont {Esarey}},
  \bibinfo {author} {\bibfnamefont {F.~J.}\ \bibnamefont {Gr\"{u}ner}}, \ and\
  \bibinfo {author} {\bibfnamefont {W.~P.}\ \bibnamefont {Leemans}}} (\bibinfo
  {year} {2012}),\ \href {https://doi.org/10.1063/1.3677358} {\bibfield
  {journal} {\bibinfo  {journal} {Phys. Plasmas}\ }\textbf {\bibinfo {volume}
  {19}},\ \bibinfo {pages} {010703}}\BibitemShut {NoStop}%
\bibitem [{\citenamefont {Schroeder}\ \emph
  {et~al.}(2013{\natexlab{a}})\citenamefont {Schroeder}, \citenamefont
  {Benedetti}, \citenamefont {Esarey}, \citenamefont {Gr\"{u}ner},\ and\
  \citenamefont {Leemans}}]{Schroeder2013}%
  \BibitemOpen
  \bibfield  {author} {\bibinfo {author} {\bibnamefont {Schroeder},
  \bibfnamefont {C~B}}, \bibinfo {author} {\bibfnamefont {C.}~\bibnamefont
  {Benedetti}}, \bibinfo {author} {\bibfnamefont {E.}~\bibnamefont {Esarey}},
  \bibinfo {author} {\bibfnamefont {F.~J.}\ \bibnamefont {Gr\"{u}ner}}, \ and\
  \bibinfo {author} {\bibfnamefont {W.~P.}\ \bibnamefont {Leemans}}} (\bibinfo
  {year} {2013}{\natexlab{a}}),\ \href {https://doi.org/10.1063/1.4803073}
  {\bibfield  {journal} {\bibinfo  {journal} {Phys. Plasmas}\ }\textbf
  {\bibinfo {volume} {20}},\ \bibinfo {pages} {056704}}\BibitemShut {NoStop}%
\bibitem [{\citenamefont {Schroeder}\ \emph
  {et~al.}(2013{\natexlab{b}})\citenamefont {Schroeder}, \citenamefont
  {Esarey}, \citenamefont {Benedetti},\ and\ \citenamefont
  {Leemans}}]{Schroeder2013a}%
  \BibitemOpen
  \bibfield  {author} {\bibinfo {author} {\bibnamefont {Schroeder},
  \bibfnamefont {C~B}}, \bibinfo {author} {\bibfnamefont {E.}~\bibnamefont
  {Esarey}}, \bibinfo {author} {\bibfnamefont {C.}~\bibnamefont {Benedetti}}, \
  and\ \bibinfo {author} {\bibfnamefont {W.~P.}\ \bibnamefont {Leemans}}}
  (\bibinfo {year} {2013}{\natexlab{b}}),\ \href
  {https://doi.org/10.1063/1.4817799} {\bibfield  {journal} {\bibinfo
  {journal} {Phys. Plasmas}\ }\textbf {\bibinfo {volume} {20}},\ \bibinfo
  {pages} {080701}}\BibitemShut {NoStop}%
\bibitem [{\citenamefont {Schroeder}\ \emph {et~al.}(2010)\citenamefont
  {Schroeder}, \citenamefont {Esarey}, \citenamefont {Geddes}, \citenamefont
  {Benedetti},\ and\ \citenamefont {Leemans}}]{Schroeder2010}%
  \BibitemOpen
  \bibfield  {author} {\bibinfo {author} {\bibnamefont {Schroeder},
  \bibfnamefont {C~B}}, \bibinfo {author} {\bibfnamefont {E.}~\bibnamefont
  {Esarey}}, \bibinfo {author} {\bibfnamefont {C.~G.~R.}\ \bibnamefont
  {Geddes}}, \bibinfo {author} {\bibfnamefont {C.}~\bibnamefont {Benedetti}}, \
  and\ \bibinfo {author} {\bibfnamefont {W.~P.}\ \bibnamefont {Leemans}}}
  (\bibinfo {year} {2010}),\ \href
  {https://doi.org/10.1103/physrevstab.13.101301} {\bibfield  {journal}
  {\bibinfo  {journal} {Phys. Rev. ST Accel. Beams}\ }\textbf {\bibinfo
  {volume} {13}},\ \bibinfo {pages} {101301}}\BibitemShut {NoStop}%
\bibitem [{\citenamefont {Schroeder}\ \emph {et~al.}(1999)\citenamefont
  {Schroeder}, \citenamefont {Whittum},\ and\ \citenamefont
  {Wurtele}}]{Schroeder1999}%
  \BibitemOpen
  \bibfield  {author} {\bibinfo {author} {\bibnamefont {Schroeder},
  \bibfnamefont {C~B}}, \bibinfo {author} {\bibfnamefont {D.~H.}\ \bibnamefont
  {Whittum}}, \ and\ \bibinfo {author} {\bibfnamefont {J.~S.}\ \bibnamefont
  {Wurtele}}} (\bibinfo {year} {1999}),\ \href
  {https://doi.org/10.1103/physrevlett.82.1177} {\bibfield  {journal} {\bibinfo
   {journal} {Phys. Rev. Lett.}\ }\textbf {\bibinfo {volume} {82}},\ \bibinfo
  {pages} {1177--1180}}\BibitemShut {NoStop}%
\bibitem [{\citenamefont {Schroeder}\ and\ \citenamefont
  {Wurtele}(2001)}]{Schroeder2001}%
  \BibitemOpen
  \bibfield  {author} {\bibinfo {author} {\bibnamefont {Schroeder},
  \bibfnamefont {C~B}}, \ and\ \bibinfo {author} {\bibfnamefont {J.~S.}\
  \bibnamefont {Wurtele}}} (\bibinfo {year} {2001}),\ in\ \href
  {https://doi.org/10.1063/1.1384389} {\emph {\bibinfo {booktitle} {{AIP} Conf.
  Proc.}}},\ Vol.\ \bibinfo {volume} {569}\ (\bibinfo  {publisher} {{AIP}})\
  pp.\ \bibinfo {pages} {616--629}\BibitemShut {NoStop}%
\bibitem [{\citenamefont {Schulte}(2016)}]{Schulte2016}%
  \BibitemOpen
  \bibfield  {author} {\bibinfo {author} {\bibnamefont {Schulte}, \bibfnamefont
  {D}}} (\bibinfo {year} {2016}),\ \href
  {https://doi.org/10.1142/s1793626816300103} {\bibfield  {journal} {\bibinfo
  {journal} {Rev. Accel. Sci. Tech.}\ }\textbf {\bibinfo {volume} {9}},\
  \bibinfo {pages} {209--233}}\BibitemShut {NoStop}%
\bibitem [{\citenamefont {Schwab}\ \emph {et~al.}(2013)\citenamefont {Schwab}
  \emph {et~al.}}]{Schwab2013}%
  \BibitemOpen
  \bibfield  {author} {\bibinfo {author} {\bibnamefont {Schwab}, \bibfnamefont
  {M~B}},  \emph {et~al.}} (\bibinfo {year} {2013}),\ \href
  {https://doi.org/10.1063/1.4829489} {\bibfield  {journal} {\bibinfo
  {journal} {Appl. Phys. Lett.}\ }\textbf {\bibinfo {volume} {103}},\ \bibinfo
  {pages} {191118}}\BibitemShut {NoStop}%
\bibitem [{\citenamefont {Schöbel}\ \emph {et~al.}(2022)\citenamefont
  {Schöbel} \emph {et~al.}}]{Schoebel2022}%
  \BibitemOpen
  \bibfield  {author} {\bibinfo {author} {\bibnamefont {Schöbel},
  \bibfnamefont {S}},  \emph {et~al.}} (\bibinfo {year} {2022}),\ \href
  {https://doi.org/10.1088/1367-2630/ac87c9} {\bibfield  {journal} {\bibinfo
  {journal} {New J. Phys.}\ }\textbf {\bibinfo {volume} {24}},\ \bibinfo
  {pages} {083034}}\BibitemShut {NoStop}%
\bibitem [{\citenamefont {Seryi}(2015)}]{Seryi2015}%
  \BibitemOpen
  \bibfield  {author} {\bibinfo {author} {\bibnamefont {Seryi}, \bibfnamefont
  {A}}} (\bibinfo {year} {2015}),\ \href {https://doi.org/10.1201/b18696}
  {\emph {\bibinfo {title} {Unifying Physics of Accelerators, Lasers and
  Plasma}}}\ (\bibinfo  {publisher} {CRC Press},\ \bibinfo {address} {Boca
  Raton, FL, United States})\BibitemShut {NoStop}%
\bibitem [{\citenamefont {Seryi}\ \emph {et~al.}(2009)\citenamefont {Seryi},
  \citenamefont {Hogan}, \citenamefont {Pei}, \citenamefont {Raubenheimer},
  \citenamefont {Tenenbaum}, \citenamefont {Katsouleas}, \citenamefont {Huang},
  \citenamefont {Joshi}, \citenamefont {Mori},\ and\ \citenamefont
  {Muggli}}]{Seryi2009}%
  \BibitemOpen
  \bibfield  {author} {\bibinfo {author} {\bibnamefont {Seryi}, \bibfnamefont
  {A}}, \bibinfo {author} {\bibfnamefont {M.}~\bibnamefont {Hogan}}, \bibinfo
  {author} {\bibfnamefont {S.}~\bibnamefont {Pei}}, \bibinfo {author}
  {\bibfnamefont {T.}~\bibnamefont {Raubenheimer}}, \bibinfo {author}
  {\bibfnamefont {P.}~\bibnamefont {Tenenbaum}}, \bibinfo {author}
  {\bibfnamefont {T.}~\bibnamefont {Katsouleas}}, \bibinfo {author}
  {\bibfnamefont {C.}~\bibnamefont {Huang}}, \bibinfo {author} {\bibfnamefont
  {C.}~\bibnamefont {Joshi}}, \bibinfo {author} {\bibfnamefont
  {W.}~\bibnamefont {Mori}}, \ and\ \bibinfo {author} {\bibfnamefont
  {P.}~\bibnamefont {Muggli}}} (\bibinfo {year} {2009}),\ in\ \href
  {https://accelconf.web.cern.ch/PAC2009/papers/we6pfp081.pdf} {\emph {\bibinfo
  {booktitle} {Proceedings of the 2009 Particle Accelerator Conf.}}},\ pp.\
  \bibinfo {pages} {2688--2690}\BibitemShut {NoStop}%
\bibitem [{\citenamefont {Shalloo}\ \emph {et~al.}(2018)\citenamefont
  {Shalloo}, \citenamefont {Arran}, \citenamefont {Corner}, \citenamefont
  {Holloway}, \citenamefont {Jonnerby}, \citenamefont {Walczak}, \citenamefont
  {Milchberg},\ and\ \citenamefont {Hooker}}]{Shalloo2018}%
  \BibitemOpen
  \bibfield  {author} {\bibinfo {author} {\bibnamefont {Shalloo}, \bibfnamefont
  {R~J}}, \bibinfo {author} {\bibfnamefont {C.}~\bibnamefont {Arran}}, \bibinfo
  {author} {\bibfnamefont {L.}~\bibnamefont {Corner}}, \bibinfo {author}
  {\bibfnamefont {J.}~\bibnamefont {Holloway}}, \bibinfo {author}
  {\bibfnamefont {J.}~\bibnamefont {Jonnerby}}, \bibinfo {author}
  {\bibfnamefont {R.}~\bibnamefont {Walczak}}, \bibinfo {author} {\bibfnamefont
  {H.~M.}\ \bibnamefont {Milchberg}}, \ and\ \bibinfo {author} {\bibfnamefont
  {S.~M.}\ \bibnamefont {Hooker}}} (\bibinfo {year} {2018}),\ \href
  {https://doi.org/10.1103/physreve.97.053203} {\bibfield  {journal} {\bibinfo
  {journal} {Phys. Rev. E}\ }\textbf {\bibinfo {volume} {97}},\ \bibinfo
  {pages} {053203}}\BibitemShut {NoStop}%
\bibitem [{\citenamefont {Shiltsev}(2011)}]{Shiltsev2011}%
  \BibitemOpen
  \bibfield  {author} {\bibinfo {author} {\bibnamefont {Shiltsev},
  \bibfnamefont {V}}} (\bibinfo {year} {2011}),\ \href {\doibase
  10.1142/s0217732311035699} {\bibfield  {journal} {\bibinfo  {journal} {Mod.
  Phys. Lett. A}\ }\textbf {\bibinfo {volume} {26}},\ \bibinfo {pages}
  {761--772}}\BibitemShut {NoStop}%
\bibitem [{\citenamefont {Shiltsev}(2020)}]{Shiltsev2020}%
  \BibitemOpen
  \bibfield  {author} {\bibinfo {author} {\bibnamefont {Shiltsev},
  \bibfnamefont {V}}} (\bibinfo {year} {2020}),\ \href {\doibase
  10.1063/pt.3.4452} {\bibfield  {journal} {\bibinfo  {journal} {Phys. Today}\
  }\textbf {\bibinfo {volume} {73}},\ \bibinfo {pages} {32--39}}\BibitemShut
  {NoStop}%
\bibitem [{\citenamefont {Shiltsev}(2022)}]{Shiltsev2022}%
  \BibitemOpen
  \bibfield  {author} {\bibinfo {author} {\bibnamefont {Shiltsev},
  \bibfnamefont {V}}} (\bibinfo {year} {2022}),\ \href {\doibase
  10.1088/1748-0221/17/05/t05010} {\bibfield  {journal} {\bibinfo  {journal}
  {J. Instrum.}\ }\textbf {\bibinfo {volume} {17}},\ \bibinfo {pages}
  {T05010}}\BibitemShut {NoStop}%
\bibitem [{\citenamefont {Shiltsev}\ and\ \citenamefont
  {Zimmermann}(2021)}]{Shiltsev2021}%
  \BibitemOpen
  \bibfield  {author} {\bibinfo {author} {\bibnamefont {Shiltsev},
  \bibfnamefont {V}}, \ and\ \bibinfo {author} {\bibfnamefont {F.}~\bibnamefont
  {Zimmermann}}} (\bibinfo {year} {2021}),\ \href
  {https://link.aps.org/doi/10.1103/RevModPhys.93.015006} {\bibfield  {journal}
  {\bibinfo  {journal} {Rev. Mod. Phys.}\ }\textbf {\bibinfo {volume} {93}},\
  \bibinfo {pages} {015006}}\BibitemShut {NoStop}%
\bibitem [{\citenamefont {Shpakov}\ \emph {et~al.}(2019)\citenamefont {Shpakov}
  \emph {et~al.}}]{Shpakov2019}%
  \BibitemOpen
  \bibfield  {author} {\bibinfo {author} {\bibnamefont {Shpakov}, \bibfnamefont
  {V}},  \emph {et~al.}} (\bibinfo {year} {2019}),\ \href
  {https://doi.org/10.1103/PhysRevLett.122.114801} {\bibfield  {journal}
  {\bibinfo  {journal} {Phys. Rev. Lett.}\ }\textbf {\bibinfo {volume} {122}},\
  \bibinfo {pages} {114801}}\BibitemShut {NoStop}%
\bibitem [{\citenamefont {Shpakov}\ \emph {et~al.}(2021)\citenamefont {Shpakov}
  \emph {et~al.}}]{Shpakov2021}%
  \BibitemOpen
  \bibfield  {author} {\bibinfo {author} {\bibnamefont {Shpakov}, \bibfnamefont
  {V}},  \emph {et~al.}} (\bibinfo {year} {2021}),\ \href
  {https://doi.org/10.1103/physrevaccelbeams.24.051301} {\bibfield  {journal}
  {\bibinfo  {journal} {Phys. Rev. Accel. Beams}\ }\textbf {\bibinfo {volume}
  {24}},\ \bibinfo {pages} {051301}}\BibitemShut {NoStop}%
\bibitem [{\citenamefont {Shvets}\ \emph {et~al.}(1996)\citenamefont {Shvets},
  \citenamefont {Wurtele}, \citenamefont {Chiou},\ and\ \citenamefont
  {Katsouleas}}]{Shvets1996}%
  \BibitemOpen
  \bibfield  {author} {\bibinfo {author} {\bibnamefont {Shvets}, \bibfnamefont
  {G}}, \bibinfo {author} {\bibfnamefont {J.~S.}\ \bibnamefont {Wurtele}},
  \bibinfo {author} {\bibfnamefont {T.~C.}\ \bibnamefont {Chiou}}, \ and\
  \bibinfo {author} {\bibfnamefont {T.~C.}\ \bibnamefont {Katsouleas}}}
  (\bibinfo {year} {1996}),\ \href {https://doi.org/10.1109/27.509999}
  {\bibfield  {journal} {\bibinfo  {journal} {IEEE T. Plasma Sci.}\ }\textbf
  {\bibinfo {volume} {24}},\ \bibinfo {pages} {351--362}}\BibitemShut {NoStop}%
\bibitem [{\citenamefont {{Silva}}\ \emph {et~al.}(2021)\citenamefont
  {{Silva}}, \citenamefont {{Amorim}}, \citenamefont {{Downer}}, \citenamefont
  {{Hogan}}, \citenamefont {{Yakimenko}}, \citenamefont {{Zgadzaj}},\ and\
  \citenamefont {{Vieira}}}]{Silva2021}%
  \BibitemOpen
  \bibfield  {author} {\bibinfo {author} {\bibnamefont {{Silva}}, \bibfnamefont
  {T}}, \bibinfo {author} {\bibfnamefont {L.~D.}\ \bibnamefont {{Amorim}}},
  \bibinfo {author} {\bibfnamefont {M.~C.}\ \bibnamefont {{Downer}}}, \bibinfo
  {author} {\bibfnamefont {M.~J.}\ \bibnamefont {{Hogan}}}, \bibinfo {author}
  {\bibfnamefont {V.}~\bibnamefont {{Yakimenko}}}, \bibinfo {author}
  {\bibfnamefont {R.}~\bibnamefont {{Zgadzaj}}}, \ and\ \bibinfo {author}
  {\bibfnamefont {J.}~\bibnamefont {{Vieira}}}} (\bibinfo {year} {2021}),\
  \href {https://doi.org/10.1103/PhysRevLett.127.104801} {\bibfield  {journal}
  {\bibinfo  {journal} {Phys. Rev. Lett.}\ }\textbf {\bibinfo {volume} {127}},\
  \bibinfo {eid} {104801}}\BibitemShut {NoStop}%
\bibitem [{\citenamefont {Silva}\ and\ \citenamefont
  {Vieira}(2023)}]{Silva2023}%
  \BibitemOpen
  \bibfield  {author} {\bibinfo {author} {\bibnamefont {Silva}, \bibfnamefont
  {T}}, \ and\ \bibinfo {author} {\bibfnamefont {J.}~\bibnamefont {Vieira}}}
  (\bibinfo {year} {2023}),\ \href
  {https://doi.org/10.1103/PhysRevAccelBeams.26.091301} {\bibfield  {journal}
  {\bibinfo  {journal} {Phys. Rev. Accel. Beams}\ }\textbf {\bibinfo {volume}
  {26}},\ \bibinfo {pages} {091301}}\BibitemShut {NoStop}%
\bibitem [{\citenamefont {Sokolov}\ and\ \citenamefont
  {Ternov}(1967)}]{Sokolov1967}%
  \BibitemOpen
  \bibfield  {author} {\bibinfo {author} {\bibnamefont {Sokolov}, \bibfnamefont
  {A~A}}, \ and\ \bibinfo {author} {\bibfnamefont {I.~M.}\ \bibnamefont
  {Ternov}}} (\bibinfo {year} {1967}),\ \href
  {https://doi.org/10.1007/BF00820300} {\bibfield  {journal} {\bibinfo
  {journal} {Sov. Phys. J.}\ }\textbf {\bibinfo {volume} {10}},\ \bibinfo
  {pages} {39--47}}\BibitemShut {NoStop}%
\bibitem [{\citenamefont {Sprangle}\ \emph {et~al.}(1990)\citenamefont
  {Sprangle}, \citenamefont {Esarey},\ and\ \citenamefont
  {Ting}}]{Sprangle1990}%
  \BibitemOpen
  \bibfield  {author} {\bibinfo {author} {\bibnamefont {Sprangle},
  \bibfnamefont {P}}, \bibinfo {author} {\bibfnamefont {E.}~\bibnamefont
  {Esarey}}, \ and\ \bibinfo {author} {\bibfnamefont {A.}~\bibnamefont {Ting}}}
  (\bibinfo {year} {1990}),\ \href
  {https://doi.org/10.1103/PhysRevLett.64.2011} {\bibfield  {journal} {\bibinfo
   {journal} {Phys. Rev. Lett.}\ }\textbf {\bibinfo {volume} {64}},\ \bibinfo
  {pages} {2011--2014}}\BibitemShut {NoStop}%
\bibitem [{\citenamefont {Steinke}\ \emph {et~al.}(2016)\citenamefont {Steinke}
  \emph {et~al.}}]{Steinke2016}%
  \BibitemOpen
  \bibfield  {author} {\bibinfo {author} {\bibnamefont {Steinke}, \bibfnamefont
  {S}},  \emph {et~al.}} (\bibinfo {year} {2016}),\ \href
  {https://doi.org/10.1038/nature16525} {\bibfield  {journal} {\bibinfo
  {journal} {Nature (London)}\ }\textbf {\bibinfo {volume} {530}},\ \bibinfo
  {pages} {190--193}}\BibitemShut {NoStop}%
\bibitem [{\citenamefont {Storey}\ \emph {et~al.}(2024)\citenamefont {Storey}
  \emph {et~al.}}]{Storey2024}%
  \BibitemOpen
  \bibfield  {author} {\bibinfo {author} {\bibnamefont {Storey}, \bibfnamefont
  {D}},  \emph {et~al.}} (\bibinfo {year} {2024}),\ \href
  {https://doi.org/10.1103/PhysRevAccelBeams.27.051302} {\bibfield  {journal}
  {\bibinfo  {journal} {Phys. Rev. Accel. Beams}\ }\textbf {\bibinfo {volume}
  {27}},\ \bibinfo {pages} {051302}}\BibitemShut {NoStop}%
\bibitem [{\citenamefont {Stupakov}(2018)}]{Stupakov2018}%
  \BibitemOpen
  \bibfield  {author} {\bibinfo {author} {\bibnamefont {Stupakov},
  \bibfnamefont {G}}} (\bibinfo {year} {2018}),\ \href
  {https://doi.org/10.1103/PhysRevAccelBeams.21.041301} {\bibfield  {journal}
  {\bibinfo  {journal} {Phys. Rev. Accel. Beams}\ }\textbf {\bibinfo {volume}
  {21}},\ \bibinfo {pages} {041301}}\BibitemShut {NoStop}%
\bibitem [{\citenamefont {Stupakov}\ \emph {et~al.}(2016)\citenamefont
  {Stupakov}, \citenamefont {Breizman}, \citenamefont {Khudik},\ and\
  \citenamefont {Shvets}}]{Stupakov2016}%
  \BibitemOpen
  \bibfield  {author} {\bibinfo {author} {\bibnamefont {Stupakov},
  \bibfnamefont {G}}, \bibinfo {author} {\bibfnamefont {B.}~\bibnamefont
  {Breizman}}, \bibinfo {author} {\bibfnamefont {V.}~\bibnamefont {Khudik}}, \
  and\ \bibinfo {author} {\bibfnamefont {G.}~\bibnamefont {Shvets}}} (\bibinfo
  {year} {2016}),\ \href {https://doi.org/10.1103/PhysRevAccelBeams.19.101302}
  {\bibfield  {journal} {\bibinfo  {journal} {Phys. Rev. Accel. Beams}\
  }\textbf {\bibinfo {volume} {19}},\ \bibinfo {pages} {101302}}\BibitemShut
  {NoStop}%
\bibitem [{\citenamefont {Su}\ \emph {et~al.}(1987)\citenamefont {Su},
  \citenamefont {Katsouleas}, \citenamefont {Dawson}, \citenamefont {Chen},
  \citenamefont {Jones},\ and\ \citenamefont {Keinigs}}]{Su1987}%
  \BibitemOpen
  \bibfield  {author} {\bibinfo {author} {\bibnamefont {Su}, \bibfnamefont
  {J~J}}, \bibinfo {author} {\bibfnamefont {T.}~\bibnamefont {Katsouleas}},
  \bibinfo {author} {\bibfnamefont {J.~M.}\ \bibnamefont {Dawson}}, \bibinfo
  {author} {\bibfnamefont {P.}~\bibnamefont {Chen}}, \bibinfo {author}
  {\bibfnamefont {M.}~\bibnamefont {Jones}}, \ and\ \bibinfo {author}
  {\bibfnamefont {R.}~\bibnamefont {Keinigs}}} (\bibinfo {year} {1987}),\ \href
  {https://doi.org/10.1109/TPS.1987.4316684} {\bibfield  {journal} {\bibinfo
  {journal} {IEEE T. Plasma Sci.}\ }\textbf {\bibinfo {volume} {15}},\ \bibinfo
  {pages} {192--198}}\BibitemShut {NoStop}%
\bibitem [{\citenamefont {Su}\ \emph {et~al.}(2023)\citenamefont {Su} \emph
  {et~al.}}]{Su2023}%
  \BibitemOpen
  \bibfield  {author} {\bibinfo {author} {\bibnamefont {Su}, \bibfnamefont
  {Q}},  \emph {et~al.}} (\bibinfo {year} {2023}),\ \href
  {https://doi.org/10.1063/5.0142940} {\bibfield  {journal} {\bibinfo
  {journal} {Phys. Plasmas}\ }\textbf {\bibinfo {volume} {30}},\ \bibinfo
  {pages} {053108}}\BibitemShut {NoStop}%
\bibitem [{\citenamefont {Sugimoto}\ \emph {et~al.}(2023)\citenamefont
  {Sugimoto}, \citenamefont {He}, \citenamefont {Iwata}, \citenamefont {Yeh},
  \citenamefont {Tangtartharakul}, \citenamefont {Arefiev},\ and\ \citenamefont
  {Sentoku}}]{Sugimoto2023}%
  \BibitemOpen
  \bibfield  {author} {\bibinfo {author} {\bibnamefont {Sugimoto},
  \bibfnamefont {K}}, \bibinfo {author} {\bibfnamefont {Y.}~\bibnamefont {He}},
  \bibinfo {author} {\bibfnamefont {N.}~\bibnamefont {Iwata}}, \bibinfo
  {author} {\bibfnamefont {I-L.}\ \bibnamefont {Yeh}}, \bibinfo {author}
  {\bibfnamefont {K.}~\bibnamefont {Tangtartharakul}}, \bibinfo {author}
  {\bibfnamefont {A.}~\bibnamefont {Arefiev}}, \ and\ \bibinfo {author}
  {\bibfnamefont {Y.}~\bibnamefont {Sentoku}}} (\bibinfo {year} {2023}),\ \href
  {https://link.aps.org/doi/10.1103/PhysRevLett.131.065102} {\bibfield
  {journal} {\bibinfo  {journal} {Phys. Rev. Lett.}\ }\textbf {\bibinfo
  {volume} {131}},\ \bibinfo {pages} {065102}}\BibitemShut {NoStop}%
\bibitem [{\citenamefont {Suk}\ \emph {et~al.}(2001)\citenamefont {Suk},
  \citenamefont {Barov}, \citenamefont {Rosenzweig},\ and\ \citenamefont
  {Esarey}}]{Suk2001}%
  \BibitemOpen
  \bibfield  {author} {\bibinfo {author} {\bibnamefont {Suk}, \bibfnamefont
  {H}}, \bibinfo {author} {\bibfnamefont {N.}~\bibnamefont {Barov}}, \bibinfo
  {author} {\bibfnamefont {J.~B.}\ \bibnamefont {Rosenzweig}}, \ and\ \bibinfo
  {author} {\bibfnamefont {E.}~\bibnamefont {Esarey}}} (\bibinfo {year}
  {2001}),\ \href {https://doi.org/10.1103/physrevlett.86.1011} {\bibfield
  {journal} {\bibinfo  {journal} {Phys. Rev. Lett.}\ }\textbf {\bibinfo
  {volume} {86}},\ \bibinfo {pages} {1011--1014}}\BibitemShut {NoStop}%
\bibitem [{\citenamefont {Tajima}\ and\ \citenamefont
  {Dawson}(1979)}]{Tajima1979}%
  \BibitemOpen
  \bibfield  {author} {\bibinfo {author} {\bibnamefont {Tajima}, \bibfnamefont
  {T}}, \ and\ \bibinfo {author} {\bibfnamefont {J.~M.}\ \bibnamefont
  {Dawson}}} (\bibinfo {year} {1979}),\ \href
  {https://doi.org/10.1103/physrevlett.43.267} {\bibfield  {journal} {\bibinfo
  {journal} {Phys. Rev. Lett.}\ }\textbf {\bibinfo {volume} {43}},\ \bibinfo
  {pages} {267--270}}\BibitemShut {NoStop}%
\bibitem [{\citenamefont {Tajima}\ \emph {et~al.}(1990)\citenamefont {Tajima}
  \emph {et~al.}}]{Tajima1990}%
  \BibitemOpen
  \bibfield  {author} {\bibinfo {author} {\bibnamefont {Tajima}, \bibfnamefont
  {T}},  \emph {et~al.}} (\bibinfo {year} {1990}),\ \href
  {https://cds.cern.ch/record/1108210/files/p235.pdf} {\bibfield  {journal}
  {\bibinfo  {journal} {Part. Accel.}\ }\textbf {\bibinfo {volume} {32}},\
  \bibinfo {pages} {235--240}}\BibitemShut {NoStop}%
\bibitem [{\citenamefont {Tang}\ \emph {et~al.}(2009)\citenamefont {Tang} \emph
  {et~al.}}]{Tang2009}%
  \BibitemOpen
  \bibfield  {author} {\bibinfo {author} {\bibnamefont {Tang}, \bibfnamefont
  {C}},  \emph {et~al.}} (\bibinfo {year} {2009}),\ \href {\doibase
  10.1016/j.nima.2009.05.088} {\bibfield  {journal} {\bibinfo  {journal} {Nucl.
  Instrum. Methods Phys. Res. A}\ }\textbf {\bibinfo {volume} {608}},\ \bibinfo
  {pages} {S70--S74}}\BibitemShut {NoStop}%
\bibitem [{\citenamefont {Terzani}\ \emph {et~al.}(2023)\citenamefont
  {Terzani}, \citenamefont {Benedetti}, \citenamefont {Bulanov}, \citenamefont
  {Schroeder},\ and\ \citenamefont {Esarey}}]{Terzani2023}%
  \BibitemOpen
  \bibfield  {author} {\bibinfo {author} {\bibnamefont {Terzani}, \bibfnamefont
  {D}}, \bibinfo {author} {\bibfnamefont {C.}~\bibnamefont {Benedetti}},
  \bibinfo {author} {\bibfnamefont {S.~S.}\ \bibnamefont {Bulanov}}, \bibinfo
  {author} {\bibfnamefont {C.~B.}\ \bibnamefont {Schroeder}}, \ and\ \bibinfo
  {author} {\bibfnamefont {E.}~\bibnamefont {Esarey}}} (\bibinfo {year}
  {2023}),\ \href {https://doi.org/10.1103/PhysRevAccelBeams.26.113401}
  {\bibfield  {journal} {\bibinfo  {journal} {Phys. Rev. Accel. Beams}\
  }\textbf {\bibinfo {volume} {26}},\ \bibinfo {pages} {113401}}\BibitemShut
  {NoStop}%
\bibitem [{\citenamefont {Th\'evenet}\ \emph {et~al.}(2019)\citenamefont
  {Th\'evenet}, \citenamefont {Lehe}, \citenamefont {Schroeder}, \citenamefont
  {Benedetti}, \citenamefont {Vay}, \citenamefont {Esarey},\ and\ \citenamefont
  {Leemans}}]{Thevenet2019}%
  \BibitemOpen
  \bibfield  {author} {\bibinfo {author} {\bibnamefont {Th\'evenet},
  \bibfnamefont {M}}, \bibinfo {author} {\bibfnamefont {R.}~\bibnamefont
  {Lehe}}, \bibinfo {author} {\bibfnamefont {C.~B.}\ \bibnamefont {Schroeder}},
  \bibinfo {author} {\bibfnamefont {C.}~\bibnamefont {Benedetti}}, \bibinfo
  {author} {\bibfnamefont {J.-L.}\ \bibnamefont {Vay}}, \bibinfo {author}
  {\bibfnamefont {E.}~\bibnamefont {Esarey}}, \ and\ \bibinfo {author}
  {\bibfnamefont {W.~P.}\ \bibnamefont {Leemans}}} (\bibinfo {year} {2019}),\
  \href {https://doi.org/10.1103/PhysRevAccelBeams.22.051302} {\bibfield
  {journal} {\bibinfo  {journal} {Phys. Rev. Accel. Beams}\ }\textbf {\bibinfo
  {volume} {22}},\ \bibinfo {pages} {051302}}\BibitemShut {NoStop}%
\bibitem [{\citenamefont {Thomas}\ and\ \citenamefont
  {Seipt}(2021)}]{Thomas2021}%
  \BibitemOpen
  \bibfield  {author} {\bibinfo {author} {\bibnamefont {Thomas}, \bibfnamefont
  {A~G~R}}, \ and\ \bibinfo {author} {\bibfnamefont {D.}~\bibnamefont {Seipt}}}
  (\bibinfo {year} {2021}),\ \href
  {https://doi.org/10.1103/physrevaccelbeams.24.104602} {\bibfield  {journal}
  {\bibinfo  {journal} {Phys. Rev. Accel. Beams}\ }\textbf {\bibinfo {volume}
  {24}},\ \bibinfo {pages} {104602}}\BibitemShut {NoStop}%
\bibitem [{\citenamefont {Thomas}\ \emph {et~al.}(2020)\citenamefont {Thomas},
  \citenamefont {H\"{u}tzen}, \citenamefont {Lehrach}, \citenamefont {Pukhov},
  \citenamefont {Ji}, \citenamefont {Wu}, \citenamefont {Geng},\ and\
  \citenamefont {B\"{u}scher}}]{Thomas2020}%
  \BibitemOpen
  \bibfield  {author} {\bibinfo {author} {\bibnamefont {Thomas}, \bibfnamefont
  {J}}, \bibinfo {author} {\bibfnamefont {A.}~\bibnamefont {H\"{u}tzen}},
  \bibinfo {author} {\bibfnamefont {A.}~\bibnamefont {Lehrach}}, \bibinfo
  {author} {\bibfnamefont {A.}~\bibnamefont {Pukhov}}, \bibinfo {author}
  {\bibfnamefont {L.}~\bibnamefont {Ji}}, \bibinfo {author} {\bibfnamefont
  {Y.}~\bibnamefont {Wu}}, \bibinfo {author} {\bibfnamefont {X.}~\bibnamefont
  {Geng}}, \ and\ \bibinfo {author} {\bibfnamefont {M.}~\bibnamefont
  {B\"{u}scher}}} (\bibinfo {year} {2020}),\ \href
  {https://doi.org/10.1103/physrevaccelbeams.23.064401} {\bibfield  {journal}
  {\bibinfo  {journal} {Phys. Rev. Accel. Beams}\ }\textbf {\bibinfo {volume}
  {23}},\ \bibinfo {pages} {064401}}\BibitemShut {NoStop}%
\bibitem [{\citenamefont {Thomas}(1927)}]{Thomas1927}%
  \BibitemOpen
  \bibfield  {author} {\bibinfo {author} {\bibnamefont {Thomas}, \bibfnamefont
  {L~H}}} (\bibinfo {year} {1927}),\ \href
  {https://doi.org/10.1080/14786440108564170} {\bibfield  {journal} {\bibinfo
  {journal} {Phil. Mag.}\ }\textbf {\bibinfo {volume} {3}},\ \bibinfo {pages}
  {1--22}}\BibitemShut {NoStop}%
\bibitem [{\citenamefont {Ting}\ \emph {et~al.}(1997)\citenamefont {Ting},
  \citenamefont {Moore}, \citenamefont {Krushelnick}, \citenamefont {Manka},
  \citenamefont {Esarey}, \citenamefont {Sprangle}, \citenamefont {Hubbard},
  \citenamefont {Burris}, \citenamefont {Fischer},\ and\ \citenamefont
  {Baine}}]{Ting1997}%
  \BibitemOpen
  \bibfield  {author} {\bibinfo {author} {\bibnamefont {Ting}, \bibfnamefont
  {A}}, \bibinfo {author} {\bibfnamefont {C.~I.}\ \bibnamefont {Moore}},
  \bibinfo {author} {\bibfnamefont {K.}~\bibnamefont {Krushelnick}}, \bibinfo
  {author} {\bibfnamefont {C.}~\bibnamefont {Manka}}, \bibinfo {author}
  {\bibfnamefont {E.}~\bibnamefont {Esarey}}, \bibinfo {author} {\bibfnamefont
  {P.}~\bibnamefont {Sprangle}}, \bibinfo {author} {\bibfnamefont
  {R.}~\bibnamefont {Hubbard}}, \bibinfo {author} {\bibfnamefont {H.~R.}\
  \bibnamefont {Burris}}, \bibinfo {author} {\bibfnamefont {R.}~\bibnamefont
  {Fischer}}, \ and\ \bibinfo {author} {\bibfnamefont {M.}~\bibnamefont
  {Baine}}} (\bibinfo {year} {1997}),\ \href {https://doi.org/10.1063/1.872332}
  {\bibfield  {journal} {\bibinfo  {journal} {Phys. Plasmas}\ }\textbf
  {\bibinfo {volume} {4}},\ \bibinfo {pages} {1889--1899}}\BibitemShut
  {NoStop}%
\bibitem [{\citenamefont {Torrado}\ \emph {et~al.}(2023)\citenamefont
  {Torrado}, \citenamefont {Lopes}, \citenamefont {Silva}, \citenamefont
  {Amoedo},\ and\ \citenamefont {Sublet}}]{Torrado2023}%
  \BibitemOpen
  \bibfield  {author} {\bibinfo {author} {\bibnamefont {Torrado}, \bibfnamefont
  {N~E}}, \bibinfo {author} {\bibfnamefont {N.~C.}\ \bibnamefont {Lopes}},
  \bibinfo {author} {\bibfnamefont {J.~F.~A.}\ \bibnamefont {Silva}}, \bibinfo
  {author} {\bibfnamefont {C.}~\bibnamefont {Amoedo}}, \ and\ \bibinfo {author}
  {\bibfnamefont {A.}~\bibnamefont {Sublet}}} (\bibinfo {year} {2023}),\ \href
  {\doibase 10.1109/tps.2023.3337314} {\bibfield  {journal} {\bibinfo
  {journal} {IEEE T. Plasma Sci.}\ }\textbf {\bibinfo {volume} {51}},\ \bibinfo
  {pages} {3619--3627}}\BibitemShut {NoStop}%
\bibitem [{\citenamefont {Tsung}\ \emph {et~al.}(2004)\citenamefont {Tsung},
  \citenamefont {Narang}, \citenamefont {Mori}, \citenamefont {Joshi},
  \citenamefont {Fonseca},\ and\ \citenamefont {Silva}}]{Tsung2004}%
  \BibitemOpen
  \bibfield  {author} {\bibinfo {author} {\bibnamefont {Tsung}, \bibfnamefont
  {F~S}}, \bibinfo {author} {\bibfnamefont {R.}~\bibnamefont {Narang}},
  \bibinfo {author} {\bibfnamefont {W.~B.}\ \bibnamefont {Mori}}, \bibinfo
  {author} {\bibfnamefont {C.}~\bibnamefont {Joshi}}, \bibinfo {author}
  {\bibfnamefont {R.~A.}\ \bibnamefont {Fonseca}}, \ and\ \bibinfo {author}
  {\bibfnamefont {L.~O.}\ \bibnamefont {Silva}}} (\bibinfo {year} {2004}),\
  \href {https://doi.org/10.1103/PhysRevLett.93.185002} {\bibfield  {journal}
  {\bibinfo  {journal} {Phys. Rev. Lett.}\ }\textbf {\bibinfo {volume} {93}},\
  \bibinfo {pages} {185002}}\BibitemShut {NoStop}%
\bibitem [{\citenamefont {Turner}\ \emph {et~al.}(2017)\citenamefont {Turner},
  \citenamefont {Biskup}, \citenamefont {Burger}, \citenamefont {Gschwendtner},
  \citenamefont {Mazzoni},\ and\ \citenamefont {Petrenko}}]{Turner2017}%
  \BibitemOpen
  \bibfield  {author} {\bibinfo {author} {\bibnamefont {Turner}, \bibfnamefont
  {M}}, \bibinfo {author} {\bibfnamefont {B.}~\bibnamefont {Biskup}}, \bibinfo
  {author} {\bibfnamefont {S.}~\bibnamefont {Burger}}, \bibinfo {author}
  {\bibfnamefont {E.}~\bibnamefont {Gschwendtner}}, \bibinfo {author}
  {\bibfnamefont {S.}~\bibnamefont {Mazzoni}}, \ and\ \bibinfo {author}
  {\bibfnamefont {A.}~\bibnamefont {Petrenko}}} (\bibinfo {year} {2017}),\
  \href {https://doi.org/10.1016/j.nima.2017.02.064} {\bibfield  {journal}
  {\bibinfo  {journal} {Nucl. Instrum. Methods Phys. Res. A}\ }\textbf
  {\bibinfo {volume} {854}},\ \bibinfo {pages} {100--106}}\BibitemShut
  {NoStop}%
\bibitem [{\citenamefont {Turner}\ \emph {et~al.}(2019)\citenamefont {Turner}
  \emph {et~al.}}]{Turner2019}%
  \BibitemOpen
  \bibfield  {author} {\bibinfo {author} {\bibnamefont {Turner}, \bibfnamefont
  {M}},  \emph {et~al.} (\bibinfo {collaboration} {AWAKE Collaboration})}
  (\bibinfo {year} {2019}),\ \href
  {https://doi.org/10.1103/physrevlett.122.054801} {\bibfield  {journal}
  {\bibinfo  {journal} {Phys. Rev. Lett.}\ }\textbf {\bibinfo {volume} {122}},\
  \bibinfo {pages} {054801}}\BibitemShut {NoStop}%
\bibitem [{\citenamefont {Turner}\ \emph {et~al.}(2020)\citenamefont {Turner}
  \emph {et~al.}}]{Turner2020}%
  \BibitemOpen
  \bibfield  {author} {\bibinfo {author} {\bibnamefont {Turner}, \bibfnamefont
  {M}},  \emph {et~al.} (\bibinfo {collaboration} {AWAKE Collaboration})}
  (\bibinfo {year} {2020}),\ \href
  {https://doi.org/10.1103/physrevaccelbeams.23.081302} {\bibfield  {journal}
  {\bibinfo  {journal} {Phys. Rev. Accel. Beams}\ }\textbf {\bibinfo {volume}
  {23}},\ \bibinfo {pages} {081302}}\BibitemShut {NoStop}%
\bibitem [{\citenamefont {Turner}\ \emph {et~al.}(2025)\citenamefont {Turner}
  \emph {et~al.}}]{Turner2025}%
  \BibitemOpen
  \bibfield  {author} {\bibinfo {author} {\bibnamefont {Turner}, \bibfnamefont
  {M}},  \emph {et~al.} (\bibinfo {collaboration} {AWAKE Collaboration})}
  (\bibinfo {year} {2025}),\ \href
  {https://doi.org/10.1103/PhysRevLett.134.155001} {\bibfield  {journal}
  {\bibinfo  {journal} {Phys. Rev. Lett.}\ }\textbf {\bibinfo {volume} {134}},\
  \bibinfo {pages} {155001}}\BibitemShut {NoStop}%
\bibitem [{\citenamefont {Tzoufras}\ \emph {et~al.}(2008)\citenamefont
  {Tzoufras}, \citenamefont {Lu}, \citenamefont {Tsung}, \citenamefont {Huang},
  \citenamefont {Mori}, \citenamefont {Katsouleas}, \citenamefont {Vieira},
  \citenamefont {Fonseca},\ and\ \citenamefont {Silva}}]{Tzoufras2008}%
  \BibitemOpen
  \bibfield  {author} {\bibinfo {author} {\bibnamefont {Tzoufras},
  \bibfnamefont {M}}, \bibinfo {author} {\bibfnamefont {W.}~\bibnamefont {Lu}},
  \bibinfo {author} {\bibfnamefont {F.~S.}\ \bibnamefont {Tsung}}, \bibinfo
  {author} {\bibfnamefont {C.}~\bibnamefont {Huang}}, \bibinfo {author}
  {\bibfnamefont {W.~B.}\ \bibnamefont {Mori}}, \bibinfo {author}
  {\bibfnamefont {T.}~\bibnamefont {Katsouleas}}, \bibinfo {author}
  {\bibfnamefont {J.}~\bibnamefont {Vieira}}, \bibinfo {author} {\bibfnamefont
  {R.~A.}\ \bibnamefont {Fonseca}}, \ and\ \bibinfo {author} {\bibfnamefont
  {L.~O.}\ \bibnamefont {Silva}}} (\bibinfo {year} {2008}),\ \href
  {https://doi.org/10.1103/physrevlett.101.145002} {\bibfield  {journal}
  {\bibinfo  {journal} {Phys. Rev. Lett.}\ }\textbf {\bibinfo {volume} {101}},\
  \bibinfo {pages} {145002}}\BibitemShut {NoStop}%
\bibitem [{\citenamefont {Tzoufras}\ \emph {et~al.}(2009)\citenamefont
  {Tzoufras}, \citenamefont {Lu}, \citenamefont {Tsung}, \citenamefont {Huang},
  \citenamefont {Mori}, \citenamefont {Katsouleas}, \citenamefont {Vieira},
  \citenamefont {Fonseca},\ and\ \citenamefont {Silva}}]{Tzoufras2009}%
  \BibitemOpen
  \bibfield  {author} {\bibinfo {author} {\bibnamefont {Tzoufras},
  \bibfnamefont {M}}, \bibinfo {author} {\bibfnamefont {W.}~\bibnamefont {Lu}},
  \bibinfo {author} {\bibfnamefont {F.~S.}\ \bibnamefont {Tsung}}, \bibinfo
  {author} {\bibfnamefont {C.}~\bibnamefont {Huang}}, \bibinfo {author}
  {\bibfnamefont {W.~B.}\ \bibnamefont {Mori}}, \bibinfo {author}
  {\bibfnamefont {T.}~\bibnamefont {Katsouleas}}, \bibinfo {author}
  {\bibfnamefont {J.}~\bibnamefont {Vieira}}, \bibinfo {author} {\bibfnamefont
  {R.~A.}\ \bibnamefont {Fonseca}}, \ and\ \bibinfo {author} {\bibfnamefont
  {L.~O.}\ \bibnamefont {Silva}}} (\bibinfo {year} {2009}),\ \href
  {https://doi.org/10.1063/1.3118628} {\bibfield  {journal} {\bibinfo
  {journal} {Phys. Plasmas}\ }\textbf {\bibinfo {volume} {16}},\ \bibinfo
  {pages} {056705}}\BibitemShut {NoStop}%
\bibitem [{\citenamefont {Ufer}\ \emph {et~al.}(2026)\citenamefont {Ufer} \emph
  {et~al.}}]{Ufer2026}%
  \BibitemOpen
  \bibfield  {author} {\bibinfo {author} {\bibnamefont {Ufer}, \bibfnamefont
  {P}},  \emph {et~al.}} (\bibinfo {year} {2026}),\ \href {\doibase
  10.1038/s42005-026-02814-1} {\bibfield  {journal} {\bibinfo  {journal}
  {Commun. Phys.}\ }\textbf {\bibinfo {volume} {9}},\ \bibinfo {pages}
  {273}}\BibitemShut {NoStop}%
\bibitem [{\citenamefont {Umstadter}\ \emph {et~al.}(1996)\citenamefont
  {Umstadter}, \citenamefont {Kim},\ and\ \citenamefont
  {Dodd}}]{Umstadter1996}%
  \BibitemOpen
  \bibfield  {author} {\bibinfo {author} {\bibnamefont {Umstadter},
  \bibfnamefont {D}}, \bibinfo {author} {\bibfnamefont {J.~K.}\ \bibnamefont
  {Kim}}, \ and\ \bibinfo {author} {\bibfnamefont {E.}~\bibnamefont {Dodd}}}
  (\bibinfo {year} {1996}),\ \href
  {https://doi.org/10.1103/physrevlett.76.2073} {\bibfield  {journal} {\bibinfo
   {journal} {Phys. Rev. Lett.}\ }\textbf {\bibinfo {volume} {76}},\ \bibinfo
  {pages} {2073--2076}}\BibitemShut {NoStop}%
\bibitem [{\citenamefont {Vaccarezza}\ \emph {et~al.}(2018)\citenamefont
  {Vaccarezza} \emph {et~al.}}]{Vaccarezza2018}%
  \BibitemOpen
  \bibfield  {author} {\bibinfo {author} {\bibnamefont {Vaccarezza},
  \bibfnamefont {C}},  \emph {et~al.}} (\bibinfo {year} {2018}),\ \href
  {https://doi.org/10.1016/j.nima.2018.01.100} {\bibfield  {journal} {\bibinfo
  {journal} {Nucl. Instrum. Methods Phys. Res. A}\ }\textbf {\bibinfo {volume}
  {909}},\ \bibinfo {pages} {314--317}}\BibitemShut {NoStop}%
\bibitem [{\citenamefont {Vafaei-Najafabadi}\ \emph {et~al.}(2014)\citenamefont
  {Vafaei-Najafabadi} \emph {et~al.}}]{VafaeiNajafabadi2014}%
  \BibitemOpen
  \bibfield  {author} {\bibinfo {author} {\bibnamefont {Vafaei-Najafabadi},
  \bibfnamefont {N}},  \emph {et~al.}} (\bibinfo {year} {2014}),\ \href
  {https://doi.org/10.1103/physrevlett.112.025001} {\bibfield  {journal}
  {\bibinfo  {journal} {Phys. Rev. Lett.}\ }\textbf {\bibinfo {volume} {112}},\
  \bibinfo {pages} {025001}}\BibitemShut {NoStop}%
\bibitem [{\citenamefont {Vafaei-Najafabadi}\ \emph {et~al.}(2016)\citenamefont
  {Vafaei-Najafabadi} \emph {et~al.}}]{VafaeiNajafabadi2016}%
  \BibitemOpen
  \bibfield  {author} {\bibinfo {author} {\bibnamefont {Vafaei-Najafabadi},
  \bibfnamefont {N}},  \emph {et~al.}} (\bibinfo {year} {2016}),\ \href
  {https://doi.org/10.1088/0741-3335/58/3/034009} {\bibfield  {journal}
  {\bibinfo  {journal} {Plasma Phys. Control. Fusion}\ }\textbf {\bibinfo
  {volume} {58}},\ \bibinfo {pages} {034009}}\BibitemShut {NoStop}%
\bibitem [{\citenamefont {Vafaei-Najafabadi}\ \emph {et~al.}(2019)\citenamefont
  {Vafaei-Najafabadi} \emph {et~al.}}]{VafaeiNajafabadi2019}%
  \BibitemOpen
  \bibfield  {author} {\bibinfo {author} {\bibnamefont {Vafaei-Najafabadi},
  \bibfnamefont {N}},  \emph {et~al.}} (\bibinfo {year} {2019}),\ \href
  {https://doi.org/10.1098/rsta.2018.0184} {\bibfield  {journal} {\bibinfo
  {journal} {Philos. Trans. R. Soc. A}\ }\textbf {\bibinfo {volume} {377}},\
  \bibinfo {pages} {20180184}}\BibitemShut {NoStop}%
\bibitem [{\citenamefont {{van Tilborg}}\ \emph {et~al.}(2015)\citenamefont
  {{van Tilborg}} \emph {et~al.}}]{vanTilborg2015}%
  \BibitemOpen
  \bibfield  {author} {\bibinfo {author} {\bibnamefont {{van Tilborg}},
  \bibfnamefont {J}},  \emph {et~al.}} (\bibinfo {year} {2015}),\ \href
  {https://doi.org/10.1103/physrevlett.115.184802} {\bibfield  {journal}
  {\bibinfo  {journal} {Phys. Rev. Lett.}\ }\textbf {\bibinfo {volume} {115}},\
  \bibinfo {pages} {184802}}\BibitemShut {NoStop}%
\bibitem [{\citenamefont {Varian}\ and\ \citenamefont
  {Varian}(1939)}]{Varian1939}%
  \BibitemOpen
  \bibfield  {author} {\bibinfo {author} {\bibnamefont {Varian}, \bibfnamefont
  {R~H}}, \ and\ \bibinfo {author} {\bibfnamefont {S.~F.}\ \bibnamefont
  {Varian}}} (\bibinfo {year} {1939}),\ \href
  {https://doi.org/10.1063/1.1707311} {\bibfield  {journal} {\bibinfo
  {journal} {J. Appl. Phys.}\ }\textbf {\bibinfo {volume} {10}},\ \bibinfo
  {pages} {321--327}}\BibitemShut {NoStop}%
\bibitem [{\citenamefont {Vay}(2007)}]{Vay2007}%
  \BibitemOpen
  \bibfield  {author} {\bibinfo {author} {\bibnamefont {Vay}, \bibfnamefont
  {J-L}}} (\bibinfo {year} {2007}),\ \href {\doibase
  10.1103/physrevlett.98.130405} {\bibfield  {journal} {\bibinfo  {journal}
  {Phys. Rev. Lett.}\ }\textbf {\bibinfo {volume} {98}},\ \bibinfo {pages}
  {130405}}\BibitemShut {NoStop}%
\bibitem [{\citenamefont {Vay}\ \emph {et~al.}(2002)\citenamefont {Vay},
  \citenamefont {Colella}, \citenamefont {McCorquodale}, \citenamefont {van
  Straalen}, \citenamefont {Friedman},\ and\ \citenamefont {Grote}}]{Vay2002}%
  \BibitemOpen
  \bibfield  {author} {\bibinfo {author} {\bibnamefont {Vay}, \bibfnamefont
  {J-L}}, \bibinfo {author} {\bibfnamefont {P.}~\bibnamefont {Colella}},
  \bibinfo {author} {\bibfnamefont {P.}~\bibnamefont {McCorquodale}}, \bibinfo
  {author} {\bibfnamefont {B.}~\bibnamefont {van Straalen}}, \bibinfo {author}
  {\bibfnamefont {A.}~\bibnamefont {Friedman}}, \ and\ \bibinfo {author}
  {\bibfnamefont {D.~P.}\ \bibnamefont {Grote}}} (\bibinfo {year} {2002}),\
  \href {https://doi.org/10.1017/S0263034602204139} {\bibfield  {journal}
  {\bibinfo  {journal} {Laser Part. Beams}\ }\textbf {\bibinfo {volume} {20}},\
  \bibinfo {pages} {569--575}}\BibitemShut {NoStop}%
\bibitem [{\citenamefont {Vay}\ \emph {et~al.}(2011)\citenamefont {Vay},
  \citenamefont {Geddes}, \citenamefont {Esarey}, \citenamefont {Schroeder},
  \citenamefont {Leemans}, \citenamefont {Cormier-Michel},\ and\ \citenamefont
  {Grote}}]{Vay2011}%
  \BibitemOpen
  \bibfield  {author} {\bibinfo {author} {\bibnamefont {Vay}, \bibfnamefont
  {J-L}}, \bibinfo {author} {\bibfnamefont {C.~G.~R.}\ \bibnamefont {Geddes}},
  \bibinfo {author} {\bibfnamefont {E.}~\bibnamefont {Esarey}}, \bibinfo
  {author} {\bibfnamefont {C.~B.}\ \bibnamefont {Schroeder}}, \bibinfo {author}
  {\bibfnamefont {W.~P.}\ \bibnamefont {Leemans}}, \bibinfo {author}
  {\bibfnamefont {E.}~\bibnamefont {Cormier-Michel}}, \ and\ \bibinfo {author}
  {\bibfnamefont {D.~P.}\ \bibnamefont {Grote}}} (\bibinfo {year} {2011}),\
  \href {\doibase 10.1063/1.3663841} {\bibfield  {journal} {\bibinfo  {journal}
  {Phys. Plasmas}\ }\textbf {\bibinfo {volume} {18}},\ \bibinfo {pages}
  {123103}}\BibitemShut {NoStop}%
\bibitem [{\citenamefont {Vay}\ and\ \citenamefont {Lehe}(2016)}]{Vay2016}%
  \BibitemOpen
  \bibfield  {author} {\bibinfo {author} {\bibnamefont {Vay}, \bibfnamefont
  {J-L}}, \ and\ \bibinfo {author} {\bibfnamefont {R.}~\bibnamefont {Lehe}}}
  (\bibinfo {year} {2016}),\ \href {https://doi.org/10.1142/s1793626816300085}
  {\bibfield  {journal} {\bibinfo  {journal} {Rev. Accel. Sci. Tech.}\ }\textbf
  {\bibinfo {volume} {9}},\ \bibinfo {pages} {165--186}}\BibitemShut {NoStop}%
\bibitem [{\citenamefont {Vay}\ \emph {et~al.}(2018)\citenamefont {Vay} \emph
  {et~al.}}]{Vay2018}%
  \BibitemOpen
  \bibfield  {author} {\bibinfo {author} {\bibnamefont {Vay}, \bibfnamefont
  {J-L}},  \emph {et~al.}} (\bibinfo {year} {2018}),\ \href
  {https://doi.org/10.1016/j.nima.2018.01.035} {\bibfield  {journal} {\bibinfo
  {journal} {Nucl. Instrum. Methods Phys. Res. A}\ }\textbf {\bibinfo {volume}
  {909}},\ \bibinfo {pages} {476--479}}\BibitemShut {NoStop}%
\bibitem [{\citenamefont {Vay}\ \emph {et~al.}(2021)\citenamefont {Vay} \emph
  {et~al.}}]{Vay2021}%
  \BibitemOpen
  \bibfield  {author} {\bibinfo {author} {\bibnamefont {Vay}, \bibfnamefont
  {J-L}},  \emph {et~al.}} (\bibinfo {year} {2021}),\ \href
  {https://doi.org/10.1063/5.0028512} {\bibfield  {journal} {\bibinfo
  {journal} {Phys. Plasmas}\ }\textbf {\bibinfo {volume} {28}},\ \bibinfo
  {pages} {023105}}\BibitemShut {NoStop}%
\bibitem [{\citenamefont {Veksler}(1956)}]{Veksler1956}%
  \BibitemOpen
  \bibfield  {author} {\bibinfo {author} {\bibnamefont {Veksler}, \bibfnamefont
  {V~I}}} (\bibinfo {year} {1956}),\ in\ \href
  {http://cds.cern.ch/record/1241563} {\emph {\bibinfo {booktitle} {CERN
  Symposium on High Energy Accelerators and Pion Physics}}}\ (\bibinfo
  {publisher} {CERN},\ \bibinfo {address} {Geneva, Switzerland})\ pp.\ \bibinfo
  {pages} {80--83}\BibitemShut {NoStop}%
\bibitem [{\citenamefont {Verboncoeur}\ \emph {et~al.}(1995)\citenamefont
  {Verboncoeur}, \citenamefont {Langdon},\ and\ \citenamefont
  {Gladd}}]{Verboncoeur1995}%
  \BibitemOpen
  \bibfield  {author} {\bibinfo {author} {\bibnamefont {Verboncoeur},
  \bibfnamefont {J~P}}, \bibinfo {author} {\bibfnamefont {A.~B.}\ \bibnamefont
  {Langdon}}, \ and\ \bibinfo {author} {\bibfnamefont {N.~T.}\ \bibnamefont
  {Gladd}}} (\bibinfo {year} {1995}),\ \href
  {https://doi.org/10.1016/0010-4655(94)00173-Y} {\bibfield  {journal}
  {\bibinfo  {journal} {Comput. Phys. Commun.}\ }\textbf {\bibinfo {volume}
  {87}},\ \bibinfo {pages} {199--211}}\BibitemShut {NoStop}%
\bibitem [{\citenamefont {Verra}\ \emph {et~al.}(2022)\citenamefont {Verra}
  \emph {et~al.}}]{Verra2022}%
  \BibitemOpen
  \bibfield  {author} {\bibinfo {author} {\bibnamefont {Verra}, \bibfnamefont
  {L}},  \emph {et~al.} (\bibinfo {collaboration} {AWAKE Collaboration})}
  (\bibinfo {year} {2022}),\ \href
  {https://doi.org/10.1103/physrevlett.129.024802} {\bibfield  {journal}
  {\bibinfo  {journal} {Phys. Rev. Lett.}\ }\textbf {\bibinfo {volume} {129}},\
  \bibinfo {pages} {024802}}\BibitemShut {NoStop}%
\bibitem [{\citenamefont {Verra}\ \emph {et~al.}(2023)\citenamefont {Verra}
  \emph {et~al.}}]{Verra2023}%
  \BibitemOpen
  \bibfield  {author} {\bibinfo {author} {\bibnamefont {Verra}, \bibfnamefont
  {L}},  \emph {et~al.}} (\bibinfo {year} {2023}),\ \href
  {https://doi.org/10.1063/5.0157391} {\bibfield  {journal} {\bibinfo
  {journal} {Phys. Plasmas}\ }\textbf {\bibinfo {volume} {30}},\ \bibinfo
  {pages} {083104}}\BibitemShut {NoStop}%
\bibitem [{\citenamefont {Verra}\ \emph
  {et~al.}(2024{\natexlab{a}})\citenamefont {Verra} \emph
  {et~al.}}]{Verra2024}%
  \BibitemOpen
  \bibfield  {author} {\bibinfo {author} {\bibnamefont {Verra}, \bibfnamefont
  {L}},  \emph {et~al.} (\bibinfo {collaboration} {AWAKE Collaboration})}
  (\bibinfo {year} {2024}{\natexlab{a}}),\ \href
  {https://link.aps.org/doi/10.1103/PhysRevE.109.055203} {\bibfield  {journal}
  {\bibinfo  {journal} {Phys. Rev. E}\ }\textbf {\bibinfo {volume} {109}},\
  \bibinfo {pages} {055203}}\BibitemShut {NoStop}%
\bibitem [{\citenamefont {Verra}\ \emph
  {et~al.}(2024{\natexlab{b}})\citenamefont {Verra} \emph
  {et~al.}}]{Verra2024b}%
  \BibitemOpen
  \bibfield  {author} {\bibinfo {author} {\bibnamefont {Verra}, \bibfnamefont
  {L}},  \emph {et~al.}} (\bibinfo {year} {2024}{\natexlab{b}}),\ \href
  {https://doi.org/10.1103/PhysRevLett.133.035001} {\bibfield  {journal}
  {\bibinfo  {journal} {Phys. Rev. Lett.}\ }\textbf {\bibinfo {volume} {133}},\
  \bibinfo {pages} {055203}}\BibitemShut {NoStop}%
\bibitem [{\citenamefont {Verra}\ \emph {et~al.}(2025)\citenamefont {Verra}
  \emph {et~al.}}]{Verra2025}%
  \BibitemOpen
  \bibfield  {author} {\bibinfo {author} {\bibnamefont {Verra}, \bibfnamefont
  {L}},  \emph {et~al.}} (\bibinfo {year} {2025}),\ \href
  {https://doi.org/10.1103/x4xl-4fnp} {\bibfield  {journal} {\bibinfo
  {journal} {Phys. Rev. E}\ }\textbf {\bibinfo {volume} {112}},\ \bibinfo
  {pages} {045205}}\BibitemShut {NoStop}%
\bibitem [{\citenamefont {Vidal}\ and\ \citenamefont
  {Cooper}(1969)}]{Vidal1969}%
  \BibitemOpen
  \bibfield  {author} {\bibinfo {author} {\bibnamefont {Vidal}, \bibfnamefont
  {C~R}}, \ and\ \bibinfo {author} {\bibfnamefont {J.}~\bibnamefont {Cooper}}}
  (\bibinfo {year} {1969}),\ \href {https://doi.org/10.1063/1.1658190}
  {\bibfield  {journal} {\bibinfo  {journal} {J. Appl. Phys.}\ }\textbf
  {\bibinfo {volume} {40}},\ \bibinfo {pages} {3370–3374}}\BibitemShut
  {NoStop}%
\bibitem [{\citenamefont {Vieira}\ and\ \citenamefont
  {Mendon\ifmmode~\mbox{\c{c}}\else \c{c}\fi{}a}(2014)}]{Vieira2014b}%
  \BibitemOpen
  \bibfield  {author} {\bibinfo {author} {\bibnamefont {Vieira}, \bibfnamefont
  {J}}, \ and\ \bibinfo {author} {\bibfnamefont {J.~T.}\ \bibnamefont
  {Mendon\ifmmode~\mbox{\c{c}}\else \c{c}\fi{}a}}} (\bibinfo {year} {2014}),\
  \href {https://doi.org/10.1103/PhysRevLett.112.215001} {\bibfield  {journal}
  {\bibinfo  {journal} {Phys. Rev. Lett.}\ }\textbf {\bibinfo {volume} {112}},\
  \bibinfo {pages} {215001}}\BibitemShut {NoStop}%
\bibitem [{\citenamefont {Vieira}\ \emph {et~al.}(2012)\citenamefont {Vieira},
  \citenamefont {Fonseca}, \citenamefont {Mori},\ and\ \citenamefont
  {Silva}}]{Vieira2012}%
  \BibitemOpen
  \bibfield  {author} {\bibinfo {author} {\bibnamefont {Vieira}, \bibfnamefont
  {J}}, \bibinfo {author} {\bibfnamefont {R.~A.}\ \bibnamefont {Fonseca}},
  \bibinfo {author} {\bibfnamefont {W.~B.}\ \bibnamefont {Mori}}, \ and\
  \bibinfo {author} {\bibfnamefont {L.~O.}\ \bibnamefont {Silva}}} (\bibinfo
  {year} {2012}),\ \href {https://doi.org/10.1103/PhysRevLett.109.145005}
  {\bibfield  {journal} {\bibinfo  {journal} {Phys. Rev. Lett.}\ }\textbf
  {\bibinfo {volume} {109}},\ \bibinfo {pages} {145005}}\BibitemShut {NoStop}%
\bibitem [{\citenamefont {Vieira}\ \emph
  {et~al.}(2014{\natexlab{a}})\citenamefont {Vieira}, \citenamefont {Fonseca},
  \citenamefont {Mori},\ and\ \citenamefont {Silva}}]{Vieira2014c}%
  \BibitemOpen
  \bibfield  {author} {\bibinfo {author} {\bibnamefont {Vieira}, \bibfnamefont
  {J}}, \bibinfo {author} {\bibfnamefont {R.~A.}\ \bibnamefont {Fonseca}},
  \bibinfo {author} {\bibfnamefont {W.~B.}\ \bibnamefont {Mori}}, \ and\
  \bibinfo {author} {\bibfnamefont {L.~O.}\ \bibnamefont {Silva}}} (\bibinfo
  {year} {2014}{\natexlab{a}}),\ \href {https://doi.org/10.1063/1.4876620}
  {\bibfield  {journal} {\bibinfo  {journal} {Phys. Plasmas}\ }\textbf
  {\bibinfo {volume} {21}},\ \bibinfo {pages} {056705}}\BibitemShut {NoStop}%
\bibitem [{\citenamefont {Vieira}\ \emph
  {et~al.}(2011{\natexlab{a}})\citenamefont {Vieira}, \citenamefont {Huang},
  \citenamefont {Mori},\ and\ \citenamefont {Silva}}]{Vieira2011b}%
  \BibitemOpen
  \bibfield  {author} {\bibinfo {author} {\bibnamefont {Vieira}, \bibfnamefont
  {J}}, \bibinfo {author} {\bibfnamefont {C.-K.}\ \bibnamefont {Huang}},
  \bibinfo {author} {\bibfnamefont {W.~B.}\ \bibnamefont {Mori}}, \ and\
  \bibinfo {author} {\bibfnamefont {L.~O.}\ \bibnamefont {Silva}}} (\bibinfo
  {year} {2011}{\natexlab{a}}),\ \href
  {https://doi.org/10.1103/physrevstab.14.071303} {\bibfield  {journal}
  {\bibinfo  {journal} {Phys. Rev. ST Accel. Beams}\ }\textbf {\bibinfo
  {volume} {14}},\ \bibinfo {pages} {071303}}\BibitemShut {NoStop}%
\bibitem [{\citenamefont {Vieira}\ \emph
  {et~al.}(2011{\natexlab{b}})\citenamefont {Vieira}, \citenamefont {Martins},
  \citenamefont {Pathak}, \citenamefont {Fonseca}, \citenamefont {Mori},\ and\
  \citenamefont {Silva}}]{Vieira2011}%
  \BibitemOpen
  \bibfield  {author} {\bibinfo {author} {\bibnamefont {Vieira}, \bibfnamefont
  {J}}, \bibinfo {author} {\bibfnamefont {S.~F.}\ \bibnamefont {Martins}},
  \bibinfo {author} {\bibfnamefont {V.~B.}\ \bibnamefont {Pathak}}, \bibinfo
  {author} {\bibfnamefont {R.~A.}\ \bibnamefont {Fonseca}}, \bibinfo {author}
  {\bibfnamefont {W.~B.}\ \bibnamefont {Mori}}, \ and\ \bibinfo {author}
  {\bibfnamefont {L.~O.}\ \bibnamefont {Silva}}} (\bibinfo {year}
  {2011}{\natexlab{b}}),\ \href
  {https://doi.org/10.1103/physrevlett.106.225001} {\bibfield  {journal}
  {\bibinfo  {journal} {Phys. Rev. Lett.}\ }\textbf {\bibinfo {volume} {106}},\
  \bibinfo {pages} {225001}}\BibitemShut {NoStop}%
\bibitem [{\citenamefont {Vieira}\ \emph {et~al.}(2016)\citenamefont {Vieira},
  \citenamefont {Mendonça},\ and\ \citenamefont {Silva}}]{Vieira2016}%
  \BibitemOpen
  \bibfield  {author} {\bibinfo {author} {\bibnamefont {Vieira}, \bibfnamefont
  {J}}, \bibinfo {author} {\bibfnamefont {J.~T.}\ \bibnamefont {Mendonça}}, \
  and\ \bibinfo {author} {\bibfnamefont {L.~O.}\ \bibnamefont {Silva}}}
  (\bibinfo {year} {2016}),\ in\ \href
  {https://aip.scitation.org/doi/abs/10.1063/1.4965655} {\emph {\bibinfo
  {booktitle} {{AIP} Conf. Proc.}}},\ Vol.\ \bibinfo {volume} {1777}\ (\bibinfo
   {publisher} {{AIP}})\ p.\ \bibinfo {pages} {070012}\BibitemShut {NoStop}%
\bibitem [{\citenamefont {Vieira}\ \emph
  {et~al.}(2014{\natexlab{b}})\citenamefont {Vieira}, \citenamefont {Mori},\
  and\ \citenamefont {Muggli}}]{Vieira2014}%
  \BibitemOpen
  \bibfield  {author} {\bibinfo {author} {\bibnamefont {Vieira}, \bibfnamefont
  {J}}, \bibinfo {author} {\bibfnamefont {W.~B.}\ \bibnamefont {Mori}}, \ and\
  \bibinfo {author} {\bibfnamefont {P.}~\bibnamefont {Muggli}}} (\bibinfo
  {year} {2014}{\natexlab{b}}),\ \href
  {https://doi.org/10.1103/physrevlett.112.205001} {\bibfield  {journal}
  {\bibinfo  {journal} {Phys. Rev. Lett.}\ }\textbf {\bibinfo {volume} {112}},\
  \bibinfo {pages} {205001}}\BibitemShut {NoStop}%
\bibitem [{\citenamefont {Villa}\ \emph {et~al.}(2023)\citenamefont {Villa}
  \emph {et~al.}}]{Villa2023}%
  \BibitemOpen
  \bibfield  {author} {\bibinfo {author} {\bibnamefont {Villa}, \bibfnamefont
  {F}},  \emph {et~al.}} (\bibinfo {year} {2023}),\ in\ \href
  {https://doi.org/10.1117/12.2668643} {\emph {\bibinfo {booktitle} {X-Ray
  Free-Electron Lasers: Advances in Source Development and Instrumentation
  VI}}}\ (\bibinfo  {publisher} {SPIE},\ \bibinfo {address} {Bellingham, WA,
  United States})\ p.\ \bibinfo {pages} {125810H}\BibitemShut {NoStop}%
\bibitem [{\citenamefont {Voss}\ and\ \citenamefont
  {Weiland}(1982)}]{Voss1982}%
  \BibitemOpen
  \bibfield  {author} {\bibinfo {author} {\bibnamefont {Voss}, \bibfnamefont
  {G~A}}, \ and\ \bibinfo {author} {\bibfnamefont {T.}~\bibnamefont {Weiland}}}
  (\bibinfo {year} {1982}),\ \href
  {https://lss.fnal.gov/conf/C8209271/p287.pdf} {}\bibinfo {type} {Tech. Rep.}\
  \bibinfo {number} {DESY-82-074}\ (\bibinfo  {institution} {DESY},\ \bibinfo
  {address} {Hamburg, Germany})\BibitemShut {NoStop}%
\bibitem [{\citenamefont {Wang}\ \emph {et~al.}(2025)\citenamefont {Wang},
  \citenamefont {Zeng}, \citenamefont {Li},\ and\ \citenamefont
  {Gao}}]{Wang2025}%
  \BibitemOpen
  \bibfield  {author} {\bibinfo {author} {\bibnamefont {Wang}, \bibfnamefont
  {J}}, \bibinfo {author} {\bibfnamefont {M.}~\bibnamefont {Zeng}}, \bibinfo
  {author} {\bibfnamefont {D.}~\bibnamefont {Li}}, \ and\ \bibinfo {author}
  {\bibfnamefont {J.}~\bibnamefont {Gao}}} (\bibinfo {year} {2025}),\ \href
  {https://doi.org/10.1103/PhysRevAccelBeams.28.031301} {\bibfield  {journal}
  {\bibinfo  {journal} {Phys. Rev. Accel. Beams}\ }\textbf {\bibinfo {volume}
  {28}},\ \bibinfo {pages} {031301}}\BibitemShut {NoStop}%
\bibitem [{\citenamefont {Wang}\ \emph {et~al.}(2002)\citenamefont {Wang} \emph
  {et~al.}}]{Wang2002}%
  \BibitemOpen
  \bibfield  {author} {\bibinfo {author} {\bibnamefont {Wang}, \bibfnamefont
  {S}},  \emph {et~al.}} (\bibinfo {year} {2002}),\ \href
  {https://doi.org/10.1103/physrevlett.88.135004} {\bibfield  {journal}
  {\bibinfo  {journal} {Phys. Rev. Lett.}\ }\textbf {\bibinfo {volume} {88}},\
  \bibinfo {pages} {135004}}\BibitemShut {NoStop}%
\bibitem [{\citenamefont {Wang}\ \emph {et~al.}(2017)\citenamefont {Wang},
  \citenamefont {Khudik}, \citenamefont {Breizman},\ and\ \citenamefont
  {Shvets}}]{2017Wang}%
  \BibitemOpen
  \bibfield  {author} {\bibinfo {author} {\bibnamefont {Wang}, \bibfnamefont
  {T}}, \bibinfo {author} {\bibfnamefont {V.}~\bibnamefont {Khudik}}, \bibinfo
  {author} {\bibfnamefont {B.}~\bibnamefont {Breizman}}, \ and\ \bibinfo
  {author} {\bibfnamefont {G.}~\bibnamefont {Shvets}}} (\bibinfo {year}
  {2017}),\ \href {https://doi.org/10.1063/1.4999629} {\bibfield  {journal}
  {\bibinfo  {journal} {Phys. Plasmas}\ }\textbf {\bibinfo {volume} {24}},\
  \bibinfo {pages} {103117}}\BibitemShut {NoStop}%
\bibitem [{\citenamefont {Wang}\ \emph
  {et~al.}(2021{\natexlab{a}})\citenamefont {Wang}, \citenamefont {Khudik},\
  and\ \citenamefont {Shvets}}]{Wang2021}%
  \BibitemOpen
  \bibfield  {author} {\bibinfo {author} {\bibnamefont {Wang}, \bibfnamefont
  {T}}, \bibinfo {author} {\bibfnamefont {V.}~\bibnamefont {Khudik}}, \ and\
  \bibinfo {author} {\bibfnamefont {G.}~\bibnamefont {Shvets}}} (\bibinfo
  {year} {2021}{\natexlab{a}}),\ \href {https://arxiv.org/abs/2110.10290} {\
  }\Eprint {http://arxiv.org/abs/2110.10290} {arXiv:2110.10290} \BibitemShut
  {NoStop}%
\bibitem [{\citenamefont {Wang}\ \emph
  {et~al.}(2021{\natexlab{b}})\citenamefont {Wang} \emph {et~al.}}]{Wang2021b}%
  \BibitemOpen
  \bibfield  {author} {\bibinfo {author} {\bibnamefont {Wang}, \bibfnamefont
  {W}},  \emph {et~al.}} (\bibinfo {year} {2021}{\natexlab{b}}),\ \href
  {https://doi.org/10.1038/s41586-021-03678-x} {\bibfield  {journal} {\bibinfo
  {journal} {Nature (London)}\ }\textbf {\bibinfo {volume} {595}},\ \bibinfo
  {pages} {516–520}}\BibitemShut {NoStop}%
\bibitem [{\citenamefont {Wang}\ \emph {et~al.}(2020)\citenamefont {Wang} \emph
  {et~al.}}]{Wang2020}%
  \BibitemOpen
  \bibfield  {author} {\bibinfo {author} {\bibnamefont {Wang}, \bibfnamefont
  {W~P}},  \emph {et~al.}} (\bibinfo {year} {2020}),\ \href
  {https://link.aps.org/doi/10.1103/PhysRevLett.125.034801} {\bibfield
  {journal} {\bibinfo  {journal} {Phys. Rev. Lett.}\ }\textbf {\bibinfo
  {volume} {125}},\ \bibinfo {pages} {034801}}\BibitemShut {NoStop}%
\bibitem [{\citenamefont {Wang}\ \emph {et~al.}(2022)\citenamefont {Wang},
  \citenamefont {Gao}, \citenamefont {An}, \citenamefont {Zhou}, \citenamefont
  {Meng}, \citenamefont {Wang}, \citenamefont {Wang}, \citenamefont {Li},
  \citenamefont {Zeng},\ and\ \citenamefont {Lu}}]{Wang2022}%
  \BibitemOpen
  \bibfield  {author} {\bibinfo {author} {\bibnamefont {Wang}, \bibfnamefont
  {X}}, \bibinfo {author} {\bibfnamefont {J.}~\bibnamefont {Gao}}, \bibinfo
  {author} {\bibfnamefont {W.}~\bibnamefont {An}}, \bibinfo {author}
  {\bibfnamefont {S.}~\bibnamefont {Zhou}}, \bibinfo {author} {\bibfnamefont
  {C.}~\bibnamefont {Meng}}, \bibinfo {author} {\bibfnamefont {D.}~\bibnamefont
  {Wang}}, \bibinfo {author} {\bibfnamefont {J.}~\bibnamefont {Wang}}, \bibinfo
  {author} {\bibfnamefont {D.}~\bibnamefont {Li}}, \bibinfo {author}
  {\bibfnamefont {M.}~\bibnamefont {Zeng}}, \ and\ \bibinfo {author}
  {\bibfnamefont {W.}~\bibnamefont {Lu}}} (\bibinfo {year} {2022}),\ \href
  {https://doi.org/10.1142/S0217751X22460034} {\bibfield  {journal} {\bibinfo
  {journal} {Int. J. Mod. Phys. A}\ }\textbf {\bibinfo {volume} {37}},\
  \bibinfo {pages} {2246003}}\BibitemShut {NoStop}%
\bibitem [{\citenamefont {Wang}\ \emph {et~al.}(2006)\citenamefont {Wang},
  \citenamefont {Ischebeck}, \citenamefont {Joshi}, \citenamefont {Muggli},\
  and\ \citenamefont {Katsouleas}}]{Wang2006}%
  \BibitemOpen
  \bibfield  {author} {\bibinfo {author} {\bibnamefont {Wang}, \bibfnamefont
  {X}}, \bibinfo {author} {\bibfnamefont {R.}~\bibnamefont {Ischebeck}},
  \bibinfo {author} {\bibfnamefont {C.}~\bibnamefont {Joshi}}, \bibinfo
  {author} {\bibfnamefont {P.}~\bibnamefont {Muggli}}, \ and\ \bibinfo {author}
  {\bibfnamefont {T.}~\bibnamefont {Katsouleas}}} (\bibinfo {year} {2006}),\
  in\ \href {https://doi.org/10.1063/1.2409185} {\emph {\bibinfo {booktitle}
  {{AIP} Conf. Proc.}}},\ Vol.\ \bibinfo {volume} {877}\ (\bibinfo  {publisher}
  {{AIP}})\ pp.\ \bibinfo {pages} {568--572}\BibitemShut {NoStop}%
\bibitem [{\citenamefont {Wang}\ \emph {et~al.}(2008)\citenamefont {Wang},
  \citenamefont {Ischebeck}, \citenamefont {Muggli}, \citenamefont
  {Katsouleas}, \citenamefont {Joshi}, \citenamefont {Mori},\ and\
  \citenamefont {Hogan}}]{Wang2008}%
  \BibitemOpen
  \bibfield  {author} {\bibinfo {author} {\bibnamefont {Wang}, \bibfnamefont
  {X}}, \bibinfo {author} {\bibfnamefont {R.}~\bibnamefont {Ischebeck}},
  \bibinfo {author} {\bibfnamefont {P.}~\bibnamefont {Muggli}}, \bibinfo
  {author} {\bibfnamefont {T.}~\bibnamefont {Katsouleas}}, \bibinfo {author}
  {\bibfnamefont {C.}~\bibnamefont {Joshi}}, \bibinfo {author} {\bibfnamefont
  {W.~B.}\ \bibnamefont {Mori}}, \ and\ \bibinfo {author} {\bibfnamefont
  {M.~J.}\ \bibnamefont {Hogan}}} (\bibinfo {year} {2008}),\ \href
  {https://link.aps.org/doi/10.1103/PhysRevLett.101.124801} {\bibfield
  {journal} {\bibinfo  {journal} {Phys. Rev. Lett.}\ }\textbf {\bibinfo
  {volume} {101}},\ \bibinfo {pages} {124801}}\BibitemShut {NoStop}%
\bibitem [{\citenamefont {Wang}\ \emph {et~al.}(2009)\citenamefont {Wang},
  \citenamefont {Muggli}, \citenamefont {Katsouleas}, \citenamefont {Joshi},
  \citenamefont {Mori}, \citenamefont {Ischebeck},\ and\ \citenamefont
  {Hogan}}]{Wang2009}%
  \BibitemOpen
  \bibfield  {author} {\bibinfo {author} {\bibnamefont {Wang}, \bibfnamefont
  {X}}, \bibinfo {author} {\bibfnamefont {P.}~\bibnamefont {Muggli}}, \bibinfo
  {author} {\bibfnamefont {T.}~\bibnamefont {Katsouleas}}, \bibinfo {author}
  {\bibfnamefont {C.}~\bibnamefont {Joshi}}, \bibinfo {author} {\bibfnamefont
  {W.~B.}\ \bibnamefont {Mori}}, \bibinfo {author} {\bibfnamefont
  {R.}~\bibnamefont {Ischebeck}}, \ and\ \bibinfo {author} {\bibfnamefont
  {M.~J.}\ \bibnamefont {Hogan}}} (\bibinfo {year} {2009}),\ \href
  {https://link.aps.org/doi/10.1103/PhysRevSTAB.12.051303} {\bibfield
  {journal} {\bibinfo  {journal} {Phys. Rev. ST Accel. Beams}\ }\textbf
  {\bibinfo {volume} {12}},\ \bibinfo {pages} {051303}}\BibitemShut {NoStop}%
\bibitem [{\citenamefont {Weingartner}\ \emph {et~al.}(2012)\citenamefont
  {Weingartner} \emph {et~al.}}]{Weingartner2012}%
  \BibitemOpen
  \bibfield  {author} {\bibinfo {author} {\bibnamefont {Weingartner},
  \bibfnamefont {R}},  \emph {et~al.}} (\bibinfo {year} {2012}),\ \href
  {https://doi.org/10.1103/physrevstab.15.111302} {\bibfield  {journal}
  {\bibinfo  {journal} {Phys. Rev. ST Accel. Beams}\ }\textbf {\bibinfo
  {volume} {15}},\ \bibinfo {pages} {111302}}\BibitemShut {NoStop}%
\bibitem [{\citenamefont {Wen}\ \emph {et~al.}(2019)\citenamefont {Wen},
  \citenamefont {Tamburini},\ and\ \citenamefont {Keitel}}]{Wen2019}%
  \BibitemOpen
  \bibfield  {author} {\bibinfo {author} {\bibnamefont {Wen}, \bibfnamefont
  {M}}, \bibinfo {author} {\bibfnamefont {M.}~\bibnamefont {Tamburini}}, \ and\
  \bibinfo {author} {\bibfnamefont {C.~H.}\ \bibnamefont {Keitel}}} (\bibinfo
  {year} {2019}),\ \href
  {https://link.aps.org/doi/10.1103/PhysRevLett.122.214801} {\bibfield
  {journal} {\bibinfo  {journal} {Phys. Rev. Lett.}\ }\textbf {\bibinfo
  {volume} {122}},\ \bibinfo {pages} {214801}}\BibitemShut {NoStop}%
\bibitem [{\citenamefont {White}\ \emph {et~al.}(2014)\citenamefont {White}
  \emph {et~al.}}]{White2014}%
  \BibitemOpen
  \bibfield  {author} {\bibinfo {author} {\bibnamefont {White}, \bibfnamefont
  {G~R}},  \emph {et~al.} (\bibinfo {collaboration} {ATF2 Collaboration})}
  (\bibinfo {year} {2014}),\ \href
  {https://doi.org/10.1103/physrevlett.112.034802} {\bibfield  {journal}
  {\bibinfo  {journal} {Phys. Rev. Lett.}\ }\textbf {\bibinfo {volume} {112}},\
  \bibinfo {pages} {034802}}\BibitemShut {NoStop}%
\bibitem [{\citenamefont {Whittum}\ \emph {et~al.}(1990)\citenamefont
  {Whittum}, \citenamefont {Sessler},\ and\ \citenamefont
  {Dawson}}]{Whittum1990}%
  \BibitemOpen
  \bibfield  {author} {\bibinfo {author} {\bibnamefont {Whittum}, \bibfnamefont
  {D~H}}, \bibinfo {author} {\bibfnamefont {A.~M.}\ \bibnamefont {Sessler}}, \
  and\ \bibinfo {author} {\bibfnamefont {J.~M.}\ \bibnamefont {Dawson}}}
  (\bibinfo {year} {1990}),\ \href
  {https://doi.org/10.1103/physrevlett.64.2511} {\bibfield  {journal} {\bibinfo
   {journal} {Phys. Rev. Lett.}\ }\textbf {\bibinfo {volume} {64}},\ \bibinfo
  {pages} {2511--2514}}\BibitemShut {NoStop}%
\bibitem [{\citenamefont {Whittum}\ \emph {et~al.}(1991)\citenamefont
  {Whittum}, \citenamefont {Sharp}, \citenamefont {Yu}, \citenamefont {Lampe},\
  and\ \citenamefont {Joyce}}]{Whittum1991}%
  \BibitemOpen
  \bibfield  {author} {\bibinfo {author} {\bibnamefont {Whittum}, \bibfnamefont
  {D~H}}, \bibinfo {author} {\bibfnamefont {W.~M.}\ \bibnamefont {Sharp}},
  \bibinfo {author} {\bibfnamefont {S.~S.}\ \bibnamefont {Yu}}, \bibinfo
  {author} {\bibfnamefont {M.}~\bibnamefont {Lampe}}, \ and\ \bibinfo {author}
  {\bibfnamefont {G.}~\bibnamefont {Joyce}}} (\bibinfo {year} {1991}),\ \href
  {https://doi.org/10.1103/physrevlett.67.991} {\bibfield  {journal} {\bibinfo
  {journal} {Phys. Rev. Lett.}\ }\textbf {\bibinfo {volume} {67}},\ \bibinfo
  {pages} {991--994}}\BibitemShut {NoStop}%
\bibitem [{\citenamefont {Wing}(2019)}]{Wing2019}%
  \BibitemOpen
  \bibfield  {author} {\bibinfo {author} {\bibnamefont {Wing}, \bibfnamefont
  {M}}} (\bibinfo {year} {2019}),\ \href
  {https://doi.org/10.1098/rsta.2018.0185} {\bibfield  {journal} {\bibinfo
  {journal} {Philos. Trans. R. Soc. A}\ }\textbf {\bibinfo {volume} {377}},\
  \bibinfo {pages} {20180185}}\BibitemShut {NoStop}%
\bibitem [{\citenamefont {Winkler}\ \emph {et~al.}(2025)\citenamefont {Winkler}
  \emph {et~al.}}]{Winkler2025}%
  \BibitemOpen
  \bibfield  {author} {\bibinfo {author} {\bibnamefont {Winkler}, \bibfnamefont
  {P}},  \emph {et~al.}} (\bibinfo {year} {2025}),\ \href {\doibase
  10.1038/s41586-025-08772-y} {\bibfield  {journal} {\bibinfo  {journal}
  {Nature (London)}\ }\textbf {\bibinfo {volume} {640}},\ \bibinfo {pages}
  {907--910}}\BibitemShut {NoStop}%
\bibitem [{\citenamefont {Wittig}\ \emph {et~al.}(2015)\citenamefont {Wittig}
  \emph {et~al.}}]{Wittig2015}%
  \BibitemOpen
  \bibfield  {author} {\bibinfo {author} {\bibnamefont {Wittig}, \bibfnamefont
  {G}},  \emph {et~al.}} (\bibinfo {year} {2015}),\ \href
  {https://doi.org/10.1103/physrevstab.18.081304} {\bibfield  {journal}
  {\bibinfo  {journal} {Phys. Rev. ST Accel. Beams}\ }\textbf {\bibinfo
  {volume} {18}},\ \bibinfo {pages} {081304}}\BibitemShut {NoStop}%
\bibitem [{\citenamefont {Wood}\ \emph {et~al.}(2024)\citenamefont {Wood} \emph
  {et~al.}}]{Wood2024}%
  \BibitemOpen
  \bibfield  {author} {\bibinfo {author} {\bibnamefont {Wood}, \bibfnamefont
  {J}},  \emph {et~al.}} (\bibinfo {year} {2024}),\ in\ \href {\doibase
  10.18429/JACOW-IPAC2024-MOPR40} {\emph {\bibinfo {booktitle} {Proceedings of
  the 15th Int. Particle Accelerator Conf.}}}\ (\bibinfo  {publisher} {JACoW},\
  \bibinfo {address} {Geneva, Switzerland})\ pp.\ \bibinfo {pages}
  {538--541}\BibitemShut {NoStop}%
\bibitem [{\citenamefont {Wood}\ \emph {et~al.}(2026)\citenamefont {Wood} \emph
  {et~al.}}]{Wood2026}%
  \BibitemOpen
  \bibfield  {author} {\bibinfo {author} {\bibnamefont {Wood}, \bibfnamefont
  {J~C}},  \emph {et~al.}} (\bibinfo {year} {2026}),\ \href {\doibase
  10.1038/s41467-026-69283-6} {\bibfield  {journal} {\bibinfo  {journal} {Nat.
  Commun.}\ }\textbf {\bibinfo {volume} {17}},\ \bibinfo {pages}
  {1588}}\BibitemShut {NoStop}%
\bibitem [{\citenamefont {Wu}\ \emph {et~al.}(2010)\citenamefont {Wu},
  \citenamefont {Tajima}, \citenamefont {Habs}, \citenamefont {Chao},\ and\
  \citenamefont {Meyer-ter Vehn}}]{Wu2010}%
  \BibitemOpen
  \bibfield  {author} {\bibinfo {author} {\bibnamefont {Wu}, \bibfnamefont
  {H-C}}, \bibinfo {author} {\bibfnamefont {T.}~\bibnamefont {Tajima}},
  \bibinfo {author} {\bibfnamefont {D.}~\bibnamefont {Habs}}, \bibinfo {author}
  {\bibfnamefont {A.~W.}\ \bibnamefont {Chao}}, \ and\ \bibinfo {author}
  {\bibfnamefont {J.}~\bibnamefont {Meyer-ter Vehn}}} (\bibinfo {year}
  {2010}),\ \href {https://doi.org/10.1103/PhysRevSTAB.13.101303} {\bibfield
  {journal} {\bibinfo  {journal} {Phys. Rev. ST Accel. Beams}\ }\textbf
  {\bibinfo {volume} {13}},\ \bibinfo {pages} {101303}}\BibitemShut {NoStop}%
\bibitem [{\citenamefont {Wu}\ \emph {et~al.}(2020)\citenamefont {Wu},
  \citenamefont {Ji}, \citenamefont {Geng}, \citenamefont {Thomas},
  \citenamefont {B\"{u}scher}, \citenamefont {Pukhov}, \citenamefont
  {H\"{u}tzen}, \citenamefont {Zhang}, \citenamefont {Shen},\ and\
  \citenamefont {Li}}]{Wu2020}%
  \BibitemOpen
  \bibfield  {author} {\bibinfo {author} {\bibnamefont {Wu}, \bibfnamefont
  {Y}}, \bibinfo {author} {\bibfnamefont {L.}~\bibnamefont {Ji}}, \bibinfo
  {author} {\bibfnamefont {X.}~\bibnamefont {Geng}}, \bibinfo {author}
  {\bibfnamefont {J.}~\bibnamefont {Thomas}}, \bibinfo {author} {\bibfnamefont
  {M.}~\bibnamefont {B\"{u}scher}}, \bibinfo {author} {\bibfnamefont
  {A.}~\bibnamefont {Pukhov}}, \bibinfo {author} {\bibfnamefont
  {A.}~\bibnamefont {H\"{u}tzen}}, \bibinfo {author} {\bibfnamefont
  {L.}~\bibnamefont {Zhang}}, \bibinfo {author} {\bibfnamefont
  {B.}~\bibnamefont {Shen}}, \ and\ \bibinfo {author} {\bibfnamefont
  {R.}~\bibnamefont {Li}}} (\bibinfo {year} {2020}),\ \href
  {https://doi.org/10.1103/physrevapplied.13.044064} {\bibfield  {journal}
  {\bibinfo  {journal} {Phys. Rev. Appl.}\ }\textbf {\bibinfo {volume} {13}},\
  \bibinfo {pages} {044064}}\BibitemShut {NoStop}%
\bibitem [{\citenamefont {Wu}\ \emph {et~al.}(2019{\natexlab{a}})\citenamefont
  {Wu} \emph {et~al.}}]{Wu2019a}%
  \BibitemOpen
  \bibfield  {author} {\bibinfo {author} {\bibnamefont {Wu}, \bibfnamefont
  {Y}},  \emph {et~al.}} (\bibinfo {year} {2019}{\natexlab{a}}),\ \href
  {https://doi.org/10.1088/1367-2630/ab2fd7} {\bibfield  {journal} {\bibinfo
  {journal} {New J. Phys.}\ }\textbf {\bibinfo {volume} {21}},\ \bibinfo
  {pages} {073052}}\BibitemShut {NoStop}%
\bibitem [{\citenamefont {Wu}\ \emph {et~al.}(2019{\natexlab{b}})\citenamefont
  {Wu} \emph {et~al.}}]{Wu2019b}%
  \BibitemOpen
  \bibfield  {author} {\bibinfo {author} {\bibnamefont {Wu}, \bibfnamefont
  {Y}},  \emph {et~al.}} (\bibinfo {year} {2019}{\natexlab{b}}),\ \href
  {https://doi.org/10.1103/physreve.100.043202} {\bibfield  {journal} {\bibinfo
   {journal} {Phys. Rev. E}\ }\textbf {\bibinfo {volume} {100}},\ \bibinfo
  {pages} {043202}}\BibitemShut {NoStop}%
\bibitem [{\citenamefont {Wu}\ \emph {et~al.}(2019{\natexlab{c}})\citenamefont
  {Wu} \emph {et~al.}}]{Wu2019c}%
  \BibitemOpen
  \bibfield  {author} {\bibinfo {author} {\bibnamefont {Wu}, \bibfnamefont
  {Y~P}},  \emph {et~al.}} (\bibinfo {year} {2019}{\natexlab{c}}),\ \href
  {https://link.aps.org/doi/10.1103/PhysRevLett.122.204804} {\bibfield
  {journal} {\bibinfo  {journal} {Phys. Rev. Lett.}\ }\textbf {\bibinfo
  {volume} {122}},\ \bibinfo {pages} {204804}}\BibitemShut {NoStop}%
\bibitem [{\citenamefont {Wu}\ \emph {et~al.}(2019{\natexlab{d}})\citenamefont
  {Wu} \emph {et~al.}}]{Wu2019d}%
  \BibitemOpen
  \bibfield  {author} {\bibinfo {author} {\bibnamefont {Wu}, \bibfnamefont
  {Y~P}},  \emph {et~al.}} (\bibinfo {year} {2019}{\natexlab{d}}),\ \href
  {https://doi.org/10.1103/PhysRevApplied.12.064011} {\bibfield  {journal}
  {\bibinfo  {journal} {Phys. Rev. Appl.}\ }\textbf {\bibinfo {volume} {12}},\
  \bibinfo {pages} {064011}}\BibitemShut {NoStop}%
\bibitem [{\citenamefont {Xu}\ \emph {et~al.}(2021)\citenamefont {Xu},
  \citenamefont {Cesar}, \citenamefont {Corde}, \citenamefont {Yakimenko},
  \citenamefont {Hogan}, \citenamefont {Joshi}, \citenamefont {Marinelli},\
  and\ \citenamefont {Mori}}]{Xu2020}%
  \BibitemOpen
  \bibfield  {author} {\bibinfo {author} {\bibnamefont {Xu}, \bibfnamefont
  {X}}, \bibinfo {author} {\bibfnamefont {D.~B.}\ \bibnamefont {Cesar}},
  \bibinfo {author} {\bibfnamefont {S.}~\bibnamefont {Corde}}, \bibinfo
  {author} {\bibfnamefont {V.}~\bibnamefont {Yakimenko}}, \bibinfo {author}
  {\bibfnamefont {M.~J.}\ \bibnamefont {Hogan}}, \bibinfo {author}
  {\bibfnamefont {C.}~\bibnamefont {Joshi}}, \bibinfo {author} {\bibfnamefont
  {A.}~\bibnamefont {Marinelli}}, \ and\ \bibinfo {author} {\bibfnamefont
  {W.~B.}\ \bibnamefont {Mori}}} (\bibinfo {year} {2021}),\ \href
  {https://link.aps.org/doi/10.1103/PhysRevLett.126.094801} {\bibfield
  {journal} {\bibinfo  {journal} {Phys. Rev. Lett.}\ }\textbf {\bibinfo
  {volume} {126}},\ \bibinfo {pages} {094801}}\BibitemShut {NoStop}%
\bibitem [{\citenamefont {Xu}\ \emph {et~al.}(2022)\citenamefont {Xu},
  \citenamefont {Li}, \citenamefont {Tsung}, \citenamefont {Miller},
  \citenamefont {Yakimenko}, \citenamefont {Hogan}, \citenamefont {Joshi},\
  and\ \citenamefont {Mori}}]{Xu2022}%
  \BibitemOpen
  \bibfield  {author} {\bibinfo {author} {\bibnamefont {Xu}, \bibfnamefont
  {X}}, \bibinfo {author} {\bibfnamefont {F.}~\bibnamefont {Li}}, \bibinfo
  {author} {\bibfnamefont {F.~S.}\ \bibnamefont {Tsung}}, \bibinfo {author}
  {\bibfnamefont {K.}~\bibnamefont {Miller}}, \bibinfo {author} {\bibfnamefont
  {V.}~\bibnamefont {Yakimenko}}, \bibinfo {author} {\bibfnamefont {M.~J.}\
  \bibnamefont {Hogan}}, \bibinfo {author} {\bibfnamefont {C.}~\bibnamefont
  {Joshi}}, \ and\ \bibinfo {author} {\bibfnamefont {W.~B.}\ \bibnamefont
  {Mori}}} (\bibinfo {year} {2022}),\ \href
  {https://doi.org/10.1038/s41467-022-30806-6} {\bibfield  {journal} {\bibinfo
  {journal} {Nat. Commun.}\ }\textbf {\bibinfo {volume} {13}},\ \bibinfo
  {pages} {3364}}\BibitemShut {NoStop}%
\bibitem [{\citenamefont {Xu}\ \emph {et~al.}(2024)\citenamefont {Xu},
  \citenamefont {Liu}, \citenamefont {Dalichaouch}, \citenamefont {Tsung},
  \citenamefont {Zhang}, \citenamefont {Huang}, \citenamefont {Hogan},
  \citenamefont {Yan}, \citenamefont {Joshi},\ and\ \citenamefont
  {Mori}}]{Xu2024}%
  \BibitemOpen
  \bibfield  {author} {\bibinfo {author} {\bibnamefont {Xu}, \bibfnamefont
  {X}}, \bibinfo {author} {\bibfnamefont {J.}~\bibnamefont {Liu}}, \bibinfo
  {author} {\bibfnamefont {T.}~\bibnamefont {Dalichaouch}}, \bibinfo {author}
  {\bibfnamefont {F.~S.}\ \bibnamefont {Tsung}}, \bibinfo {author}
  {\bibfnamefont {Z.}~\bibnamefont {Zhang}}, \bibinfo {author} {\bibfnamefont
  {Z.}~\bibnamefont {Huang}}, \bibinfo {author} {\bibfnamefont {M.~J.}\
  \bibnamefont {Hogan}}, \bibinfo {author} {\bibfnamefont {X.}~\bibnamefont
  {Yan}}, \bibinfo {author} {\bibfnamefont {C.}~\bibnamefont {Joshi}}, \ and\
  \bibinfo {author} {\bibfnamefont {W.~B.}\ \bibnamefont {Mori}}} (\bibinfo
  {year} {2024}),\ \href {https://doi.org/10.1103/PhysRevAccelBeams.27.011301}
  {\bibfield  {journal} {\bibinfo  {journal} {Phys. Rev. Accel. Beams}\
  }\textbf {\bibinfo {volume} {27}},\ \bibinfo {pages} {011301}}\BibitemShut
  {NoStop}%
\bibitem [{\citenamefont {Xu}\ \emph {et~al.}(2017)\citenamefont {Xu},
  \citenamefont {Li}, \citenamefont {An}, \citenamefont {Dalichaouch},
  \citenamefont {Yu}, \citenamefont {Lu}, \citenamefont {Joshi},\ and\
  \citenamefont {Mori}}]{Xu2017}%
  \BibitemOpen
  \bibfield  {author} {\bibinfo {author} {\bibnamefont {Xu}, \bibfnamefont
  {X~L}}, \bibinfo {author} {\bibfnamefont {F.}~\bibnamefont {Li}}, \bibinfo
  {author} {\bibfnamefont {W.}~\bibnamefont {An}}, \bibinfo {author}
  {\bibfnamefont {T.~N.}\ \bibnamefont {Dalichaouch}}, \bibinfo {author}
  {\bibfnamefont {P.}~\bibnamefont {Yu}}, \bibinfo {author} {\bibfnamefont
  {W.}~\bibnamefont {Lu}}, \bibinfo {author} {\bibfnamefont {C.}~\bibnamefont
  {Joshi}}, \ and\ \bibinfo {author} {\bibfnamefont {W.~B.}\ \bibnamefont
  {Mori}}} (\bibinfo {year} {2017}),\ \href
  {https://doi.org/10.1103/physrevaccelbeams.20.111303} {\bibfield  {journal}
  {\bibinfo  {journal} {Phys. Rev. Accel. Beams}\ }\textbf {\bibinfo {volume}
  {20}},\ \bibinfo {pages} {111303}}\BibitemShut {NoStop}%
\bibitem [{\citenamefont {Xu}\ \emph {et~al.}(2016)\citenamefont {Xu} \emph
  {et~al.}}]{Xu2016}%
  \BibitemOpen
  \bibfield  {author} {\bibinfo {author} {\bibnamefont {Xu}, \bibfnamefont
  {X~L}},  \emph {et~al.}} (\bibinfo {year} {2016}),\ \href
  {https://doi.org/10.1103/physrevlett.116.124801} {\bibfield  {journal}
  {\bibinfo  {journal} {Phys. Rev. Lett.}\ }\textbf {\bibinfo {volume} {116}},\
  \bibinfo {pages} {124801}}\BibitemShut {NoStop}%
\bibitem [{\citenamefont {Yakimenko}\ \emph {et~al.}(2006)\citenamefont
  {Yakimenko}, \citenamefont {Babzien}, \citenamefont {Kallos}, \citenamefont
  {Kimura}, \citenamefont {Kusche},\ and\ \citenamefont
  {Muggli}}]{Yakimenko2006}%
  \BibitemOpen
  \bibfield  {author} {\bibinfo {author} {\bibnamefont {Yakimenko},
  \bibfnamefont {V}}, \bibinfo {author} {\bibfnamefont {M.}~\bibnamefont
  {Babzien}}, \bibinfo {author} {\bibfnamefont {E.~K.}\ \bibnamefont {Kallos}},
  \bibinfo {author} {\bibfnamefont {W.~D.}\ \bibnamefont {Kimura}}, \bibinfo
  {author} {\bibfnamefont {K.}~\bibnamefont {Kusche}}, \ and\ \bibinfo {author}
  {\bibfnamefont {P.}~\bibnamefont {Muggli}}} (\bibinfo {year} {2006}),\ in\
  \href {https://jacow.org/f06/papers/TUPPH072.pdf} {\emph {\bibinfo
  {booktitle} {Proceedings of FEL 2006}}}\ (\bibinfo  {publisher} {JACoW},\
  \bibinfo {address} {Geneva, Switzerland})\ pp.\ \bibinfo {pages}
  {481--484}\BibitemShut {NoStop}%
\bibitem [{\citenamefont {Yakimenko}\ \emph {et~al.}(2003)\citenamefont
  {Yakimenko} \emph {et~al.}}]{Yakimenko2003}%
  \BibitemOpen
  \bibfield  {author} {\bibinfo {author} {\bibnamefont {Yakimenko},
  \bibfnamefont {V}},  \emph {et~al.}} (\bibinfo {year} {2003}),\ \href
  {https://doi.org/10.1103/physrevlett.91.014802} {\bibfield  {journal}
  {\bibinfo  {journal} {Phys. Rev. Lett.}\ }\textbf {\bibinfo {volume} {91}},\
  \bibinfo {pages} {014802}}\BibitemShut {NoStop}%
\bibitem [{\citenamefont {Yakimenko}\ \emph {et~al.}(2019)\citenamefont
  {Yakimenko} \emph {et~al.}}]{Yakimenko2019}%
  \BibitemOpen
  \bibfield  {author} {\bibinfo {author} {\bibnamefont {Yakimenko},
  \bibfnamefont {V}},  \emph {et~al.}} (\bibinfo {year} {2019}),\ \href
  {https://doi.org/10.1103/PhysRevAccelBeams.22.101301} {\bibfield  {journal}
  {\bibinfo  {journal} {Phys. Rev. Accel. Beams}\ }\textbf {\bibinfo {volume}
  {22}},\ \bibinfo {pages} {101301}}\BibitemShut {NoStop}%
\bibitem [{\citenamefont {Yi}\ \emph {et~al.}(2013)\citenamefont {Yi},
  \citenamefont {Khudik}, \citenamefont {Siemon},\ and\ \citenamefont
  {Shvets}}]{Yi2013}%
  \BibitemOpen
  \bibfield  {author} {\bibinfo {author} {\bibnamefont {Yi}, \bibfnamefont
  {S~A}}, \bibinfo {author} {\bibfnamefont {V.}~\bibnamefont {Khudik}},
  \bibinfo {author} {\bibfnamefont {C.}~\bibnamefont {Siemon}}, \ and\ \bibinfo
  {author} {\bibfnamefont {G.}~\bibnamefont {Shvets}}} (\bibinfo {year}
  {2013}),\ \href {https://doi.org/10.1063/1.4775774} {\bibfield  {journal}
  {\bibinfo  {journal} {Phys. Plasmas}\ }\textbf {\bibinfo {volume} {20}},\
  \bibinfo {pages} {013108}}\BibitemShut {NoStop}%
\bibitem [{\citenamefont {Yu}\ \emph {et~al.}(2014)\citenamefont {Yu},
  \citenamefont {Schroeder}, \citenamefont {Li}, \citenamefont {Benedetti},
  \citenamefont {Chen}, \citenamefont {Weng}, \citenamefont {Sheng},\ and\
  \citenamefont {Esarey}}]{Yu2014}%
  \BibitemOpen
  \bibfield  {author} {\bibinfo {author} {\bibnamefont {Yu}, \bibfnamefont
  {L-L}}, \bibinfo {author} {\bibfnamefont {C.~B.}\ \bibnamefont {Schroeder}},
  \bibinfo {author} {\bibfnamefont {F.-Y.}\ \bibnamefont {Li}}, \bibinfo
  {author} {\bibfnamefont {C.}~\bibnamefont {Benedetti}}, \bibinfo {author}
  {\bibfnamefont {M.}~\bibnamefont {Chen}}, \bibinfo {author} {\bibfnamefont
  {S.-M.}\ \bibnamefont {Weng}}, \bibinfo {author} {\bibfnamefont {Z.-M.}\
  \bibnamefont {Sheng}}, \ and\ \bibinfo {author} {\bibfnamefont
  {E.}~\bibnamefont {Esarey}}} (\bibinfo {year} {2014}),\ \href
  {https://doi.org/10.1063/1.4903536} {\bibfield  {journal} {\bibinfo
  {journal} {Phys. Plasmas}\ }\textbf {\bibinfo {volume} {21}},\ \bibinfo
  {pages} {120702}}\BibitemShut {NoStop}%
\bibitem [{\citenamefont {Zeng}\ \emph {et~al.}(2020)\citenamefont {Zeng},
  \citenamefont {{Martinez de la Ossa}},\ and\ \citenamefont
  {Osterhoff}}]{Zeng2020}%
  \BibitemOpen
  \bibfield  {author} {\bibinfo {author} {\bibnamefont {Zeng}, \bibfnamefont
  {M}}, \bibinfo {author} {\bibfnamefont {A.}~\bibnamefont {{Martinez de la
  Ossa}}}, \ and\ \bibinfo {author} {\bibfnamefont {J.}~\bibnamefont
  {Osterhoff}}} (\bibinfo {year} {2020}),\ \href
  {https://doi.org/10.1088/1367-2630/abc9ee} {\bibfield  {journal} {\bibinfo
  {journal} {New J. Phys.}\ }\textbf {\bibinfo {volume} {22}},\ \bibinfo
  {pages} {123003}}\BibitemShut {NoStop}%
\bibitem [{\citenamefont {Zeng}\ and\ \citenamefont {Seto}(2021)}]{Zeng2021}%
  \BibitemOpen
  \bibfield  {author} {\bibinfo {author} {\bibnamefont {Zeng}, \bibfnamefont
  {M}}, \ and\ \bibinfo {author} {\bibfnamefont {K.}~\bibnamefont {Seto}}}
  (\bibinfo {year} {2021}),\ \href {https://doi.org/10.1088/1367-2630/ac12fa}
  {\bibfield  {journal} {\bibinfo  {journal} {New J. Phys.}\ }\textbf {\bibinfo
  {volume} {23}},\ \bibinfo {pages} {075008}}\BibitemShut {NoStop}%
\bibitem [{\citenamefont {Zgadzaj}\ \emph {et~al.}(2020)\citenamefont {Zgadzaj}
  \emph {et~al.}}]{Zgadzaj2020}%
  \BibitemOpen
  \bibfield  {author} {\bibinfo {author} {\bibnamefont {Zgadzaj}, \bibfnamefont
  {R}},  \emph {et~al.}} (\bibinfo {year} {2020}),\ \href
  {https://doi.org/10.1038/s41467-020-18490-w} {\bibfield  {journal} {\bibinfo
  {journal} {Nat. Commun.}\ }\textbf {\bibinfo {volume} {11}},\ \bibinfo
  {pages} {4753}}\BibitemShut {NoStop}%
\bibitem [{\citenamefont {Zha}\ \emph {et~al.}(2016)\citenamefont {Zha},
  \citenamefont {Latina}, \citenamefont {Grudiev}, \citenamefont {{De
  Michele}}, \citenamefont {Solodko}, \citenamefont {Wuensch}, \citenamefont
  {Schulte}, \citenamefont {Adli}, \citenamefont {Lipkowitz},\ and\
  \citenamefont {Yocky}}]{Zha2016}%
  \BibitemOpen
  \bibfield  {author} {\bibinfo {author} {\bibnamefont {Zha}, \bibfnamefont
  {H}}, \bibinfo {author} {\bibfnamefont {A.}~\bibnamefont {Latina}}, \bibinfo
  {author} {\bibfnamefont {A.}~\bibnamefont {Grudiev}}, \bibinfo {author}
  {\bibfnamefont {G.}~\bibnamefont {{De Michele}}}, \bibinfo {author}
  {\bibfnamefont {A.}~\bibnamefont {Solodko}}, \bibinfo {author} {\bibfnamefont
  {W.}~\bibnamefont {Wuensch}}, \bibinfo {author} {\bibfnamefont
  {D.}~\bibnamefont {Schulte}}, \bibinfo {author} {\bibfnamefont
  {E.}~\bibnamefont {Adli}}, \bibinfo {author} {\bibfnamefont {N.}~\bibnamefont
  {Lipkowitz}}, \ and\ \bibinfo {author} {\bibfnamefont {G.~S.}\ \bibnamefont
  {Yocky}}} (\bibinfo {year} {2016}),\ \href
  {https://doi.org/10.1103/physrevaccelbeams.19.011001} {\bibfield  {journal}
  {\bibinfo  {journal} {Phys. Rev. Accel. Beams}\ }\textbf {\bibinfo {volume}
  {19}},\ \bibinfo {pages} {011001}}\BibitemShut {NoStop}%
\bibitem [{\citenamefont {Zhang}\ \emph {et~al.}(2019)\citenamefont {Zhang},
  \citenamefont {Huang}, \citenamefont {Marsh}, \citenamefont {Xu},
  \citenamefont {Li}, \citenamefont {Hogan}, \citenamefont {Yakimenko},
  \citenamefont {Corde}, \citenamefont {Mori},\ and\ \citenamefont
  {Joshi}}]{Zhang2019}%
  \BibitemOpen
  \bibfield  {author} {\bibinfo {author} {\bibnamefont {Zhang}, \bibfnamefont
  {C}}, \bibinfo {author} {\bibfnamefont {C.-K.}\ \bibnamefont {Huang}},
  \bibinfo {author} {\bibfnamefont {K.~A.}\ \bibnamefont {Marsh}}, \bibinfo
  {author} {\bibfnamefont {X.~L.}\ \bibnamefont {Xu}}, \bibinfo {author}
  {\bibfnamefont {F.}~\bibnamefont {Li}}, \bibinfo {author} {\bibfnamefont
  {M.}~\bibnamefont {Hogan}}, \bibinfo {author} {\bibfnamefont
  {V.}~\bibnamefont {Yakimenko}}, \bibinfo {author} {\bibfnamefont
  {S.}~\bibnamefont {Corde}}, \bibinfo {author} {\bibfnamefont {W.~B.}\
  \bibnamefont {Mori}}, \ and\ \bibinfo {author} {\bibfnamefont
  {C.}~\bibnamefont {Joshi}}} (\bibinfo {year} {2019}),\ \href
  {https://doi.org/10.1103/physrevaccelbeams.22.111301} {\bibfield  {journal}
  {\bibinfo  {journal} {Phys. Rev. Accel. Beams}\ }\textbf {\bibinfo {volume}
  {22}},\ \bibinfo {pages} {111301}}\BibitemShut {NoStop}%
\bibitem [{\citenamefont {Zhang}\ \emph {et~al.}(2024)\citenamefont {Zhang}
  \emph {et~al.}}]{Zhang2024}%
  \BibitemOpen
  \bibfield  {author} {\bibinfo {author} {\bibnamefont {Zhang}, \bibfnamefont
  {C}},  \emph {et~al.}} (\bibinfo {year} {2024}),\ \href
  {https://dx.doi.org/10.1088/1361-6587/ad1ae4} {\bibfield  {journal} {\bibinfo
   {journal} {Plasma Phys. Control. Fusion}\ }\textbf {\bibinfo {volume}
  {66}},\ \bibinfo {pages} {025013}}\BibitemShut {NoStop}%
\bibitem [{\citenamefont {Zhang}\ \emph {et~al.}(2025)\citenamefont {Zhang}
  \emph {et~al.}}]{Zhang2025}%
  \BibitemOpen
  \bibfield  {author} {\bibinfo {author} {\bibnamefont {Zhang}, \bibfnamefont
  {C}},  \emph {et~al.}} (\bibinfo {year} {2025}),\ \href
  {https://doi.org/10.1038/s41467-025-65742-8} {\bibfield  {journal} {\bibinfo
  {journal} {Nat. Commun.}\ }\textbf {\bibinfo {volume} {16}},\ \bibinfo
  {pages} {10719}}\BibitemShut {NoStop}%
\bibitem [{\citenamefont {Zhang}\ \emph {et~al.}(2026)\citenamefont {Zhang}
  \emph {et~al.}}]{Zhang2026}%
  \BibitemOpen
  \bibfield  {author} {\bibinfo {author} {\bibnamefont {Zhang}, \bibfnamefont
  {C}},  \emph {et~al.}} (\bibinfo {year} {2026}),\ \href {\doibase
  10.48550/arxiv.2607.08069} {}\Eprint {http://arxiv.org/abs/2607.08069}
  {arXiv:2607.08069} \BibitemShut {NoStop}%
\bibitem [{\citenamefont {Zhao}\ \emph {et~al.}(2020)\citenamefont {Zhao},
  \citenamefont {Lehe}, \citenamefont {Myers}, \citenamefont {Th{\'{e}}venet},
  \citenamefont {Huebl}, \citenamefont {Schroeder},\ and\ \citenamefont
  {Vay}}]{Zhao2020}%
  \BibitemOpen
  \bibfield  {author} {\bibinfo {author} {\bibnamefont {Zhao}, \bibfnamefont
  {Y}}, \bibinfo {author} {\bibfnamefont {R.}~\bibnamefont {Lehe}}, \bibinfo
  {author} {\bibfnamefont {A.}~\bibnamefont {Myers}}, \bibinfo {author}
  {\bibfnamefont {M.}~\bibnamefont {Th{\'{e}}venet}}, \bibinfo {author}
  {\bibfnamefont {A.}~\bibnamefont {Huebl}}, \bibinfo {author} {\bibfnamefont
  {C.~B.}\ \bibnamefont {Schroeder}}, \ and\ \bibinfo {author} {\bibfnamefont
  {J.-L.}\ \bibnamefont {Vay}}} (\bibinfo {year} {2020}),\ \href
  {https://doi.org/10.1063/5.0023776} {\bibfield  {journal} {\bibinfo
  {journal} {Phys. Plasmas}\ }\textbf {\bibinfo {volume} {27}},\ \bibinfo
  {pages} {113105}}\BibitemShut {NoStop}%
\bibitem [{\citenamefont {Zhao}\ \emph {et~al.}(2024)\citenamefont {Zhao} \emph
  {et~al.}}]{Zhao2024}%
  \BibitemOpen
  \bibfield  {author} {\bibinfo {author} {\bibnamefont {Zhao}, \bibfnamefont
  {Y}},  \emph {et~al.}} (\bibinfo {year} {2024}),\ \href
  {https://doi.org/10.1063/5.0206378} {\bibfield  {journal} {\bibinfo
  {journal} {Phys. Plasmas}\ }\textbf {\bibinfo {volume} {31}},\ \bibinfo
  {pages} {063106}}\BibitemShut {NoStop}%
\bibitem [{\citenamefont {Zhou}\ \emph {et~al.}(2007)\citenamefont {Zhou} \emph
  {et~al.}}]{Zhou2007}%
  \BibitemOpen
  \bibfield  {author} {\bibinfo {author} {\bibnamefont {Zhou}, \bibfnamefont
  {M}},  \emph {et~al.}} (\bibinfo {year} {2007}),\ in\ \href
  {https://doi.org/10.1109/PAC.2007.4440669} {\emph {\bibinfo {booktitle}
  {Proceedings of the 2007 Particle Accelerator Conf.}}}\ (\bibinfo
  {publisher} {IEEE},\ \bibinfo {address} {New York, NY, United States})\ pp.\
  \bibinfo {pages} {3064--3066}\BibitemShut {NoStop}%
\bibitem [{\citenamefont {Zhou}\ \emph {et~al.}(2025)\citenamefont {Zhou},
  \citenamefont {Ding}, \citenamefont {An}, \citenamefont {Su}, \citenamefont
  {Hua}, \citenamefont {Li}, \citenamefont {Mori}, \citenamefont {Joshi},\ and\
  \citenamefont {Lu}}]{Zhou2025}%
  \BibitemOpen
  \bibfield  {author} {\bibinfo {author} {\bibnamefont {Zhou}, \bibfnamefont
  {S}}, \bibinfo {author} {\bibfnamefont {S.}~\bibnamefont {Ding}}, \bibinfo
  {author} {\bibfnamefont {W.}~\bibnamefont {An}}, \bibinfo {author}
  {\bibfnamefont {Q.}~\bibnamefont {Su}}, \bibinfo {author} {\bibfnamefont
  {J.}~\bibnamefont {Hua}}, \bibinfo {author} {\bibfnamefont {F.}~\bibnamefont
  {Li}}, \bibinfo {author} {\bibfnamefont {W.~B.}\ \bibnamefont {Mori}},
  \bibinfo {author} {\bibfnamefont {C.}~\bibnamefont {Joshi}}, \ and\ \bibinfo
  {author} {\bibfnamefont {W.}~\bibnamefont {Lu}}} (\bibinfo {year} {2025}),\
  \href {https://doi.org/10.34133/research.0878} {\bibfield  {journal}
  {\bibinfo  {journal} {Research}\ }\textbf {\bibinfo {volume} {8}},\ \bibinfo
  {pages} {0878}}\BibitemShut {NoStop}%
\bibitem [{\citenamefont {Zhou}\ \emph {et~al.}(2021)\citenamefont {Zhou},
  \citenamefont {Hua}, \citenamefont {An}, \citenamefont {Mori}, \citenamefont
  {Joshi}, \citenamefont {Gao},\ and\ \citenamefont {Lu}}]{Zhou2021}%
  \BibitemOpen
  \bibfield  {author} {\bibinfo {author} {\bibnamefont {Zhou}, \bibfnamefont
  {S}}, \bibinfo {author} {\bibfnamefont {J.}~\bibnamefont {Hua}}, \bibinfo
  {author} {\bibfnamefont {W.}~\bibnamefont {An}}, \bibinfo {author}
  {\bibfnamefont {W.~B.}\ \bibnamefont {Mori}}, \bibinfo {author}
  {\bibfnamefont {C.}~\bibnamefont {Joshi}}, \bibinfo {author} {\bibfnamefont
  {J.}~\bibnamefont {Gao}}, \ and\ \bibinfo {author} {\bibfnamefont
  {W.}~\bibnamefont {Lu}}} (\bibinfo {year} {2021}),\ \href
  {https://doi.org/10.1103/physrevlett.127.174801} {\bibfield  {journal}
  {\bibinfo  {journal} {Phys. Rev. Lett.}\ }\textbf {\bibinfo {volume} {127}},\
  \bibinfo {pages} {174801}}\BibitemShut {NoStop}%
\bibitem [{\citenamefont {Zhou}\ \emph {et~al.}(2022)\citenamefont {Zhou},
  \citenamefont {Hua}, \citenamefont {Lu}, \citenamefont {An}, \citenamefont
  {Su}, \citenamefont {Mori},\ and\ \citenamefont {Joshi}}]{Zhou2022a}%
  \BibitemOpen
  \bibfield  {author} {\bibinfo {author} {\bibnamefont {Zhou}, \bibfnamefont
  {S}}, \bibinfo {author} {\bibfnamefont {J.}~\bibnamefont {Hua}}, \bibinfo
  {author} {\bibfnamefont {W.}~\bibnamefont {Lu}}, \bibinfo {author}
  {\bibfnamefont {W.}~\bibnamefont {An}}, \bibinfo {author} {\bibfnamefont
  {Q.}~\bibnamefont {Su}}, \bibinfo {author} {\bibfnamefont {W.~B.}\
  \bibnamefont {Mori}}, \ and\ \bibinfo {author} {\bibfnamefont
  {C.}~\bibnamefont {Joshi}}} (\bibinfo {year} {2022}),\ \href
  {https://doi.org/10.1103/physrevaccelbeams.25.091303} {\bibfield  {journal}
  {\bibinfo  {journal} {Phys. Rev. Accel. Beams}\ }\textbf {\bibinfo {volume}
  {25}},\ \bibinfo {pages} {091303}}\BibitemShut {NoStop}%
\bibitem [{\citenamefont {Zyngier}(1977)}]{Zyngier1977}%
  \BibitemOpen
  \bibfield  {author} {\bibinfo {author} {\bibnamefont {Zyngier}, \bibfnamefont
  {H}}} (\bibinfo {year} {1977}),\ \href
  {https://inspirehep.net/literature/122210} {}\bibinfo {type} {Tech. Rep.}\
  \bibinfo {number} {LAL-77/35}\ (\bibinfo  {institution} {LAL},\ \bibinfo
  {address} {Orsay, France})\BibitemShut {NoStop}%
\end{thebibliography}
\end{document}